\documentclass[a4paper,10pt]{extarticle}
\pdfoutput=1 

\usepackage{jheppub} 
\usepackage{hyperref}
\hypersetup{
    bookmarks=true,         
    unicode=false,          
    pdftoolbar=true,        
   pdfmenubar=true,        
    pdffitwindow=false,     
    pdfstartview={FitH},    
    pdftitle={My title},    
    pdfauthor={Author},     
    pdfsubject={Subject},   
    pdfcreator={Creator},   
    pdfproducer={Producer}, 
    pdfkeywords={keyword1} {key2} {key3}, 
    pdfnewwindow=true,      
    colorlinks=true,       
    linkcolor=red,          
    citecolor=purple,        
    filecolor=magenta,      
    urlcolor=blue,           
    linktocpage=true
}
\usepackage{graphics}
\usepackage{rotating}
\usepackage{subfigure}
\usepackage{pstricks}
\usepackage{color}
\usepackage{amsfonts}
\usepackage{mathrsfs,amsmath}
\usepackage{epsfig}
\usepackage{amsmath,amssymb,amsthm,graphicx,latexsym}
\usepackage{ulem}
\renewcommand{\arraystretch}{2}

\newcommand{\be}{\begin{equation}}
\newcommand{\ee}{\end{equation}}
\newcommand{\bea}{\begin{eqnarray}}
\newcommand{\eea}{\end{eqnarray}}
\newcommand{\bml}{\begin{subequations}}
\newcommand{\eml}{\end{subequations}}
\newcommand{\bfig}{\begin{figure}}
\newcommand{\efig}{\end{figure}}

\newcommand{\slashed}{\hspace{-1.1ex}/}

\newcommand{\bmat}{\begin{pmatrix}}
\newcommand{\emat}{\end{pmatrix}}
\begin{document}
	$~~~~~~~~~~~~~~~~~~~~~~~~~~~~~~~~~~~~~~~~~~~~~~~~~~~~~~~~~~~~~~~~~~~~~~~~~~~~~~~~~~~~~~~~~~~~~~~~~~~~~~~~~~~~~~~~~~~~~~~$\textcolor{red}{\bf \Large TIFR/TH/16-19}
	\title{\textsc{\fontsize{47}{90}\selectfont \sffamily \bfseries
	\textcolor{purple}{Bell violation in the Sky}}}

	\author[a]{Sayantan Choudhury,
		\footnote{\textcolor{purple}{\bf Presently working as a Visiting (Post-Doctoral) fellow at DTP, TIFR, Mumbai, Alternative
	E-mail: sayanphysicsisi@gmail.com}. ${}^{}$}}
	\author[b,c]{Sudhakar Panda
}
	\author[d]{and Rajeev Singh}
	
	\affiliation[a]{Department of Theoretical Physics, Tata Institute of Fundamental Research, Colaba, Mumbai - 400005, India.
	}
	\affiliation[b]{Institute of Physics, Sachivalaya Marg, Bhubaneswar, Odisha - 751005, India.
	}
	\affiliation[c]{Homi Bhabha National Institute, Training School Complex,
	Anushakti Nagar, Mumbai-400085, India.
	}
	\affiliation[d]{Department of Physics, Savitribai Phule Pune University, Pune - 411007, India.
	}
	\emailAdd{sayantan@theory.tifr.res.in, panda@iopb.res.in, rajeevsingh240291@gmail.com }

	\abstract{In this work, we have studied the possibility of setting up Bell's inequality violating experiment
 in the context of cosmology, based on the basic principles of quantum mechanics. First we 
 start with the physical motivation of implementing the Bell's inequality violation in the context of cosmology. Then to set up 
 the cosmological Bell violating test experiment we introduce a model independent theoretical framework using which we have studied 
 the creation of new massive particles by implementing the WKB approximation method for the scalar 
 fluctuations in presence of additional time dependent mass contribution in the cosmological perturbation theory.
 Here for completeness we compute total number density and energy density of the newly created particles 
 in terms of Bogoliubov coefficients using WKB approximation method. 
 Next using the background scalar fluctuation in presence of new time dependent mass contribution, we explicitly 
 compute the expression for the one point and two point correlation functions. Furthermore, using the results for one point function 
 we introduce a new theoretical cosmological parameter which can be expressed in
 terms of the other known inflationary observables and can also be treated as a future theoretical probe   
 to break the degeneracy amongst various models of inflation. Additionally, we also fix the scale of 
 inflation in a model independent way without any prior knowledge of 
 primordial gravitational waves. Also using the input from newly introduced
 cosmological parameter, we finally give a theoretical estimate for the tensor-to-scalar ratio in
 a model independent way. Next, we also comment on the technicalities of measurements from isospin breaking interactions and the future prospects of newly 
 introduced massive particles in cosmological Bell violating test experiment. Further, we cite a precise example of this 
 set up applicable in the context of string theory motivated axion monodromy model. Then we comment on the explicit role of decoherence effect and high spin on cosmological Bell violating test experiment.
 In fine, we provide a theoretical bound on the heavy particle mass parameter for scalar fields, graviton and other high spin fields from our proposed setup.}


	\maketitle
	\flushbottom
	\textcolor{blue}{{\it ``That one body may act upon another at a distance through a vacuum without
the mediation of anything else....is to me so great an absurdity, that I
believe no man, who has in philosophical matters a competent faculty for thinking,
can ever fall into'' }--}\textcolor{violet}{\bf Sir Isaac Newton}
	\section{Introduction}
	
In the year  \textcolor{red}{\bf 1935}, \textcolor{violet}{\bf Einstein, Podolsky and Rosen (EPR)} in ref.~\cite{epr:1935ja} mentioned that,
\textcolor{blue}{{\it ``If,  without  in  any  way  disturbing  a  system,  we  can  predict  with  certainty  (i.e.,  with
probability equal to unity) the value of a physical quantity, then there exists an element of
physical reality corresponding to this physical quantity''}}. This work also claimed that
quantum mechanics cannot be a complete theoretical framework, therefore there has to be
some element exists using which it is not possible to describe within the basic principles of quantum mechanics. 
Furthermore the authors also added that, \textcolor{blue}{{\it ``While we have thus shown that the wave function does not provide a complete description
of the physical reality, we left open the question of whether or not such a description exists.
We believe, however, that such a theory is possible''}}. Based on all such statements one can ask a question regarding the existence of all such missing elements 
in quantum physics theory.

Later \textcolor{violet}{\bf J. S. Bell} introduced the existence of \textcolor{blue}{\it``hidden''} variables which directly implies
that in spin correlation measurements the measurable probabilities
must satisfy the proposed \textcolor{blue}{\it Bell's inequality} \cite{Bell:1964kc} within the framework of quantum mechanics.
For completeness here we also mention some of the remarkable works in the area of quantum mechanics proposed upto Bell test experiment:
\begin{itemize}
\item \textcolor{red}{\bf 1927} \textcolor{blue}{{\it Copenhagen interpretation of Quantum Mechanics (Bohr, Heisenberg)}},
\item  \textcolor{red}{\bf 1935} \textcolor{blue}{{\it Einstein-Podolsky-Rosen (EPR) paradox}},
\item  \textcolor{red}{\bf 1952} \textcolor{blue}{{\it De Broglie-Bohm nonlocal hidden variable theory (Bohmian Mechanics)}},
\item  \textcolor{red}{\bf 1964} \textcolor{blue}{{\it Bell’s Theorem on local hidden variables}},
\item  \textcolor{red}{\bf 1972} \textcolor{blue}{{\it First experimental Bell test (Freedman and Clauser)}}.
\end{itemize}
Later the actual version of the Bell's inequality have been proved incorrect
by many experiments performed till date, which in turn proves 
that the nature is non-local and hence all the particles can interact
with each other without bothering about the underlying interaction scale and the corresponding distance (length scale) between all of them.
This underlying principle of violation of Bell's inequality is thoroughly
used in our work to setup the cosmological experiment and to study some of the unexplored important 
features in the context of early universe.

 It is a very well known fact that our 
present understanding of the large scale structure formation of universe is that,
it is actually originated from
the small scale perturbations and once the universe became matter dominated then gravitational effects 
mimics its role in cosmological evolution, which we observe today through various cosmological observations.
For the formation of the structure due to gravitational instability of what we
observe today, there has to be pre-existing small fluctuations on physical
length scales. In the model of Big Bang it is almost impossible to
produce fluctuations in any arbitrary length scale, so in such a case we put these small
perturbations by hand. The proper physical explanation for
these small scale perturbations is that these perturbations
arises due to density fluctuations in the inflationary epoch \cite{juan:2015ja,Mukhanov:1981ja,Hawking:1981ja,Starobinsky:1982ja,bardeen:1983jj}, which have a quantum mechanical origin.

In the context of modern cosmology, as we know one of the main basic idea is that there occurred
an event namely epoch in the very early time of the universe where the universe is vacuum
dominated matter or radiation. Therefore during this era the scale
factor grew almost exponentially in time. We can also understand why the observable
universe is homogeneous and isotropic if this quasi exponential expansion occurred
in the very early age of universe. This epoch is commonly known as inflation. 
This theory was first introduced by \textcolor{violet}{\bf A. Guth} in ref.~\cite{Alan:1981ja}.
Primordial density perturbation is actually the vacuum fluctuation
which survived after the period of inflation which may be the most possible reason for
the large scale structure formation of our universe and CMB anisotropy. In the present context we are
primarily interested in the specific type of inflation theory which removes the shortcomings of
standard Big Bang theory, also helps us to the get the mostly favored possible explanation
of the homogeneity and isotropic of CMB and to construct a Bell's inequality violating cosmological setup. Therefore the inflationary paradigm predicts that the
origin of large scale structure,
which we actually observe is nothing but the outcome of quantum mechanical
fluctuations after the inflationary period. Such quantum fluctuations make the inflationary paradigm consistent with various cosmological observations 
compared to the other classical statistical fluctuations appearing in the present context by following the same epoch \cite{Mukhanov:1981ja,Hawking:1981ja,Starobinsky:1982ja,bardeen:1983jj,Alan:1981ja}.
Here it is important to note that, in case of classical statistical approach frictional force acts as a external source using which inflaton energy is converted to the
other forms of energy and finally produce fluctuations. Now further using this information one can compute,
also compare and constrain two and three point correlation functions from 
quantum fluctuations and classical statistical fluctuations and check the consistency relations from any higher point correlation functions.
Here additionally it is important to note that, during the quantum mechanical interpretation of the required fluctuations, highly entangled quantum mechanical wave function of the universe plays a significant role.
Due to this fact, quantum fluctuations can be theoretically demonstrated as well as implemented in the
context of primordial cosmology, iff we can
perform a Bell's inequality violating cosmological experiment using the highly quantum mechanical
entangled wave function
of the universe defined in the inflationary period. Throughout this paper we will develop a theoretical
setup to address various fundamental questions related to general aspect of Bell's inequality
violation and 
also study the various unexplored physical consequences
from cosmological Bell's inequality violating experiment. Now to describe the theoretical framework and background methodology in detail it is important to mention that, in the context of quantum mechanics, Bell test experiment is described by the measurement of two non-commutating physical operators which are associated with 
two distinctive locations in the space-time. Similarly using this similar analogy in the context of primordial cosmology, one can also perform similar cosmological observations on two spatially separated as well as causally disconnected places upto the epoch of reheating.
In case of cosmological observations one can able to measure the numerical values of various cosmological observables (along with cosmic variance), which can be computed from scalar curvature fluctuation. 
Apart from the success in its observational ground it is important to point that for all such observations it is impossible to measure the 
value of associated canonically conjugate momentum. Consequently, for these observables it is impossible to measure the imprints of two non-commuting operators in the context of primordial cosmology. This directly 
implies that due to this serious drawback in the underlying structural setup it is not at all possible to setup a Bell's inequality violating experimental setup in the context of cosmology. But to make a further strong 
conclusive statement regarding this issue one needs to 
investigate for decoherence effect and its impact in cosmological observation \cite{Burgess:2006jn,Nelson:2016kjm,Dpaas:1996st,Fcldn:2005st,Marti:2007st,Blhu:1995st,Rlaflamme:1990st,
Prokopec:2007st,Gdmoore:2007st,Holmann:2008st,franco:2011st}. If the cosmological observables are
satisfying the basic requirements of decoherence effect then it is possible to perform 
measurements from two exactly commuting cosmological observables and one can able to design a Bell's inequality violating cosmological experimental setup. In the context of quantum mechanics, to design such experimental setup
one needs to perform repeated measurement on the same object (here it is the same quantum state) and in such a physical situation one can justify the appearance of each and every measurement through a single quantum state. Using the same idea 
one can also design a cosmological experimental setup in the present context. In the context of cosmology, one can similarly consider two spatially separated portions in the full sky which exactly 
mimics the role of performing repeated cosmological Bell's inequality violating experiment
via the same quantum mechanical state. Due to this here one can choose the appropriate and required properties of two spatially separated portions in the full sky to setup Bell's inequality violating experimental setup in cosmology.
Most importantly it is important to mention here that if it is possible to connect direct a link between these mentioned non-commuting cosmological observables and a classical probability distribution function originated from 
inflationary paradigm then it is surely possible to setup a Bell's inequality violating cosmological experimental setup.

In this work we have addressed the following important points through which it is possible to understand the underlying framework and consequences from the proposed Bell's inequality violating experimental setup in the context of cosmology.
These issues are:
\begin{itemize}
 \item Setting up cosmological Bell's inequality violating experiment in presence of new heavy fields within the framework of inflation where these heavy fields are the additional field content, appearing along with the inflaton field.
 We have shown that the time dependent mass profile for such heavy fields play a significant role to setup Bell's inequality violating experiment.
 
 \item Explicit role of one point and two point correlation functions, which play significant role to quantify the effect of Bell's inequality violation in presence of significant heavy field mass profile. 
 
 \item Particle creation mechanism of all such heavy fields for different time dependent mass profiles which are responsible for Bell's inequality violation in cosmological setup.
 
 \item The exact connection between all such heavy fields and axion fields as appearing in the context of monodromy model in string theory. 
 
 \item Specific role of isospin breaking phenomenological interactions for heavy fields during the Bell's inequality violating experimental measurement.
 
 \item Exact role of high spin for heavy particles to determine the particle creation and quantify the amount of Bell's inequality violation in cosmological setup.
 
 \item To give a generic mass bound on the scalar heavy fields and high spin heavy fields within a model independent framework of inflationary paradigm. 
 For this purpose we use Effective Field Theory (EFT) framework for inflation \cite{lopez:2012jj,Behbahani:2011it,Choudhury:2014wsa,Choudhury:2015eua,
 Choudhury:2015pqa,Cheung:2007st,noumi:2013jj} in the present context.
 
 \item To identify the connection between scale of inflation or more precisely the exact theory of inflation and amount of Bell's inequality violation in proposed cosmological experimental setup.
 
 \item To give a model independent quantification for primordial gravitational waves through tensor-to-scalar ratio from inflation with the help of the amount of Bell's inequality violation in cosmology.
 If we have any prior knowledge of the amount of Bell's inequality violation in the cosmological setup then using this model independent relation one can put stringent constraint on various inflationary models. If it is not possible to 
 quantify the amount of Bell's inequality violation from any other experimental probe and
 if one can able to measure the value of tensor-to-scalar from future observational probes, subsequently it is possible to quantify the amount of Bell's inequality violation in cosmology with the help of this proposed model independent relation.
 
 \item To study the exact role of initial conditions or choice of inflationary vacuum to violate Bell's inequality in the context of de Sitter and quasi de Sitter cosmological setup.
 
 \item Proposed a specific form of cosmological observable within the framework of inflationary paradigm through which the effect of Bell's inequality violation can be explicitly quantified~\footnote{In ref.~\cite{juan:2015ja} the author have also 
 mentioned this possibility in the context of baroque inflationary model where one can perform the cosmological Bell's inequality violating experiment. In this paper we explore other possibilities in detail by proposing various 
 time dependent mass profiles for the heavy fields for arbitrary choice of initial conditions or choice of vacuum. Hence we will quote the results for Bunch Davies vacuum and $\alpha$ vacuum for the sake of 
 completeness. Also in our paper we provide an explicit form of the new inflationary observable through which one can quantify the effect of Bell's inequality violation in
 cosmological setup.}. Also 
 expressed various known inflationary observables in terms of this newly proposed observable. Here it is important to note that, this conversion is only possible if the heavy fields are massive compared to the Hubble scale 
 and follow a profile as mentioned earlier.
\end{itemize}
Now before going to the further details let us mention the underlying assumptions clearly to understand the background setup for this paper:
\begin{enumerate}
 \item UV cut-off of the effective theory is given by the scale $\Lambda_{UV}$. For our purpose we fix $\Lambda_{UV}=M_p$, where $M_p$ is the
reduced Planck mass. 
 \item Inflaton and the heavy fields are minimally coupled to the Einstein gravity sector.
 
 \item Effective sound speed $c_{S}\neq 1$. Within EFT it is always $c_{S}\leq 1$. For canonical slow roll models $c_S=1$ and for other cases $c_S<1$.
 
 \item Various choices for initial conditions are taken into account during our computation. We first derive the results for arbitrary choice of vacuum and then quote the results for Bunch Davies, $\alpha$ and special type of vacuum.
 
 \item To express the scale of inflation in terms of the amount of Bell's inequality violation in cosmological experimental setup we assume that slow-roll prescription perfectly holds good in the EFT sector. Consequently we have used 
 the consistency conditions which are applicable for slow roll to find out the expression for tensor-to-scalar ratio in terms of the Bell's inequality violating observable. For example, we use here $r=16\epsilon c_{S}$. But without assuming any slow-roll one can find out the 
 expression for the first Hubble slow roll parameter $\epsilon=-\dot{H}/H^2$ in terms of the Bell's inequality violating observable within the framework of EFT.
 
 \item For the computation of Bogoliubov co-efficients we have introduced cut-off in conformal time scale to collect the regularized finite analytical contribution for different time dependent mass profile.
 Consequently the rest of the parameters derived from Bogoliubov co-efficients i.e. reflection and transmission coefficients, number density and energy density follow the same approximation during massive particle creation.
 
 \item To use the analogy with axion monodromy model in the context of string theory we neglect the effect of back-reaction and restricted upto the mass term in the effective potential. 
 This helps us to perfectly identify the analogy between heavy fields and axion.
 
 \item We use approximated WKB solutions to quantify the particle creation for different arbitrary time dependent mass profile for heavy fields as it is not always possible to compute the exact mode functions for the heavy fields 
 in Fourier space by exactly solving the equation of motion for the heavy fields. Some of the cases we provide exact solution where the time dependence in the mass parameter is slowly varying. We use these results to compute the
 one point and two point correlation functions in the present context.
 
 \item To study the role of arbitrary spin fields with spin ${\cal S}>2$ in Bell's inequality violation we assume that the dynamics of all such fields is similar with the scalar field and graviton. 
\end{enumerate}
\begin{table*}
\centering
\footnotesize\begin{tabular}{|c|c|c|
}
\hline
\hline
{\bf Properties}& {\bf Relativistic Quantum Theory} & {\bf Cosmology}
\\
\hline\hline\hline
 {\bf Importance}&Theory of Entanglement &  Important hidden features in
 \\ &came into picture &the context of early universe can be known
\\
\hline
{\bf Fluctuation} & Helps to produce & Helps to produce\\
&virtual particles(pairs & hot and cold spots in CMB
\\ &of particle and antiparticle) & \\
\hline
 {\bf Assumptions}&Concepts of locality and reality  &  Slow-roll prescription 
\\ 
\hline
 {\bf Decoherence } &Provides reasons for  & Primordial non gaussianity 
\\ &the collapse of wave function &can be enhanced\\
\hline
{\bf Applications }  &Quantum information, computing  &   Origin of Large Scale Structure formation.  
\\&and many more &

\\
\hline
\hline
\hline
\end{tabular}
\caption{ Table showing the connection between Relativistic Quantum Theory and Cosmology
in context to Bell's inequality violation.}\label{fig:cosmoscv}
\vspace{.4cm}
\end{table*}
\begin{figure*}[htb]
\centering
{
    \includegraphics[width=17.2cm,height=19cm] {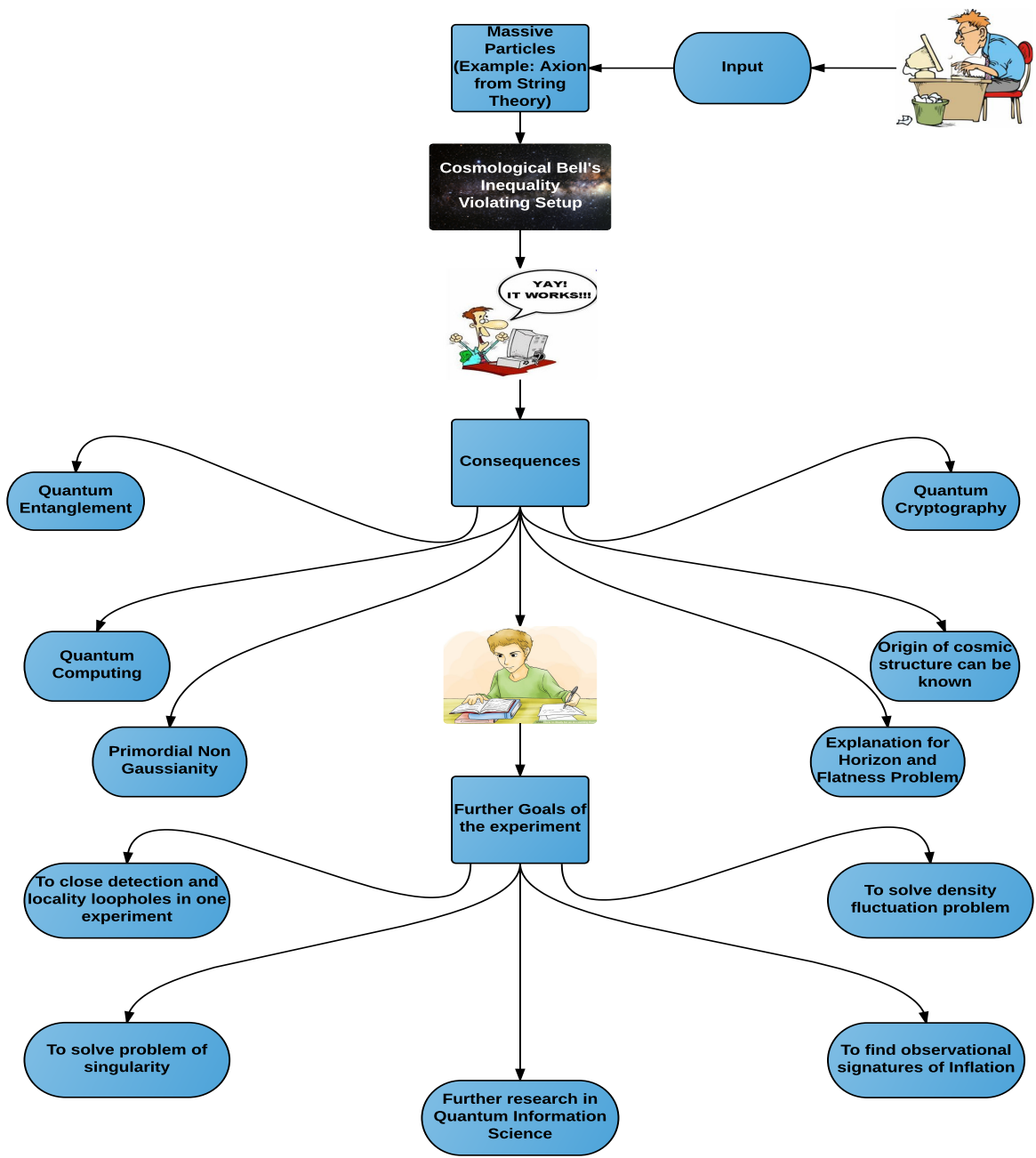}
}
\caption{Flow chart of the Bell's inequality violating cosmological setup.}
\label{sqz2cv}
\end{figure*}
\begin{figure*}[htb]
\centering
{
    \includegraphics[width=17.2cm,height=19cm] {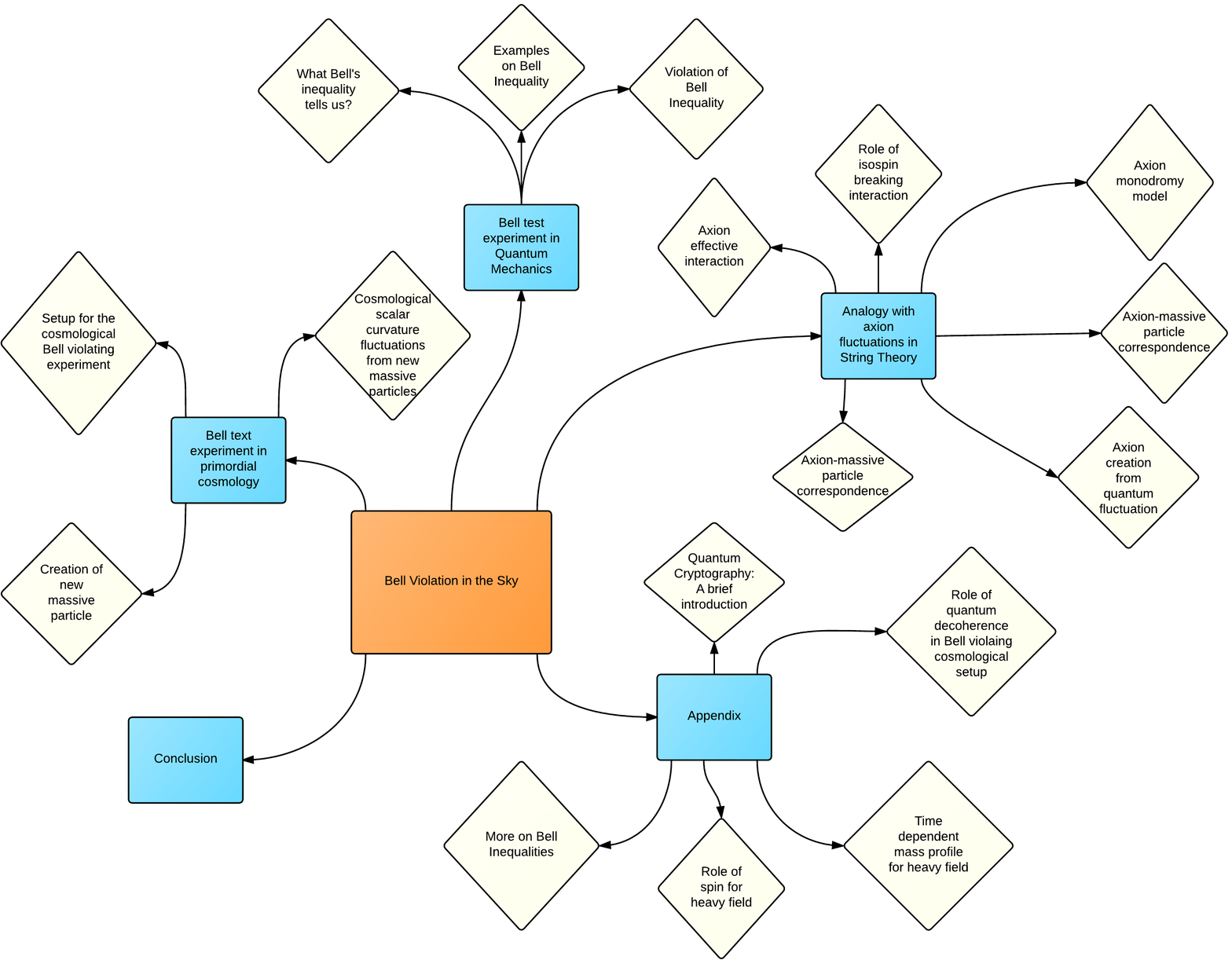}
}
\caption{Flow chart of the basic structural setup of this paper.}
\label{sqz2}
\end{figure*}
In table~(\ref{fig:cosmoscv}), we show the connection between relativistic quantum theory and cosmology
in context to Bell's inequality violation. In fig.~(\ref{sqz2cv}) and fig.~(\ref{sqz2}), we have schematically shown the flow chart of the Bell's inequality violating cosmological setup and 
basic structural setup of the present paper which we have discussed in detail
as follows:
\begin{itemize}
 \item \underline{\textcolor{violet}{\bf Section }\ref{sec2}}: Here we review Bell's inequality in quantum mechanics and its implications. For this we review the proof of Bell's inequality followed by an example of Bell's inequality with spin system.
Further we discuss briefly the violation of Bell's inequality in quantum mechanics. Hence we provide the explanation for such violation and the consequences which finally give rise to new physical concepts like quantum entanglement.

\item  \underline{\textcolor{violet}{\bf Section } \ref{sec3}}: Here in subsection \ref{sec3a} we briefly discuss about the setup for Bell's inequality violating test experiment in the context of primordial cosmology. Then we study creation of new massive particles
as introduced in the context of inflationary paradigm for various choice of time dependent mass profile in subsection \ref{sec3b}. We also present the calculation for the three limiting
situations-{\bf (1)} $m\approx H$, {\bf (2)} $m>>H$ and {\bf (3)} $m<<H$. Now to describe a very small 
fraction of particle creation after inflation we need to find the Bogoliubov coefficient $\beta$ in FLRW space-time, which characterizes the amount of mixing between the two types of WKB
solutions. Therefore we provide detailed mathematical calculations to find the Bogoliubov coefficient $\beta$ for each of the different cases. 
Using the results for Bogoliubov co-efficients we further calculated reflection and transmission co-efficients, number density and energy density of the created particles for
various mass profiles for two equivalent representations. Since the exact analytical expression for the integrals involved in all of these parameters are not always computable, we use the approximation 
in three physical sub regions. Here we provide the results for three specific cases:-\begin{enumerate}
                                                                              \item  $|kc_{S}\eta|=c_{S}k/aH<<1$ (\textcolor{blue}{super horizon}),
                                                                              \item $|kc_{S}\eta|=c_{S}k/aH\approx 1$ (\textcolor{blue}{horizon crossing}),
                                                                              \item  $|kc_{S}\eta|=c_{S}k/aH>>1$ (\textcolor{blue}{sub horizon}).
                                                                             \end{enumerate}
Further in subsection \ref{sec3c} we study cosmological scalar curvature fluctuations in presence of new massive particles for arbitrary choice of initial condition and also for any arbitrary time dependent mass profile. Here 
we explicitly derive the expression for one point and two point correlation function using in-in formalism. Then we quote the results 
for the three limiting
situations-{\bf (1)} $m\approx H$, {\bf (2)} $m>>H$ and {\bf (3)} $m<<H$ in \textcolor{blue}{super horizon}, \textcolor{blue}{sub horizon} and \textcolor{blue}{horizon crossing}.
Here we introduce a new cosmological observable which captures the effect of Bell's inequality violation in cosmology.
Further we express the scale of inflation in terms of the amount of Bell's inequality violation in cosmology experimental setup. Additionally we derive
 a model independent expression for first Hubble slow roll parameter $\epsilon=-\dot{H}/H^2$ and tensor-to-scalar ratio in terms of the Bell's inequality violating observable
 within the framework of EFT. Additionally, in subsection \ref{sec3c} we give an estimate of inflaton mass parameter $m_{inf}/H$~\footnote{Here $m_{inf}$ is the mass of inflaton field and $H$ is the Hubble scale.}. Further 
 we consider a very special phenomenological case, where inflaton mass is comparable with the new particle mass parameter $m_{inf}\approx m$ and using this we provide an estimate of 
 heavy field mass parameter $m/H$~\footnote{Here $m$ is the mass of heavy field and $H$ is the Hubble scale.} which is an important ingredient
to violate Bell's inequality within cosmological setup.

\item  \underline{\textcolor{violet}{\bf Section } \ref{sec4}}: In subsection \ref{sec4a} we give an example of axion model with time dependent decay constant as appearing in the context of string theory. Hence in the next subsection \ref{sec4b} we mention the effective axion interaction of axion fields.
Now to give a analogy between the newly introduced massive particle and the axion we further discuss the creation of axion in early universe in subsection \ref{sec4c}.
Further in subsection \ref{sec4d} and \ref{sec4e} we establish the one to one correspondence between heavy field and axion by comparing the particle creation mechanism, one and two point correlation functions.
Additionally, in subsection \ref{sec4e} we give an estimate of axion mass parameter
$m_{axion}/f_{a}H$~\footnote{Here $m_{axion}$ is the axion mass, $f_{a}$ is the time dependent decay constant for axion and $H$ is the Hubble scale.}which is an important ingredient
to violate Bell's inequality within cosmological setup. Finally, in subsection \ref{sec4f} we discuss the specific role of 
 isospin breaking phenomenological interaction for axion type of heavy fields to measure the effect of Bell's inequality violation in primordial cosmology.
 \item \underline{\textcolor{violet}{\bf Section } \ref{sec5}}: Here we conclude with future prospects from this present work.
 
 \item  \underline{\textcolor{violet}{\bf Appendix} \ref{sec6}}: In appendix \ref{sec6a} we explicitly show the role of quantum decoherence in cosmological setup to violate Bell's inequality. Additionally here we also 
 mention a possibility to enhance the value of primordial non-Gaussianity from Bell's inequality violating setup in presence of massive time dependent field profile. Further in appendix \ref{sec6b} we discuss the role of 
 three specific time dependent mass profile for producing massive particles and to generate quantum fluctuations. Further, in appendix \ref{sec6b} we discuss the role of arbitrary spin heavy field to violate Bell's inequality. 
 Here we provide a bound on the mass parameter for massive scalar with spin ${\cal S}=0$, axion with spin ${\cal S}=0$, graviton with spin ${\cal S}=2$ and for particles with high spin ${\cal S}>2$ in \textcolor{blue}{horizon crossing}, \textcolor{blue}{super horizon}
 and \textcolor{blue}{sub horizon} regime. Then we provide the extended class of Bell's inequality, called \textcolor{blue}{\it CHSH inequality}. Finally, we give a very brief discussion on \textcolor{blue}{\it quantum cryptography} related to the present 
 topic of the paper.

\end{itemize}

\section{Bell test experiment in Quantum Mechanics}
\label{sec2}
\subsection{What Bell's inequality tells us?}
\label{sec2a}

In ref.~\cite{epr:1935ja} authors first demonstrated that quantum theory is incomplete by the help of \textcolor{blue}{\it EPR (Einstein Podolsky Rosen) paradox}.
According to Einstein's theory of special relativity, we know that speed of light is the fastest that we can get. Indeed it was the discrepancy between the predictions
of relativity and quantum theory concerning the correlations between events in space-like separated regions that led Albert Einstein, Boris Podolsky and Nathan Rosen to point out an effect, known as EPR, where one part
of entangled quantum systems appears to influence another at the same instant.
To Einstein, Podolsky and Rosen quantum theory gave only an incomplete account of physical reality.
As special theory of relativity \cite{Einstein:1905ve} says that nothing can go faster than light or in other words speed of light is the fastest we can get, they believed that the correlations in measurement
outcomes of experiments which measures both members of particles which are highly separated and
entangled could be explained by hypothesizing that separated particles are not
entangled rather had fixed values of all their
measurable attributes from the outset. Hence the outcomes of experiment must be determined by  \textcolor{blue}{\it``hidden variables''}.

Later in ref.~\cite{Bell:1964kc}, \textcolor{violet}{\bf John Stewart Bell} showed that \textcolor{blue}{\it ``In a theory in which parameters are added to quantum mechanics to determine the results of
individual measurements, without changing the statistical predictions, there must be a mechanism whereby the setting of one measuring device can influence the reading of another instrument, however remote. (Bell 1987, p. 20.)''}.
He also showed that there was the difference between the predictions of any hidden variable theory and predictions of quantum theory. Bell's article refers to ref.~\cite{epr:1935ja} that challenged the completeness of quantum theory.
In that paper, Bell started his theory with two assumptions which were:
\begin{enumerate}
 \item \textcolor{blue}{\it concept of reality} (real properties of microscopic objects determines the results of quantum mechanical experiments),
 \item \textcolor{blue}{\it concept of locality} (that reality in one place is not affected by experiments done at the same time at a distant place).
\end{enumerate}
Using these two assumptions  Bell derived an important result, which is known as  \textcolor{blue}{\it`Bell's Inequality'}.
Bell proved with his inequality that \textcolor{blue}{\it ``no local hidden variable theory is compatible with quantum
mechanics.''} The following example will develop some physical intuition for Bell inequality which is clearly explained in the following steps:
\begin{itemize}
\item What would we observe if an experiment is performed on a set of pairs of polarization measurements?
For simplicity let's say that the pair of photons exist in an entangled state such that both polarizations are same but are otherwise unknown when they are measured.
\item Let's call our experimentalists Aace and Bace, and let's say that they agree to place their polarizers in the same direction. Thus the angle between their polarizers is  $0^{o}$. What would they see? Since the entangled particles are correlated, every time Alice observes `vertical', Bace also observes the same i.e. `vertical'. And every time Aace sees `horizontal', Bace sees `horizontal'. The percentage that they agree mutually on the result is $100$.
\item Now lets rotate the polarizer of Bace by $90^{o}$. Now when Bace sees `vertical', Aace observes `horizontal'. And when they perform polarization measurements on respective pairs of correlated photons, their results will be anti-correlated. Every time Bace sees `vertical', Aace sees `horizontal' and vice versa. The percentage they agree on the results is $0$.
\item Now suppose Bace rotates his polarizer back towards Aace's vertical so that polarizer of the Bace makes an angle to Aace's vertical. Now Aace measures her photon to be `vertical'.
Thus the twin photon will also be `vertical' (in Aace frame).
To Bace, the photon he receives will appear to be in superposition of his `horizontal' and `vertical' orientations.
\item Hence the result of Bace's polarization measurement is uncertain, sometimes when Bace measures a photon that Aace observes as `vertical' Bace will also sees `vertical' too. But at other times when Bace measures a photon Aace sees as `vertical', but Bace sees `horizontal'. As a result the percentage they agree is between $0$ and $100$. The exact percentage depends on the angle between their polarizers.
\end{itemize}

\subsection{1st example on Bell's inequality}
\label{sec2b}
\begin{itemize}

\item Chace's idea was to test the theory of locality of Einstein by using the
properties of the correlations between measurement outcomes obtained by experimenters Aace and Bace.
Now suppose Aace is in Mumbai with three coins with its head or tails facing upwards,
but Aace can't tell which side is up as he is blind folded and also there is a black cloth on each of
the coin.
\item When Aace uncover one coin, suddenly other two coins disappear. Therefore probability of getting either
head or tails is same.
\item Similarly his friend Bace (in Calcutta) has same type of coins and does the same experiment. He too have the
same probability of getting either head or tails.
\item Both of them repeats their experiments again and again
to find out the correlation between their coins. Therefore they found out that whenever they uncover their
coins with the same label, that is, first, second or third, they both got head(H) or tail(T).
\item They did their experiments number of times to be sure but they got their coins correlated each time.
But Aace wants to find out two coins in one turn, but he can't as when he uncover one coin the other
coins suddenly disappears.
\item So when he talked to Bace, he told that if he (Bace) uncovers second coin and tell him what he got,
then Aace will certainly know what he will get if he (Aace) uncovers second coin without uncovering it.
Then he can uncover first coin and hence by this way he will get to know the results of two coins.
\item But Aace got one doubt which is, if Bace uncovers second coin, his first and third coins disappeared
and he himself uncovers first coin and remaining two coin disappeared, but there is no way to find out
when they actually uncover the second coin.
When Bace uncovers his coin, it does not have any influence on Aace's coin.
In fact what Bace finds by uncovering his coin, it reveals some information about the coin of
Aace.
\item They went to their friend Chace to clarify their doubt. He told Bace to uncover his one coin and assume
to know for sure what Aace will find when he uncovers his own coin without Aace disturbing his coin.
Therefore there has to be some variables which are hidden that specify the condition of Aace's coins.
And if we can anyhow know those hidden variables, then we will be able to know the value of Aace's coins.
\item Chace told that there has to be some probability distribution that specify the condition
on the three coins of Aace and it must not be negative and its sum is one.
Aace can not uncover all his coins, therefore he will not be able to measure the probability distribution.
But with the help of Bace, he can uncover any two coins as Bace suggested.
\item After doing Bell experiment they found out that correlations found by them violates Bell's inequality.

\end{itemize}

\subsection{2nd example on Bell's inequality}

\begin{figure*}[htb]
\centering
{
    \includegraphics[width=9.2cm,height=9cm] {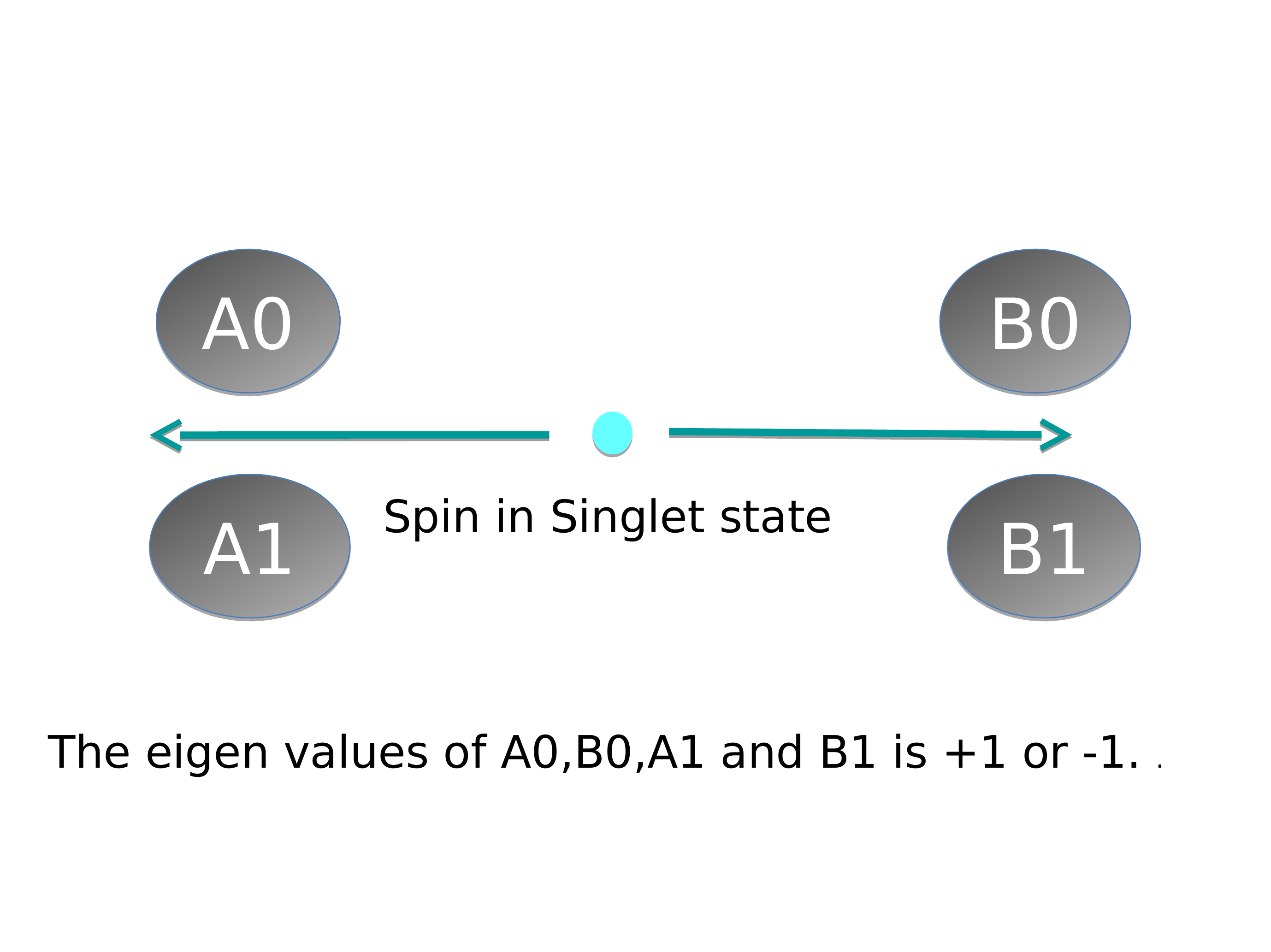}
}
\caption{Schematic diagram of Bell's inequality example for a spin system.}
\label{spinc2}
\end{figure*}
See fig~(\ref{spinc2}) for the representative setup for the spin system.
Here the operators which are $A_0$, $B_0$, $A_1$ and $B_1$ correspond to measuring the spin and their eigenvalues are $\pm$ 1,
we have to choose the value of operators as:
\bea A_0 = \bf{n}_{0}.{\bf\sigma},\\
A_1 = \bf{n}_{1}.{\bf\sigma},\\
B_0 = \bf{n}_{0}.{\bf\sigma},\\
B_1 = \bf{n}_{1}.{\bf\sigma}.\eea

Therefore assuming the other variable which is
\begin{equation}\label{qwe}
\langle R \rangle = \langle A_0 B_0 \rangle + \langle A_1 B_0 \rangle + \langle A_0 B_1 \rangle - \langle A_1 B_1 \rangle
\end{equation}

According to classical theory of hidden variable, \be \vert \langle R \rangle \rvert \leq 2.\ee
But in quantum mechanics, the expectation value of R can be found bigger.
By squaring the Eq.~(\ref{qwe}) one can show that:
\begin{equation}
R^2 = 4 + [A_1,A_0][B_1,B_0] \Rightarrow \lvert \langle R \rangle \rvert >2
\end{equation}
making $\lvert \langle R \rangle \rvert$ larger than $2$, which violates Bell's inequality. The question now arises is how to draw above conclusion, choosing:
\bea A_0 &=& \bf{x}.\bf{\sigma},\\
A_1 &=& \bf{y}.\bf{\sigma},\\
B_0 &=& sin~ \theta (\bf{x}.\bf{\sigma}) + cos~ \theta (\bf{y}.\bf{\sigma}),\\ 
B_1 &=& cos~ \theta(\bf{x}.\bf{\sigma}) - sin~ \theta (\bf{y}.\bf{\sigma})\eea
we get the extra $\sqrt{2}$ factor for the maximal violation i.e.
\be \lvert \langle R \rangle \rvert >2\sqrt{2}.\ee

\subsection{Review on the proof of Bell's inequality}
\label{sec2d}
Bell's inequality gives a general condition, that hold for any local deterministic hidden variable theory.
Let us consider two spin half particles and we define two functions which are P$\left(a,\lambda\right)$
and Q$\left(b,\lambda\right)$ which give results of the spin measurements on particle $1$ in the direction
of `a' and on particle $2$ in the direction of `b' respectively. Here these functions depend on the parameter $\lambda$, which is a hidden variable.

Therefore we have
\bea P(a,\lambda) &=& \pm 1,\\ Q(b,\lambda) &=& \pm 1\eea
Now we want to calculate average value of the product of two components $P(a,\lambda)$ and $Q(b,\lambda)$
\begin{equation}
A(a,b) = \int \rho(\lambda) P(a,\lambda) Q(b,\lambda) d\lambda
\end{equation}
where $\rho\left(\lambda\right)$ is the probability distribution of $\lambda$.
Since \be A(d,d) = -1\forall d\ee 
where the detectors are perfectly aligned and the results are perfectly anti-correlated.
It means
\begin{equation}
P(d,\lambda) = -Q(d,\lambda) \forall \lambda
\end{equation}
Therefore 
\begin{equation}
A(a,b) = -\int \rho(\lambda) P(a,\lambda) Q(b,\lambda) d\lambda
\end{equation}
\begin{equation}
A(a,b)-A(a,c) = - \int \left[\rho(\lambda) P(a,\lambda) P(b,\lambda) - P(a,\lambda) P(c,\lambda)\right]
\end{equation}
where `c' is taken as unit vector.

Since 
\be \left(P\left(b,\lambda\right)\right)^2 = 1\ee

Then, 
\begin{equation}
A\left(a,b\right) - A\left(a,c\right) = - \int \rho\left(\lambda\right)\left[1-P\left(b,\lambda\right) P\left(c,\lambda\right)\right]P\left(a,\lambda\right)P\left(b,\lambda\right) d\lambda
\end{equation}

Now we know that \be P\left(a,\lambda\right) = \pm 1\ee

Therefore we can write,
\bea
-1\leqslant  P\left(a,\lambda\right)P\left(b,\lambda\right)&\leqslant& 1,\\
\rho\left(\lambda\right)\left[1-P\left(b,\lambda\right)P\left(c,\lambda\right)\right] &\geqslant & 0
\eea
Hence,
\bea
\lvert A\left(a,b\right) - A\left(a,c\right)\rvert\leqslant \int \rho\left(\lambda\right)\left[1-P\left(b,\lambda\right)P\left(c,\lambda\right)\right]d\lambda
\eea
or equivalently
\begin{equation}
 \lvert A\left(a,b\right) - A\left(a,c\right) \rvert \leqslant 1 + A\left(b,c\right)
\end{equation}
The above equation is the Bell Inequality.
\subsection{Bell's inequality in a spin system}
\label{sec2e}
In accordance to Chace, there has to be a variable $\lambda$, which is hidden to both Aace and Bace, 
and $\lambda$ describes spins of particles.
Chace sets \be S\left(\bf{\hat{p}},\lambda\right) = \pm 1\ee indicating
the sign of the projection of the spin in $\bf{\hat{p}}$ direction.
We know that the total spin is zero, therefore `S' will give opposite spins
of particles 1 and 2, due to the conservation of angular momentum.
For Chace, $\lambda$ can take any value, but it will be fixed if the
initial state is set up. We can find the value of $\lambda$, from
the probability distribution $P\left(\lambda\right)$
where, \begin{equation}
\int d\lambda ~P\left(\lambda\right) = 1
\end{equation}
Aace and Bace can measure the projection of spin in $\bf{\hat{d_1}}$ and $\bf{\hat{d_2}}$ direction respectively.
While doing measurements consecutively, correlation function is
\bea
\langle{({\bf s_{1}.\hat{d_1}}) ({\bf s_{2}.\hat{d_2}})}\rangle = - \langle
({\bf s_{1}.\hat{d_1}}) ({\bf s_{1}.\hat{d_2}})\rangle &=& -\frac{1}{4} \int d\lambda ~
P(\lambda) S(\bf{\hat{d_1}},\lambda) S(\bf{\hat{d_2}},\lambda)\nonumber\\
&=& -\frac{1}{4} {\bf{\hat{d_1}.\hat{d_2}}} + \frac{i}{2}(\bf{\hat{d_1}}\times{\bf \hat{d_2}}).\langle\bf{s_1}\rangle.
\eea
Now since here \be \langle\bf{s_1}\rangle = 0\ee one can finally write the following expression:
\bea
\langle{\bf{(s_{1}.\hat{d_1}) (s_{2}.\hat{d_2})}}\rangle &=& -\frac{1}{4} \bf{\hat{d_1}.\hat{d_2}}\nonumber\\
&=& \langle{\bf{(s_{1}.\hat{d_2}) (s_{2}.\hat{d_3})}}\rangle 
+ \frac{1}{4} \int d\lambda~ P\left(\lambda\right) S({\bf{\hat{d_2}}},\lambda) 
\left[S({\bf{\hat{d_3}}},\lambda) - S({\bf{\hat{d_1}}},\lambda)\right].
\eea
Now here we use the following constraint, \be S^2 ({\bf{\hat{d_3}}},\lambda) = 1\ee and consequently we get:
\bea
\langle{\bf{(s_{1}.\hat{d_1}) (s_{2}.\hat{d_2})}}\rangle 
- \langle{\bf{(s_{1}.\hat{d_2}) (s_{2}.\hat{d_3})}}\rangle = \frac{1}{4} \int d\lambda~P(\lambda) 
S({\bf{\hat{d_2}}},\lambda) S({\bf{\hat{d_3}}},\lambda) 
\left[1 - S({\bf{\hat{d_1}}},\lambda) S({\bf{\hat{d_3}}},\lambda)\right].~~~~~
\eea
This directly implies that:
\bea
\lvert\langle{\bf{(s_{1}.\hat{d_1}) (s_{2}.\hat{d_2})}}\rangle - 
\langle{\bf{(s_{1}.\hat{d_2}) (s_{2}.\hat{d_3})}}\rangle\rvert 
\leq \frac{1}{4} \int d\lambda~P(\lambda) 
\left[1 - S({\bf{\hat{d_1}}},\lambda) S({\bf{\hat{d_3}}},\lambda)\right].
\eea
Hence Bell's inequality follows from the theory of hidden variable and one can write:
\begin{equation}
\lvert\langle{\bf{(s_{1}.\hat{d_1}) (s_{2}.\hat{d_2})}}\rangle 
- \langle{\bf{(s_{1}.\hat{d_2}) (s_{2}.\hat{d_3})}}\rangle\rvert \leq 
\frac{1}{4} + \langle{\bf{(s_{1}.\hat{d_1}) (s_{2}.\hat{d_3})}}\rangle
\end{equation}
If we choose \bea {\bf{\hat{d_1}.\hat{d_2}}} &=& 0,\\ 
{\bf{\hat{d_3}}} &=& {\bf{\hat{d_1}}}\cos \theta + {\bf{\hat{d_2}}}\sin \theta,\eea then one can write the following expression:
\bea
\langle{\bf{(s_{1}.\hat{d_2})(s_{2}.\hat{d_3})}}\rangle &=&
-\frac{1}{4} \sin \theta,\\ \langle{\bf{(s_{1}.\hat{d_1})(s_{2}.\hat{d_3})}}\rangle &=& -\frac{1}{4} \cos \theta
\eea
Hence we get:
\bea
\lvert\langle{\bf{(s_{1}.\hat{d_1}) (s_{2}.\hat{d_2})}}\rangle 
- \langle{\bf{(s_{1}.\hat{d_2}) (s_{2}.\hat{d_3})}}\rangle\rvert &=& \frac{1}{4} \lvert \sin \theta \rvert,\\
\frac{1}{4} + \langle{\bf{(s_{1}.\hat{d_1})(s_{2}.\hat{d_3})}}\rangle = \frac{1}{4} \left(1-\cos \theta\right)
\eea
Here it is important to note that according to Bell's inequality, the following quantity $I(\theta)$ is negative i.e.
\bea
I\left(\theta\right) &=& \lvert\langle{\bf{(s_{1}.\hat{d_1}) (s_{2}.\hat{d_2})}}\rangle 
- \langle{\bf{(s_{1}.\hat{d_2}) (s_{2}.\hat{d_3})}}\rangle\rvert+\frac{1}{4}
+ \langle{\bf{(s_{1}.\hat{d_1})(s_{2}.\hat{d_3})}}\rangle\nonumber \\
&=&\frac{1}{4} \left[ \lvert \sin \theta \rvert
+ \cos \theta - 1 \right]<0.
\eea
But as an exception for the range $\theta < \lvert \theta \rvert < \frac{\pi}{2}$, the quantity $I\left(\theta\right)>0$.
\subsection{Violation of Bell Inequality}
\label{sec2f}
If reality is \textcolor{blue}{\it`local'} then Bell's inequality must hold regardless of the angles at which polarization detectors are set.
The first actual Bell test was done using Freedman's inequality in ref.~\cite{Freedman:1972zza}.
The delay in experiment was due to the inability to build perfect polarization detectors and to coordinate closed timed
measurements that no speed of light could make it from one photon to the other within the duration of pair of measurements.
The results of above experiment confirmed the violation of Bell's inequality. Hence the inequality is wrong.
However only assumption we used was \textcolor{blue}{\it`concept of locality'}. 

\subsubsection{ {\bf Case I:} Explanation from earlier experiments}
After the proof presented in ref.~\cite{Freedman:1972zza}, many experiments were done such as:
\begin{itemize}
 \item \underline{\textcolor{violet}{\bf Aspect (1982)}\cite{Aspect:1981zz,Aspect:1982fx},
\textcolor{violet}{\bf Tittel and Geneva group (1998) }\cite{Tittel:1998ja}, \textcolor{violet}{\bf Rowe (2001)} \cite{Rowe:2001ja}}:
These experiments are performed to close the detection loophole, 

\item \underline{\textcolor{violet}{\bf Groblacher (2007) }\cite{Groblacher:2001jb}}: Test of
Leggett-type non-local realist theories,
 
 \item  \underline{\textcolor{violet}{\bf Salart (2008)} \cite{Salart:2008jc}}: Separation in
a Bell Test,
 
 \item  \underline{\textcolor{violet}{\bf Ansmann (2009)} \cite{Ansmann:2009jd}}:  Overcoming the detection loophole in
solid state,
 
 \item \underline{\textcolor{violet}{\bf Christensen (2013)} \cite{Christensen:2013je}}:  Overcoming
the detection loophole for photons,
 
 \item \underline{\textcolor{violet}{\bf Hensen (2015)} \cite{Hensen:2015ccp}}:  A loophole-free
Bell test and many others,

\item \underline{\textcolor{violet}{\bf Giustina (2015)} \cite{Giustina:2015jf}, \textcolor{violet}{\bf Shalm (2015)} \cite{Shalm:2015jj}}: Recently 
performed Loophole-free Bell tests with photons which provide strong
experimental proof for non-local reality.
\end{itemize}

\subsubsection{ {\bf Case II:} Explanation from recent experiments}
In view of local realism concept, physical properties of objects exist independently
of measurement and physical influences cannot exceed the speed of light as we already know.
Even though the previous experiments supported the predictions of quantum
theory, yet every experiment requires assumptions which will provide
loopholes for a local realist explanation. Therefore in this experiment
they reported a Bell test that closes the most significant of these
loopholes at the same instant of time. They used photons which are
entangled in nature, rapid setting generation, and
superconducting detectors with very efficiency and then observed violation of Bell inequality.
Every time particles interact with one another their quantum states
tend to entangle. Hence when one member of the pair is being
measured then the other member behaves as if it is also being
measured, and thus acquires a definite state.
\subsubsection{ {\bf Case III:} Explanation for entanglement}
We deduced from the violation of Bell's inequality that hidden variables theory is incorrect,
therefore let us consider an experiment where large number of measurements are
 done on the spin of particles then the outcome should follow Bell's inequality but that doesn't happen.
 There are many experimental evidence which proves the violation of Bell's inequality, but those experiments have loophole
problems that is the results of the measurements are correlated with each other which 
means we cannot measure properties simultaneously.

The following example will explain the entanglement in simple way in some steps:
\begin{itemize}
\item Let's say we have two particle states having same mass, spin
and also no forces acting on both the particles.

\item Let $\bf{p_1}$
and $\bf{m_1}$ and $\bf{p_2}$ and $\bf{m_2}$ be the position
and momentum of first and second particle respectively. Therefore
for the two particle system the basis states will be
$|\bf{p_1}\rangle \otimes \bf{p_2}\rangle$. But we should
have states labeled by center of mass momentum, i.e.
\bea {\bf{M}} &=& {\bf{m_1}} + {\bf{m_2}},\\ \bf{p} &=& {\bf{p_1}} + {\bf{p_2}},\eea
therefore, unitary transformation to the basis is $\bf{|M,p\rangle}$

\item Now for instance Aace and Bace set up the two-particle system
where initially $\bf{M}$ is 0 i.e. $|0,\bf{p_0}\rangle$.

\item Now Aace makes a measurement on the momentum of first particle
and found the accurate outcome to be $\bf{m_1}$ ,then,
\be {\bf{m_2}} = -{\bf{m_1}}.\ee There are large uncertainties in the
positions of the two particles but \be {\bf{p}} = {\bf{p_0}} + {2t\bf{m_1}}{\rm 1/mass}.\ee 

\item When the two particles are very far from each other, then Bace accurately
measures the position of the second particle which is $\bf{p_2}$. But we
do not have any idea about the accurate values of position and momentum
of both the particles. When Bace measures $\bf{p_2}$, it makes momentum
of the second particle $\bf{m_2}$ uncertain, making $\bf{m_1}$ uncertain
instantaneously, this will occur even if the distance is very large.

\item Hence this spooky action at a distance is known as \textcolor{blue}{\it quantum entanglement}.
\end{itemize}

\section{Bell test experiment in primordial cosmology}
\label{sec3}
\subsection{Setup for the cosmological Bell violating experiment}
\label{sec3a}
Metric of a uniform, spatially flat ($k=0$), FLRW space-time is given by,
\begin{equation}
ds^2 = -dt^2 + a\left(t\right)^2 d{\bf x}^2  = a^2\left(\eta\right) \left[-d\eta^2 + d{\bf x}^2 \right]
\end{equation}
\begin{figure*}[htb]
\centering
{
    \includegraphics[width=14.2cm,height=12cm] {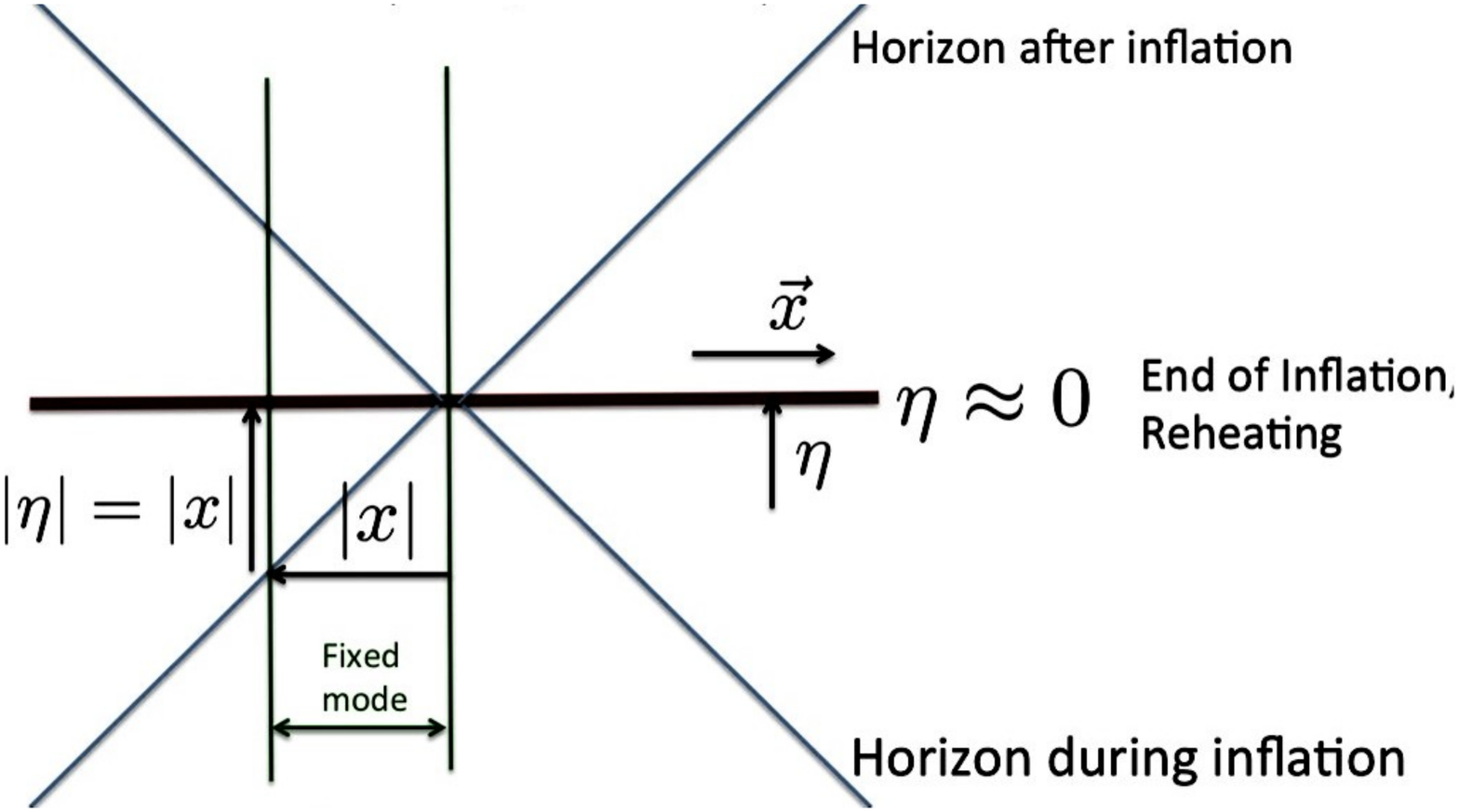}
}
\caption[Optional caption for list of figures]{ Schematic diagram of the evolution of the Universe.} 
\label{figc}
\end{figure*}
where t is proper time and $\eta$ is conformal time defined as:
\be\label{fg1} \eta=\int \frac{dt}{a(t)}.\ee
The conformal time described here is negative (so that we can make scale factor positive) and ranges from
from $-\infty$ to $0$. Here $a(t)$ is the scale factor which characterize the Hubble parameter $\frac{\dot{a}}{a}\approx H(t)$.
During the inflationary period the scale factor grows exponentially ($a(t)\propto e^{Ht}$), just like de Sitter and quasi de Sitter
space and consequently the Hubble parameter $H(t)$ varying slowly. Using this fact in Eq~(\ref{fg1}),
during inflation the scale factor can be expressed in terms of the conformal time $\eta$ as:
\be\begin{array}{lll}\label{kg1}
 \displaystyle a(\eta) =\left\{\begin{array}{ll}
                    \displaystyle   -\frac{1}{H\eta}~~~~ &
 \mbox{\small {\bf for ~dS}}  \\ 
	\displaystyle  -\frac{1}{H\eta}\left(1+\epsilon\right)~~~~ & \mbox{\small {\bf for~ qdS}}.
          \end{array}
\right.
\end{array}\ee
where $\epsilon$ is the Hubble slow-roll parameter defined in Eq~(\ref{rt1}). But for simplicity one can neglect the contribution from $\epsilon$ in the 
leading order for quasi de-Sitter case as it is sufficiently small in the slow-roll regime. For our computation henceforth 
we will follow this assumption. Additionally it is important to note that for de Sitter and quasi de Sitter case the relation between conformal time $\eta$ and physical time $t$ can be expressed 
through the following expression:
\be t=-\frac{1}{H}\ln(-H\eta),\ee
which we will use throughout the paper. Within this setup inflation ends when the conformal time $\eta\sim 0$, as clearly depicted in figure~\ref{figc}.

Here it can be easily shown that the role of quantum mechanics plays very significant role to produce spatially dependent fluctuations in terms of the scalar fields.
We know that according to the theory of inflation, in the early universe, quantum mechanical effects are responsible
for primordial fluctuations.
But it is interesting to know that the fluctuations we have observed today is completely classical in nature.
It is a very well known fact that in the context of inflation 
all such fluctuations become classical as they exit the horizon and inside the horizon all of them are quantum. In this discussion the fluctuations are characterized by the following quantity, known as the curvature perturbation:
\be \zeta=-\frac{H}{\dot{\bar{\rho}}}\delta\rho~~~~\overrightarrow{\bf Fourier~transform}~~~~ \zeta_{k}=-\frac{H}{\dot{\bar{\phi_{0}}}M_p}\phi_{k},\ee
where for each values of $k$ in Fourier space it represent harmonic oscillator. Now in FLRW background one can compute the following commutator:
\be \left[\zeta,\dot{\zeta}\right]\propto a^{-3},\ee
where $\dot{\zeta}$ is the canonically conjugate variable of $\zeta$.
Further this expression can be re-expressed after Fourier transformation as:
\be\begin{array}{lll}
 \displaystyle \left|k^3\left[\zeta_{\bf k},\dot{\zeta}_{\bf k}\right]\right|\sim k^3\left|\left[\phi_{\bf k}(\eta),\eta\partial_{\eta}\phi_{\bf k}\right]\right|\approx\left\{\begin{array}{ll}
                    \displaystyle  H^3 (\eta k)^{3}~~~~ &
 \mbox{\small {\bf for ~dS}}  \\ 
	\displaystyle  H^3 (\eta k)^{3}(1-3\epsilon)~~~~ & \mbox{\small {\bf for~ qdS}}.
          \end{array}
\right.
\end{array}\ee
which becomes zero as $\eta k\rightarrow 0$, at the end of inflation. This is the signature of Bell inequality violation in the context of primordial cosmological setup. Most importantly,
after inflation when reheating phenomena occurs one can write down a classical measure or more precisely a classical probability distribution function of fluctuation $\zeta(x)$ as:
\be \rho[\zeta(x)]=|\Psi[\zeta(x)]|^2 \Rightarrow\rho[\phi(x)]=\mu[\phi(x)].\ee
Here $|\Psi[\zeta(x)]|^2$ or equivalently $\mu[\phi(x)]$ represent the classical probability distribution~\footnote{For multifield case one needs to take the contribution from 
the isocurvature fluctuation as well. In such a physical situation the classical probability distribution function of curvature fluctuation $\zeta(x)$ and isocurvature fluctuation $\chi(x)$ can be written as:
\be \rho[\zeta(x),\chi(x)]=|\Psi[\zeta(x),\chi(x)]|^2 \Rightarrow\rho[\phi_{i}(x),\phi_{j}(x)]=\mu[\phi_{i}(x),\phi_{j}(x)],\ee where $i$ and $j$ stands for number field contents in multifield scenario.},
which is nothing but the state of the universe at the spatial hyper surface where reheating phenomena occurs. In the present context, all the fluctuations can be treated as distribution of classical random variables.
Additionally it is important to note that, here due to commutivity of $\zeta$ and $\dot{\zeta}$ at the end of inflation for above mentioned non-commutative observables it is not at all possible to distinguish $|\Psi[\zeta(x)]|^2$ from classical probability 
distribution function $\rho[\zeta(x)]$. Now here one can also calculate spread in the canonically conjugate variable $\dot{\zeta}_{\bf k}$ of scalar curvature fluctuation $\zeta_{\bf k}$ in Fourier space as:
\be \sqrt{\langle |\dot{\zeta}_{\bf k}|^2 \rangle} \propto \tilde{c}^2_{S}(\eta k)^{2},\ee
where in the present context the effective sound speed $c_{S}$ is defined as:
\be c_{S}=\tilde{c}_{S}\sqrt{1+\frac{\bar{M}^3_{1}}{\epsilon H M^2_p}},\ee
where $\tilde{c}_{S}$ is the actual sound speed in the absence of all effective interactions and in our discussion it is defined as:
\be \tilde{c}_{S}=\frac{1}{\sqrt{1-\frac{2M^4_2}{\dot{H}M^2_p}}}.\ee
Here $\bar{M}^3_{1}$ and $M_2$ are the time dependent coefficients of specific type of effective interactions introduced in ref.~\cite{Cheung:2007st}.
In the slow roll regime, $\frac{\bar{M}^3_{1}}{\epsilon H M^2_p}$ is a very slowly varying function with respect to time and can be treated as a 
constant for our discussion. Here it is important to note that, after horizon crossing modes become classical in nature and in that case the spread becomes zero as $\eta k\rightarrow 0$, at the end of inflation. 
As a result we do not able to measure the canonically conjugate variable through various cosmological observations. But in the present context of discussion from the computed classical probability 
distribution function $\rho[\zeta(x)]$ one cannot comment on the exact measurement procedure on a quantum state. This type of phenomena is commonly studied in the context of quantum mechanical decoherence in which 
to setup a measuring device one needs to introduce a coupling between additional 
environment and long wavelength cosmological perturbations in the present context of discussion. Decoherence phenomena in quantum mechanics is guided by the dynamical behaviour of the phase factor appearing in the expression for the quantum mechanical wave function  $\Psi[\zeta(x)]$. On the other hand, here one can say that $|\Psi[\zeta(x)]|^2$ 
is connected with the correlation functions in cosmological perturbation theory. Now as $|\Psi[\zeta(x)]|^2$ cannot be distinguishable from classical probability 
distribution function $\rho[\zeta(x)]$, one can easily identify this quantity with the post inflationary correlations functions in context of primordial cosmology. Let us mention few possibilities in the following for system environment
interactions and associated couplings which are commonly used to study the phenomena of quantum decoherence during inflationary epoch:
\begin{enumerate}
 \item Gravitational waves \cite{Hawking:1981ja,Starobinsky:1982ja,Alan:1981ja,Linde:1982st,Albs:1982st,Ahgsyp:1982mn,Radam:2015st,
Pade:2015st,CKDP:2009st,Dpaas:1996st,aapfmj:1994so,Lpgyv:1989st,Lpgu:1990st,ahgu:1985st,
ckdp:1998st,Fcldn:2005st,Marti:2007st,Blhu:1995st,Sakagami:1988st,Rlaflamme:1990st,
Prokopec:2007st,Gdmoore:2007st,holman:2015st,Holmann:2008st,franco:2011st,feldman:1993st,
itzhaki:2007jj,Panda:2007ie,Baumann:2007ah,Baumann:2007np},
 \item Effects of multifield components and associated isocurvature perturbation \cite{Hawking:1981ja,Starobinsky:1982ja,Alan:1981ja,Linde:1982st,Albs:1982st,Ahgsyp:1982mn,Radam:2015st,
Pade:2015st,CKDP:2009st,Dpaas:1996st,aapfmj:1994so,Lpgyv:1989st,Lpgu:1990st,ahgu:1985st,
ckdp:1998st,Fcldn:2005st,Marti:2007st,Blhu:1995st,Sakagami:1988st,Rlaflamme:1990st,
Prokopec:2007st,Gdmoore:2007st,holman:2015st,Holmann:2008st,franco:2011st,feldman:1993st,
itzhaki:2007jj,Panda:2007ie,Baumann:2007ah,Baumann:2007np},
 \item Interaction between short and long wavelength fluctuations in cosmological perturbations \cite{Hawking:1981ja,Starobinsky:1982ja,Alan:1981ja,Linde:1982st,Albs:1982st,Ahgsyp:1982mn,Radam:2015st,
Pade:2015st,CKDP:2009st,Dpaas:1996st,aapfmj:1994so,Lpgyv:1989st,Lpgu:1990st,ahgu:1985st,
ckdp:1998st,Fcldn:2005st,Marti:2007st,Blhu:1995st,Sakagami:1988st,Rlaflamme:1990st,
Prokopec:2007st,Gdmoore:2007st,holman:2015st,Holmann:2008st,franco:2011st,feldman:1993st,
itzhaki:2007jj,Panda:2007ie,Baumann:2007ah,Baumann:2007np},
 \item Contribution from the self interaction between inflatons \cite{Hawking:1981ja,Starobinsky:1982ja,Alan:1981ja,Linde:1982st,Albs:1982st,Ahgsyp:1982mn,Radam:2015st,
Pade:2015st,CKDP:2009st,Dpaas:1996st,aapfmj:1994so,Lpgyv:1989st,Lpgu:1990st,ahgu:1985st,
ckdp:1998st,Fcldn:2005st,Marti:2007st,Blhu:1995st,Sakagami:1988st,Rlaflamme:1990st,
Prokopec:2007st,Gdmoore:2007st,holman:2015st,Holmann:2008st,franco:2011st,feldman:1993st,
itzhaki:2007jj,Panda:2007ie,Baumann:2007ah,Baumann:2007np}.
\end{enumerate}
More generically, such interactions with the additional environment can be expressed in FLRW background as:
\be\begin{array}{lll}
 \displaystyle  H_{int}=\int d^3x~a^3~\zeta({\bf x})~G({\bf x})=\left\{\begin{array}{ll}
                    -\displaystyle  \int d^3x~\frac{\zeta({\bf x})~G({\bf x})}{H^3\eta^3}~~~~ &
 \mbox{\small {\bf for ~dS}}  \\ 
	-\displaystyle  \int d^3x~\frac{\zeta({\bf x})~G({\bf x})}{H^3\eta^3}(1+2\epsilon)~~~~ & \mbox{\small {\bf for~ qdS}},
          \end{array}
\right.
\end{array}\ee
where $\zeta({\bf x})$ signifies the scalar curvature fluctuation and $G({\bf x})$ characterizes source function for high frequency fluctuation in real position space. Additionally it is important to note that, as the 
approximate time translational symmetry and nearly scale invariant feature is maintained in the primordial power spectrum for scalar modes the dynamical behaviour of decoherence phenomena is same in all momentum scales. 
For more details on this crucial aspect see Appendix \ref{sec6a}.
Additionally, in the present one can interpret $\zeta({\bf x})$ as the Goldstone modes that is appearing from the breaking of time translational symmetry in the de Sitter and quasi de Sitter cosmological background.
This is exactly equivalent to spontaneous symmetry breaking mechanism applicable in the context of gauge theory \cite{Cheung:2007st}.

\subsection{Creation of new massive particle}
\label{sec3b}
The classical time dependence of the inflation leads to a time dependent mass $m(\eta)$. The equation of motion for the massive field is~\footnote{In case of scalar curvature fluctuation equation of motion for inflaton field 
looks like exactly similar to the heavy field case and in that case we need to replace heavy particle mass term $m(\eta)$ with the inflaton mass term $m_{inf}$. In most of the computations one assumes that $m_{inf}<<H$. But we keep this term intact 
and will explicitly show that this contribution is necessarily required to explain the observed data for inflation. Here for scalar curvature fluctuation equation of motion for inflaton field can be written as:\bea
h''_k + \left\{c^2_{S}k^2 + \left(\frac{m^2_{inf}}{H^2} - 2\right) \frac{1}{\eta^2} \right\} h_k &=& 0~~~~~~~{\bf for~ dS}\\ 
h''_k + \left\{c^2_{S}k^2 + \left(\frac{m^2_{inf}}{H^2} - \left[\nu^2-\frac{1}{4}\right]
\right) \frac{1}{\eta^2} \right\} h_k &=& 0~~~~~~~{\bf for~ qdS}.
\eea
The most general solution of the mode function for de Sitter and quasi de Sitter case can be written as:
\be\begin{array}{lll}\label{yu2xxxx}
 \displaystyle h_k (\eta) =\left\{\begin{array}{ll}
                    \displaystyle   \sqrt{-\eta}\left[C_1  H^{(1)}_{\sqrt{\frac{9}{4}-\frac{m^2_{inf}}{H^2}}} \left(-kc_{S}\eta\right) 
+ C_2 H^{(2)}_{\sqrt{\frac{9}{4}-\frac{m^2_{inf}}{H^2}}} \left(-kc_{S}\eta\right)\right]~~~~ &
 \mbox{\small {\bf for ~dS}}  \\ 
	\displaystyle \sqrt{-\eta}\left[C_1  H^{(1)}_{\sqrt{\nu^2-\frac{m^2_{inf}}{H^2}}} \left(-kc_{S}\eta\right) 
+ C_2 H^{(2)}_{\sqrt{\nu^2-\frac{m^2_{inf}}{H^2}}} \left(-kc_{S}\eta\right)\right]~~~~ & \mbox{\small {\bf for~ qdS}}.
          \end{array}
\right.
\end{array}\ee
Here $C_{1}$ and $C_{2}$ are the arbitrary integration constants and the numerical value depend on the choice of the 
initial condition or more precisely the vacuum.}:
\bea
h''_k + \left\{c^2_{S}k^2 + \left(\frac{m^2}{H^2} - 2\right) \frac{1}{\eta^2} \right\} h_k &=& 0~~~~~~~{\bf for~ dS}\\ 
h''_k + \left\{c^2_{S}k^2 + \left(\frac{m^2}{H^2} - \left[\nu^2-\frac{1}{4}\right]
\right) \frac{1}{\eta^2} \right\} h_k &=& 0~~~~~~~{\bf for~ qdS}.
\eea
where in the quasi de Sitter case the parameter $\nu$ can be written as:
\be\label{we} \nu= \frac{3}{2}+\epsilon+\frac{\eta}{2}+\frac{s}{2},\ee
where $\epsilon$ and $\eta$ are the Hubble slow-roll parameter defined as:
\bea \label{rt1}\epsilon &=& -\frac{\dot{H}}{H^2},\\
\eta \label{rt2}&=& \frac{\dot{\epsilon}}{H\epsilon},\\
s\label{rt3}&=&\frac{\dot{c_{S}}}{Hc_{S}}.\eea
In the slow-roll regime of inflation $\epsilon<<1$ and $|\eta|<<1$ and at the end of inflation 
sow-roll condition breaks when any of the criteria satisfy, (1) $\epsilon=1$ or $|\eta|=1$, (2) $\epsilon=1=|\eta|$. 

The most general solution of the mode function for de Sitter and quasi de Sitter case can be written as:
\be\begin{array}{lll}\label{yu2}
 \displaystyle h_k (\eta) =\left\{\begin{array}{ll}
                    \displaystyle   \sqrt{-\eta}\left[C_1  H^{(1)}_{\sqrt{\frac{9}{4}-\frac{m^2}{H^2}}} \left(-kc_{S}\eta\right) 
+ C_2 H^{(2)}_{\sqrt{\frac{9}{4}-\frac{m^2}{H^2}}} \left(-kc_{S}\eta\right)\right]~~~~ &
 \mbox{\small {\bf for ~dS}}  \\ 
	\displaystyle \sqrt{-\eta}\left[C_1  H^{(1)}_{\sqrt{\nu^2-\frac{m^2}{H^2}}} \left(-kc_{S}\eta\right) 
+ C_2 H^{(2)}_{\sqrt{\nu^2-\frac{m^2}{H^2}}} \left(-kc_{S}\eta\right)\right]~~~~ & \mbox{\small {\bf for~ qdS}}.
          \end{array}
\right.
\end{array}\ee
Here $C_{1}$ and $C_{2}$ are the arbitrary integration constants and the numerical value depend on the choice of the 
initial condition or more precisely the vacuum. In the present context apart from the arbitrary vacuum we consider the following 
choice of the vacuum for the computation:
\begin{enumerate}
 \item {\bf Bunch Davies vacuum:} In this case we choose $C_{1}=\sqrt{\frac{\pi}{2}}$ and $C_{2}=0$.
 \item {\bf $\alpha$ vacuum Type-I:} In this case we choose $C_{1}=\cosh\alpha$ and $C_{2}=e^{i\delta}\sinh\alpha$. 
 Here $\delta$ is a phase factor.
 \item {\bf $\alpha$ vacuum Type-II:} In this case we choose $C_{1}=N_{\alpha}$ and $C_{2}=N_{\alpha}~e^{\alpha}$. 
 Here $N_{\alpha}=\frac{1}{\sqrt{1-e^{\alpha+\alpha^{*}}}}$.
 \item {\bf Special vacuum:} In this case we choose $C_{1}=C_{2}=C$.
\end{enumerate}

Here it is important to mention that the argument in the Hankel function for the solution of the $h_k$ takes the following 
values in different regime~\footnote{In case of inflaton field, argument of the Hankel function involves the inflaton mass $m_{inf}$ term as given by: \be\begin{array}{lll}\label{infmassgo}\small
 \displaystyle \underline{\rm \bf For~dS:}~~~~~~~~~\sqrt{\frac{9}{4}-\frac{m^2_{inf}}{H^2}} \approx\left\{\begin{array}{ll}
                    \displaystyle  \frac{\sqrt{5}}{2}~~~~ &
 \mbox{\small {\bf for ~$m_{inf}\approx H$}}  \\ 
	\displaystyle \frac{3}{2}~~~~ & \mbox{\small {\bf for ~$m_{inf}<< H$}}.
          \end{array}
\right.
\end{array}\ee
\be\begin{array}{lll}\label{infmassgo2}\small
 \displaystyle \underline{\rm \bf For~qdS:}~~~~~~~~~\sqrt{\nu^2-\frac{m^2_{inf}}{H^2}} \approx\left\{\begin{array}{ll}
                    \displaystyle  \sqrt{\nu^2-1}~~~~ &
 \mbox{\small {\bf for ~$m_{inf}\approx H$}}  \\ 
	\displaystyle \nu~~~~ & \mbox{\small {\bf for ~$m_{inf}<< H$}}.
          \end{array}
\right.
\end{array}\ee
Here inflation mass is always $m_{inf}<< H$ or $m_{inf}\approx H$, as the mass scale of the inflaton cannot be larger the scale of inflation itself.}:
\be\begin{array}{lll}\label{yu2dfvq}\small
 \displaystyle \underline{\rm \bf For~dS:}~~~~~~~~~\sqrt{\frac{9}{4}-\frac{m^2}{H^2}} \approx\left\{\begin{array}{ll}
                    \displaystyle  \frac{\sqrt{5}}{2}~~~~ &
 \mbox{\small {\bf for ~$m\approx H$}}  \\ 
	\displaystyle \frac{3}{2}~~~~ & \mbox{\small {\bf for ~$m<< H$}}\\ 
	\displaystyle i\sqrt{\Upsilon^2-\frac{9}{4}}~~~~ & \mbox{\small {\bf for ~$m>> H$}}.
          \end{array}
\right.
\end{array}\ee
\be\begin{array}{lll}\label{yu2dfve}\small
 \displaystyle \underline{\rm \bf For~qdS:}~~~~~~~~~\sqrt{\nu^2-\frac{m^2}{H^2}} \approx\left\{\begin{array}{ll}
                    \displaystyle  \sqrt{\nu^2-1}~~~~ &
 \mbox{\small {\bf for ~$m\approx H$}}  \\ 
	\displaystyle \nu~~~~ & \mbox{\small {\bf for ~$m<< H$}}\\ 
	\displaystyle i\sqrt{\Upsilon^2-\nu^2}~~~~ & \mbox{\small {\bf for ~$m>> H$}}.
          \end{array}
\right.
\end{array}\ee
Here we set $m=\Upsilon H$, where the parameter $\Upsilon>>1$ for $m>>H$ case.
In the present context we are interested in the following cases for both de Sitter and quasi de Sitter solution which we will follow throughout the rest of the discussion in this paper:
\begin{enumerate}
 \item \underline{\textcolor{violet}{\bf Case I:}}~$m\approx H$, in which we treat the mass scale of the heavy fields is comparable with the inflationary scale. This is a special case where we treat $m/H$ is a constant 
 parameter for the sake of simplicity. In this case the particle production of heavy fields deal with non-local effects. 
 But only changing the structure of effective Lagrangian it is not at all possible to explain the characteristic of non-local effects in the present context.
 \item \underline{\textcolor{violet}{\bf Case II:}}~$m >>H$, in which we treat the mass scale of the heavy fields is much higher compared to the
 the inflationary scale. This is another special case where we treat $m/H$ is a constant 
 parameter for the sake of simplicity. In this case one can interpret that such heavy fields belongs
 to the hidden sector. In this case we can integrate them from the theory and finally they generate an effective field theory of light inflaton fields.
 As we don't know anything about the UV complete theory of inflation it is not possible
 to detect all such heavy contributions.
 \item \underline{\textcolor{violet}{\bf Case III:}}~$m<<H$, in which we treat the mass scale of the heavy fields is much smaller 
 compared to the inflationary scale. In this case one can neglect the contributions from all such fields in the mode
 equation for scalar fluctuations. This situation is exactly similar to the inflationary framework as the mode function for the scalar fluctuation are 
 exactly same and in such a physical situation these extra dynamical fields serves the purpose of inflaton. One can interpret 
 this situation by using two field scenario or inflaton-curvaton scenario in the present context. Here it is important to mention that this specific scenario does not give rise to the violation of cosmological Bell's 
 inequality. We have quoted the results for completeness, which gives the information about the particle production during inflation, where the effect of the heavy particle mass is negligibly small compared to the scale of inflation or background cosmological Hubble scale.
 
 \item \underline{\textcolor{violet}{\bf Case IV:}}~\\We also take the following phenomenological cases for the conformal
 time dependent parametrization on mass parameter:\\
 
 \textcolor{red}{\bf A.} $\displaystyle m= \sqrt{\gamma\left(\frac{\eta}{\eta_0} - 1\right)^2 + \delta}~H$, 
 where $\gamma$, $\delta$ and $\eta_{0}$ are fixed parameters of the model. This is very special model
 using which one can explicitly study the specific amount and significant signatures of Bell
 violation in primordial cosmological setup. In ref.~\cite{juan:2015ja} it is first proposed
 to study the Baroque model of the universe to study the violation of cosmological Bell inequalities.\\
   
 \textcolor{red}{\bf B.} $\displaystyle m= \frac{m_0}{\sqrt{2}}\sqrt{\left[1
 -\tanh\left(\frac{\rho}{H}\ln(-H\eta)\right)\right]}$, where $\rho$ and $m_{0}$ are fixed parameters of the model. 
 This is a model for the heavy particle mass which was earlier used to study the phenomena of 
 quantum critical quench and thermalization in the context of Conformal Field Theory (CFT). In case of quantum quench $m_{0}$ is known as 
 quench parameter. See ref \cite{Mandal:2015kxi,Das:2016lla,Das:2014hqa,Das:2014jna,Das:2014lda,Basu:2013soa,Basu:2012gg,Mandal:2013id} for more details in this direction. 
 In this context we are 
 interested in this specific type of mass parametrization as the corresponding equivalent version of 
 Schr$\ddot{o}$dinger quantum mechanical problem can easily solvable. Here this can be treated as another 
 model to explain the parametrization of heavy particle mass parameter.\\
 
 \textcolor{red}{\bf C.} $\displaystyle 
 m= m_0~{\rm sech}\left(\frac{\rho}{H}\ln(-H\eta)\right) $, 
 where $\rho$ and $m_{0}$ are fixed parameters of the model. This is another model for the heavy particle mass which was also earlier used to study the phenomena of 
 quantum critical quench and thermalization in the context of Conformal Field Theory (CFT). As mentioned earlier in case of quantum quench $m_{0}$ is known as 
 quench parameter. See ref \cite{Mandal:2015kxi,Das:2016lla,Das:2014hqa,Das:2014jna,Das:2014lda,Basu:2013soa,Basu:2012gg,Mandal:2013id} for more details in this direction. Here this can be treated as another 
 model to explain the parametrization of heavy particle mass parameter. In the last part of this paper we have shown that the axion decay constant profile in string theory is exactly 
 mimics the same behavior as presented in this context.
\end{enumerate}

\begin{figure*}[htb]
\centering
\subfigure[Heavy field mass profile for A with $\gamma=1$ and $\delta=1$.]{
    \includegraphics[width=7.2cm,height=8cm] {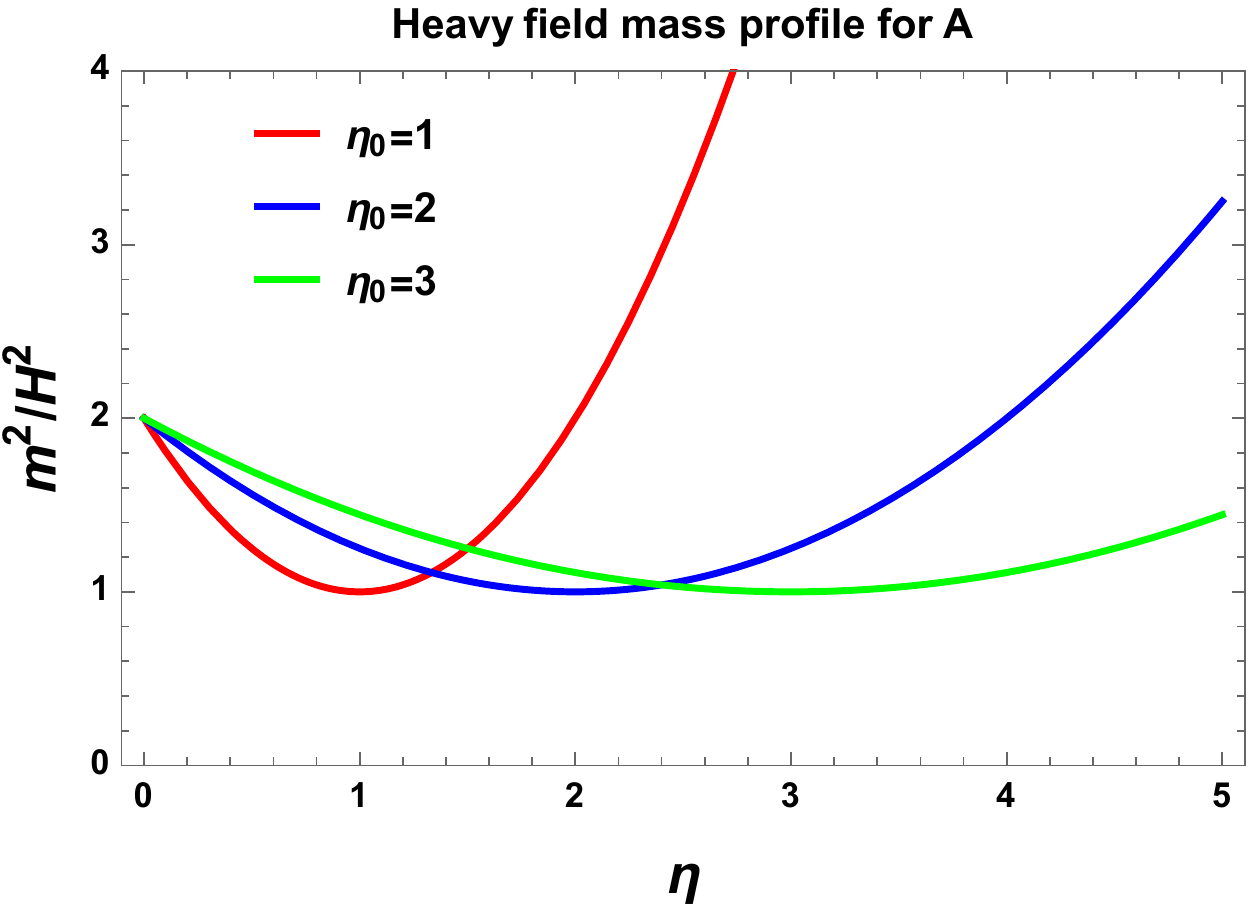}
    \label{fig1a}
}
\subfigure[Heavy field mass profile for B with $\frac{m^{2}_{0}}{2H^2}=1$.]{
    \includegraphics[width=7.2cm,height=8cm] {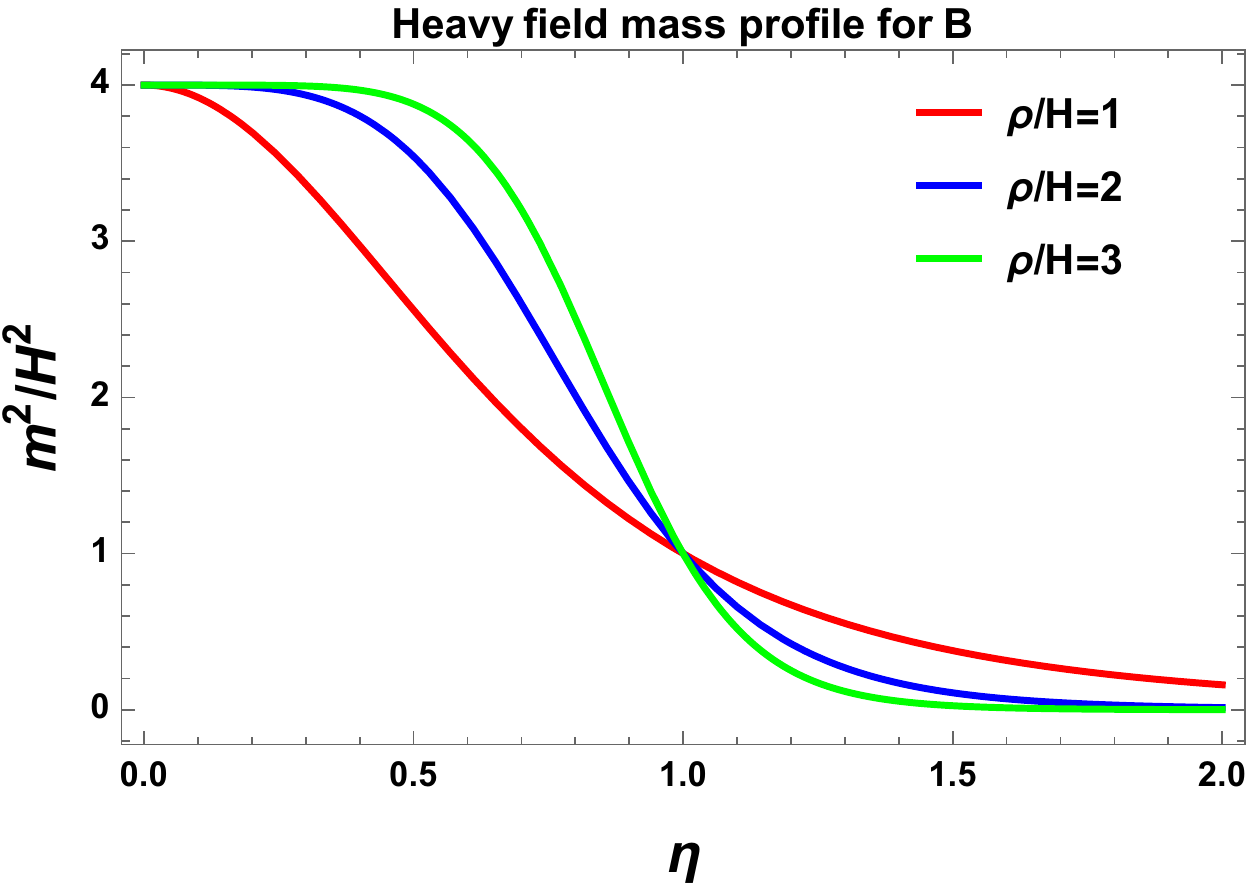}
    \label{fig2b}
}
\subfigure[Heavy field mass profile for C with $\frac{m^{2}_{0}}{H^2}=1$.]{
    \includegraphics[width=10.2cm,height=8cm] {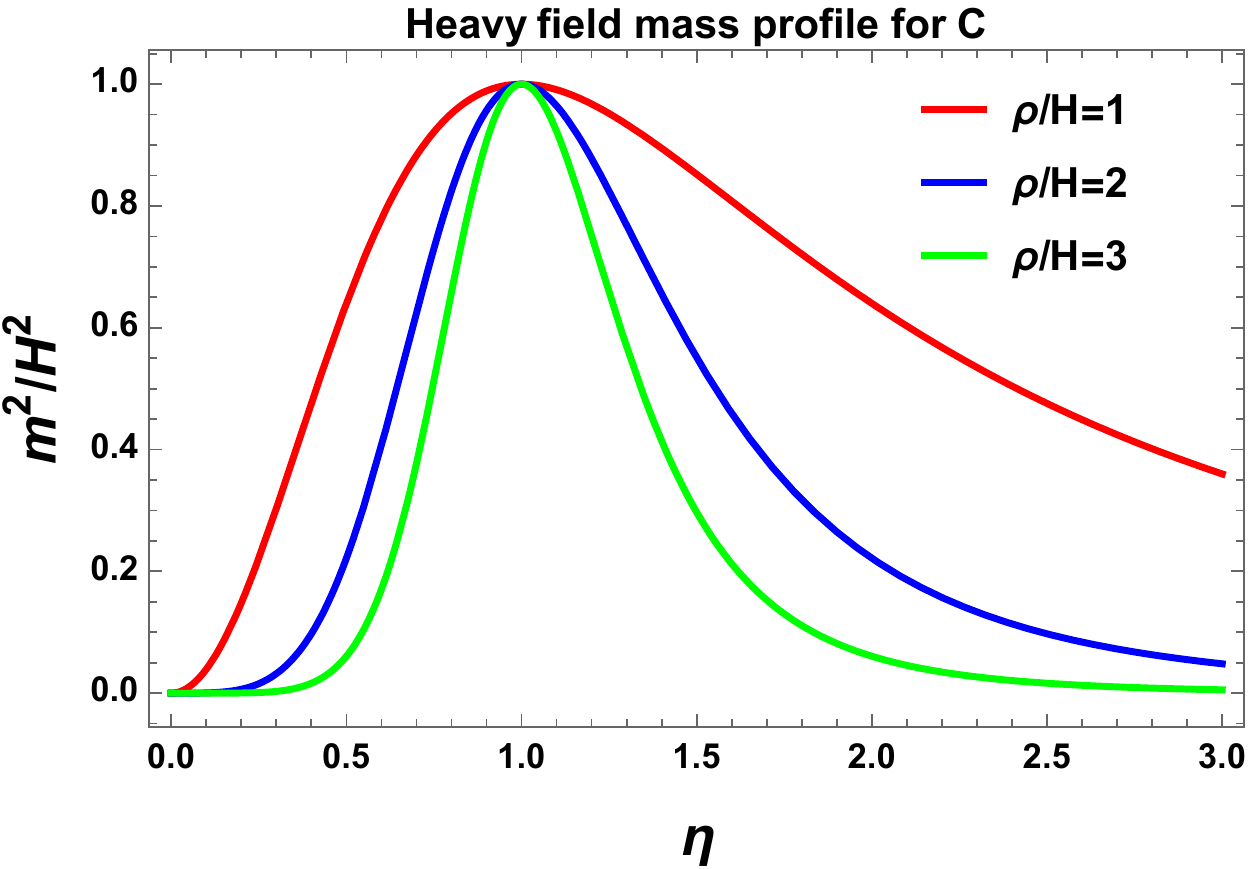}
    \label{fig3c}
}
\caption[Optional caption for list of figures]{Behaviour of the heavy field mass profile with $\eta$.} 
\label{fza}
\end{figure*}

\begin{figure*}[htb]
\centering
\subfigure[Heavy field mass profile for A with $\gamma=1$ and $\delta=1$.]{
    \includegraphics[width=7.2cm,height=7cm] {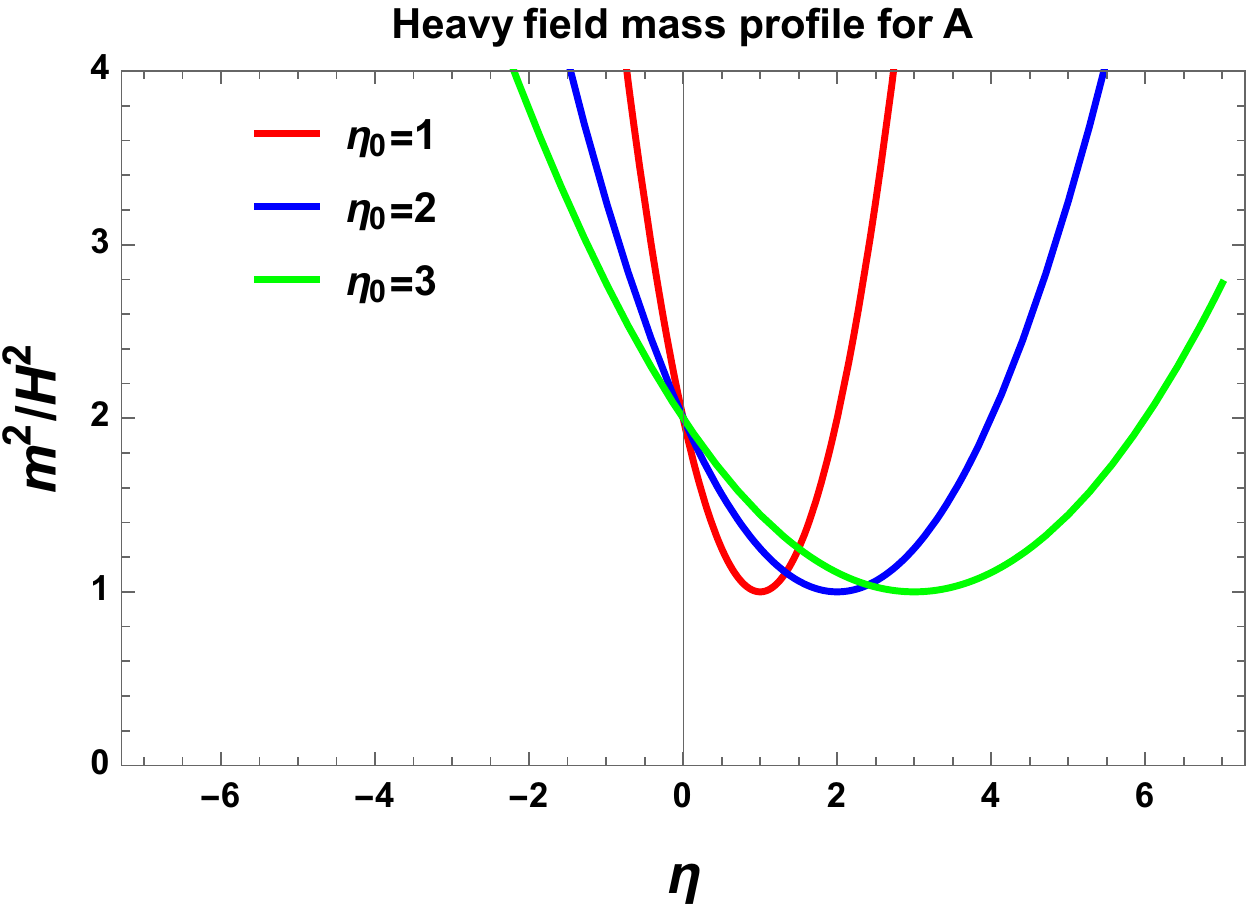}
    \label{fig1a}
}
\subfigure[Heavy field mass profile for B with $\frac{m^{2}_{0}}{2H^2}=1$.]{
    \includegraphics[width=7.2cm,height=7cm] {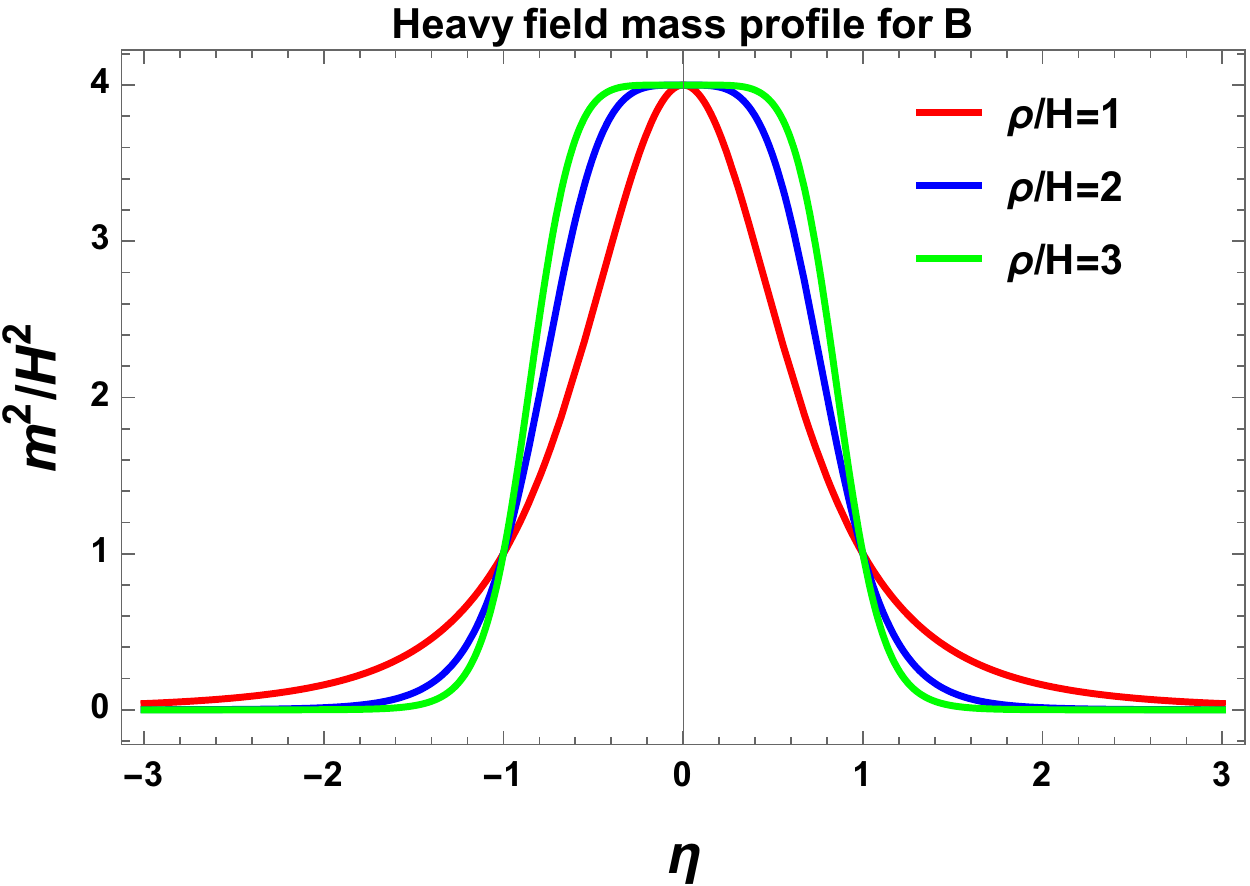}
    \label{fig2b}
}
\subfigure[Heavy field mass profile for C with $\frac{m^{2}_{0}}{H^2}=1$.]{
    \includegraphics[width=10.2cm,height=7cm] {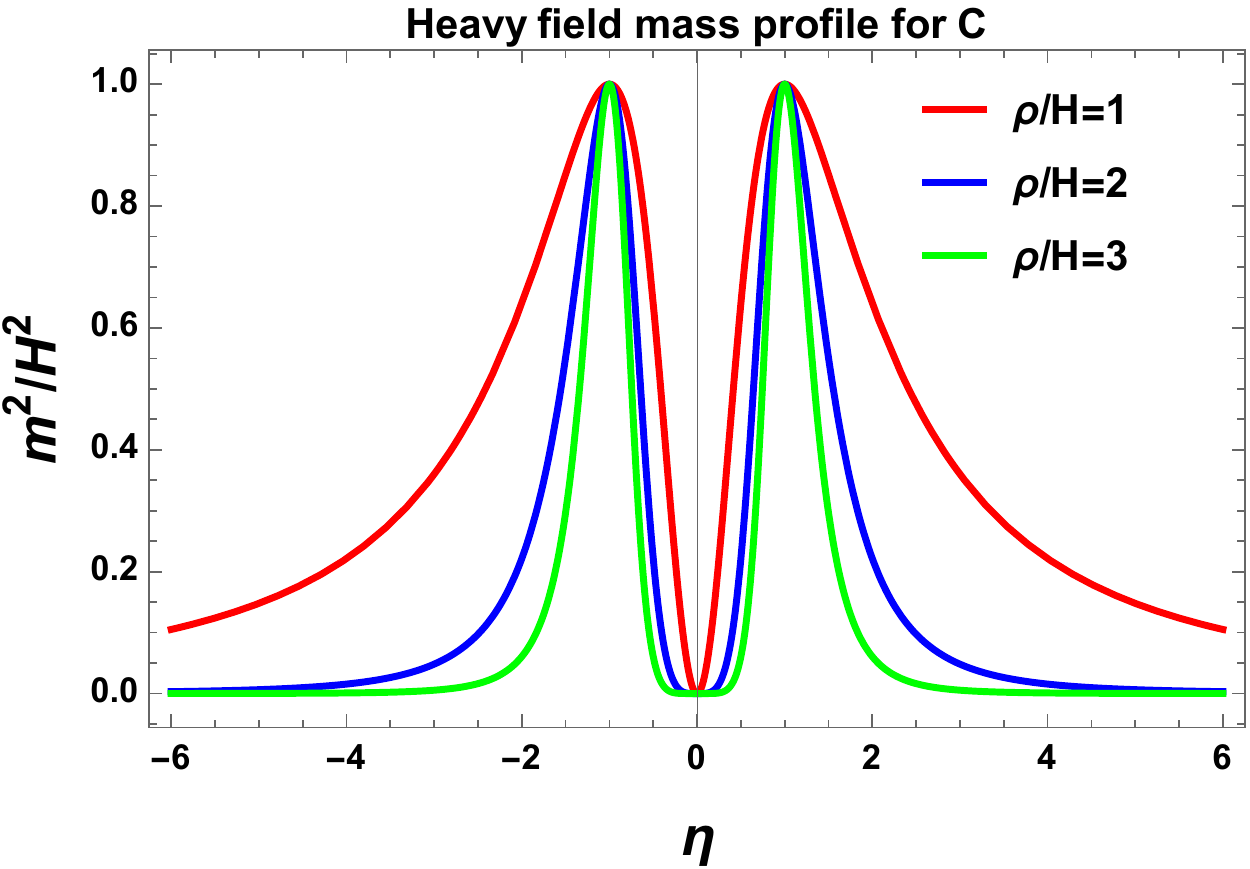}
    \label{fig3c}
}
\caption[Optional caption for list of figures]{Behaviour of the heavy field mass profile for all range of $\eta$.} 
\label{fza}
\end{figure*}
 For the most general solution as stated in Eq~(\ref{yu2}) one can consider the following limiting physical 
 situations:
 \begin{enumerate}
  \item \underline{\textcolor{blue}{\bf Superhorizon regime:}} $|kc_{S}\eta|<<1$ or equivalently  $|kc_{S}\eta|\rightarrow 0$,
  \item \underline{\textcolor{blue}{\bf Horizon crossing:}} $|kc_{S}\eta|= 1$ or equivalently  $|kc_{S}\eta|\approx 1-\Delta$ with $\Delta\rightarrow 0$,
  \item \underline{\textcolor{blue}{\bf Subhorizon regime:}} $|kc_{S}\eta|>>1$ or equivalently $|kc_{S}\eta|\rightarrow -\infty$.
 \end{enumerate}
 Consequently for the arbitrary choice of the initial condition or vacuum we get the following results:
\begin{eqnarray}
\label{sol1c1}\lim_{kc_{S}\eta\rightarrow -\infty}H^{(1,2)}_{\Lambda}(-kc_{S}\eta)&=&\pm
\sqrt{\frac{2}{\pi}}\frac{1}{\sqrt{-kc_{S}\eta}}e^{ \mp ikc_{S}\eta}
e^{\mp\frac{i\pi}{2}\left(\Lambda+\frac{1}{2}\right)},\\
\label{sol1cc}\lim_{kc_{S}\eta\rightarrow 0}H^{(1,2)}_{\Lambda}(-kc_{S}\eta)&=&\pm
\frac{i}{\pi}\Gamma(\Lambda)\left(-\frac{kc_{S}\eta}{2}\right)^{-\Lambda}.
\end{eqnarray}
where the parameter $\Lambda$ is defined as~\footnote{For inflaton field we have:\be\begin{array}{lll}\label{lalaas}\small
 \displaystyle \Lambda=\left\{\begin{array}{ll}
                    \displaystyle \sqrt{\frac{9}{4}-\frac{m^2_{inf}}{H^2}} ~~~~ &
 \mbox{\small {\bf for ~dS}}  \\ 
	\displaystyle \sqrt{\nu^2-\frac{m^2_{inf}}{H^2}} ~~~~ & \mbox{\small {\bf for~ qdS}}.
          \end{array}
\right.
\end{array}\ee}:
 \be\begin{array}{lll}\label{yu2dfvqdfdfdf10}\small
 \displaystyle \Lambda=\left\{\begin{array}{ll}
                    \displaystyle \sqrt{\frac{9}{4}-\frac{m^2}{H^2}} ~~~~ &
 \mbox{\small {\bf for ~dS}}  \\ 
	\displaystyle \sqrt{\nu^2-\frac{m^2}{H^2}} ~~~~ & \mbox{\small {\bf for~ qdS}}.
          \end{array}
\right.
\end{array}\ee
One can also consider the following approximations to simplify the final derived form of the solution for arbitrary vacuum with $|kc_{S}\eta|= 1$ or equivalently  $|kc_{S}\eta|\approx 1-\Delta$ with $\Delta\rightarrow 0$ case:
\begin{enumerate}
 \item We start with the {\it Laurent expansion} of the Gamma function:
       \bea \label{r1zax} \Gamma(\Lambda) &=& \frac{1}{\Lambda}-\gamma+\frac{1}{2}\left(\gamma^2+\frac{\pi^2}{6}\right)\Lambda-\frac{1}{6}\left(\gamma^3+\frac{\gamma \pi^2}{2}+2\zeta(3)\right)\Lambda^2 +{\cal O}(\Lambda^3)
       \nonumber\\
      &=&\footnotesize\left\{\begin{array}{lll}
                    \displaystyle  
                   \frac{1}{\left(\sqrt{\frac{9}{4}-\frac{m^2}{H^2}}\right)}-\gamma+\frac{1}{2}
                   \left(\gamma^2+\frac{\pi^2}{6}\right)\left(\sqrt{\frac{9}{4}-\frac{m^2}{H^2}}\right)
                   -\frac{1}{6}\left(\gamma^3+\frac{\gamma \pi^2}{2}+2\zeta(3)\right)
                   \left(\sqrt{\frac{9}{4}-\frac{m^2}{H^2}}\right)^2 +\cdots\,,~~~~ &
 \mbox{\small {\bf for~dS}}  \\
 \displaystyle  
  \frac{1}{ \left\{\sqrt{\nu^2-\frac{m^2}{H^2}}\right\}}
  -\gamma 
  +\frac{1}{2}\left(\gamma^2+\frac{\pi^2}{6}\right) \left\{\sqrt{\nu^2-\frac{m^2}{H^2}}\right\}
  -\frac{1}{6}\left(\gamma^3+\frac{\gamma \pi^2}{2}+2\zeta(3)\right) \left\{\sqrt{\nu^2-\frac{m^2}{H^2}}\right\}^2
  +\cdots \,,~~~ &
 \mbox{\small {\bf for ~qdS}}.
          \end{array}
\right.
\eea
where $\gamma$ is known as the Euler Mascheroni constant and $\zeta(3)$ characterizing
the Reimann zeta function of 
       order $3$ originating in the expansion of the gamma function. 
\item In this case the solution Hankel functions of first and second kind can re-expressed into the following simplified form  as:
\begin{eqnarray}
\label{sol1cccc}\lim_{|kc_{S}\eta|\approx 1-\Delta(\rightarrow 0)}H^{(1,2)}_{\Lambda}(-kc_{S}\eta)&=&\pm
\frac{i}{\pi}\left[ \frac{1}{\Lambda}-\gamma+\frac{1}{2}\left(\gamma^2+\frac{\pi^2}{6}\right)\Lambda\right.\\ &&\nonumber\left.
~~~~~~~~~~~~~~-\frac{1}{6}\left(\gamma^3+\frac{\gamma \pi^2}{2}+2\zeta(3)\right)\Lambda^2 +\cdots\right]\left(\frac{1+\Delta}{2}\right)^{-\Lambda}.
\end{eqnarray}
\end{enumerate}
After taking $kc_{S}\eta\rightarrow -\infty$, $kc_{S}\eta\rightarrow 0$ and $|kc_{S}\eta|\approx 1-\Delta(\rightarrow 0)$ limit the most general 
solution as stated in Eq~(\ref{yu2}) can be recast as:
\bea\label{yu2}
 \displaystyle h_k (\eta) &\stackrel{|kc_{S}\eta|\rightarrow-\infty}{=}&
 \sqrt{\frac{2}{\pi kc_{S}}}\left[C_1  e^{- ikc_{S}\eta}
e^{-\frac{i\pi}{2}\left(\Lambda+\frac{1}{2}\right)} 
+ C_2 e^{ ikc_{S}\eta}
e^{\frac{i\pi}{2}\left(\Lambda+\frac{1}{2}\right)}\right]\\ &=&\left\{\begin{array}{ll}
                    \displaystyle   \sqrt{\frac{2}{\pi kc_{S}}}\left[C_1  e^{ -ikc_{S}\eta}
e^{-\frac{i\pi}{2}\left(\sqrt{\frac{9}{4}-\frac{m^2}{H^2}}+\frac{1}{2}\right)} 
+ C_2 e^{ ikc_{S}\eta}
e^{\frac{i\pi}{2}\left(\sqrt{\frac{9}{4}-\frac{m^2}{H^2}}+\frac{1}{2}\right)}\right]~~~~ &
 \mbox{\small {\bf for ~dS}}  \nonumber\\ 
	\displaystyle \sqrt{\frac{2}{\pi kc_{S}}}\left[C_1  e^{ -ikc_{S}\eta}
e^{-\frac{i\pi}{2}\left(\sqrt{\nu^2-\frac{m^2}{H^2}}+\frac{1}{2}\right)} 
+ C_2 e^{ ikc_{S}\eta}
e^{\frac{i\pi}{2}\left(\sqrt{\nu^2-\frac{m^2}{H^2}}+\frac{1}{2}\right)}\right]~~~~ & \mbox{\small {\bf for~ qdS}}.
          \end{array}
\right.\nonumber\eea 
\bea
\label{yu2}
 \displaystyle h_k (\eta) &\stackrel{|kc_{S}\eta|\rightarrow 0}{=}&\frac{i}{\pi}\sqrt{-\eta}\Gamma(\Lambda)\left(-\frac{kc_{S}\eta}{2}\right)^{-\Lambda}\left[C_1   
- C_2 \right]\\ &=&\left\{\begin{array}{ll}
                    \displaystyle  \frac{i\sqrt{-\eta}}{\pi}\Gamma\left(\sqrt{\frac{9}{4}-\frac{m^2}{H^2}}\right)
                    \left(-\frac{kc_{S}\eta}{2}\right)^{-\sqrt{\frac{9}{4}-\frac{m^2}{H^2}}}\left[C_1   
- C_2 \right]~ &
 \mbox{\small {\bf for ~dS}} \nonumber \\ 
	\displaystyle\frac{i\sqrt{-\eta}}{\pi}\Gamma\left(\sqrt{\nu^2-\frac{m^2}{H^2}}\right)\left(-\frac{kc_{S}\eta}{2}\right)^{-\sqrt{\nu^2-\frac{m^2}{H^2}}}\left[C_1   
- C_2 \right]~ & \mbox{\small {\bf for~ qdS}}.
          \end{array}
\right.
\eea
\bea
\label{yu2}
 \displaystyle h_k (\eta) &\stackrel{|kc_{S}\eta|\approx 1-\Delta(\rightarrow 0)}{=}&\frac{i}{\pi}\sqrt{-\eta}\left[ \frac{1}{\Lambda}-\gamma+\frac{1}{2}\left(\gamma^2+\frac{\pi^2}{6}\right)\Lambda\right.\\ &&\nonumber\left.
-\frac{1}{6}\left(\gamma^3+\frac{\gamma \pi^2}{2}+2\zeta(3)\right)\Lambda^2 +\cdots\right]\left(\frac{1+\Delta}{2}\right)^{-\Lambda}\left[C_1   
- C_2 \right]\\ &=&\left\{\begin{array}{ll}
                    \displaystyle  \frac{i}{\pi}\sqrt{-\eta}\left[ \frac{1}{\left(\sqrt{\frac{9}{4}-\frac{m^2}{H^2}}\right)}-\gamma+\frac{1}{2}
                   \left(\gamma^2+\frac{\pi^2}{6}\right)\left(\sqrt{\frac{9}{4}-\frac{m^2}{H^2}}\right)\right.\\ \displaystyle \left.
                   \displaystyle~~~~~~~~-\frac{1}{6}\left(\gamma^3+\frac{\gamma \pi^2}{2}+2\zeta(3)\right)
                   \left(\sqrt{\frac{9}{4}-\frac{m^2}{H^2}}\right)^2 +\cdots\right]\\ \displaystyle ~~~~~~~~~~~~\times\left(\frac{1+\Delta}{2}\right)^{-\sqrt{\frac{9}{4}-\frac{m^2}{H^2}}}\left[C_1   
- C_2 \right]~&
 \mbox{\small {\bf for ~dS}} \nonumber \\ 
	\displaystyle\frac{i}{\pi}\sqrt{-\eta}\left[ \frac{1}{ \left\{\sqrt{\nu^2-\frac{m^2}{H^2}}\right\}}
  -\gamma  \displaystyle +\frac{1}{2}\left(\gamma^2+\frac{\pi^2}{6}\right) \left\{\sqrt{\nu^2-\frac{m^2}{H^2}}\right\}
  \right.\\ \displaystyle \left. \displaystyle~-\frac{1}{6}\left(\gamma^3+\frac{\gamma \pi^2}{2}+2\zeta(3)\right) \left\{\sqrt{\nu^2-\frac{m^2}{H^2}}\right\}^2
  +\cdots\right]\\ \displaystyle ~~~~~~~~~~~~\times\left(\frac{1+\Delta}{2}\right)^{-\sqrt{\nu^2-\frac{m^2}{H^2}}}\left[C_1   
- C_2 \right]~ & \mbox{\small {\bf for~ qdS}}.
          \end{array}
\right.
\eea
In the next subsections we use all these limiting results for the previously mentioned cases- (1) $m\approx H$, (2) $m>>H$, (3) $m<<H$.
 Here we can think of a physical 
condition where the WKB approximation is valid (approximately) for the solution for the 
mode function $h_k$. Here we provide the solution for the fluctuations by exactly solving the equation of motion for the heavy fields, where we assume that time variation in heavy field mass parameter is very slow. For arbitrary time dependence case 
it is only possible depending on the complexity of the mathematical structure of the heavy field mass parameter $m(\eta)$.
In the standard WKB approximation the total solution can be recast in the following form:
\bea\label{df3}
h_k (\eta)&=& \left[D_{1}u_{k}(\eta) + D_{2} \bar{u}_{k}(\eta)\right],\eea
where $D_{1}$ and and $D_{2}$ are two arbitrary integration constants, which depend on the 
choice of the initial condition during WKB approximation at early and late time scale. In our discussion two arbitrary integration constants $D_{1}$ and and $D_{2}$ can be identified with the 
Bogoliubov co-efficient in momentum space~\footnote{Here one can chose another convention for the Bogoliubov co-efficient in momentum space as given by:\bea D_{1} &=& \alpha(k),\\
     D_{2} &=& \beta(k).\eea But for our computation we will follow other convention for the Bogoliubov co-efficient in momentum space stated in Eq~(\ref{opopi}).}:
\bea\label{opopi} D_{1} &=& \beta(k),\\
     D_{2} &=& \alpha(k).\eea
In the present context $u_{k}(\eta)$ and $\bar{u}_{k}(\eta)$ are defined as:
\bea\label{solpzxzx}
 \displaystyle\small u_{k}(\eta) &=&
                    \displaystyle   \frac{1}{\sqrt{2p(\eta)}}
\exp\left[i\int^{\eta} d\eta^{\prime} p(\eta^{'})\right]\\
\label{solazxzx}
 \displaystyle\small \bar{u}_{k}(\eta) &=&
                    \displaystyle   \frac{1}{\sqrt{2p(\eta)}}
\exp\left[-i\int^{\eta} d\eta^{\prime} p(\eta^{'})\right]
\eea
where we have written the total solution for the mode $h_k$ in terms of 
two linearly independent solutions. Here in the most generalized situation
the new conformal time dependent factor $p(\eta)$ is defined as:
\be\begin{array}{lll}\label{soladfasx2}
 \displaystyle p(\eta) =\footnotesize\left\{\begin{array}{ll}
                    \displaystyle   \sqrt{\left\{c^{2}_{S}k^2 + \left(\frac{m^2}{H^2} - 2\right) \frac{1}{\eta^2} \right\}}~~~~ &
 \mbox{\small {\bf for ~dS}}  \\ 
	\displaystyle  \sqrt{\left\{c^{2}_{S}k^2 + \left(\frac{m^2}{H^2} - \left[\nu^2-\frac{1}{4}\right]
\right) \frac{1}{\eta^2} \right\}}~~~~ & \mbox{\small {\bf for~ qdS}}.
          \end{array}
\right.
\end{array}\ee
which we use thoroughly in our computation. Here it is important to mention the expressions for the controlling 
factor $p(\eta)$ in different regime of solution:
\be\begin{array}{lll}\label{yu2dfvqasx3}\small
 \displaystyle \underline{\rm \bf For~dS:}~~~~~~~~~p(\eta) \approx\footnotesize\left\{\begin{array}{ll}
                    \displaystyle  \sqrt{\left\{c^{2}_{S}k^2 - \frac{1}{\eta^2} \right\}}~~~~ &
 \mbox{\small {\bf for ~$m\approx H$}}  \\ 
	\displaystyle \sqrt{\left\{c^{2}_{S}k^2 - \frac{2}{\eta^2} \right\}}~~~~ & \mbox{\small {\bf for ~$m<< H$}}\\ 
	\displaystyle \sqrt{\left\{c^{2}_{S}k^2 + \left(\Upsilon^2-2\right)\frac{1}{\eta^2} \right\}}~~~~
	& \mbox{\small {\bf for ~$m>> H$}}\\ 
	\displaystyle \sqrt{\left\{c^{2}_{S}k^2 + \left(\gamma\left(\frac{\eta}{\eta_0} - 1\right)^2 +\delta - 2\right) \frac{1}{\eta^2} \right\}}~~~~
	& \mbox{\small {\bf for ~$m\approx \sqrt{\gamma\left(\frac{\eta}{\eta_0} - 1\right)^2 + \delta}~H$}}\\ 
	\displaystyle \sqrt{\left\{c^{2}_{S}k^2 + \left(\frac{m^2_0}{2H^2}\left[1
 -\tanh\left(\frac{\rho}{H}\ln(-H\eta)\right)\right] - 2\right) \frac{1}{\eta^2} \right\}}~~~~
	& \mbox{\small {\bf for ~$\displaystyle m= \frac{m_0}{\sqrt{2}}\sqrt{\left[1
 -\tanh\left(\frac{\rho}{H}\ln(-H\eta)\right)\right]}$}}\\ 
	\displaystyle \sqrt{\left\{c^{2}_{S}k^2 + \left(\frac{m^2_0}{H^2}{\rm sech}^2\left(\frac{\rho}{H}\ln(-H\eta)\right) - 2\right) \frac{1}{\eta^2} \right\}}~~~~
	& \mbox{\small {\bf for ~$\displaystyle 
 m= m_0~{\rm sech}\left(\frac{\rho}{H}\ln(-H\eta)\right) $}}.
          \end{array}
\right.
\end{array}\ee
\\
\be\begin{array}{lll}\label{yu2dfveasx4}\small
 \displaystyle \underline{\rm \bf For~qdS:}~~~p(\eta) \approx\footnotesize\left\{\begin{array}{ll}
                    \displaystyle  \sqrt{\left\{c^{2}_{S}k^2 - \left[\nu^2-\frac{5}{4}\right]
\frac{1}{\eta^2} \right\}} &
 \mbox{\small {\bf for ~$m\approx H$}}  \\ 
	\displaystyle \sqrt{\left\{c^{2}_{S}k^2 - \left[\nu^2-\frac{1}{4}\right]
\frac{1}{\eta^2} \right\}} & \mbox{\small {\bf for ~$m<< H$}}\\ 
	\displaystyle \sqrt{\left\{c^{2}_{S}k^2 + \left(\Upsilon^2 - \left[\nu^2-\frac{1}{4}\right]
\right) \frac{1}{\eta^2} \right\}} & \mbox{\small {\bf for ~$m>> H$}}\\ 
	\displaystyle \sqrt{\left\{c^{2}_{S}k^2 + \left(\gamma\left(\frac{\eta}{\eta_0} - 1\right)^2 +\delta 
	- \left[\nu^2-\frac{1}{4}\right]\right) \frac{1}{\eta^2} \right\}}
	& \mbox{\small {\bf for ~$m\approx \sqrt{\gamma\left(\frac{\eta}{\eta_0} - 1\right)^2 + \delta}~H$}}\\ 
	\displaystyle \sqrt{\left\{c^{2}_{S}k^2 + \left(\frac{m^2_0}{2H^2}\left[1
 -\tanh\left(\frac{\rho}{H}\ln(-H\eta)\right)\right] - \left[\nu^2-\frac{1}{4}\right]\right) \frac{1}{\eta^2} \right\}}
	& \mbox{\small {\bf for ~$\displaystyle m= \frac{m_0}{\sqrt{2}}\sqrt{\left[1
 -\tanh\left(\frac{\rho}{H}\ln(-H\eta)\right)\right]}$}}\\ 
	\displaystyle \sqrt{\left\{c^{2}_{S}k^2 + \left(\frac{m^2_0}{H^2}{\rm sech}^2\left(\frac{\rho}{H}\ln(-H\eta)\right) - \left[\nu^2-\frac{1}{4}\right]\right) \frac{1}{\eta^2} \right\}}~~~~
	& \mbox{\small {\bf for ~$\displaystyle 
 m= m_0~{\rm sech}\left(\frac{\rho}{H}\ln(-H\eta)\right) $}}.
          \end{array}
\right.
\end{array}\ee

It is important to note that, if in the present discussion 
the WKB approximation were exactly valid, then for the EFT driven present cosmological
setup no particle creation occur. Now to describe a very small 
fraction of particle creation after inflation in the present context 
we start with a Bogoliubov coefficient $\beta$ 
in FLRW space time, which characterizes the amount of mixing between the two types of WKB approximated solutions. 
Here it is important to mention that, in the \textcolor{blue}{sub Hubble region} ($|kc_{S}\eta|>>1$) Bogoliubov coefficient $\beta$
is small and consequently the representative probability
distribution $P(x)$ for the relative comoving distance $x$ between the two pairs peaks at the comoving length scale given by,
$x \sim |\eta_{\rm pair}|$ i.e. $\frac{dP(x)}{dx}|_{x \sim |\eta_{\rm pair}|}=0$, $\frac{d^2P(x)}{dx^2}|_{x \sim |\eta_{\rm pair}|}<0$ and $P(x \sim |\eta_{\rm pair}|)=P_{max}$. When the typical comoving distance $x$ 
is of the order of the time $\eta_{\rm pair}$, all the pair is created within the present EFT setup.
It is important to mention that the general formula for the Bogoliubov coefficient $\beta$ in Fourier 
space is given by the following approximation:
\be\begin{array}{lll}\label{soladfasx}
 \displaystyle \beta(k) =
                    \displaystyle   \int^{0}_{-\infty}d\eta~\frac{\left(p^{'}(\eta)\right)^2}{
                    4p^3(\eta)}\exp\left[2i\int^{\eta}_{-\infty}
d\eta^{'}p(\eta^{'})\right]
\end{array}\ee
One can use another equivalent way to define the the Bogoliubov coefficient $\beta$ in Fourier 
space by implementing instantaneous Hamiltonian diagonalization method in the present
context \cite{lopez:2012jj,Behbahani:2011it,Choudhury:2014wsa,Choudhury:2015eua,
 Choudhury:2015pqa,Cheung:2007st,noumi:2013jj}. 
Using this diagonalized representation the regularized Bogoliubov coefficient $\beta$ in Fourier 
space can be written as:
\be\begin{array}{lll}\label{soladfasxsd3}
 \displaystyle \beta_{diag}(k;\tau,\tau^{'}) =
                    \displaystyle   \int^{\tau}_{\tau^{'}}d\eta~\frac{p^{'}(\eta)}{
                    2p(\eta)}\exp\left[-2i\int^{\eta}
d\eta^{'}p(\eta^{'})\right]
\end{array}\ee
where $\tau$ and $\tau^{'}$ introduced as the conformal time regulator in the present context.
We will also derive the expressions using Eq~(\ref{soladfasxsd3}) in the next three subsections.
In the next three subsection we will explicitly discuss three physical possibilities which captures the effect of 
massive particles in our computation.

In this context one can compare the dynamical equations for scalar mode fluctuations with the well known 
Schr$\ddot{o}$dinger scattering problem in one spatial dimension as given by~\footnote{Here we set $h/2\pi=1$.}:
\bea \left[-\frac{1}{2m}\frac{d^2}{dx^2}+V(x)-E\right]\Psi(x)=0,\eea 
where the following identification exists between quantum mechanical Schr$\ddot{o}$dinger equation and cosmological 
dynamical equations for scalar mode fluctuations:
\bea \eta \Longrightarrow  t=-\frac{1}{H}\ln(-H\eta)&\Longleftrightarrow& x,\nonumber\\
h_{k}(\eta)\Longrightarrow h_{k}(t)&\Longleftrightarrow& \Psi(x),\nonumber\\
p^2(\eta)\Longrightarrow p^2(t)= \footnotesize\left\{\begin{array}{ll}
                    \displaystyle  \left\{c^{2}_{S}k^2 + \left(\frac{m^2}{H^2} - 2\right) H^2~e^{2Ht} \right\}~~~~ &
 \mbox{\small {\bf for ~dS}}  \\ 
	\displaystyle  \left\{c^{2}_{S}k^2 + \left(\frac{m^2}{H^2} - \left[\nu^2-\frac{1}{4}\right]
\right)  H^2~e^{2Ht} \right\}~~~~ & \mbox{\small {\bf for~ qdS}}.
          \end{array}
\right.&\Longleftrightarrow& 2m\left[E-V(t)\right].~~~~~
\eea
Here the signature of $p^{2}(t)$ in Schr$\ddot{o}$dinger quantum mechanics signify the following physical 
situations:
\begin{itemize}
 \item If $p^{2}(t)>0$ then it corresponds to the propagation over the barrier for $E>V(t)$.
 \item If $p^{2}(t)<0$ then it corresponds to tunneling solution for $E<V(t)$.
\end{itemize}
Most importantly if we use the analogy between Schr$\ddot{o}$dinger quantum mechanics and cosmology then one can write:
\bea
 \displaystyle V(t)&=&\left\{\begin{array}{ll}
                    \displaystyle -\frac{1}{2m}\left(\frac{m^2}{H^2} - 2\right) H^2~e^{2Ht} ~~~~ &
 \mbox{\small {\bf for ~dS}}  \\ 
	\displaystyle -\frac{1}{2m}\left(\frac{m^2}{H^2} - \left[\nu^2-\frac{1}{4}\right]
\right)  H^2~e^{2Ht} ~~~~ & \mbox{\small {\bf for~ qdS}}.
          \end{array}
\right.\\
E&=&\frac{1}{2m}c^{2}_{S}k^2.
\eea
\bea
 \displaystyle V(t)&=&\left\{\begin{array}{ll}
                    \displaystyle   \frac{H^2}{2m}e^{2Ht} ~~~~ &
 \mbox{\small {\bf for ~$m\approx H$}}  \\  
	\displaystyle -\frac{1}{2m}\left(\Upsilon^2 - 2\right) H^2~e^{2Ht} ~~~~ & \mbox{\small {\bf for~ $m>> H$}} \\ 
	\displaystyle  \frac{H^2}{m}e^{2Ht}~~~~ & \mbox{\small {\bf for~ $m<< H$}} \\
	\displaystyle -\frac{1}{2m}\left[\gamma\left(\frac{e^{-Ht}}{H\eta_0}+1\right)^2+\delta-2\right]H^2~e^{2Ht}  ~~~~ & \mbox{\small {\bf for~ $m\approx \sqrt{\gamma\left(\frac{\eta}{\eta_0} - 1\right)^2 + \delta}~H$}} \\ 
	\displaystyle -\frac{1}{2m}\left[\frac{m^2_0}{2H^2}\left[1
 +\tanh\left(\rho t\right)\right]-2\right]H^2~e^{2Ht} ~~~~ & \mbox{\small {\bf for~ $\displaystyle m= \frac{m_0}{\sqrt{2}}\sqrt{\left[1
 -\tanh\left(\frac{\rho}{H}\ln(-H\eta)\right)\right]}$}} \\ 
	\displaystyle -\frac{1}{2m}\left[\frac{m^2_0}{H^2}{\rm sech}^2\left(\rho t\right)-2\right]H^2~e^{2Ht}  ~~~~ & \mbox{\small {\bf for~ $\displaystyle 
 m= m_0~{\rm sech}\left(\frac{\rho}{H}\ln(-H\eta)\right) $}}.
          \end{array}
\right.
\eea
\bea
 \displaystyle V(t)&=&\left\{\begin{array}{ll}
                    \displaystyle  \frac{\left[\nu^2-\frac{5}{4}\right]H^2}{2m}e^{2Ht} ~~~~ &
 \mbox{\small {\bf for ~$m\approx H$}}  \\ 
	\displaystyle -\frac{1}{2m}\left(\Upsilon^2  -\left[\nu^2-\frac{1}{4}\right]\right) H^2~e^{2Ht} ~~~~ & \mbox{\small {\bf for~ $m>> H$}} \\ 
	\displaystyle \frac{\left[\nu^2-\frac{1}{4}\right]H^2}{2m}e^{2Ht} ~~~~ & \mbox{\small {\bf for~ $m<< H$}} \\ 
	\displaystyle -\frac{1}{2m}\left[\gamma\left(\frac{e^{-Ht}}{H\eta_0}+1\right)^2+\delta-\left[\nu^2-\frac{1}{4}\right]\right]H^2~e^{2Ht}  ~~~~ & \mbox{\small {\bf for~ $m\approx \sqrt{\gamma\left(\frac{\eta}{\eta_0} - 1\right)^2 + \delta}~H$}} \\ 
	\displaystyle -\frac{1}{2m}\left[\frac{m^2_0}{2H^2}\left[1
 +\tanh\left(\rho t\right)\right]-\left[\nu^2-\frac{1}{4}\right]\right]H^2~e^{2Ht} ~~~~ & \mbox{\small {\bf for~ $\displaystyle m= \frac{m_0}{\sqrt{2}}\sqrt{\left[1
 -\tanh\left(\frac{\rho}{H}\ln(-H\eta)\right)\right]}$}} \\ 
	\displaystyle -\frac{1}{2m}\left[\frac{m^2_0}{H^2}{\rm sech}^2\left(\rho t\right)-\left[\nu^2-\frac{1}{4}\right]\right]H^2~e^{2Ht} ~~~~ & \mbox{\small {\bf for~ $\displaystyle 
 m= m_0~{\rm sech}\left(\frac{\rho}{H}\ln(-H\eta)\right) $}}.
          \end{array}
\right.
\eea
Now if we assume that in the past field has the structure $\Psi_{past}(t)=e^{ip(t)t}$, in the future the solution is given by,
$\Psi_{future}(t)=\alpha ~e^{ip(t)t}+\beta~e^{-ip(t)t}$, due to tunneling. Here $\alpha$ and $\beta$ are the Bogoliubov
co-efficients in the 
present context of discussion. This correspond to the particle creation with probability $P\propto |\beta|^2$
~\footnote{Here it is important to note that for $p^{2}(t)>0$ case we also get some amount of 
scattering over the barrier.}. 

In the context of primordial cosmology one can also study the particle creation mechanism following the same prescription 
in Schr$\ddot{o}$dinger quantum mechanics. In case of cosmology the past field has the
pseudo-nomr structure and this could be identified with the left-moving wave $\Psi_{L}=e^{-ip(t)t}$ and 
in the future the solution is given by,
$\Psi_{LR}=\alpha ~e^{-ip(t)t}+\beta~e^{ip(t)t}$, which can be interpreted as the mixture of left-moving and right-moving wave. 
Consequently, the Bogoliubov co-efficients $\alpha$ and $\beta$ are related to the refection and transmission 
co-efficients ${\cal R}$ and ${\cal T}$ via the following identifications:
\bea \alpha &=& \frac{1}{{\cal T}},\\
\beta &=& \frac{{\cal R}}{{\cal T}}.\eea
In this context the Bogoliubov co-efficients $\alpha$ and $\beta$ satisfies the normalization condition:
\bea\label{sdq1} |\alpha|^2 -|\beta|^2 &=& 1,\eea
which implies the following well known conservation law:
\bea\label{sdq2} |{\cal R}|^2+|{\cal T}|^2 &=& 1,\eea
applicable in the context of Schr$\ddot{o}$dinger quantum mechanics.

Further using the expressions for Bogoliubov co-efficient $\beta$ in two different representations as mentioned in
Eq~(\ref{soladfasx}) and Eq~(\ref{soladfasxsd3}), and substituting them in Eq~(\ref{sdq1}) we get the following
expressions for the Bogoliubov co-efficient $\alpha$ in two different representations as given by:
\be\begin{array}{lll}\label{soladfasxv1}
 \displaystyle \alpha(k) =
                    \displaystyle \sqrt{\left[ 1+\left| \int^{0}_{-\infty}d\eta~\frac{\left(p^{'}(\eta)\right)^2}{
                    4p^3(\eta)}\exp\left[2i\int^{\eta}_{-\infty}
d\eta^{'}p(\eta^{'})\right]\right|^2\right]}~e^{i\phi}
\end{array}\ee
\be\begin{array}{lll}\label{soladfasxsd3v2}
 \displaystyle \alpha_{diag}(k;\tau,\tau^{'}) =
                    \displaystyle   \sqrt{\left[ 1+\left| \int^{\tau}_{\tau^{'}}d\eta~\frac{p^{'}(\eta)}{
                    2p(\eta)}\exp\left[-2i\int^{\eta}
d\eta^{'}p(\eta^{'})\right]\right|^2\right]}~e^{i\phi_{diag}}
\end{array}\ee
where $\phi$ and $\phi_{diag}$ are the associated phase factors in two different representations.
Further using the expressions for Bogoliubov co-efficient $\alpha$ in two different representations as mentioned in
Eq~(\ref{soladfasxv1}) and Eq~(\ref{soladfasxsd3v2}), and substituting them in Eq~(\ref{sdq2}) we get the following
expressions for the reflection and transmission co-efficient in two different representations as given by:
\be\begin{array}{lll}\label{soladfasxv11}
 \displaystyle {\cal R} =\frac{\beta}{\alpha}=
                    \displaystyle \frac{\displaystyle\int^{0}_{-\infty}d\eta~\frac{\left(p^{'}(\eta)\right)^2}{
                    4p^3(\eta)}\exp\left[2i\int^{\eta}_{-\infty}
d\eta^{'}p(\eta^{'})\right]}{\sqrt{\displaystyle\left[ 1+\left| \int^{0}_{-\infty}d\eta~\frac{\left(p^{'}(\eta)\right)^2}{
                    4p^3(\eta)}\exp\left[2i\int^{\eta}_{-\infty}
d\eta^{'}p(\eta^{'})\right]\right|^2\right]}}~e^{-i\phi},
\end{array}\ee
\be\begin{array}{lll}\label{soladfasxv22}
 \displaystyle {\cal T} =\frac{1}{\alpha}=
                    \displaystyle \frac{~e^{-i\phi}}{\sqrt{\displaystyle\left[ 1+\left| \int^{0}_{-\infty}d\eta~\frac{\left(p^{'}(\eta)\right)^2}{
                    4p^3(\eta)}\exp\left[2i\int^{\eta}_{-\infty}
d\eta^{'}p(\eta^{'})\right]\right|^2\right]}},
\end{array}\ee
and 
\be\begin{array}{lll}\label{soladfasxv33}
 \displaystyle {\cal R}_{diag}(k;\tau,\tau^{'}) =\frac{\beta_{diag}(k;\tau,\tau^{'})}{\alpha_{diag}(k;\tau,\tau^{'})}=
                    \displaystyle \frac{\displaystyle \int^{\tau}_{\tau^{'}}d\eta~\frac{p^{'}(\eta)}{
                    2p(\eta)}\exp\left[-2i\int^{\eta}
d\eta^{'}p(\eta^{'})\right]}{\sqrt{\displaystyle\left[ 1+\left| \int^{\tau}_{\tau^{'}}d\eta~\frac{p^{'}(\eta)}{
                    2p(\eta)}\exp\left[-2i\int^{\eta}
d\eta^{'}p(\eta^{'})\right]\right|^2\right]}}~e^{-i\phi_{diag}},
\end{array}\ee
\be\begin{array}{lll}\label{soladfasxv44}
 \displaystyle {\cal T}_{diag}(k;\tau,\tau^{'}) =\frac{1}{\alpha_{diag}(k;\tau,\tau^{'})}=
                    \displaystyle \frac{~e^{-i\phi_{diag}}}{\sqrt{\displaystyle\left[ 1+\left| \int^{\tau}_{\tau^{'}}d\eta~\frac{p^{'}(\eta)}{
                    2p(\eta)}\exp\left[-2i\int^{\eta}
d\eta^{'}p(\eta^{'})\right]\right|^2\right]}}.
\end{array}\ee
Next the expression for the
number of produced particles at time $\tau$ can be calculated in the two representations using 
from the following formula as:
\bea {\cal N}(\tau,\tau^{'})&=&\frac{1}{(2\pi a)^3}\int d^{3}{\bf k}~|\beta(k,\tau,\tau^{'})|^2
\nonumber\\
&=& \frac{1}{(2\pi a)^3}\int d^{3}{\bf k}~\left|\displaystyle\int^{0}_{-\infty}d\eta~\frac{\left(p^{'}(\eta)\right)^2}{
                    4p^3(\eta)}\exp\left[2i\int^{\eta}_{-\infty}
d\eta^{'}p(\eta^{'})\right]\right|^2.\eea
\bea {\cal N}_{diag}(\tau,\tau^{'})&=&\frac{1}{(2\pi a)^3}\int d^{3}{\bf k}~|\beta_{diag}(\tau,\tau^{'})|^2\nonumber\\
&=& \frac{1}{(2\pi a)^3}\int d^{3}{\bf k}~\left|\displaystyle \int^{\tau}_{\tau^{'}}d\eta~\frac{p^{'}(\eta)}{
                    2p(\eta)}\exp\left[-2i\int^{\eta}
d\eta^{'}p(\eta^{'})\right]\right|^2.\eea
Finally, one can define the total energy density of the produced particles using the following expression:
\bea \rho(\tau,\tau^{'})&=&\frac{1}{(2\pi a)^3a}\int d^{3}{\bf k}~p(\tau)~|\beta(k,\tau,\tau^{'})|^2\nonumber\\ 
&=&\frac{1}{(2\pi a)^3a}\int d^{3}{\bf k}~p(\tau)~\left|\displaystyle\int^{0}_{-\infty}d\eta~\frac{\left(p^{'}(\eta)\right)^2}{
                    4p^3(\eta)}\exp\left[2i\int^{\eta}_{-\infty}
d\eta^{'}p(\eta^{'})\right]\right|^2,\eea
\bea \rho_{diag}(\tau,\tau^{'},\eta^{'})&=&\frac{1}{(2\pi a)^3a}
\int d^{3}{\bf k}~p(\tau)~|\beta_{diag}(k,\tau,\tau^{'})|^2\nonumber\\ 
&=&\frac{1}{(2\pi a)^3a}
\int d^{3}{\bf k}~p(\tau)~\left|\displaystyle \int^{\tau}_{\tau^{'}}d\eta~\frac{p^{'}(\eta)}{
                    2p(\eta)}\exp\left[-2i\int^{\eta}
d\eta^{'}p(\eta^{'})\right]\right|^2.\eea
\subsubsection{\bf Case I: $m \approx H$}
Equation of motion for the massive field is:
\bea
h''_k + \left\{c^{2}_{S}k^2 - \frac{1}{\eta^2} \right\} h_k &=& 0~~~~~~~{\bf for~ dS}\\\nonumber\\
h''_k + \left\{c^{2}_{S}k^2 - \left[\nu^2-\frac{5}{4}\right]
\frac{1}{\eta^2} \right\} h_k &=& 0~~~~~~~{\bf for~ qdS}.
\eea
The solution for the mode function for de Sitter and quasi de Sitter space can be expressed as: 
\be\begin{array}{lll}\label{yu12}
 \displaystyle h_k (\eta) =\left\{\begin{array}{ll}
                    \displaystyle   \sqrt{-\eta}\left[C_1  H^{(1)}_{\sqrt{5}/2} \left(-kc_{S}\eta\right) 
+ C_2 H^{(2)}_{\sqrt{5}/2} \left(-kc_{S}\eta\right)\right]~~~~ &
 \mbox{\small {\bf for ~dS}}  \\ 
	\displaystyle \sqrt{-\eta}\left[C_1  H^{(1)}_{\sqrt{\nu^2-1}} \left(-kc_{S}\eta\right) 
+ C_2 H^{(2)}_{\sqrt{\nu^2-1}} \left(-kc_{S}\eta\right)\right]~~~~ & \mbox{\small {\bf for~ qdS}}.
          \end{array}
\right.
\end{array}\ee
where $C_{1}$ and and $C_{2}$ are two arbitrary integration constant, which depend on the 
choice of the initial condition.

After taking $kc_{S}\eta\rightarrow -\infty$, $kc_{S}\eta\rightarrow 0$ and $|kc_{S}\eta|\approx 1-\Delta(\rightarrow 0)$ limit the most general 
solution as stated in Eq~(\ref{yu12}) can be recast as:
\bea\label{yu2}
 \displaystyle h_k (\eta) &\stackrel{|kc_{S}\eta|\rightarrow-\infty}{=}&\left\{\begin{array}{ll}
                    \displaystyle   \sqrt{\frac{2}{\pi kc_{S}}}\left[C_1  e^{ -ikc_{S}\eta}
e^{-\frac{i\pi}{2}\left(\frac{\sqrt{5}+1}{2}\right)} 
+ C_2 e^{ ikc_{S}\eta}
e^{\frac{i\pi}{2}\left(\frac{\sqrt{5}+1}{2}\right)}\right]~~~~ &
 \mbox{\small {\bf for ~dS}} \\ 
	\displaystyle \sqrt{\frac{2}{\pi kc_{S}}}\left[C_1  e^{ -ikc_{S}\eta}
e^{-\frac{i\pi}{2}\left(\sqrt{\nu^2-1}+\frac{1}{2}\right)} 
+ C_2 e^{ ikc_{S}\eta}
e^{\frac{i\pi}{2}\left(\sqrt{\nu^2-1}+\frac{1}{2}\right)}\right]~~~~ & \mbox{\small {\bf for~ qdS}}.
          \end{array}
\right.\eea 
\bea
\label{yu2}
 \displaystyle h_k (\eta) &\stackrel{|kc_{S}\eta|\rightarrow 0}{=}&\left\{\begin{array}{ll}
                    \displaystyle  \frac{i\sqrt{-\eta}}{\pi}\Gamma\left(\frac{\sqrt{5}}{2}\right)
                    \left(-\frac{kc_{S}\eta}{2}\right)^{-\frac{\sqrt{5}}{2}}\left[C_1   
- C_2 \right]~ &
 \mbox{\small {\bf for ~dS}}  \\ 
	\displaystyle\frac{i\sqrt{-\eta}}{\pi}\Gamma\left(\sqrt{\nu^2-1}\right)\displaystyle
	\left(-\frac{kc_{S}\eta}{2}\right)^{-\sqrt{\nu^2-1}}\left[C_1   
- C_2 \right]~ & \mbox{\small {\bf for~ qdS}}.
          \end{array}
\right.
\eea
\bea
\label{yu2}
 \displaystyle h_k (\eta) &\stackrel{|kc_{S}\eta|\approx 1-\Delta(\rightarrow 0)}{=}&\left\{\begin{array}{ll}
                    \displaystyle  \frac{i}{\pi}\sqrt{-\eta}\left[ \frac{2}{\sqrt{5}}-\gamma+\frac{\sqrt{5}}{4}
                   \left(\gamma^2+\frac{\pi^2}{6}\right)\right.\\ \displaystyle \left.
                   \displaystyle~~~~~~~~-\frac{5}{24}\left(\gamma^3+\frac{\gamma \pi^2}{2}+2\zeta(3)\right)
                    +\cdots\right]\\ \displaystyle ~~~~~~~~~~~~\times\left(\frac{1+\Delta}{2}\right)^{-\frac{\sqrt{5}}{2}}\left[C_1   
- C_2 \right]~&
 \mbox{\small {\bf for ~dS}}  \\ 
	\displaystyle\frac{i}{\pi}\sqrt{-\eta}\left[ \frac{1}{ \left\{\sqrt{\nu^2-1}\right\}}
  -\gamma  \displaystyle +\frac{1}{2}\left(\gamma^2+\frac{\pi^2}{6}\right) \left\{\sqrt{\nu^2-1}\right\}
  \right.\\ \displaystyle \left. \displaystyle~-\frac{1}{6}\left(\gamma^3+\frac{\gamma \pi^2}{2}+2\zeta(3)\right) \left\{\sqrt{\nu^2-1}\right\}^2
  +\cdots\right]\\ \displaystyle ~~~~~~~~~~~~\times\left(\frac{1+\Delta}{2}\right)^{-\sqrt{\nu^2-1}}\left[C_1   
- C_2 \right]~ & \mbox{\small {\bf for~ qdS}}.
          \end{array}
\right.
\eea
Next we assume that the WKB approximation is approximately valid for all times for the solution for the mode function $h_{k}$.
In the standard WKB approximation the total solution can be recast in the following form:
\bea\label{df13}
h_k (\eta)&=& \left[D_{1}u_{k}(\eta) + D_{2} \bar{u}_{k}(\eta)\right],\eea
where $D_{1}$ and and $D_{2}$ are two arbitrary integration constant, which depend on the 
choice of the initial condition during WKB approximation at early and late time scale.
In the present context $u_{k}(\eta)$ and $\bar{u}_{k}(\eta)$ are defined as:
\be\begin{array}{lll}\label{solp}
 \displaystyle\small u_{k}(\eta) =
 \left\{\begin{array}{ll}
                    \displaystyle   \frac{1}{\sqrt{2\sqrt{c^{2}_{S}k^2-\frac{1}{\eta^2}}}}
\exp\left[i\int^{\eta} d\eta^{\prime} \sqrt{c^{2}_{S}k^2-\frac{1}{\eta^{'2}}}\right]\\
= \displaystyle\frac{1}{\sqrt{2\sqrt{c^{2}_{S}k^2-\frac{1}{\eta^2}}}} 
\exp\left[i\left(\eta\sqrt{c^{2}_{S}k^2-\frac{1}{\eta^{2}}}+tan^{-1}\left[\frac{1}{\eta\sqrt{c^{2}_{S}k^2-\frac{1}{\eta^{2}}}}\right]
\right)\right]~~~~ &
 \mbox{\small {\bf for ~dS}}  \\ 
	\displaystyle \frac{1}{\sqrt{2\sqrt{c^{2}_{S}k^2-\frac{\left[\nu^2-\frac{5}{4}\right]}{\eta^2}}}}
\exp\left[i\int^{\eta} d\eta^{\prime} \sqrt{c^{2}_{S}k^2-\frac{\left[\nu^2-\frac{5}{4}\right]}{\eta^{'2}}}\right]\\
= \displaystyle\frac{1}{\sqrt{2\sqrt{c^{2}_{S}k^2-\frac{\left[\nu^2-\frac{5}{4}\right]}{\eta^2}}}} 
\exp\left[i\left(\eta\sqrt{c^{2}_{S}k^2-\frac{\left[\nu^2-\frac{5}{4}\right]}{\eta^{2}}}
\right.\right.\\ \left.\left.~~~~~~~~~~~~\displaystyle +\sqrt{\nu^2-\frac{5}{4}}~tan^{-1}\left[\frac{\sqrt{\nu^2-\frac{5}{4}}}{\eta\sqrt{c^{2}_{S}k^2-\frac{\left[\nu^2-\frac{5}{4}\right]}{\eta^{2}}}}\right]
\right)\right]~~~~ & \mbox{\small {\bf for~ qdS}}.
          \end{array}
\right.
\end{array}\ee
\\
\be\begin{array}{lll}\label{sola}
 \displaystyle\small \bar{u}_{k}(\eta) =
 \left\{\begin{array}{ll}
                    \displaystyle   \frac{1}{\sqrt{2\sqrt{c^{2}_{S}k^2-\frac{1}{\eta^2}}}}
\exp\left[-i\int^{\eta} d\eta^{\prime} \sqrt{c^{2}_{S}k^2-\frac{1}{\eta^{'2}}}\right]\\
= \displaystyle\frac{1}{\sqrt{2\sqrt{c^{2}_{S}k^2-\frac{1}{\eta^2}}}} 
\exp\left[-i\left(\eta\sqrt{c^{2}_{S}k^2-\frac{1}{\eta^{2}}}+tan^{-1}\left[\frac{1}{\eta\sqrt{c^{2}_{S}k^2-\frac{1}{\eta^{2}}}}\right]
\right)\right]~~~~ &
 \mbox{\small {\bf for ~dS}}  \\ 
	\displaystyle \frac{1}{\sqrt{2\sqrt{c^{2}_{S}k^2-\frac{\left[\nu^2-\frac{5}{4}\right]}{\eta^2}}}}
\exp\left[-i\int^{\eta} d\eta^{\prime} \sqrt{c^{2}_{S}k^2-\frac{\left[\nu^2-\frac{5}{4}\right]}{\eta^{'2}}}\right]\\
= \displaystyle\frac{1}{\sqrt{2\sqrt{c^{2}_{S}k^2-\frac{\left[\nu^2-\frac{5}{4}\right]}{\eta^2}}}} 
\exp\left[-i\left(\eta\sqrt{c^{2}_{S}k^2-\frac{\left[\nu^2-\frac{5}{4}\right]}{\eta^{2}}}
\right.\right.\\ \left.\left.~~~~~~~~~~~~\displaystyle +\sqrt{\nu^2-\frac{5}{4}}~tan^{-1}\left[\frac{\sqrt{\nu^2-\frac{5}{4}}}{\eta\sqrt{c^{2}_{S}k^2-\frac{\left[\nu^2-\frac{5}{4}\right]}{\eta^{2}}}}\right]
\right)\right]~~~~ & \mbox{\small {\bf for~ qdS}}.
          \end{array}
\right.
\end{array}\ee
where we have written the total solution for the mode $h_k$ in terms of 
two linearly independent solutions. Here it is important to note that the both of the 
solutions are hermitian conjugate of each other. If in the present context the exact solution of the mode $h_k$ 
is expanded with respect to these two linearly independent solutions then particle creation is absent in our EFT 
setup. In the present context correctness of 
WKB approximation is guarantee at very early and very late time scales. In this discussion $u_{k}(\eta)$
is valid at very early time scale and $\bar{u}_{k}(\eta)$ perfectly works in the late time scale.

Now we will explicitly check that the exactness of the above mentioned WKB result derived in Eq~(\ref{df13}) 
with the actual solution of the mode function as presented in Eq~(\ref{yu12}). As mentioned earlier 
in FLRW space-time in Fourier space Bogoliubov coefficient $\beta(k)$ measures this exactness for a given 
setup. The particle creation mechanism and its exact amount
is described by finding the Bogoliubov coefficient $\beta(k)$ in Fourier space
which in principle measures the exact 
amount of late times solution $u_{k}(\eta)$, if in the present context we exactly 
start with the early time scale solution $u_{k}(\eta)$. In our present computation 
we consider a physical situation where the
WKB approximation is correct up to the leading order throughout the cosmological evolution in time scale.
In the present context the Bogoliubov coefficient $\beta(k)$ in Fourier space
can be computed approximately using the following regularized integral:
\be\begin{array}{lll}\label{soladf1}
 \displaystyle \beta(k,\tau,\tau^{'},\eta^{'}) =
 \left\{\begin{array}{ll}
                    \displaystyle   \int^{\tau}_{\tau^{'}}d\eta~\frac{1}{
                    4\eta^{6}\left(c^{2}_{S}k^2-\frac{1}{\eta^{2}}\right)^{\frac{5}{2}}}\exp\left[2i\int^{\eta}_{\eta^{'}}
d\eta^{''}\sqrt{c^{2}_{S}k^2-\frac{1}{\eta^{''2}}}\right]&
 \mbox{\small {\bf for ~dS}}  \\ 
	\displaystyle   \int^{\tau}_{\tau^{'}}d\eta~\frac{\left[\nu^2-\frac{5}{4}\right]^2}{
                    4\eta^{6}\left(c^{2}_{S}k^2-\frac{\left[\nu^2-\frac{5}{4}\right]}{\eta^{2}}\right)^{\frac{5}{2}}}
                    \exp\left[2i\int^{\eta}_{\eta^{'}}
d\eta^{''}\sqrt{c^{2}_{S}k^2-\frac{\left[\nu^2-\frac{5}{4}\right]}{\eta^{''2}}}\right]& \mbox{\small {\bf for~ qdS}}.
          \end{array}
\right.
\end{array}\ee
which is not exactly analytically computable. To study the behaviour of this integral we consider here three 
consecutive physical situations-$|kc_{S}\eta|<<1$, $|kc_{S}\eta|\approx 1-\Delta(\rightarrow 0)$ and $|kc_{S}\eta|>>1$ for de Sitter and quasi de Sitter case. 
In three cases we have:
\be\begin{array}{lll}\label{yu2dfvqasx3}\small
 \displaystyle \underline{\rm \bf For~dS:}~~~~~~~~~\sqrt{\left\{c^{2}_{S}k^2 - \frac{1}{\eta^2} \right\}} \approx
 \left\{\begin{array}{ll}
                    \displaystyle  \frac{i}{\eta}~~~~ &
 \mbox{\small {\bf for ~$|kc_{S}\eta|<<1$}}  \\ 
	\displaystyle \frac{i\sqrt{2\Delta}}{\eta}~~~~ & \mbox{\small
	{\bf for ~$|kc_{S}\eta|\approx 1-\Delta(\rightarrow 0)$}}\\ 
	\displaystyle kc_{S}~~~~ & \mbox{\small {\bf for ~$|kc_{S}\eta|>>1$}}.
          \end{array}
\right.
\\
\small
 \displaystyle \underline{\rm \bf For~qdS:}~~~~~~~~~\sqrt{\left\{c^{2}_{S}k^2 - \left[\nu^2-\frac{5}{4}\right]
\frac{1}{\eta^2} \right\}} \approx
\left\{\begin{array}{ll}
                    \displaystyle \frac{i\sqrt{\left[\nu^2-\frac{5}{4}\right]}}{\eta} ~~~~ &
 \mbox{\small {\bf for ~$|kc_{S}\eta|<<1$}}  \\ 
	\displaystyle \frac{i\sqrt{2\Delta+\left[\nu^2-\frac{9}{4}\right]}}{\eta}~~~~ 
	& \mbox{\small {\bf for ~$|kc_{S}\eta|\approx 1-\Delta(\rightarrow 0)$}}\\ 
	\displaystyle kc_{S}~~~~ & \mbox{\small {\bf for ~$|kc_{S}\eta|>>1$}}.
          \end{array}
\right.
\end{array}\ee
and further using this result Bogoliubov coefficient $\beta(k)$ in Fourier space can be expressed as:
\bea &&\underline{\bf For~dS:}\\
\small\beta(k,\tau,\tau^{'},\eta^{'})&=& 
\left\{\begin{array}{ll}
                    \displaystyle  \frac{\eta^{'2}}{
                    8i}\left[\frac{1}{\tau^{'2}}-\frac{1}{\tau^2}\right]
                     &
 \mbox{\small {\bf for ~$|kc_{S}\eta|<<1$}}  \\ \\
 \displaystyle   \frac{\eta^{'2\sqrt{2\Delta}}}{
                    8i\left(2\Delta\right)^{3}}\left[\frac{1}{\tau^{'~2\sqrt{2\Delta}}}-\frac{1}{\tau^{2\sqrt{2\Delta}}}
                    \right] &
 \mbox{\small {\bf for ~$|kc_{S}\eta|\approx 1-\Delta(\rightarrow 0)$}}  \\ \\
	\displaystyle \left[i\frac{\text{Ei}(2 ikc_{S} \eta)e^{-2ikc_{S}\eta^{'}}}{15}\right.\\ \left.\displaystyle-\frac{e^{2 i k c_{S}(\eta-\eta^{'})} }{120 (c_{S}k)^5 \eta^5}\left(
	4 (c_{S}k)^4 \eta^4-2 i (c_{S}k)^3 \eta^3\right.\right.\\ \left.\left.-2 (c_{S}k)^2 \eta^2+3 i c_{S}k \eta+6\right)\right]^{\tau}_{\tau^{'}}& \mbox{\small {\bf for ~$|kc_{S}\eta|>>1$}}.
          \end{array}
\right.\nonumber\eea
\bea
&&\underline{\bf For~qdS:}\\
\small\beta(k,\tau,\tau^{'},\eta^{'})&=&
\left\{\begin{array}{ll}
                     \frac{\eta^{'2\sqrt{\nu^2-\frac{5}{4}}}}{
                    8i\left(\nu^2-\frac{5}{4}\right)}\left[\frac{1}{\tau^{'2\sqrt{\nu^2-\frac{5}{4}}}}-\frac{1}{\tau^{2\sqrt{\nu^2-\frac{5}{4}}}}\right]&
 \mbox{\small {\bf for ~$|kc_{S}\eta|<<1$}}  \\ \\
  \frac{\eta^{'2\sqrt{2\Delta+\nu^2-\frac{9}{4}}}\left[\nu^2-\frac{5}{4}\right]^2}{
                    8i\left(2\Delta+\nu^2-\frac{9}{4}\right)^3}
                    \left[\frac{1}{\tau^{'2\sqrt{2\Delta+\nu^2-\frac{9}{4}}}}
                    -\frac{1}{\tau^{2\sqrt{2\Delta+\nu^2-\frac{9}{4}}}}\right]
                    &
 \mbox{\small {\bf for ~$|kc_{S}\eta|\approx 1-\Delta(\rightarrow 0)$}} \\ \\
	\displaystyle \left[\nu^2-\frac{5}{4}\right]^2 \left[i\frac{\text{Ei}(2 ikc_{S} \eta)e^{-2ikc_{S}\eta^{'}}}{15}\right.\\ \left.\displaystyle-\frac{e^{2 i k c_{S}(\eta-\eta^{'})} }{120 (c_{S}k)^5 \eta^5}\left(
	4 (c_{S}k)^4 \eta^4-2 i (c_{S}k)^3 \eta^3\right.\right.\\ \left.\left.-2 (c_{S}k)^2 \eta^2+3 i c_{S}k \eta+6\right)\right]^{\tau}_{\tau^{'}} & \mbox{\small {\bf for ~$|kc_{S}\eta|>>1$}}.
          \end{array}
\right.\nonumber
\eea
In all the situation described for de Sitter and quasi de Sitter case here 
the magnitude of the Bogoliubov coefficient $|\beta(k)|$ in Fourier space is considerably small. Specifically it is important 
to point out here that for the case when $|kc_{S}\eta|>>1$ the value
of the Bogoliubov coefficient $\beta(k)$ in Fourier space
is even smaller as the WKB approximated solution is strongly consistent for all time scales. On the other hand near the 
vicinity of the conformal time scale $\eta\sim \eta_{pair}$ for $|kc_{S}\eta_{pair}|<<1$
the WKB approximated solution is less strongly valid and to validate the solution at this time scale 
one can neglect the momentum $k$ dependence in the Bogoliubov coefficient $\beta(k)$ in Fourier space.
Here $|\eta_{pair}|$ characterizes the relative
separation between the created particles.

As mentioned earlier here one can use another equivalent way
to define the the Bogoliubov coefficient $\beta$ in Fourier 
space by implementing instantaneous Hamiltonian
diagonalization method to interpret the results. 
Using this diagonalized representation the
regularized Bogoliubov coefficient $\beta$ in Fourier 
space can be written as:
\be
\ee
where $\tau$ and $\tau^{'}$ introduced as the conformal time regulator in the present context.
In this case as well the Bogoliubov coefficient is not exactly analytically computable. 
To study the behaviour of this integral we consider here three similar
consecutive physical situations for de Sitter and quasi de Sitter case as discussed earlier.
\bea &&\underline{\bf For~dS:}\\
\small\beta_{diag}(k;\tau,\tau^{'}) &=& 
\left\{\begin{array}{ll}
                    \displaystyle  \frac{1}{
                    4}\left[\tau^{'2}-\tau^2\right]
                     &
 \mbox{\small {\bf for ~$|kc_{S}\eta|<<1$}}  \\ \\
 \displaystyle   \frac{1}{
                    4\left(2\Delta\right)^{3/2}}\left[\tau^{'~2\sqrt{2\Delta}}-\tau^{2\sqrt{2\Delta}}
                    \right] &
 \mbox{\small {\bf for ~$|kc_{S}\eta|\approx 1-\Delta(\rightarrow 0)$}}  \\ \\
	\displaystyle \left[\frac{e^{-2 i kc_{S} \eta} \left(2ikc_{S}\eta-1\right)}{4 (c_{S}k)^2 \eta^2}-\text{Ei}(-2 i kc_{S} \eta)\right]^{\tau}_{\tau^{'}} & \mbox{\small {\bf for ~$|kc_{S}\eta|>>1$}}.
          \end{array}
\right.\eea
\bea
&&\underline{\bf For~qdS:}\\
\small\beta_{diag}(k;\tau,\tau^{'}) &=& 
\left\{\begin{array}{ll}
                    \displaystyle   \frac{1}{
                    4\left(\nu^2-\frac{5}{4}\right)}\left[\tau^{'2\sqrt{\nu^2-\frac{5}{4}}}-\tau^{2\sqrt{\nu^2-\frac{5}{4}}}\right] &
 \mbox{\small {\bf for ~$|kc_{S}\eta|<<1$}}  \\ \\
 \displaystyle   \frac{\left[\nu^2-\frac{5}{4}\right]}{
                    4\left(2\Delta+\nu^2-\frac{9}{4}\right)^{3/2}}
                    \left[\tau^{'2\sqrt{2\Delta+\nu^2-\frac{9}{4}}}
                    -\tau^{2\sqrt{2\Delta+\nu^2-\frac{9}{4}}}\right]
                    &
 \mbox{\small {\bf for ~$|kc_{S}\eta|\approx 1-\Delta(\rightarrow 0)$}} \\ \\
	\displaystyle \left[\nu^2-\frac{5}{4}\right]^2\left[\frac{e^{-2 i kc_{S} \eta} \left(2ikc_{S}\eta-1\right)}{4 (c_{S}k)^2 \eta^2}-\text{Ei}(-2 i kc_{S} \eta)\right]^{\tau}_{\tau^{'}} & \mbox{\small {\bf for ~$|kc_{S}\eta|>>1$}}.
          \end{array}
\right.\nonumber
\eea

Further using the regularized expressions for Bogoliubov co-efficient $\beta$ in two different representations as mentioned in
Eq~(\ref{soladf1}) and Eq~(\ref{soladfasxsd3cv1}), and substituting them in Eq~(\ref{sdq1}) we get the following
regularized expressions for the Bogoliubov co-efficient $\alpha$ in two different representations as given by:
\be%
\ee
where $\phi$ and $\phi_{diag}$ are the associated phase factors in two different representations.
Here the results are not exactly analytically computable. To study the behaviour of this integral we consider here three 
consecutive physical situations-$|kc_{S}\eta|<<1$, $|kc_{S}\eta|\approx 1-\Delta(\rightarrow 0)$ and
$|kc_{S}\eta|>>1$ for de Sitter and quasi de Sitter case. 
\bea &&\underline{\bf For~dS:}\\
\small\alpha(k,\tau,\tau^{'},\eta^{'})&=& 
\left\{\begin{array}{ll}
                    \displaystyle  \sqrt{\left[ 1+ \left|\frac{\eta^{'2}}{
                    8i}\left[\frac{1}{\tau^{'2}}-\frac{1}{\tau^2}\right]\right|^2\right]}~e^{i\phi}
                     &
 \mbox{\small {\bf for ~$|kc_{S}\eta|<<1$}}  \\ \\
 \displaystyle   \sqrt{\left[ 1+ \left| \frac{\eta^{'2\sqrt{2\Delta}}}{
                    8i\left(2\Delta\right)^{3}}\left[\frac{1}{\tau^{'~2\sqrt{2\Delta}}}-\frac{1}{\tau^{2\sqrt{2\Delta}}}
                    \right]\right|^2\right]}~e^{i\phi}  &
 \mbox{\small {\bf for ~$|kc_{S}\eta|\approx 1-\Delta(\rightarrow 0)$}}  \\ \\
	\displaystyle 
\left[1+\left|\left[i\frac{\text{Ei}(2 ikc_{S} \eta)e^{-2ikc_{S}\eta^{'}}}{15}\right.\right.\right.\\ \left.\left.\left.  
\displaystyle-\frac{e^{2 i k c_{S}(\eta-\eta^{'})} }{120 (c_{S}k)^5 \eta^5}\left(
	4 (c_{S}k)^4 \eta^4-2 i (c_{S}k)^3 \eta^3\right.\right.\right.\right.\\ \left.\left.\left.\left. 
	-2 (c_{S}k)^2 \eta^2+3 i c_{S}k \eta+6\right)\right]^{\tau}_{\tau^{'}}\right|^2\right]^{1/2}~e^{i\phi}& \mbox{\small {\bf for ~$|kc_{S}\eta|>>1$}}.
          \end{array}
\right.\nonumber\eea
\bea
&&\underline{\bf For~qdS:}\\
\small\alpha(k,\tau,\tau^{'},\eta^{'})&=& 
\left\{\begin{array}{ll}
                     \sqrt{\left[ 1+ \left| \frac{\eta^{'2\sqrt{\nu^2-\frac{5}{4}}}}{
                    8i\left(\nu^2-\frac{5}{4}\right)}\left[\frac{1}{\tau^{'2\sqrt{\nu^2-\frac{5}{4}}}}
                    -\frac{1}{\tau^{2\sqrt{\nu^2-\frac{5}{4}}}}\right]\right|^2\right]}~e^{i\phi}&
 \mbox{\small {\bf for ~$|kc_{S}\eta|<<1$}}  \\ \\
  \sqrt{\left[ 1+ \left| \frac{\eta^{'2\sqrt{2\Delta+\nu^2-\frac{9}{4}}}\left[\nu^2-\frac{5}{4}\right]^2}{
                    8i\left(2\Delta+\nu^2-\frac{9}{4}\right)^3}
                    \left[\frac{1}{\tau^{'2\sqrt{2\Delta+\nu^2-\frac{9}{4}}}}
                    -\frac{1}{\tau^{2\sqrt{2\Delta+\nu^2-\frac{9}{4}}}}\right]\right|^2\right]}~e^{i\phi}

                    &
 \mbox{\small {\bf for ~$|kc_{S}\eta|\approx 1-\Delta(\rightarrow 0)$}} \\ \\
	\displaystyle \left[1+\left|\left[\nu^2-\frac{5}{4}\right]^2 \left[i\frac{\text{Ei}(2 ikc_{S} \eta)e^{-2ikc_{S}\eta^{'}}}{15}\right.\right.\right.\\ \left.\left.\left.\displaystyle-\frac{e^{2 i k c_{S}(\eta-\eta^{'})} }{120 (c_{S}k)^5 \eta^5}\left(
	4 (c_{S}k)^4 \eta^4-2 i (c_{S}k)^3 \eta^3\right.\right.\right.\right.\\ \left.\left.\left.\left.-2 (c_{S}k)^2 \eta^2
	+3 i c_{S}k \eta+6\right)\right]^{\tau}_{\tau^{'}}\right|^2\right]^{1/2}~e^{i\phi} & \mbox{\small {\bf for ~$|kc_{S}\eta|>>1$}}.
          \end{array}
\right.\nonumber
\eea
\bea &&\underline{\bf For~dS:}\\
\small\alpha_{diag}(k;\tau,\tau^{'}) &=& 
\left\{\begin{array}{ll}
               \sqrt{\left[ 1+ \left|  \frac{1}{
                    4}\left[\tau^{'2}-\tau^2\right]\right|^2\right]}~e^{i\phi}\nonumber
                     &
 \mbox{\small {\bf for ~$|kc_{S}\eta|<<1$}}  \\ \\
    \sqrt{\left[ 1+ \left| \frac{1}{
                    4\left(2\Delta\right)^{3/2}}\left[\tau^{'~2\sqrt{2\Delta}}-\tau^{2\sqrt{2\Delta}}
                    \right]\right|^2\right]}~e^{i\phi} &
 \mbox{\small {\bf for ~$|kc_{S}\eta|\approx 1-\Delta(\rightarrow 0)$}}  \\ \\
	 \sqrt{\left[ 1+ \left| \left[\frac{e^{-2 i kc_{S} \eta} 
	\left(2ikc_{S}\eta-1\right)}{4 (c_{S}k)^2 \eta^2}-\text{Ei}(-2 i kc_{S} \eta)
	\right]^{\tau}_{\tau^{'}}\right|^2\right]}~e^{i\phi} & \mbox{\small {\bf for ~$|kc_{S}\eta|>>1$}}.
          \end{array}
\right.\eea
\bea
&&\underline{\bf For~qdS:}\\
\small\alpha_{diag}(k;\tau,\tau^{'}) &=& 
\left\{\begin{array}{ll}
                  \sqrt{\left[ 1+ \left|  \frac{1}{
                    4\left(\nu^2-\frac{5}{4}\right)}\left[\tau^{'2\sqrt{\nu^2-\frac{5}{4}}}
                    -\tau^{2\sqrt{\nu^2-\frac{5}{4}}}\right]\right|^2\right]}~e^{i\phi} &
 \mbox{\small {\bf for ~$|kc_{S}\eta|<<1$}}  \\ \\
     \sqrt{\left[ 1+ \left|  \frac{\left[\nu^2-\frac{5}{4}\right]}{
                    4\left(2\Delta+\nu^2-\frac{9}{4}\right)^{3/2}}
                    \left[\tau^{'2\sqrt{2\Delta+\nu^2-\frac{9}{4}}}
                    -\tau^{2\sqrt{2\Delta+\nu^2-\frac{9}{4}}}\right]\right|^2\right]}~e^{i\phi}
                    &
 \mbox{\small {\bf for ~$|kc_{S}\eta|\approx 1-\Delta(\rightarrow 0)$}} \\ \\
	  \sqrt{\left[ 1+ \left|  \left[\nu^2-\frac{5}{4}\right]^2
	\left[\frac{e^{-2 i kc_{S} \eta} \left(2ikc_{S}\eta-1\right)}{4 (c_{S}k)^2 \eta^2}
	-\text{Ei}(-2 i kc_{S} \eta)\right]^{\tau}_{\tau^{'}}\right|^2\right]}~e^{i\phi} & \mbox{\small {\bf for ~$|kc_{S}\eta|>>1$}}.
          \end{array}
\right.\nonumber
\eea
Further using the expressions for Bogoliubov co-efficient $\alpha$ in two different representations
and substituting them in Eq~(\ref{sdq2}) we get the following
expressions for the reflection and transmission co-efficient in two different representations 
for three consecutive physical situations-$|kc_{S}\eta|<<1$, $|kc_{S}\eta|\approx 1-\Delta(\rightarrow 0)$ and
$|kc_{S}\eta|>>1$ for de Sitter and quasi de Sitter case as given by:
\bea &&\underline{\bf For~dS:}\\
{\cal R}&=& \footnotesize\left\{\begin{array}{ll}
                    \displaystyle  \frac{\frac{\eta^{'2}}{
                    8i}\left[\frac{1}{\tau^{'2}}-\frac{1}{\tau^2}\right]}{\sqrt{\left[ 1+ \left|\frac{\eta^{'2}}{
                    8i}\left[\frac{1}{\tau^{'2}}-\frac{1}{\tau^2}\right]\right|^2\right]}}~e^{-i\phi}
                     &
 \mbox{\small {\bf for ~$|kc_{S}\eta|<<1$}}  \\ 
 \displaystyle   \frac{ \frac{\eta^{'2\sqrt{2\Delta}}}{
                    8i\left(2\Delta\right)^{3}}\left[\frac{1}{\tau^{'~2\sqrt{2\Delta}}}-\frac{1}{\tau^{2\sqrt{2\Delta}}}
                    \right]}{\sqrt{\left[ 1+ \left| \frac{\eta^{'2\sqrt{2\Delta}}}{
                    8i\left(2\Delta\right)^{3}}\left[\frac{1}{\tau^{'~2\sqrt{2\Delta}}}-\frac{1}{\tau^{2\sqrt{2\Delta}}}
                    \right]\right|^2\right]}}~e^{-i\phi}  &
 \mbox{\small {\bf for ~$|kc_{S}\eta|\approx 1-\Delta(\rightarrow 0)$}}  \\ 
	\displaystyle 
\left[i\frac{\text{Ei}(2 ikc_{S} \eta)e^{-2ikc_{S}\eta^{'}}}{15}
-\frac{e^{2 i k c_{S}(\eta-\eta^{'})} }{120 (c_{S}k)^5 \eta^5}\left(
	4 (c_{S}k)^4 \eta^4-2 i (c_{S}k)^3 \eta^3\right.\right.\\ \left.\left.-2 (c_{S}k)^2 \eta^2+3 i c_{S}k \eta+6\right)\right]^{\tau}_{\tau^{'}}/\left[1+\left|\left[i\frac{\text{Ei}(2 ikc_{S} \eta)e^{-2ikc_{S}\eta^{'}}}{15}\right.\right.\right.\\ \left.\left.\left.  
\displaystyle-\frac{e^{2 i k c_{S}(\eta-\eta^{'})} }{120 (c_{S}k)^5 \eta^5}\left(
	4 (c_{S}k)^4 \eta^4-2 i (c_{S}k)^3 \eta^3\right.\right.\right.\right.\\ \left.\left.\left.\left. 
	-2 (c_{S}k)^2 \eta^2+3 i c_{S}k \eta+6\right)\right]^{\tau}_{\tau^{'}}\right|^2\right]^{1/2}~e^{-i\phi}& \mbox{\small {\bf for ~$|kc_{S}\eta|>>1$}}.
          \end{array}
\right.\nonumber\eea

\bea &&\underline{\bf For~dS:}\\
{\cal T}&=&\footnotesize\left\{\begin{array}{ll}
                    \displaystyle  \frac{1}{\sqrt{\left[ 1+ \left|\frac{\eta^{'2}}{
                    8i}\left[\frac{1}{\tau^{'2}}-\frac{1}{\tau^2}\right]\right|^2\right]}}~e^{-i\phi}
                     &
 \mbox{\small {\bf for ~$|kc_{S}\eta|<<1$}}  \\ 
 \displaystyle   \frac{1}{\sqrt{\left[ 1+ \left| \frac{\eta^{'2\sqrt{2\Delta}}}{
                    8i\left(2\Delta\right)^{3}}\left[\frac{1}{\tau^{'~2\sqrt{2\Delta}}}-\frac{1}{\tau^{2\sqrt{2\Delta}}}
                    \right]\right|^2\right]}}~e^{-i\phi}  &
 \mbox{\small {\bf for ~$|kc_{S}\eta|\approx 1-\Delta(\rightarrow 0)$}}  \\ 
	\displaystyle 
1/\left[1+\left|\left[i\frac{\text{Ei}(2 ikc_{S} \eta)e^{-2ikc_{S}\eta^{'}}}{15}\right.\right.\right.\\ \left.\left.\left.  
\displaystyle-\frac{e^{2 i k c_{S}(\eta-\eta^{'})} }{120 (c_{S}k)^5 \eta^5}\left(
	4 (c_{S}k)^4 \eta^4-2 i (c_{S}k)^3 \eta^3\right.\right.\right.\right.\\ \left.\left.\left.\left. 
	-2 (c_{S}k)^2 \eta^2+3 i c_{S}k \eta+6\right)\right]^{\tau}_{\tau^{'}}\right|^2\right]^{1/2}~e^{-i\phi}& \mbox{\small {\bf for ~$|kc_{S}\eta|>>1$}}.
          \end{array}
\right.\nonumber\eea

\bea &&\underline{\bf For~qdS:}\\
{\cal R}&=&\footnotesize \left\{\begin{array}{ll}
                    \displaystyle  \frac{ \frac{\eta^{'2\sqrt{\nu^2-\frac{5}{4}}}}{
                    8i\left(\nu^2-\frac{5}{4}\right)}\left[\frac{1}{\tau^{'2\sqrt{\nu^2-\frac{5}{4}}}}-\frac{1}{\tau^{2\sqrt{\nu^2-\frac{5}{4}}}}\right]}{\sqrt{\left[ 1+ \left|\frac{\eta^{'2}}{
                    8i}\left[\frac{1}{\tau^{'2}}-\frac{1}{\tau^2}\right]\right|^2\right]}}~e^{-i\phi}
                     &
 \mbox{\small {\bf for ~$|kc_{S}\eta|<<1$}}  \\ 
 \displaystyle   \frac{  \frac{\eta^{'2\sqrt{2\Delta+\nu^2-\frac{9}{4}}}\left[\nu^2-\frac{5}{4}\right]^2}{
                    8i\left(2\Delta+\nu^2-\frac{9}{4}\right)^3}
                    \left[\frac{1}{\tau^{'2\sqrt{2\Delta+\nu^2-\frac{9}{4}}}}
                    -\frac{1}{\tau^{2\sqrt{2\Delta+\nu^2-\frac{9}{4}}}}\right]}{\sqrt{\left[ 1+ \left| \frac{\eta^{'2\sqrt{2\Delta}}}{
                    8i\left(2\Delta\right)^{3}}\left[\frac{1}{\tau^{'~2\sqrt{2\Delta}}}-\frac{1}{\tau^{2\sqrt{2\Delta}}}
                    \right]\right|^2\right]}}~e^{-i\phi}  &
 \mbox{\small {\bf for ~$|kc_{S}\eta|\approx 1-\Delta(\rightarrow 0)$}}  \\ 
	\displaystyle 
\left[\nu^2-\frac{5}{4}\right]^2 \left[i\frac{\text{Ei}(2 ikc_{S} \eta)e^{-2ikc_{S}\eta^{'}}}{15}
-\frac{e^{2 i k c_{S}(\eta-\eta^{'})} }{120 (c_{S}k)^5 \eta^5}\left(
	4 (c_{S}k)^4 \eta^4-2 i (c_{S}k)^3 \eta^3\right.\right.\\ \left.\left.-2 (c_{S}k)^2 \eta^2+3 i c_{S}k \eta+6\right)\right]^{\tau}_{\tau^{'}}/\left[1+\left|\left[i\frac{\text{Ei}(2 ikc_{S} \eta)e^{-2ikc_{S}\eta^{'}}}{15}\right.\right.\right.\\ \left.\left.\left.  
\displaystyle-\frac{e^{2 i k c_{S}(\eta-\eta^{'})} }{120 (c_{S}k)^5 \eta^5}\left(
	4 (c_{S}k)^4 \eta^4-2 i (c_{S}k)^3 \eta^3\right.\right.\right.\right.\\ \left.\left.\left.\left. 
	-2 (c_{S}k)^2 \eta^2+3 i c_{S}k \eta+6\right)\right]^{\tau}_{\tau^{'}}\right|^2\right]^{1/2}~e^{-i\phi}& \mbox{\small {\bf for ~$|kc_{S}\eta|>>1$}}.
          \end{array}
\right.\nonumber\eea

\bea &&\underline{\bf For~qdS:}\\
{\cal T}&=&\footnotesize\left\{\begin{array}{ll}
                    \displaystyle  \frac{1}{\sqrt{\left[ 1+ \left| \frac{\eta^{'2\sqrt{\nu^2-\frac{5}{4}}}}{
                    8i\left(\nu^2-\frac{5}{4}\right)}\left[\frac{1}{\tau^{'2\sqrt{\nu^2-\frac{5}{4}}}}
                    -\frac{1}{\tau^{2\sqrt{\nu^2-\frac{5}{4}}}}\right]\right|^2\right]}}~e^{-i\phi}
                     &
 \mbox{\small {\bf for ~$|kc_{S}\eta|<<1$}}  \\ 
 \displaystyle   \frac{1}{\sqrt{\left[ 1+ \left| \frac{\eta^{'2\sqrt{2\Delta+\nu^2-\frac{9}{4}}}\left[\nu^2-\frac{5}{4}\right]^2}{
                    8i\left(2\Delta+\nu^2-\frac{9}{4}\right)^3}
                    \left[\frac{1}{\tau^{'2\sqrt{2\Delta+\nu^2-\frac{9}{4}}}}
                    -\frac{1}{\tau^{2\sqrt{2\Delta+\nu^2-\frac{9}{4}}}}\right]\right|^2\right]}}~e^{-i\phi}  &
 \mbox{\small {\bf for ~$|kc_{S}\eta|\approx 1-\Delta(\rightarrow 0)$}}  \\ 
	\displaystyle 
1/\left[1+\left|\left[\nu^2-\frac{5}{4}\right]^2 \left[i\frac{\text{Ei}(2 ikc_{S} \eta)e^{-2ikc_{S}\eta^{'}}}{15}
-\frac{e^{2 i k c_{S}(\eta-\eta^{'})} }{120 (c_{S}k)^5 \eta^5}\left(
	4 (c_{S}k)^4 \eta^4-2 i (c_{S}k)^3 \eta^3\right.\right.\right.\right.\\ \left.\left.\left.\left.-2 (c_{S}k)^2 \eta^2
	+3 i c_{S}k \eta+6\right)\right]^{\tau}_{\tau^{'}}\right|^2\right]^{1/2}~e^{-i\phi}& \mbox{\small {\bf for ~$|kc_{S}\eta|>>1$}}.
          \end{array}
\right.\nonumber\eea
and 
\bea &&\underline{\bf For~dS:}\\
\small{\cal R}_{diag}(k;\tau,\tau^{'})&=& \footnotesize\left\{\begin{array}{ll}
                    \displaystyle  \frac{\frac{1}{
                    4}\left[\tau^{'2}-\tau^2\right]}{\sqrt{\left[ 1+ \left|\frac{\eta^{'2}}{
                    8i}\left[\frac{1}{\tau^{'2}}-\frac{1}{\tau^2}\right]\right|^2\right]}}~e^{-i\phi}
                     &
 \mbox{\small {\bf for ~$|kc_{S}\eta|<<1$}}  \\ 
 \displaystyle   \frac{ \frac{1}{
                    4\left(2\Delta\right)^{3/2}}\left[\tau^{'~2\sqrt{2\Delta}}-\tau^{2\sqrt{2\Delta}}
                    \right]}{  \sqrt{\left[ 1+ \left| \frac{1}{
                    4\left(2\Delta\right)^{3/2}}\left[\tau^{'~2\sqrt{2\Delta}}-\tau^{2\sqrt{2\Delta}}
                    \right]\right|^2\right]}}~e^{-i\phi}  &
 \mbox{\small {\bf for ~$|kc_{S}\eta|\approx 1-\Delta(\rightarrow 0)$}}  \\ 
	\displaystyle 
\frac{\left[\frac{e^{-2 i kc_{S} \eta} \left(2ikc_{S}\eta-1\right)}{4 (c_{S}k)^2 \eta^2}
-\text{Ei}(-2 i kc_{S} \eta)\right]^{\tau}_{\tau^{'}}}{\sqrt{\left[ 1+ \left| \left[\frac{e^{-2 i kc_{S} \eta} 
	\left(2ikc_{S}\eta-1\right)}{4 (c_{S}k)^2 \eta^2}-\text{Ei}(-2 i kc_{S} \eta)
	\right]^{\tau}_{\tau^{'}}\right|^2\right]}}~e^{-i\phi}& \mbox{\small {\bf for ~$|kc_{S}\eta|>>1$}}.
          \end{array}
\right.\nonumber\eea

\bea &&\underline{\bf For~dS:}\\
\small{\cal T}_{diag}(k;\tau,\tau^{'})&=& \footnotesize\left\{\begin{array}{ll}
                    \displaystyle  \frac{1}{ \sqrt{\left[ 1+ \left|  \frac{1}{
                    4}\left[\tau^{'2}-\tau^2\right]\right|^2\right]}}~e^{-i\phi}
                     &
 \mbox{\small {\bf for ~$|kc_{S}\eta|<<1$}}  \\ 
 \displaystyle   \frac{1}{ \sqrt{\left[ 1+ \left| \frac{1}{
                    4\left(2\Delta\right)^{3/2}}\left[\tau^{'~2\sqrt{2\Delta}}-\tau^{2\sqrt{2\Delta}}
                    \right]\right|^2\right]}}~e^{-i\phi}  &
 \mbox{\small {\bf for ~$|kc_{S}\eta|\approx 1-\Delta(\rightarrow 0)$}}  \\ 
	\displaystyle 
\frac{1}{\sqrt{\left[ 1+ \left| \left[\frac{e^{-2 i kc_{S} \eta} 
	\left(2ikc_{S}\eta-1\right)}{4 (c_{S}k)^2 \eta^2}-\text{Ei}(-2 i kc_{S} \eta)
	\right]^{\tau}_{\tau^{'}}\right|^2\right]}}~e^{-i\phi}& \mbox{\small {\bf for ~$|kc_{S}\eta|>>1$}}.
          \end{array}
\right.\nonumber\eea

\bea &&\underline{\bf For~qdS:}\\
\small{\cal R}_{diag}(k;\tau,\tau^{'}) 
&=& \footnotesize\left\{\begin{array}{ll}
                    \displaystyle  \frac{\frac{1}{
                    4\left(\nu^2-\frac{5}{4}\right)}\left[\tau^{'2\sqrt{\nu^2-\frac{5}{4}}}-\tau^{2\sqrt{\nu^2-\frac{5}{4}}}\right]}{\sqrt{\left[ 1+ \left|  \frac{1}{
                    4\left(\nu^2-\frac{5}{4}\right)}\left[\tau^{'2\sqrt{\nu^2-\frac{5}{4}}}
                    -\tau^{2\sqrt{\nu^2-\frac{5}{4}}}\right]\right|^2\right]}}~e^{-i\phi}
                     &
 \mbox{\small {\bf for ~$|kc_{S}\eta|<<1$}}  \\ 
 \displaystyle   \frac{  \frac{\left[\nu^2-\frac{5}{4}\right]}{
                    4\left(2\Delta+\nu^2-\frac{9}{4}\right)^{3/2}}
                    \left[\tau^{'2\sqrt{2\Delta+\nu^2-\frac{9}{4}}}
                    -\tau^{2\sqrt{2\Delta+\nu^2-\frac{9}{4}}}\right]}{\sqrt{\left[ 1+ \left|  \frac{\left[\nu^2-\frac{5}{4}\right]}{
                    4\left(2\Delta+\nu^2-\frac{9}{4}\right)^{3/2}}
                    \left[\tau^{'2\sqrt{2\Delta+\nu^2-\frac{9}{4}}}
                    -\tau^{2\sqrt{2\Delta+\nu^2-\frac{9}{4}}}\right]\right|^2\right]}}~e^{-i\phi}  &
 \mbox{\small {\bf for ~$|kc_{S}\eta|\approx 1-\Delta(\rightarrow 0)$}}  \\ 
	\displaystyle 
\frac{\left[\nu^2-\frac{5}{4}\right]^2\left[\frac{e^{-2 i kc_{S} \eta} \left(2ikc_{S}\eta-1\right)}{4 (c_{S}k)^2 \eta^2}-\text{Ei}(-2 i kc_{S} \eta)\right]^{\tau}_{\tau^{'}}}{ \sqrt{\left[ 1+ \left|  \left[\nu^2-\frac{5}{4}\right]^2
	\left[\frac{e^{-2 i kc_{S} \eta} \left(2ikc_{S}\eta-1\right)}{4 (c_{S}k)^2 \eta^2}
	-\text{Ei}(-2 i kc_{S} \eta)\right]^{\tau}_{\tau^{'}}\right|^2\right]}}~e^{-i\phi}& \mbox{\small {\bf for ~$|kc_{S}\eta|>>1$}}.
          \end{array}
\right.\nonumber\eea

\bea &&\underline{\bf For~qdS:}\\
\small{\cal T}_{diag}(k;\tau,\tau^{'}) &=&\footnotesize\left\{\begin{array}{ll}
                    \displaystyle  \frac{1}{ \sqrt{\left[ 1+ \left|  \frac{1}{
                    4\left(\nu^2-\frac{5}{4}\right)}\left[\tau^{'2\sqrt{\nu^2-\frac{5}{4}}}
                    -\tau^{2\sqrt{\nu^2-\frac{5}{4}}}\right]\right|^2\right]}}~e^{-i\phi}
                     &
 \mbox{\small {\bf for ~$|kc_{S}\eta|<<1$}}  \\ 
 \displaystyle   \frac{1}{ \sqrt{\left[ 1+ \left|  \frac{\left[\nu^2-\frac{5}{4}\right]}{
                    4\left(2\Delta+\nu^2-\frac{9}{4}\right)^{3/2}}
                    \left[\tau^{'2\sqrt{2\Delta+\nu^2-\frac{9}{4}}}
                    -\tau^{2\sqrt{2\Delta+\nu^2-\frac{9}{4}}}\right]\right|^2\right]}}~e^{-i\phi}  &
 \mbox{\small {\bf for ~$|kc_{S}\eta|\approx 1-\Delta(\rightarrow 0)$}}  \\ 
	\displaystyle 
\frac{1}{ \sqrt{\left[ 1+ \left|  \left[\nu^2-\frac{5}{4}\right]^2
	\left[\frac{e^{-2 i kc_{S} \eta} \left(2ikc_{S}\eta-1\right)}{4 (c_{S}k)^2 \eta^2}
	-\text{Ei}(-2 i kc_{S} \eta)\right]^{\tau}_{\tau^{'}}\right|^2\right]}}~e^{-i\phi}& \mbox{\small {\bf for ~$|kc_{S}\eta|>>1$}}.
          \end{array}
\right.\nonumber\eea
 Next the expression for the
number of produced particles at time $\tau$ in the two representations can be calculated for de~Sitter and quasi de sitter as:
\be%
\ee
which is not exactly analytically computable. To study the behaviour of this integral we consider here three 
consecutive physical situations-$|kc_{S}\eta|<<1$, $|kc_{S}\eta|\approx 1-\Delta(\rightarrow 0)$ and $|kc_{S}\eta|>>1$ for de Sitter and quasi de Sitter case. 
In three cases we have:
\bea &&\underline{\bf For~dS:}\\
\small{\cal N}(\tau,\tau^{'},\eta^{'})&=& \footnotesize\left\{\begin{array}{ll}
                    \displaystyle  \frac{\eta^{'4} V}{
                    64(2\pi a)^3}\left[\frac{1}{\tau^{'2}}-\frac{1}{\tau^2}\right]^2
                    ~~&
 \mbox{\small {\bf for ~$|kc_{S}\eta|<<1$}}  \\ 
 \displaystyle  \frac{\eta^{'4\sqrt{2\Delta}}V}{
                    64(2\pi a)^3\left(2\Delta\right)^{6}}\left[\frac{1}{\tau^{'~2\sqrt{2\Delta}}}-\frac{1}{\tau^{2\sqrt{2\Delta}}}
                    \right]^2 &
 \mbox{\small {\bf for ~$|kc_{S}\eta|\approx 1-\Delta(\rightarrow 0)$}}  \\ 
	 \frac{4\pi}{(2\pi a)^3}\int^{\infty}_{0}dk~k^2\left|\left\{\frac{i~e^{-2ikc_{S}\eta^{'}}}{15}\left\{\text{Ei}(2 ikc_{S} \tau)-\text{Ei}(2 ik c_{S}\tau^{'})\right\}
	\right.\right.\\\left.\left.-\frac{e^{-2 ikc_{S}\eta^{'}}}{120k^5}\left[\frac{ e^{2ikc_{S}\tau}\left(
	4 (c_{S}k)^4 \tau^4-2 i(c_{S}k)^3 \tau^3-2 (c_{S}k)^2 \tau^2+3 ikc_{S} \tau+6\right)}{\tau^5}\right.\right.\right.\\\left.\left.\left.
	~~-\frac{ e^{2kc_{S}\tau}\left(
	4 (c_{S}k)^4 \tau^{'4}-2 i(c_{S}k)^3 \tau^{'3}-2 (c_{S}k)^2 \tau^{'2}+3 ikc_{S} \tau^{'}+6\right)}{\tau^{'5}}\right]\right\}\right|^2~~~ & \mbox{\small {\bf for ~$|kc_{S}\eta|>>1$}}.
          \end{array}
\right.\nonumber\eea
\bea
&&\underline{\bf For~qdS:}\\
\small{\cal N}(\tau,\tau^{'},\eta^{'})&=&\footnotesize\left\{\begin{array}{ll}
                    \displaystyle  \frac{\eta^{'4\sqrt{\nu^2-\frac{5}{4}}}V}{
                    64(2\pi a)^3\left(\nu^2-\frac{5}{4}\right)^2}\left[\frac{1}{\tau^{'2\sqrt{\nu^2-\frac{5}{4}}}}-\frac{1}{\tau^{2\sqrt{\nu^2-\frac{5}{4}}}}\right]^2~~ &
 \mbox{\small {\bf for ~$|kc_{S}\eta|<<1$}}  \\ 
 \displaystyle   \frac{\eta^{'4\sqrt{2\Delta+\nu^2-\frac{9}{4}}}V \left[\nu^2-\frac{5}{4}\right]^4}{
                    64(2\pi a)^3\left(2\Delta+\nu^2-\frac{9}{4}\right)^6}
                    \left[\frac{1}{\tau^{'2\sqrt{2\Delta+\nu^2-\frac{9}{4}}}}
                    -\frac{1}{\tau^{2\sqrt{2\Delta+\nu^2-\frac{9}{4}}}}\right]^2
                    ~~&
 \mbox{\small {\bf for ~$|kc_{S}\eta|\approx 1-\Delta(\rightarrow 0)$}} \\ 
	\displaystyle \frac{4\pi}{(2\pi a)^3}\left[\nu^2-\frac{5}{4}\right]^4 \int^{\infty}_{0}dk~k^2 
	\left|\left\{\frac{i~e^{-2ikc_{S}\eta^{'}}}{15}\left\{\text{Ei}(2 ikc_{S} \tau)-\text{Ei}(2 ik c_{S}\tau^{'})\right\}
	\right.\right.\\\left.\left.\displaystyle -\frac{e^{-2 ikc_{S}\eta^{'}}}{120k^5}\left[\frac{ e^{2ikc_{S}\tau}\left(
	4 (c_{S}k)^4 \tau^4-2 i(c_{S}k)^3 \tau^3-2 (c_{S}k)^2 \tau^2+3 ikc_{S} \tau+6\right)}{\tau^5}\right.
	\right.\right.\\\left.\left.\left. \displaystyle 
	~~-\frac{ e^{2kc_{S}\tau}\left(
	4 (c_{S}k)^4 \tau^{'4}-2 i(c_{S}k)^3 \tau^{'3}-2 (c_{S}k)^2 \tau^{'2}+3 ikc_{S} \tau^{'}+6\right)}{\tau^{'5}}\right]\right\}\right|^2~~ & \mbox{\small {\bf for ~$|kc_{S}\eta|>>1$}}.~~~~~~~~
          \end{array}
\right.\nonumber
\eea
\bea &&\underline{\bf For~dS:}\\
\small{\cal N}_{diag}(\tau,\tau^{'}) &=& \footnotesize\left\{\begin{array}{ll}
                    \displaystyle  \frac{V}{
                    16(2\pi a)^3}\left[\tau^{'2}-\tau^2\right]^2
                    ~~~~ &
 \mbox{\small {\bf for ~$|kc_{S}\eta|<<1$}}  \\ 
 \displaystyle  \frac{V}{
                    16(2\pi a)^3\left(2\Delta\right)^{3}}\left[\tau^{'~2\sqrt{2\Delta}}-\tau^{2\sqrt{2\Delta}}
                    \right]^2~~~~ &
 \mbox{\small {\bf for ~$|kc_{S}\eta|\approx 1-\Delta(\rightarrow 0)$}}  \\ \nonumber
	\displaystyle \frac{4\pi}{(2\pi a)^3}\int^{\infty}_{0}dk~k^2\left|\left[\left\{\frac{e^{-2 i k c_{S}\tau} \left(2ikc_{S}\tau-1\right)}{4 (c_{S}k)^2 \tau^2}
	-\frac{e^{-2 i k c_{S}\tau^{'}} \left(2ikc_{S}\tau^{'}-1\right)}{4 (c_{S}k)^2 \tau^{'2}}\right\}
	\right.\right.\\ \left.\left.~~~~~~~~~-\left\{\text{Ei}(-2 i k c_{S}\tau)-\text{Ei}(-2 i k c_{S}\tau^{'})\right\}\right]\right|^2~~~~~~~~ & \mbox{\small {\bf for ~$|kc_{S}\eta|>>1$}}.~~~~~~~~
          \end{array}
\right.\eea 
\bea
&&\underline{\bf For~qdS:}\\
\small{\cal N}_{diag}(\tau,\tau^{'}) &=& \footnotesize\left\{\begin{array}{ll}
                    \displaystyle   \frac{V}{
                    16(2\pi a)^3\left(\nu^2-\frac{5}{4}\right)^2}\left[\tau^{'2\sqrt{\nu^2-\frac{5}{4}}}-\tau^{2\sqrt{\nu^2-\frac{5}{4}}}\right]^2 &
 \mbox{\small {\bf for ~$|kc_{S}\eta|<<1$}}  \\ 
 \displaystyle  \frac{V\left[\nu^2-\frac{5}{4}\right]^2}{
                    16(2\pi a)^3\left(2\Delta+\nu^2-\frac{9}{4}\right)^{3}}
                    \left[\tau^{'2\sqrt{2\Delta+\nu^2-\frac{9}{4}}}
                    -\tau^{2\sqrt{2\Delta+\nu^2-\frac{9}{4}}}\right]^2
                    &
 \mbox{\small {\bf for ~$|kc_{S}\eta|\approx 1-\Delta(\rightarrow 0)$}} \\ 
	\displaystyle \frac{4\pi}{(2\pi a)^3}\left[\nu^2-\frac{5}{4}\right]^4\int^{\infty}_{0}dk~k^2\left|\left[\left\{\frac{e^{-2 i k c_{S}\tau}
	\left(2ikc_{S}\tau-1\right)}{4 (c_{S}k)^2 \tau^2}\right.\right.\right.\\ \left.\left.\left.\displaystyle~~~~~~~~~~~~~~~~~~-\frac{e^{-2 i k c_{S}\tau^{'}} \left(2ikc_{S}\tau^{'}-1\right)}{4 (c_{S}k)^2 \tau^{'2}}\right\}
	\right.\right.\\ \left.\left.~~~~~~~~~-\left\{\text{Ei}(-2 i k c_{S}\tau)-\text{Ei}(-2 i k c_{S}\tau^{'})\right\}\right]\right|^2 & \mbox{\small {\bf for ~$|kc_{S}\eta|>>1$}}.
          \end{array}
\right.\nonumber
\eea
Throughout the discussion of total number of particle production we have introduced a symbol $V$ defined as:
\be V=\int d^3 {\bf k},\ee
which physically signifies the total finite volume in momentum space within which the produced particles are occupied. 

Finally one can define the total energy density of the produced particles using the following expression:
\be%
\ee
which is not exactly analytically computable. To study the behaviour of this integral we consider here three 
consecutive physical situations-$|kc_{S}\eta|<<1$, $|kc_{S}\eta|\approx 1-\Delta(\rightarrow 0)$ and $|kc_{S}\eta|>>1$ for de Sitter and quasi de Sitter case. 
In three cases we have:
\bea &&\underline{\bf For~dS:}\\
\small\rho(\tau,\tau^{'},\eta^{'})&=& \footnotesize\left\{\begin{array}{ll}
                    \displaystyle  \frac{\eta^{'4}J}{
                    64(2\pi a)^3a }\left[\frac{1}{\tau^{'2}}-\frac{1}{\tau^2}\right]^2
                    ~~&
 \mbox{\small {\bf for ~$|kc_{S}\eta|<<1$}}  \\ 
 \displaystyle  \frac{\eta^{'4\sqrt{2\Delta}}J}{
                    64(2\pi a)^3a\left(2\Delta\right)^{6}}\left[\frac{1}{\tau^{'~2\sqrt{2\Delta}}}-\frac{1}{\tau^{2\sqrt{2\Delta}}}
                    \right]^2~~~~ &
 \mbox{\small {\bf for ~$|kc_{S}\eta|\approx 1-\Delta(\rightarrow 0)$}}  \\ \nonumber
	 \frac{4\pi c_{S}}{(2\pi a)^3a}\int^{\infty}_{0}dk~k^3
	\left|\left\{\frac{i~e^{-2ikc_{S}\eta^{'}}}{15}\left\{\text{Ei}(2 ik c_{S}\tau)-\text{Ei}(2 ik c_{S}\tau^{'})\right\}
	\right.\right.\\\left.\left.-\frac{e^{-2 ikc_{S}\eta^{'}}}{120k^5}\left[\frac{ e^{2ikc_{S}\tau}\left(
	4 (c_{S}k)^4 \tau^4-2 i(c_{S}k)^3 \tau^3-2 (c_{S}k)^2 \tau^2+3 i(c_{S}k) \tau+6\right)}{\tau^5}
	\right.\right.\right.\\\left.\left.\left. 
	~~-\frac{ e^{2kc_{S}\tau^{'}}\left(
	4 (c_{S}k)^4 \tau^{'4}-2 i(c_{S}k)^3 
	\tau^{'3}-2 k^2 \tau^{'2}+3 ik \tau^{'}+6\right)}{\tau^{'5}}\right]\right\}\right|^2 & \mbox{\small {\bf for ~$|kc_{S}\eta|>>1$}}.~~~~~~~~
          \end{array}
\right.\nonumber\eea
\bea
&&\underline{\bf For~qdS:}\\
\small\rho(\tau,\tau^{'},\eta^{'})&=&\footnotesize\left\{\begin{array}{ll}
                    \displaystyle  \frac{\eta^{'4\sqrt{\nu^2-\frac{5}{4}}}J}{
                    64(2\pi a)^3a\left(\nu^2-\frac{5}{4}\right)^{2}}\left[\frac{1}{\tau^{'2\sqrt{\nu^2-\frac{5}{4}}}}-\frac{1}{\tau^{2\sqrt{\nu^2-\frac{5}{4}}}}\right]^2~~ &
 \mbox{\small {\bf for ~$|kc_{S}\eta|<<1$}}  \\ 
 \displaystyle   \frac{\eta^{'4\sqrt{2\Delta+\nu^2-\frac{9}{4}}}J \left[\nu^2-\frac{5}{4}\right]^4}{
                    64(2\pi a)^3a\left(2\Delta+\nu^2-\frac{9}{4}\right)^{6}}
                    \left[\frac{1}{\tau^{'2\sqrt{2\Delta+\nu^2-\frac{9}{4}}}}
                    -\frac{1}{\tau^{2\sqrt{2\Delta+\nu^2-\frac{9}{4}}}}\right]^2
                    ~~&
 \mbox{\small {\bf for ~$|kc_{S}\eta|\approx 1-\Delta(\rightarrow 0)$}} \\ 
	\displaystyle \frac{4\pi c_{S}}{(2\pi a)^3a}\left[\nu^2-\frac{5}{4}\right]^4 \int^{\infty}_{0}dk~k^3
	  \left|\left\{\frac{i~e^{-2ikc_{S}\eta^{'}}}{15}\left\{\text{Ei}(2 ik c_{S}\tau)-\text{Ei}(2 ik c_{S}\tau^{'})\right\}
	\right.\right.\\\left.\left.-\frac{e^{-2 ikc_{S}\eta^{'}}}{120k^5}\left[\frac{ e^{2ikc_{S}\tau}\left(
	4 (c_{S}k)^4 \tau^4-2 i(c_{S}k)^3 \tau^3-2 (c_{S}k)^2 \tau^2+3 i(c_{S}k) \tau+6\right)}{\tau^5}
	\right.\right.\right.\\\left.\left.\left. 
	~~-\frac{ e^{2kc_{S}\tau^{'}}\left(
	4 (c_{S}k)^4 \tau^{'4}-2 i(c_{S}k)^3 \tau^{'3}-2 k^2 \tau^{'2}+3 ik \tau^{'}+6\right)}{\tau^{'5}}\right]\right\}\right|^2~~ & \mbox{\small {\bf for ~$|kc_{S}\eta|>>1$}}.~~~~~~~~
          \end{array}
\right.\nonumber
\eea
\bea &&\underline{\bf For~dS:}\\
\small\rho_{diag}(\tau,\tau^{'})&=&\footnotesize \left\{\begin{array}{ll}
                    \displaystyle  \frac{J}{
                    16(2\pi a)^3a}\left[\tau^{'2}-\tau^2\right]^2
                    ~~&
 \mbox{\small {\bf for ~$|kc_{S}\eta|<<1$}}  \\ 
 \displaystyle  \frac{J}{
                    16(2\pi a)^3a\left(2\Delta\right)^{3}}\left[\tau^{'~2\sqrt{2\Delta}}-\tau^{2\sqrt{2\Delta}}
                    \right]^2~~~~ &
 \mbox{\small {\bf for ~$|kc_{S}\eta|\approx 1-\Delta(\rightarrow 0)$}}  \\ 
	\displaystyle \frac{4\pi c_{S}}{(2\pi a)^3a}\int^{\infty}_{0}dk~k^3\left|\frac{e^{-2ikc_{S}\tau}(2ikc_{S}\tau-1)}{4(c_{S}k\tau)^2}-\frac{e^{-2ikc_{S}\tau^{'}}(2ikc_{S}\tau^{'}-1)}{4(c_{S}k\tau^{'})^2}
	\right.\\ \left. +\left\{\text{Ei}(-2 ik c_{S}\tau)-\text{Ei}(-2 ik c_{S}\tau^{'})\right\}\right|^2 & \mbox{\small {\bf for ~$|kc_{S}\eta|>>1$}}.~~~~~~~~
          \end{array}
\right.\nonumber\eea
\bea
&&\underline{\bf For~qdS:}\\
\small\rho_{diag}(\tau,\tau^{'},\eta^{'})&=&\footnotesize\left\{\begin{array}{ll}
                    \displaystyle   \frac{J}{
                    16(2\pi a)^3a}\left[\tau^{'2}-\tau^2\right]^2~~ &
 \mbox{\small {\bf for ~$|kc_{S}\eta|<<1$}}  \\ 
 \displaystyle   \frac{J}{
                    16(2\pi a)^3a\left(2\Delta\right)^{3}}\left[\tau^{'~2\sqrt{2\Delta}}-\tau^{2\sqrt{2\Delta}}
                    \right]^2
                    ~~&
 \mbox{\small {\bf for ~$|kc_{S}\eta|\approx 1-\Delta(\rightarrow 0)$}} \\ 
	\displaystyle \frac{4\pi c_{S}}{(2\pi a)^3a}\left[\nu^2-\frac{5}{4}\right]^4\int^{\infty}_{0}dk~k^3\left|\frac{e^{-2ikc_{S}\tau}(2ikc_{S}\tau-1)}{4(c_{S}k\tau)^2}-\frac{e^{-2ikc_{S}\tau^{'}}(2ikc_{S}\tau^{'}-1)}{4(c_{S}k\tau^{'})^2}
	\right.\\ \left. +\left\{\text{Ei}(-2 ik c_{S}\tau)-\text{Ei}(-2 ik c_{S}\tau^{'})\right\}\right|^2~~ & \mbox{\small {\bf for ~$|kc_{S}\eta|>>1$}}.~~~~~~~~
          \end{array}
\right.\nonumber
\eea
\begin{figure*}[htb]
\centering
\subfigure[]{
    \includegraphics[width=7.2cm,height=6cm] {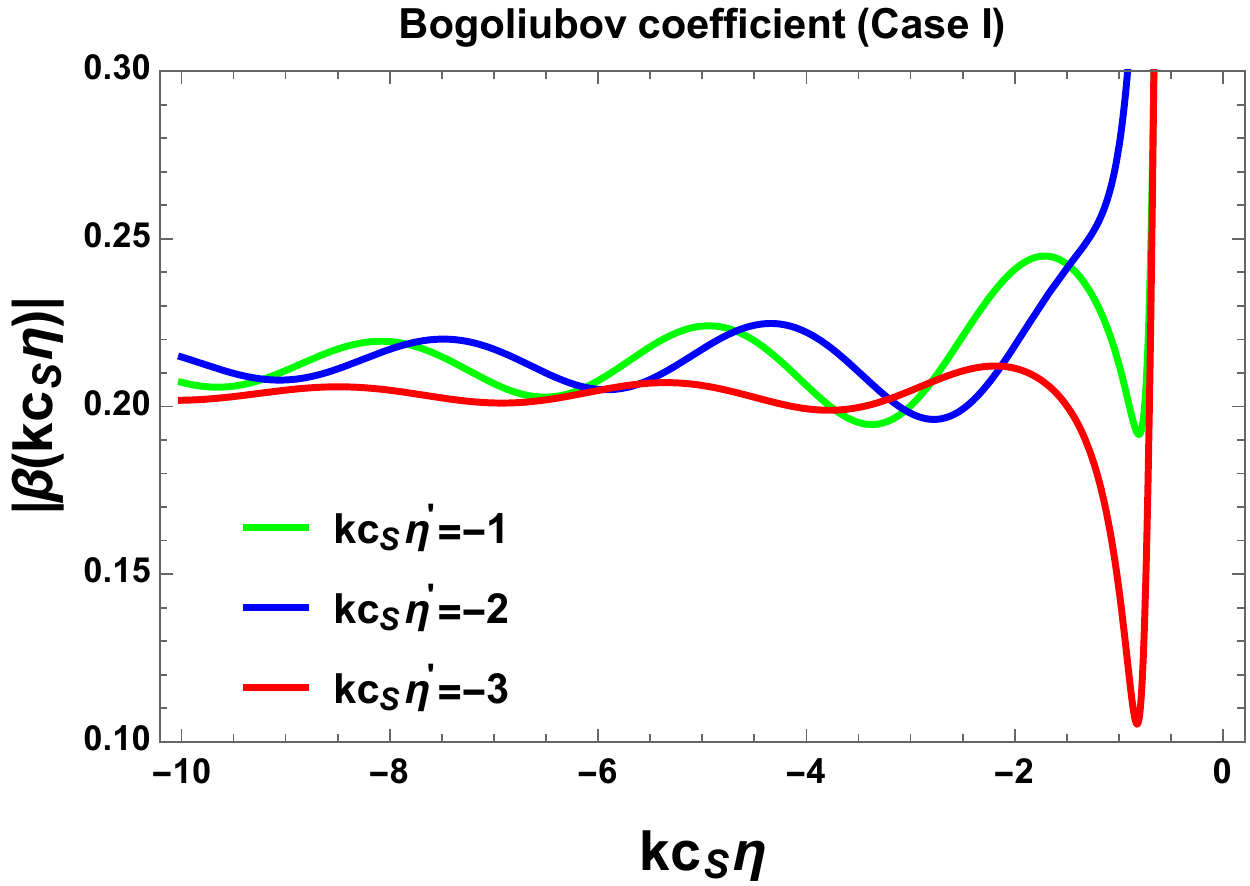}
    \label{fig1xcc1}
}
\subfigure[]{
    \includegraphics[width=7.2cm,height=6cm] {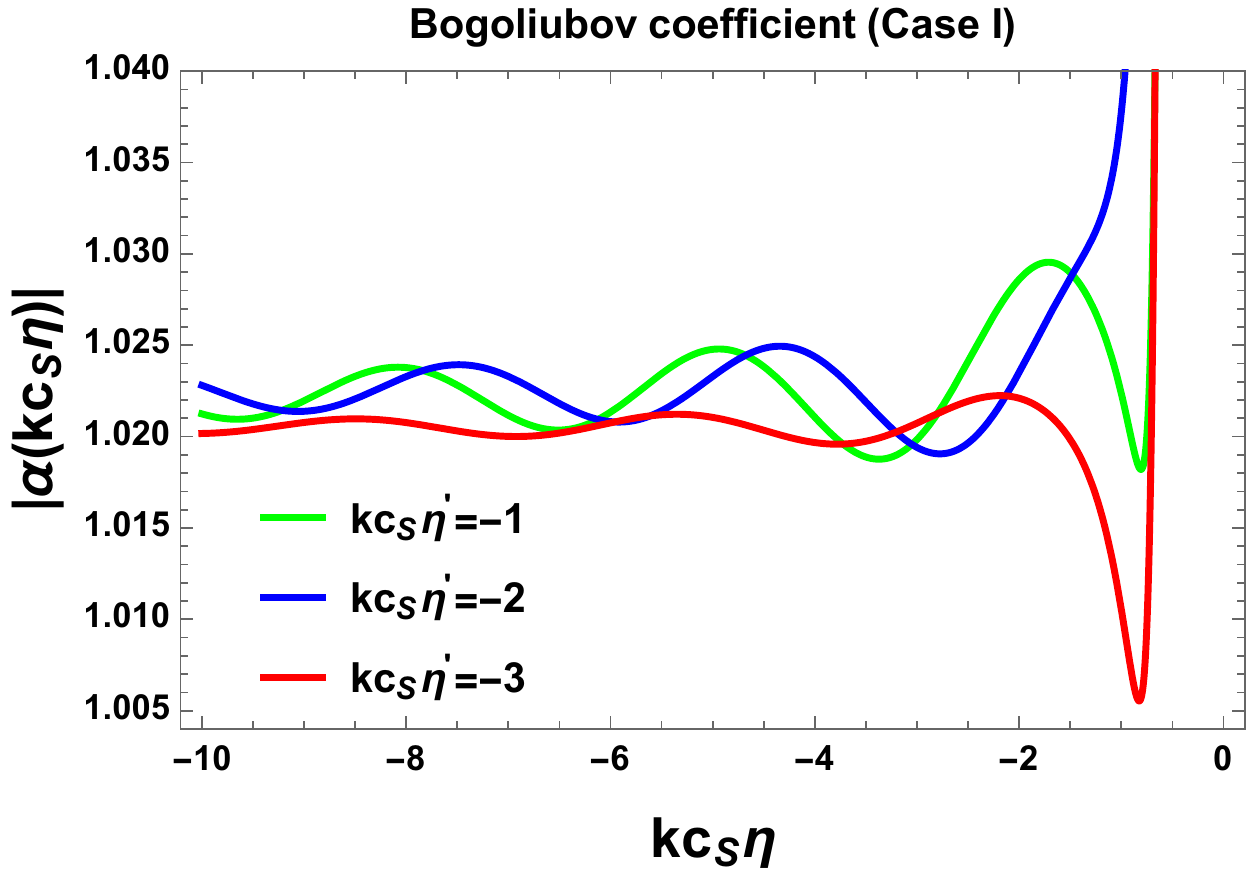}
    \label{fig2xcc2}
}
\subfigure[]{
    \includegraphics[width=7.2cm,height=6cm] {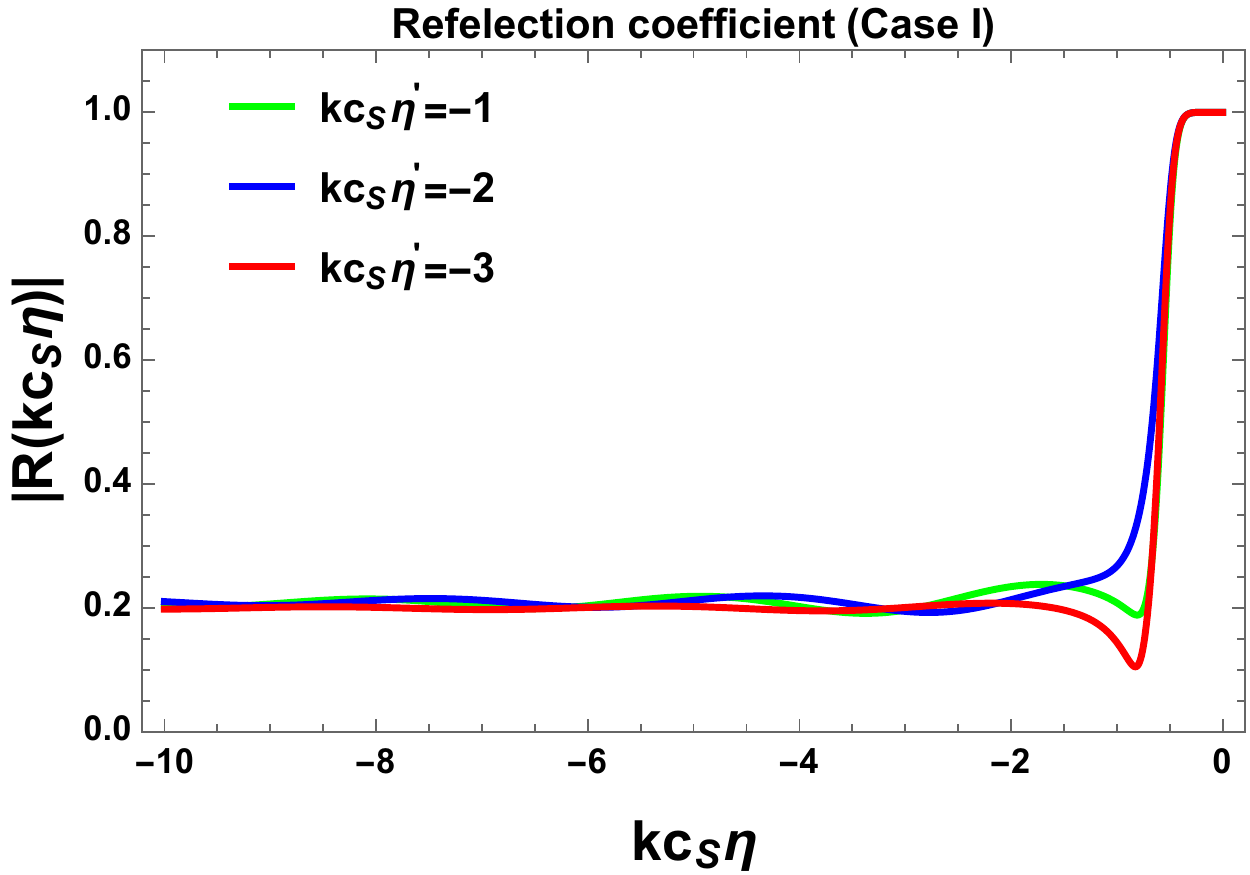}
    \label{fig3xcc3}
}
\subfigure[]{
    \includegraphics[width=7.2cm,height=6cm] {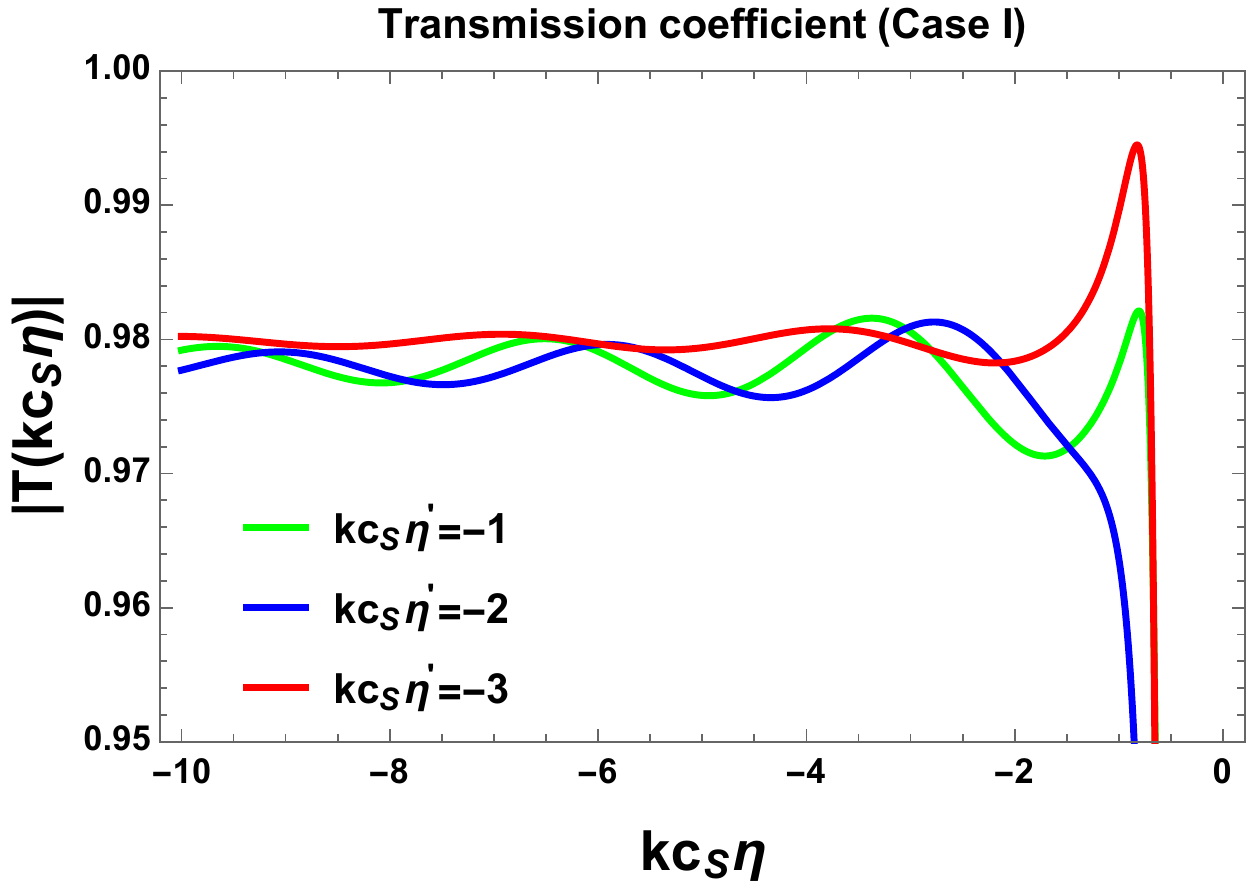}
    \label{fig4xcc4}
}
\subfigure[]{
    \includegraphics[width=7.2cm,height=6.1cm] {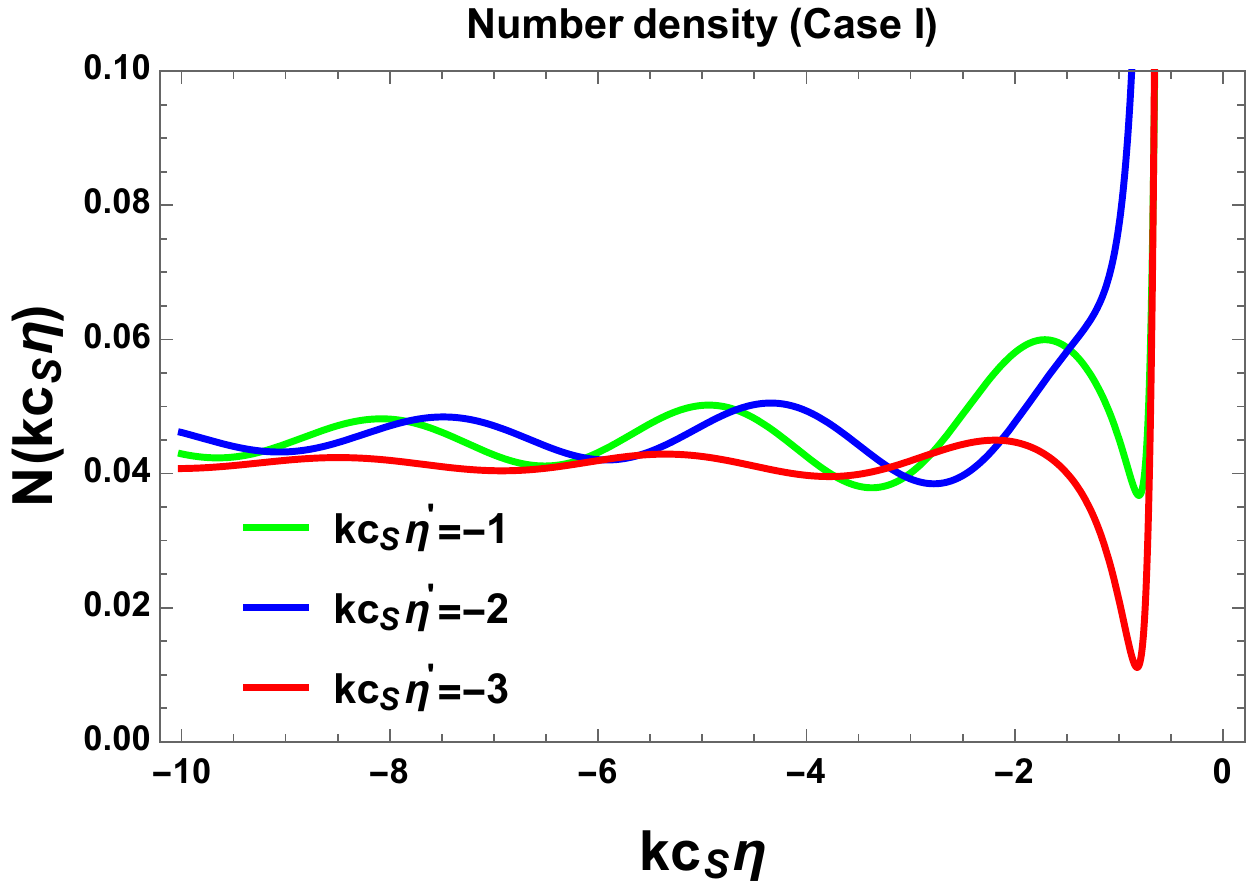}
    \label{fig3xcc5}
}
\subfigure[]{
    \includegraphics[width=7.2cm,height=6.1cm] {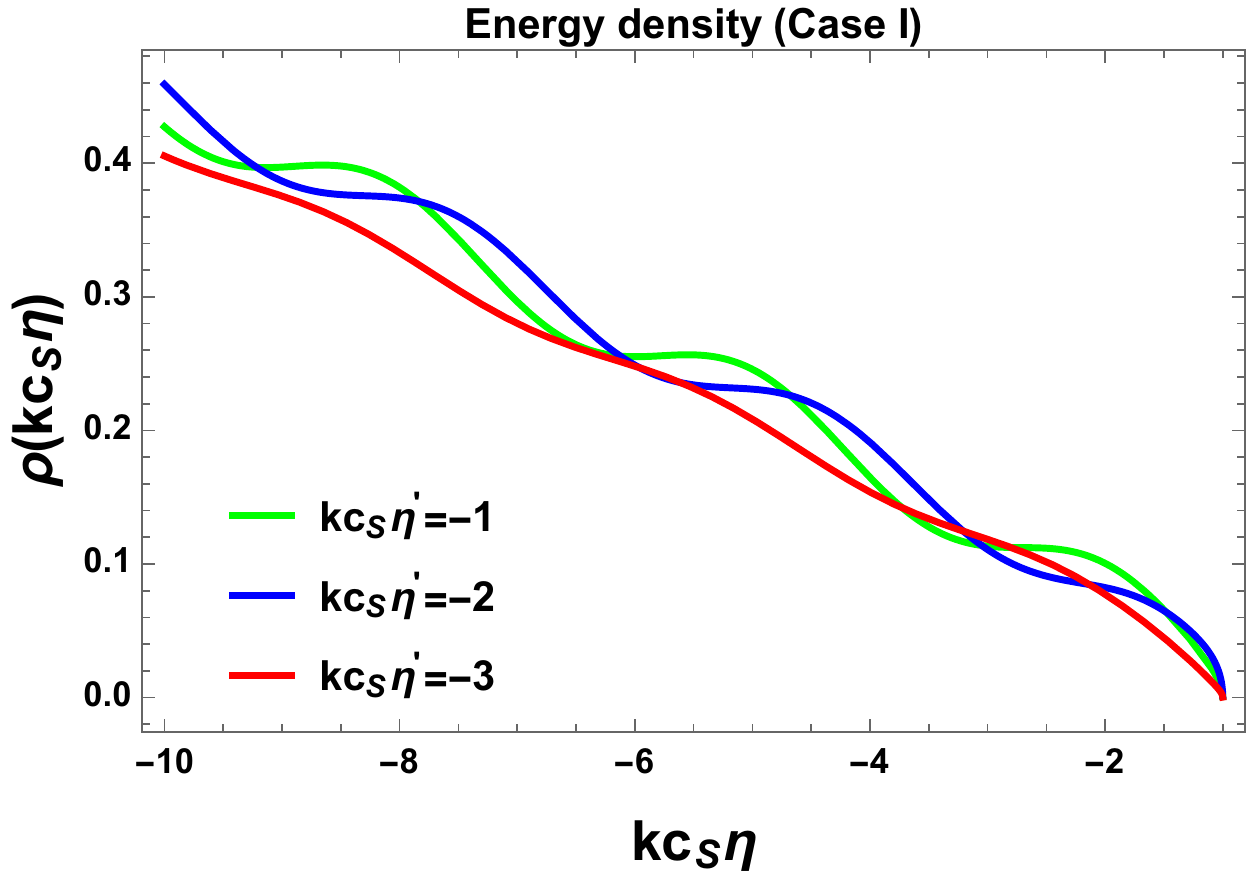}
    \label{fig4xcc6}
}
\caption[Optional caption for list of figures]{Particle creation profile for {\bf Case I}.} 
\label{bogg1}
\end{figure*}
\begin{figure*}[htb]
\centering
\subfigure[]{
    \includegraphics[width=7.2cm,height=6cm] {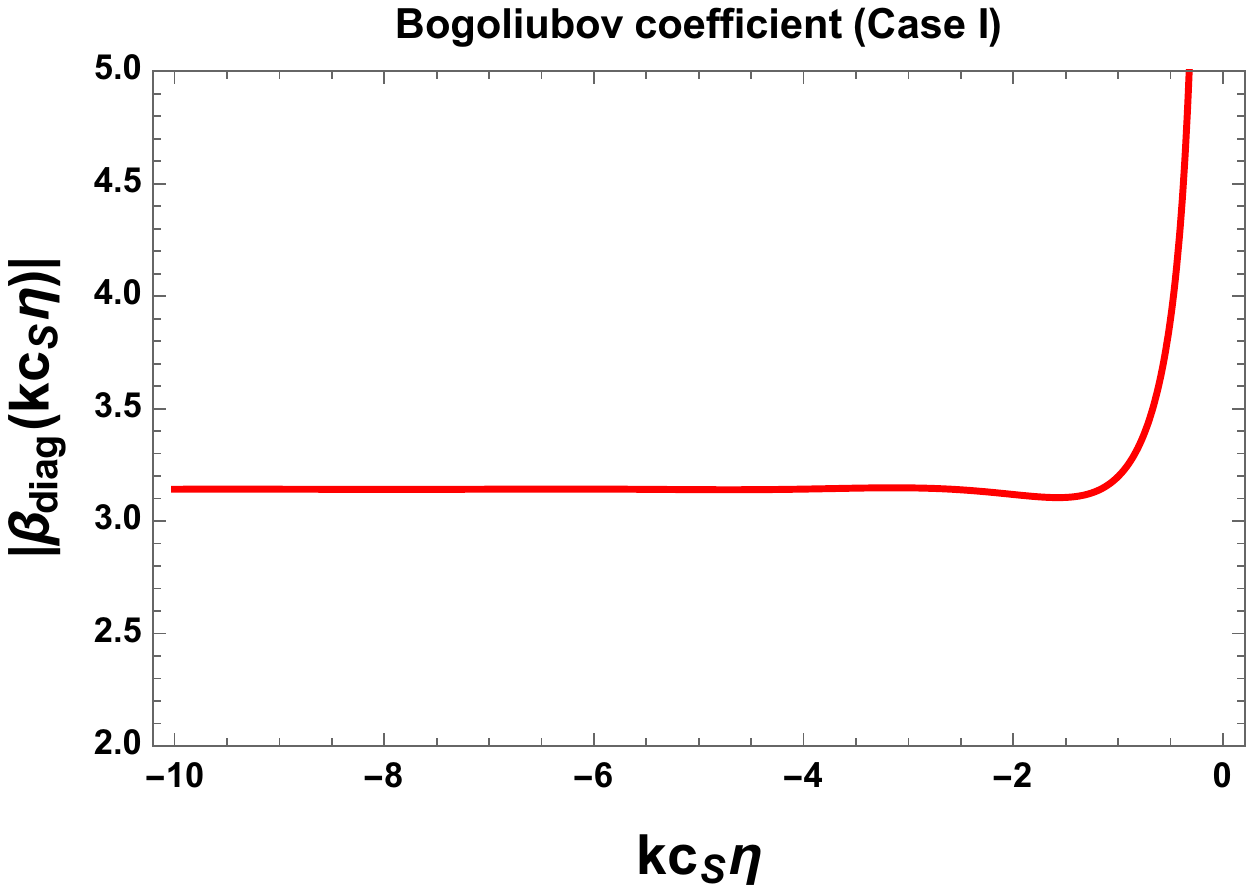}
    \label{fig1xcc}
}
\subfigure[]{
    \includegraphics[width=7.2cm,height=6cm] {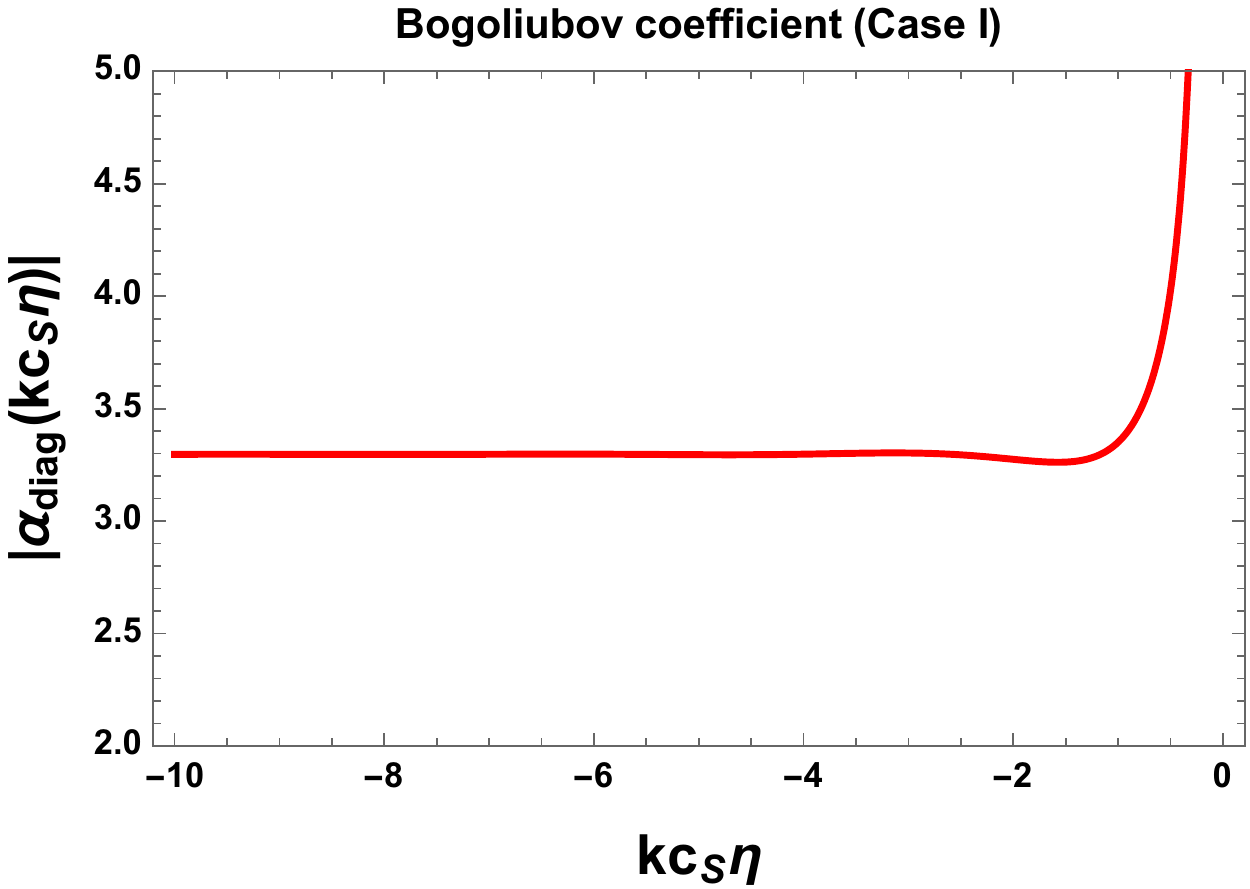}
    \label{fig2xcc}
}
\subfigure[]{
    \includegraphics[width=7.2cm,height=6cm] {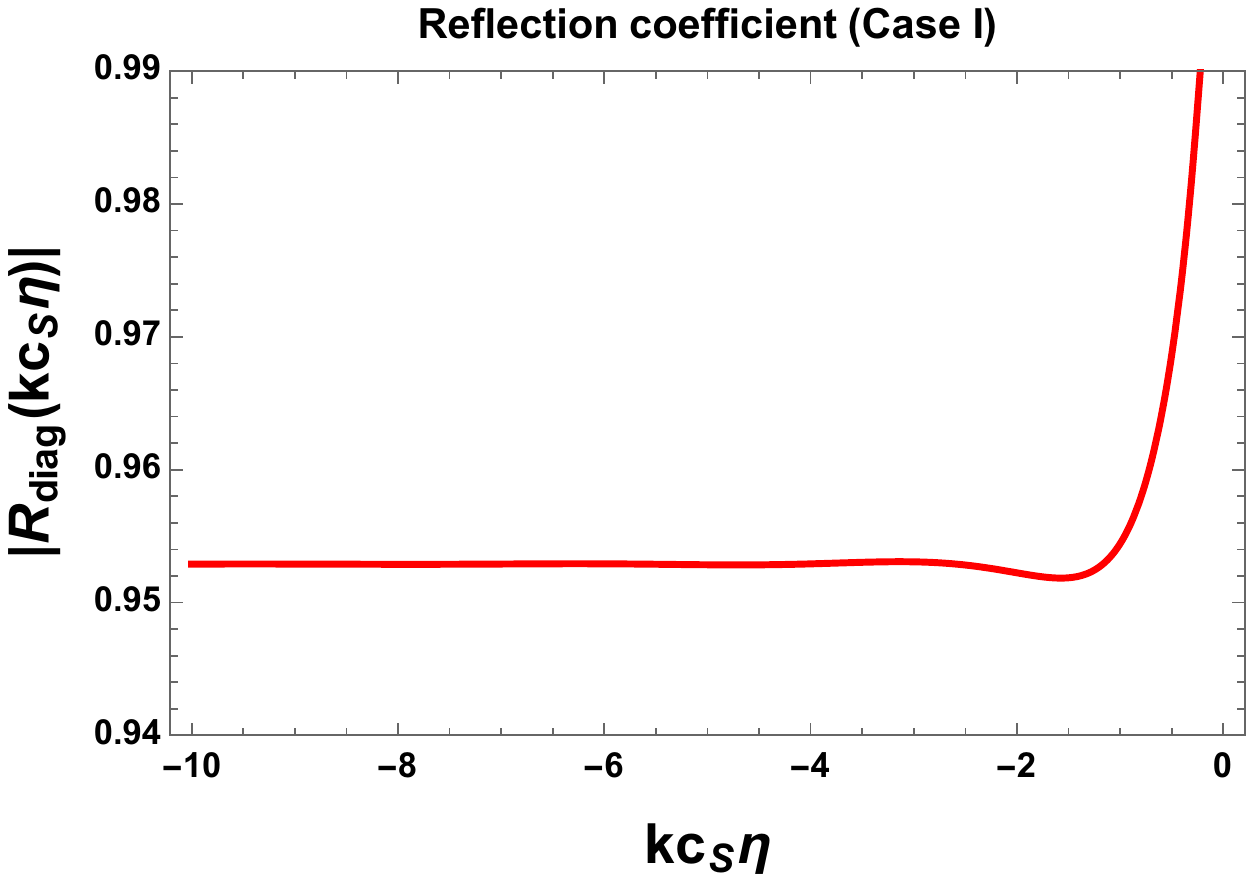}
    \label{fig3xcc}
}
\subfigure[]{
    \includegraphics[width=7.2cm,height=6cm] {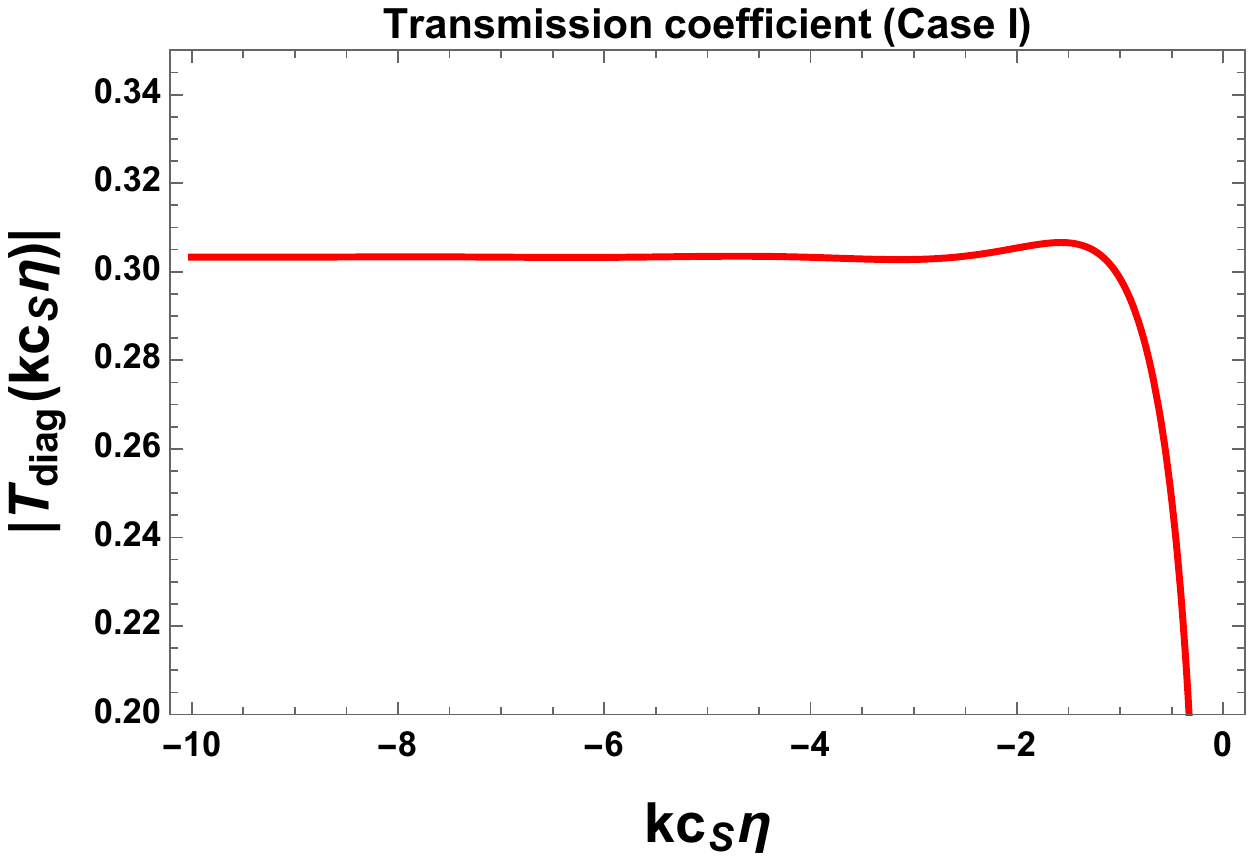}
    \label{fig4xcc}
}
\subfigure[]{
    \includegraphics[width=7.2cm,height=6.1cm] {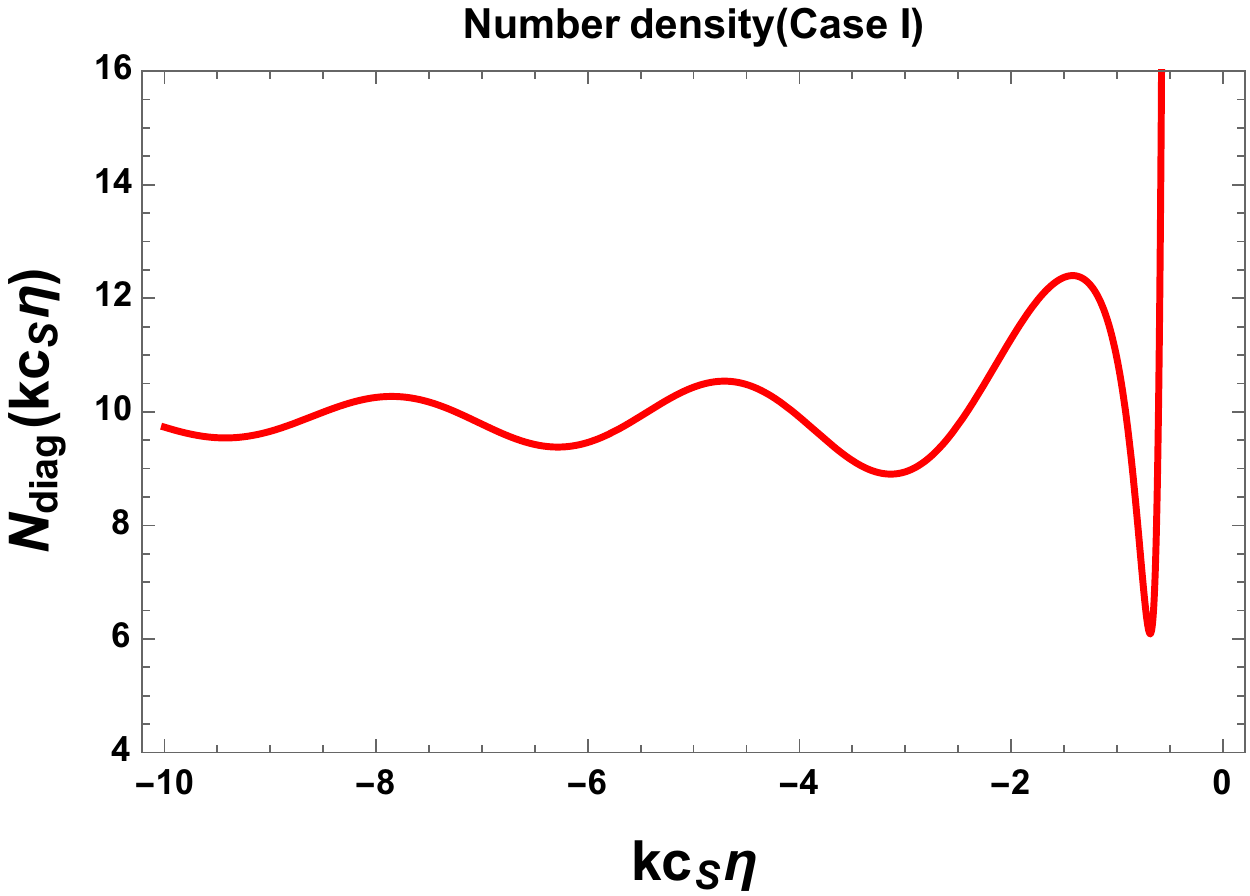}
    \label{fig3xcc}
}
\subfigure[]{
    \includegraphics[width=7.2cm,height=6.1cm] {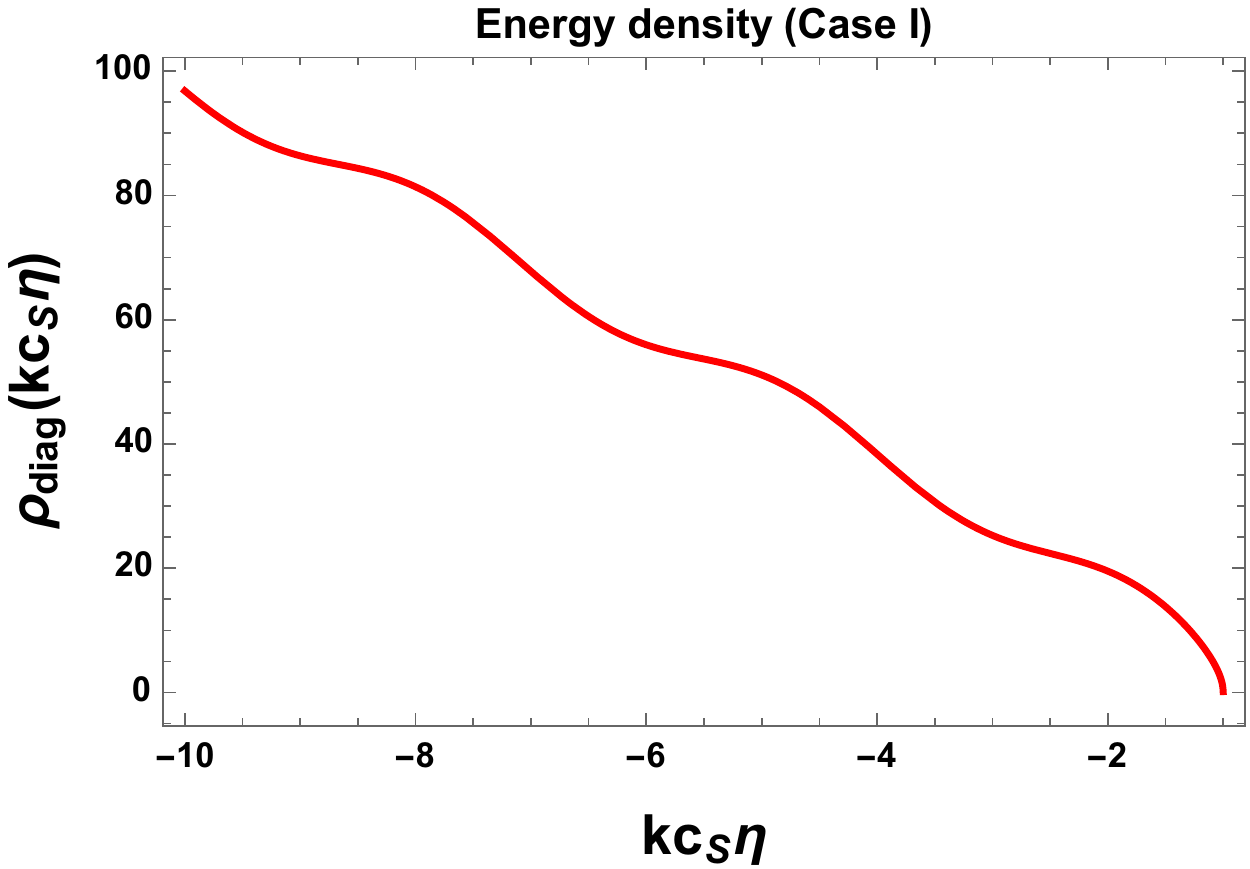}
    \label{fig4xcc}
}
\caption[Optional caption for list of figures]{Particle creation profile for {\bf Case I} in diagonalized representation.} 
\label{bogg2}
\end{figure*}
Throughout the discussion of total energy density of the produced particles 
we have introduced a symbol $J$ defined as:
\be\begin{array}{lll}\label{kghjk}
 \displaystyle J=\int d^3 {\bf k}~p(\tau)=
 \left\{\begin{array}{ll}
                    \displaystyle   \int d^3 {\bf k}~\sqrt{c^2_{S}k^2-\frac{1}{\tau^2}}~~~~ &
 \mbox{\small {\bf for ~dS}}  \\ 
	\displaystyle  \int d^3 {\bf k}~\sqrt{c^2_{S}k^2-\frac{\left[\nu^2-\frac{5}{4}\right]}{\tau^2}}~~~~ & \mbox{\small {\bf for~ qdS}}.
          \end{array}
\right.
\end{array}\ee
which physically signifies the total finite volume weighted by $p(\eta)$ 
in momentum space within which the produced particles are occupied. 

To study the behaviour of this integral we consider here three 
consecutive physical situations-$|kc_{S}\eta|<<1$, $|kc_{S}\eta|\approx 1-\Delta(\rightarrow 0)$ and $|kc_{S}\eta|>>1$ for de Sitter and quasi de Sitter case. 
In three cases we have:
\bea &&\underline{\bf For~dS:}\\
\small J&=& 
\left\{\begin{array}{ll}
                    \displaystyle  \int d^3 {\bf k}~\frac{1}{\tau}=\frac{V}{\tau}
                    ~~&
 \mbox{\small {\bf for ~$|kc_{S}\eta|<<1$}}  \\ 
 \displaystyle  \int d^3 {\bf k}~\frac{\sqrt{2\Delta}}{\tau}=\frac{\sqrt{2\Delta}~V}{\tau}~~~~ &
 \mbox{\small {\bf for ~$|kc_{S}\eta|\approx 1-\Delta(\rightarrow 0)$}}  \\ 
	\displaystyle\int d^3 {\bf k}~kc_{S} & \mbox{\small {\bf for ~$|kc_{S}\eta|>>1$}}.~~~~~~~~
          \end{array}
\right.\nonumber\eea
\bea
&&\underline{\bf For~qdS:}\\
\small J&=&
\left\{\begin{array}{ll}
                    \displaystyle  \int d^3 {\bf k}~\frac{\sqrt{\left[\nu^2-\frac{5}{4}\right]}}{\tau}=\frac{V\sqrt{\left[\nu^2-\frac{5}{4}\right]}}{\tau}~~ &
 \mbox{\small {\bf for ~$|kc_{S}\eta|<<1$}}  \\ 
 \displaystyle   \int d^3 {\bf k}~\frac{\sqrt{2\Delta +\left[\nu^2-\frac{9}{4}\right]}}{\tau}=\frac{\sqrt{2\Delta +\left[\nu^2-\frac{9}{4}\right]}~V}{\tau}
                    ~~&
 \mbox{\small {\bf for ~$|kc_{S}\eta|\approx 1-\Delta(\rightarrow 0)$}} \\ 
	\displaystyle \int d^3 {\bf k}~kc_{S}~~ & \mbox{\small {\bf for ~$|kc_{S}\eta|>>1$}}.~~~~~~~~
          \end{array}
\right.\nonumber
\eea
In fig.~(\ref{bogg1}) and fig.~(\ref{bogg2}), we have explicitly shown the particle creation profile for {\bf Case I} for two representations.
\subsubsection{\bf Case II: $m>>H$}
Here we set $m=\Upsilon H$, where the parameter $\Upsilon>>1$ in this case. Here the equation of motion for the massive field is:
\bea
h''_k + \left\{c^{2}_{S}k^2 + \frac{\Upsilon^2-2}{\eta^2} \right\} h_k &=& 0~~~~~~~{\bf for~ dS}\\\nonumber \\
h''_k + \left\{c^{2}_{S}k^2 + 
\frac{\left[\Upsilon^2-\left(\nu^2-\frac{1}{4}\right)\right]}{\eta^2} \right\} h_k &=& 0~~~~~~~{\bf for~ qdS}.
\eea
The solution for the mode function for de Sitter and quasi de Sitter space can be expressed as: 
\be\begin{array}{lll}\label{yu212}
 \displaystyle h_k (\eta) =
 \left\{\begin{array}{ll}
                    \displaystyle   \sqrt{-\eta}\left[C_1  H^{(1)}_{i\sqrt{\Upsilon^2-\frac{9}{4}}} \left(-kc_{S}\eta\right) 
+ C_2 H^{(2)}_{i\sqrt{\Upsilon^2-\frac{9}{4}}} \left(-kc_{S}\eta\right)\right]~~~~ &
 \mbox{\small {\bf for ~dS}}  \\ 
	\displaystyle \sqrt{-\eta}\left[C_1  H^{(1)}_{i\sqrt{\Upsilon^2-\nu^2}} \left(-kc_{S}\eta\right) 
+ C_2 H^{(2)}_{i\sqrt{\Upsilon^2-\nu^2}} \left(-kc_{S}\eta\right)\right]~~~~ & \mbox{\small {\bf for~ qdS}}.
          \end{array}
\right.
\end{array}\ee
where $C_{1}$ and and $C_{2}$ are two arbitrary integration constant, which depend on the 
choice of the initial condition. 

After taking $kc_{S}\eta\rightarrow -\infty$, $kc_{S}\eta\rightarrow 0$ and $|kc_{S}\eta|\approx 1$ limit the most general 
solution as stated in Eq~(\ref{yu212}) can be recast as:
\bea\label{yu2}
 \displaystyle h_k (\eta) &\stackrel{|kc_{S}\eta|\rightarrow-\infty}{=}&\footnotesize\left\{\begin{array}{ll}
                    \displaystyle   \sqrt{\frac{2}{\pi kc_{S}}}\left[C_1  e^{ -ikc_{S}\eta}
e^{-\frac{i\pi}{2}\left(i\sqrt{\Upsilon^2-\frac{9}{4}}+\frac{1}{2}\right)} 
+ C_2 e^{ ikc_{S}\eta}
e^{\frac{i\pi}{2}\left(i\sqrt{\Upsilon^2-\frac{9}{4}}+\frac{1}{2}\right)}\right]~~~~ &
 \mbox{\small {\bf for ~dS}} \\ 
	\displaystyle \sqrt{\frac{2}{\pi kc_{S}}}\left[C_1  e^{ -ikc_{S}\eta}
e^{-\frac{i\pi}{2}\left(i\sqrt{\Upsilon^2-\nu^2}+\frac{1}{2}\right)} 
+ C_2 e^{ ikc_{S}\eta}
e^{\frac{i\pi}{2}\left(i\sqrt{\Upsilon^2-\nu^2}+\frac{1}{2}\right)}\right]~~~~ & \mbox{\small {\bf for~ qdS}}.
          \end{array}
\right.\eea 
\bea
\label{yu2}
 \displaystyle h_k (\eta) &\stackrel{|kc_{S}\eta|\rightarrow 0}{=}&\footnotesize\left\{\begin{array}{ll}
                    \displaystyle  \frac{i\sqrt{-\eta}}{\pi}\Gamma\left(i\sqrt{\Upsilon^2-\frac{9}{4}}\right)
                    \left(-\frac{kc_{S}\eta}{2}\right)^{-i\sqrt{\Upsilon^2-\frac{9}{4}}}\left[C_1   
- C_2 \right]~ &
 \mbox{\small {\bf for ~dS}} \\ 
	\displaystyle\frac{i\sqrt{-\eta}}{\pi}\Gamma\left(i\sqrt{\Upsilon^2-\nu^2}\right)\left(-\frac{kc_{S}\eta}{2}\right)^{-i\sqrt{\Upsilon^2-\nu^2}}\left[C_1   
- C_2 \right]~ & \mbox{\small {\bf for~ qdS}}.
          \end{array}
\right.
\eea
\bea
\label{yu2}
 \displaystyle h_k (\eta) &\stackrel{|kc_{S}\eta|\approx 1-\Delta(\rightarrow 0)}{=}&\footnotesize\left\{\begin{array}{ll}
                    \displaystyle  \frac{i}{\pi}\sqrt{-\eta}\left[ \frac{1}{\left(i\sqrt{\Upsilon^2-\frac{9}{4}}\right)}-\gamma+\frac{i}{2}
                   \left(\gamma^2+\frac{\pi^2}{6}\right)\left(\sqrt{\Upsilon^2-\frac{9}{4}}\right)\right.\\ \displaystyle \left.
                   \displaystyle~~~~~~~~+\frac{1}{6}\left(\gamma^3+\frac{\gamma \pi^2}{2}+2\zeta(3)\right)
                   \left(\sqrt{\Upsilon^2-\frac{9}{4}}\right)^2 +\cdots\right]\\ \displaystyle ~~~~~~~~~~~~\times
                   \left(\frac{1+\Delta}{2}\right)^{-i\sqrt{\Upsilon^2-\frac{9}{4}}}\left[C_1   
- C_2 \right]~&
 \mbox{\small {\bf for ~dS}} \\ 
	\displaystyle\frac{i}{\pi}\sqrt{-\eta}\left[ \frac{1}{ \left\{i\sqrt{\Upsilon^2-\nu^2}\right\}}
  -\gamma  \displaystyle +\frac{i}{2}\left(\gamma^2+\frac{\pi^2}{6}\right) \left\{\sqrt{\Upsilon^2-\nu^2}\right\}
  \right.\\ \displaystyle \left. \displaystyle~+\frac{1}{6}\left(\gamma^3+\frac{\gamma \pi^2}{2}+2\zeta(3)\right) \left\{\sqrt{\Upsilon^2-\nu^2}\right\}^2
  +\cdots\right]\\ \displaystyle ~~~~~~~~~~~~\times\left(\frac{1+\Delta}{2}\right)^{-i\sqrt{\Upsilon^2-\nu^2}}\left[C_1   
- C_2 \right]~ & \mbox{\small {\bf for~ qdS}}.
          \end{array}
\right.
\eea
In the standard WKB approximation the total solution can be recast in the following form:
\bea\label{df313}
h_k (\eta)&=& \left[D_{1}u_{k}(\eta) + D_{2} \bar{u}_{k}(\eta)\right],\eea
where $D_{1}$ and and $D_{2}$ are two arbitrary integration constant, which depend on the 
choice of the initial condition during WKB approximation at early and late time scale.
In the present context $u_{k}(\eta)$ and $\bar{u}_{k}(\eta)$ are defined as:
\be\begin{array}{lll}\label{solp}
 \tiny u_{k}(\eta) =\footnotesize\left\{\begin{array}{ll}
                    \displaystyle   \frac{1}{\sqrt{2\sqrt{c^{2}_{S}k^2 + \frac{\Upsilon^2-2}{\eta^2}}}}
\exp\left[i\int^{\eta} d\eta^{\prime} \sqrt{c^{2}_{S}k^2 + \frac{\Upsilon^2-2}{\eta^{'2}}}\right]\\
= \displaystyle\frac{1}{\sqrt{2\sqrt{c^{2}_{S}k^2 + \frac{\Upsilon^2-2}{\eta^2}}}} 
\exp\left[i\left(\eta\sqrt{c^{2}_{S}k^2 + \frac{\Upsilon^2-2}{\eta^2}}
+\sqrt{\Upsilon^2-2}\ln\left[\frac{2}{\eta\sqrt{\Upsilon^2-2}}+\frac{2\sqrt{c^{2}_{S}k^2 + \frac{\Upsilon^2-2}{\eta^{2}}}}{(\Upsilon^2-2)}\right]
\right)\right]~~~~ &
 \mbox{\small {\bf for ~dS}}  \\ 
	\displaystyle \frac{1}{\sqrt{2\sqrt{c^{2}_{S}k^2 + 
\frac{\left[\Upsilon^2-\left(\nu^2-\frac{1}{4}\right)\right]}{\eta^2}}}}
\exp\left[i\int^{\eta} d\eta^{\prime} \sqrt{c^{2}_{S}k^2 + 
\frac{\left[\Upsilon^2-\left(\nu^2-\frac{1}{4}\right)\right]}{\eta^{'2}}}\right]\\
= \displaystyle\frac{1}{\sqrt{2\sqrt{c^{2}_{S}k^2 + 
\frac{\left[\Upsilon^2-\left(\nu^2-\frac{1}{4}\right)\right]}{\eta^2}}}} 
\exp\left[i\left(\eta\sqrt{c^{2}_{S}k^2 + 
\frac{\left[\Upsilon^2-\left(\nu^2-\frac{1}{4}\right)\right]}{\eta^2}}
\right.\right.\\ \left.\left. \displaystyle +\sqrt{\Upsilon^2-\left(\nu^2-\frac{1}{4}\right)}\ln\left[\frac{2}{\eta\sqrt{\Upsilon^2-\left(\nu^2-\frac{1}{4}\right)}}
+\frac{2\sqrt{c^{2}_{S}k^2 + \frac{\Upsilon^2-\left(\nu^2-\frac{1}{4}\right)}{\eta^{2}}}}{(\Upsilon^2-\left(\nu^2-\frac{1}{4}\right))}\right]
\right)\right]~~~~ & \mbox{\small {\bf for~ qdS}}.
          \end{array}
\right.
\end{array}\ee
\be\begin{array}{lll}\label{sola}
 \tiny \bar{u}_{k}(\eta) =\footnotesize\left\{\begin{array}{ll}
                    \displaystyle   \frac{1}{\sqrt{2\sqrt{c^{2}_{S}k^2 + \frac{\Upsilon^2-2}{\eta^2}}}}
\exp\left[-i\int^{\eta} d\eta^{\prime} \sqrt{c^{2}_{S}k^2 + \frac{\Upsilon^2-2}{\eta^{'2}}}\right]\\
= \displaystyle\frac{1}{\sqrt{2\sqrt{c^{2}_{S}k^2 + \frac{\Upsilon^2-2}{\eta^2}}}} 
\exp\left[-i\left(\eta\sqrt{c^{2}_{S}k^2 + \frac{\Upsilon^2-2}{\eta^2}}
+\sqrt{\Upsilon^2-2}\ln\left[\frac{2}{\eta\sqrt{\Upsilon^2-2}}+\frac{2\sqrt{c^{2}_{S}k^2 + \frac{\Upsilon^2-2}{\eta^{2}}}}{(\Upsilon^2-2)}\right]
\right)\right]~~~~ &
 \mbox{\small {\bf for ~dS}}  \\
	\displaystyle \frac{1}{\sqrt{2\sqrt{c^{2}_{S}k^2 + 
\frac{\left[\Upsilon^2-\left(\nu^2-\frac{1}{4}\right)\right]}{\eta^2}}}}
\exp\left[-i\int^{\eta} d\eta^{\prime} \sqrt{c^{2}_{S}k^2 + 
\frac{\left[\Upsilon^2-\left(\nu^2-\frac{1}{4}\right)\right]}{\eta^{'2}}}\right]\\
= \displaystyle\frac{1}{\sqrt{2\sqrt{c^{2}_{S}k^2 + 
\frac{\left[\Upsilon^2-\left(\nu^2-\frac{1}{4}\right)\right]}{\eta^2}}}} 
\exp\left[-i\left(\eta\sqrt{c^{2}_{S}k^2 + 
\frac{\left[\Upsilon^2-\left(\nu^2-\frac{1}{4}\right)\right]}{\eta^2}}
\right.\right.\\ \left.\left. \displaystyle +\sqrt{\Upsilon^2-\left(\nu^2-\frac{1}{4}\right)}\ln\left[\frac{2}{\eta\sqrt{\Upsilon^2-\left(\nu^2-\frac{1}{4}\right)}}
+\frac{2\sqrt{c^{2}_{S}k^2 + \frac{\Upsilon^2-\left(\nu^2-\frac{1}{4}\right)}{\eta^{2}}}}{(\Upsilon^2-\left(\nu^2-\frac{1}{4}\right))}\right]
\right)\right]~~~~ & \mbox{\small {\bf for~ qdS}}.
          \end{array}
\right.
\end{array}\ee
where we have written the total solution for the mode $h_k$ in terms of 
two linearly independent solutions. Here it is important to note that the both of the 
solutions are hermitian conjugate of each other. If in the present context the exact solution of the mode $h_k$ 
is expanded with respect to these two linearly independent solutions then particle creation is absent in our EFT 
setup. In the present context correctness of 
WKB approximation is guarantee at very early and very late time scales. In this discussion $u_{k}(\eta)$
is valid at very early time scale and $\bar{u}_{k}(\eta)$ perfectly works in the late time scale.

Now we will explicitly check that the exactness of the above mentioned WKB result derived in Eq~(\ref{df313}) 
with the actual solution of the mode function as presented in Eq~(\ref{yu212}). As mentioned earlier 
in FLRW space-time in Fourier space Bogoliubov coefficient $\beta(k)$ measures this exactness for a given 
setup. The particle creation mechanism and its exact amount
is described by finding the Bogoliubov coefficient $\beta(k)$ in Fourier space
which in principle measures the exact 
amount of late times solution $u_{k}(\eta)$, if in the present context we exactly 
start with the early time scale solution $u_{k}(\eta)$. In our present computation 
we consider a physical situation where the
WKB approximation is correct up to the leading order throughout the cosmological evolution in time scale.
In the present context the Bogoliubov coefficient $\beta(k)$ in Fourier space
can be computed approximately using the following regularized integral:
\be\begin{array}{lll}\label{soladf12}
 \displaystyle \beta(k,\tau,\tau^{'},\eta^{'}) =
 \left\{\begin{array}{ll}
                    \displaystyle   \int^{\tau}_{\tau^{'}}d\eta~\frac{(\Upsilon^2-2)^2\exp\left[2i\int^{\eta}_{\eta^{'}}
d\eta^{''}\sqrt{c^{2}_{S}k^2 + \frac{\Upsilon^2-2}{\eta^{''2}}}\right]}{
                    4\eta^{6}\left(c^{2}_{S}k^2 + \frac{\Upsilon^2-2}{\eta^2}\right)^{\frac{5}{2}}}&
 \mbox{\small {\bf for ~dS}}  \\ 
	\displaystyle   \int^{\tau}_{\tau^{'}}d\eta~\frac{\left(\Upsilon^2-\left[\nu^2-\frac{1}{4}\right]\right)^2\exp\left[2i\int^{\eta}_{\eta^{'}}
d\eta^{''}\sqrt{c^{2}_{S}k^2 + 
\frac{\left[\Upsilon^2-\left(\nu^2-\frac{1}{4}\right)\right]}{\eta^{''2}}}\right]}{
                    4\eta^{6}\left(c^{2}_{S}k^2 + 
\frac{\left[\Upsilon^2-\left(\nu^2-\frac{1}{4}\right)\right]}{\eta^2}\right)^{\frac{5}{2}}}& \mbox{\small {\bf for~ qdS}}.
          \end{array}
\right.
\end{array}\ee
which is not exactly analytically computable. To study the behaviour of this integral we consider here three 
consecutive physical situations-$|kc_{S}\eta|<<1$, $|kc_{S}\eta|\approx 1-\Delta(\rightarrow 0)$ and $|kc_{S}\eta|>>1$ for de Sitter and quasi de Sitter case. 
In three cases we have:
\be\begin{array}{lll}\label{yu2dfvqasx3}\small
 \displaystyle \underline{\rm \bf For~dS:}~~~~~~\sqrt{\left\{c^{2}_{S}k^2 + \frac{\Upsilon^2-2}{\eta^2}\right\}} \approx
 \left\{\begin{array}{ll}
                    \displaystyle  \frac{\sqrt{\Upsilon^2-2}}{\eta}~~~~ &
 \mbox{\small {\bf for ~$|kc_{S}\eta|<<1$}}  \\ 
	\displaystyle \frac{\sqrt{\Upsilon^2-2\Delta-1}}{\eta}~~~~ & \mbox{\small {\bf for ~$|kc_{S}\eta|\approx 1-\Delta(\rightarrow 0)$}}\\ 
	\displaystyle kc_{S}~~~~ & \mbox{\small {\bf for ~$|kc_{S}\eta|>>1$}}.
          \end{array}
\right.
\\
\small
 \displaystyle \underline{\rm \bf For~qdS:}~~~~~\sqrt{\left\{c^{2}_{S}k^2 + 
\frac{\left[\Upsilon^2-\left(\nu^2-\frac{1}{4}\right)\right]}{\eta^2}\right\}} \approx
\left\{\begin{array}{ll}
                    \displaystyle \frac{\sqrt{\Upsilon^2-\left(\nu^2-\frac{1}{4}\right)}}{\eta} ~~~~ &
 \mbox{\small {\bf for ~$|kc_{S}\eta|<<1$}}  \\ 
	\displaystyle \frac{\sqrt{\Upsilon^2-2\Delta-\left(\nu^2-\frac{5}{4}\right)}}{\eta}~~~~ & \mbox{\small {\bf for ~$|kc_{S}\eta|\approx 1-\Delta(\rightarrow 0)$}}\\ 
	\displaystyle kc_{S}~~~~ & \mbox{\small {\bf for ~$|kc_{S}\eta|>>1$}}.
          \end{array}
\right.
\end{array}\ee
and further using this result Bogoliubov coefficient $\beta(k)$ in Fourier space can be expressed as:
\bea &&\underline{\bf For~dS:}\\
\small\beta(k,\tau,\tau^{'},\eta^{'})&=& \footnotesize\left\{\begin{array}{ll}
                    \displaystyle  \frac{\left[\tau^{2i\sqrt{\Upsilon^2-2}}-\tau^{'2i\sqrt{\Upsilon^2-2}}\right]}{
                    8i(\Upsilon^2-2)\eta^{'2i\sqrt{\Upsilon^2-2}}}
                     &
 \mbox{\small {\bf for ~$|kc_{S}\eta|<<1$}}  \\ 
 \displaystyle   \frac{(\Upsilon^2-2)^2\left[\tau^{2i\sqrt{\Upsilon^2-2\Delta-1}}-\tau^{'2i\sqrt{\Upsilon^2-2\Delta-1}}\right]}{
                    8i(\Upsilon^2-2\Delta-1)^3\eta^{'2i\sqrt{\Upsilon^2-2\Delta-1}}} &
 \mbox{\small {\bf for ~$|kc_{S}\eta|\approx 1-\Delta(\rightarrow 0)$}}  \\ 
	\displaystyle (\Upsilon^2-2)^2\left[i\frac{\text{Ei}(2 ikc_{S} \eta)e^{-2ikc_{S}\eta^{'}}}{15}
	-\frac{e^{2 i k c_{S}(\eta-\eta^{'})} }{120 (c_{S}k)^5 \eta^5}\left(
	4 (c_{S}k)^4 \eta^4-2 i (c_{S}k)^3 \eta^3\right.\right.\\ \left.\left.-2 (c_{S}k)^2 \eta^2+3 i c_{S}k \eta+6\right)\right]^{\tau}_{\tau^{'}}& \mbox{\small {\bf for ~$|kc_{S}\eta|>>1$}}.
          \end{array}
\right.\nonumber\eea
\bea
&&\underline{\bf For~qdS:}\\
\small\beta(k,\tau,\tau^{'},\eta^{'})&=& \footnotesize\left\{\begin{array}{ll}
                     \displaystyle \frac{\left[\tau^{2i\sqrt{\Upsilon^2-\left(\nu^2-\frac{1}{4}\right)}}-\tau^{'2i\sqrt{\Upsilon^2-\left(\nu^2-\frac{1}{4}\right)}}\right]}{
                    8i\left[\Upsilon^2-\left(\nu^2-\frac{1}{4}\right)\right]\eta^{'2i\sqrt{\Upsilon^2-\left(\nu^2-\frac{1}{4}\right)}}}
                     &
 \mbox{\small {\bf for ~$|kc_{S}\eta|<<1$}}  \\ 
   \displaystyle \frac{\left[\Upsilon^2-\left(\nu^2-\frac{1}{4}\right)\right]^2\left[\tau^{2i\sqrt{\Upsilon^2-2\Delta-\left(\nu^2-\frac{5}{4}\right)}}-\tau^{'2i\sqrt{\Upsilon^2-2\Delta-\left(\nu^2-\frac{5}{4}\right)}}\right]}{
                    8i\left[\Upsilon^2-2\Delta-\left(\nu^2-\frac{5}{4}\right)\right]^3\eta^{'2i\sqrt{\Upsilon^2-2\Delta-\left(\nu^2-\frac{5}{4}\right)}}} &
 \mbox{\small {\bf for ~$|kc_{S}\eta|\approx 1-\Delta(\rightarrow 0)$}}  \\ 
	 \left[\Upsilon^2-\left(\nu^2-\frac{1}{4}\right)\right]^2\left[i\frac{\text{Ei}(2 ikc_{S} \eta)e^{-2ikc_{S}\eta^{'}}}{15}
	 -\frac{e^{2 i k c_{S}(\eta-\eta^{'})} }{120 (c_{S}k)^5 \eta^5}\left(
	4 (c_{S}k)^4 \eta^4-2 i (c_{S}k)^3 \eta^3\right.\right.\\ \left.\left.-2 (c_{S}k)^2 \eta^2+3 i c_{S}k \eta+6\right)\right]^{\tau}_{\tau^{'}}& \mbox{\small {\bf for ~$|kc_{S}\eta|>>1$}}.
          \end{array}
\right.\nonumber\eea
In all the situation described for de Sitter and quasi de Sitter case here 
the magnitude of the Bogoliubov coefficient $|\beta(k)|$ in Fourier space is considerably small. Specifically it is important 
to point out here that for the case when $|kc_{S}\eta|>>1$ the value
of the Bogoliubov coefficient $\beta(k)$ in Fourier space
is even smaller as the WKB approximated solution is strongly consistent for all time scales. On the other hand near the 
vicinity of the conformal time scale $\eta\sim \eta_{pair}$ for $|kc_{S}\eta_{pair}|<<1$
the WKB approximated solution is less strongly valid and to validate the solution at this time scale 
one can neglect the momentum $k$ dependence in the Bogoliubov coefficient $\beta(k)$ in Fourier space.
Here $|\eta_{pair}|$ characterizes the relative
separation between the created particles.

As mentioned earlier here one can use another equivalent way
to define the the Bogoliubov coefficient $\beta$ in Fourier 
space by implementing instantaneous Hamiltonian
diagonalization method to interpret the results. 
Using this diagonalized representation the
regularized Bogoliubov coefficient $\beta$ in Fourier 
space can be written as:
\be
\ee
where $\tau$ and $\tau^{'}$ introduced as the conformal time regulator in the present context.
In this case as well the Bogoliubov coefficient is not exactly analytically computable. 
To study the behaviour of this integral we consider here three similar
consecutive physical situations for de Sitter and quasi de Sitter case as discussed earlier.
\bea &&\underline{\bf For~dS:}\\
\small\beta_{diag}(k;\tau,\tau^{'}) &=& 
\left\{\begin{array}{ll}
                    \displaystyle  \frac{1}{
                    4i\sqrt{\Upsilon^2-2}}\left[\frac{1}{\tau^{'2i\sqrt{\Upsilon^2-2}}}-
                    \frac{1}{\tau^{2i\sqrt{\Upsilon^2-2}}}\right]
                     &
 \mbox{\small {\bf for ~$|kc_{S}\eta|<<1$}}  \\ \\
 \displaystyle    \frac{(\Upsilon^2-2)}{
                    4i(\Upsilon^2-2\Delta-1)^{3/2}}\left[\frac{1}{\tau^{'2i\sqrt{\Upsilon^2-2\Delta-1}}}-
                    \frac{1}{\tau^{2i\sqrt{\Upsilon^2-2\Delta-1}}}\right] &
 \mbox{\small {\bf for ~$|kc_{S}\eta|\approx 1-\Delta(\rightarrow 0)$}}  \\ \\
	\displaystyle (\Upsilon^2-2)\left[\frac{e^{-2 i kc_{S} \eta} \left(2ikc_{S}\eta-1\right)}{4 (c_{S}k)^2 \eta^2}-\text{Ei}(-2 i kc_{S} \eta)\right]^{\tau}_{\tau^{'}} & \mbox{\small {\bf for ~$|kc_{S}\eta|>>1$}}.
          \end{array}
\right.\nonumber\\\nonumber\\
&&\underline{\bf For~qdS:}\\
\small\beta_{diag}(k;\tau,\tau^{'}) &=&
\left\{\begin{array}{ll}
                    \displaystyle  \frac{1}{
                    4i\sqrt{\Upsilon^2-\left(\nu^2-\frac{1}{4}\right)}}\left[\frac{1}{\tau^{'2i\sqrt{\Upsilon^2-\left(\nu^2-\frac{1}{4}\right)}}}-
                    \frac{1}{\tau^{2i\sqrt{\Upsilon^2-\left(\nu^2-\frac{1}{4}\right)}}}\right]
                     &
 \mbox{\small {\bf for ~$|kc_{S}\eta|<<1$}}  \\ \\
  \frac{\left[\Upsilon^2-\left(\nu^2-\frac{1}{4}\right)\right]}{
                    4i\left[\Upsilon^2-2\Delta-\left(\nu^2-\frac{5}{4}\right)\right]^{3/2}}\left[\frac{1}{\tau^{'2i\sqrt{\Upsilon^2-2\Delta-\left(\nu^2-\frac{5}{4}\right)}}}-
                    \frac{1}{\tau^{2i\sqrt{\Upsilon^2-2\Delta-\left(\nu^2-\frac{5}{4}\right)}}}\right] &
 \mbox{\small {\bf for ~$|kc_{S}\eta|\approx 1-\Delta(\rightarrow 0)$}}  \\ \\
	 \left[\Upsilon^2-\left(\nu^2-\frac{1}{4}\right)\right]\left[\frac{e^{-2 i kc_{S} \eta} \left(2ikc_{S}\eta-1\right)}{4 (c_{S}k)^2 \eta^2}-\text{Ei}(-2 i kc_{S} \eta)\right]^{\tau}_{\tau^{'}} & \mbox{\small {\bf for ~$|kc_{S}\eta|>>1$}}.
          \end{array}
\right.\nonumber
\eea

Further using the regularized expressions for Bogoliubov co-efficient $\beta$ in two different representations as mentioned in
Eq~(\ref{soladf12}) and Eq~(\ref{soladfasxsd3cv12}), and
substituting them in Eq~(\ref{sdq1}) we get the following
regularized expressions for the Bogoliubov co-efficient $\alpha$ in two different representations as given by:
\be%
\ee

where $\phi$ and $\phi_{diag}$ are the associated phase factors in two different representations.
Here the results are not exactly analytically computable. To study the behaviour of this integral we consider here three 
consecutive physical situations-$|kc_{S}\eta|<<1$, $|kc_{S}\eta|\approx 1-\Delta(\rightarrow 0)$ and
$|kc_{S}\eta|>>1$ for de Sitter and quasi de Sitter case. 
\bea &&\underline{\bf For~dS:}\\
\small\alpha(k,\tau,\tau^{'},\eta^{'})&=& 
\left\{\begin{array}{ll}
                    \displaystyle   \sqrt{\left[ 1+ \left|\frac{\left[\tau^{2i\sqrt{\Upsilon^2-2}}-\tau^{'2i\sqrt{\Upsilon^2-2}}\right]}{
                    8i(\Upsilon^2-2)\eta^{'2i\sqrt{\Upsilon^2-2}}}\right|^2\right]}~e^{i\phi}
                     &
 \mbox{\small {\bf for ~$|kc_{S}\eta|<<1$}}  \\ \\
 \displaystyle    \sqrt{\left[ 1+ \left|\frac{(\Upsilon^2-2)^2\left[\tau^{2i\sqrt{\Upsilon^2-2\Delta-1}}-\tau^{'2i\sqrt{\Upsilon^2-2\Delta-1}}\right]}{
                    8i(\Upsilon^2-2\Delta-1)^3\eta^{'2i\sqrt{\Upsilon^2-2\Delta-1}}}\right|^2\right]}~e^{i\phi} &
 \mbox{\small {\bf for ~$|kc_{S}\eta|\approx 1-\Delta(\rightarrow 0)$}}  \\ \\
	\displaystyle \left[ 1+ \left|(\Upsilon^2-2)^2\left[i\frac{\text{Ei}(2 ikc_{S} \eta)e^{-2ikc_{S}\eta^{'}}}{15}
	-\frac{e^{2 i k c_{S}(\eta-\eta^{'})} }{120 (c_{S}k)^5 \eta^5}\left(
	4 (c_{S}k)^4 \eta^4\right.\right.\right.\right.\\ \left.\left.\left.\left.-2 i (c_{S}k)^3 \eta^3-2 (c_{S}k)^2 
	\eta^2+3 i c_{S}k \eta+6\right)\right]^{\tau}_{\tau^{'}}\right|^2\right]^{1/2}~e^{i\phi} & \mbox{\small {\bf for ~$|kc_{S}\eta|>>1$}}.
          \end{array}
\right.\nonumber\eea
\bea
&&\underline{\bf For~qdS:}\\
\small\alpha(k,\tau,\tau^{'},\eta^{'})&=& 
\left\{\begin{array}{ll}
                     \displaystyle \sqrt{\left[ 1+ \left| \frac{\left[\tau^{2i\sqrt{\Upsilon^2-\left(\nu^2-\frac{1}{4}\right)}}-\tau^{'2i\sqrt{\Upsilon^2-\left(\nu^2-\frac{1}{4}\right)}}\right]}{
                    8i\left[\Upsilon^2-\left(\nu^2-\frac{1}{4}\right)\right]
                    \eta^{'2i\sqrt{\Upsilon^2-\left(\nu^2-\frac{1}{4}\right)}}}\right|^2\right]}~e^{i\phi}
                     &
 \mbox{\small {\bf for ~$|kc_{S}\eta|<<1$}}  \\ \\
 \sqrt{\left[ 1+ \left|\frac{\left[\Upsilon^2-\left(\nu^2-\frac{1}{4}\right)\right]^2\left[\tau^{2i\sqrt{\Upsilon^2-2\Delta-\left(\nu^2-\frac{5}{4}\right)}}-\tau^{'2i\sqrt{\Upsilon^2-2\Delta-\left(\nu^2-\frac{5}{4}\right)}}\right]}{
                    8i\left[\Upsilon^2-2\Delta-\left(\nu^2-\frac{5}{4}\right)\right]^3
                    \eta^{'2i\sqrt{\Upsilon^2-2\Delta-\left(\nu^2-\frac{5}{4}\right)}}}\right|^2\right]}~e^{i\phi} &
 \mbox{\small {\bf for ~$|kc_{S}\eta|\approx 1-\Delta(\rightarrow 0)$}}  \\ \\
	 \left[ 1+ \left|\left[\Upsilon^2-\left(\nu^2-\frac{1}{4}\right)\right]^2\left[i\frac{\text{Ei}(2 ikc_{S} 
	 \eta)e^{-2ikc_{S}\eta^{'}}}{15}\right.\right.\right.\\ \left.\left.\left.
	 -\frac{e^{2 i k c_{S}(\eta-\eta^{'})} }{120 (c_{S}k)^5 \eta^5}\left(
	4 (c_{S}k)^4 \eta^4-2 i (c_{S}k)^3 \eta^3
	-2 (c_{S}k)^2 
	\eta^2+3 i c_{S}k \eta+6\right)\right]^{\tau}_{\tau^{'}}\right|^2\right]^{1/2}e^{i\phi}& \mbox{\small {\bf for ~$|kc_{S}\eta|>>1$}}.
          \end{array}
\right.\nonumber\eea

\bea &&\underline{\bf For~dS:}\\
\small\alpha_{diag}(k;\tau,\tau^{'}) &=& 
\left\{\begin{array}{ll}
            \displaystyle   \sqrt{\left[ 1+ \left| \frac{1}{
                    4i\sqrt{\Upsilon^2-2}}\left[\frac{1}{\tau^{'2i\sqrt{\Upsilon^2-2}}}-
                    \frac{1}{\tau^{2i\sqrt{\Upsilon^2-2}}}\right]\right|^2\right]}~e^{i\phi}\nonumber
                     &
 \mbox{\small {\bf for ~$|kc_{S}\eta|<<1$}}  \\ \\
  \displaystyle  \sqrt{\left[ 1+ \left|\frac{(\Upsilon^2-2)}{
                    4i(\Upsilon^2-2\Delta-1)^{3/2}}\left[\frac{1}{\tau^{'2i\sqrt{\Upsilon^2-2\Delta-1}}}-
                    \frac{1}{\tau^{2i\sqrt{\Upsilon^2-2\Delta-1}}}\right]\right|^2\right]}~e^{i\phi} &
 \mbox{\small {\bf for ~$|kc_{S}\eta|\approx 1-\Delta(\rightarrow 0)$}}  \\ \\
	\displaystyle \sqrt{\left[ 1+ \left| (\Upsilon^2-2)\left[\frac{e^{-2 i kc_{S} \eta} \left(2ikc_{S}\eta-1\right)}{4 (c_{S}k)^2 \eta^2}-\text{Ei}(-2 i kc_{S} \eta)\right]^{\tau}_{\tau^{'}}\right|^2\right]}~e^{i\phi} & \mbox{\small {\bf for ~$|kc_{S}\eta|>>1$}}.
          \end{array}
\right.\eea
\bea
&&\underline{\bf For~qdS:}\\
\small\alpha_{diag}(k;\tau,\tau^{'}) &=&
\left\{\begin{array}{ll}
               \displaystyle   \sqrt{\left[ 1+ \left|  \frac{1}{
                    4i\sqrt{\Upsilon^2-\left(\nu^2-\frac{1}{4}\right)}}\left[\frac{1}{\tau^{'2i\sqrt{\Upsilon^2-\left(\nu^2-\frac{1}{4}\right)}}}-
                    \frac{1}{\tau^{2i\sqrt{\Upsilon^2-\left(\nu^2-\frac{1}{4}\right)}}}\right]\right|^2\right]}~e^{i\phi} &
 \mbox{\small {\bf for ~$|kc_{S}\eta|<<1$}}  \\ \\
     \sqrt{\left[ 1+ \left|  \frac{\left[\Upsilon^2-\left(\nu^2-\frac{1}{4}\right)\right]}{
                    4i\left[\Upsilon^2-2\Delta-\left(\nu^2-\frac{5}{4}\right)\right]^{3/2}}\left[\frac{1}{\tau^{'2i\sqrt{\Upsilon^2-2\Delta-\left(\nu^2-\frac{5}{4}\right)}}}-
                    \frac{1}{\tau^{2i\sqrt{\Upsilon^2-2\Delta-\left(\nu^2-\frac{5}{4}\right)}}}\right]\right|^2\right]}~e^{i\phi}
                    &
 \mbox{\small {\bf for ~$|kc_{S}\eta|\approx 1-\Delta(\rightarrow 0)$}} \\ \\
	  \sqrt{\left[ 1+ \left|  \left[\Upsilon^2-\left(\nu^2-\frac{1}{4}\right)\right]\left[\frac{e^{-2 i kc_{S} \eta} \left(2ikc_{S}\eta-1\right)}{4 (c_{S}k)^2 \eta^2}-\text{Ei}(-2 i kc_{S} \eta)\right]^{\tau}_{\tau^{'}}\right|^2\right]}~e^{i\phi} & \mbox{\small {\bf for ~$|kc_{S}\eta|>>1$}}.
          \end{array}
\right.\nonumber
\eea
Further using the expressions for Bogoliubov co-efficient $\alpha$ in two different representations
and substituting them in Eq~(\ref{sdq2}) we get the following
expressions for the reflection and transmission co-efficient in two different representations 
for three consecutive physical situations-$|kc_{S}\eta|<<1$, $|kc_{S}\eta|\approx 1-\Delta(\rightarrow 0)$ and
$|kc_{S}\eta|>>1$ for de Sitter and quasi de Sitter case as given by:
\bea &&\underline{\bf For~dS:}\\
{\cal R}&=& \left\{\begin{array}{ll}
                    \displaystyle  \frac{\frac{\left[\tau^{2i\sqrt{\Upsilon^2-2}}-\tau^{'2i\sqrt{\Upsilon^2-2}}\right]}{
                    8i(\Upsilon^2-2)\eta^{'2i\sqrt{\Upsilon^2-2}}}}{ \sqrt{\left[ 1+ \left|\frac{\left[\tau^{2i\sqrt{\Upsilon^2-2}}-\tau^{'2i\sqrt{\Upsilon^2-2}}\right]}{
                    8i(\Upsilon^2-2)\eta^{'2i\sqrt{\Upsilon^2-2}}}\right|^2\right]}}~e^{-i\phi}
                     &
 \mbox{\small {\bf for ~$|kc_{S}\eta|<<1$}}  \\ \\
 \displaystyle   \frac{ \frac{(\Upsilon^2-2)^2\left[\tau^{2i\sqrt{\Upsilon^2-2\Delta-1}}-\tau^{'2i\sqrt{\Upsilon^2-2\Delta-1}}\right]}{
                    8i(\Upsilon^2-2\Delta-1)^3\eta^{'2i\sqrt{\Upsilon^2-2\Delta-1}}}}{\sqrt{\left[ 1+ \left|\frac{(\Upsilon^2-2)^2\left[\tau^{2i\sqrt{\Upsilon^2-2\Delta-1}}-\tau^{'2i\sqrt{\Upsilon^2-2\Delta-1}}\right]}{
                    8i(\Upsilon^2-2\Delta-1)^3\eta^{'2i\sqrt{\Upsilon^2-2\Delta-1}}}\right|^2\right]}}~e^{-i\phi}  &
 \mbox{\small {\bf for ~$|kc_{S}\eta|\approx 1-\Delta(\rightarrow 0)$}}  \\ \\
	\displaystyle 
(\Upsilon^2-2)^2\left[i\frac{\text{Ei}(2 ikc_{S} \eta)e^{-2ikc_{S}\eta^{'}}}{15}
-\frac{e^{2 i k c_{S}(\eta-\eta^{'})} }{120 (c_{S}k)^5 \eta^5}\left(
	4 (c_{S}k)^4 \eta^4-2 i (c_{S}k)^3 \eta^3\right.\right.\\ \left.\left.-2 (c_{S}k)^2 
	\eta^2+3 i c_{S}k \eta+6\right)\right]^{\tau}_{\tau^{'}}/\left[ 1+ \left|(\Upsilon^2-2)^2\left[i\frac{\text{Ei}(2 ikc_{S} \eta)e^{-2ikc_{S}\eta^{'}}}{15}\right.\right.\right.\\ \left.\left.\left.\displaystyle-\frac{e^{2 i k c_{S}(\eta-\eta^{'})} }{120 (c_{S}k)^5 \eta^5}\left(
	4 (c_{S}k)^4 \eta^4-2 i (c_{S}k)^3 \eta^3
	-2 (c_{S}k)^2 
	\eta^2+3 i c_{S}k \eta+6\right)\right]^{\tau}_{\tau^{'}}\right|^2\right]^{1/2}~e^{-i\phi}& \mbox{\small {\bf for ~$|kc_{S}\eta|>>1$}}.
          \end{array}
\right.\nonumber\eea

\bea &&\underline{\bf For~dS:}\\
{\cal T}&=&\left\{\begin{array}{ll}
                    \displaystyle  \frac{1}{ \sqrt{\left[ 1+ \left|\frac{\left[\tau^{2i\sqrt{\Upsilon^2-2}}-\tau^{'2i\sqrt{\Upsilon^2-2}}\right]}{
                    8i(\Upsilon^2-2)\eta^{'2i\sqrt{\Upsilon^2-2}}}\right|^2\right]}}~e^{-i\phi}
                     &
 \mbox{\small {\bf for ~$|kc_{S}\eta|<<1$}}  \\ \\
 \displaystyle   \frac{1}{\sqrt{\left[ 1+ \left|\frac{(\Upsilon^2-2)^2\left[\tau^{2i\sqrt{\Upsilon^2-2\Delta-1}}-\tau^{'2i\sqrt{\Upsilon^2-2\Delta-1}}\right]}{
                    8i(\Upsilon^2-2\Delta-1)^3\eta^{'2i\sqrt{\Upsilon^2-2\Delta-1}}}\right|^2\right]}}~e^{-i\phi}  &
 \mbox{\small {\bf for ~$|kc_{S}\eta|\approx 1-\Delta(\rightarrow 0)$}}  \\ \\
	\displaystyle 
1/\left[ 1+ \left|(\Upsilon^2-2)^2\left[i\frac{\text{Ei}(2 ikc_{S} \eta)e^{-2ikc_{S}\eta^{'}}}{15}
-\frac{e^{2 i k c_{S}(\eta-\eta^{'})} }{120 (c_{S}k)^5 \eta^5}\left(
	4 (c_{S}k)^4 \eta^4-2 i (c_{S}k)^3 \eta^3\right.\right.\right.\right.\\ \left.\left.\left.\left.-2 (c_{S}k)^2 
	\eta^2+3 i c_{S}k \eta+6\right)\right]^{\tau}_{\tau^{'}}\right|^2\right]^{1/2}~e^{-i\phi}& \mbox{\small {\bf for ~$|kc_{S}\eta|>>1$}}.
          \end{array}
\right.\nonumber\eea

\bea &&\underline{\bf For~qdS:}\\
{\cal R}&=& \left\{\begin{array}{ll}
                    \displaystyle  \frac{ \frac{\left[\tau^{2i\sqrt{\Upsilon^2-\left(\nu^2-\frac{1}{4}\right)}}-\tau^{'2i\sqrt{\Upsilon^2-\left(\nu^2-\frac{1}{4}\right)}}\right]}{
                    8i\left[\Upsilon^2-\left(\nu^2-\frac{1}{4}\right)\right]
                    \eta^{'2i\sqrt{\Upsilon^2-\left(\nu^2-\frac{1}{4}\right)}}}}{\sqrt{\left[ 1+ \left| \frac{\left[\tau^{2i\sqrt{\Upsilon^2-\left(\nu^2-\frac{1}{4}\right)}}-\tau^{'2i\sqrt{\Upsilon^2-\left(\nu^2-\frac{1}{4}\right)}}\right]}{
                    8i\left[\Upsilon^2-\left(\nu^2-\frac{1}{4}\right)\right]
                    \eta^{'2i\sqrt{\Upsilon^2-\left(\nu^2-\frac{1}{4}\right)}}}\right|^2\right]}}~e^{-i\phi}
                     &
 \mbox{\small {\bf for ~$|kc_{S}\eta|<<1$}}  \\ \\
 \displaystyle   \frac{  \frac{\left[\Upsilon^2-\left(\nu^2-\frac{1}{4}\right)\right]^2\left[\tau^{2i\sqrt{\Upsilon^2-2\Delta-\left(\nu^2-\frac{5}{4}\right)}}-\tau^{'2i\sqrt{\Upsilon^2-2\Delta-\left(\nu^2-\frac{5}{4}\right)}}\right]}{
                    8i\left[\Upsilon^2-2\Delta-\left(\nu^2-\frac{5}{4}\right)\right]^3
                    \eta^{'2i\sqrt{\Upsilon^2-2\Delta-\left(\nu^2-\frac{5}{4}\right)}}}}{\sqrt{\left[ 1+ \left|\frac{\left[\Upsilon^2-\left(\nu^2-\frac{1}{4}\right)\right]^2\left[\tau^{2i\sqrt{\Upsilon^2-2\Delta-\left(\nu^2-\frac{5}{4}\right)}}-\tau^{'2i\sqrt{\Upsilon^2-2\Delta-\left(\nu^2-\frac{5}{4}\right)}}\right]}{
                    8i\left[\Upsilon^2-2\Delta-\left(\nu^2-\frac{5}{4}\right)\right]^3
                    \eta^{'2i\sqrt{\Upsilon^2-2\Delta-\left(\nu^2-\frac{5}{4}\right)}}}\right|^2\right]}}~e^{-i\phi}  &
 \mbox{\small {\bf for ~$|kc_{S}\eta|\approx 1-\Delta(\rightarrow 0)$}}  \\ \\
	\displaystyle 
\left[\Upsilon^2-\left(\nu^2-\frac{1}{4}\right)\right]^2\left[i\frac{\text{Ei}(2 ikc_{S} 
	 \eta)e^{-2ikc_{S}\eta^{'}}}{15}
	 -\frac{e^{2 i k c_{S}(\eta-\eta^{'})} }{120 (c_{S}k)^5 \eta^5}\left(
	4 (c_{S}k)^4 \eta^4-2 i (c_{S}k)^3 \eta^3\right.\right.\\
\end{array}
\right.\nonumber\eea

\bea &&\underline{\bf For~qdS:}\\
{\cal T} &=&\footnotesize\left\{\begin{array}{ll}
                    \displaystyle  \frac{1}{\sqrt{\left[ 1+ \left| \frac{\left[\tau^{2i\sqrt{\Upsilon^2-\left(\nu^2-\frac{1}{4}\right)}}-\tau^{'2i\sqrt{\Upsilon^2-\left(\nu^2-\frac{1}{4}\right)}}\right]}{
                    8i\left[\Upsilon^2-\left(\nu^2-\frac{1}{4}\right)\right]
                    \eta^{'2i\sqrt{\Upsilon^2-\left(\nu^2-\frac{1}{4}\right)}}}\right|^2\right]}}~e^{-i\phi}
                     &
 \mbox{\small {\bf for ~$|kc_{S}\eta|<<1$}}  \\ \\
 \displaystyle   \frac{1}{\sqrt{\left[ 1+ \left|\frac{\left[\Upsilon^2-\left(\nu^2-\frac{1}{4}\right)\right]^2\left[\tau^{2i\sqrt{\Upsilon^2-2\Delta-\left(\nu^2-\frac{5}{4}\right)}}-\tau^{'2i\sqrt{\Upsilon^2-2\Delta-\left(\nu^2-\frac{5}{4}\right)}}\right]}{
                    8i\left[\Upsilon^2-2\Delta-\left(\nu^2-\frac{5}{4}\right)\right]^3
                    \eta^{'2i\sqrt{\Upsilon^2-2\Delta-\left(\nu^2-\frac{5}{4}\right)}}}\right|^2\right]}}~e^{-i\phi}  &
 \mbox{\small {\bf for ~$|kc_{S}\eta|\approx 1-\Delta(\rightarrow 0)$}}  \\ \\
	\displaystyle 
1/\left[ 1+ \left|\left[\Upsilon^2-\left(\nu^2-\frac{1}{4}\right)\right]^2\left[i\frac{\text{Ei}(2 ikc_{S} 
	 \eta)e^{-2ikc_{S}\eta^{'}}}{15}
	 -\frac{e^{2 i k c_{S}(\eta-\eta^{'})} }{120 (c_{S}k)^5 \eta^5}\left(
	4 (c_{S}k)^4 \eta^4-2 i (c_{S}k)^3 \eta^3\right.\right.\right.\right.\\ \left.\left.\left.\left.-2 (c_{S}k)^2 
	\eta^2+3 i c_{S}k \eta+6\right)\right]^{\tau}_{\tau^{'}}\right|^2\right]^{1/2}~e^{-i\phi}& \mbox{\small {\bf for ~$|kc_{S}\eta|>>1$}}.
          \end{array}
\right.\nonumber\eea
and 
\bea &&\underline{\bf For~dS:}\\
{\cal R}_{diag}(k;\tau,\tau^{'}) &=& \left\{\begin{array}{ll}
                    \displaystyle  \frac{\frac{1}{
                    4i\sqrt{\Upsilon^2-2}}\left[\frac{1}{\tau^{'2i\sqrt{\Upsilon^2-2}}}-
                    \frac{1}{\tau^{2i\sqrt{\Upsilon^2-2}}}\right]}{\sqrt{\left[ 1+ \left| \frac{1}{
                    4i\sqrt{\Upsilon^2-2}}\left[\frac{1}{\tau^{'2i\sqrt{\Upsilon^2-2}}}-
                    \frac{1}{\tau^{2i\sqrt{\Upsilon^2-2}}}\right]\right|^2\right]}}~e^{-i\phi}
                     &
 \mbox{\small {\bf for ~$|kc_{S}\eta|<<1$}}  \\ 
 \displaystyle   \frac{ \frac{(\Upsilon^2-2)}{
                    4i(\Upsilon^2-2\Delta-1)^{3/2}}\left[\frac{1}{\tau^{'2i\sqrt{\Upsilon^2-2\Delta-1}}}-
                    \frac{1}{\tau^{2i\sqrt{\Upsilon^2-2\Delta-1}}}\right]}{  \sqrt{\left[ 1+ \left|\frac{(\Upsilon^2-2)}{
                    4i(\Upsilon^2-2\Delta-1)^{3/2}}\left[\frac{1}{\tau^{'2i\sqrt{\Upsilon^2-2\Delta-1}}}-
                    \frac{1}{\tau^{2i\sqrt{\Upsilon^2-2\Delta-1}}}\right]\right|^2\right]}}~e^{-i\phi}  &
 \mbox{\small {\bf for ~$|kc_{S}\eta|\approx 1-\Delta(\rightarrow 0)$}}  \\ 
	\displaystyle 
\frac{(\Upsilon^2-2)\left[\frac{e^{-2 i kc_{S} \eta} \left(2ikc_{S}\eta-1\right)}{4 (c_{S}k)^2 \eta^2}-\text{Ei}(-2 i kc_{S} \eta)\right]^{\tau}_{\tau^{'}}}{\sqrt{\left[ 1+ \left| (\Upsilon^2-2)\left[\frac{e^{-2 i kc_{S} \eta} \left(2ikc_{S}\eta-1\right)}{4 (c_{S}k)^2 \eta^2}-\text{Ei}(-2 i kc_{S} \eta)\right]^{\tau}_{\tau^{'}}\right|^2\right]}}~e^{-i\phi}& \mbox{\small {\bf for ~$|kc_{S}\eta|>>1$}}.
          \end{array}
\right.\nonumber\eea

\bea &&\underline{\bf For~dS:}\\
{\cal T}_{diag}(k;\tau,\tau^{'}) &=& \left\{\begin{array}{ll}
                    \displaystyle  \frac{1}{\sqrt{\left[ 1+ \left| \frac{1}{
                    4i\sqrt{\Upsilon^2-2}}\left[\frac{1}{\tau^{'2i\sqrt{\Upsilon^2-2}}}-
                    \frac{1}{\tau^{2i\sqrt{\Upsilon^2-2}}}\right]\right|^2\right]}}~e^{-i\phi}
                     &
 \mbox{\small {\bf for ~$|kc_{S}\eta|<<1$}}  \\ 
 \displaystyle   \frac{1}{ \sqrt{\left[ 1+ \left|\frac{(\Upsilon^2-2)}{
                    4i(\Upsilon^2-2\Delta-1)^{3/2}}\left[\frac{1}{\tau^{'2i\sqrt{\Upsilon^2-2\Delta-1}}}-
                    \frac{1}{\tau^{2i\sqrt{\Upsilon^2-2\Delta-1}}}\right]\right|^2\right]}}~e^{-i\phi}  &
 \mbox{\small {\bf for ~$|kc_{S}\eta|\approx 1-\Delta(\rightarrow 0)$}}  \\ 
	\displaystyle 
\frac{1}{\sqrt{\left[ 1+ \left| (\Upsilon^2-2)\left[\frac{e^{-2 i kc_{S} \eta} \left(2ikc_{S}\eta-1\right)}{4 (c_{S}k)^2 \eta^2}-\text{Ei}(-2 i kc_{S} \eta)\right]^{\tau}_{\tau^{'}}\right|^2\right]}}~e^{-i\phi}& \mbox{\small {\bf for ~$|kc_{S}\eta|>>1$}}.
          \end{array}
\right.\nonumber\eea

\bea &&\underline{\bf For~qdS:}\\
\small{\cal R}_{diag}(k;\tau,\tau^{'})
&=& \footnotesize\left\{\begin{array}{ll}
                    \displaystyle  \frac{ \frac{1}{
                    4i\sqrt{\Upsilon^2-\left(\nu^2-\frac{1}{4}\right)}}\left[\frac{1}{\tau^{'2i\sqrt{\Upsilon^2-\left(\nu^2-\frac{1}{4}\right)}}}-
                    \frac{1}{\tau^{2i\sqrt{\Upsilon^2-\left(\nu^2-\frac{1}{4}\right)}}}\right]}{\sqrt{\left[ 1+ \left|  \frac{1}{
                    4i\sqrt{\Upsilon^2-\left(\nu^2-\frac{1}{4}\right)}}\left[\frac{1}{\tau^{'2i\sqrt{\Upsilon^2-\left(\nu^2-\frac{1}{4}\right)}}}-
                    \frac{1}{\tau^{2i\sqrt{\Upsilon^2-\left(\nu^2-\frac{1}{4}\right)}}}\right]\right|^2\right]}}~e^{-i\phi}
                     &
 \mbox{\small {\bf for ~$|kc_{S}\eta|<<1$}}  \\ 
 \displaystyle   \frac{  \frac{\left[\Upsilon^2-\left(\nu^2-\frac{1}{4}\right)\right]}{
                    4i\left[\Upsilon^2-2\Delta-\left(\nu^2-\frac{5}{4}\right)\right]^{3/2}}\left[\frac{1}{\tau^{'2i\sqrt{\Upsilon^2-2\Delta-\left(\nu^2-\frac{5}{4}\right)}}}-
                    \frac{1}{\tau^{2i\sqrt{\Upsilon^2-2\Delta-\left(\nu^2-\frac{5}{4}\right)}}}\right]}{\sqrt{\left[ 1+ \left|  \frac{\left[\Upsilon^2-\left(\nu^2-\frac{1}{4}\right)\right]}{
                    4i\left[\Upsilon^2-2\Delta-\left(\nu^2-\frac{5}{4}\right)\right]^{3/2}}\left[\frac{1}{\tau^{'2i\sqrt{\Upsilon^2-2\Delta-\left(\nu^2-\frac{5}{4}\right)}}}-
                    \frac{1}{\tau^{2i\sqrt{\Upsilon^2-2\Delta-\left(\nu^2-\frac{5}{4}\right)}}}\right]\right|^2\right]}}~e^{-i\phi}  &
 \mbox{\small {\bf for ~$|kc_{S}\eta|\approx 1-\Delta(\rightarrow 0)$}}  \\
\end{array}
\right.\nonumber\eea

\bea &&\underline{\bf For~qdS:}\\
\small{\cal T}_{diag}(k;\tau,\tau^{'}) &=& \footnotesize\left\{\begin{array}{ll}
                    \displaystyle  \frac{1}{\sqrt{\left[ 1+ \left|  \frac{1}{
                    4i\sqrt{\Upsilon^2-\left(\nu^2-\frac{1}{4}\right)}}\left[\frac{1}{\tau^{'2i\sqrt{\Upsilon^2-\left(\nu^2-\frac{1}{4}\right)}}}-
                    \frac{1}{\tau^{2i\sqrt{\Upsilon^2-\left(\nu^2-\frac{1}{4}\right)}}}\right]\right|^2\right]}}~e^{-i\phi}
                     &
 \mbox{\small {\bf for ~$|kc_{S}\eta|<<1$}}  \\ 
 \displaystyle   \frac{1}{ \sqrt{\left[ 1+ \left|  \frac{\left[\Upsilon^2-\left(\nu^2-\frac{1}{4}\right)\right]}{
                    4i\left[\Upsilon^2-2\Delta-\left(\nu^2-\frac{5}{4}\right)\right]^{3/2}}\left[\frac{1}{\tau^{'2i\sqrt{\Upsilon^2-2\Delta-\left(\nu^2-\frac{5}{4}\right)}}}-
                    \frac{1}{\tau^{2i\sqrt{\Upsilon^2-2\Delta-\left(\nu^2-\frac{5}{4}\right)}}}\right]\right|^2\right]}}~e^{-i\phi}  &
 \mbox{\small {\bf for ~$|kc_{S}\eta|\approx 1-\Delta(\rightarrow 0)$}}  \\ 
	\displaystyle 
\frac{1}{ \sqrt{\left[ 1+ \left|  \left[\Upsilon^2-\left(\nu^2-\frac{1}{4}\right)\right]\left[\frac{e^{-2 i kc_{S} \eta} \left(2ikc_{S}\eta-1\right)}{4 (c_{S}k)^2 \eta^2}-\text{Ei}(-2 i kc_{S} \eta)\right]^{\tau}_{\tau^{'}}\right|^2\right]}}~e^{-i\phi}& \mbox{\small {\bf for ~$|kc_{S}\eta|>>1$}}.
          \end{array}
\right.\nonumber\eea

 Next the expression for the
number of produced particles at time $\tau$ can be calculated from the following expression:
\be%
\ee
which is not exactly analytically computable. To study the behaviour of this integral we consider here three 
consecutive physical situations-$|kc_{S}\eta|<<1$, $|kc_{S}\eta|\approx 1-\Delta(\rightarrow 0)$ and $|kc_{S}\eta|>>1$ for de Sitter and quasi de Sitter case. 
In three cases we have:
\bea &&\underline{\bf For~dS:}\\
\small{\cal N}(\tau,\tau^{'},\eta^{'})&=& \left\{\begin{array}{ll}
                    \displaystyle  \frac{V}{
                    64(2\pi a)^3(\Upsilon^2-2)^2}\left|\frac{\left[\tau^{2i\sqrt{\Upsilon^2-2}}-\tau^{'2i\sqrt{\Upsilon^2-2}}\right]}{
                    \eta^{'2i\sqrt{\Upsilon^2-2}}}\right|^2
                    ~~&
 \mbox{\small {\bf for ~$|kc_{S}\eta|<<1$}}  \\ 
 \displaystyle  \frac{V(\Upsilon^2-2)^4}{
                    64(2\pi a)^3(\Upsilon^2-2\Delta-1)^6}\left| \frac{\left[\tau^{2i\sqrt{\Upsilon^2-2\Delta-1}}-\tau^{'2i\sqrt{\Upsilon^2-2\Delta-1}}\right]}{
                    \eta^{'2i\sqrt{\Upsilon^2-2\Delta-1}}}
                    \right|^2 &
 \mbox{\small {\bf for ~$|kc_{S}\eta|\approx 1-\Delta(\rightarrow 0)$}}  \\ 
	\displaystyle \frac{4\pi(\Upsilon^2-2)^2}{(2\pi a)^3}\int^{\infty}_{0}dk~k^2\left|\left\{\frac{i~e^{-2ikc_{S}\eta^{'}}}{15}\left\{\text{Ei}(2 ikc_{S} \tau)-\text{Ei}(2 ik c_{S}\tau^{'})\right\}
	\right.\right.\\\left.\left.-\frac{e^{-2 ikc_{S}\eta^{'}}}{120k^5}\left[\frac{ e^{2ikc_{S}\tau}\left(
	4 (c_{S}k)^4 \tau^4-2 i(c_{S}k)^3 \tau^3-2 (c_{S}k)^2 \tau^2+3 ikc_{S} \tau+6\right)}{\tau^5}\right.\right.\right.\\\left.\left.\left.
	~~-\frac{ e^{2kc_{S}\tau}\left(
	4 (c_{S}k)^4 \tau^{'4}-2 i(c_{S}k)^3 \tau^{'3}-2 (c_{S}k)^2 \tau^{'2}+3 ikc_{S} \tau^{'}+6\right)}{\tau^{'5}}\right]\right\}\right|^2~~~ & \mbox{\small {\bf for ~$|kc_{S}\eta|>>1$}}.
          \end{array}%
\right.\nonumber
\eea
where $V$ is defined in the earlier subsection. 

Finally one can define the total energy density of the produced particles using the following expression:
\be%
\ee
which is not exactly analytically computable. To study the behaviour of this integral we consider here three 
consecutive physical situations-$|kc_{S}\eta|<<1$, $|kc_{S}\eta|\approx 1-\Delta(\rightarrow 0)$ and $|kc_{S}\eta|>>1$ for de Sitter and quasi de Sitter case. 
In three cases we have:
\bea &&\underline{\bf For~dS:}\\
\small\rho(\tau,\tau^{'},\eta^{'})&=&\footnotesize \left\{%
\right.\nonumber
\eea
\begin{figure*}[htb]
\centering
\subfigure[]{
    \includegraphics[width=7.2cm,height=6cm] {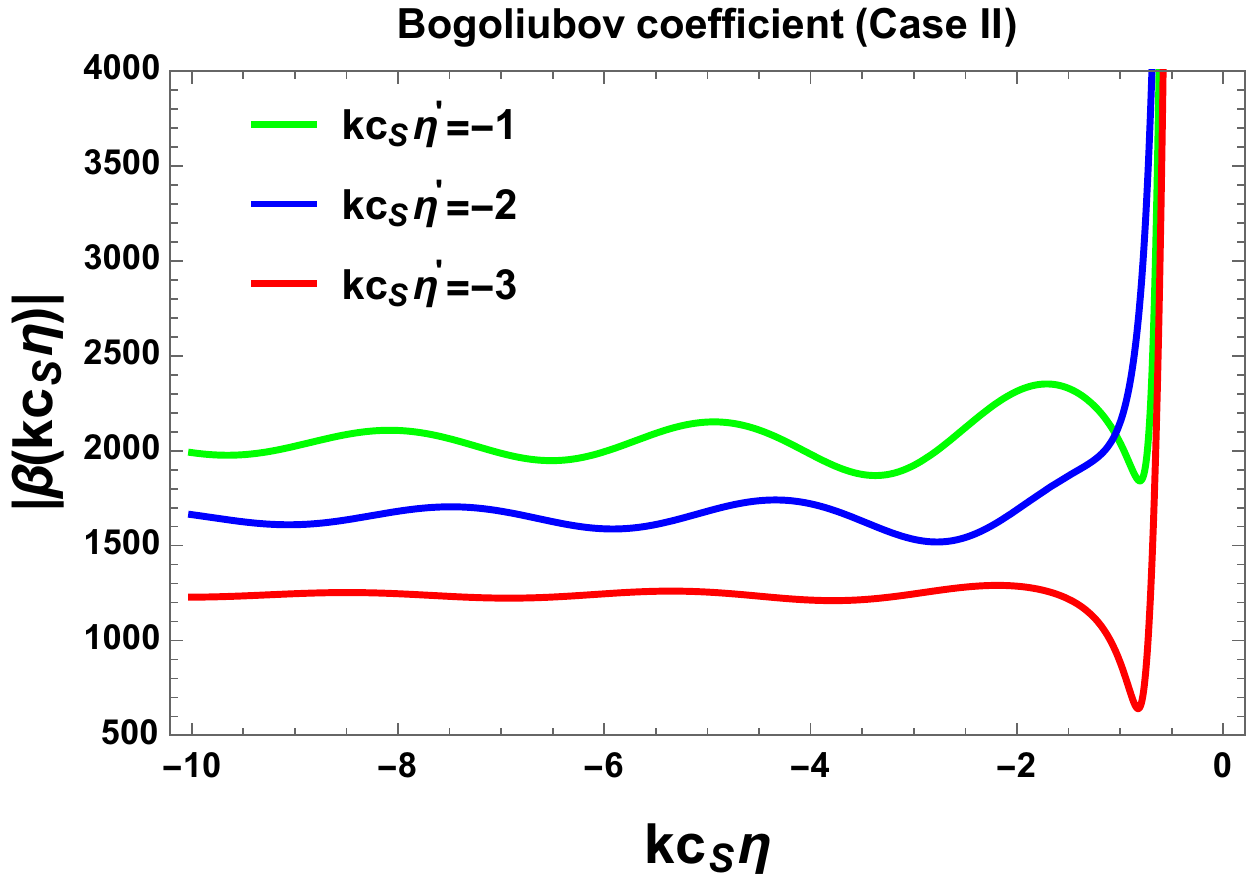}
    \label{fig1xcca}
}
\subfigure[]{
    \includegraphics[width=7.2cm,height=6cm] {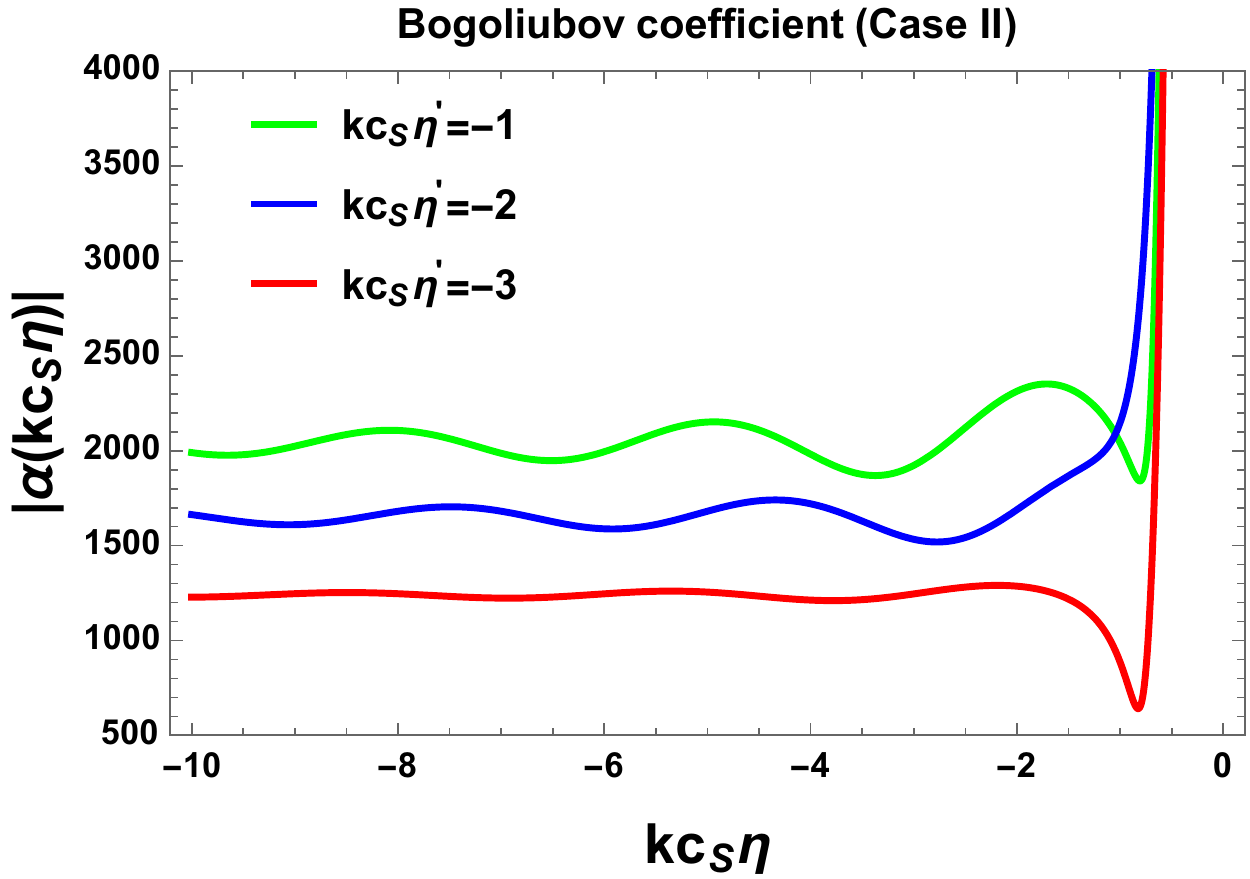}
    \label{fig2xccb}
}
\subfigure[]{
    \includegraphics[width=7.2cm,height=6cm] {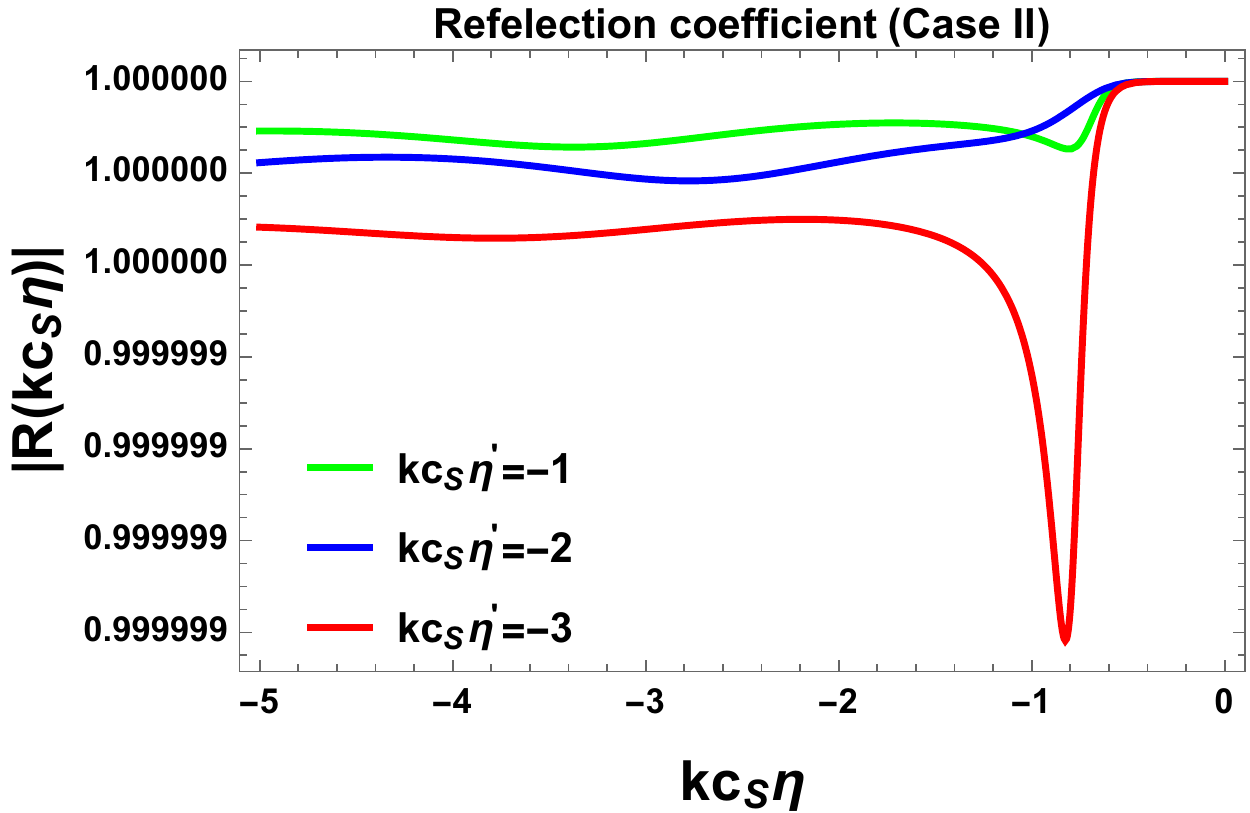}
    \label{fig3xccc}
}
\subfigure[]{
    \includegraphics[width=7.2cm,height=6cm] {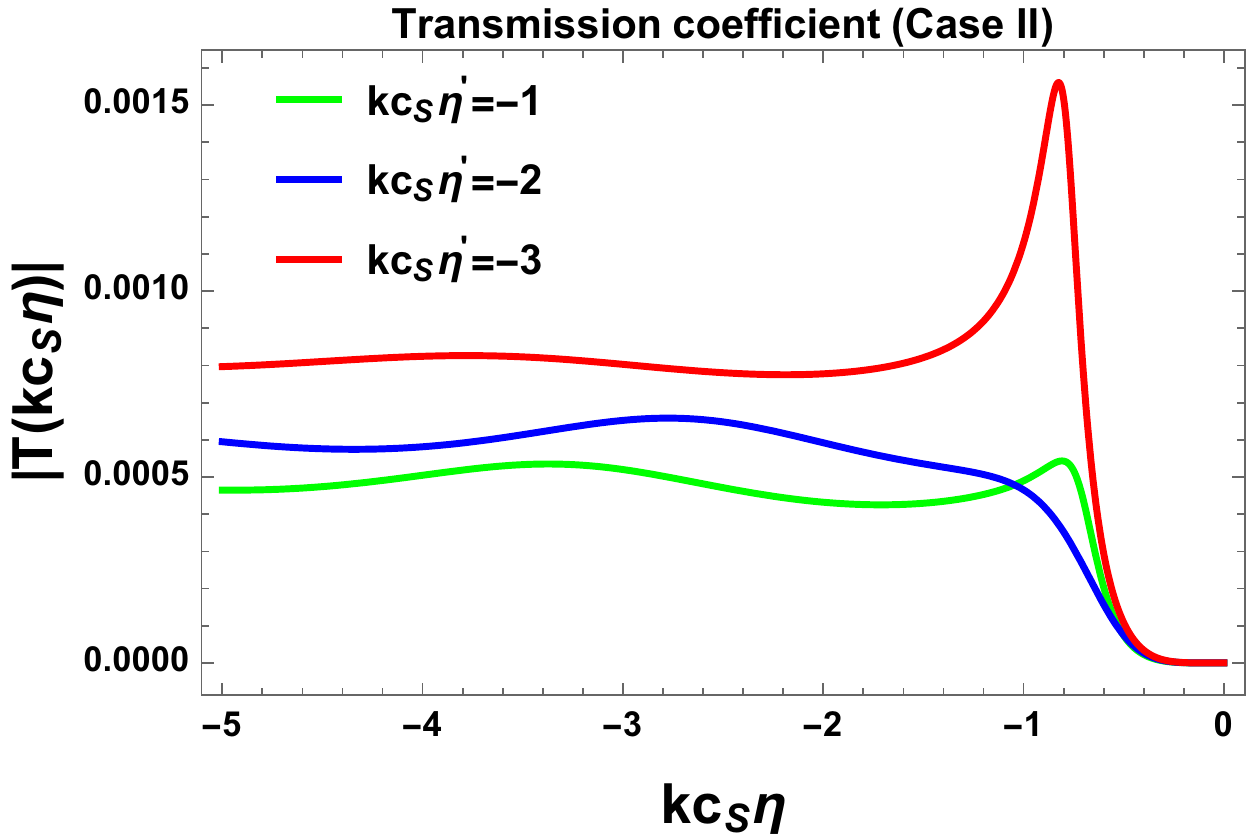}
    \label{fig4xccd}
}
\subfigure[]{
    \includegraphics[width=7.2cm,height=6.1cm] {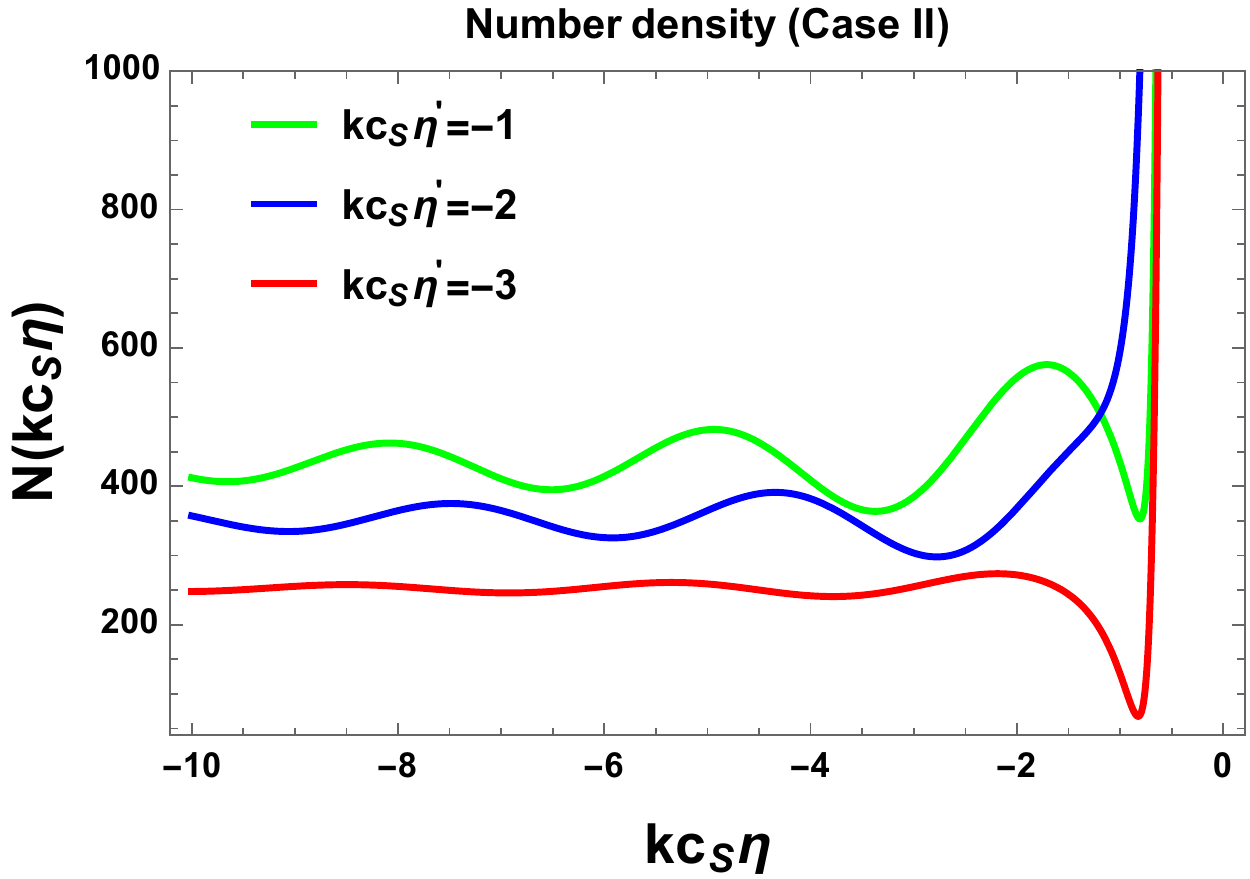}
    \label{fig3xcce}
}
\subfigure[]{
    \includegraphics[width=7.2cm,height=6.1cm] {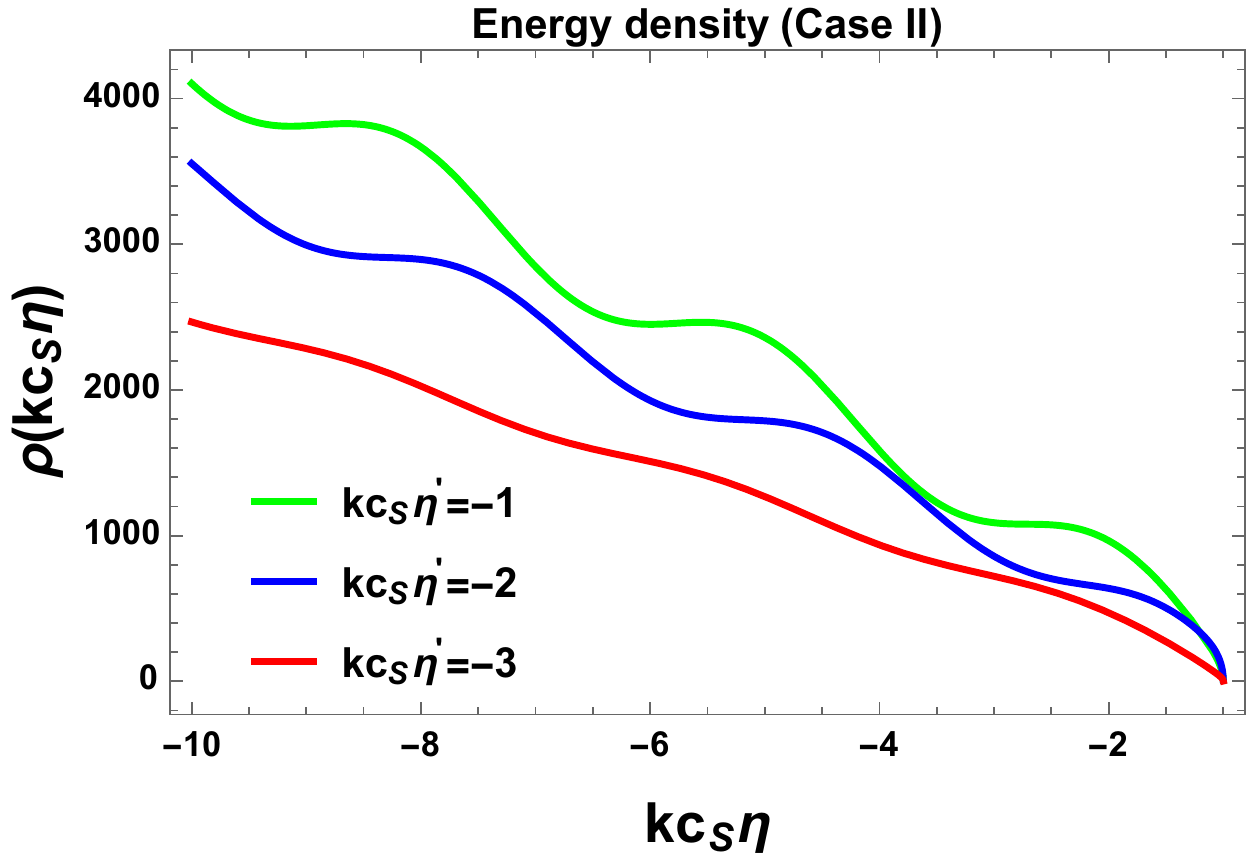}
    \label{fig4xccf}
}
\caption[Optional caption for list of figures]{Particle creation profile for {\bf Case II}.} 
\label{bog3}
\end{figure*}
\begin{figure*}[htb]
\centering
\subfigure[]{
    \includegraphics[width=7.2cm,height=6cm] {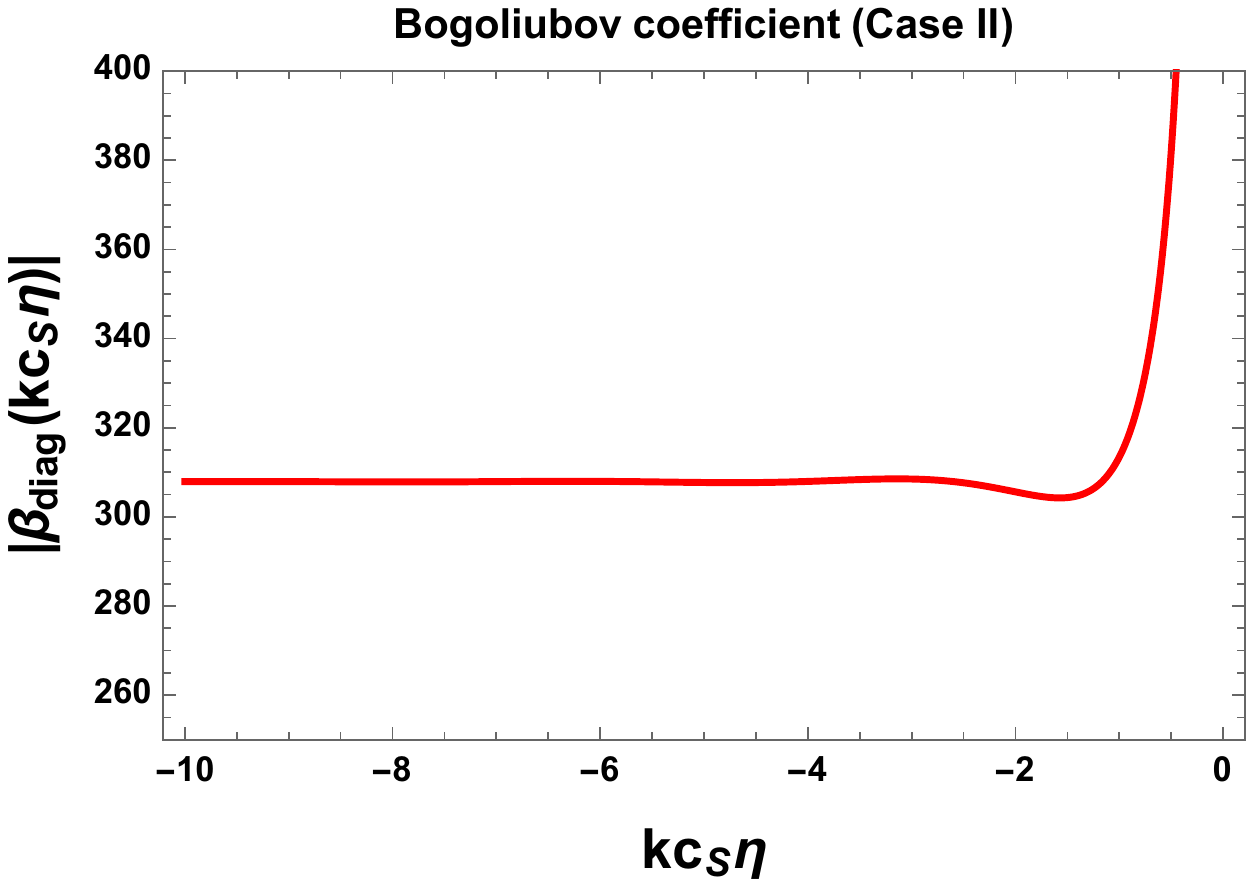}
    \label{fig1xccg}
}
\subfigure[]{
    \includegraphics[width=7.2cm,height=6cm] {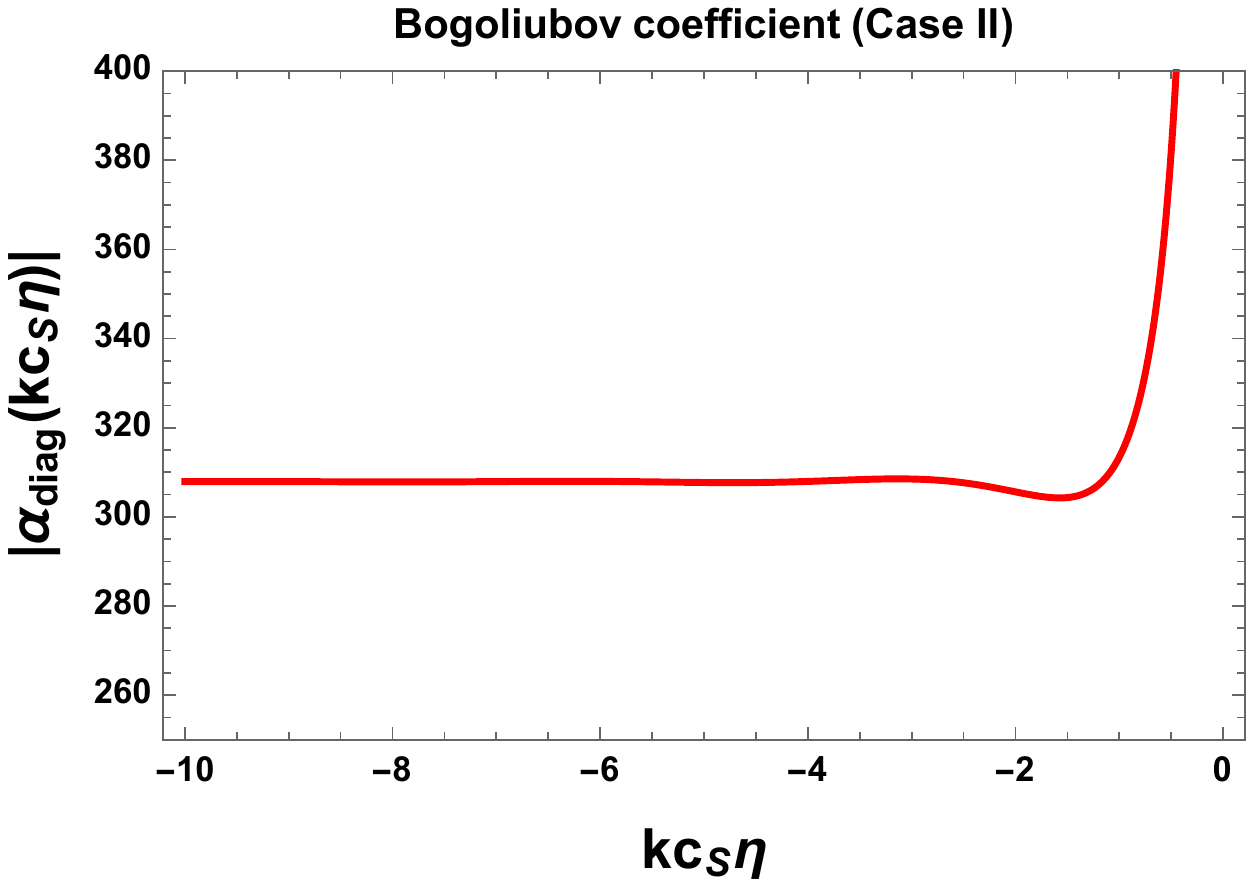}
    \label{fig2xcch}
}
\subfigure[]{
    \includegraphics[width=7.2cm,height=6cm] {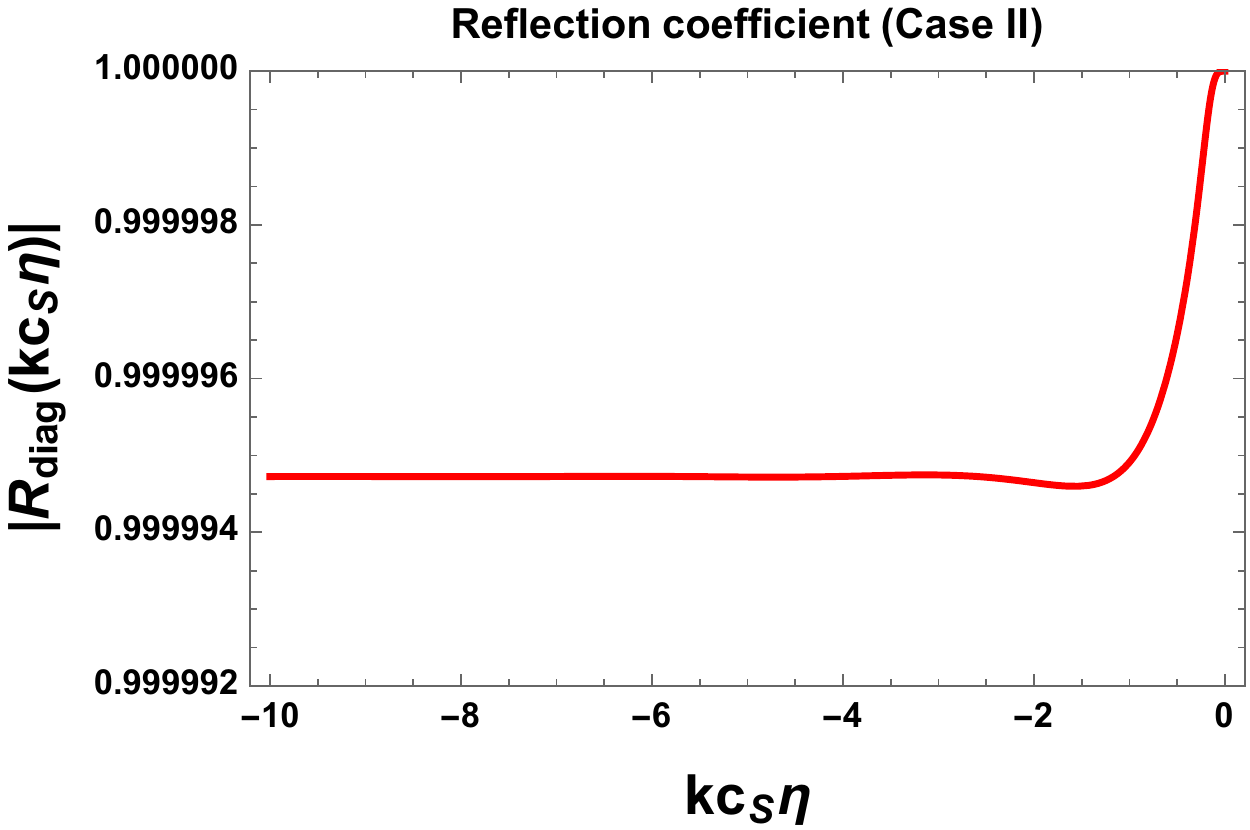}
    \label{fig3xcci}
}
\subfigure[]{
    \includegraphics[width=7.2cm,height=6cm] {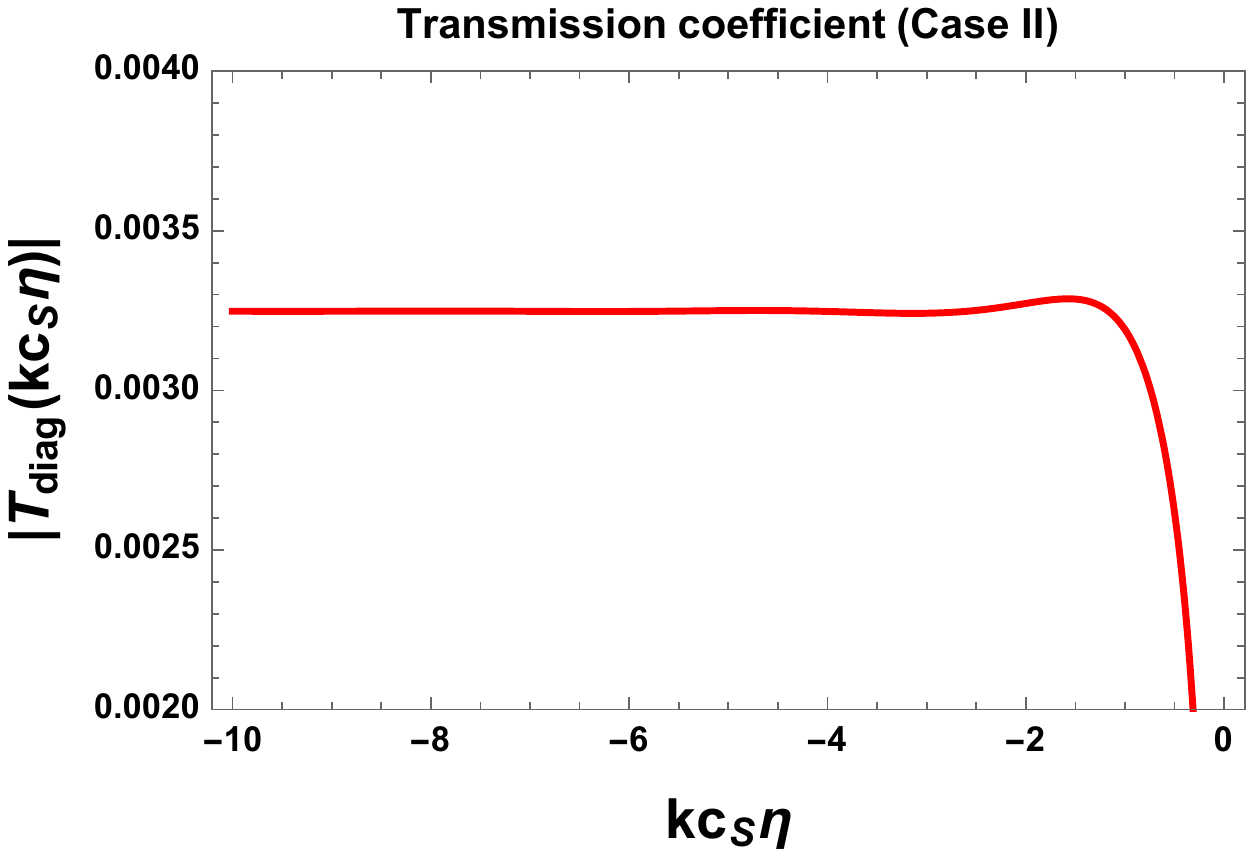}
    \label{fig4xccj}
}
\subfigure[]{
    \includegraphics[width=7.2cm,height=6.1cm] {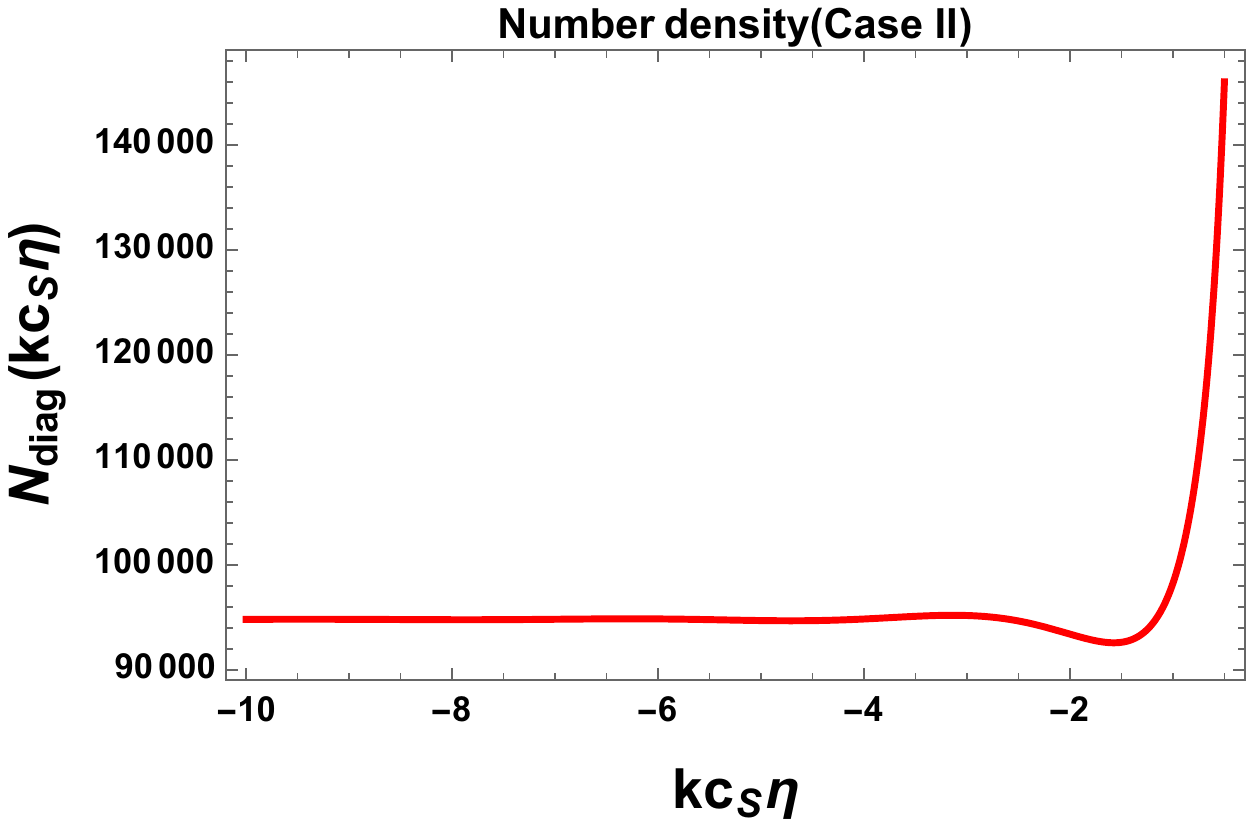}
    \label{fig3xcck}
}
\subfigure[]{
    \includegraphics[width=7.2cm,height=6.1cm] {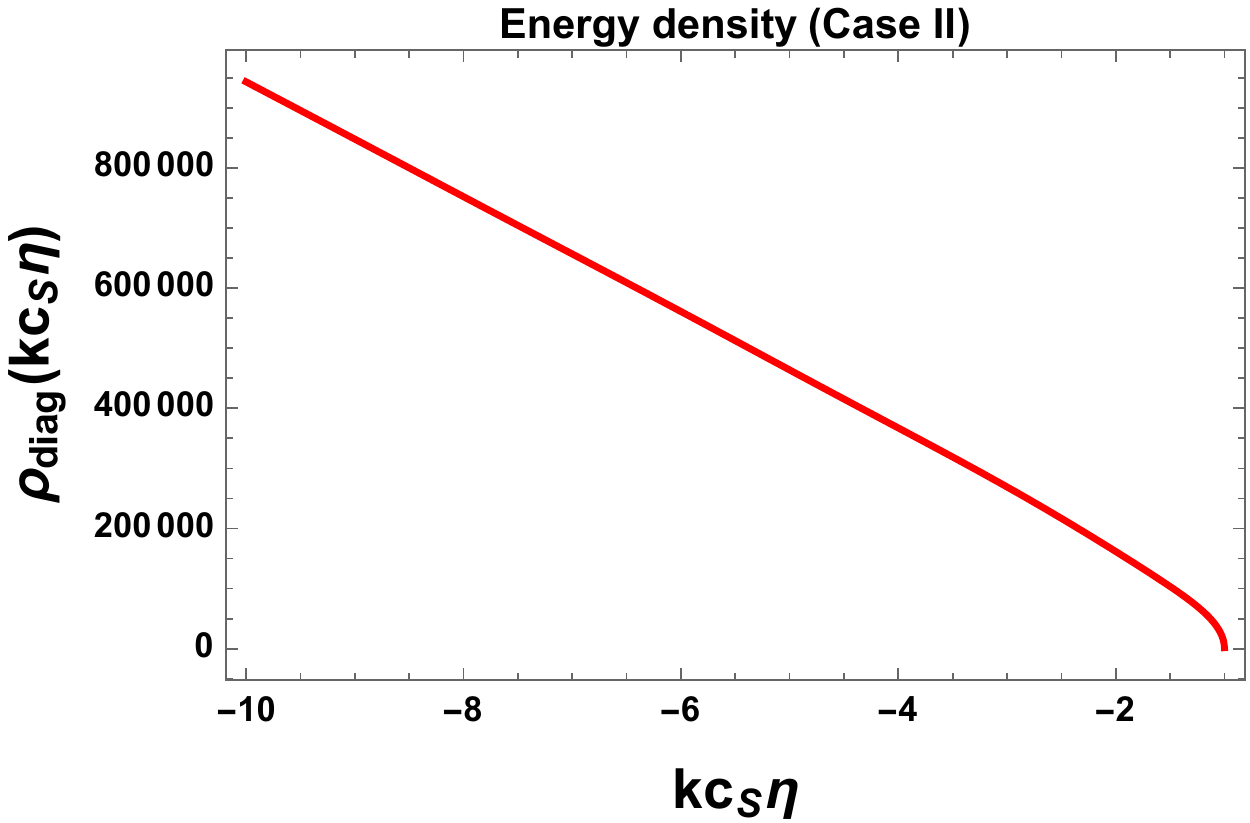}
    \label{fig4xccl}
}
\caption[Optional caption for list of figures]{Particle creation profile for {\bf Case II} in diagonalized representation.} 
\label{bog4}
\end{figure*}
Throughout the discussion of total energy density of the produced particles 
we have introduced a symbol $J$ defined as:
\be\begin{array}{lll}\label{kghjk}
 \displaystyle J=\int d^3 {\bf k}~p(\tau)=\left\{\begin{array}{ll}
                    \displaystyle   \int d^3 {\bf k}~\sqrt{c^{2}_{S}k^2 + \frac{\Upsilon^2-2}{\tau^2}}~~~~ &
 \mbox{\small {\bf for ~dS}}  \\ 
	\displaystyle  \int d^3 {\bf k}~\sqrt{c^{2}_{S}k^2 + 
\frac{\left[\Upsilon^2-\left(\nu^2-\frac{1}{4}\right)\right]}{\tau^2}}~~~~ & \mbox{\small {\bf for~ qdS}}.
          \end{array}
\right.
\end{array}\ee
which physically signifies the total finite volume weighted by $p(\eta)$ 
in momentum space within which the produced particles are occupied. 

To study the behaviour of this integral we consider here three 
consecutive physical situations-$|kc_{S}\eta|<<1$, $|kc_{S}\eta|\approx 1-\Delta(\rightarrow 0)$ and $|kc_{S}\eta|>>1$ for de Sitter and quasi de Sitter case. 
In three cases we have:
\bea &&\underline{\bf For~dS:}\\
\small J&=&\left\{\begin{array}{ll}
                    \displaystyle  \int d^3 {\bf k}~ \frac{\sqrt{\Upsilon^2-2}}{\tau}=\frac{V\sqrt{\Upsilon^2-2}}{\tau}
                    ~~&
 \mbox{\small {\bf for ~$|kc_{S}\eta|<<1$}}  \\ 
 \displaystyle  \int d^3 {\bf k}~\frac{\sqrt{\Upsilon^2-2\Delta-1}}{\tau}=\frac{\sqrt{\Upsilon^2-2\Delta-1}~V}{\tau}~~~~ &
 \mbox{\small {\bf for ~$|kc_{S}\eta|\approx 1-\Delta(\rightarrow 0)$}}  \\ 
	\displaystyle\int d^3 {\bf k}~kc_{S} & \mbox{\small {\bf for ~$|kc_{S}\eta|>>1$}}.~~~~~~~~
          \end{array}
\right.\nonumber\eea
\bea
&&\underline{\bf For~qdS:}\\
\small J&=&\left\{\begin{array}{ll}
                    \displaystyle  \int d^3 {\bf k}~\frac{\sqrt{\Upsilon^2-\left(\nu^2-\frac{1}{4}\right)}}{\tau}=\frac{V\sqrt{\Upsilon^2-\left(\nu^2-\frac{1}{4}\right)}}{\tau}~~ &
 \mbox{\small {\bf for ~$|kc_{S}\eta|<<1$}}  \\ 
 \displaystyle   \int d^3 {\bf k}~\frac{\sqrt{\Upsilon^2-2\Delta-\left(\nu^2-\frac{5}{4}\right)}}{\tau}=\frac{\sqrt{\Upsilon^2-2\Delta-\left(\nu^2-\frac{5}{4}\right)}~V}{\tau}
                    ~~&
 \mbox{\small {\bf for ~$|kc_{S}\eta|\approx 1-\Delta(\rightarrow 0)$}} \\ 
	\displaystyle \int d^3 {\bf k}~kc_{S}~~ & \mbox{\small {\bf for ~$|kc_{S}\eta|>>1$}}.~~~~~~~~
          \end{array}
\right.\nonumber
\eea
In fig.~(\ref{bog3}) and fig.~(\ref{bog4}), we have explicitly shown the particle creation profile for {\bf Case II} for two representations.
\subsubsection{\bf Case III: $m<<H$}
Here we set $m<<H$ for the computation. Here the equation of motion for the field with mass $m<<H$ is given by:
\bea
h''_k + \left\{c^{2}_{S}k^2 - \frac{2}{\eta^2} \right\} h_k &=& 0~~~~~~~{\bf for~ dS}\\\nonumber \\
h''_k + \left\{c^{2}_{S}k^2- 
\frac{\left(\nu^2-\frac{1}{4}\right)}{\eta^2} \right\} h_k &=& 0~~~~~~~{\bf for~ qdS}.
\eea
The solution for the mode function for de Sitter and quasi de Sitter space can be expressed as: 
\be\begin{array}{lll}\label{yu223}
 \displaystyle h_k (\eta) =\left\{\begin{array}{ll}
                    \displaystyle   \sqrt{-\eta}\left[C_1  H^{(1)}_{3/2} \left(-kc_{S}\eta\right) 
+ C_2 H^{(2)}_{3/2} \left(-kc_{S}\eta\right)\right]~~~~ &
 \mbox{\small {\bf for ~dS}}  \\ 
	\displaystyle \sqrt{-\eta}\left[C_1  H^{(1)}_{\nu} \left(-kc_{S}\eta\right) 
+ C_2 H^{(2)}_{\nu} \left(-kc_{S}\eta\right)\right]~~~~ & \mbox{\small {\bf for~ qdS}}.
          \end{array}
\right.
\end{array}\ee
where $C_{1}$ and and $C_{2}$ are two arbitrary integration constant, which depend on the 
choice of the initial condition.

After taking $kc_{S}\eta\rightarrow -\infty$, $kc_{S}\eta\rightarrow 0$ and $|kc_{S}\eta|\approx 1$ limit the most general 
solution as stated in Eq~(\ref{yu223}) can be recast as:
\bea\label{yu2}
 \displaystyle h_k (\eta) &\stackrel{|kc_{S}\eta|\rightarrow-\infty}{=}&\left\{\begin{array}{ll}
                    \displaystyle   -\sqrt{\frac{2}{\pi kc_{S}}}\left[C_1  e^{ -ikc_{S}\eta}
+ C_2 e^{ ikc_{S}\eta}\right]~~~~ &
 \mbox{\small {\bf for ~dS}}  \\ 
	\displaystyle \sqrt{\frac{2}{\pi kc_{S}}}\left[C_1  e^{- ikc_{S}\eta}
e^{-\frac{i\pi}{2}\left(\nu+\frac{1}{2}\right)} 
+ C_2 e^{ ikc_{S}\eta}
e^{\frac{i\pi}{2}\left(\nu+\frac{1}{2}\right)}\right]~~~~ & \mbox{\small {\bf for~ qdS}}.
          \end{array}
\right.\eea 
\bea
\label{yu2}
 \displaystyle h_k (\eta) &\stackrel{|kc_{S}\eta|\rightarrow 0}{=}&\left\{\begin{array}{ll}
                    \displaystyle  \frac{i\sqrt{-\eta}}{2\sqrt{\pi}}
                    \left(-\frac{kc_{S}\eta}{2}\right)^{-\frac{3}{2}}\left[C_1   
- C_2 \right]~ &
 \mbox{\small {\bf for ~dS}}  \\ 
	\displaystyle\frac{i\sqrt{-\eta}}{\pi}\Gamma\left(\nu\right)\left(-\frac{kc_{S}\eta}{2}\right)^{-\nu}\left[C_1   
- C_2 \right]~ & \mbox{\small {\bf for~ qdS}}.
          \end{array}
\right.
\eea
\bea
\label{yu2}
 \displaystyle h_k (\eta) &\stackrel{|kc_{S}\eta|\approx 1-\Delta(\rightarrow 0)}{=}&\left\{\begin{array}{ll}
                    \displaystyle  \frac{i}{\pi}\sqrt{-\eta}\left[ \frac{2}{3}-\gamma+\frac{3}{4}
                   \left(\gamma^2+\frac{\pi^2}{6}\right)\right.\\ \displaystyle \left.
                   \displaystyle~~~~~~~~-\frac{9}{24}\left(\gamma^3+\frac{\gamma \pi^2}{2}+2\zeta(3)\right)
                   +\cdots\right]\\ \displaystyle ~~~~~~~~~~~~\times\left(\frac{1+\Delta}{2}\right)^{-\frac{3}{2}}\left[C_1   
- C_2 \right]~&
 \mbox{\small {\bf for ~dS}}  \\ 
	\displaystyle\frac{i}{\pi}\sqrt{-\eta}\left[ \frac{1}{\nu}
  -\gamma  \displaystyle +\frac{\nu}{2}\left(\gamma^2+\frac{\pi^2}{6}\right) 
  \right.\\ \displaystyle \left. \displaystyle~-\frac{\nu^2}{6}\left(\gamma^3+\frac{\gamma \pi^2}{2}+2\zeta(3)\right) 
  +\cdots\right]\\ \displaystyle ~~~~~~~~~~~~\times\left(\frac{1+\Delta}{2}\right)^{-\nu}\left[C_1   
- C_2 \right]~ & \mbox{\small {\bf for~ qdS}}.
          \end{array}
\right.
\eea
In the standard WKB approximation the total solution can be recast in the following form:
\bea\label{df323}
h_k (\eta)&=& \left[D_{1}u_{k}(\eta) + D_{2} \bar{u}_{k}(\eta)\right],\eea
where $D_{1}$ and and $D_{2}$ are two arbitrary integration constant, which depend on the 
choice of the initial condition during WKB approximation at early and late time scale.
In the present context $u_{k}(\eta)$ and $\bar{u}_{k}(\eta)$ are defined as:
\be\begin{array}{lll}\label{solp}
\small u_{k}(\eta) =\footnotesize\left\{\begin{array}{ll}
                    \displaystyle   \frac{1}{\sqrt{2\sqrt{c^{2}_{S}k^2 - \frac{2}{\eta^2} }}}
\exp\left[i\int^{\eta} d\eta^{\prime} \sqrt{c^{2}_{S}k^2 - \frac{2}{\eta^{'2}}}\right]\\
= \displaystyle\frac{1}{\sqrt{2\sqrt{c^{2}_{S}k^2 - \frac{2}{\eta^2}}}} 
\exp\left[i\left(\eta\sqrt{c^{2}_{S}k^2 - \frac{2}{\eta^2}} 
+\sqrt{2}{\rm tan}^{-1}\left[\frac{\sqrt{2}}{\eta\sqrt{c^{2}_{S}k^2 - \frac{2}{\eta^{2}}}}\right]
\right)\right]~~~~ &
 \mbox{\small {\bf for ~dS}}  \\ 
	\displaystyle \frac{1}{\sqrt{2\sqrt{c^{2}_{S}k^2 - 
\frac{\left(\nu^2-\frac{1}{4}\right)}{\eta^2}}}}
\exp\left[i\int^{\eta} d\eta^{\prime} \sqrt{c^{2}_{S}k^2 - 
\frac{\left(\nu^2-\frac{1}{4}\right)}{\eta^{'2}}}\right]\\
= \displaystyle\frac{1}{\sqrt{2\sqrt{c^{2}_{S}k^2 - 
\frac{\left(\nu^2-\frac{1}{4}\right)}{\eta^2}}}} 
\exp\left[i\left(\eta\sqrt{c^{2}_{S}k^2 - 
\frac{\left(\nu^2-\frac{1}{4}\right)}{\eta^2}}
\right.\right.\\ \left.\left. \displaystyle +\sqrt{\nu^2-\frac{1}{4}}{\rm tan}^{-1}\left[\frac{\sqrt{\left(\nu^2-\frac{1}{4}\right)}}{\eta\sqrt{c^{2}_{S}k^2 - \frac{\left(\nu^2-\frac{1}{4}\right)}{\eta^{2}}}}\right]
\right)\right]~~~~ & \mbox{\small {\bf for~ qdS}}.
          \end{array}
\right.
\end{array}\ee
\\
\be\begin{array}{lll}\label{sola}
 \small \bar{u}_{k}(\eta) =\footnotesize\left\{\begin{array}{ll}
                    \displaystyle   \frac{1}{\sqrt{2\sqrt{c^{2}_{S}k^2 - \frac{2}{\eta^2} }}}
\exp\left[-i\int^{\eta} d\eta^{\prime} \sqrt{c^{2}_{S}k^2 - \frac{2}{\eta^{'2}}}\right]\\
= \displaystyle\frac{1}{\sqrt{2\sqrt{c^{2}_{S}k^2 - \frac{2}{\eta^2}}}} 
\exp\left[-i\left(\eta\sqrt{c^{2}_{S}k^2 - \frac{2}{\eta^2}} \right.\right.\\ \left.\left. \displaystyle 
+\sqrt{2}{\rm tan}^{-1}\left[\frac{\sqrt{2}}{\eta\sqrt{c^{2}_{S}k^2 - \frac{2}{\eta^{2}}}}\right]
\right)\right]~~~~ &
 \mbox{\small {\bf for ~dS}}  \\  
	\displaystyle \frac{1}{\sqrt{2\sqrt{c^{2}_{S}k^2 - 
\frac{\left(\nu^2-\frac{1}{4}\right)}{\eta^2}}}}
\exp\left[-i\int^{\eta} d\eta^{\prime} \sqrt{c^{2}_{S}k^2 - 
\frac{\left(\nu^2-\frac{1}{4}\right)}{\eta^{'2}}}\right]\\
= \displaystyle\frac{1}{\sqrt{2\sqrt{c^{2}_{S}k^2 - 
\frac{\left(\nu^2-\frac{1}{4}\right)}{\eta^2}}}} 
\exp\left[-i\left(\eta\sqrt{c^{2}_{S}k^2 - 
\frac{\left(\nu^2-\frac{1}{4}\right)}{\eta^2}}
\right.\right.\\ \left.\left. \displaystyle +\sqrt{\nu^2-\frac{1}{4}}{\rm tan}^{-1}\left[\frac{\sqrt{\left(\nu^2-\frac{1}{4}\right)}}{\eta\sqrt{c^{2}_{S}k^2 - \frac{\left(\nu^2-\frac{1}{4}\right)}{\eta^{2}}}}\right]
\right)\right]~~~~ & \mbox{\small {\bf for~ qdS}}.
          \end{array}
\right.
\end{array}\ee
where we have written the total solution for the mode $h_k$ in terms of 
two linearly independent solutions. Here it is important to note that the both of the 
solutions are hermitian conjugate of each other. If in the present context the exact solution of the mode $h_k$ 
is expanded with respect to these two linearly independent solutions then particle creation is absent in our EFT 
setup. In the present context correctness of 
WKB approximation is guarantee at very early and very late time scales. In this discussion $u_{k}(\eta)$
is valid at very early time scale and $\bar{u}_{k}(\eta)$ perfectly works in the late time scale.

Now we will explicitly check that the exactness of the above mentioned WKB result derived in Eq~(\ref{df323}) 
with the actual solution of the mode function as presented in Eq~(\ref{yu223}). As mentioned earlier 
in FLRW space-time in Fourier space Bogoliubov coefficient $\beta(k)$ measures this exactness for a given 
setup. The particle creation mechanism and its exact amount
is described by finding the Bogoliubov coefficient $\beta(k)$ in Fourier space
which in principle measures the exact 
amount of late times solution $u_{k}(\eta)$, if in the present context we exactly 
start with the early time scale solution $u_{k}(\eta)$. In our present computation 
we consider a physical situation where the
WKB approximation is correct up to the leading order throughout the cosmological evolution in time scale.
In the present context the Bogoliubov coefficient $\beta(k)$ in Fourier space
can be computed approximately using the following regularized integral:
\be\begin{array}{lll}\label{soladfbnbn123}
 \displaystyle \beta(k,\tau,\tau^{'},\eta^{'}) =\left\{\begin{array}{ll}
                    \displaystyle   \int^{\tau}_{\tau^{'}}d\eta~\frac{\exp\left[2i\int^{\eta}_{\eta^{'}}
d\eta^{''}\sqrt{c^{2}_{S}k^2 - \frac{2}{\eta^{''2}}}\right]}{
                    \eta^{6}\left(c^{2}_{S}k^2 - \frac{2}{\eta^2}\right)^{\frac{5}{2}}}&
 \mbox{\small {\bf for ~dS}}  \\ 
	\displaystyle   \int^{\tau}_{\tau^{'}}d\eta~\frac{\left[\nu^2-\frac{1}{4}\right]^2
	\exp\left[2i\int^{\eta}_{\eta^{'}}
d\eta^{''}\sqrt{c^{2}_{S}k^2 - 
\frac{\left(\nu^2-\frac{1}{4}\right)}{\eta^{''2}}}\right]}{
                    4\eta^{6}\left(c^{2}_{S}k^2 - 
\frac{\left(\nu^2-\frac{1}{4}\right)}{\eta^2}\right)^{\frac{5}{2}}} & \mbox{\small {\bf for~ qdS}}.
          \end{array}
\right.
\end{array}\ee
which is not exactly analytically computable. To study the behaviour of this integral we consider here three 
consecutive physical situations-$|kc_{S}\eta|<<1$, $|kc_{S}\eta|\approx 1-\Delta(\rightarrow 0)$ and $|kc_{S}\eta|>>1$ for de Sitter and quasi de Sitter case. 
In three cases we have:
\be\begin{array}{lll}\label{yu2dfvqasx3}\small
 \displaystyle \underline{\rm \bf For~dS:}~~~~~~\sqrt{\left\{c^{2}_{S}k^2 - \frac{2}{\eta^2}\right\}}
 \approx\left\{\begin{array}{ll}
                    \displaystyle  \frac{i\sqrt{2}}{\eta}~~~~ &
 \mbox{\small {\bf for ~$|kc_{S}\eta|<<1$}}  \\ 
	\displaystyle \frac{i\sqrt{2\Delta+1}}{\eta}~~~~ & \mbox{\small {\bf for ~$|kc_{S}\eta|\approx 1-\Delta(\rightarrow 0)$}}\\ 
	\displaystyle kc_{S}~~~~ & \mbox{\small {\bf for ~$|kc_{S}\eta|>>1$}}.
          \end{array}
\right.
\end{array}
\ee
\be\begin{array}{lll}\label{yu2dfvqasx3vvb}
\small
 \displaystyle \underline{\rm \bf For~qdS:}~~~~~\sqrt{\left\{c^{2}_{S}k^2 - 
\frac{\left(\nu^2-\frac{1}{4}\right)}{\eta^2}\right\}} \approx\left\{\begin{array}{ll}
                    \displaystyle \frac{i\sqrt{\left(\nu^2-\frac{1}{4}\right)}}{\eta} ~~~~ &
 \mbox{\small {\bf for ~$|kc_{S}\eta|<<1$}}  \\ 
	\displaystyle \frac{i\sqrt{2\Delta+\left(\nu^2-\frac{5}{4}\right)}}{\eta}
	~~~~ & \mbox{\small {\bf for ~$|kc_{S}\eta|\approx 1-\Delta(\rightarrow 0)$}}\\ 
	\displaystyle kc_{S}~~~~ & \mbox{\small {\bf for ~$|kc_{S}\eta|>>1$}}.
          \end{array}
\right.
\end{array}\ee
and further using this result Bogoliubov coefficient $\beta(k)$ in Fourier space can be expressed as:
\bea &&\underline{\bf For~dS:}\\
\small\beta(k,\tau,\tau^{'},\eta^{'})&=&\left\{\begin{array}{ll}
                    \displaystyle  \frac{\eta^{'2\sqrt{2}}}{
                    16i}\left[\frac{1}{\tau^{'2\sqrt{2}}}
                    -\frac{1}{\tau^{2\sqrt{2}}}\right]
                     &
 \mbox{\small {\bf for ~$|kc_{S}\eta|<<1$}}  \\ \\
   \displaystyle \frac{\eta^{'2\sqrt{2\Delta +1}}}{
                    2i(2\Delta+1)^3}\left[\frac{1}{\tau^{'2\sqrt{2\Delta +1}}}-\frac{1}{\tau^{2\sqrt{2\Delta +1}}}\right] &
 \mbox{\small {\bf for ~$|kc_{S}\eta|\approx 1-\Delta(\rightarrow 0)$}}  \\ \\
	\displaystyle 4\left[i\frac{\text{Ei}(2 ikc_{S} \eta)e^{-2ikc_{S}\eta^{'}}}{15}\right.\\ \left.\displaystyle-\frac{e^{2 i k c_{S}(\eta-\eta^{'})} }{120 (c_{S}k)^5 \eta^5}\left(
	4 (c_{S}k)^4 \eta^4-2 i (c_{S}k)^3 \eta^3\right.\right.\\ \left.\left.-2 (c_{S}k)^2 \eta^2+3 i c_{S}k \eta+6\right)\right]^{\tau}_{\tau^{'}}& \mbox{\small {\bf for ~$|kc_{S}\eta|>>1$}}.
          \end{array}
\right.\nonumber\eea
\bea
&&\underline{\bf For~qdS:}\\
\small\beta(k,\tau,\tau^{'},\eta^{'})&=& 
\left\{\begin{array}{ll}
                 \displaystyle   \frac{\eta^{'2\sqrt{\nu^2-\frac{1}{4}}}}{
                    8i\left(\nu^2-\frac{1}{4}\right)}
                    \left[\frac{1}{\tau^{'2\sqrt{\nu^2-\frac{1}{4}}}}-\frac{1}{\tau^{2\sqrt{\nu^2-\frac{1}{4}}}}\right]
                     &
 \mbox{\small {\bf for ~$|kc_{S}\eta|<<1$}}  \\ \\
 \displaystyle\frac{\left(\nu^2-\frac{1}{4}\right)^2\eta^{'2\sqrt{2\Delta +\left(\nu^2-\frac{5}{4}\right)}}}{
                    8i\left[2\Delta +\left(\nu^2-\frac{5}{4}\right)\right]^3
                    }\left[\frac{1}{\tau^{'2\sqrt{2\Delta +\left(\nu^2-\frac{5}{4}\right)}}}
                    -\frac{1}{\tau^{2\sqrt{2\Delta +\left(\nu^2-\frac{5}{4}\right)}}}\right] &
 \mbox{\small {\bf for ~$|kc_{S}\eta|\approx 1-\Delta(\rightarrow 0)$}}  \\ \\
	\displaystyle \left(\nu^2-\frac{1}{4}\right)^2\left[i\frac{\text{Ei}(2 ikc_{S} \eta)e^{-2ikc_{S}\eta^{'}}}{15}\right.\\ \left.\displaystyle-\frac{e^{2 i k c_{S}(\eta-\eta^{'})} }{120 (c_{S}k)^5 \eta^5}\left(
	4 (c_{S}k)^4 \eta^4-2 i (c_{S}k)^3 \eta^3\right.\right.\\ \left.\left.-2 (c_{S}k)^2 \eta^2+3 i c_{S}k \eta+6\right)\right]^{\tau}_{\tau^{'}}& \mbox{\small {\bf for ~$|kc_{S}\eta|>>1$}}.
          \end{array}
\right.\nonumber\eea
In all the situation described for de Sitter and quasi de Sitter case here 
the magnitude of the Bogoliubov coefficient $|\beta(k)|$ in Fourier space is considerably small. Specifically it is important 
to point out here that for the case when $|kc_{S}\eta|>>1$ the value
of the Bogoliubov coefficient $\beta(k)$ in Fourier space
is even smaller as the WKB approximated solution is strongly consistent for all time scales. On the other hand near the 
vicinity of the conformal time scale $\eta\sim \eta_{pair}$ for $|kc_{S}\eta_{pair}|<<1$
the WKB approximated solution is less strongly valid and to validate the solution at this time scale 
one can neglect the momentum $k$ dependence in the Bogoliubov coefficient $\beta(k)$ in Fourier space.
Here $|\eta_{pair}|$ characterizes the relative
separation between the created particles. 

As mentioned earlier here one can use another equivalent way
to define the the Bogoliubov coefficient $\beta$ in Fourier 
space by implementing instantaneous Hamiltonian
diagonalization method to interpret the results. 
Using this diagonalized representation the
regularized Bogoliubov coefficient $\beta$ in Fourier 
space can be written as:
\be
\ee
where $\tau$ and $\tau^{'}$ introduced as the conformal time regulator in the present context.
In this case as well the Bogoliubov coefficient is not exactly analytically computable. 
To study the behaviour of this integral we consider here three similar
consecutive physical situations for de Sitter and quasi de Sitter case as discussed earlier.
\bea &&\underline{\bf For~dS:}\\
\small\beta_{diag}(k;\tau,\tau^{'}) &=&\left\{\begin{array}{ll}
                    \displaystyle  \frac{1}{
                    4\sqrt{2}}\left[\tau^{2\sqrt{2}}-
                    \tau^{'2\sqrt{2}}\right]
                     &
 \mbox{\small {\bf for ~$|kc_{S}\eta|<<1$}}  \\ \\ 
 \displaystyle    \frac{1}{
                    2i(2\Delta +1)^{3/2}}\left[\tau^{2\sqrt{2\Delta+1}}-
                    \tau^{'2\sqrt{2\Delta+1}}\right] &
 \mbox{\small {\bf for ~$|kc_{S}\eta|\approx 1-\Delta(\rightarrow 0)$}}  \\ \\ 
	\displaystyle -2\left[\frac{e^{-2 i kc_{S} \eta} \left(2ikc_{S}\eta-1\right)}{4 (c_{S}k)^2 \eta^2}-\text{Ei}(-2 i kc_{S} \eta)\right]^{\tau}_{\tau^{'}} & \mbox{\small {\bf for ~$|kc_{S}\eta|>>1$}}.
          \end{array}%
\right.\nonumber
\eea

Further using the regularized expressions for Bogoliubov co-efficient $\beta$ in two different representations as mentioned in
Eq~(\ref{soladfbnbn123}) and Eq~(\ref{soladfasxsd3cv123}), and substituting them in Eq~(\ref{sdq1}) we get the following
regularized expressions for the Bogoliubov co-efficient $\alpha$ in two different representations as given by:
\be%
\ee
where $\phi$ and $\phi_{diag}$ are the associated phase factors in two different representations.
Here the results are not exactly analytically computable. To study the behaviour of this integral we consider here three 
consecutive physical situations-$|kc_{S}\eta|<<1$, $|kc_{S}\eta|\approx 1-\Delta(\rightarrow 0)$ and
$|kc_{S}\eta|>>1$ for de Sitter and quasi de Sitter case. 
\bea &&\underline{\bf For~dS:}\\
\small\alpha(k,\tau,\tau^{'},\eta^{'})&=& \left\{%
\right.\nonumber
\eea
Further using the expressions for Bogoliubov co-efficient $\alpha$ in two different representations
and substituting them in Eq~(\ref{sdq2}) we get the following
expressions for the reflection and transmission co-efficient in two different representations 
for three consecutive physical situations-$|kc_{S}\eta|<<1$, $|kc_{S}\eta|\approx 1-\Delta(\rightarrow 0)$ and
$|kc_{S}\eta|>>1$ for de Sitter and quasi de Sitter case as given by:
\bea &&\underline{\bf For~dS:}\\
{\cal R}&=&\left\{%
\right.\nonumber\eea

 Next the expression for the
number of produced particles at time $\tau$ can be calculated from the following formula:
\be%
\ee
which is not exactly analytically computable. To study the behaviour of this integral we consider here three 
consecutive physical situations-$|kc_{S}\eta|<<1$, $|kc_{S}\eta|\approx 1-\Delta(\rightarrow 0)$ and $|kc_{S}\eta|>>1$ for de Sitter and quasi de Sitter case. 
In three cases we have:
\bea &&\underline{\bf For~dS:}\\
{\cal N}(\tau,\tau^{'},\eta^{'})&=& \footnotesize\left\{%
\right.\nonumber
\eea
where $V$ is introduced earlier. 

Finally one can define the total energy density of the produced particles using the following expression:
\be%
\ee
which is not exactly analytically computable. To study the behaviour of this integral we consider here three 
consecutive physical situations-$|kc_{S}\eta|<<1$, $|kc_{S}\eta|\approx 1-\Delta(\rightarrow 0)$ and $|kc_{S}\eta|>>1$ for de Sitter and quasi de Sitter case. 
In three cases we have:
\bea &&\underline{\bf For~dS:}\\
\small\rho(\tau,\tau^{'},\eta^{'})&=& \footnotesize\left\{%
\right.\nonumber
\eea 
\begin{figure*}[htb]
\centering
\subfigure[]{
    \includegraphics[width=7.2cm,height=6cm] {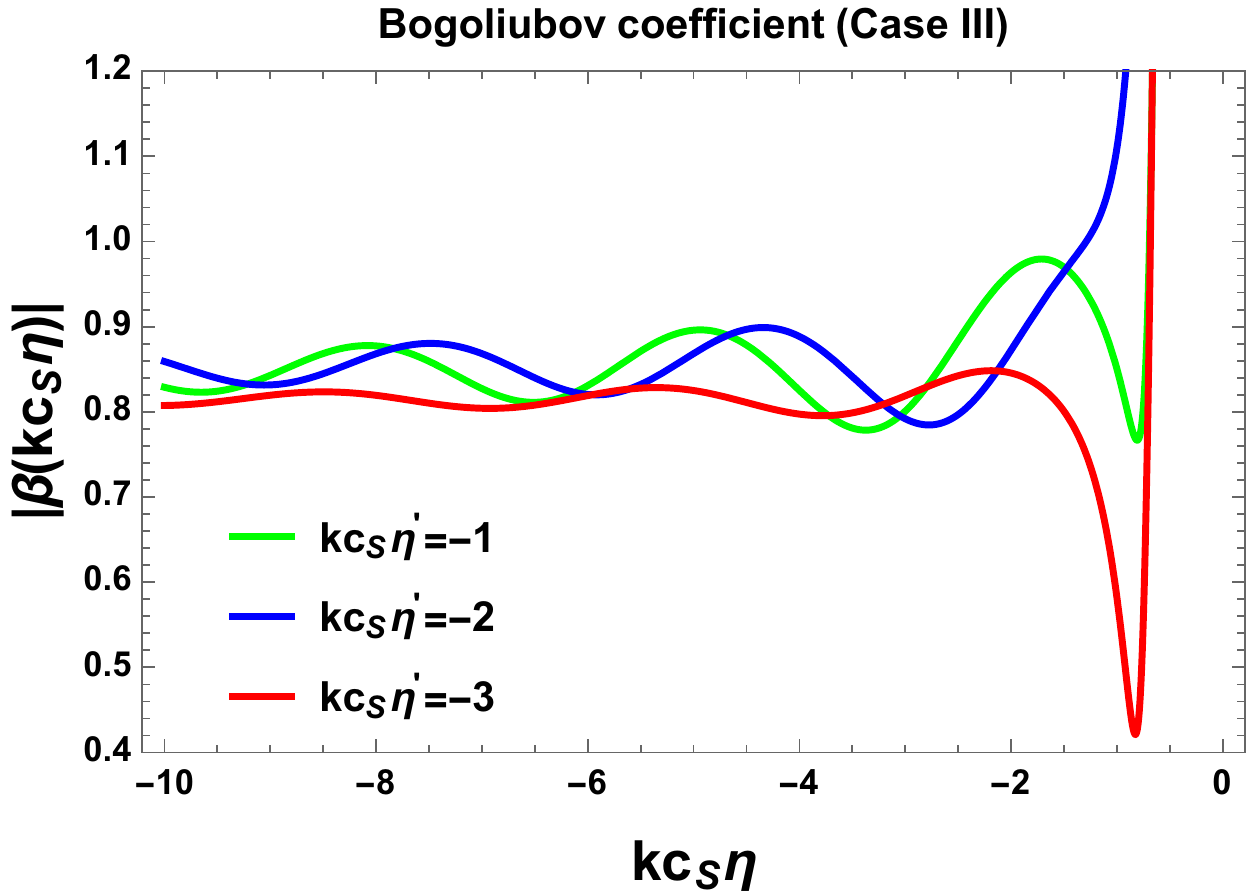}
    \label{fig1xcc}
}
\subfigure[]{
    \includegraphics[width=7.2cm,height=6cm] {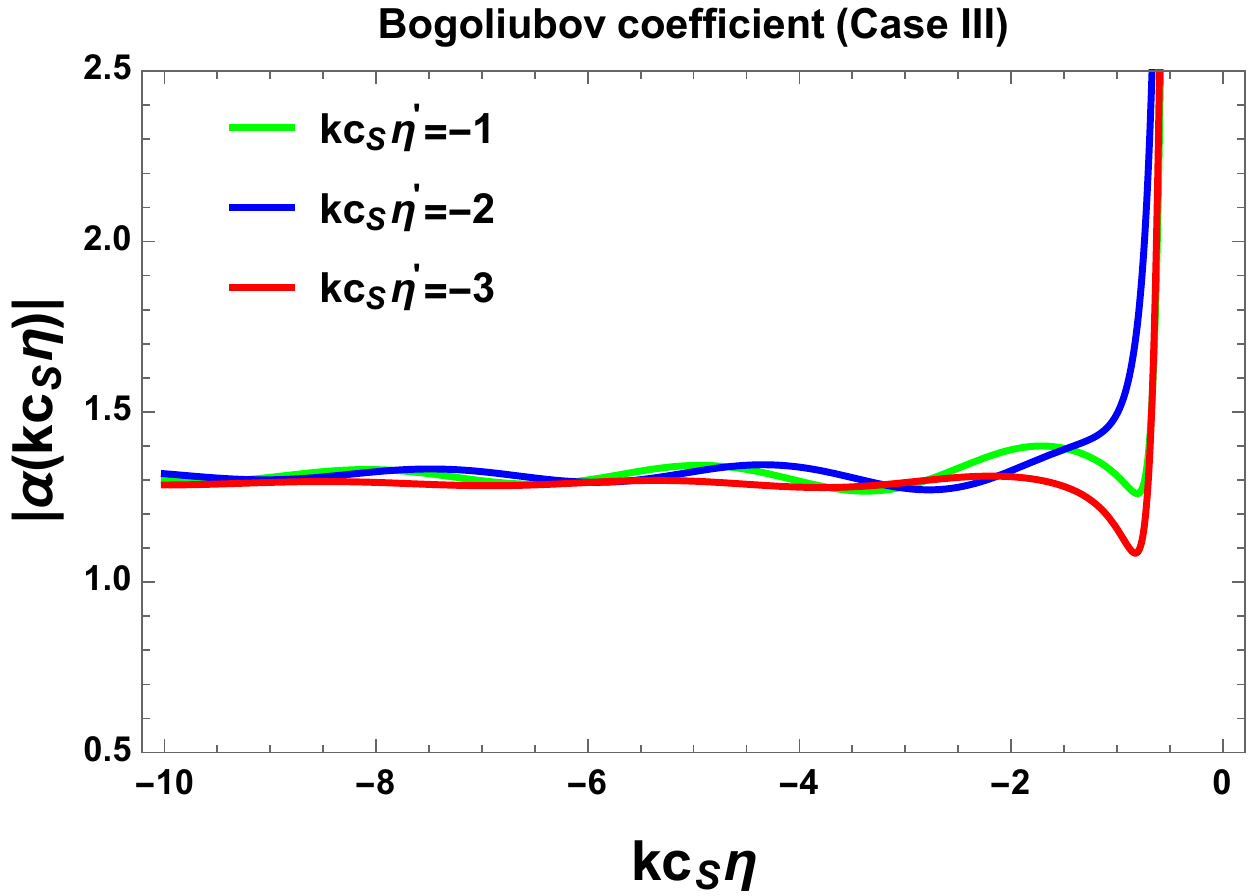}
    \label{fig2xcc}
}
\subfigure[]{
    \includegraphics[width=7.2cm,height=6cm] {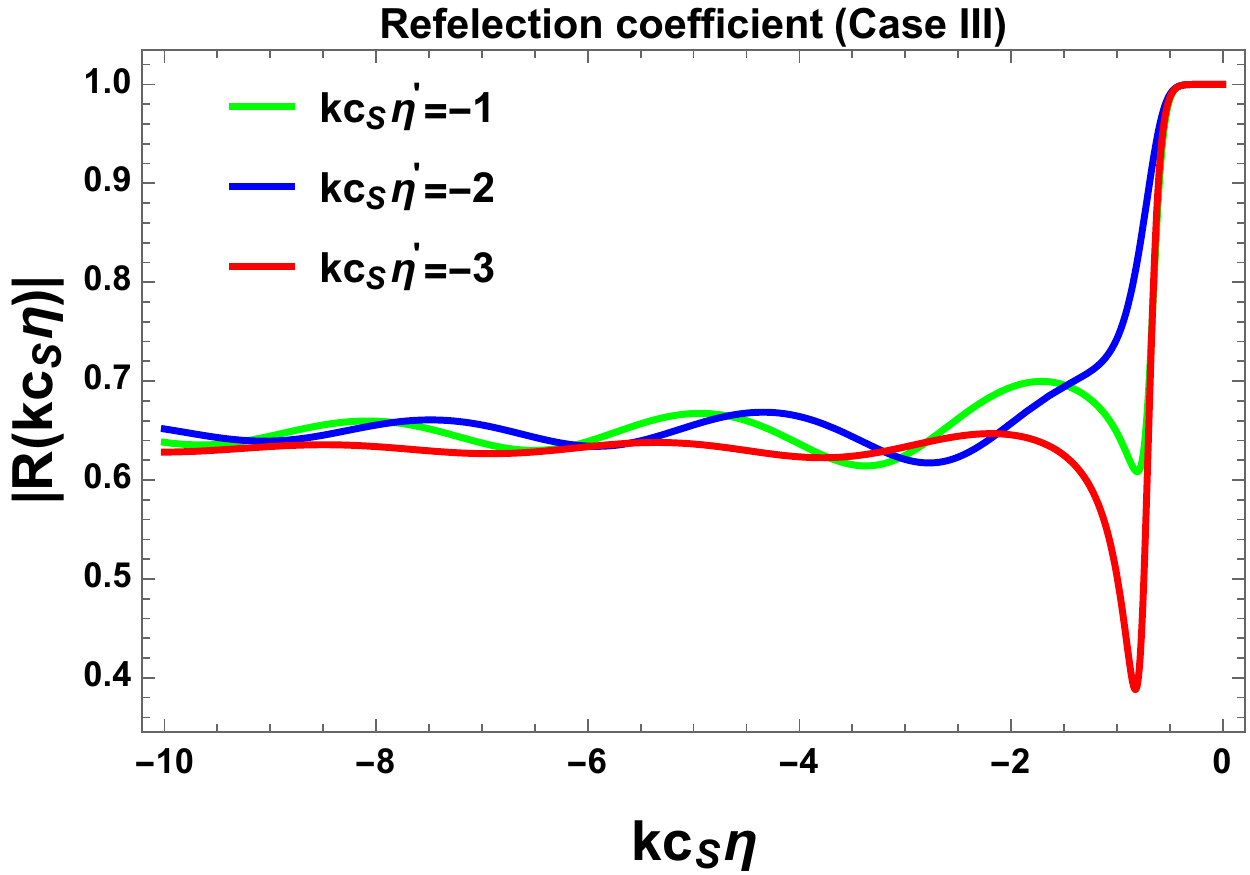}
    \label{fig3xcc}
}
\subfigure[]{
    \includegraphics[width=7.2cm,height=6cm] {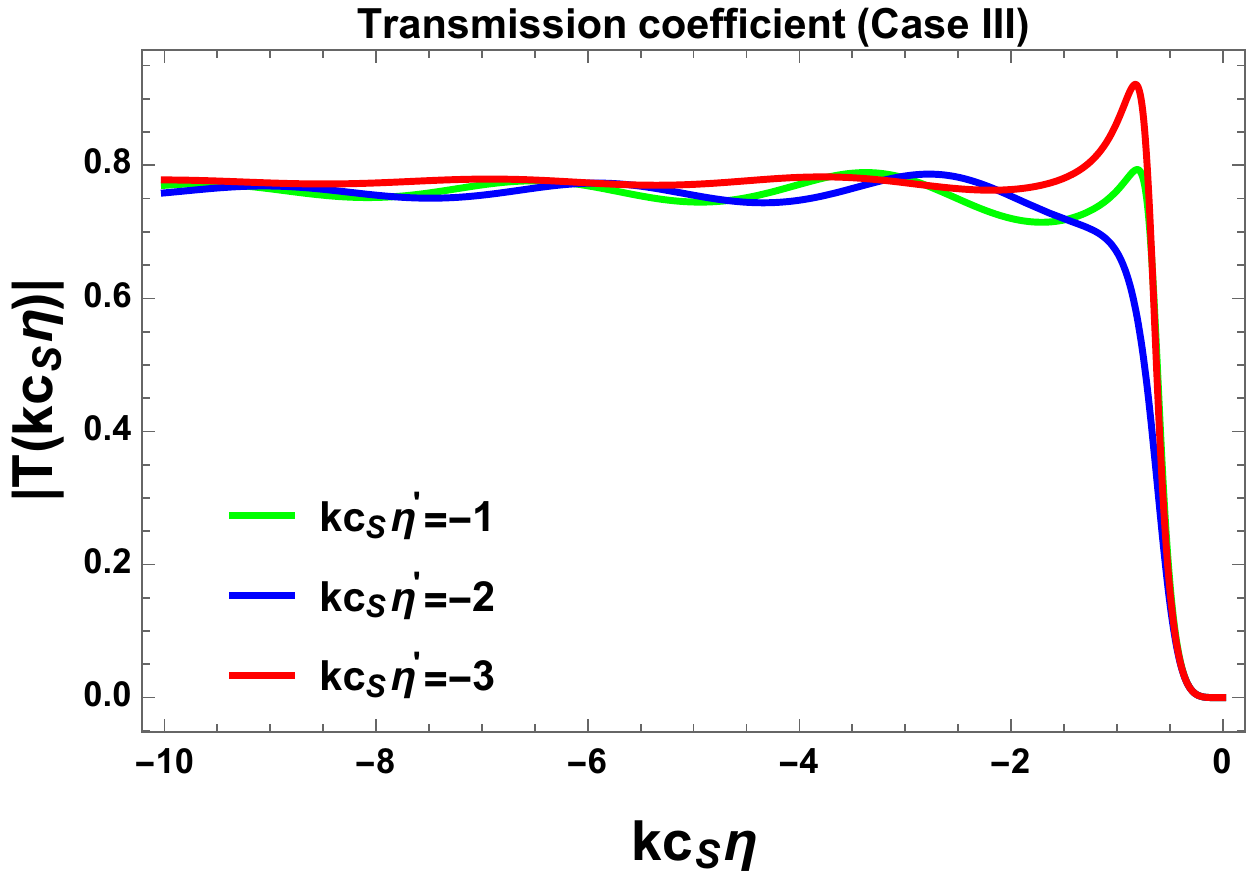}
    \label{fig4xcc}
}
\subfigure[]{
    \includegraphics[width=7.2cm,height=6.1cm] {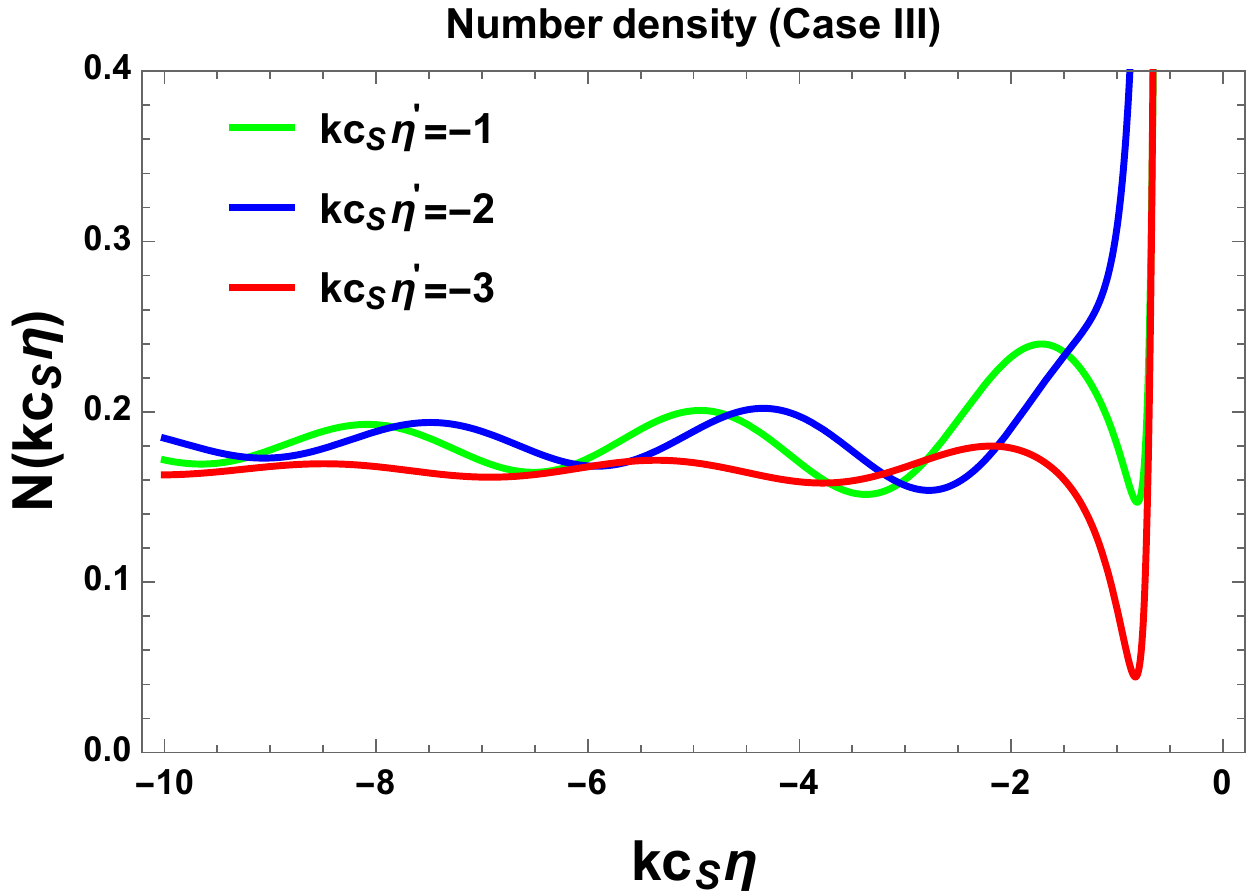}
    \label{fig3xcc}
}
\subfigure[]{
    \includegraphics[width=7.2cm,height=6.1cm] {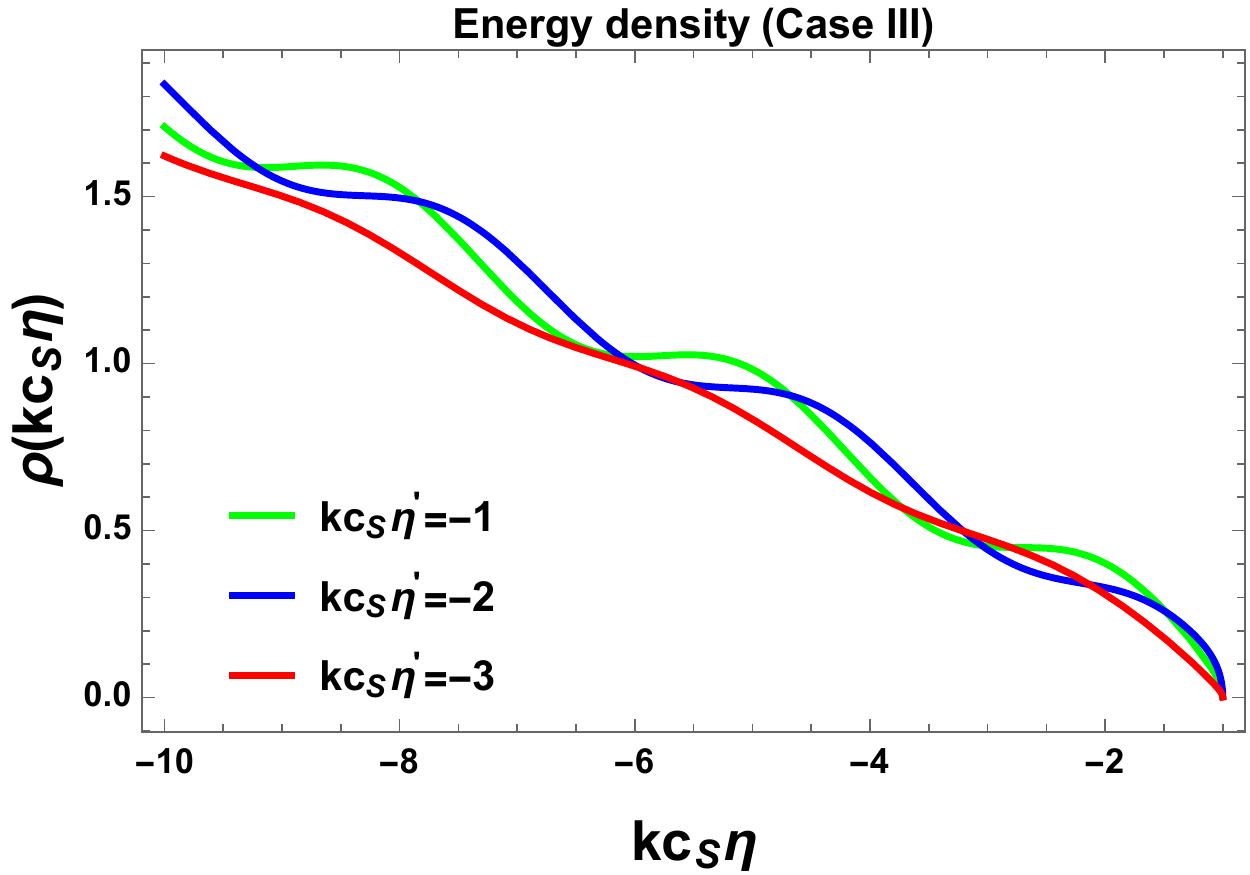}
    \label{fig4xcc}
}
\caption[Optional caption for list of figures]{Particle creation profile for {\bf Case III}.} 
\label{bog1v3}
\end{figure*}
\begin{figure*}[htb]
\centering
\subfigure[]{
    \includegraphics[width=7.2cm,height=6cm] {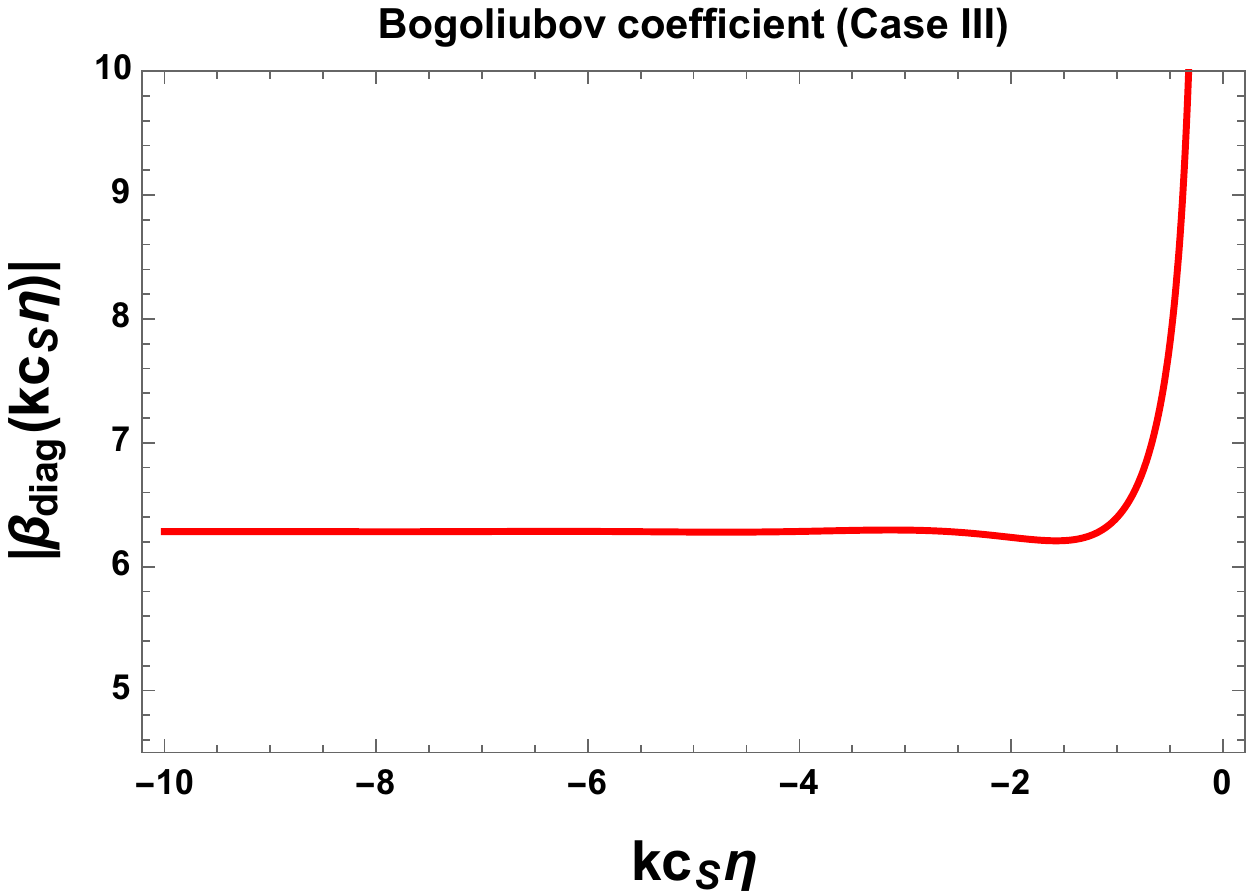}
    \label{fig1xcc}
}
\subfigure[]{
    \includegraphics[width=7.2cm,height=6cm] {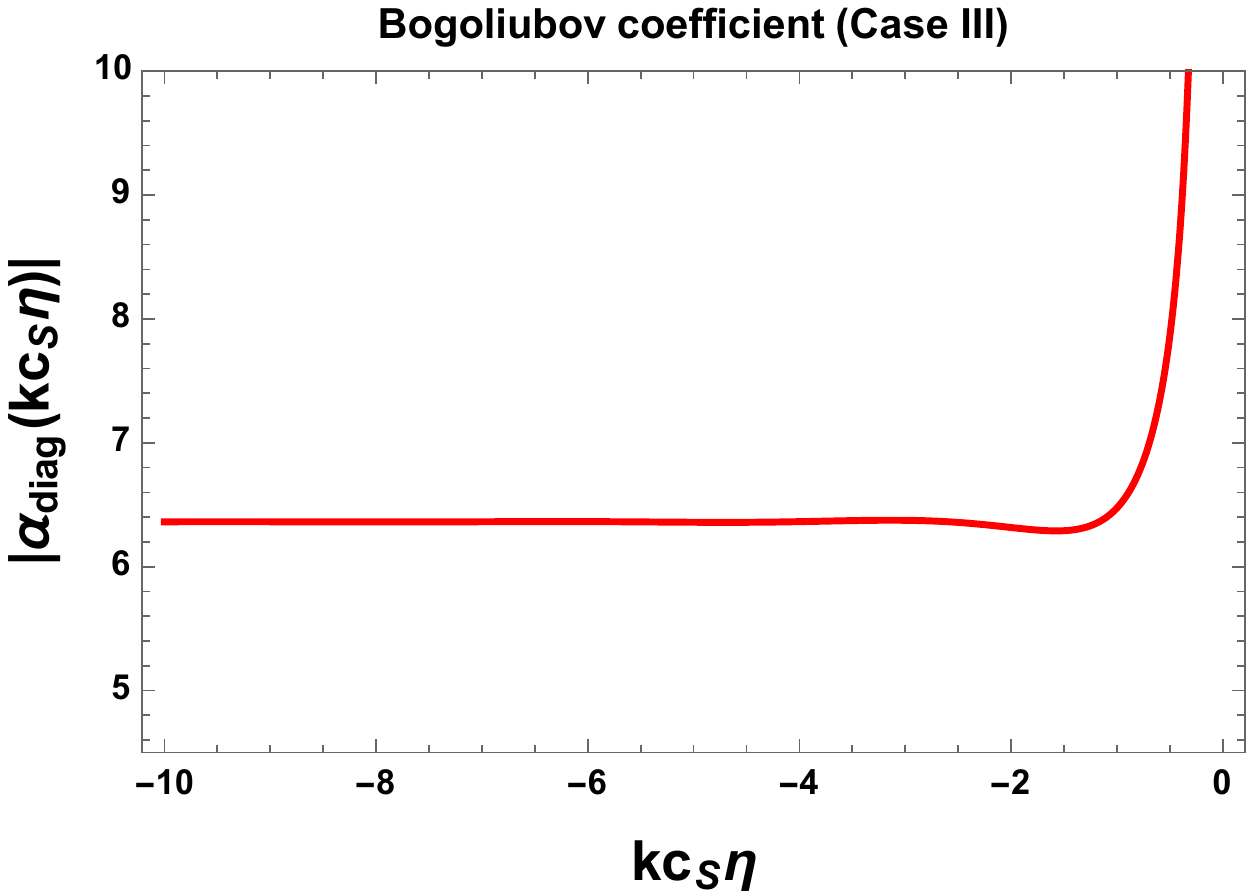}
    \label{fig2xcc}
}
\subfigure[]{
    \includegraphics[width=7.2cm,height=6cm] {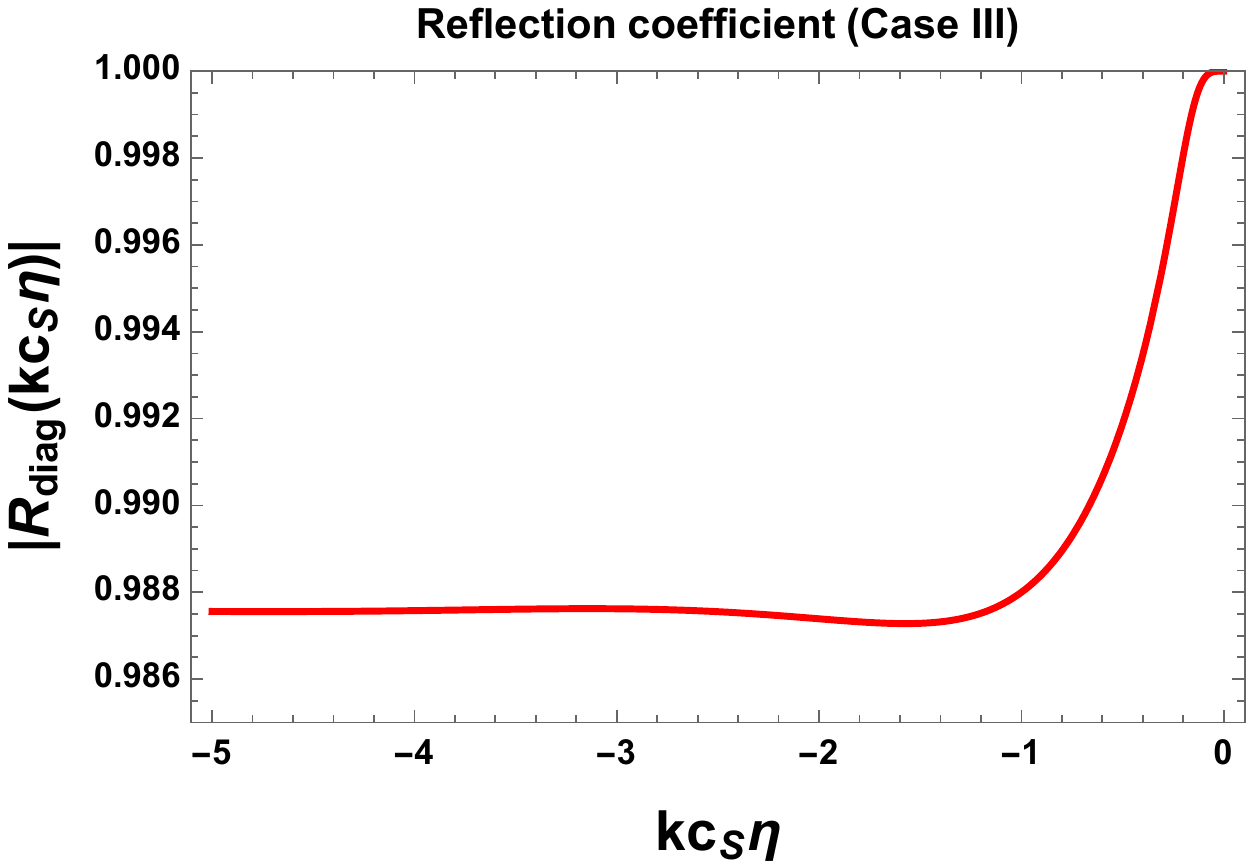}
    \label{fig3xcc}
}
\subfigure[]{
    \includegraphics[width=7.2cm,height=6cm] {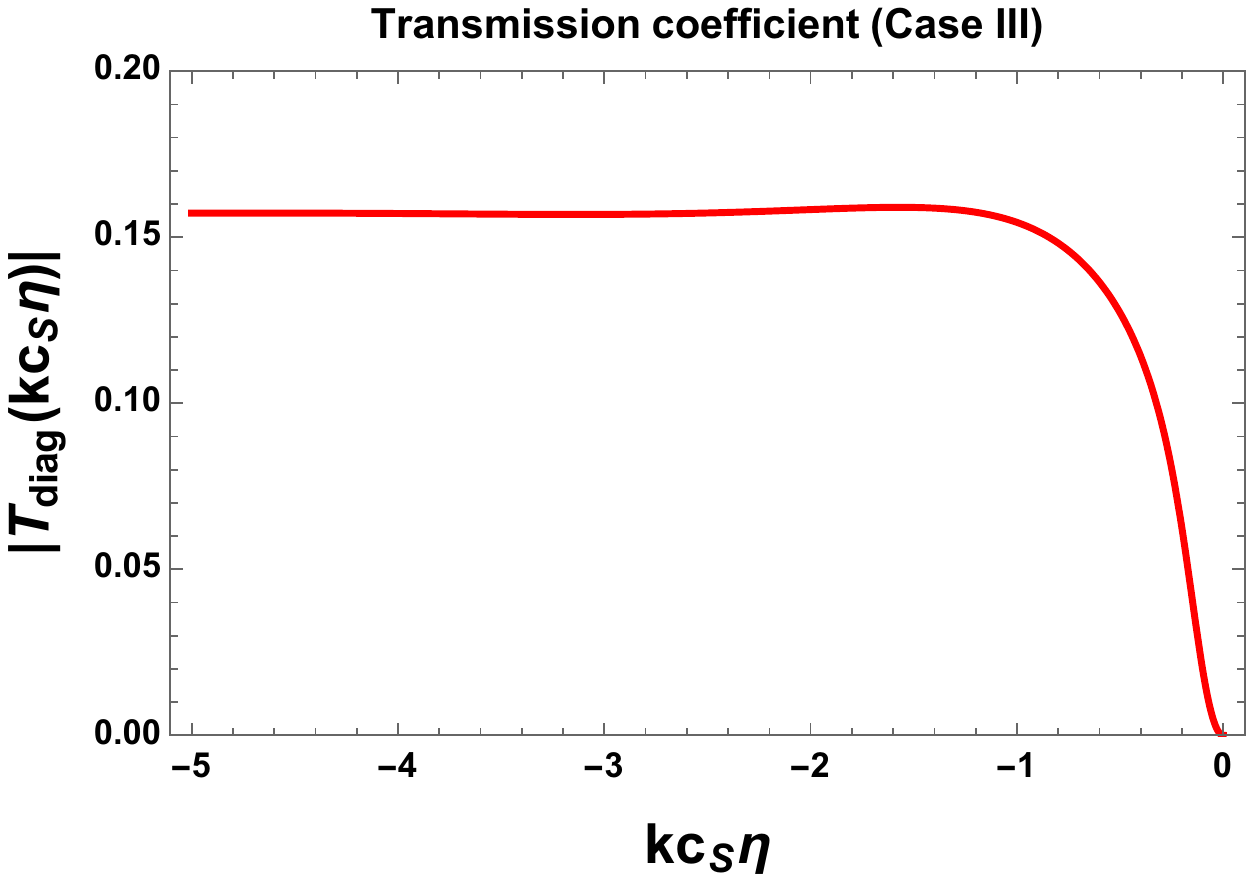}
    \label{fig4xcc}
}
\subfigure[]{
    \includegraphics[width=7.2cm,height=6.1cm] {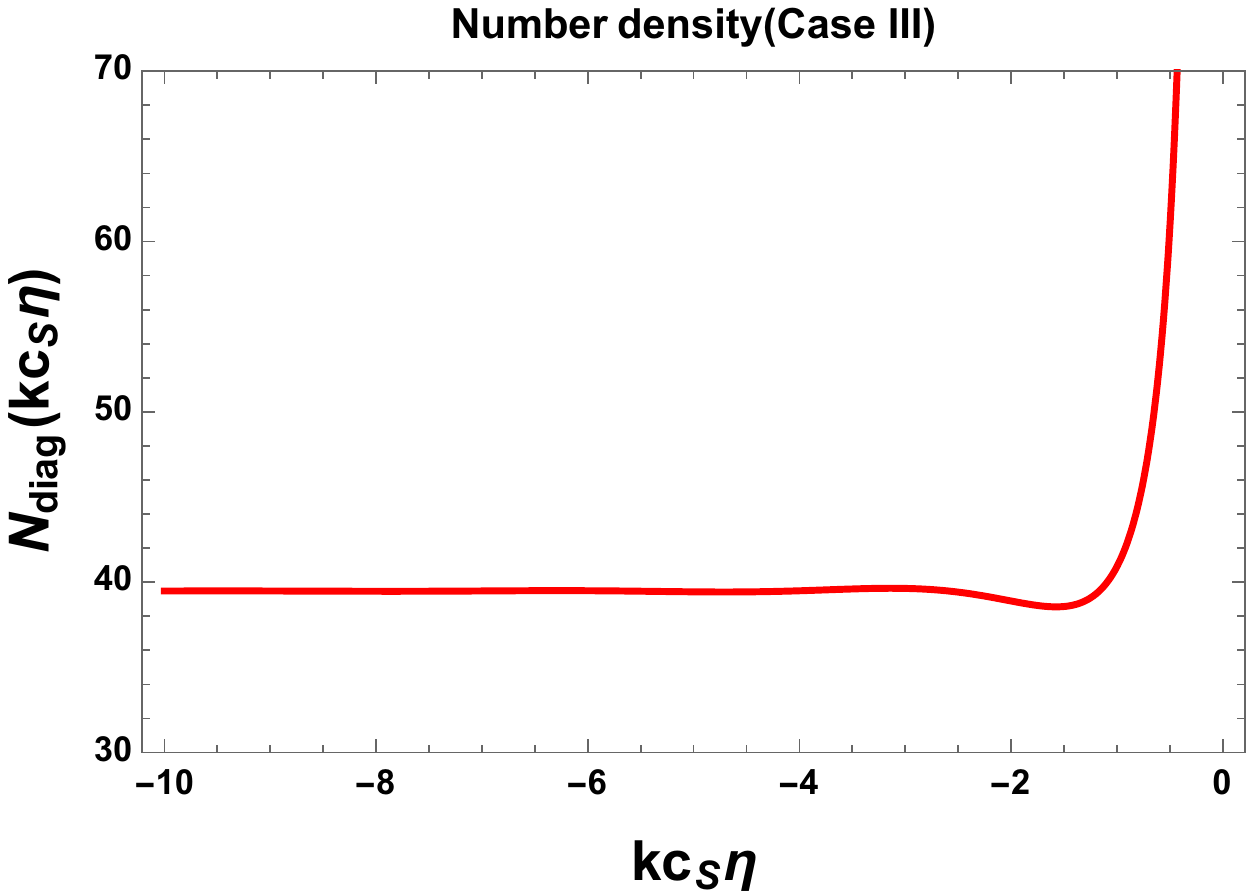}
    \label{fig3xcc}
}
\subfigure[]{
    \includegraphics[width=7.2cm,height=6.1cm] {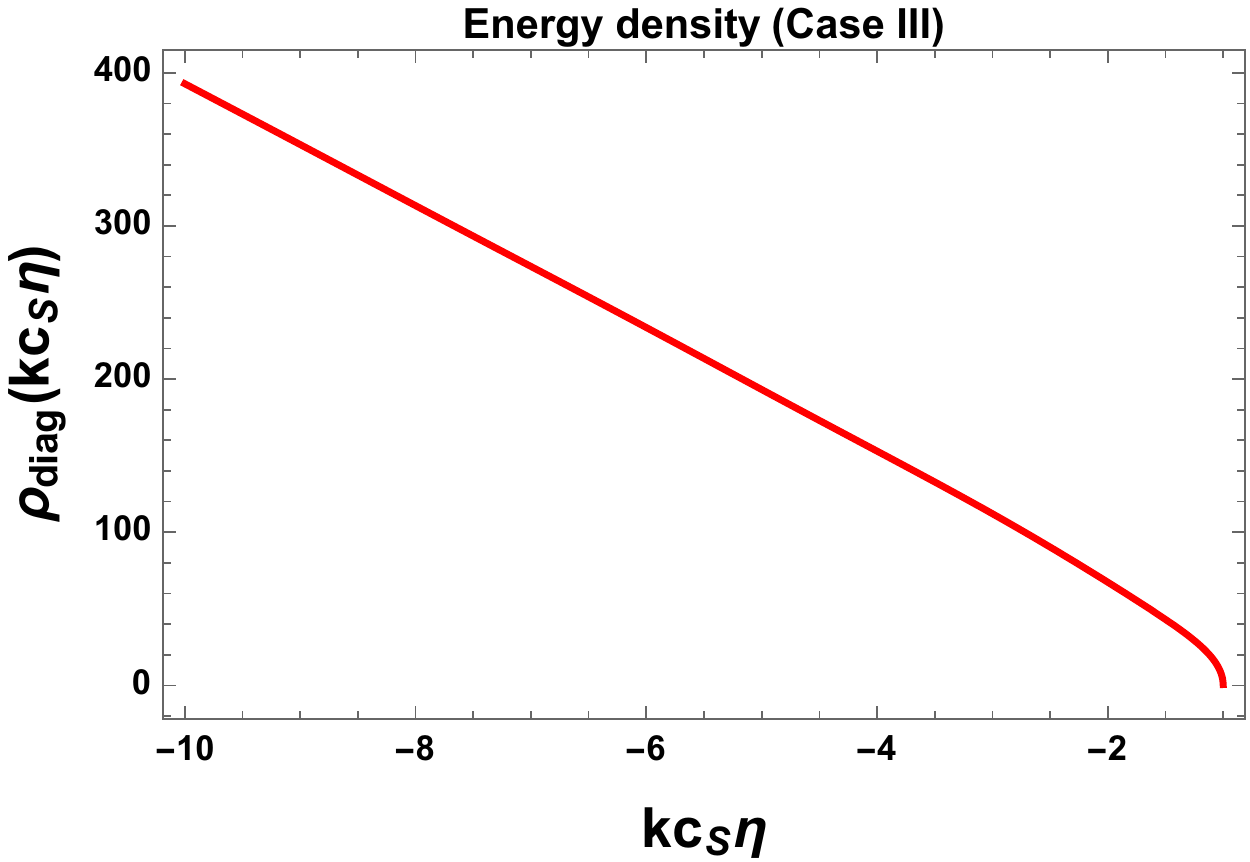}
    \label{fig4xcc}
}
\caption[Optional caption for list of figures]{Particle creation profile for {\bf Case III} in diagonalized representation.} 
\label{bog2v3}
\end{figure*}
Throughout the discussion of total energy density of the produced particles 
we have introduced a symbol $J$ defined as:
\be\begin{array}{lll}\label{kghjk}
 \displaystyle J=\int d^3 {\bf k}~p(\tau)=\left\{\begin{array}{ll}
                    \displaystyle   \int d^3 {\bf k}~\sqrt{c^{2}_{S}k^2 - \frac{2}{\tau^2}}~~~~ &
 \mbox{\small {\bf for ~dS}}  \\ 
	\displaystyle  \int d^3 {\bf k}~\sqrt{c^{2}_{S}k^2 - 
\frac{\left(\nu^2-\frac{1}{4}\right)}{\tau^2}}~~~~ & \mbox{\small {\bf for~ qdS}}.
          \end{array}
\right.
\end{array}\ee
which physically signifies the total finite volume weighted by $p(\eta)$ 
in momentum space within which the produced particles are occupied. 

To study the behaviour of this integral we consider here three 
consecutive physical situations-$|kc_{S}\eta|<<1$, $|kc_{S}\eta|\approx 1-\Delta(\rightarrow 0)$ and $|kc_{S}\eta|>>1$ for de Sitter and quasi de Sitter case. 
In three cases we have:
\bea &&\underline{\bf For~dS:}\\
\small J&=& \left\{\begin{array}{ll}
                    \displaystyle  \int d^3 {\bf k}~ \frac{\sqrt{2}}{\tau}=\frac{V\sqrt{2}}{\tau}
                    ~~&
 \mbox{\small {\bf for ~$|kc_{S}\eta|<<1$}}  \\ 
 \displaystyle  \int d^3 {\bf k}~\frac{\sqrt{2\Delta+1}}{\tau}=\frac{\sqrt{2\Delta+1}~V}{\tau}~~~~ &
 \mbox{\small {\bf for ~$|kc_{S}\eta|\approx 1-\Delta(\rightarrow 0)$}}  \\ 
	\displaystyle\int d^3 {\bf k}~kc_{S} & \mbox{\small {\bf for ~$|kc_{S}\eta|>>1$}}.~~~~~~~~
          \end{array}
\right.\nonumber\eea
\bea
&&\underline{\bf For~qdS:}\\
\small J&=&\left\{\begin{array}{ll}
                    \displaystyle  \int d^3 {\bf k}~\frac{\sqrt{\left(\nu^2-\frac{1}{4}\right)}}{\tau}=\frac{V\sqrt{\left(\nu^2-\frac{1}{4}\right)}}{\tau}~~ &
 \mbox{\small {\bf for ~$|kc_{S}\eta|<<1$}}  \\ 
 \displaystyle   \int d^3 {\bf k}~\frac{\sqrt{2\Delta+\left(\nu^2-\frac{5}{4}\right)}}{\tau}=\frac{\sqrt{2\Delta+\left(\nu^2-\frac{5}{4}\right)}~V}{\tau}
                    ~~&
 \mbox{\small {\bf for ~$|kc_{S}\eta|\approx 1-\Delta(\rightarrow 0)$}} \\ 
	\displaystyle \int d^3 {\bf k}~kc_{S}~~ & \mbox{\small {\bf for ~$|kc_{S}\eta|>>1$}}.~~~~~~~~
          \end{array}
\right.\nonumber
\eea
In fig.~(\ref{bog1v3}) and fig.~(\ref{bog2v3}), we have explicitly shown the particle creation profile for {\bf Case III} for two representations.
\subsection{Cosmological scalar curvature fluctuations from new massive particles}
\label{sec3c}
To describe the effect of the massive particles on the scalar curvature fluctuations here we start with the second order action derived in the framework of effective field theory as:
\begin{equation}\label{eq1jh}
S = S_{1}+S_{2}
\end{equation}
where $S_{1}$ and $S_{2}$ is given by:
\bea
S_{1}&=&{\frac{1}{2}} \int d\eta d^3x \frac{2 \epsilon M^2_{p}}{\tilde{c}^2_{S}H^2} \left[\frac{(\partial_\eta \zeta)^2 
- c^2_{S}(\partial_i \zeta)^2}{\eta^2}-\frac{m^2_{inf}}{H^2\eta^2}\right],\\
S_{2}&=&- \int \frac{d\eta}{\tilde{c}_{S}H} m(\eta) \partial_\eta \zeta(\eta,{\bf x}=0).\eea
Here the action contains the following crucial information:
\begin{itemize}
 \item The term $S_{2}$ contains the effect of massive particle as explicitly the mass factor appears here. In the most generalized picture the 
 mass parameter $m(\eta)$ is function of conformal time $\eta$. Additionally it is important to 
 note that, the term $S_{2}$ in the effective action for curvature fluctuation represents specific 
 interaction term in which inflaton field is interacting with the heavy fields. This implies that the time dependent 
 coupling $m(\eta)$ mimics the role of coupling constant in the present context. 

 \item In inflationary mass term we have neglected the other contributions from the effective potential. This is a complicated 
 inflationary model as it contains both inflaton and heavy field with a time dependent coupling $m(\eta)$. For this 
 reason initially we keep the mass contribution from the inflationary sector. But it is important to note that, in the most 
 simpler inflationary models one can neglect the mass contribution as well, 
 because in all those cases $m_{inf}<<H$ approximation is valid. But due to the presence of mass term of the inflaton equation of motion also modified and 
 this will further appear in the solutions as well.
 \item Here $c_{S}$ is the effective sound speed parameter and $\tilde{c}_{S}$ is the actual sound speed as introduced in the previous section of this paper.
 \item When all the effective field theoretic interactions are absent in that case both $c_{S}\sim \tilde{c}_{S}=1$ and one can get back the results for canonical slow-roll models.
 \item On the other hand when the previously mentioned effective field theoretic interactions are switched on within the present description, one can accommodate the non-canonical as well 
 as non-minimal interactions. In that case both $c_{S}$ and $\tilde{c}_{S}$ are less than
 unity and in such a situation one can always constraint the sound speed parameter as well the strength 
 of the effective field theoretic interactions using observational probes (Planck 2015 data).
 \item In case of canonical interactions one can easily compare the present setup with effective time varying mass 
 parameter with the axions with time varying decay constant.
 \item For $m<<H$ case the last term in the above mentioned effective action is absent and in that case the reduced form 
 of the action will able to explain the effective field of inflation in presence of previously mentioned non-trivial 
 effective interactions. Once we switch off all such interactions the above action mimics the case for single field 
 slow-roll inflation.
 \item Here first we derive the results for arbitrary parametrization of $m/H$ and then discuss about the results for $m\approx H$,
 $m<<H$, $m>>H$ cases. Also we derive the results for arbitrary choice of initial conditions. Then we discuss the results for 
 Buch Davies vacuum, $\alpha$ vacuum and another specific choice of vacuum which we explicitly discussed in this section.
\end{itemize}
To extract further informations from Eq~(\ref{eq1jh}), first of all one needs to write down the second order 
action by applying Fourier transform. For this the Fourier transform of the curvature perturbation $\zeta(\eta,{\bf x})$ is defined as:
\bea \zeta(\eta,{\bf x})&=&\int \frac{d^{3}k}{(2\pi)^3}\zeta_{\bf k}(\eta)\exp(i{\bf k}.{\bf x}),\eea
where $\zeta_{\bf k}(\eta)$ is the time dependent part of the curvature fluctuation after Fourier transform and can be expressed 
in terms of the normalized time dependent scalar mode function $h_{\bf{k}}(\eta)$ as:
\begin{equation}\label{sdf}
\zeta_{\bf{k}}(\eta) = \frac{h_{\bf{k}}(\eta)}{zM_{p}} = \frac{h\left(\eta , \bf{k}\right) a \left(\bf{k}\right) 
+ h^* \left(\eta ,-\bf{k}\right) a^\dagger \left(-\bf{k}\right)}{zM_{p}}
\end{equation}
where $z$ is the Mukhanov Sasaki variable defined as:
\be z=\frac{a\sqrt{2\epsilon}}{\tilde{c}_{S}}\ee
and $a({\bf k})$ and $a^{\dagger}({\bf k})$ are the creation and annihilation operator satisfies
the following commutation relations:
\bea \left[a({\bf k}),a^{\dagger}(-{\bf k}^{'})\right]&=&(2\pi)^3\delta^{3}({\bf k}+{\bf k}^{'}),\\ 
\left[a({\bf k}),a({\bf k}^{'})\right]&=&0,\\ 
\left[a^{\dagger}({\bf k}),a^{\dagger}({\bf k}^{'})\right]&=&0.\eea
Additionally it is important to mention here that the exact solution and its WKB approximated results for the time dependent scalar mode function $h_{\bf{k}}(\eta)$ are explicitly derived in the previous section.

Presently our prime objective is to compute the VEV of the curvature fluctuation in momentum space in presence of the mass contribution $S_{2}$ with respect to the arbitrary choice of vacuum, 
which leads to important contribution to the Bell's inequalities or violation in the context of primordial cosmology.
Using the interaction picture the one point function of the curvature fluctuation in momentum space can be expressed as:  
\begin{equation}\label{opio}
\langle\zeta_k(\eta=0)\rangle = -i \int\limits_{-\infty}^0 {d\eta}~ a(\eta)~ \langle0|\left[\zeta_{\bf{k}} \left(0\right) , H_{int} \left(\eta\right)\right]|0\rangle,
\end{equation}
where $a(\eta)$ is the scale factor defined in the earlier section in terms of Hubble parameter $H$ and conformal time scale $\eta$. In the interaction picture the Hamiltonian can written as:
\begin{equation}\label{hamil}
H_{int}(\eta) = - \frac{m}{\tilde{c}_{S}H} \partial_\eta \zeta(\eta,{\bf x}=0)
\end{equation}
which gives the primary information to compute the explicit expression for the one point function or more precisely the Bell's inequality violation in the present context.
After applying Fourier transform in Eq~(\ref{hamil}) we get the following expression:
\begin{equation}\label{uytd}
H_{int} = -\int \frac{d^3k}{(2\pi)^3\tilde{c}_{S}zM_p} \frac{m}{H} \left[{h}^\prime \left(\eta , \bf{k}\right) a\left(\bf{k}\right) + {h^{\dagger}}^\prime\left(\eta,-\bf{k}\right) a^\dagger 
\left(-\bf{k}\right)\right] 
\end{equation}
and further substituting Eq~(\ref{uytd}) in Eq~(\ref{opio}) finally we get:
\be\begin{array}{llll}\label{dfin}
\displaystyle \langle\zeta_{\bf{k}}(\eta=0)\rangle = -i \int\limits_{-\infty}^0 {d\eta}~ 
\frac{a(\eta)}{z^2M^2_p}
\frac{m}{\tilde{c}_{S}} \left({h}_{\bf{k}}
\left(0\right) {h^\dagger}^{\prime}_{\bf{k}}\left(\eta\right) - 
{h^\dagger}_{-\bf{k}}\left(0\right) {h}^{\prime}_{-\bf{k}} \left(\eta\right)\right)
\end{array}\ee
where $h_{\bf k}(\eta)$ is the exact solution of the mode function as explicitly computed
in the earlier section of this paper. Now sometimes it happens that the exact solution of mode function is not 
exactly defined at $\eta=0$ point. To avoid such complexity in the present computation for the sake of clarity 
here we introduce a Infra-Red (IR) cut-off regulator $\xi$ in the conformal time integral and consequently Eq~(\ref{dfin}) can be 
recast as:
\be\begin{array}{llll}\label{dfinv2}
\displaystyle \langle\zeta_{\bf{k}}(\eta=\xi\rightarrow 0)\rangle = -i \lim_{kc_{S}\xi\rightarrow 0}\int\limits_{-\infty}^{\xi} {d\eta}
~ \frac{a(\eta)}{z^2M^2_p} \frac{m}{\tilde{c}_{S}}\left({h}_{\bf{k}}
\left(\xi\right) {h^\dagger}^{\prime}_{\bf{k}}\left(\eta\right) - 
{h^\dagger}_{-\bf{k}}\left(\xi\right) {h}^{\prime}_{-\bf{k}} \left(\eta\right)\right).
\end{array}\ee
Further substituting the explicit form of the scalar mode functions computed from the 
exact solution we get the following generalized expression for 
the one point function of the curvature fluctuation in momentum space:
\be\begin{array}{llll}\label{dfinv3}
\displaystyle \langle\zeta_{\bf{k}}(\eta=\xi\rightarrow 0)\rangle = -i \lim_{kc_{S}\xi\rightarrow 0}
\int\limits_{-\infty}^{\xi} {d\eta}
~ \frac{a(\eta)}{z^2M^2_p} \frac{m}{\tilde{c}_{S}} \sum^{2}_{i=1}\sum^{2}_{j=1}C^{*}_{i}C_{j}{\cal A}_{ij}(\eta,k).
\end{array}\ee
where the conformal time dependent functions ${\cal A}_{ij}\forall i,j=1,2$ in momentum space is defined as:
\bea {\cal A}_{11}(\eta,k)&=& \sqrt{\xi\eta}\left[H^{(1)}_{\Lambda}(-kc_{S}\xi)H^{(1)*'}_{\Lambda}(-kc_{S}\eta)
-H^{(1)*}_{\Lambda}(kc_{S}\xi)H^{(1)'}_{\Lambda}(kc_{S}\eta)\right]\nonumber\\
 &&~~~~~~~~~-\frac{1}{2}\sqrt{\frac{\xi}{\eta}}\left[H^{(1)}_{\Lambda}(-kc_{S}\xi)H^{(1)*}_{\Lambda}(-kc_{S}\eta)
-H^{(1)*}_{\Lambda}(kc_{S}\xi)H^{(1)}_{\Lambda}(kc_{S}\eta)\right],\\
{\cal A}_{22}(\eta,k)&=& \sqrt{\xi\eta}\left[H^{(2)}_{\Lambda}(-kc_{S}\xi)H^{(2)*'}_{\Lambda}(-kc_{S}\eta)
-H^{(2)*}_{\Lambda}(kc_{S}\xi)H^{(2)'}_{\Lambda}(kc_{S}\eta)\right]\nonumber\\
 &&~~~~~~~~~-\frac{1}{2}\sqrt{\frac{\xi}{\eta}}\left[H^{(2)}_{\Lambda}(-kc_{S}\xi)H^{(2)*}_{\Lambda}(-kc_{S}\eta)
-H^{(2)*}_{\Lambda}(kc_{S}\xi)H^{(2)}_{\Lambda}(kc_{S}\eta)\right],\\
{\cal A}_{12}(\eta,k)&=& \sqrt{\xi\eta}\left[H^{(1)}_{\Lambda}(-kc_{S}\xi)H^{(2)*'}_{\Lambda}(-kc_{S}\eta)
-H^{(1)*}_{\Lambda}(kc_{S}\xi)H^{(2)'}_{\Lambda}(kc_{S}\eta)\right]\nonumber\\
 &&~~~~~~~~~-\frac{1}{2}\sqrt{\frac{\xi}{\eta}}\left[H^{(1)}_{\Lambda}(-kc_{S}\xi)H^{(2)*}_{\Lambda}(-kc_{S}\eta)
-H^{(1)*}_{\Lambda}(kc_{S}\xi)H^{(2)}_{\Lambda}(kc_{S}\eta)\right],\\
{\cal A}_{21}(\eta,k)&=& \sqrt{\xi\eta}\left[H^{(2)}_{\Lambda}(-kc_{S}\xi)H^{(1)*'}_{\Lambda}(-kc_{S}\eta)
-H^{(2)*}_{\Lambda}(kc_{S}\xi)H^{(1)'}_{\Lambda}(kc_{S}\eta)\right]\nonumber\\
 &&~~~~~~~~~-\frac{1}{2}\sqrt{\frac{\xi}{\eta}}\left[H^{(2)}_{\Lambda}(-kc_{S}\xi)H^{(1)*}_{\Lambda}(-kc_{S}\eta)
-H^{(2)*}_{\Lambda}(kc_{S}\xi)H^{(1)}_{\Lambda}(kc_{S}\eta)\right],
 \eea
the parameter $\Lambda$ is defined in Eq~(\ref{yu2dfvqdfdfdf10}) for dS and quasi-dS case. Here after taking the limit 
$kc_{S}\xi\rightarrow 0$ the conformal time dependent functions ${\cal A}_{ij}\forall i,j=1,2$ in momentum space can be recast as:
\bea \lim_{kc_{S}\xi\rightarrow 0}{\cal A}_{11}(\eta,k)&=& \frac{i}{\pi}\Gamma(\Lambda)
\sqrt{\eta\xi}\left[\left(-\frac{kc_{S}\xi}{2}\right)^{-\Lambda}H^{(1)*'}_{\Lambda}(-kc_{S}\eta)
+\left(\frac{kc_{S}\xi}{2}\right)^{-\Lambda}H^{(1)'}_{\Lambda}(kc_{S}\eta)\right]\nonumber\\
 &&~~~-\frac{i\sqrt{\xi}}{2\pi\sqrt{\eta}}\Gamma(\Lambda)\left[\left(-\frac{kc_{S}\xi}{2}\right)^{-\Lambda}H^{(1)*}_{\Lambda}(-kc_{S}\eta)
+\left(\frac{kc_{S}\xi}{2}\right)^{-\Lambda}H^{(1)}_{\Lambda}(kc_{S}\eta)\right],~~~~\\
\lim_{kc_{S}\xi\rightarrow 0}{\cal A}_{22}(\eta,k)&=& -\frac{i}{\pi}\Gamma(\Lambda)
\sqrt{\eta\xi}\left[\left(-\frac{kc_{S}\xi}{2}\right)^{-\Lambda}H^{(2)*'}_{\Lambda}(-kc_{S}\eta)
+\left(\frac{kc_{S}\xi}{2}\right)^{-\Lambda}H^{(2)'}_{\Lambda}(kc_{S}\eta)\right]\nonumber\\
 &&~~~+\frac{i\sqrt{\xi}}{2\pi\sqrt{\eta}}\Gamma(\Lambda)\left[\left(-\frac{kc_{S}\xi}{2}\right)^{-\Lambda}H^{(2)*}_{\Lambda}(-kc_{S}\eta)
+\left(\frac{kc_{S}\xi}{2}\right)^{-\Lambda}H^{(2)}_{\Lambda}(kc_{S}\eta)\right],~~~~\\
\lim_{kc_{S}\xi\rightarrow 0}{\cal A}_{12}(\eta,k)&=& \frac{i}{\pi}\Gamma(\Lambda)\sqrt{\eta\xi}
\left[\left(-\frac{kc_{S}\xi}{2}\right)^{-\Lambda}H^{(2)*'}_{\Lambda}(-kc_{S}\eta)
+\left(\frac{kc_{S}\xi}{2}\right)^{-\Lambda}H^{(2)'}_{\Lambda}(kc_{S}\eta)\right]\nonumber\\
 &&~~~-\frac{i\sqrt{\xi}}{2\pi\sqrt{\eta}}\Gamma(\Lambda)\left[\left(-\frac{kc_{S}\xi}{2}\right)^{-\Lambda}H^{(2)*}_{\Lambda}(-kc_{S}\eta)
+\left(\frac{kc_{S}\xi}{2}\right)^{-\Lambda}H^{(2)}_{\Lambda}(kc_{S}\eta)\right],~~~~\\
\lim_{kc_{S}\xi\rightarrow 0}{\cal A}_{21}(\eta,k)&=& -\frac{i}{\pi}\Gamma(\Lambda)
\sqrt{\eta\xi}\left[\left(-\frac{kc_{S}\xi}{2}\right)^{-\Lambda}H^{(1)*'}_{\Lambda}(-kc_{S}\eta)
+\left(\frac{kc_{S}\xi}{2}\right)^{-\Lambda}H^{(1)'}_{\Lambda}(kc_{S}\eta)\right]\nonumber\\
 &&~~~+\frac{i\sqrt{\xi}}{2\pi\sqrt{\eta}}\Gamma(\Lambda)\left[\left(-\frac{kc_{S}\xi}{2}\right)^{-\Lambda}H^{(1)*}_{\Lambda}(-kc_{S}\eta)
+\left(\frac{kc_{S}\xi}{2}\right)^{-\Lambda}H^{(1)}_{\Lambda}(kc_{S}\eta)\right].
 \eea
 For further simplification we consider here two limiting cases $|kc_{S}\eta|\rightarrow -\infty$,
 $|kc_{S}\eta|\rightarrow 0$ and $|kc_{S}\eta|\approx 1$
 which are physically acceptable in the present context. First of we consider the
 results for $|kc_{S}\eta|\rightarrow -\infty$. In this case we get:
 \bea\label{dfinv4bnbncv}
\displaystyle \langle\zeta_{\bf{k}}(\eta=0)\rangle_{|kc_{S}\eta|\rightarrow -\infty} &=& \frac{2}{M^2_{p}\pi}
\int\limits_{-\infty}^{0} {d\eta}
~\frac{\tilde{c}_{S}H}{a(\eta)\epsilon k c_{S}} \frac{m}{H}\left[|C_{2}|^2 
e^{-ikc_{S}\eta}-|C_{1}|^2 e^{ikc_{S}\eta}\right.\nonumber \\ &&\left. 
\displaystyle~~~~~~~~~~~-i\left(C^{*}_{1}C_{2}e^{i\pi\left(\Lambda+\frac{1}{2}\right)}
+C_{1}C^{*}_{2}e^{-i\pi\left(\Lambda+\frac{1}{2}\right)}\right)\sin {\it kc_{S}\eta}\right]\\&=&\left\{\begin{array}{ll}
                    \displaystyle  -\frac{1}{M^2_{p}}
\int\limits_{-\infty}^{0} {d\eta}
~\frac{\tilde{c}_{S}H}{a(\eta)\epsilon k c_{S}} \frac{m}{H}e^{ikc_{S}\eta}~~~~ &
 \mbox{\small {\bf for ~Bunch Davies vacua}}  \\ 
	\displaystyle  \frac{2}{M^2_{p}\pi}
\int\limits_{-\infty}^{0} {d\eta}
~\frac{\tilde{c}_{S}H}{a(\eta)\epsilon  k c_{S}} \frac{m}{H}\left[\sinh^2\alpha~ 
e^{ikc_{S}\eta}-\cosh^2\alpha~ e^{-ikc_{S}\eta}\right.\nonumber \\ \left. 
\displaystyle~~~~~~~~~~~+i~\sinh2\alpha\cos\left({\it \pi\left(\Lambda+\frac{1}{2}\right)}+\delta\right)\sin{\it 
kc_{S}\eta}\right]
                    ~~~~ & \mbox{\small {\bf for~$\alpha$~vacua~Type-I}}\\ 
	\displaystyle  \frac{2|N_{\alpha}|^2}{M^2_{p}\pi}
\int\limits_{-\infty}^{0} {d\eta}
~\frac{\tilde{c}_{S}H}{a(\eta)\epsilon  k c_{S}} \frac{m}{H}\left[e^{\alpha+\alpha^{*}} 
e^{ikc_{S}\eta}-e^{-ikc_{S}\eta}\right.\nonumber \\ \left. 
\displaystyle~~~~~~~~~~~+i\left(e^{\alpha}e^{i\pi\left(\Lambda+\frac{1}{2}\right)}
+e^{\alpha^{*}}e^{-i\pi\left(\Lambda+\frac{1}{2}\right)}\right)\sin{\it 
kc_{S}\eta}\right]
                    ~~~~ & \mbox{\small {\bf for~$\alpha$~vacua~Type-II}}\\ 
	\displaystyle  \frac{8i|C|^2}{M^2_{p}\pi}
\int\limits_{-\infty}^{0} {d\eta}
~\frac{\tilde{c}_{S}H}{a(\eta)\epsilon  k c_{S}} \frac{m}{H}\sin{\it 
kc_{S}\eta}\cos^2{\it \frac{\pi}{2}\left(\Lambda+\frac{1}{2}\right)}
                    ~~~~ & \mbox{\small {\bf for~special~vacua}}.
          \end{array}
\right.
\eea
On the other hand for $|kc_{S}\eta|\rightarrow 0$ we get the following simplified expression:
\bea\label{dfinv4bnbnxcxcx}
\displaystyle \langle\zeta_{\bf{k}}(\eta=\xi\rightarrow 0)\rangle_{|kc_{S}\eta|\rightarrow 0} &=& 
\frac{4\pi\sqrt{\xi}}{M^2_{p}}\left(C^{*}_{1}C_{2}
+C_{1}C^{*}_{2}-|C_{1}|^2-|C_{2}|^2\right)
\int\limits_{-\infty}^{\xi} {d\eta}
~\frac{\tilde{c}^2_{S}H}{a(\eta)(2\pi)^3\epsilon\sqrt{-\eta}} \frac{m}{H\tilde{c}_{S}}\\ &&\displaystyle~~~~~~~~~~~~~~~~ 
\left(\Lambda-\frac{1}{2}\right)\left[\left(-\frac{kc_{S}\eta}{2}\right)^{-\Lambda}\left(-\frac{kc_{S}\xi}{2}\right)^{-\Lambda}
+\left(\frac{kc_{S}\eta}{2}\right)^{-\Lambda}\left(\frac{kc_{S}\xi}{2}\right)^{-\Lambda}\right]\nonumber
\\&=&\frac{I_{\xi}}{(2\pi)^3}\times\left\{\begin{array}{ll}
                    \displaystyle  -\frac{2\pi^2\sqrt{\xi}}{M^2_{p}}~~~~ &
 \mbox{\small {\bf for ~Bunch Davies vacua}}  \\ 
	\displaystyle  \frac{4\pi\sqrt{\xi}}{M^2_{p}}\left(\cos\delta\sinh 2\alpha-\cosh^2\alpha-\sinh^2\alpha\right)
                    ~~~~ & \mbox{\small {\bf for~$\alpha$~vacua~Type-I}}\\ 
	\displaystyle  \frac{4\pi\sqrt{\xi}|N_{\alpha}|^2}{M^2_{p}}\left(e^{\alpha}+e^{\alpha^{*}}
-e^{\alpha+\alpha^{*}}-1\right)
\nonumber
                    ~~~~ & \mbox{\small {\bf for~$\alpha$~vacua~Type-II}}\\ 
	\displaystyle  0
~~~ & \mbox{\small {\bf for~special~vacua}}.
          \end{array}
\right.
\eea
where the integral $I_{\xi}$ is defined as:
\bea I_{\xi}&=&\int\limits_{-\infty}^{\xi} {d\eta}
~\frac{\tilde{c}^2_{S}H}{a(\eta)(2\pi)^3\epsilon\sqrt{-\eta}} \frac{m}{H\tilde{c}_{S}}
\left(\Lambda-\frac{1}{2}\right)\left[\left(-\frac{kc_{S}\eta}{2}\right)^{-\Lambda}\left(-\frac{kc_{S}\xi}{2}\right)^{-\Lambda}
+\left(\frac{kc_{S}\eta}{2}\right)^{-\Lambda}\left(\frac{kc_{S}\xi}{2}\right)^{-\Lambda}\right].~~~~~~~~~~~~\eea
Finally the other hand for $|kc_{S}\eta|\approx 1$ we get the following simplified expression:
\bea\label{dfinv4bnbnxcxcx}
\displaystyle \langle\zeta_{\bf{k}}(\eta=\xi\rightarrow 0)\rangle_{|kc_{S}\eta|\approx 1} &=& 
\frac{4\pi\sqrt{\xi}}{M^2_{p}}\left(C^{*}_{1}C_{2}
+C_{1}C^{*}_{2}-|C_{1}|^2-|C_{2}|^2\right)
\int\limits_{-\infty}^{\xi} {d\eta}
~\frac{\tilde{c}^2_{S}H}{a(\eta)(2\pi)^3\epsilon\sqrt{-\eta}} \frac{m}{H\tilde{c}_{S}}\\ &&\displaystyle~~~~~~~~~~~~~~~~ 
\left(\Lambda-\frac{1}{2}\right)\left(\frac{1}{2}\right)^{-\Lambda}\left[\left(-\frac{kc_{S}\xi}{2}\right)^{-\Lambda}
+\left(-1\right)^{-\Lambda}\left(\frac{kc_{S}\xi}{2}\right)^{-\Lambda}\right]\nonumber
\\&=&J_{\xi}\times\left\{\begin{array}{ll}
                    \displaystyle  -\frac{2\pi^2\sqrt{\xi}}{M^2_{p}}~~~~ &
 \mbox{\small {\bf for ~Bunch Davies vacua}}  \\ 
	\displaystyle  \frac{4\pi\sqrt{\xi}}{M^2_{p}}\left(\cos\delta\sinh 2\alpha-\cosh^2\alpha-\sinh^2\alpha\right)
                    ~~~~ & \mbox{\small {\bf for~$\alpha$~vacua~Type-I}}\\ 
	\displaystyle  \frac{4\pi\sqrt{\xi}|N_{\alpha}|^2}{M^2_{p}}\left(e^{\alpha}+e^{\alpha^{*}}
-e^{\alpha+\alpha^{*}}-1\right)
\nonumber
                    ~~~~ & \mbox{\small {\bf for~$\alpha$~vacua~Type-II}}\\ 
	\displaystyle  0
~~~ & \mbox{\small {\bf for~special~vacua}}.
          \end{array}
\right.
\eea
where the integral $J_{\xi}$ is defined as:
\bea J_{\xi}&=&\int\limits_{-\infty}^{\xi} {d\eta}
~\frac{\tilde{c}^2_{S}H}{a(\eta)(2\pi)^3\epsilon\sqrt{-\eta}} \frac{m}{H\tilde{c}_{S}}
\left(\Lambda-\frac{1}{2}\right)\left(\frac{1}{2}\right)^{-\Lambda}
\left[\left(-\frac{kc_{S}\xi}{2}\right)^{-\Lambda}
+\left(-1\right)^{-\Lambda}\left(\frac{kc_{S}\xi}{2}\right)^{-\Lambda}\right].~~~~~~~~~~~~\eea
Now to analyze the behaviour of the expectation value of scalar curvature perturbation in position space 
we need to take the Fourier transform of the expectation value of scalar curvature perturbation already computed 
in momentum space. For the most generalized solution we get the following result:
\be\begin{array}{llll}\label{dfinsd1}
\displaystyle \langle\zeta({\bf x},\eta=0)\rangle = -i \int \frac{d^{3}k}{(2\pi)^3}\int\limits_{-\infty}^0 {d\eta}~ \frac{a(\eta)}{z^2M^2_p}
\frac{m}{\tilde{c}_{S}} e^{i{\bf k}.{\bf x}}\left({h}_{\bf{k}}
\left(0\right) {h^\dagger}^{\prime}_{\bf{k}}\left(\eta\right) - 
{h^\dagger}_{-\bf{k}}\left(0\right) {h}^{\prime}_{-\bf{k}} \left(\eta\right)\right)
\end{array}\ee
where $h_{\bf k}(\eta)$ is the exact solution of the mode function as explicitly computed
in the earlier section of this paper. Following the previous methodology here we also
introduce a Infra-Red (IR) cut-off regulator $\xi$ in the conformal time integral and 
consequently Eq~(\ref{dfinsd1}) can be 
recast in the following form as:
\be\begin{array}{llll}\label{dfinv2}
\displaystyle \langle\zeta({\bf x},\eta=0)\rangle = -i \lim_{kc_{S}\xi\rightarrow 0}\int \frac{d^{3}k}{(2\pi)^3}\int\limits_{-\infty}^{\xi} {d\eta}
~ \frac{a(\eta)}{z^2M^2_p} \frac{m}{H\tilde{c}_{S}}e^{i{\bf k}.{\bf x}}\left({h}_{\bf{k}}
\left(\xi\right) {h^\dagger}^{\prime}_{\bf{k}}\left(\eta\right) - 
{h^\dagger}_{-\bf{k}}\left(\xi\right) {h}^{\prime}_{-\bf{k}} \left(\eta\right)\right).
\end{array}\ee
Further substituting the explicit form of the scalar mode functions computed from the 
exact solution we get the following simplified expression for 
the one point function of the curvature fluctuation in position space:
\be\begin{array}{llll}\label{dfinv3}
\displaystyle \langle\zeta({\bf x},\eta=0)\rangle = -i \lim_{kc_{S}\xi\rightarrow 0}\int \frac{d^{3}k}{(2\pi)^3}\int\limits_{-\infty}^{\xi} {d\eta}
~ \frac{a(\eta)}{z^2M^2_p} \frac{m}{\tilde{c}_{S}} \sum^{2}_{i=1}\sum^{2}_{j=1}C^{*}_{i}C_{j}{\cal A}_{ij}e^{i{\bf k}.{\bf x}}.
\end{array}\ee
where the conformal time dependent functions ${\cal A}_{ij}\forall i,j=1,2$ in momentum
space is already defined earlier.

Similarly in the position space the representative expressions for the expectation value of the 
scalar curvature perturbation along with three
limiting cases $|kc_{S}\eta|\rightarrow -\infty$, $|kc_{S}\eta|\rightarrow 0$ and $|kc_{S}\eta|\approx 1$
are given by:
\bea\label{dfinv4bnbncv}
\displaystyle \langle\zeta({\bf x},\eta=0)\rangle_{|kc_{S}\eta|\rightarrow -\infty} &\approx& 
-\frac{2H\tilde{c}_{S}}{M^2_{p}\epsilon\pi c_{S}}
\left[|C_{2}|^2 O_{1}
-|C_{1}|^2 O_{2}\right.\\ &&\left.\nonumber~~~~~~~~~~~~~~~~~~~~~~-i\left(C^{*}_{1}C_{2}e^{i\pi\left(\Lambda+\frac{1}{2}\right)}
+C_{1}C^{*}_{2}e^{-i\pi\left(\Lambda+\frac{1}{2}\right)}\right)O_{3}\right]\\&=&\left\{\begin{array}{ll}
                  \displaystyle     \frac{H\tilde{c}_{S}}{M^2_{p}\epsilon c_{S}}
O_{2}\nonumber &
 \mbox{\small {\bf for ~Bunch Davies}}  \\ 
	 \displaystyle  -\frac{2H\tilde{c}_{S}}{M^2_{p}\epsilon\pi c_{S}}
\left[\sinh^2\alpha~ 
O_{1}-\cosh^2\alpha~O_{2}\right.\nonumber \\ \left.  \displaystyle
-i~\sinh2\alpha\cos\left({\it \pi\left(\Lambda+\frac{1}{2}\right)}+\delta\right)O_{3}\right]
                     & \mbox{\small {\bf for~$\alpha$~vacua~Type-I}}\\ 
	 \displaystyle  -\frac{2|N_{\alpha}|^2\tilde{c}_{S}H}{M^2_{p}\epsilon\pi c_{S}}
\left[e^{\alpha+\alpha^{*}} 
O_{1}-O_{2}\right.\nonumber \\ \left.  \displaystyle
-i\left(e^{\alpha}e^{i\pi\left(\Lambda+\frac{1}{2}\right)}
+e^{\alpha^{*}}e^{-i\pi\left(\Lambda+\frac{1}{2}\right)}\right)O_{3}\right]
                     & \mbox{\small {\bf for~$\alpha$~vacua~Type-II}}\\ 
	 \displaystyle - \frac{8i H|C|^2\tilde{c}_{S}}{M^2_{p}\epsilon\pi c_{S}}
\cos^2{\it \frac{\pi}{2}\left(\Lambda+\frac{1}{2}\right)}O_{3}
                    & \mbox{\small {\bf for~special~vacua}}.
          \end{array}
\right.
\eea
\bea\label{dfinv4bnbnxcxcx}
\small\displaystyle \langle\zeta({\bf x},\eta=\xi\rightarrow 0)\rangle_{|kc_{S}\eta|\rightarrow 0} &=& 
\frac{H\sqrt{\xi}\tilde{c}_{S}}{2M^2_{p}\epsilon\pi^2}\left(C^{*}_{1}C_{2}
+C_{1}C^{*}_{2}-|C_{1}|^2-|C_{2}|^2\right)
O^{\xi\theta}_{4}
\\&=&O^{\xi\theta}_{4}\times\left\{\begin{array}{ll}
                    \displaystyle  -\frac{2H\pi^2\sqrt{\xi}\tilde{c}_{S}}{M^2_{p}\epsilon}~~~~ &
 \mbox{\small {\bf for ~Bunch Davies vacua}}  \\ 
	\displaystyle  \frac{H\sqrt{\xi}\tilde{c}_{S}}{2M^2_{p}\pi^2\epsilon}\left(\cos\delta\sinh 2\alpha-\cosh^2\alpha-\sinh^2\alpha\right)
                    ~~~~ & \mbox{\small {\bf for~$\alpha$~vacua~Type-I}}\\ 
	\displaystyle  \frac{H\sqrt{\xi}|N_{\alpha}|^2\tilde{c}_{S}}{2M^2_{p}\pi^2\epsilon}\left(e^{\alpha}+e^{\alpha^{*}}
-e^{\alpha+\alpha^{*}}-1\right)
\nonumber
                    ~~~~ & \mbox{\small {\bf for~$\alpha$~vacua~Type-II}}\\ 
	\displaystyle  0
~~~ & \mbox{\small {\bf for~special~vacua}}.
          \end{array}
\right.
\eea
\bea\label{dfinv4bnbnxcxcx}
\small\displaystyle \langle\zeta({\bf x},\eta=\xi\rightarrow 0)\rangle_{|kc_{S}\eta|\approx 1} &=& 
\frac{H\sqrt{\xi}\tilde{c}_{S}}{2M^2_{p}\epsilon\pi^2}\left(C^{*}_{1}C_{2}
+C_{1}C^{*}_{2}-|C_{1}|^2-|C_{2}|^2\right)
O^{\xi\theta}_{5}
\\&=&O^{\xi\theta}_{5}\times\left\{\begin{array}{ll}
                    \displaystyle  -\frac{2H\pi^2\sqrt{\xi}\tilde{c}_{S}}{M^2_{p}\epsilon}~~~~ &
 \mbox{\small {\bf for ~Bunch Davies vacua}}  \\ 
	\displaystyle  \frac{H\sqrt{\xi}\tilde{c}_{S}}{2M^2_{p}\pi^2\epsilon}\left(\cos\delta\sinh 2\alpha-\cosh^2\alpha-\sinh^2\alpha\right)
                    ~~~~ & \mbox{\small {\bf for~$\alpha$~vacua~Type-I}}\\ 
	\displaystyle  \frac{H\sqrt{\xi}|N_{\alpha}|^2\tilde{c}_{S}}{2M^2_{p}\pi^2\epsilon}\left(e^{\alpha}+e^{\alpha^{*}}
-e^{\alpha+\alpha^{*}}-1\right)
\nonumber
                    ~~~~ & \mbox{\small {\bf for~$\alpha$~vacua~Type-II}}\\ 
	\displaystyle  0
~~~ & \mbox{\small {\bf for~special~vacua}}.
          \end{array}
\right.
\eea
where we introduce the following integrals $O_{1}, O_{2}, O_{3}, O^{\xi\theta}_{4}, O^{\xi\theta}_{5}$ are given by:
\bea O_{1}&=& \int \frac{d^{3}k}{(2\pi)^3}~e^{i{\bf k}.{\bf x}}\int\limits_{-\infty}^{0} {d\eta}~
~\frac{\eta}{k}~m(\eta)
e^{-ikc_{S}\eta},\\ 
 O_{2}&=&  \int \frac{d^{3}k}{(2\pi)^3}~e^{i{\bf k}.{\bf x}}\int\limits_{-\infty}^{0} {d\eta}
~\frac{\eta}{k}~m(\eta)~e^{ikc_{S}\eta},\\ 
 O_{3}&=& \int \frac{d^{3}k}{(2\pi)^3}~e^{i{\bf k}.{\bf x}}\int\limits_{-\infty}^{0} {d\eta}
~\frac{\eta}{k}~m(\eta)\sin {\it kc_{S}\eta},\\
 O^{\xi\theta}_{4}&=&\int \frac{d^{3}k}{(2\pi)^3}~e^{i{\bf k}.{\bf x}}\int\limits_{-\infty}^{\xi} {d\eta}
~\frac{1}{a(\eta)\sqrt{-\eta}} \frac{m}{H}
\left(\Lambda-\frac{1}{2}\right)
\left[\left(-\frac{kc_{S}\eta}{2}\right)^{-\Lambda}\left(-\frac{kc_{S}\xi}{2}\right)^{-\Lambda}
+\left(\frac{kc_{S}\eta}{2}\right)^{-\Lambda}\left(\frac{kc_{S}\xi}{2}\right)^{-\Lambda}\right],~~~~~~~\\
 O^{\xi\theta}_{5}&=&\int \frac{d^{3}k}{(2\pi)^3}~e^{i{\bf k}.{\bf x}}\int\limits_{-\infty}^{\xi} {d\eta}
~\frac{1}{a(\eta)\sqrt{-\eta}} \frac{m}{H}
\left(\Lambda-\frac{1}{2}\right)\left(\frac{1}{2}\right)^{-\Lambda}
\left[\left(-\frac{kc_{S}\xi}{2}\right)^{-\Lambda}
+\left(-1\right)^{-\Lambda}\left(\frac{kc_{S}\xi}{2}\right)^{-\Lambda}\right].~~\eea
 Now to compute these momentum integrals we follow the following sets of assumptions:
 \begin{enumerate}
  \item We choose spherical polar coordinate for the computation of momentum volume integral.
  
 \item  We take two situations where ${\bf k}$ and ${\bf x}$ are parallel and having an angle $\Theta$ in between them. For the 
  first case \be {\bf k}.{\bf x}=kx\ee and for the second case we have \be {\bf k}.{\bf x}=kx\cos\Theta,\ee where the range of the 
  angular parameter is lying within the window $\Theta_{1}\leq \Theta\leq \Theta_{2}$, where 
  $\Theta_{1}$ and $\Theta_{2}$ are two cut-off in the angular coordinate which are introduced 
  to regularize the momentum integrals in the present context. Additionally it is also important to note that for the first 
  case the volume element for the momentum integration is given by $d^{3}k=4\pi~k^2dk$  and for the second case 
  the volume element for the momentum integration is considered as $ d^{3}k=k^2\sin\Theta dk d\Theta d\phi,$ where 
  $\phi$ is called the azimuthal coordinate and lying within the window $0< \phi<2\pi$.
  
  \item Last but not the least, to perform volume integration in momentum space for two point function 
  we need to introduce a momentum IR cut-off at \be k_{IR}=\frac{1}{L_{IR}}.\ee
 \end{enumerate}
 For Case I and Case II we get the following results: 
\bea {\bf For~Case~I:}\nonumber\\
O_{1}&=& \frac{1}{2\pi^2}\int^{\infty}_{0} dk~\int\limits_{-\infty}^{0} {d\eta}~\eta k~m(\eta)~e^{ik\left(x-c_{S}\eta\right)}
=-\frac{i}{2\pi^2c^2_{S}}m(\eta=-|{\bf x}|/c_{S}),\\ 
 O_{2}&=&  \frac{1}{2\pi^2}\int^{\infty}_{0} dk~\int\limits_{-\infty}^{0} {d\eta}~\eta k~m(\eta)~e^{ik\left(x+c_{S}\eta\right)}
=\frac{i}{2\pi^2c^2_{S}}m(\eta=-|{\bf x}|/c_{S}),\\ 
 O_{3}&=&  \frac{1}{2\pi^2}\int^{\infty}_{0} dk\int\limits_{-\infty}^{0} {d\eta}
~\eta k~m(\eta)~e^{ikx}\sin {\it kc_{S}\eta}=\frac{1}{2\pi^2c^2_{S}}m(\eta=-|{\bf x}|/c_{S}),\\
 O^{\xi\theta}_{4}&=& \frac{1}{2\pi^2}\int^{\infty}_{0} dk~k^2~e^{ikx}\int\limits_{-\infty}^{\xi} {d\eta}
~\frac{\tilde{c}^2_{S}H}{a(\eta)\epsilon\sqrt{-\eta}} \frac{m}{H\tilde{c}_{S}}
\left(\Lambda-\frac{1}{2}\right)\\
\nonumber&&~~~~~~~~~~~~~~~~~~~\left[\left(-\frac{kc_{S}\eta}{2}\right)^{-\Lambda}\left(-\frac{kc_{S}\xi}{2}\right)^{-\Lambda}
+\left(\frac{kc_{S}\eta}{2}\right)^{-\Lambda}\left(\frac{kc_{S}\xi}{2}\right)^{-\Lambda}\right],~~~~~~~~~~~~\\
 O^{\xi\theta}_{5}&=& \frac{1}{2\pi^2}\int^{\infty}_{0} dk~e^{ikx}~k^{2-\Lambda}\int\limits_{-\infty}^{\xi} {d\eta}
~\frac{\tilde{c}^2_{S}H}{a(\eta)\epsilon\sqrt{-\eta}} \frac{m}{H\tilde{c}_{S}}
\left(\Lambda-\frac{1}{2}\right)\left(\frac{1}{2}\right)^{-\Lambda}\\
\nonumber&&~~~~~~~~~~~~~~~~~~~\left[\left(-\frac{c_{S}\xi}{2}\right)^{-\Lambda}
+\left(-1\right)^{-\Lambda}\left(\frac{c_{S}\xi}{2}\right)^{-\Lambda}\right].~~~~~~~~~~~~\eea
 \bea \footnotesize{\bf For~Case~II:}\nonumber\\
 O_{1}&=&  \frac{1}{4\pi^2}\int^{\infty}_{0} dk~\int\limits_{-\infty}^{0} {d\eta}\int^{\Theta_{2}}_{\Theta_{1}}d\Theta~\eta k~m(\eta)
 ~e^{ik\left(x\cos\Theta-c_{S}\eta\right)}
=-\int^{\Theta_{2}}_{\Theta_{1}}d\Theta\frac{im(\eta=-\frac{|{\bf x}|\cos\Theta}{c_{S}})}{4\pi^2c^2_{S}}\cos\Theta,~~~~~~~~~\\ 
 O_{2}&=& \frac{1}{4\pi^2}\int^{\infty}_{0} dk~\int\limits_{-\infty}^{0} {d\eta}\int^{\Theta_{2}}_{\Theta_{1}}d\Theta~\eta k~m(\eta)~
 e^{ik\left(x\cos\Theta+c_{S}\eta\right)}
=\int^{\Theta_{2}}_{\Theta_{1}}d\Theta\frac{im(\eta=-\frac{|{\bf x}|\cos\Theta}{c_{S}})}{4\pi^2c^2_{S}}\cos\Theta,\\ 
 O_{3}&=& \frac{1}{4\pi^2}\int^{\infty}_{0} dk\int\limits_{-\infty}^{0} {d\eta}\int^{\Theta_{2}}_{\Theta_{1}}d\Theta
~\eta k~m(\eta)~e^{ikx\cos\Theta}\sin {\it kc_{S}\eta}=\int^{\Theta_{2}}_{\Theta_{1}}d\Theta\frac{m(\eta=-\frac{|{\bf x}|\cos\Theta}{c_{S}})}{4\pi^2c^2_{S}}\cos\Theta,\\
 O^{\xi\theta}_{4}&=& \frac{1}{4\pi^2}\int^{\infty}_{0} dk~\int\limits_{-\infty}^{\xi} {d\eta}~\int^{\Theta_{2}}_{\Theta_{1}}~d\Theta~k^2~e^{ikx\cos\Theta}
~\frac{\tilde{c}^2_{S}H}{a(\eta)\epsilon\sqrt{-\eta}} \frac{m}{H\tilde{c}_{S}}
\left(\Lambda-\frac{1}{2}\right)\\
\nonumber&&~~~~~~~~~~~~~~~~~~~\left[\left(-\frac{kc_{S}\eta}{2}\right)^{-\Lambda}\left(-\frac{kc_{S}\xi}{2}\right)^{-\Lambda}
+\left(\frac{kc_{S}\eta}{2}\right)^{-\Lambda}\left(\frac{kc_{S}\xi}{2}\right)^{-\Lambda}\right],~~~~~~~~~~~~\\
 O^{\xi\theta}_{5}&=&\frac{1}{4\pi^2}\int^{\infty}_{0} dk~\int^{\Theta_{2}}_{\Theta_{1}}~d\Theta~e^{ikx\cos \Theta}~k^{2-\Lambda}\int\limits_{-\infty}^{\xi} {d\eta}
~\frac{\tilde{c}^2_{S}H}{a(\eta)\epsilon\sqrt{-\eta}} \frac{m}{H\tilde{c}_{S}}
\left(\Lambda-\frac{1}{2}\right)\left(\frac{1}{2}\right)^{-\Lambda}\\
\nonumber&&~~~~~~~~~~~~~~~~~~~\left[\left(-\frac{c_{S}\xi}{2}\right)^{-\Lambda}
+\left(-1\right)^{-\Lambda}\left(\frac{c_{S}\xi}{2}\right)^{-\Lambda}\right].~~~~~~~~~~~~\eea
 where $\Theta_{1}$ and $\Theta_{2}$ plays the role of angular regulator in the present context.
 
 Now our objective is to compute the expression for the 
 two point correlation function from scalar curvature perturbation. Following the previously mentioned 
 computational technique of in-in formalism, which is commonly known as the Swinger-Keyldish formalism here we 
 compute the expression for the two point correlation function from scalar curvature perturbation.
 Using the interaction picture the two point correlation function
 of the curvature fluctuation in momentum space can be expressed as:  
\begin{equation}\label{opiovbvb}
\langle\zeta_{\bf k}(\eta)\zeta_{\bf q}(\eta)\rangle = (2\pi)^3 \delta^{3}({\bf k}+{\bf q})\frac{2\pi^2}{k^3}
\Delta_{\zeta}(k),
\end{equation}
where the primordial power spectrum for scalar mode at any arbitrary momentum scale can be written as:
\bea\label{dfinv4bnbncv}
\displaystyle \Delta_{\zeta}(k) &=& \frac{k^3|h_{\bf k}|^2}{2\pi^2 z^2M^2_p}\\
&=&\frac{(-k\eta \tilde{c}_{S})^3 H^2}{4\tilde{c}_{S}\epsilon \pi^2 M^2_p}\sum^{2}_{i=1}\sum^{2}_{j=1} C^{*}_{i}C_{j} 
U_{ij}(\eta,k)\nonumber.
\eea
where $U_{ij}\forall i,j=1,2$ are defined as:
\bea U_{11}(\eta,k)&=& H^{(1)*}_{\Lambda} \left(-kc_{S}\eta\right)H^{(1)}_{\Lambda} \left(-kc_{S}\eta\right),\\
U_{12}(\eta,k)&=& H^{(1)*}_{\Lambda} \left(-kc_{S}\eta\right)H^{(2)}_{\Lambda} \left(-kc_{S}\eta\right),\\
U_{21}(\eta,k)&=& H^{(2)*}_{\Lambda} \left(-kc_{S}\eta\right)H^{(1)}_{\Lambda} \left(-kc_{S}\eta\right),\\
U_{22}(\eta,k)&=& H^{(2)*}_{\Lambda} \left(-kc_{S}\eta\right)H^{(2)}_{\Lambda} \left(-kc_{S}\eta\right).\eea
For further simplification we consider here three limiting cases $|kc_{S}\eta|\rightarrow -\infty$,
$|kc_{S}\eta|\rightarrow 0$ and $|kc_{S}\eta|\approx 1$,
 which are physically acceptable in the present context. First of we consider the
 results for $|kc_{S}\eta|\rightarrow -\infty$. In this case we get:
 \begin{equation}\label{opiovbvb}
\langle\zeta_{\bf k}(\eta)\zeta_{\bf q}(\eta)\rangle_{|kc_{S}\eta|\rightarrow -\infty} =
(2\pi)^3 \delta^{3}({\bf k}+{\bf q})\frac{2\pi^2}{k^3}
\left[\Delta_{\zeta}(k)\right]_{{|kc_{S}\eta|\rightarrow -\infty}},
\end{equation}
where the primordial power spectrum for scalar mode at $|kc_{S}\eta|\rightarrow -\infty$ can be written as:
 \bea\label{dfinv4bnbncv}
\displaystyle \left[\Delta_{\zeta}(k)\right]_{{|kc_{S}\eta|\rightarrow -\infty}} &\approx& 
\frac{H^2}{2\epsilon M^2_{p}}\frac{k^2\eta^2\tilde{c}^2_{S}}{c_{S}\pi^3}
\left[|C_{2}|^2+|C_{1}|^2\right.\\ &&\left.~~~~~~~\nonumber+\left(C^{*}_{1}C_{2}e^{2ikc_{S}\eta}e^{i\pi\left(\Lambda+\frac{1}{2}\right)}
+C_{1}C^{*}_{2}e^{-2ikc_{S}\eta}e^{-i\pi\left(\Lambda+\frac{1}{2}\right)}\right)\right]\\&=&\left\{\begin{array}{ll}
                   \displaystyle    \frac{H^2}{2\epsilon M^2_{p}c_{S}}\frac{(-k\eta\tilde{c}_{S})^2}{2\pi^2}\nonumber &
 \mbox{\small {\bf for ~Bunch Davies}}  \\ 
	 \displaystyle  \frac{H^2}{2\epsilon M^2_{p}c_{S}}\frac{(-k\tilde{c}_{S}\eta)^2}{\pi^3}
\left[\sinh^2\alpha +\cosh^2\alpha
\right.\\ \left. \displaystyle ~~~~~~~~~~~~~+\sinh2\alpha\cos{\it \left(2kc_{S}\eta+\pi\left(\Lambda+\frac{1}{2}\right)+\delta\right)}\right]
                     & \mbox{\small {\bf for~$\alpha$~vacua~Type-I}}\\ 
	  \displaystyle \frac{H^2}{2\epsilon M^2_{p}c_{S}}\frac{4(-k\tilde{c}_{S}\eta)^2 |N_{\alpha}|^2}{\pi^3}
\cos^2{\it \left(kc_{S}\eta+\frac{\pi}{2}\left(\Lambda+\frac{1}{2}\right)-i\frac{\alpha}{2}\right)}
                     & \mbox{\small {\bf for~$\alpha$~vacua~Type-II}}\\ 
	 \displaystyle \frac{H^2}{2\epsilon M^2_{p}c_{S}}\frac{4(-k\tilde{c}_{S}\eta)^2}{\pi^3}
\cos^2{\it \left(kc_{S}\eta+\frac{\pi}{2}\left(\Lambda+\frac{1}{2}\right)\right)}
                    & \mbox{\small {\bf for~special~vacua}}.
          \end{array}
\right.
\eea
Next we consider the
 results for $|kc_{S}\eta|\rightarrow 0$. In this case we get:
 \begin{equation}\label{opiovbvb}
\langle\zeta_{\bf k}(\eta)\zeta_{\bf q}(\eta)\rangle_{|kc_{S}\eta|\rightarrow 0} =
(2\pi)^3 \delta^{3}({\bf k}+{\bf q})\frac{2\pi^2}{k^3}
\left[\Delta_{\zeta}(k)\right]_{{|kc_{S}\eta|\rightarrow 0}},
\end{equation}
where the primordial power spectrum for scalar mode at $|kc_{S}\eta|\rightarrow 0$ can be written as:
 \bea\label{dfinv4bnbncv}
\displaystyle \left[\Delta_{\zeta}(k)\right]_{{|kc_{S}\eta|\rightarrow 0}} &\approx& 
\frac{H^2}{2\epsilon M^2_{p}}\frac{(-k\eta\tilde{c}_{S})^3}{2\tilde{c}_{S}\pi^4}\Gamma^2(\Lambda)\left(-\frac{kc_{S}\eta}{2}\right)^{-2\Lambda}
\left[|C_{2}|^2+|C_{1}|^2-\left(C^{*}_{1}C_{2}
+C_{1}C^{*}_{2}\right)\right]\\&=&\left\{\begin{array}{ll}
                    \displaystyle   \frac{H^2}{2\epsilon M^2_{p}}\frac{(-k\eta{c}_{S})^{3-2\Lambda}}{2^{2(2-\Lambda)}
                      \pi^2}\frac{\tilde{c}^2_{S}}{c^3_{S}}\left|\frac{\Gamma(\Lambda)}{\Gamma\left(\frac{3}{2}\right)}\right|^2
                     \nonumber &
 \mbox{\small {\bf for ~Bunch Davies}}  \\ 
	 \displaystyle \frac{H^2}{2\epsilon M^2_{p}}\frac{(-k\eta{c}_{S})^{3-2\Lambda}}{2^{(3-2\Lambda)}
                      \pi^3}\frac{\tilde{c}^2_{S}}{c^3_{S}}\left|\frac{\Gamma(\Lambda)}{\Gamma\left(\frac{3}{2}\right)}\right|^2
\left[\sinh^2\alpha +\cosh^2\alpha
-\sinh2\alpha\cos{\it \delta}\right]
                     & \mbox{\small {\bf for~$\alpha$~vacua~Type-I}}\\ 
	 \displaystyle \frac{H^2}{2\epsilon M^2_{p}}\frac{(-k\eta{c}_{S})^{3-2\Lambda}|N_{\alpha}|^2}{2^{(1-2\Lambda)}
                      \pi^3}\frac{\tilde{c}^2_{S}}{c^3_{S}}\left|\frac{\Gamma(\Lambda)}
                      {\Gamma\left(\frac{3}{2}\right)}\right|^2
\sin^2\frac{\alpha}{2}
                     & \mbox{\small {\bf for~$\alpha$~vacua~Type-II}}\\ 
	 0
                    & \mbox{\small {\bf for~special~vacua}}.
          \end{array}
\right.
\eea
Finally we consider the
 results for $|kc_{S}\eta|\approx 1$. In this case we get:
 \begin{equation}\label{opiovbvb}
\langle\zeta_{\bf k}(\eta=0)\zeta_{\bf q}(\eta=0)\rangle_{|kc_{S}\eta|\approx 1} =
(2\pi)^3 \delta^{3}({\bf k}+{\bf q})\frac{2\pi^2}{k^3}
\left[\Delta_{\zeta}(k)\right]_{{|kc_{S}\eta|\approx 1}},
\end{equation}
where the primordial power spectrum for scalar mode at $|kc_{S}\eta|\approx 1$ can be written as:
 \bea\label{dfinv4bnbncv}
\displaystyle \left[\Delta_{\zeta}(k)\right]_{{|kc_{S}\eta|\approx 1}} &\approx& 
\frac{H^2}{2\epsilon M^2_{p}}\frac{1}{2^{2(1-\Lambda)}\pi^3}\frac{\tilde{c}^2_{S}}{c^3_{S}}\left|\frac{\Gamma(\Lambda)}{\Gamma
\left(\frac{3}{2}\right)}\right|^2
\left[|C_{2}|^2+|C_{1}|^2-\left(C^{*}_{1}C_{2}
+C_{1}C^{*}_{2}\right)\right]\\&=&\left\{\begin{array}{ll}
                   \displaystyle    \frac{H^2}{2\epsilon M^2_{p}}\frac{1}{2^{2(2-\Lambda)}
                      \pi^2}\frac{\tilde{c}^2_{S}}{c^3_{S}}\left|\frac{\Gamma(\Lambda)}{\Gamma\left(\frac{3}{2}\right)}\right|^2
                     \nonumber &
 \mbox{\small {\bf for ~Bunch Davies}}  \\ 
	 \displaystyle \frac{H^2}{2\epsilon M^2_{p}}\frac{1}{2^{(3-2\Lambda)}
                      \pi^3}\frac{\tilde{c}^2_{S}}{c^3_{S}}\left|\frac{\Gamma(\Lambda)}{\Gamma\left(\frac{3}{2}\right)}\right|^2
\left[\sinh^2\alpha +\cosh^2\alpha
\right.\\ \left. \displaystyle ~~~~~~~~~~~~~~~~~~~~~~~~~~~~~~~~~~~~~-\sinh2\alpha\cos{\it \delta}\right]
                     & \mbox{\small {\bf for~$\alpha$~vacua~Type-I}}\\ 
	 \displaystyle \frac{H^2}{2\epsilon M^2_{p}}\frac{|N_{\alpha}|^2}{2^{(1-2\Lambda)}
                      \pi^3}\frac{\tilde{c}^2_{S}}{c^3_{S}}\left|\frac{\Gamma(\Lambda)}
                      {\Gamma\left(\frac{3}{2}\right)}\right|^2
\sin^2\frac{\alpha}{2}
                     & \mbox{\small {\bf for~$\alpha$~vacua~Type-II}}\\ 
	 0
                    & \mbox{\small {\bf for~special~vacua}}.
          \end{array}
\right.
\eea
Now to analyze the behaviour of the two point correlation function of scalar curvature perturbation in position space 
we need to take the Fourier transform of the two point correlation function of scalar curvature perturbation already computed 
in momentum space. For the most generalized solution we get the following result:
\be\begin{array}{llll}\label{dfinv3}
\displaystyle \langle\zeta({\bf x},\eta)\zeta({\bf y},\eta)\rangle = \int \frac{d^{3}k}{(2\pi)^3 }
~e^{i{\bf k}.({\bf x}-{\bf y})}\frac{(-\eta\tilde{c}_{S})^3}{2\tilde{c}_{S}\epsilon M^2_p}\sum^{2}_{i=1}\sum^{2}_{j=1}C^{*}_{i}C_{j}U_{ij}.
\end{array}\ee
where the conformal time dependent functions $U_{ij}\forall i,j=1,2$ in momentum
space is already defined earlier.

Similarly in the position space the representative expressions for the expectation value of the 
scalar curvature perturbation along with 
limiting case $|kc_{S}\eta|\rightarrow -\infty$ is given by:
 \bea\label{dfinv4bnbncv}
\displaystyle \langle\zeta({\bf x},\eta)\zeta({\bf y},\eta)\rangle_{|kc_{S}\eta|\rightarrow -\infty}&\approx& 
\frac{1}{4\pi^4}\frac{H^2}{2\epsilon M^2_{p}}\frac{\eta^2\tilde{c}^2_{S}}{c_{S}}
\left[\left(|C_{2}|^2+|C_{1}|^2\right)J_{1}\right.\\ &&\left.~~~~~~~\nonumber+\left(C^{*}_{1}C_{2}
e^{i\pi\left(\Lambda+\frac{1}{2}\right)}J_{2}
+C_{1}C^{*}_{2}e^{-i\pi\left(\Lambda+\frac{1}{2}\right)}J_{3}\right)\right]\\&=&\left\{\begin{array}{ll}
                   \displaystyle    \frac{H^2}{2\epsilon M^2_{p}c_{S}}\frac{(\eta\tilde{c}_{S})^2}{(2\pi)^3}J_{1}\nonumber &
 \mbox{\small {\bf for ~Bunch Davies}}  \\ \\
	 \displaystyle  \frac{H^2}{2\epsilon M^2_{p}c_{S}}\frac{(\tilde{c}_{S}\eta)^2}{4\pi^4}
\left[\left(\sinh^2\alpha +\cosh^2\alpha\right)J_{1}
\right.\\ \left. \displaystyle ~~~~~~~~~~~~~+\frac{1}{2}\sinh2\alpha\left( e^{
i\left(\pi\left(\Lambda+\frac{1}{2}\right)+\delta\right)}J_{2}+e^{
-i\left(\pi\left(\Lambda+\frac{1}{2}\right)+\delta\right)}\right)J_{3}\right]
                     & \mbox{\small {\bf for~$\alpha$~vacua~Type-I}}\\ \\
	  \displaystyle \frac{H^2}{2\epsilon M^2_{p}c_{S}}\frac{(\tilde{c}_{S}\eta)^2 |N_{\alpha}|^2}{2\pi^4}
\left[J_{1}+\left(e^{i\pi\left(\Lambda+\frac{1}{2}\right)}e^{\alpha}J_{2}+
e^{-i\pi\left(\Lambda+\frac{1}{2}\right)}e^{-\alpha}J_{3}\right)\right]
                     & \mbox{\small {\bf for~$\alpha$~vacua~Type-II}}\\ \\
	 \displaystyle \frac{H^2}{2\epsilon M^2_{p}c_{S}}\frac{(\tilde{c}_{S}\eta)^2}{2\pi^4}
\left[J_{1}+\left(e^{i\pi\left(\Lambda+\frac{1}{2}\right)}e^{\alpha}J_{2}+
e^{-i\pi\left(\Lambda+\frac{1}{2}\right)}e^{-\alpha}J_{3}\right)\right]
                    & \mbox{\small {\bf for~special~vacua}}.
          \end{array}
\right.
\eea
where the momentum integrals $J_{1}, J_{2}$ and $J_{3}$ are defined in the following:
\bea  J_{1}&=&\int d^3k~\frac{e^{i{\bf k}.({\bf x}-{\bf y})}}{k},\\
J_{2}&=&\int d^3k~\frac{e^{i{\bf k}.({\bf x}-{\bf y})}}{k}~e^{2ikc_{S}\eta},\\
J_{3}&=&\int d^3k~\frac{e^{i{\bf k}.({\bf x}-{\bf y})}}{k}~e^{-2ikc_{S}\eta}.\eea
To compute this integrals we follow the same assumptions as mentioned for the computation of 
VEV of curvature perturbation in position space. Here we have the following results:
\bea  {\bf For~~Case~ I:}~~~~~\nonumber\\
J_{1}&=&\int^{\infty}_{0} dk~k~e^{ik(|{\bf x}-{\bf y}|)}=-\frac{4\pi}{|{\bf x}-{\bf y}|^2},\\
J_{2}&=&\int^{\infty}_{0} dk~k~e^{ik(|{\bf x}-{\bf y}|+2c_{S}\eta)}=-\frac{4\pi}{(|{\bf x}-{\bf y}|+2c_{S}\eta)^2},\\
J_{3}&=&\int^{\infty}_{0} dk~k~e^{ik(|{\bf x}-{\bf y}|-2c_{S}\eta)}=-\frac{4\pi}{(|{\bf x}-{\bf y}|-2c_{S}\eta)^2}.\eea
\bea  {\bf For~~Case~ II:}~~~~~\nonumber\\
J_{1}&=&\int^{\infty}_{0} dk\int^{\Theta_{2}}_{\Theta_{1}}~d\Theta~k~e^{ikx\cos\Theta}=\frac{2\pi(\tan \Theta_{1} -\tan \Theta_{2})}{|{\bf x}-{\bf y}|^2},\\
J_{2}&=&\int^{\infty}_{0} dk\int^{\Theta_{2}}_{\Theta_{1}}~d\Theta~k~e^{ik(x\cos\Theta+2c_{S}\eta)}\nonumber\\
&=&\frac{2\pi|{\bf x}-{\bf y}|}{(|{\bf x}-{\bf y}|^2-4c^2_{S}\eta^2)}
\left\{\left[\frac{\sin \Theta_{1}}{(|{\bf x}-{\bf y}|\cos \Theta_{1}+2c_{S}\eta)}
-\frac{\sin \Theta_{2}}{(|{\bf x}-{\bf y}|\cos \Theta_{2}+2c_{S}\eta)}\right]\right.\nonumber\\&& \left.
~~~~~~~~~~~~+\frac{4c_{S}\eta}{\sqrt{|{\bf x}-{\bf y}|^2-4c^2_{S}\eta^2}}\left[{\rm tanh}^{-1}
\left(\frac{(2c_{S}\eta-|{\bf x}-{\bf y}|){\rm tan}\frac{\Theta_{1}}{2}}{\sqrt{|{\bf x}-{\bf y}|^2-4c^2_{S}\eta^2}}
\right)\right.\right.\nonumber\\&& \left.\left.~~~~~~~~~~~~~~~~~~~~~~~~~~~-{\rm tanh}^{-1}
\left(\frac{(2c_{S}\eta-|{\bf x}-{\bf y}|){\rm tan}\frac{\Theta_{2}}{2}}{\sqrt{|{\bf x}-{\bf y}|^2-4c^2_{S}\eta^2}}
\right)\right]\right\},\\
J_{3}&=&\int^{\infty}_{0} dk\int^{\Theta_{2}}_{\Theta_{1}}~d\Theta~k~e^{ik(x\cos\Theta-2c_{S}\eta)}\nonumber\\
&=&\frac{2\pi|{\bf x}-{\bf y}|}{(|{\bf x}-{\bf y}|^2-4c^2_{S}\eta^2)}
\left\{\left[\frac{\sin \Theta_{1}}{(|{\bf x}-{\bf y}|\cos \Theta_{1}-2c_{S}\eta)}
-\frac{\sin \Theta_{2}}{(|{\bf x}-{\bf y}|\cos \Theta_{2}-2c_{S}\eta)}\right]\right.\nonumber\\&& \left.
~~~~~~~~~~~~+\frac{4c_{S}\eta}{\sqrt{|{\bf x}-{\bf y}|^2-4c^2_{S}\eta^2}}\left[{\rm tanh}^{-1}
\left(\frac{(2c_{S}\eta+|{\bf x}-{\bf y}|){\rm tan}\frac{\Theta_{1}}{2}}{\sqrt{|{\bf x}-{\bf y}|^2-4c^2_{S}\eta^2}}
\right)\right.\right.\nonumber\\&& \left.\left.~~~~~~~~~~~~~~~~~~~~~~~~~~~-{\rm tanh}^{-1}
\left(\frac{(2c_{S}\eta+|{\bf x}-{\bf y}|){\rm tan}\frac{\Theta_{2}}{2}}{\sqrt{|{\bf x}-{\bf y}|^2-4c^2_{S}\eta^2}}
\right)\right]\right\}.\eea
Here by setting ${\bf y}=0$ one can derive the results for $\langle\zeta({\bf x},\eta)\zeta(0,\eta)\rangle$
at conformal time scale $\eta$ with $|kc_{S}\eta|\rightarrow -\infty$.

Next we consider the
 results for $|kc_{S}\eta|\rightarrow 0$. In this case we get:
 \bea\label{dfinv4bnbncv}
\displaystyle \langle\zeta({\bf x},\eta)\zeta({\bf y},\eta)\rangle_{|kc_{S}\eta|\rightarrow 0}&\approx& 
\frac{H^2}{2\epsilon M^2_{p}}\frac{(-\eta c_{S})^{3-2\Lambda}}{2^{2(2-\Lambda)}\pi^4}\frac{\tilde{c}^2_{S}}{c^3_{S}}\left|\frac{\Gamma(\Lambda)}{\Gamma
\left(\frac{3}{2}\right)}\right|^2
\left[(|C_{2}|^2+|C_{1}|^2)-\left(C^{*}_{1}C_{2}
+C_{1}C^{*}_{2}\right)\right]K_{I}\\&=& K_{I}\times\left\{\begin{array}{ll}
                \displaystyle       \frac{H^2}{2\epsilon M^2_{p}}\frac{(-\eta{c}_{S})^{3-2\Lambda}}{2^{2(3-\Lambda)}
                      \pi^3}\frac{\tilde{c}^2_{S}}{c^3_{S}}\left|\frac{\Gamma(\Lambda)}{\Gamma\left(\frac{3}{2}\right)}\right|^2
                     \nonumber &
 \mbox{\small {\bf for ~Bunch Davies}}  \\ 
	 \displaystyle \frac{H^2}{2\epsilon M^2_{p}}\frac{(-\eta{c}_{S})^{3-2\Lambda}}{2^{(5-2\Lambda)}
                      \pi^4}\frac{\tilde{c}^2_{S}}{c^3_{S}}\left|\frac{\Gamma(\Lambda)}{\Gamma\left(\frac{3}{2}\right)}\right|^2
\left[\sinh^2\alpha +\cosh^2\alpha
\right.\\ \left. \displaystyle ~~~~~~~~~~~~~~~~~~~~~~~~~~~~~~~~~~~~~
-\sinh2\alpha\cos{\it \delta}\right]
                     & \mbox{\small {\bf for~$\alpha$~vacua~Type-I}}\\ 
	 \displaystyle \frac{H^2}{2\epsilon M^2_{p}}\frac{(-\eta{c}_{S})^{3-2\Lambda}|N_{\alpha}|^2}{2^{(3-2\Lambda)}
                      \pi^4}\frac{\tilde{c}^2_{S}}{c^3_{S}}\left|\frac{\Gamma(\Lambda)}
                      {\Gamma\left(\frac{3}{2}\right)}\right|^2
\sin^2\frac{\alpha}{2}
                     & \mbox{\small {\bf for~$\alpha$~vacua~Type-II}}\\ 
	 0
                    & \mbox{\small {\bf for~special~vacua}}.
          \end{array}
\right.
\eea
where the momentum integrals $K_{I}$ is defined in the following:
\bea  K_{I}&=&\int~d^3k~k^{1-2\Lambda}~e^{i{\bf k}.({\bf x}-{\bf y})}.\eea
To compute this integral we follow the same assumptions as mentioned for the computation of 
VEV of curvature perturbation in position space. Here we have the following results:
\bea  {\bf For~ Case~I:}~~~~K_{I}&=&4\pi\int^{\infty}_{0}dk~k^{3-2\Lambda}~e^{ik|{\bf x}-{\bf y}|}
=4\pi\left(\frac{i}{|{\bf x}-{\bf y}|}\right)^{3-2\Lambda}\Gamma(3-2\Lambda).\eea
\bea  {\bf For~ Case~II:}~~~~K_{I}&=&2\pi\int^{\infty}_{0}dk~\int^{\Theta_{2}}_{\Theta_{2}}d\Theta~
k^{3-2\Lambda}~e^{ik|{\bf x}-{\bf y}|\cos \Theta}\nonumber\\
&=&2\pi\frac{\Gamma (4-2 \Lambda ) 
(-i)^{2 \Lambda }}{(3-2\Lambda)|{\bf x}-{\bf y}|^{4-2\Lambda}}\left[( \sec \Theta_{1})^{3-2\Lambda} 
\, _2F_1\left(\frac{1}{2},\Lambda -\frac{3}{2};\Lambda -\frac{1}{2};\cos ^2\Theta_{1}\right) \nonumber \right.\\ && \left. 
~~~~~~~~~~~~~~~~~~-( \sec \Theta_{2})^{3-2\Lambda} 
\, _2F_1\left(\frac{1}{2},\Lambda -\frac{3}{2};\Lambda -\frac{1}{2};\cos ^2\Theta_{2}\right) \right].\eea
Here by setting ${\bf y}=0$ one can derive the results for $\langle\zeta({\bf x},\eta)\zeta(0,\eta)\rangle$
at conformal time scale $\eta$ with $|kc_{S}\eta|\rightarrow 0$.
 
Finally we consider the
 results for $|kc_{S}\eta|\approx 1$. In this case we get:
  \bea\label{dfinv4bnbncv}
\displaystyle \langle\zeta({\bf x},\eta=0)\zeta({\bf y},\eta=0)\rangle_{|kc_{S}\eta|\approx 1}&\approx& 
\frac{H^2}{2\epsilon M^2_{p}}\frac{1}{2^{2(2-\Lambda)}\pi^4}\frac{\tilde{c}^2_{S}}{c^3_{S}}\left|\frac{\Gamma(\Lambda)}{\Gamma
\left(\frac{3}{2}\right)}\right|^2
\left[(|C_{2}|^2+|C_{1}|^2)-\left(C^{*}_{1}C_{2}
+C_{1}C^{*}_{2}\right)\right]Z_{I}\\&=& Z_{I}\times\left\{\begin{array}{ll}
                      \displaystyle \frac{H^2}{2\epsilon M^2_{p}}\frac{1}{2^{2(3-\Lambda)}
                      \pi^3}\frac{\tilde{c}^2_{S}}{c^3_{S}}\left|\frac{\Gamma(\Lambda)}{\Gamma\left(\frac{3}{2}\right)}\right|^2
                     \nonumber &
 \mbox{\small {\bf for ~Bunch Davies}}  \\ 
	 \displaystyle \frac{H^2}{2\epsilon M^2_{p}}\frac{1}{2^{(5-2\Lambda)}
                      \pi^4}\frac{\tilde{c}^2_{S}}{c^3_{S}}\left|\frac{\Gamma(\Lambda)}{\Gamma\left(\frac{3}{2}\right)}\right|^2
\left[\sinh^2\alpha +\cosh^2\alpha
\right.\\ \left.  \displaystyle~~~~~~~~~~~~~~~~~~~~~~~~~~~~~~~~~~~~~
-\sinh2\alpha\cos{\it \delta}\right]
                     & \mbox{\small {\bf for~$\alpha$~vacua~Type-I}}\\ 
	  \displaystyle\frac{H^2}{2\epsilon M^2_{p}}\frac{|N_{\alpha}|^2}{2^{(3-2\Lambda)}
                      \pi^4}\frac{\tilde{c}^2_{S}}{c^3_{S}}\left|\frac{\Gamma(\Lambda)}
                      {\Gamma\left(\frac{3}{2}\right)}\right|^2
\sin^2\frac{\alpha}{2}
                     & \mbox{\small {\bf for~$\alpha$~vacua~Type-II}}\\ 
	 0
                    & \mbox{\small {\bf for~special~vacua}}.
          \end{array}
\right.
\eea
where the momentum integrals $Z_{I}$ is defined in the following:
\bea  Z_{I}&=&\int~d^3k~\frac{e^{i{\bf k}.({\bf x}-{\bf y})}}{k^3}.\eea
To compute this integral we follow the same assumptions as mentioned for the computation of 
VEV of curvature perturbation in position space. Additionally we introduce a infrared cut-off $L_{IR}$ to regularize the integral.
Here we have the following results:
\bea  {\bf For~ Case~I:}~~~~Z_{I}&=&4\pi\int^{\infty}_{1/L_{IR}}dk~\frac{e^{ik|{\bf x}-{\bf y}|}}{k}
\approx 4\pi\left\{\ln\left(\frac{L_{IR}}{|{\bf x}-{\bf y}|}\right)-\gamma_{E}\right\}.\eea
\bea  {\bf For~ Case~II:}~~~~Z_{I}&=&2\pi\int^{\infty}_{1/L_{IR}}dk~\int^{\Theta_{2}}_{\Theta_{1}}d\Theta~
\frac{e^{ik|{\bf x}-{\bf y}|\cos \Theta}}{k}\nonumber\\
&=&4\pi\left[(\Theta_{2}-\Theta_{1})\left\{\ln\left(\frac{L_{IR}}{|{\bf x}-{\bf y}|}\right)
+\ln\left(\frac{\cos \Theta_{2}}{\cos \Theta_{1}}\right)-\gamma_{E}
\right.\right.\\ && \left.\left.+{\rm Ei}\left(\frac{i|{\bf x}-{\bf y}|\cos \Theta_{2}}{L_{IR}}\right)
-{\rm Ei}\left(\frac{i|{\bf x}-{\bf y}|\cos \Theta_{2}}{L_{IR}}\right)\right\}
-\int^{\Theta_{2}}_{\Theta_{1}}d\Theta~{\rm Ei}\left(\frac{i|{\bf x}-{\bf y}|\cos \Theta}{L_{IR}}\right) \right].\nonumber\eea
Here by setting ${\bf y}=0$ one can derive the results for $\langle\zeta({\bf x},\eta=0)\zeta(0,\eta=0)\rangle$
at conformal time scale $\eta$ with $|kc_{S}\eta|\rightarrow -\infty$.

Now our prime objective is following: 
\begin{itemize}
 \item to derive a exact connection between the computed 
 VEV and two point function of the scalar curvature perturbation in presence of mass parameter $m(\eta)$,
 \item to derive the exact connection between the VEV of the curvature perturbation with the real cosmological 
 observables,
 \item to put additional constraint on the primordial cosmological setup in presence of new mass parameter $m(\eta)$
 within the prescription of EFT, 
 \item to check the future possibility of detecting VEV of the curvature perturbation in the cosmological experiments 
 and to put stringent bound on the Wilsonian operators of the background EFT framework,
 \item to give a theoretical understanding of the new cosmological quantity i.e. VEV of the curvature perturbation 
 in presence of mass parameter $m(\eta)$, using which it may be possible to break the degeneracy between various cosmological 
 parameters and once can able to discriminate between various models in primordial cosmology~\footnote{Implementing 
 the present techniques it is also possible to derive a direct connection between the primordial non-Gaussianity 
 computed from the three point and one point scalar fluctuations for the newly
 introduced mass parameter $m(\eta)$. We will explore this 
 possibility in near future in great detail within the context of EFT.}.
\end{itemize}
To establish such a clear 
connection in momentum space we write down the following sets of new consistency relations in primordial cosmology
for the three limiting cases $|kc_{S}\eta|\rightarrow -\infty$, $|kc_{S}\eta|\rightarrow 0$
and $|kc_{S}\eta|\approx 1$:
\bea\langle \zeta_{\bf k}(\eta=0) \rangle_{|kc_{S}\eta|
\rightarrow -\infty}&=&\langle \zeta_{\bf k}(\eta=0)\zeta_{\bf q}(\eta=0)
\rangle^{'}_{|kc_{S}\eta|\rightarrow -\infty}\times\frac{2 k }{\eta^2 \tilde{c}_{S}H}{\cal I}_{1},\\
\langle \zeta_{\bf k}(\eta=\xi\rightarrow 0) \rangle_{|kc_{S}\eta|
\rightarrow 0}&=&\langle \zeta_{\bf k}(\eta=\xi\rightarrow 0)\zeta_{\bf q}(\eta=\xi\rightarrow 0)
\rangle^{'}_{|kc_{S}\eta|\rightarrow 0}\nonumber\\
&&~~~~~\times\frac{2^{2-2\Lambda} k^{2\Lambda}c^{2\Lambda}_{S}}{(-\eta)^{3-2\Lambda} H\pi}
\left|\frac{\Gamma\left(\frac{3}{2}\right)}{\Gamma\left(\Lambda\right)}\right|^2{\cal I}_{2},\\
\langle \zeta_{\bf k}(\eta=0) \rangle_{|kc_{S}\eta|
\approx 1}&=&\langle \zeta_{\bf k}(\eta=0)\zeta_{\bf q}(\eta=0)
\rangle^{'}_{|kc_{S}\eta|\approx 1}\nonumber\\
&&~~~~~\times\frac{2^{2-2\Lambda} k^{3}c^{3}_{S}}{ H\pi}
\left|\frac{\Gamma\left(\frac{3}{2}\right)}{\Gamma\left(\Lambda\right)}\right|^2{\cal I}_{3},
\eea
where $\langle \zeta_{\bf k}(\eta=0)\zeta_{\bf q}(\eta=0)
\rangle^{'}_{|kc_{S}\eta|\rightarrow -\infty}$, $\langle \zeta_{\bf k}(\eta=0)\zeta_{\bf q}(\eta=0)
\rangle^{'}_{|kc_{S}\eta|\rightarrow 0}$ and $\langle \zeta_{\bf k}(\eta=0)\zeta_{\bf q}(\eta=0)
\rangle^{'}_{|kc_{S}\eta|\approx 1}$ are defined as:
\bea \langle \zeta_{\bf k}(\eta=0)\zeta_{\bf q}(\eta=0)
\rangle^{'}_{|kc_{S}\eta|\rightarrow -\infty}&=&\frac{\langle \zeta_{\bf k}(\eta=0)\zeta_{\bf q}(\eta=0)
\rangle_{|kc_{S}\eta|\rightarrow -\infty}}{(2\pi)^3\delta^{3}({\bf k}+{\bf q})},\\ 
\langle \zeta_{\bf k}(\eta=0)\zeta_{\bf q}(\eta=0)
\rangle^{'}_{|kc_{S}\eta|\rightarrow 0}&=&\frac{\langle \zeta_{\bf k}(\eta=0)\zeta_{\bf q}(\eta=0)
\rangle_{|kc_{S}\eta|\rightarrow 0}}{(2\pi)^3\delta^{3}({\bf k}+{\bf q})},\\
\langle \zeta_{\bf k}(\eta=0)\zeta_{\bf q}(\eta=0)
\rangle^{'}_{|kc_{S}\eta|\approx 1}&=&\frac{\langle \zeta_{\bf k}(\eta=0)\zeta_{\bf q}(\eta=0)
\rangle_{|kc_{S}\eta|\approx 1}}{(2\pi)^3\delta^{3}({\bf k}+{\bf q})}.
\eea
Let us now define a new cosmological observable $\hat{ O}_{obs}$ in the three limiting cases $|kc_{S}\eta|\rightarrow -\infty$, $|kc_{S}\eta|\rightarrow 0$
and $|kc_{S}\eta|\approx 1$ as:
\bea \hat{ O}_{obs}&\stackrel{|kc_{S}\eta|\rightarrow -\infty}{=}&\frac{2\tilde{c}_{S}}{(-k\eta \tilde{c}_{S})^2H}{\cal I}_{1},\\ 
\hat{ O}_{obs}&\stackrel{|kc_{S}\eta|\rightarrow 0}{=}&\frac{2^{3-2\Lambda} c^{3}_{S}\pi
}{(-k\eta c_{S})^{3-2\Lambda} H}
\left|\frac{\Gamma\left(\frac{3}{2}\right)}{\Gamma\left(\Lambda\right)}\right|^2{\cal I}_{2},\\ 
\hat{ O}_{obs}&\stackrel{|kc_{S}\eta|\approx 1}{=}&\frac{2^{3-2\Lambda} c^{3}_{S} \pi }{ H}
\left|\frac{\Gamma\left(\frac{3}{2}\right)}{\Gamma\left(\Lambda\right)}\right|^2{\cal I}_{3},\eea
where ${\cal I}_{1}, {\cal I}_{2}$ and ${\cal I}_{3}$ are defined as:
\be\begin{array}{lll}\label{dfinv4bnbncv}
 \displaystyle{\cal I}_{1} = -\frac{
\int\limits_{-\infty}^{0} {d\eta}
~\frac{\eta}{k}m\left[|C_{2}|^2 
e^{-ikc_{S}\eta}-|C_{1}|^2 e^{ikc_{S}\eta} 
-i\left(C^{*}_{1}C_{2}e^{i\pi\left(\Lambda+\frac{1}{2}\right)}
+C_{1}C^{*}_{2}e^{-i\pi\left(\Lambda+\frac{1}{2}\right)}\right)\sin {\it kc_{S}\eta}\right]}
{\left[|C_{2}|^2+|C_{1}|^2
+\left(C^{*}_{1}C_{2}e^{2ikc_{S}\eta}e^{i\pi\left(\Lambda+\frac{1}{2}\right)}
+C_{1}C^{*}_{2}e^{-2ikc_{S}\eta}e^{-i\pi\left(\Lambda+\frac{1}{2}\right)}\right)\right]}
\\=\left\{\begin{array}{ll}
                    \displaystyle 
\int\limits_{-\infty}^{0} {d\eta}
~\frac{\eta}{k} m~e^{ikc_{S}\eta} &
 \mbox{\small {\bf for ~Bunch Davies}}  \\ 
	\displaystyle  -\frac{
\int\limits_{-\infty}^{0} {d\eta}
~\frac{\eta}{k} m\left[\sinh^2\alpha~ 
e^{ikc_{S}\eta}-\cosh^2\alpha~ e^{-ikc_{S}\eta}
+i~\sinh2\alpha\cos\left({\it \pi\left(\Lambda+\frac{1}{2}\right)}+\delta\right)\sin{\it 
kc_{S}\eta}\right]}{\left[\sinh^2\alpha +\cosh^2\alpha
+\sinh2\alpha\cos{\it \left(2kc_{S}\eta+\pi\left(\Lambda+\frac{1}{2}\right)+\delta\right)}\right]}
                     & \mbox{\small {\bf for~$\alpha$~vacua~Type-I}}\\ 
	\displaystyle  -\frac{
\int\limits_{-\infty}^{0} {d\eta}
~\frac{\eta}{k} m\left[e^{\alpha+\alpha^{*}} 
e^{ikc_{S}\eta}-e^{-ikc_{S}\eta}
+i\left(e^{\alpha}e^{i\pi\left(\Lambda+\frac{1}{2}\right)}
+e^{\alpha^{*}}e^{-i\pi\left(\Lambda+\frac{1}{2}\right)}\right)\sin{\it 
kc_{S}\eta}\right]}{4
\cos^2{\it \left(kc_{S}\eta+\frac{\pi}{2}\left(\Lambda+\frac{1}{2}\right)-i\frac{\alpha}{2}\right)}}
                     & \mbox{\small {\bf for~$\alpha$~vacua~Type-II}}\\ 
	\displaystyle  -\frac{i\pi^2
\int\limits_{-\infty}^{0} {d\eta}
~\frac{\eta}{k} m\sin{\it 
kc_{S}\eta}\cos^2{\it \frac{\pi}{2}\left(\Lambda+
\frac{1}{2}\right)}}{\cos^2{\it \left(kc_{S}\eta+\frac{\pi}{2}\left(\Lambda+\frac{1}{2}\right)\right)}}
                     & \mbox{\small {\bf for~special~vacua}}.
          \end{array}
\right.
\end{array}\ee
\bea\label{dfinv4bnbncv}
 \displaystyle{\cal I}_{2} &=& \sqrt{\xi}\int\limits_{-\infty}^{\xi} {d\eta}
~\sqrt{-\eta}~m~
\left(\Lambda-\frac{1}{2}\right)\left[\left(-\frac{kc_{S}\eta}{2}\right)^{-\Lambda}\left(-\frac{kc_{S}\xi}{2}\right)^{-\Lambda}
+\left(\frac{kc_{S}\eta}{2}\right)^{-\Lambda}\left(\frac{kc_{S}\xi}{2}\right)^{-\Lambda}\right],~~~~~~~~\eea
\bea\label{dfinv4bnbncv}
 \displaystyle{\cal I}_{3} &=& \sqrt{\xi}\int\limits_{-\infty}^{\xi} {d\eta}
~\sqrt{-\eta}~m~
\left(\Lambda-\frac{1}{2}\right)2^{1+\Lambda}\left(\frac{kc_{S}\xi}{2}\right)^{-\Lambda}
\left(-1\right)^{-\Lambda},~~~~~~~~\eea
Next we write down the expression for new cosmological observable $\hat{ O}_{obs}$ 
in terms of all other known cosmological 
\bea\label{fgdf}\hat{ O}_{obs}&\stackrel{|kc_{S}\eta|\approx 1}{=}&\frac{2^{n_{\zeta}-1} c^{3}_{S} \pi }{ H}
\left|\frac{\Gamma\left(\frac{3}{2}\right)}{\Gamma\left(\frac{4-n_{\zeta}}{2}\right)}\right|^2{\cal I}_{3}\left(\Lambda=
\frac{4-n_{\zeta}}{2}\right),\eea
where spectral tilt for scalar fluctuations is given by: 
\be\begin{array}{ll}\label{dfgs} n_{\zeta}-1\equiv\left(\frac{d\ln \Delta_{\zeta}(k)}{d\ln k}\right)_{|kc_{S}\eta|\approx 1}
\approx 3-2\Lambda=\left\{\begin{array}{ll}
                    \displaystyle 3-2\sqrt{\frac{9}{4}-\frac{m^2_{inf}}{H^2}}
 &
 \mbox{\small {\bf for ~de~Sitter}}  \\ 
	\displaystyle  3-2\sqrt{\nu^2-\frac{m^2_{inf}}{H^2}}
                     & \mbox{\small {\bf for~quasi~de~Sitter}}.
          \end{array}
\right.
\end{array}\ee 
One can approximately consider a phenomenological situation where inflaton and the new particles are exactly identical. Technically this means both of them have the comparable mass i.e. $m_{inf}\approx m$.
In that situation spectral tilt for scalar fluctuations is given by: 
\be\begin{array}{ll}\label{dfgs} n_{\zeta}-1\equiv\left(\frac{d\ln \Delta_{\zeta}(k)}{d\ln k}\right)_{|kc_{S}\eta|\approx 1}
\approx \left\{\begin{array}{ll}
                    \displaystyle 3-2\sqrt{\frac{9}{4}-\frac{m^2}{H^2}}
 &
 \mbox{\small {\bf for ~de~Sitter}}  \\ 
	\displaystyle  3-2\sqrt{\nu^2-\frac{m^2}{H^2}}
                     & \mbox{\small {\bf for~quasi~de~Sitter}}.
          \end{array}
\right.
\end{array}\ee 
As the value of scalar spectral tilt $n_{\zeta}$ is known from observation, one can easily give the estimate of the
value of the ratio of the mass parameter $m$ with Hubble scale during inflation $H$ as:
\be\begin{array}{ll}\label{dfgs1} \displaystyle \left|\frac{m_{inf}}{H}\right|=\left\{\begin{array}{ll}
                    \displaystyle \left|\sqrt{\frac{9}{4}-\frac{(4-n_{\zeta})^2}{4}}\right|
 &
 \mbox{\small {\bf for ~de~Sitter}}  \\ 
	\displaystyle   \left|\sqrt{\nu^2-\frac{(4-n_{\zeta})^2}{4}}\right|
                     & \mbox{\small {\bf for~quasi~de~Sitter}}.
          \end{array}
\right.
\end{array}\ee 
In the approximated situation as mentioned earlier one can similarly give an estimate of the
value of the ratio of the mass parameter $m$ with Hubble scale during inflation $H$ as:
\be\begin{array}{ll}\label{esti} \displaystyle \left|\frac{m}{H}\right|=\left\{\begin{array}{ll}
                    \displaystyle \left|\sqrt{\frac{9}{4}-\frac{(4-n_{\zeta})^2}{4}}\right|
 &
 \mbox{\small {\bf for ~de~Sitter}}  \\ 
	\displaystyle   \left|\sqrt{\nu^2-\frac{(4-n_{\zeta})^2}{4}}\right|
                     & \mbox{\small {\bf for~quasi~de~Sitter}}.
          \end{array}
\right.
\end{array}\ee 
Additionally, in this context the integral ${\cal I}_{3}\left(\Lambda=
\frac{4-n_{\zeta}}{2}\right)$ is defined in terms of the scalar spectral tilt as:
\bea\label{dfinv4bnbncv}
 \displaystyle{\cal I}_{3} &=& \sqrt{\xi}\int\limits_{-\infty}^{\xi} {d\eta}
~\sqrt{-\eta}~m~
\left(\frac{3-n_{\zeta}}{2}\right)2^{3-\frac{n_{\zeta}}{2}}\left(\frac{kc_{S}\xi}{2}\right)^{\frac{n_{\zeta}-1}{2}}
\left(-1\right)^{\frac{n_{\zeta}-1}{2}}.~~~~~~~~\eea 
Now our main claim is that, if in near future with sufficient statistical accuracy one can measure the new cosmological parameter 
$\hat{ O}_{obs}$ then in that case using Eq~(\ref{fgdf}) one can further write the expression for the scale of inflation 
in presence of new particle with mass $m$ as:
\bea\label{fgdf2}H&=&\frac{2^{n_{\zeta}-1} c^{3}_{S} \pi}{\hat{ O}_{obs} }
\left|\frac{\Gamma\left(\frac{3}{2}\right)}{\Gamma\left(\frac{4-n_{\zeta}}{2}\right)}\right|^2{\cal I}_{3}\left(\Lambda=
\frac{4-n_{\zeta}}{2}\right),\eea
where all the observables are computed at the horizon crossing $|kc_{S}\eta|\approx 1$ and the temporal 
IR cut-off $\xi$ 
should be fixed at the pivot scale $k_{*}$ where $|kc_{S}\eta|\approx 1$ condition is additionally satisfied. Here the main advantage of this expression is that we don't need any prior 
knowledge or any equivalent information from tensor-to-scalar 
ratio $r$. This implies that even we don't able to detect the signatures of 
primordial gravitational waves through tensor-to-scalar 
ratio $r$ it is possible to quantity exactly the scale of inflationary paradigm and comment on the 
new physics associated with this scenario. Additional advantage of such relation is the computation is 
applicable for any models of inflation as it is based on background EFT framework.

Next we express the scale of inflation in terms of the amplitude of scalar power spectrum and the 
scalar spectral tilt as:
\bea\label{dinf}
\displaystyle H
 &\approx& \sqrt{\epsilon  }
\frac{2^{\frac{n_{\zeta}-1}{2}}\pi^{3/2}\sqrt{\left[\Delta_{\zeta}(k)\right]_{{|kc_{S}\eta|\approx 1}}}}{\sqrt{\left[|C_{2}|^2+|C_{1}|^2-\left(C^{*}_{1}C_{2}
+C_{1}C^{*}_{2}\right)\right]}}\frac{c^{3/2}_{S}}{\tilde{c}_{S}}\left|\frac{\Gamma
\left(\frac{3}{2}\right)}{\Gamma\left(\frac{4-n_{\zeta}}{2}\right)}\right|~M_p\\&=&\left\{\begin{array}{ll}
                     \displaystyle \sqrt{\epsilon  }~~
2^{\frac{n_{\zeta}}{2}}\pi\sqrt{\left[\Delta_{\zeta}(k)\right]_{{|kc_{S}\eta|\approx 1}}}\frac{c^{3/2}_{S}}{\tilde{c}_{S}}\left|\frac{\Gamma
\left(\frac{3}{2}\right)}{\Gamma\left(\frac{4-n_{\zeta}}{2}\right)}\right|~M_p
                     \nonumber &
 \mbox{\small {\bf for ~Bunch Davies}}  \\ 
	\displaystyle \sqrt{\epsilon  }
\frac{2^{\frac{n_{\zeta}-1}{2}}\pi^{3/2}\sqrt{\left[\Delta_{\zeta}(k)\right]_{{|kc_{S}\eta|\approx 1}}}}{\sqrt{\left[\sinh^2\alpha +\cosh^2\alpha
-\sinh2\alpha\cos{\it \delta}\right]}}\frac{c^{3/2}_{S}}{\tilde{c}_{S}}\left|\frac{\Gamma
\left(\frac{3}{2}\right)}{\Gamma\left(\frac{4-n_{\zeta}}{2}\right)}\right|~M_p
                     & \mbox{\small {\bf for~$\alpha$~vacua~Type-I}}\\ 
	\displaystyle \sqrt{\epsilon  }
\frac{2^{\frac{n_{\zeta}-2}{2}}\pi^{3/2}\sqrt{\left[\Delta_{\zeta}(k)\right]_{{|kc_{S}\eta|\approx 1}}}}{|N_{\alpha}|}\frac{c^{3/2}_{S}}{\tilde{c}_{S}}\left|\frac{\Gamma
\left(\frac{3}{2}\right)}{\Gamma\left(\frac{4-n_{\zeta}}{2}\right)}\right|~{\rm cosec}\frac{\alpha}{2}~M_p

                     & \mbox{\small {\bf for~$\alpha$~vacua~Type-II}}\\ 
	\displaystyle \rightarrow \infty

                     & \mbox{\small {\bf for~special~vacua}}.
          \end{array}
\right.
\eea
where we don't need any information from the VEV of the scalar fluctuation.
Before going to the further details also here it is important to 
mention that in Eq~(\ref{fgdf2}) and Eq~(\ref{dinf}) and the following parameters are known from Planck 2015 data set
within $2\sigma$ C.L.:
\begin{enumerate}
 \item Amplitude of the scalar power spectrum:\\
 \be \ln(10^{10}\left[\Delta_{\zeta}(k)\right]_{{|kc_{S}\eta|\approx 1}})=3.089\pm 0.036,\ee
 \item Scalar spectral tilt:\\
 \be n_{\zeta}=0.9569\pm 0.0077,\ee
 \item Sound speed:\\
 \be 0.02< \tilde{c}_{S}<1.\ee
\end{enumerate}
As using both Eq~(\ref{fgdf2}) and Eq~(\ref{dinf}) one can predict the scale of inflation, here one can compare both of them to 
compute the value of first slow roll parameter $\epsilon$ in model independent way in the background of EFT framework. After doing 
this here we get the following expression for slow-roll parameter $\epsilon$ as given by:
\bea\label{dinfx}
\displaystyle \epsilon
 &\approx& \frac{2^{n_{\zeta}-1} c^{3}_{S} \tilde{c}^2_{S}\left[|C_{2}|^2+|C_{1}|^2-\left(C^{*}_{1}C_{2}
+C_{1}C^{*}_{2}\right)\right] }{\left[\Delta_{\zeta}(k)\right]_{{|kc_{S}\eta|\approx 1}}\hat{ O}^2_{obs} M^2_p \pi}
\left|\frac{\Gamma\left(\frac{3}{2}\right)}{\Gamma\left(\frac{4-n_{\zeta}}{2}\right)}\right|^2{\cal I}^2_{3}\left(\Lambda=
\frac{4-n_{\zeta}}{2}\right)\\&=&\left\{\begin{array}{ll}
                     \displaystyle \frac{2^{n_{\zeta}-2} c^{3}_{S} \tilde{c}^2_{S}}{\left[\Delta_{\zeta}(k)\right]_{{|kc_{S}\eta|\approx 1}}\hat{ O}^2_{obs} M^2_p}
\left|\frac{\Gamma\left(\frac{3}{2}\right)}{\Gamma\left(\frac{4-n_{\zeta}}{2}\right)}\right|^2{\cal I}^2_{3}\left(\Lambda=
\frac{4-n_{\zeta}}{2}\right)
                     \nonumber &
 \mbox{\small {\bf for ~Bunch Davies}}  \\ 
	\displaystyle \frac{2^{n_{\zeta}-1} c^{3}_{S} \tilde{c}^2_{S}\left[\sinh^2\alpha +\cosh^2\alpha
-\sinh2\alpha\cos{\it \delta}\right] }{\left[\Delta_{\zeta}(k)\right]_{{|kc_{S}\eta|\approx 1}}\hat{ O}^2_{obs} M^2_p \pi}
\left|\frac{\Gamma\left(\frac{3}{2}\right)}{\Gamma\left(\frac{4-n_{\zeta}}{2}\right)}\right|^2{\cal I}^2_{3}\left(\Lambda=
\frac{4-n_{\zeta}}{2}\right)
                     & \mbox{\small {\bf for~$\alpha$~vacua~Type-I}}\\ 
	\displaystyle \frac{2^{n_{\zeta}+1} c^{3}_{S} \tilde{c}^2_{S}|N_{\alpha}|^2 }{\left[\Delta_{\zeta}(k)\right]_{{|kc_{S}\eta|\approx 1}}\hat{ O}^2_{obs} M^2_p \pi}
\left|\frac{\Gamma\left(\frac{3}{2}\right)}{\Gamma\left(\frac{4-n_{\zeta}}{2}\right)}\right|^2{\cal I}^2_{3}\left(\Lambda=
\frac{4-n_{\zeta}}{2}\right)\sin^2\frac{\alpha}{2}
& \mbox{\small {\bf for~$\alpha$~vacua~Type-II}}\\ 
	\displaystyle \rightarrow 0
& \mbox{\small {\bf for~special~vacua}}.
          \end{array}
\right.
\eea
On the other hand within the framework of EFT using the slow-roll parameter $\epsilon$ 
one can write down a consistency relation for tensor-to-scalar ratio $r$ 
in presence of sound speed $\tilde{c}_{S}$ as:
\be\label{dinf3} r=
\frac{\left[\Delta_{h}(k)\right]_{{|kc_{S}\eta|\approx 1}}}{\left[\Delta_{\zeta}(k)\right]_{{|kc_{S}\eta|\approx 1}}}
=16 \epsilon \tilde{c}_{S},\ee
where $\left[\Delta_{h}(k)\right]_{{|kc_{S}\eta|\approx 1}}$ 
and $\left[\Delta_{\zeta}(k)\right]_{{|kc_{S}\eta|\approx 1}}$ signify the amplitude of tensor and scalar fluctuations. 
Further using Eq~(\ref{dinf3}) in Eq~(\ref{dinfx}) we get the following expression for 
the tensor-to-scalar ratio and the amplitude of tensor fluctuation in a model independent fashion as:
\bea\label{dinfx}
\displaystyle r
 &\approx& \frac{2^{n_{\zeta}+3} c^{3}_{S} \tilde{c}^3_{S}\left[|C_{2}|^2+|C_{1}|^2-\left(C^{*}_{1}C_{2}
+C_{1}C^{*}_{2}\right)\right] }{\left[\Delta_{\zeta}(k)\right]_{{|kc_{S}\eta|\approx 1}}\hat{ O}^2_{obs} M^2_p \pi}
\left|\frac{\Gamma\left(\frac{3}{2}\right)}{\Gamma\left(\frac{4-n_{\zeta}}{2}\right)}\right|^2{\cal I}^2_{3}\left(\Lambda=
\frac{4-n_{\zeta}}{2}\right)\\&=&\left\{\begin{array}{ll}
                     \displaystyle \frac{2^{n_{\zeta}+2} c^{3}_{S} \tilde{c}^3_{S}}{\left[\Delta_{\zeta}(k)\right]_{{|kc_{S}\eta|\approx 1}}\hat{ O}^2_{obs} M^2_p}
\left|\frac{\Gamma\left(\frac{3}{2}\right)}{\Gamma\left(\frac{4-n_{\zeta}}{2}\right)}\right|^2{\cal I}^2_{3}\left(\Lambda=
\frac{4-n_{\zeta}}{2}\right)
                     \nonumber &
 \mbox{\small {\bf for ~Bunch Davies}}  \\ 
	\displaystyle \frac{2^{n_{\zeta}+3} c^{3}_{S} \tilde{c}^3_{S}\left[\sinh^2\alpha +\cosh^2\alpha
-\sinh2\alpha\cos{\it \delta}\right] }{\left[\Delta_{\zeta}(k)\right]_{{|kc_{S}\eta|\approx 1}}\hat{ O}^2_{obs} M^2_p \pi}
\left|\frac{\Gamma\left(\frac{3}{2}\right)}{\Gamma\left(\frac{4-n_{\zeta}}{2}\right)}\right|^2{\cal I}^2_{3}\left(\Lambda=
\frac{4-n_{\zeta}}{2}\right)
                     & \mbox{\small {\bf for~$\alpha$~vacua~Type-I}}\\ 
	\displaystyle \frac{2^{n_{\zeta}+5} c^{3}_{S} \tilde{c}^3_{S}|N_{\alpha}|^2 }{\left[\Delta_{\zeta}(k)\right]_{{|kc_{S}\eta|\approx 1}}\hat{ O}^2_{obs} M^2_p \pi}
\left|\frac{\Gamma\left(\frac{3}{2}\right)}{\Gamma\left(\frac{4-n_{\zeta}}{2}\right)}\right|^2{\cal I}^2_{3}\left(\Lambda=
\frac{4-n_{\zeta}}{2}\right)\sin^2\frac{\alpha}{2}
& \mbox{\small {\bf for~$\alpha$~vacua~Type-II}}\\ 
	\displaystyle \rightarrow 0
& \mbox{\small {\bf for~special~vacua}}.
          \end{array}
\right.
\eea
and 
\bea\label{dinfx}
\displaystyle \left[\Delta_{h}(k)\right]_{{|kc_{S}\eta|\approx 1}}
 &\approx& \frac{2^{n_{\zeta}+3} c^{3}_{S} \tilde{c}^3_{S}\left[|C_{2}|^2+|C_{1}|^2-\left(C^{*}_{1}C_{2}
+C_{1}C^{*}_{2}\right)\right] }{\hat{ O}^2_{obs} M^2_p \pi}\nonumber\\ &&~~~~~~~~~~~~~~~~~~~~~~~~~~\times
\left|\frac{\Gamma\left(\frac{3}{2}\right)}{\Gamma\left(\frac{4-n_{\zeta}}{2}\right)}\right|^2{\cal I}^2_{3}\left(\Lambda=
\frac{4-n_{\zeta}}{2}\right)\\&=&\left\{\begin{array}{ll}
                     \displaystyle \frac{2^{n_{\zeta}+2} c^{3}_{S} \tilde{c}^3_{S}}{\hat{ O}^2_{obs} M^2_p}
\left|\frac{\Gamma\left(\frac{3}{2}\right)}{\Gamma\left(\frac{4-n_{\zeta}}{2}\right)}\right|^2{\cal I}^2_{3}\left(\Lambda=
\frac{4-n_{\zeta}}{2}\right)
                     \nonumber &
 \mbox{\small {\bf for ~Bunch Davies}}  \\ 
	\displaystyle \frac{2^{n_{\zeta}+3} c^{3}_{S} \tilde{c}^3_{S}\left[\sinh^2\alpha +\cosh^2\alpha
-\sinh2\alpha\cos{\it \delta}\right] }{\hat{ O}^2_{obs} M^2_p \pi}
\nonumber\\ \displaystyle~~~~~~~~~~~~~~~~~~~~~~~~~~\times\left|\frac{\Gamma\left(\frac{3}{2}\right)}{\Gamma\left(\frac{4-n_{\zeta}}{2}\right)}\right|^2{\cal I}^2_{3}\left(\Lambda=
\frac{4-n_{\zeta}}{2}\right)
                     & \mbox{\small {\bf for~$\alpha$~vacua~Type-I}}\\ 
	\displaystyle \frac{2^{n_{\zeta}+5} c^{3}_{S} \tilde{c}^3_{S}\pi^5|N_{\alpha}|^2 }{\hat{ O}^2_{obs} M^2_p \pi}
\left|\frac{\Gamma\left(\frac{3}{2}\right)}{\Gamma\left(\frac{4-n_{\zeta}}{2}\right)}\right|^2{\cal I}^2_{3}\left(\Lambda=
\frac{4-n_{\zeta}}{2}\right)\sin^2\frac{\alpha}{2}
& \mbox{\small {\bf for~$\alpha$~vacua~Type-II}}\\ 
	\displaystyle \rightarrow 0
& \mbox{\small {\bf for~special~vacua}}.
          \end{array}
\right.
\eea
Here in summary from the derived results we get the following information:
\begin{itemize}
 \item Scale of inflation and the associated new physics can be predicted without the prior knowledge of primordial 
 gravity waves in a model independent way. Here we only need to measure the mass of the particles participating in evolution of universe and for
 serve this purpose we need to measure the value of the VEV of the scalar fluctuations or the one point function or more precisely a new cosmological observable 
 depicted by $\hat{O}_{obs}$ as introduced in the context of EFT. Additionally, it is important to mention here that the final analytical expression for the scale of inflation 
 is independent of the choice of any initial condition during inflation. 
 \item Here one can give an estimate of scaled mass parameter $m_{inf}/H$ and $m/H$ in terms of scalar spectral tilt $n_{\zeta}$ for 
 de-Sitter and quasi de-Sitter case.
 \item One can also compute the expression for the scale of inflation in therms of known inflationary observables 
 for a proper choice of initial condition. 
 \item Further if we demand that the scale of inflation computed from both the techniques should give 
 unique result then by comparing both of the expressions we derive the analytical model independent expression for the first 
 slow-roll parameter $\epsilon$ within the framework of EFT.
 \item Further using the consistency relations valid in the context of inflation one can further derive the expression 
 for both tensor-to-scalar ratio $r$ and the amplitude of the tensor fluctuations in a model independent way.
 \item In case of single field slow-roll models of inflation the amount of non-Gaussianity is proportional to the 
 first slow-roll parameter $\epsilon$ or more precisely with the primordial gravity waves through tensor-to-scalar
 ratio $r$. So using the prescribed methodology mentioned in this paper one can easily derive
 the model independent expression for the non-Gaussian amplitude in terms of the time dependent mass parameter within 
 the framework of EFT. We will report soon on these aspects.
 \item It is important to mention here that if we use the constraint on scalar spectral tilt as obtained from Planck 2015 data we get the following $2\sigma$ bound on the magnitude of the 
 mass parameter of the inflaton field:
 \bea
 &&\underline{\bf For ~de~Sitter:}\nonumber\\
 && 0.23<\left|\frac{m_{inf}}{H}\right|_{kc_{S}\eta\approx 1}<0.28,\\
 &&\underline{\bf For~quasi~de~Sitter:}\nonumber\\
 && 0.23\times\left|\sqrt{1-56.18\left(\epsilon+\frac{\eta}{2}+\frac{s}{2}\right)}\right|<
 \left|\frac{m_{inf}}{H}\right|_{kc_{S}\eta\approx 1}<0.28\times\left|\sqrt{1-39.06\left(\epsilon+\frac{\eta}{2}+\frac{s}{2}\right)}\right|.~~~~\eea 
 and for the approximated situation where $m_{inf}\approx m$ we get the following $2\sigma$ bound on the magnitude of the 
 mass parameter of the
 new heavy particles:
 \bea
 &&\underline{\bf For ~de~Sitter:}\nonumber\\
 && 0.23<\left|\frac{m}{H}\right|_{kc_{S}\eta\approx 1}<0.28,\\
 &&\underline{\bf For~quasi~de~Sitter:}\nonumber\\
 && 0.23\times\left|\sqrt{1-56.18\left(\epsilon+\frac{\eta}{2}+\frac{s}{2}\right)}\right|<
 \left|\frac{m}{H}\right|_{kc_{S}\eta\approx 1}<0.28\times\left|\sqrt{1-39.06\left(\epsilon+\frac{\eta}{2}+\frac{s}{2}\right)}\right|.~~~~\eea 
 Importance of the obtained bound on the mass parameter of the new heavy particles can be justified in a following manner:
 \begin{enumerate}
  \item If the contribution from the inflaton field mass term or in the special case if the heavy massive field is very negligible then in that situation for de-Sitter case we get the feature of exact scale invariance of the primordial power spectrum.
  But as the various observational probes confirms the fact that primordial power spectrum for scalar fluctuations are not exactly scale invariant, it is required to use quasi de Sitter approximation
  in the present context. As in the special case mass of the heavy field is negligibly small, this implies that the amount of Bell violation is also negligibly small.
  \item From the above mentioned bound it is clear that to get nearly scale invariant primordial power spectrum non-negligible contribution in the inflaton field mass term and for the special case 
  heavy field mass is necessarily required in de Sitter and quasi de Sitter 
  both the cases. Most importantly in de Sitter case also we get the nearly scale invariant feature in the primordial power spectrum in this case. The non-negligible mass contribution from the obtained bound 
  also implies that the amount of Bell violation in the primordial universe is not negligibly small. 
 \end{enumerate}

 \item On the other hand if in near future any observational probe can measure the value of tensor-to-scalar ratio and primordial non-Gaussianity then also there is a possibility to give
 an estimate of the newly introduced massive particle and 
 the new cosmological observable 
 depicted by $\hat{O}_{obs}$. This will surely quantify the amount of Bell violation in early universe cosmology.
\end{itemize}

\subsubsection{{\bf Case I:} $m\approx H$}
 For further simplification we consider here $m\approx H$ along with two limiting cases
 $|kc_{S}\eta|\rightarrow -\infty$, $|kc_{S}\eta|\rightarrow 0$ and $|kc_{S}\eta|\approx 1$
 which are physically acceptable in the present context. 
 \bea\label{dfinv4bnbncv}
\displaystyle \langle\zeta_{\bf{k}}(\eta=0)\rangle_{|kc_{S}\eta|\rightarrow -\infty} &\approx& \tiny\tiny
-\frac{16\pi^2H\tilde{c}_{S}}{M^2_{p}\epsilon c_{S}}
\left[|C_{2}|^2 I_{1}
-|C_{1}|^2 I_{2}-i\left(C^{*}_{1}C_{2}e^{i\pi\left(\Lambda+\frac{1}{2}\right)}
+C_{1}C^{*}_{2}e^{-i\pi\left(\Lambda+\frac{1}{2}\right)}\right)I_{3}\right]~~~~~~~~~~~~\\&=&\left\{\begin{array}{ll}
                      \frac{8\pi^3H\tilde{c}_{S}}{M^2_{p}\epsilon c_{S}}
I_{2}\nonumber &
 \mbox{\small {\bf for ~Bunch Davies vacua}}  \\ 
	  -\frac{16\pi^2H\tilde{c}_{S}}{M^2_{p}\epsilon c_{S}}
\left[\sinh^2\alpha~ 
I_{1}-\cosh^2\alpha~I_{2}\right.\nonumber \\ \left. 
\displaystyle~~~~~~~~~~~
-i~\sinh2\alpha\cos\left({\it \pi\left(\Lambda+\frac{1}{2}\right)}+\delta\right)I_{3}\right]
                     & \mbox{\small {\bf for~$\alpha$~vacua~Type-I}}\\ 
	  -\frac{16\pi^2|N_{\alpha}|^2H\tilde{c}_{S}}{M^2_{p}\epsilon c_{S}}
\left[e^{\alpha+\alpha^{*}} 
I_{1}-I_{2}\right.\nonumber \\ \left. 
\displaystyle~~~~~~~~~~~
-i\left(e^{\alpha}e^{i\pi\left(\Lambda+\frac{1}{2}\right)}
+e^{\alpha^{*}}e^{-i\pi\left(\Lambda+\frac{1}{2}\right)}\right)I_{3}\right]
                     & \mbox{\small {\bf for~$\alpha$~vacua~Type-II}}\\ 
	 - \frac{64i\pi^2|C|^2H\tilde{c}_{S}}{M^2_{p}\epsilon c_{S}}
\cos^2{\it \frac{\pi}{2}\left(\Lambda+\frac{1}{2}\right)}I_{3}
                    & \mbox{\small {\bf for~special~vacua}}.
          \end{array}
\right.
\eea
\bea\label{dfinv4bnbnxcxcx}
\small\displaystyle \langle\zeta_{\bf{k}}(\eta=\xi\rightarrow 0)\rangle_{|kc_{S}\eta|\rightarrow 0} &=& 
\frac{4\pi\sqrt{\xi}H\tilde{c}_{S}}{M^2_{p}\epsilon}\left(C^{*}_{1}C_{2}
+C_{1}C^{*}_{2}-|C_{1}|^2-|C_{2}|^2\right)
I^{\xi\theta_{1}}_{4}
\\&=&I^{\xi\theta_{1}}_{4}\times\left\{\begin{array}{ll}
                    \displaystyle  -\frac{2\pi^2\sqrt{\xi}H\tilde{c}_{S}}{M^2_{p}\epsilon}~~~~ &
 \mbox{\small {\bf for ~Bunch Davies vacua}}  \\ 
	\displaystyle  \frac{4\pi\sqrt{\xi}H\tilde{c}_{S}}{M^2_{p}\epsilon}\left(\cos\delta\sinh 2\alpha-\cosh^2\alpha-\sinh^2\alpha\right)
                    ~~~~ & \mbox{\small {\bf for~$\alpha$~vacua~Type-I}}\\ 
	\displaystyle  \frac{4\pi\sqrt{\xi}|N_{\alpha}|^2H\tilde{c}_{S}}{M^2_{p}\epsilon}\left(e^{\alpha}+e^{\alpha^{*}}
-e^{\alpha+\alpha^{*}}-1\right)
\nonumber
                    ~~~~ & \mbox{\small {\bf for~$\alpha$~vacua~Type-II}}\\ 
	\displaystyle  0
~~~ & \mbox{\small {\bf for~special~vacua}}.
          \end{array}
\right.
\eea
\bea\label{dfinv4bnbnxcxcx}
\small\displaystyle \langle\zeta_{\bf{k}}(\eta=\xi\rightarrow 0)\rangle_{|kc_{S}\eta|\approx 1} &=& 
\frac{4\pi\sqrt{\xi}H\tilde{c}_{S}}{M^2_{p}\epsilon}\left(C^{*}_{1}C_{2}
+C_{1}C^{*}_{2}-|C_{1}|^2-|C_{2}|^2\right)
I^{\xi\theta_{2}}_{5}
\\&=&I^{\xi\theta_{2}}_{5}\times\left\{\begin{array}{ll}
                    \displaystyle  -\frac{2\pi^2\sqrt{\xi}H\tilde{c}_{S}}{M^2_{p}\epsilon}~~~~ &
 \mbox{\small {\bf for ~Bunch Davies vacua}}  \\ 
	\displaystyle  \frac{4\pi\sqrt{\xi}H\tilde{c}_{S}}{M^2_{p}\epsilon}\left(\cos\delta\sinh 2\alpha-\cosh^2\alpha-\sinh^2\alpha\right)
                    ~~~~ & \mbox{\small {\bf for~$\alpha$~vacua~Type-I}}\\ 
	\displaystyle  \frac{4\pi\sqrt{\xi}|N_{\alpha}|^2H\tilde{c}_{S}}{M^2_{p}\epsilon}\left(e^{\alpha}+e^{\alpha^{*}}
-e^{\alpha+\alpha^{*}}-1\right)
\nonumber
                    ~~~~ & \mbox{\small {\bf for~$\alpha$~vacua~Type-II}}\\ 
	\displaystyle  0
~~~ & \mbox{\small {\bf for~special~vacua}}.
          \end{array}
\right.
\eea
where we introduce the following integrals $I_{1}, I_{2}, I_{3}, I^{\xi\theta_{1}}_{4}, I^{\xi\theta_{2}}_{4}$ as given by:
\bea I_{1}&=& \frac{1}{(2\pi)^3}\int\limits_{-\infty}^{0} {d\eta}~
~\frac{\eta}{k}~H~
e^{-ikc_{S}\eta},\\ 
 I_{2}&=& \frac{1}{(2\pi)^3}\int\limits_{-\infty}^{0} {d\eta}
~\frac{\eta}{k}~H~e^{ikc_{S}\eta},\\ 
 I_{3}&=& \frac{1}{(2\pi)^3}\int\limits_{-\infty}^{0} {d\eta}
~\frac{\eta}{k}~H~\sin {\it kc_{S}\eta},\\
 I^{\xi\theta_{1}}_{4}&=&-i\frac{H}{(2\pi)^3}\frac{\left(\Lambda-\frac{1}{2}\right)}{\left(\Lambda-\frac{3}{2}\right)}\left[1+(-1)^{-2\Lambda}\right]
 \left(\frac{kc_{S}}{2}\right)^{-2\Lambda}\xi^{-\Lambda}\left[\xi^{\frac{3}{2}-\Lambda}+(-1)^{\frac{1}{2}-\Lambda}\theta^{\frac{3}{2}-\Lambda}_{1}\right],
 ~~~~~~~~~~~~\\
 I^{\xi\theta_{2}}_{5}&=&-i\frac{H}{(2\pi)^3}\frac{\left(\Lambda-\frac{1}{2}\right)}{\left(\Lambda-\frac{3}{2}\right)}\left[1+(-1)^{-2\Lambda}\right]
 \left(-\frac{1}{2\theta_{2}}\right)^{-2\Lambda}\xi^{-\Lambda}\left[\xi^{\frac{3}{2}-\Lambda}+(-1)^{\frac{1}{2}-\Lambda}\theta^{\frac{3}{2}-\Lambda}_{2}\right]~~~~~~~~~~~~\eea
and here we introduce a UV cut-off regulator $\theta_{1}$ and $\theta_{2}$ in $I^{\xi\theta_{1}}_{4}$ and $I^{\xi\theta_{2}}_{4}$ respectively. Additionally, it is important to note that 
here the parameter $\Lambda$ is given by~\footnote{For the special case where inflaton mass are comparable with the new particle then for $m_{inf}\approx m\approx H$ case we have:\be\begin{array}{lll}\label{y5}\small
 \displaystyle \Lambda\approx\left\{\begin{array}{ll}
                    \displaystyle \frac{\sqrt{5}}{2}~~~~ &
 \mbox{\small {\bf for ~dS}}  \\ 
	\displaystyle \sqrt{\nu^2-1} ~~~~ & \mbox{\small {\bf for~ qdS}}.
          \end{array}
\right.
\end{array}\ee}:
 \be\begin{array}{lll}\label{y6}\small
     \underline{{\bf For}~ m_{inf}<<H:}~~~~~~~~~~~~~~
     \displaystyle \Lambda\approx\left\{\begin{array}{ll}
                    \displaystyle \frac{3}{2}~~~~ &
 \mbox{\small {\bf for ~dS}}  \\ 
	\displaystyle \nu ~~~~ & \mbox{\small {\bf for~ qdS}}.
          \end{array}
\right.
\end{array}\ee
\be\begin{array}{lll}\label{y7}\small
     \underline{{\bf For}~ m_{inf}\approx H:}~~~~~~~~~~~~~~
     \displaystyle \Lambda\approx\left\{\begin{array}{ll}
                    \displaystyle \frac{\sqrt{5}}{2}~~~~ &
 \mbox{\small {\bf for ~dS}}  \\ 
	\displaystyle \sqrt{\nu^2-1} ~~~~ & \mbox{\small {\bf for~ qdS}}.
          \end{array}
\right.
\end{array}\ee
Similarly in the position space the representative expressions for the expectation value of the 
scalar curvature perturbation for $m\approx H$ along with three limiting cases $|kc_{S}\eta|\rightarrow -\infty$,
$|kc_{S}\eta|\rightarrow 0$ and $|kc_{S}\eta|\approx 1$
are given by:
\be\begin{array}{lll}\label{dfinv4bnbncv}
\tiny \langle\zeta({\bf x},\eta=0)\rangle_{|kc_{S}\eta|\rightarrow -\infty} \approx 
-\frac{2H\tilde{c}_{S}}{M^2_{p}\epsilon\pi c_{S}}
\left[|C_{2}|^2 B_{1}
-|C_{1}|^2 B_{2}-i\left(C^{*}_{1}C_{2}e^{i\pi\left(\Lambda+\frac{1}{2}\right)}
+C_{1}C^{*}_{2}e^{-i\pi\left(\Lambda+\frac{1}{2}\right)}\right)B_{3}\right]\\~~~~~~~~~~~~~~~=\left\{\begin{array}{ll}
                      \frac{H\tilde{c}_{S}}{M^2_{p}\epsilon c_{S}}
B_{2}\nonumber &
 \mbox{\small {\bf for ~Bunch Davies}}  \\ 
	  -\frac{2H\tilde{c}_{S}}{M^2_{p}\epsilon\pi c_{S}}
\left[\sinh^2\alpha~ 
B_{1}-\cosh^2\alpha~B_{2}
-i~\sinh2\alpha\cos\left({\it \pi\left(\Lambda+\frac{1}{2}\right)}+\delta\right)B_{3}\right]
                     & \mbox{\small {\bf for~$\alpha$~vacua~Type-I}}\\ 
	  -\frac{2|N_{\alpha}|^2H\tilde{c}_{S}}{M^2_{p}\epsilon\pi c_{S}}
\left[e^{\alpha+\alpha^{*}} 
B_{1}-B_{2}
-i\left(e^{\alpha}e^{i\pi\left(\Lambda+\frac{1}{2}\right)}
+e^{\alpha^{*}}e^{-i\pi\left(\Lambda+\frac{1}{2}\right)}\right)B_{3}\right]
                     & \mbox{\small {\bf for~$\alpha$~vacua~Type-II}}\\ 
	 - \frac{8i|C|^2H\tilde{c}_{S}}{M^2_{p}\epsilon\pi c_{S}}
\cos^2{\it \frac{\pi}{2}\left(\Lambda+\frac{1}{2}\right)}B_{3}
                    & \mbox{\small {\bf for~special~vacua}}.
          \end{array}
\right.
\end{array}\ee
\bea\label{dfinv4bnbnxcxcx}
\small\displaystyle \langle\zeta({\bf x},\eta=\xi\rightarrow 0)\rangle_{|kc_{S}\eta|\rightarrow 0} &=& 
\frac{\sqrt{\xi}H\tilde{c}_{S}}{2M^2_{p}\epsilon\pi^2}\left(C^{*}_{1}C_{2}
+C_{1}C^{*}_{2}-|C_{1}|^2-|C_{2}|^2\right)
B^{\xi\theta_{1}}_{4}
\\&=&B^{\xi\theta_{1}}_{4}\times\left\{\begin{array}{ll}
                    \displaystyle  -\frac{2\pi^2\sqrt{\xi}H\tilde{c}_{S}}{M^2_{p}\epsilon}~~~~ &
 \mbox{\small {\bf for ~Bunch Davies vacua}}  \\ 
	\displaystyle  \frac{\sqrt{\xi}H\tilde{c}_{S}}{2M^2_{p}\pi^2\epsilon}\left(\cos\delta\sinh 2\alpha-\cosh^2\alpha-\sinh^2\alpha\right)
                    ~~~~ & \mbox{\small {\bf for~$\alpha$~vacua~Type-I}}\\ 
	\displaystyle  \frac{\sqrt{\xi}|N_{\alpha}|^2H\tilde{c}_{S}}{2M^2_{p}\pi^2\epsilon}\left(e^{\alpha}+e^{\alpha^{*}}
-e^{\alpha+\alpha^{*}}-1\right)
\nonumber
                    ~~~~ & \mbox{\small {\bf for~$\alpha$~vacua~Type-II}}\\ 
	\displaystyle  0
~~~ & \mbox{\small {\bf for~special~vacua}}.
          \end{array}
\right.
\eea
\bea\label{dfinv4bnbnxcxcx}
\small\displaystyle \langle\zeta({\bf x},\eta=\xi\rightarrow 0)\rangle_{|kc_{S}\eta|\approx 1} &=& 
\frac{\sqrt{\xi}H\tilde{c}_{S}}{2M^2_{p}\epsilon\pi^2}\left(C^{*}_{1}C_{2}
+C_{1}C^{*}_{2}-|C_{1}|^2-|C_{2}|^2\right)
B^{\xi\theta_{2}}_{5}
\\&=&B^{\xi\theta_{2}}_{5}\times\left\{\begin{array}{ll}
                    \displaystyle  -\frac{2\pi^2\sqrt{\xi}H\tilde{c}_{S}}{M^2_{p}\epsilon}~~~~ &
 \mbox{\small {\bf for ~Bunch Davies vacua}}  \\ 
	\displaystyle  \frac{\sqrt{\xi}H\tilde{c}_{S}}{2M^2_{p}\pi^2\epsilon}\left(\cos\delta\sinh 2\alpha-\cosh^2\alpha-\sinh^2\alpha\right)
                    ~~~~ & \mbox{\small {\bf for~$\alpha$~vacua~Type-I}}\\ 
	\displaystyle  \frac{\sqrt{\xi}|N_{\alpha}|^2H\tilde{c}_{S}}{2M^2_{p}\pi^2\epsilon}\left(e^{\alpha}+e^{\alpha^{*}}
-e^{\alpha+\alpha^{*}}-1\right)
\nonumber
                    ~~~~ & \mbox{\small {\bf for~$\alpha$~vacua~Type-II}}\\ 
	\displaystyle  0
~~~ & \mbox{\small {\bf for~special~vacua}}.
          \end{array}
\right.
\eea
where we introduce the following integrals $B_{1}, B_{2}, B_{3}, B^{\xi\theta}_{4}, B^{\xi\theta}_{5}$ as given by:
\bea B_{1}&=& \frac{1}{(2\pi)^3}\int d^{3}k~e^{i{\bf k}.{\bf x}}\int\limits_{-\infty}^{0} {d\eta}~
~\frac{\eta}{k}~H~
e^{-ikc_{S}\eta},\\ 
 B_{2}&=&  \frac{1}{(2\pi)^3}\int d^{3}k~e^{i{\bf k}.{\bf x}}\int\limits_{-\infty}^{0} {d\eta}
~\frac{\eta}{k}~H~e^{ikc_{S}\eta},\\ 
 B_{3}&=& \frac{1}{(2\pi)^3}\int d^{3}k~e^{i{\bf k}.{\bf x}}\int\limits_{-\infty}^{0} {d\eta}
~\frac{\eta}{k}~H~\sin {\it kc_{S}\eta},\\
 B^{\xi\theta_{1}}_{4}&=&-i\frac{H}{(2\pi)^3}\frac{\left(\Lambda-\frac{1}{2}\right)}{\left(\Lambda-\frac{3}{2}\right)}
 \left[1+(-1)^{-2\Lambda}\right]
 \left(\frac{c_{S}}{2}\right)^{-2\Lambda}\xi^{-\Lambda}\left[\xi^{\frac{3}{2}-\Lambda}+(-1)^{\frac{1}{2}-\Lambda}
 \theta^{\frac{3}{2}-\Lambda}_{1}\right]\int d^{3}k~e^{i{\bf k}.{\bf x}}k^{-2\Lambda},~~~~~~~~~~~~\\
 B^{\xi\theta_{2}}_{5}&=&-i\frac{H}{(2\pi)^3}\frac{\left(\Lambda-\frac{1}{2}\right)}{\left(\Lambda-\frac{3}{2}\right)}\left[1+(-1)^{-2\Lambda}\right]
 \left(-\frac{1}{2\theta_{2}}\right)^{-2\Lambda}\xi^{-\Lambda}\left[\xi^{\frac{3}{2}-\Lambda}+(-1)^{\frac{1}{2}-\Lambda}
 \theta^{\frac{3}{2}-\Lambda}_{2}\right]\int d^{3}k~e^{i{\bf k}.{\bf x}}~~~~~~~~~~~~\eea
 Now using the previously mentioned assumptions to compute the momentum integral 
 for Case I and Case II we get the following results:
\bea {\bf For~Case~I:}\nonumber\\
B_{1}&=& \frac{1}{2\pi^2}\int^{\infty}_{0} dk~\int\limits_{-\infty}^{0} {d\eta}~\eta k~H~e^{ik\left(x-c_{S}\eta\right)}
=-\frac{iH\left(\eta=-|{\bf x}|/c_{S}\right)}{2\pi^2c^2_{S}},\\ 
 B_{2}&=& \frac{1}{2\pi^2}\int^{\infty}_{0} dk~\int\limits_{-\infty}^{0} {d\eta}~\eta k~H~e^{ik\left(x+c_{S}\eta\right)}
=\frac{iH\left(\eta=-|{\bf x}|/c_{S}\right)}{2\pi^2c^2_{S}},\\ 
 B_{3}&=& \frac{1}{2\pi^2}\int^{\infty}_{0} dk\int\limits_{-\infty}^{0} {d\eta}
~\eta k~H~e^{ikx}\sin {\it kc_{S}\eta}=\frac{H\left(\eta=-|{\bf x}|/c_{S}\right) }{2\pi^2c^2_{S}},\\
 B^{\xi\theta_{1}}_{4}&=&-i\frac{H\Gamma(3-2\Lambda)}{(2\pi)^3}\frac{\left(\Lambda-\frac{1}{2}\right)}{\left(\Lambda-\frac{3}{2}\right)}\left[1+(-1)^{-2\Lambda}\right]
 \left(\frac{c_{S}}{2}\right)^{-2\Lambda}\xi^{-\Lambda}\left[\xi^{\frac{3}{2}-\Lambda}+(-1)^{\frac{1}{2}-\Lambda}
 \theta^{\frac{3}{2}-\Lambda}\right](-ix)^{2\Lambda-3},~~~~~~~~~~~~\\
 B^{\xi\theta_{2}}_{5}&=&-2\frac{H}{(2\pi)^3}\frac{\left(\Lambda-\frac{1}{2}\right)}{\left(\Lambda-\frac{3}{2}\right)}\left[1+(-1)^{-2\Lambda}\right]
 \left(-\frac{1}{2\theta_{2}}\right)^{-2\Lambda}\xi^{-\Lambda}\left[\xi^{\frac{3}{2}-\Lambda}+(-1)^{\frac{1}{2}-\Lambda}
 \theta^{\frac{3}{2}-\Lambda}\right]x^{-3}.\eea
 \bea {\bf For~Case~II:}\nonumber\\
 B_{1}&=& \frac{1}{4\pi^2}\int^{\infty}_{0} dk~\int\limits_{-\infty}^{0} {d\eta}\int^{\Theta_{2}}_{\Theta_{1}}d\Theta~\eta k~H
 ~e^{ik\left(x\cos\Theta+c_{S}\eta\right)}
=-\int^{\Theta_{2}}_{\Theta_{1}}d\Theta\frac{iH(\eta=-\frac{|{\bf x}|\cos\Theta}{c_{S}})}{4\pi^2c^2_{S}}\cos\Theta,~~~~~~~~~~~~~\\ 
 B_{2}&=& \frac{1}{4\pi^2}\int^{\infty}_{0} dk~\int\limits_{-\infty}^{0} {d\eta}\int^{\Theta_{2}}_{\Theta_{1}}d\Theta
 ~\eta k~e^{ik\left(x\cos\Theta+c_{S}\eta\right)}
=\int^{\Theta_{2}}_{\Theta_{1}}d\Theta\frac{iH(\eta=-\frac{|{\bf x}|\cos\Theta}{c_{S}})}{4\pi^2c^2_{S}}\cos\Theta,\\ 
 B_{3}&=& \frac{1}{4\pi^2}\int^{\infty}_{0} dk\int\limits_{-\infty}^{0} {d\eta}\int^{\Theta_{2}}_{\Theta_{1}}d\Theta
~\eta k~e^{ikx\cos\Theta}\sin {\it kc_{S}\eta}
=\int^{\Theta_{2}}_{\Theta_{1}}d\Theta\frac{H(\eta=-\frac{|{\bf x}|\cos\Theta}{c_{S}})}{4\pi^2c^2_{S}}\cos\Theta,\\
 B^{\xi\theta_{1}}_{4}&=&-i\frac{H}{(2\pi)^3}\frac{\left(\Lambda-\frac{1}{2}\right)}{\left(\Lambda-\frac{3}{2}\right)}\left[1+(-1)^{-2\Lambda}\right]
 \left(\frac{c_{S}}{2}\right)^{-2\Lambda}\xi^{-\Lambda}\left[\xi^{\frac{3}{2}-\Lambda}+(-1)^{\frac{1}{2}-\Lambda}
 \theta^{\frac{3}{2}-\Lambda}_{1}\right]Q_{1},~~~~~~~~~~~~\\
 B^{\xi\theta_{2}}_{5}&=&-i\frac{H}{(2\pi)^3}\frac{\left(\Lambda-\frac{1}{2}\right)}{\left(\Lambda-\frac{3}{2}\right)}\left[1+(-1)^{-2\Lambda}\right]
 \left(-\frac{1}{2\theta_{2}}\right)^{-2\Lambda}\xi^{-\Lambda}\left[\xi^{\frac{3}{2}-\Lambda}+(-1)^{\frac{1}{2}-\Lambda}
 \theta^{\frac{3}{2}-\Lambda}_{2}\right]Q_{2}.~~~~~~~~~~~~\eea
 where $\Theta_{1}$ and $\Theta_{2}$ plays the role of angular regulator in the present context and $Q_{1}$ and $Q_{2}$ 
 are defined as:
 \bea Q_{1}&=&\int^{\pi}_{0}\int^{\infty}_{0}d\Theta~dk~e^{ikx\cos\Theta}~k^{2-2\Lambda}\nonumber\\
 &=& \nonumber\frac{i \sqrt{\pi } (-i x)^{2 \Lambda} \sec (2 \pi  \Lambda) \Gamma (3-2 \Lambda) }{8 x^3 \Gamma^2 (2-\Lambda)
 \Gamma (2 \Lambda-2)
 \Gamma \left(2 \Lambda-\frac{3}{2}\right)}\left[\pi ^{3/2} 4^\Lambda \Gamma \left(2 \Lambda-\frac{3}{2}\right)\nonumber\right.\\ && \left. +4  \left(e^{4 i \pi  \Lambda}-1\right)
 \, _2F_1\left(\frac{1}{2},2 \Lambda-2;2 \Lambda-\frac{3}{2};-1\right) 
 \Gamma^2 (2-\Lambda) \Gamma^2 (2 \Lambda-2)\right],\\ 
 Q_{2}&=&\int^{\pi}_{0}\int^{\infty}_{0}d\Theta~dk~e^{ikx\cos\Theta}~k^{2}\nonumber\\
 &=& \frac{i }{x^3 }\left[\ln\left[
 \frac{\tan\left(\frac{\pi}{4}+\frac{\Theta_{2}}{2}\right)}{\tan\left(\frac{\pi}{4}
 +\frac{\Theta_{1}}{2}\right)}\right]+\sec\Theta_{1}\tan\Theta_{1}-\sec\Theta_{2}\tan\Theta_{2}\right].\eea
 Let us now write down the expressions for newly introduced cosmological observable $\hat{ O}_{obs}$ for $m\approx H$ 
 case in the three limiting cases $|kc_{S}\eta|\rightarrow -\infty$, $|kc_{S}\eta|\rightarrow 0$
and $|kc_{S}\eta|\approx 1$ as given by:
\bea \left[\hat{ O}_{obs}\right]_{m\approx H}&\stackrel{|kc_{S}\eta|\rightarrow -\infty}{=}&
\frac{2\tilde{c}_{S}}{(-k\eta \tilde{c}_{S})^2H}{\cal I}^{I}_{1},\\ 
\left[\hat{ O}_{obs}\right]_{m\approx H}&\stackrel{|kc_{S}\eta|\rightarrow 0}{=}&\frac{2^{3-2\Lambda} c^{3}_{S}\pi
}{(-k\eta c_{S})^{3-2\Lambda} H}
\left|\frac{\Gamma\left(\frac{3}{2}\right)}{\Gamma\left(\Lambda\right)}\right|^2{\cal I}^{I}_{2},\\ 
\left[\hat{ O}_{obs}\right]_{m\approx H}&\stackrel{|kc_{S}\eta|\approx 1}{=}&\frac{2^{3-2\Lambda} c^{3}_{S} \pi}{ H}
\left|\frac{\Gamma\left(\frac{3}{2}\right)}{\Gamma\left(\Lambda\right)}\right|^2{\cal I}^{I}_{3},\eea
where for $m\approx H$ case the integrals ${\cal I}^{I}_{1}, {\cal I}^{I}_{2}$ and ${\cal I}^{I}_{3}$ are given by the following expressions:
\be\begin{array}{lll}\label{dfinv4bnbncv}
 \displaystyle{\cal I}^{I}_{1} = -H\frac{
\left[|C_{2}|^2 I_{1}-|C_{1}|^2 I_{2} 
-i\left(C^{*}_{1}C_{2}e^{i\pi\left(\Lambda+\frac{1}{2}\right)}
+C_{1}C^{*}_{2}e^{-i\pi\left(\Lambda+\frac{1}{2}\right)}\right)I_{3}\right]}
{\left[|C_{2}|^2+|C_{1}|^2
+\left(C^{*}_{1}C_{2}e^{2ikc_{S}\eta}e^{i\pi\left(\Lambda+\frac{1}{2}\right)}
+C_{1}C^{*}_{2}e^{-2ikc_{S}\eta}e^{-i\pi\left(\Lambda+\frac{1}{2}\right)}\right)\right]}
\\=\left\{\begin{array}{ll}
                    \displaystyle 
HI_{2} &
 \mbox{\small {\bf for ~Bunch Davies}}  \\ \\
	\displaystyle  -H\frac{
\left[\sinh^2\alpha~ 
I_{1}-\cosh^2\alpha~ I_{2}
+i~\sinh2\alpha\cos\left({\it \pi\left(\Lambda+\frac{1}{2}\right)}+\delta\right)I_{3}\right]}{\left[\sinh^2\alpha +\cosh^2\alpha
+\sinh2\alpha\cos{\it \left(2kc_{S}\eta+\pi\left(\Lambda+\frac{1}{2}\right)+\delta\right)}\right]}
                     & \mbox{\small {\bf for~$\alpha$~vacua~Type-I}}\\ \\
	\displaystyle  -H\frac{
\left[e^{\alpha+\alpha^{*}} 
I_{1}-I_{2}
+i\left(e^{\alpha}e^{i\pi\left(\Lambda+\frac{1}{2}\right)}
+e^{\alpha^{*}}e^{-i\pi\left(\Lambda+\frac{1}{2}\right)}\right)I_{3}\right]}{4
\cos^2{\it \left(kc_{S}\eta+\frac{\pi}{2}\left(\Lambda+\frac{1}{2}\right)-i\frac{\alpha}{2}\right)}}
                     & \mbox{\small {\bf for~$\alpha$~vacua~Type-II}}\\ \\
	\displaystyle  -H\frac{i\pi^2
I_{3}\cos^2{\it \frac{\pi}{2}\left(\Lambda+
\frac{1}{2}\right)}}{\cos^2{\it \left(kc_{S}\eta+\frac{\pi}{2}\left(\Lambda+\frac{1}{2}\right)\right)}}
                     & \mbox{\small {\bf for~special~vacua}}.
          \end{array}
\right.
\end{array}\ee
\bea\label{dfinv4bnbncv}
 \displaystyle{\cal I}_{2} &=& \sqrt{\xi}HI^{\xi\theta_{1}}_{4},~~~~~~~~~~~~~~~~~~~~~~~~~~~~~~~~~~~~~~~~~~\eea
\bea\label{dfinv4bnbncv}
 \displaystyle{\cal I}_{3} &=& \sqrt{\xi}HI^{\xi\theta_{2}}_{5}.~~~~~~~~~~~~~~~~~~~~~~~~~~~~~~~~~~~~~~~~~~\eea
 The results obtained in this section implies that if we take $m\approx H$ then it may be possible to measure the effect of Bell violation in the context of primordial cosmology, specifically for the inflationary paradigm.
 In this case the scale of inflation is comparable of the order of the mass parameter $m$. In such a case to get unique prediction of the scale of inflation 
 and as well as the nature of the new particle by knowing the impact of one point function or the newly defined observable. Additionally it is important to note that as the results in this case is dependent on 
 temporal cut-off scale $\xi$, we need to choose this parameter in such a way that the obtained results are consistent with the numerical value of all other inflationary observables as recently observed by Planck.
\subsubsection{{\bf Case II:} $m>> H$}
For further simplification we consider here $m\approx \Upsilon H$ with $\Upsilon>>1$ and also consider three
limiting cases $|kc_{S}\eta|\rightarrow -\infty$, $|kc_{S}\eta|\rightarrow 0$  and $|kc_{S}\eta|\approx 1$
 which are physically acceptable in the present context. 
 \bea\label{dfinv4bnbncv}
\displaystyle \langle\zeta_{\bf{k}}(\eta=0)\rangle_{|kc_{S}\eta|\rightarrow -\infty} &\approx& 
-\frac{16\Upsilon\pi^2H\tilde{c}_{S}}{M^2_{p}\epsilon}
\left[|C_{2}|^2 I_{1}
-|C_{1}|^2 I_{2}\right. \\ \nonumber && \left. ~~~~~~~~~~~~-i\left(C^{*}_{1}C_{2}e^{i\pi\left(\Lambda+\frac{1}{2}\right)}
+C_{1}C^{*}_{2}e^{-i\pi\left(\Lambda+\frac{1}{2}\right)}\right)I_{3}\right]\\&=&\left\{\begin{array}{ll}
                      \frac{8\Upsilon\pi^3H\tilde{c}_{S}}{M^2_{p}\epsilon}
I_{2} &
 \mbox{\small {\bf for ~Bunch Davies}}  \\ 
	  -\frac{16\Upsilon\pi^2H\tilde{c}_{S}}{M^2_{p}\epsilon}
\left[\sinh^2\alpha~ 
I_{1}-\cosh^2\alpha~I_{2}\right.\nonumber \\ \left. 
\displaystyle~~~~~~~~~~~
-i~\sinh2\alpha\cos\left({\it \pi\left(\Lambda+\frac{1}{2}\right)}+\delta\right)I_{3}\right]
                     & \mbox{\small {\bf for~$\alpha$~vacua~Type-I}}\\ 
	  -\frac{16\Upsilon\pi^2|N_{\alpha}|^2H\tilde{c}_{S}}{M^2_{p}\epsilon}
\left[e^{\alpha+\alpha^{*}} 
I_{1}-I_{2}\right.\nonumber \\ \left. 
\displaystyle~~~~~~~~~~~
-i\left(e^{\alpha}e^{i\pi\left(\Lambda+\frac{1}{2}\right)}
+e^{\alpha^{*}}e^{-i\pi\left(\Lambda+\frac{1}{2}\right)}\right)I_{3}\right]\nonumber
                     & \mbox{\small {\bf for~$\alpha$~vacua~Type-II}}\\ 
	 - \frac{64i\Upsilon\pi^2|C|^2H\tilde{c}_{S}}{M^2_{p}\epsilon}
\cos^2{\it \frac{\pi}{2}\left(\Lambda+\frac{1}{2}\right)}I_{3}
                    & \mbox{\small {\bf for~special~vacua}}.
          \end{array}
\right.
\eea
\bea\label{dfinv4bnbnxcxcx}
\small\displaystyle \langle\zeta_{\bf{k}}(\eta=\xi\rightarrow 0)\rangle_{|kc_{S}\eta|\rightarrow 0} &=& 
\frac{4\Upsilon\pi\sqrt{\xi}H\tilde{c}_{S}}{M^2_{p}\epsilon}\left(C^{*}_{1}C_{2}
+C_{1}C^{*}_{2}-|C_{1}|^2-|C_{2}|^2\right)
I^{\xi\theta_{1}}_{4}
\\&=&I^{\xi\theta_{1}}_{4}\times\left\{\begin{array}{ll}
                    \displaystyle  -\frac{2\Upsilon\pi^2\sqrt{\xi}H\tilde{c}_{S}}{M^2_{p}\epsilon}~~~~ &
 \mbox{\small {\bf for ~Bunch Davies vacua}}  \\ 
	\displaystyle  \frac{4\Upsilon\pi\sqrt{\xi}H\tilde{c}_{S}}{M^2_{p}\epsilon}\left(\cos\delta\sinh 2\alpha-\cosh^2\alpha-\sinh^2\alpha\right)
                    ~~~~ & \mbox{\small {\bf for~$\alpha$~vacua~Type-I}}\\ 
	\displaystyle  \frac{4\Upsilon\pi\sqrt{\xi}|N_{\alpha}|^2H\tilde{c}_{S}}{M^2_{p}\epsilon}\left(e^{\alpha}+e^{\alpha^{*}}
-e^{\alpha+\alpha^{*}}-1\right)
\nonumber
                    ~~~~ & \mbox{\small {\bf for~$\alpha$~vacua~Type-II}}\\ 
	\displaystyle  0
~~~ & \mbox{\small {\bf for~special~vacua}}.
          \end{array}
\right.
\eea
\bea\label{dfinv4bnbnxcxcx}
\small\displaystyle \langle\zeta_{\bf{k}}(\eta=\xi\rightarrow 0)\rangle_{|kc_{S}\eta|\approx 1} &=& 
\frac{4\Upsilon\pi\sqrt{\xi}H\tilde{c}_{S}}{M^2_{p}\epsilon}\left(C^{*}_{1}C_{2}
+C_{1}C^{*}_{2}-|C_{1}|^2-|C_{2}|^2\right)
I^{\xi\theta_{2}}_{5}
\\&=&I^{\xi\theta_{2}}_{5}\times\left\{\begin{array}{ll}
                    \displaystyle  -\frac{2\Upsilon\pi^2\sqrt{\xi}H\tilde{c}_{S}}{M^2_{p}\epsilon}~~~~ &
 \mbox{\small {\bf for ~Bunch Davies vacua}}  \\ 
	\displaystyle  \frac{4\Upsilon\pi\sqrt{\xi}H\tilde{c}_{S}}{M^2_{p}\epsilon}\left(\cos\delta\sinh 2\alpha-\cosh^2\alpha-\sinh^2\alpha\right)
                    ~~~~ & \mbox{\small {\bf for~$\alpha$~vacua~Type-I}}\\ 
	\displaystyle  \frac{4\Upsilon\pi\sqrt{\xi}|N_{\alpha}|^2H\tilde{c}_{S}}{M^2_{p}\epsilon}\left(e^{\alpha}+e^{\alpha^{*}}
-e^{\alpha+\alpha^{*}}-1\right)
\nonumber
                    ~~~~ & \mbox{\small {\bf for~$\alpha$~vacua~Type-II}}\\ 
	\displaystyle  0
~~~ & \mbox{\small {\bf for~special~vacua}}.
          \end{array}
\right.
\eea
where the integrals $I_{1}, I_{2}, I_{3}, I^{\xi\theta_{1}}_{4}, I^{\xi\theta_{2}}_{5}$ are defined earler.
Additionally, it is important to note that 
here the parameter $\Lambda$ is given by~\footnote{For the case $m>>H$ the approximation $m_{inf}\approx m$ is not valid as the mass scale of the inflaton cannot be larger than the scale of inflation.}:
 \be\begin{array}{lll}\label{y8}\small
     \underline{{\bf For}~ m_{inf}<<H:}~~~~~~~~~~~~~~
     \displaystyle \Lambda\approx\left\{\begin{array}{ll}
                    \displaystyle \frac{3}{2}~~~~ &
 \mbox{\small {\bf for ~dS}}  \\ 
	\displaystyle \nu ~~~~ & \mbox{\small {\bf for~ qdS}}.
          \end{array}
\right.
\end{array}\ee
\be\begin{array}{lll}\label{y9}\small
     \underline{{\bf For}~ m_{inf}\approx H:}~~~~~~~~~~~~~~
     \displaystyle \Lambda\approx\left\{\begin{array}{ll}
                    \displaystyle \frac{\sqrt{5}}{2}~~~~ &
 \mbox{\small {\bf for ~dS}}  \\ 
	\displaystyle \sqrt{\nu^2-1} ~~~~ & \mbox{\small {\bf for~ qdS}}.
          \end{array}
\right.
\end{array}\ee
Similarly in the position space the representative expressions for the expectation value of the 
scalar curvature perturbation along with two limiting cases $|kc_{S}\eta|\rightarrow -\infty$ and $|kc_{S}\eta|\rightarrow 0$ 
are given by:
\bea\label{dfinv4bnbncv}
\displaystyle \langle\zeta({\bf x},\eta=0)\rangle_{|kc_{S}\eta|\rightarrow -\infty} &\approx& 
-\frac{2\Upsilon H\tilde{c}_{S}}{M^2_{p}\epsilon\pi}
\left[|C_{2}|^2 B_{1}
-|C_{1}|^2 B_{2}\right.\\ \nonumber && \left. ~~~~~~~~~~~~-i\left(C^{*}_{1}C_{2}e^{i\pi\left(\Lambda+\frac{1}{2}\right)}
+C_{1}C^{*}_{2}e^{-i\pi\left(\Lambda+\frac{1}{2}\right)}\right)B_{3}\right]\\&=&\left\{\begin{array}{ll}
                      \frac{H\Upsilon\tilde{c}_{S}}{M^2_{p}\epsilon}
B_{2}\nonumber &
 \mbox{\small {\bf for ~Bunch Davies}}  \\ 
	  -\frac{2H\Upsilon\tilde{c}_{S}}{M^2_{p}\epsilon\pi}
\left[\sinh^2\alpha~ 
B_{1}-\cosh^2\alpha~B_{2}\right.\nonumber \\ \left. 
\displaystyle~~~~~~~~~~~
-i~\sinh2\alpha\cos\left({\it \pi\left(\Lambda+\frac{1}{2}\right)}+\delta\right)B_{3}\right]
                     & \mbox{\small {\bf for~$\alpha$~vacua~Type-I}}\\ 
	  -\frac{2|N_{\alpha}|^2H\tilde{c}_{S}}{M^2_{p}\epsilon\pi}
\left[e^{\alpha+\alpha^{*}} 
B_{1}-B_{2}\right.\nonumber \\ \left. 
\displaystyle~~~~~~~~~~~
-i\left(e^{\alpha}e^{i\pi\left(\Lambda+\frac{1}{2}\right)}
+e^{\alpha^{*}}e^{-i\pi\left(\Lambda+\frac{1}{2}\right)}\right)B_{3}\right]
                     & \mbox{\small {\bf for~$\alpha$~vacua~Type-II}}\\ 
	 - \frac{8i\Upsilon|C|^2H\tilde{c}_{S}}{M^2_{p}\epsilon\pi}
\cos^2{\it \frac{\pi}{2}\left(\Lambda+\frac{1}{2}\right)}B_{3}
                    & \mbox{\small {\bf for~special~vacua}}.
          \end{array}
\right.
\eea
\bea\label{dfinv4bnbnxcxcx}
\small\displaystyle \langle\zeta({\bf x},\eta=\xi\rightarrow 0)\rangle_{|kc_{S}\eta|\rightarrow 0} &=& 
\frac{\Upsilon\sqrt{\xi}H\tilde{c}_{S}}{2M^2_{p}\epsilon\pi^2}\left(C^{*}_{1}C_{2}
+C_{1}C^{*}_{2}-|C_{1}|^2-|C_{2}|^2\right)
B^{\xi\theta_{1}}_{4}
\\&=&B^{\xi\theta_{1}}_{4}\times\left\{\begin{array}{ll}
                    \displaystyle  -\frac{2\Upsilon\pi^2\sqrt{\xi}H\tilde{c}_{S}}{M^2_{p}\epsilon}~~~~ &
 \mbox{\small {\bf for ~Bunch Davies vacua}}  \\ 
	\displaystyle  \frac{\Upsilon\sqrt{\xi}H\tilde{c}_{S}}{2 M^2_{p}\pi^2\epsilon}\left(\cos\delta\sinh 2\alpha-\cosh^2\alpha-\sinh^2\alpha\right)
                    ~~~~ & \mbox{\small {\bf for~$\alpha$~vacua~Type-I}}\\ 
	\displaystyle  \frac{\Upsilon\sqrt{\xi}|N_{\alpha}|^2H\tilde{c}_{S}}{2M^2_{p}\pi^2\epsilon}\left(e^{\alpha}+e^{\alpha^{*}}
-e^{\alpha+\alpha^{*}}-1\right)
\nonumber
                    ~~~~ & \mbox{\small {\bf for~$\alpha$~vacua~Type-II}}\\ 
	\displaystyle  0
~~~ & \mbox{\small {\bf for~special~vacua}}.
          \end{array}
\right.
\eea
\bea\label{dfinv4bnbnxcxcx}
\small\displaystyle \langle\zeta({\bf x},\eta=\xi\rightarrow 0)\rangle_{|kc_{S}\eta|\approx 1} &=& 
\frac{\Upsilon\sqrt{\xi}H\tilde{c}_{S}}{2M^2_{p}\epsilon\pi^2}\left(C^{*}_{1}C_{2}
+C_{1}C^{*}_{2}-|C_{1}|^2-|C_{2}|^2\right)
B^{\xi\theta_{2}}_{5}
\\&=&B^{\xi\theta_{2}}_{5}\times\left\{\begin{array}{ll}
                    \displaystyle  -\frac{2\Upsilon\pi^2\sqrt{\xi}H\tilde{c}_{S}}{M^2_{p}\epsilon}~~~~ &
 \mbox{\small {\bf for ~Bunch Davies vacua}}  \\ 
	\displaystyle  \frac{\Upsilon\sqrt{\xi}H\tilde{c}_{S}}{2 M^2_{p}\pi^2\epsilon}\left(\cos\delta\sinh 2\alpha-\cosh^2\alpha-\sinh^2\alpha\right)
                    ~~~~ & \mbox{\small {\bf for~$\alpha$~vacua~Type-I}}\\ 
	\displaystyle  \frac{\Upsilon\sqrt{\xi}|N_{\alpha}|^2H\tilde{c}_{S}}{2M^2_{p}\pi^2\epsilon}\left(e^{\alpha}+e^{\alpha^{*}}
-e^{\alpha+\alpha^{*}}-1\right)
\nonumber
                    ~~~~ & \mbox{\small {\bf for~$\alpha$~vacua~Type-II}}\\ 
	\displaystyle  0
~~~ & \mbox{\small {\bf for~special~vacua}}.
          \end{array}
\right.
\eea
where the integrals $B_{1}, B_{2}, B_{3}, B^{\xi\theta_{1}}_{4}, B^{\xi\theta_{2}}_{5}$ are defined earlier.

Let us now write down the expressions for newly introduced cosmological observable $\hat{ O}_{obs}$ for $m\approx H$ 
 case in the three limiting cases $|kc_{S}\eta|\rightarrow -\infty$, $|kc_{S}\eta|\rightarrow 0$
and $|kc_{S}\eta|\approx 1$ as given by:
\bea \left[\hat{ O}_{obs}\right]_{m>> H}&\stackrel{|kc_{S}\eta|\rightarrow -\infty}{=}&
\frac{2\tilde{c}_{S}}{(-k\eta \tilde{c}_{S})^2H}{\cal I}^{I}_{1},\\ 
\left[\hat{ O}_{obs}\right]_{m>> H}&\stackrel{|kc_{S}\eta|\rightarrow 0}{=}&\frac{2^{3-2\Lambda} c^{3}_{S}\pi
}{(-k\eta c_{S})^{3-2\Lambda} H}
\left|\frac{\Gamma\left(\frac{3}{2}\right)}{\Gamma\left(\Lambda\right)}\right|^2{\cal I}^{I}_{2},\\ 
\left[\hat{ O}_{obs}\right]_{m>> H}&\stackrel{|kc_{S}\eta|\approx 1}{=}&\frac{2^{3-2\Lambda} c^{3}_{S} \pi}{ H}
\left|\frac{\Gamma\left(\frac{3}{2}\right)}{\Gamma\left(\Lambda\right)}\right|^2{\cal I}^{I}_{3},\eea
where for $m\approx H$ case the integrals ${\cal I}^{I}_{1}, {\cal I}^{I}_{2}$ and ${\cal I}^{I}_{3}$ are given by the following expressions:
\be\begin{array}{lll}\label{dfinv4bnbncv}
 \displaystyle{\cal I}^{I}_{1} = -\Upsilon H\frac{
\left[|C_{2}|^2 I_{1}-|C_{1}|^2 I_{2} 
-i\left(C^{*}_{1}C_{2}e^{i\pi\left(\Lambda+\frac{1}{2}\right)}
+C_{1}C^{*}_{2}e^{-i\pi\left(\Lambda+\frac{1}{2}\right)}\right)I_{3}\right]}
{\left[|C_{2}|^2+|C_{1}|^2
+\left(C^{*}_{1}C_{2}e^{2ikc_{S}\eta}e^{i\pi\left(\Lambda+\frac{1}{2}\right)}
+C_{1}C^{*}_{2}e^{-2ikc_{S}\eta}e^{-i\pi\left(\Lambda+\frac{1}{2}\right)}\right)\right]}
\\=\left\{\begin{array}{ll}
                    \displaystyle 
\Upsilon H I_{2} &
 \mbox{\small {\bf for ~Bunch Davies}}  \\ 
	\displaystyle  -\Upsilon H\frac{
\left[\sinh^2\alpha~ 
I_{1}-\cosh^2\alpha~ I_{2}
+i~\sinh2\alpha\cos\left({\it \pi\left(\Lambda+\frac{1}{2}\right)}+\delta\right)I_{3}\right]}{\left[\sinh^2\alpha +\cosh^2\alpha
+\sinh2\alpha\cos{\it \left(2kc_{S}\eta+\pi\left(\Lambda+\frac{1}{2}\right)+\delta\right)}\right]}
                     & \mbox{\small {\bf for~$\alpha$~vacua~Type-I}}\\ 
	\displaystyle  -\Upsilon H\frac{
\left[e^{\alpha+\alpha^{*}} 
I_{1}-I_{2}
+i\left(e^{\alpha}e^{i\pi\left(\Lambda+\frac{1}{2}\right)}
+e^{\alpha^{*}}e^{-i\pi\left(\Lambda+\frac{1}{2}\right)}\right)I_{3}\right]}{4
\cos^2{\it \left(kc_{S}\eta+\frac{\pi}{2}\left(\Lambda+\frac{1}{2}\right)-i\frac{\alpha}{2}\right)}}
                     & \mbox{\small {\bf for~$\alpha$~vacua~Type-II}}\\ 
	\displaystyle  -\Upsilon H\frac{i\pi^2
I_{3}\cos^2{\it \frac{\pi}{2}\left(\Lambda+
\frac{1}{2}\right)}}{\cos^2{\it \left(kc_{S}\eta+\frac{\pi}{2}\left(\Lambda+\frac{1}{2}\right)\right)}}
                     & \mbox{\small {\bf for~special~vacua}}.
          \end{array}
\right.
\end{array}\ee
\bea\label{dfinv4bnbncv}
 \displaystyle{\cal I}_{2} &=& \sqrt{\xi}\Upsilon HI^{\xi\theta_{1}}_{4},~~~~~~~~~~~~~~~~~~~~~~~~~~~~~~~~~~~~~~~~~~\\
 \displaystyle{\cal I}_{3} &=& \sqrt{\xi}\Upsilon HI^{\xi\theta_{2}}_{5}.~~~~~~~~~~~~~~~~~~~~~~~~~~~~~~~~~~~~~~~~~~\eea
The results obtained in this section implies that if we take $m>>H$ then it may be possible to measure the effect of Bell violation in the context of primordial cosmology, specifically for the inflationary paradigm. 
 In such a case to get unique prediction of the scale of inflation and associated new physics we need to measure the impact of one point function or the newly defined observable and in such a case we don't need to 
 wait for future observations for primordial gravitational waves and primordial non-Gaussianity to comment on the unique scale of inflation in a model independent way. But as the results in this case dependent on 
 temporal cut-off scale $\xi$, we need to choose this parameter in such a way that the obtained results are consistent with the numerical value of all other inflationary observables as recently observed by Planck.
\subsubsection{{\bf Case III:} $m<< H$}
For further simplification we consider here $m<<H$ with three limiting
cases $|kc_{S}\eta|\rightarrow -\infty$, $|kc_{S}\eta|\rightarrow 0$ and $|kc_{S}\eta|\approx 1$ 
 which are physically acceptable in the present context. 
 \bea\label{dfinv4bnbncv}
\displaystyle \langle\zeta_{\bf{k}}(\eta=0)\rangle_{|kc_{S}\eta|\rightarrow -\infty} &\approx& 0,\\
\small\displaystyle \langle\zeta_{\bf{k}}(\eta=\xi\rightarrow 0)\rangle_{|kc_{S}\eta|\rightarrow 0} &\approx & 0,\\
\displaystyle \langle\zeta_{\bf{k}}(\eta=\xi\rightarrow 0)\rangle_{|kc_{S}\eta|\approx 1} &\approx & 0.
\eea
 Additionally, it is important to note that 
here the parameter $\Lambda$ is given by~\footnote{For the special case where inflaton mass are comparable with the new particle then for $m_{inf}\approx m<< H$ case we have:\be\begin{array}{lll}\label{y10}\small
 \displaystyle \Lambda\approx\left\{\begin{array}{ll}
                    \displaystyle \frac{3}{2}~~~~ &
 \mbox{\small {\bf for ~dS}}  \\ 
	\displaystyle \nu ~~~~ & \mbox{\small {\bf for~ qdS}}.
          \end{array}
\right.
\end{array}\ee}:
 \be\begin{array}{lll}\label{y11}\small
     \underline{{\bf For}~ m_{inf}<<H:}~~~~~~~~~~~~~~
     \displaystyle \Lambda\approx\left\{\begin{array}{ll}
                    \displaystyle \frac{3}{2}~~~~ &
 \mbox{\small {\bf for ~dS}}  \\ 
	\displaystyle \nu ~~~~ & \mbox{\small {\bf for~ qdS}}.
          \end{array}
\right.
\end{array}\ee
\be\begin{array}{lll}\label{y12}\small
     \underline{{\bf For}~ m_{inf}\approx H:}~~~~~~~~~~~~~~
     \displaystyle \Lambda\approx\left\{\begin{array}{ll}
                    \displaystyle \frac{\sqrt{5}}{2}~~~~ &
 \mbox{\small {\bf for ~dS}}  \\ 
	\displaystyle \sqrt{\nu^2-1} ~~~~ & \mbox{\small {\bf for~ qdS}}.
          \end{array}
\right.
\end{array}\ee
After taking Fourier transform in position space we get:
\bea\label{dfinv4bnbncv}
\displaystyle \langle\zeta ({\bf x},\eta=0)\rangle_{|kc_{S}\eta|\rightarrow -\infty} &\approx& 0,\\
\small\displaystyle \langle\zeta ({\bf x},\eta=\xi\rightarrow 0)\rangle_{|kc_{S}\eta|\rightarrow 0} &\approx & 0,\\
\displaystyle \langle\zeta ({\bf x},\eta=\xi\rightarrow 0)\rangle_{|kc_{S}\eta|\approx 1} &\approx & 0.
\eea
Let us now define a new cosmological observable $\hat{ O}_{obs}$ in the three limiting cases $|kc_{S}\eta|\rightarrow -\infty$, $|kc_{S}\eta|\rightarrow 0$
and $|kc_{S}\eta|\approx 1$ as:
\bea \hat{ O}_{obs}&\stackrel{|kc_{S}\eta|\rightarrow -\infty}{=}&\frac{2\tilde{c}_{S}}{(-k\eta \tilde{c}_{S})^2H}{\cal I}_{1}\approx 0,\\ 
\hat{ O}_{obs}&\stackrel{|kc_{S}\eta|\rightarrow 0}{=}&\frac{2^{3-2\Lambda} c^{3}_{S}\pi
}{(-k\eta c_{S})^{3-2\Lambda} H}
\left|\frac{\Gamma\left(\frac{3}{2}\right)}{\Gamma\left(\Lambda\right)}\right|^2{\cal I}_{2}\approx 0,\\ 
\hat{ O}_{obs}&\stackrel{|kc_{S}\eta|\approx 1}{=}&\frac{2^{3-2\Lambda} c^{3}_{S} \pi}{ H}
\left|\frac{\Gamma\left(\frac{3}{2}\right)}{\Gamma\left(\Lambda\right)}\right|^2{\cal I}_{3}\approx 0,\eea
where for $m<<H$ case ${\cal I}_{1}, {\cal I}_{2}$ and ${\cal I}_{3}$ are defined as:
\bea
 \displaystyle{\cal I}_{1} &\approx& 0,\\
 \displaystyle{\cal I}_{2} &\approx & 0,~~~~~~~~~~~~~~~~~~~~~~~~~~~~~~~~~~~~~~~~~~\\
 \displaystyle{\cal I}_{3} &\approx & 0.~~~~~~~~~~~~~~~~~~~~~~~~~~~~~~~~~~~~~~~~~~\eea
 The results obtained in this section implies that if we take $m<<H$ then it is not possible to measure the effect of Bell violation in the context of primordial cosmology, specifically for the inflationary paradigm. 
 In such a case to get unique prediction of the scale of inflation and associated new physics we need to wait for future observations for primordial gravitational waves and primordial non-Gaussianity.

\section{Specific example: Analogy with axion fluctuations in String Theory}
\label{sec4}
In this section we discuss about the axion fluctuations originated from string theory and its 
exact connection with the present topic of discussion in this paper. Here we discuss about the background string theoretic framework and 
its four dimensional effective field theory version which will participate in the axion fluctuations in primordial cosmology.
See ref.~\cite{silverstein:2008jj,Sandip:2004jj,Witten:2013pra,Svrcek:2006yi,Beasley:2005iu,Beasley:2005hr,
Witten:2002wb,Witten:2001zr,Witten:1995gx,Witten:1995zh,
Witten:1995ex,Witten:1993ed,Choudhury:2015hvr,Panda:2010uq} for further details in this direction.
\subsection{Axion monodromy model}
\label{sec4a}
Let us start our discussion with the canonically normalized string theory originated axion action:
\bea S_{axion}&=& \int d^{4}x \sqrt{-g}\left[-\frac{1}{2}(\partial \phi)^2 -V(\phi)\right],\eea
where $\phi$ is the axion field and the corresponding potential from string theory can be expressed as:
\bea\label{ax} V(\phi)&=&\mu^3\phi+\Lambda^4_{C}\cos\left(\frac{\phi}{f_{a}}\right)\nonumber\\
&=&\mu^3\left[\phi+bf_{a}\cos\left(\frac{\phi}{f_{a}}\right)\right],\eea
where we define a new parameter $b$ as:
\bea b&=& \frac{\Lambda^4_{C}}{\mu^3 f_{a}}.\eea
Here it is important to mention that, the linear part of the axion potential as appearing in Eq~(\ref{ax}) 
has been derived in the context of string theory in ref.~\cite{Panda:2010uq}. Where as the cosine part of the axion potential has 
its origin in non-perturbative aspects in string theory \cite{Svrcek:2006yi}.

One can also express the axion action by introducing a dimensionless axion field, $\phi=a~f_{a}$ as:
\bea S_{axion}&=& \int d^{4}x \sqrt{-g}\left[-\frac{f^2_{a}}{2}(\partial a)^2 -U(a)\right],\eea
where the effective axion potential can be recast as:
\bea U(a)=V(af_{a})&=&\mu^3af_{a}+\Lambda^4_{C}\cos a\nonumber\\
&=&\mu^3f_{a}\left[a+b\cos a\right].\eea
\begin{figure*}[htb]
\centering
\subfigure[Various parts of potential with $b=2$]{
    \includegraphics[width=7.2cm,height=7cm] {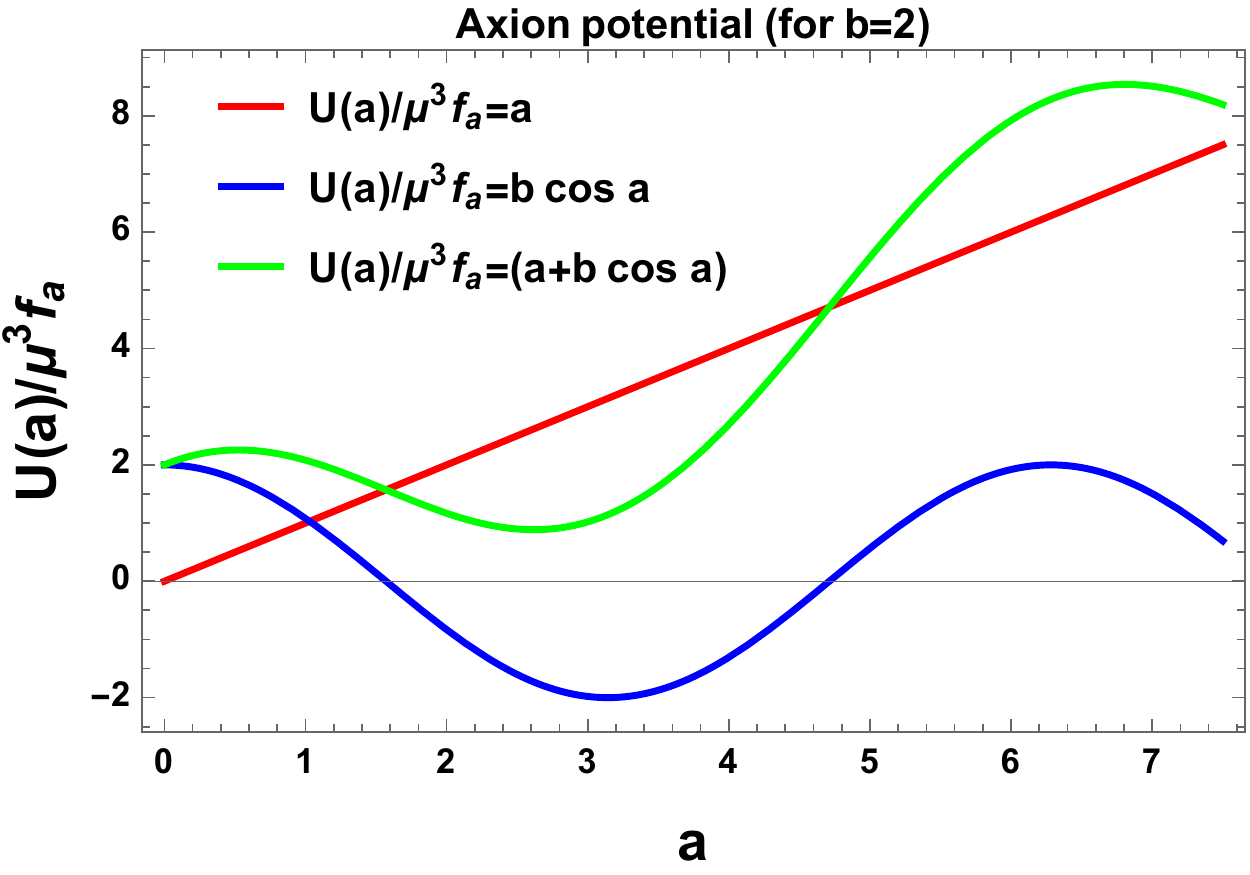}
    \label{fig1}
}
\subfigure[Total potential with $b=1,2,3$]{
    \includegraphics[width=7.2cm,height=7cm] {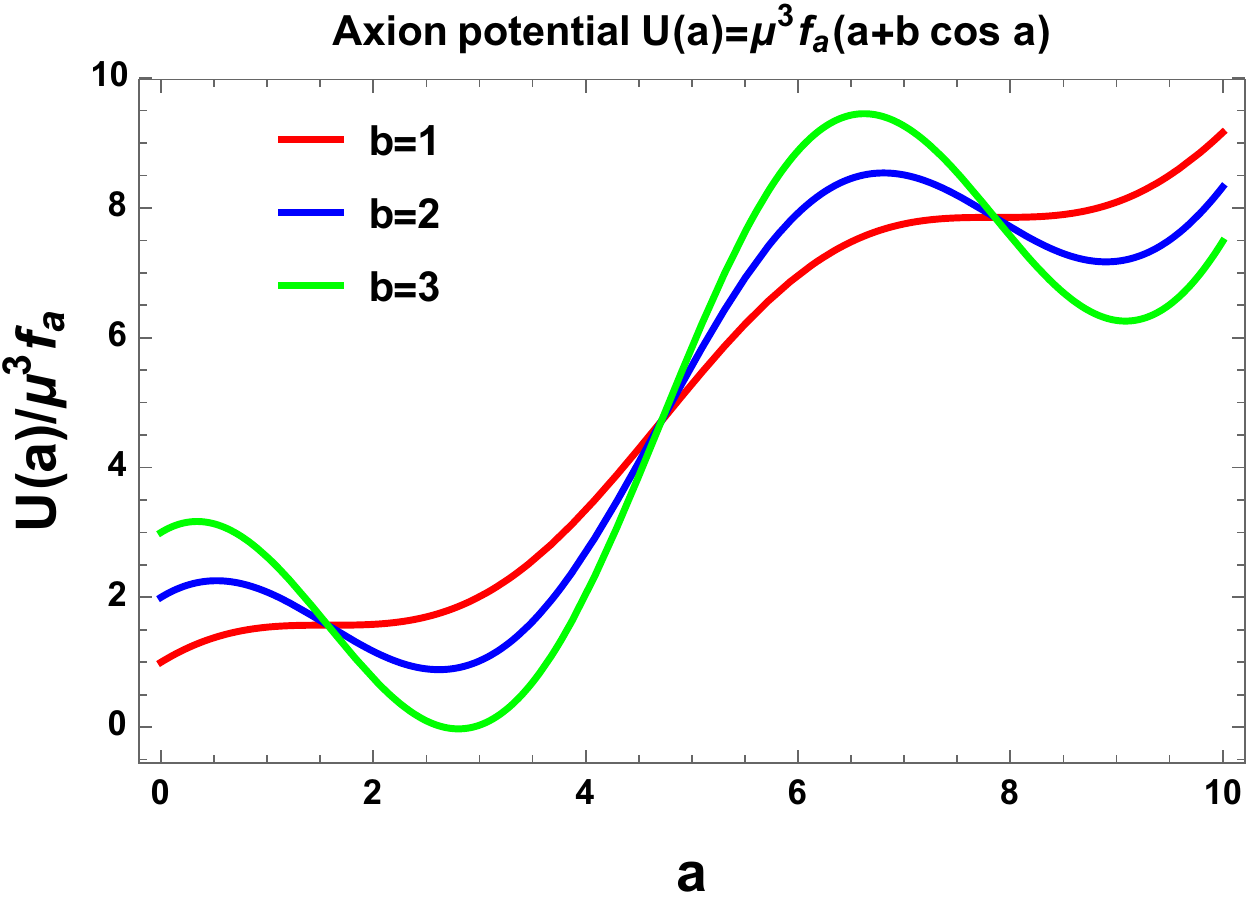}
    \label{fig2}
}
\caption[Optional caption for list of figures]{Behaviour of the axion effective potential.} 
\label{fza}
\end{figure*}
Further introducing conformal time in this computation axion action can be recast as:
\bea S_{axion}&=& \int d\eta~d^{3}x \left[\frac{f^2_{a}(\eta)}{2H^2}\frac{\left[(\partial_{\eta}a)^2-(\partial_{i}a)^2\right]}{\eta^2} -\frac{U(a)}{H^4 \eta^4}\right],\eea
where only mass contribution for axion field will contribute to the fluctuations and other part can be treated as back-reaction effect which one can neglect due to its smallness.

Additionally, it is important to mention the following information regarding the axion action and the representative theoretical setup:
\begin{itemize}
 \item In this case the scale $\Lambda_{C}$ is characterized by:
       \be\label{sdfg} \Lambda_{C}=\sqrt{m_{SUSY}M_{p}}~e^{-cS_{inst}},\ee
       where $S_{inst}$ characterize the action of the instanton which gives rise to the effective 
       potential in the present context, $c$ signifies a constant factor which is of the order of unity,
       $m_{SUSY}$ represents the supersymmetry breaking mass scale and $M_p$
       is the reduced Planck mass, defined as:
       \be M_p=\frac{L^3}{\sqrt{\alpha^{'}}g_{s}}.\ee
       Substituting this expression in eq~(\ref{sdfg}) we finally get:
       \be\label{sdfg1} \Lambda_{C}=\sqrt{\frac{m_{SUSY} L^3}{\sqrt{\alpha^{'}}g_{s}}}~e^{-cS_{inst}}.\ee
       
 \item Here the overall scale of the effective potential is given by:
        \be V_{0}=\mu^3 f_a=\frac{1}{\alpha^{'2}g_{s}}e^{4A_0}+\frac{R^2}{\alpha^{'}L^4}m^4_{SUSY}e^{2A_0}.\ee
 where $e^{A_0}$ is the warp factor at the bottom portion of the throat, $\alpha^{'}$ is the Regge slope parameter, 
 $g_{s}$ is the string coupling constant, $L^6$ is the volume factor in string units and $R$ approximately characterizes 
 the radius of the AdS like (Klebanov-Strassler) throat region in which the 5 brane and antibranes are placed.
 \item Here the second term in the effective potential has a periodicity of $2\pi f_{a}$ and maintains the shift symmetry 
 $\phi\rightarrow \phi+2\pi f_{a}$. Here $f_{a}$ characterizes the decay constant. On the other hand, the first term in the effective potential breaks the shift symmetry.
 This implies that the total effective potential breaks the shift symmetry in the present context.
 
 \item Additionally, the total scale of the axion effective potential is determined by the term $\mu^3$ after 
 introducing the additional parameter $b$. In the present context in terms of string parameters the factor $b$ 
 can be recast as`:
 \be b=\frac{\Lambda^4_C}{\mu^3 f_a}=\left(\frac{m^2_{SUSY}M^2_{p}~e^{-4cS_{inst}}}{\frac{1}{\alpha^{'2}g_{s}}e^{4A_0}
 +\frac{R^2}{\alpha^{'}L^4}m^4_{SUSY}e^{2A_0}}\right)=\left(\frac{m^2_{SUSY}\frac{L^6}{\alpha^{'}g^2_{s}}~e^{-4cS_{inst}}}{\frac{1}{\alpha^{'2}g_{s}}e^{4A_0}
 +\frac{R^2}{\alpha^{'}L^4}m^4_{SUSY}e^{2A_0}}\right).\ee
 Here the warp factor $e^{A_0}$ is given by:
 \be e^{A_0}=\left(\frac{\Lambda_{C}}{m_{SUSY}}\right)^2\frac{L}{R}\sqrt{g_{s}\alpha^{'}}
 =\frac{L^4}{m_{SUSY}R}\sqrt{\frac{\alpha^{'}}{g_{s}}}~e^{-2cS_{inst}}.\ee
 Substituting the explicit expression for the warp factor in $b$ we finally get:
\bea b=\frac{\Lambda^4_C}{\mu^3 f_a}&=&\left(\frac{m^2_{SUSY}\frac{L^6}{\alpha^{'}
g^2_{s}}~e^{-4cS_{inst}}}{\frac{L^{16}}{m^4_{SUSY}g^3_{s}R^4}e^{-8cS_{inst}}
 +\frac{L^4}{g_{s}}m^2_{SUSY}e^{-4cS_{inst}}}\right)\nonumber\\
 &=&\left(\frac{\frac{L^2}{\alpha^{'}
g_{s}}}{1+\frac{L^{12}}{m^6_{SUSY}g^2_{s}R^4}e^{-4cS_{inst}}
 }\right).\eea
 Additionally, the warped down string scale at the bottom the throat is given by:
 \be M_{s}=\frac{e^{A_{0}}}{\sqrt{\alpha^{'}}}=\left(\frac{\Lambda_{C}}{m_{SUSY}}\right)^2\frac{L}{R}\sqrt{g_{s}}
 =\frac{L^4}{m_{SUSY}R\sqrt{g_{s}}}~e^{-2cS_{inst}}.\ee
 \item Most importantly if one can treat the axion decay constant $f_{a}$ is background conformal 
 time scale dependent or if one interpret this to be parameter of string theory then in our present discussion 
 of the paper it exactly mimics the role of slow-roll parameter $\epsilon$ rescaled with sound speed $c_{S}$ of the new particle. Most importantly, one can treat 
 such new particles to be axion field originated from string theory.
 \item In this context we also assume that the axion decay constant $f_{a}$ is inflaton field dependent and this is perfectly consistent with the fact that $f_{a}$ is background conformal 
 time scale dependent.
 \item In the present context we also assume that the axion decay constant $f_{a}$ initially becomes large compared the the Hubble scale during inflation i.e. $f_{a}>>H$ and becomes smaller compared to the Hubble scale i.e. 
 $f_{a}<<H$ during some time interval, a few
e-foldings after the massive new particles were
created. Then it becomes large again. Here due to the increase in the value of the axion decay constant $f_{a}$ one can suppress the effect of quantum fluctuations at shorter distance scale. As a result
in the next setup we get an admissible value for the magnitude of the dimensionless axion field $a=\phi/f_{a}$ at the location of the each created particle from axion. As $f_{a}$ mimics the role of mass parameter $m$ in the present context all the earlier computation for four cases of the choice of mass parameters are valid here.
\item In this computation we additionally setup the the cosmological detector settings or more precisely the decider variables are adjusted in such a way that it creates an axion field with fluctuations at a
characteristic scale controlled by the mutual effects from $\Lambda^4_{C}$ and $\mu^3$ as introduced earlier. Here the effect from $\Lambda^4_{C}$ becomes dominant in the early universe and decides the scale of inflation. On
the other hand the effect from the term $\mu^3$ becomes larger during late time acceleration of our universe. In the present context, more precisely the energy scales $\Lambda^4_{C}$ and $\mu^3$ serve the purpose of 
cosmological constant at early and late times of the evolution of the universe. 
\item The quantum fluctuations in the dimensionless stringy axionic field $a=\phi/f_{a}$ exactly mimics the role of scalar (curvature) perturbation $\zeta$ as introduced in the earlier section. Additionally it 
is important to note that the quantum fluctuations in the dimensionless stringy axionic field $a$ are larger at distance scale corresponding to a particular conformal time $x_{dist}\sim |\eta_{c}|$, and for distances
smaller than $x<x_{dist}\sim |\eta_{c}|$ the effect of quantum fluctuations are also smaller. For this reason, one can interpret such a distance or the corresponding time scale 
to be the critical one in the present context. 

\item During the computation detector settings or more precisely the decider variables are chosen in such a way that they will appear locally around each massive particle. Most importantly it is important to mention here that
here we also assume that during this process detector settings or more precisely the decider variables are independent from the environment of the other massive particle pair in the present context. In our case 
axion plays the role of such decider variable.

\item In case of axion the cosmological Bell test experimental setup is prepared in such a way that it survives after inflationary epoch upto very late times. After the end of inflation the oscillating and periodic part of the 
potential $U(a)\sim \Lambda^{4}_{C}\cos a$ gives the mass contribution to the axion field and due to the specific structure of the potential, axion field oscillates at later times after the end of inflation. This additionally implies 
that in the present context the axion field contributes to explain the dark matter content of the universe. Moreover, apart from explaining fluctuation appearing from curvature perturbation in the context of axion isocurvature fluctuations also contribute to
the dark matter. But according to the present day observations, we cannot able to see such fluctuations at all and here one can interpret that the contribution from the isocurvature
fluctuations become highly suppressed in cosmological perturbations to explain the existence of dark matter. But such fluctuations are theoretically useful to
determine the initial field value of the dimensionless axion field $a=\phi/f_{a}$. 

\item The mass contribution to the axion field appearing from the periodic part of the 
potential $U(a)\sim \Lambda^{4}_{C}\cos a$ is exactly equivalent to the conformal time scale dependent mass parameter $m(\eta)$ in the present context of discussion.
\end{itemize}
\subsection{Axion effective interaction}
\label{sec4b}
Before going to the further details let us analyze the structure of the effective axionic potential. If we consider the total 
effective potential then the potential has extrema at $a=a_{0}$ given by the following constraint condition:
\bea U^{'}(a=a_{0})=0 \Longrightarrow a_{0}=\frac{\phi_{0}}{f_{a}}=\sin^{-1}\left(\frac{1}{b}\right)
=\sin^{-1}\left(\frac{\mu^3 f_{a}}{\Lambda^4_{C}}\right).\eea
Here $'$ signifies derivatives with respect to axion field $a$.
In this context the time dependent effective mass of the dimensionless axion field at extrema $a=a_{0}=\phi_{0}/f_{a}$
is computed from the total axionic effective potential $U(a)$ as:
\bea m^2_{axion}=U^{''}(a=a_{0})=-\mu^3 f_{a}b \cos\left(\sin^{-1}\left(\frac{1}{b}\right)\right)=
-\Lambda^4_{C} \cos\left(\sin^{-1}\left(\frac{\mu^3 f_{a}}{\Lambda^4_{C}}\right)\right).
\eea
To get the minima/maxima at the location $a=a_{0}$ we get the following set of possible constraint from the 
total effective potential:
\begin{itemize}
 \item For maxima we need: \be U^{''}(a=a_{0})<0\ee and this implies that in such a situation the axion mass term 
 \be m^2_{axion}<0,\ee which gives us tachyonic type of instability. Consequently as always $\cos\left(\sin^{-1}\left(\frac{1}{b}\right)\right)>0$ then here we get the following constraint condition on the axion model parameters:
 \bea \mu^3 f_{a}b=\Lambda^{4}_{C} >0\eea
 This is possible:
 \begin{enumerate}
  \item If $\mu^3 f_{a}>0$ and $b>0$ or precisely $\Lambda^4_{C}>0$.
  \item If $\mu^3 f_{a}<0$ and $b<0$ or precisely $\Lambda^4_{C}>0$.
 \end{enumerate}

 \item For minima we need: \be U^{''}(a=a_{0})>0\ee and this implies that in such a situation the axion mass term 
 \be m^2_{axion}>0,\ee which avoids tachyonic type of instability and makes the analysis more consistent in the present context. Consequently as always $\cos\left(\sin^{-1}\left(\frac{1}{b}\right)\right)>0$ then here we get the following constraint condition on the axion model parameters:
 \bea \mu^3 f_{a}b=\Lambda^{4}_{C} <0\eea
 This is possible:
 \begin{enumerate}
  \item If $\mu^3 f_{a}>0$ and $b<0$ or precisely $\Lambda^4_{C}<0$.
  \item If $\mu^3 f_{a}<0$ and $b>0$ or precisely $\Lambda^4_{C}<0$.
 \end{enumerate}
\end{itemize}
On the other hand if we consider the fact that during inflation and in later stages oscillating part of the 
effective axionic potential contributes larger compared to the linear contribution then for such periodic structure of 
the effective potential we get the extrema at $a=a_{0}$ given by the following constraint condition:
\bea U^{'}(a=a_{0})=0 \Longrightarrow a_{0}=\frac{\phi_{0}}{f_{a}}=m \pi,\eea
where $m \subset {\bf Z}$. In this context the time dependent effective mass of the dimensionless axion field at extrema $a=a_{0}=\phi_{0}/f_{a}$
is computed from the total axionic effective potential $U(a)$ as:
\bea m^2_{axion}=U^{''}(a=a_{0})=-\mu^3 f_{a}b \cos\left(m \pi\right)=
-\mu^3 f_{a}b (-1)^{m}.
\eea
To get the minima/maxima at the location $a=a_{0}$ we get the following set of possible constraint from the 
total effective potential:
\begin{itemize}
 \item For maxima we need: \be U^{''}(a=a_{0})<0\ee and this implies that in such a situation the axion mass term 
 \be m^2_{axion}<0,\ee which gives us tachyonic type of instability. Consequently we get the following constraint condition on the axion model parameters:
 \bea \mu^3 f_{a}b (-1)^{m}=\Lambda^{4}_{C}(-1)^{m}>0\eea
 This is possible:
 \begin{enumerate}
  \item If $\mu^3 f_{a}>0$, $b>0$ or precisely $\Lambda^4_{C}>0$ and $(-1)^{m}>0$ i.e. $m$ is even integer number.
  \item If $\mu^3 f_{a}>0$, $b<0$ or precisely $\Lambda^4_{C}<0$ and $(-1)^{m}<0$ i.e. $m$ is odd integer number.
  \item If $\mu^3 f_{a}<0$, $b>0$ or precisely $\Lambda^4_{C}<0$ and $(-1)^{m}<0$ i.e. $m$ is odd integer number.
  \item If $\mu^3 f_{a}<0$, $b<0$ or precisely $\Lambda^4_{C}>0$ and $(-1)^{m}>0$ i.e. $m$ is even integer number.
 \end{enumerate}

 \item For minima we need: \be U^{''}(a=a_{0})>0\ee and this implies that in such a situation the axion mass term 
 \be m^2_{axion}>0,\ee which avoids tachyonic type of instability and makes the analysis more consistent in the present context. Consequently we get the following constraint condition on the axion model parameters:
 \bea \mu^3 f_{a}b (-1)^{m}=\Lambda^{4}_{C}(-1)^{m}<0\eea
 This is possible:
 \begin{enumerate}
  \item If $\mu^3 f_{a}>0$, $b>0$ or precisely $\Lambda^4_{C}>0$ and $(-1)^{m}<0$ i.e. $m$ is odd integer number.
  \item If $\mu^3 f_{a}>0$, $b<0$ or precisely $\Lambda^4_{C}<0$ and $(-1)^{m}>0$ i.e. $m$ is even integer number.
  \item If $\mu^3 f_{a}<0$, $b>0$ or precisely $\Lambda^4_{C}<0$ and $(-1)^{m}>0$ i.e. $m$ is even integer number.
  \item If $\mu^3 f_{a}<0$, $b<0$ or precisely $\Lambda^4_{C}>0$ and $(-1)^{m}<0$ i.e. $m$ is odd integer number.
 \end{enumerate}
\end{itemize}
Now to comment on the effect of very small back reaction one can compute the following terms at $a=a_{0}$:
\bea U(a=a_{0})=\left\{\begin{array}{ll}
                    \displaystyle \mu^3f_{a}\left[\sin^{-1}\left(\frac{1}{b}\right)+b\cos \left(\sin^{-1}\left(\frac{1}{b}\right)\right)\right]
                  ~~~~ &
 \mbox{\small {\bf for ~total~$U(a)$}}  \\ 
	\displaystyle \mu^3f_{a}\left[m \pi+b(-1)^{m}\right]~~~~ & \mbox{\small {\bf for~osc.~$U(a)$}}.
          \end{array}
\right.\eea
\bea U^{'''}(a=a_{0})=\left\{\begin{array}{ll}
                    \displaystyle \mu^3 f_{a}~~~~ &
 \mbox{\small {\bf for ~total~$U(a)$}}  \\ 
	\displaystyle 0 ~~~~ & \mbox{\small {\bf for~osc.~$U(a)$}}.
          \end{array}
\right.\eea
\bea U^{''''}(a=a_{0})=\lambda_{self}=\left\{\begin{array}{ll}
                    \displaystyle \mu^3 f_{a}b\cos \left(\sin^{-1}\left(\frac{1}{b}\right)\right)~~~~ &
 \mbox{\small {\bf for ~total~$U(a)$}}  \\ 
	\displaystyle \mu^3 f_{a}b(-1)^{m} ~~~~ & \mbox{\small {\bf for~osc.~$U(a)$}}.
          \end{array}
\right.\eea
Here we restrict upto fourth derivative terms to make the potential re-normalizable.
Now if we claim that $\mu^{3}f_{a}$ is small then one can neglect $U(a=a_{0})$, $U^{'''}(a=a_{0})$ and self interaction term $U^{''''}(a=a_{0})$ due to small back reaction. Consequently if we take the Taylor expansion of the 
axion potential around $a=a_{0}$ we get:
\be U(a)\approx \frac{1}{2}m^{2}_{axion}(a-a_{0})^2.\ee
\subsection{Axion creation from quantum fluctuation}
\label{sec4c}
Now to study the effects of fluctuations explicitly let us first write down the equation of motion corresponding to axion field as given by:
\be \partial_{\eta}\left(\frac{f^2_{a}}{H^2\eta^2}(\partial_{\eta}\bar{a})\right)
-\frac{f^2_{a}}{H^2\eta^2}(\partial^2_{i}\bar{a})+\frac{m^{2}_{axion}}{H^4\eta^4}\bar{a}=0,\ee
where we use the fact that $f_{a}/H$ is a conformal time dependent factor and $\bar{a}$ is defined as, $\bar{a}=a-a_{0}$. Further taking the following ansatz for Fourier transformation:
\be \bar{a}=\bar{a}(\eta,{\bf x})=\int\frac{d^3k}{(2\pi)^3}\bar{a}_{\bf k}(\eta)\exp(i{\bf k}.{\bf x}),\ee
in momentum space the equation of motion can be recast as:
\be \partial^{2}_{\eta}\vartheta_{\bf k}+
\left(k^2-\frac{\partial^2_{\eta}\left(\frac{f^2_{a}}{H^2\eta^2}\right)}{\left(\frac{f^2_{a}}{H^2\eta^2}\right)}
+\frac{m^{2}_{axion}}{f^2_{a}H^2\eta^2}\right)\vartheta_{\bf k}=0,\ee
where introduce a new variable $\vartheta_{\bf k}$ defined as:
\be \vartheta_{\bf k} = \frac{f^2_{a}}{H^2\eta^2M^2_p}\bar{a}_{\bf k}.\ee
Here it is important to mention here that to solve the above mentioned mode equation exactly or using WKB approximation method 
for axion fluctuation we need to assume some specific structural form of the conformal time scale dependent axion decay constant.
Here we take the following parametric profile for the conformal time scale dependent axion decay constant:
\bea f_a= \sqrt{100-\frac{80}{1+\left(\ln\frac{\eta}{\eta_{c}}\right)^2}}~H.\eea
\begin{figure*}[htb]
\centering
\subfigure[Axion decay constant profile for $\eta_{c}=-1,-2,-3$.]{
    \includegraphics[width=7.2cm,height=6cm] {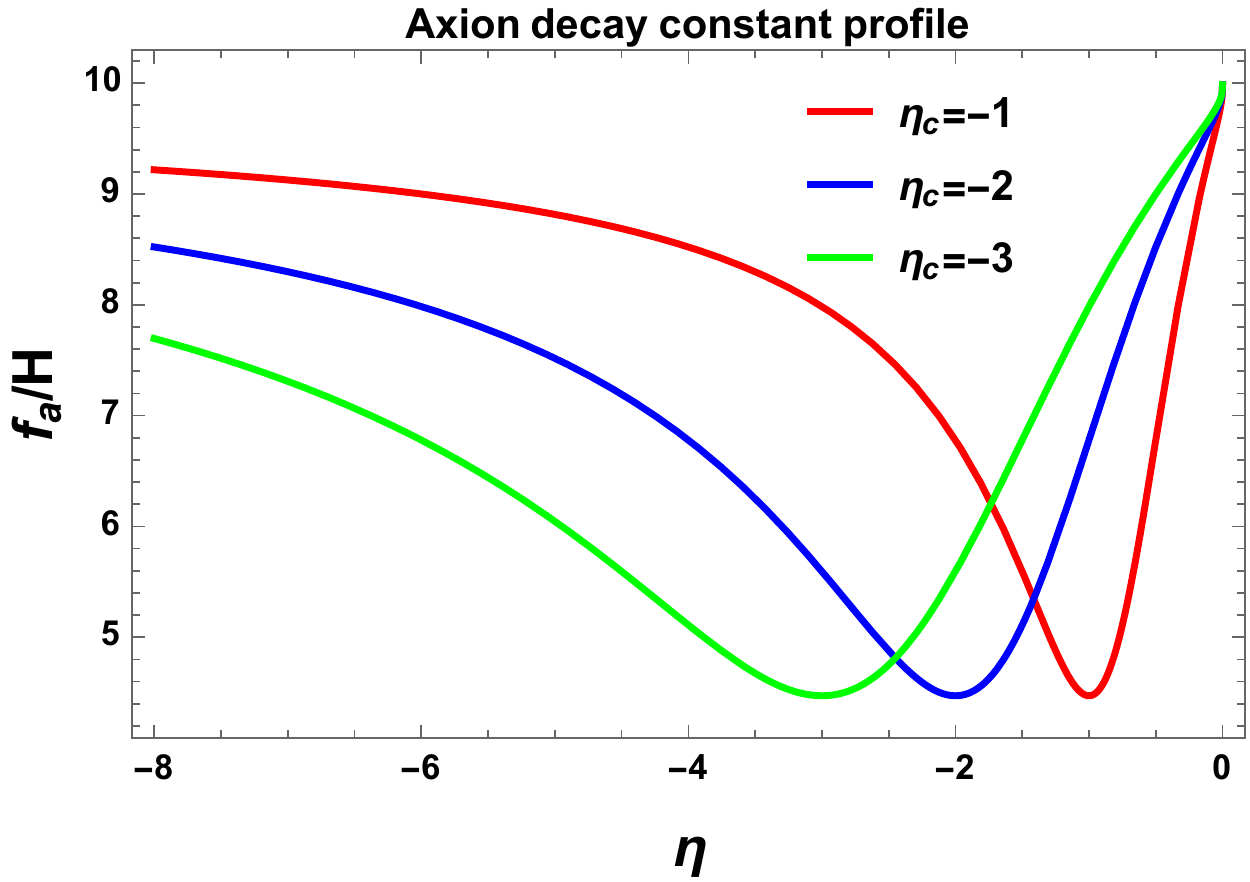}
    \label{fig1}
}
\subfigure[Axion decay constant profile for $\eta_{c}=1,2,3$.]{
    \includegraphics[width=7.2cm,height=6cm] {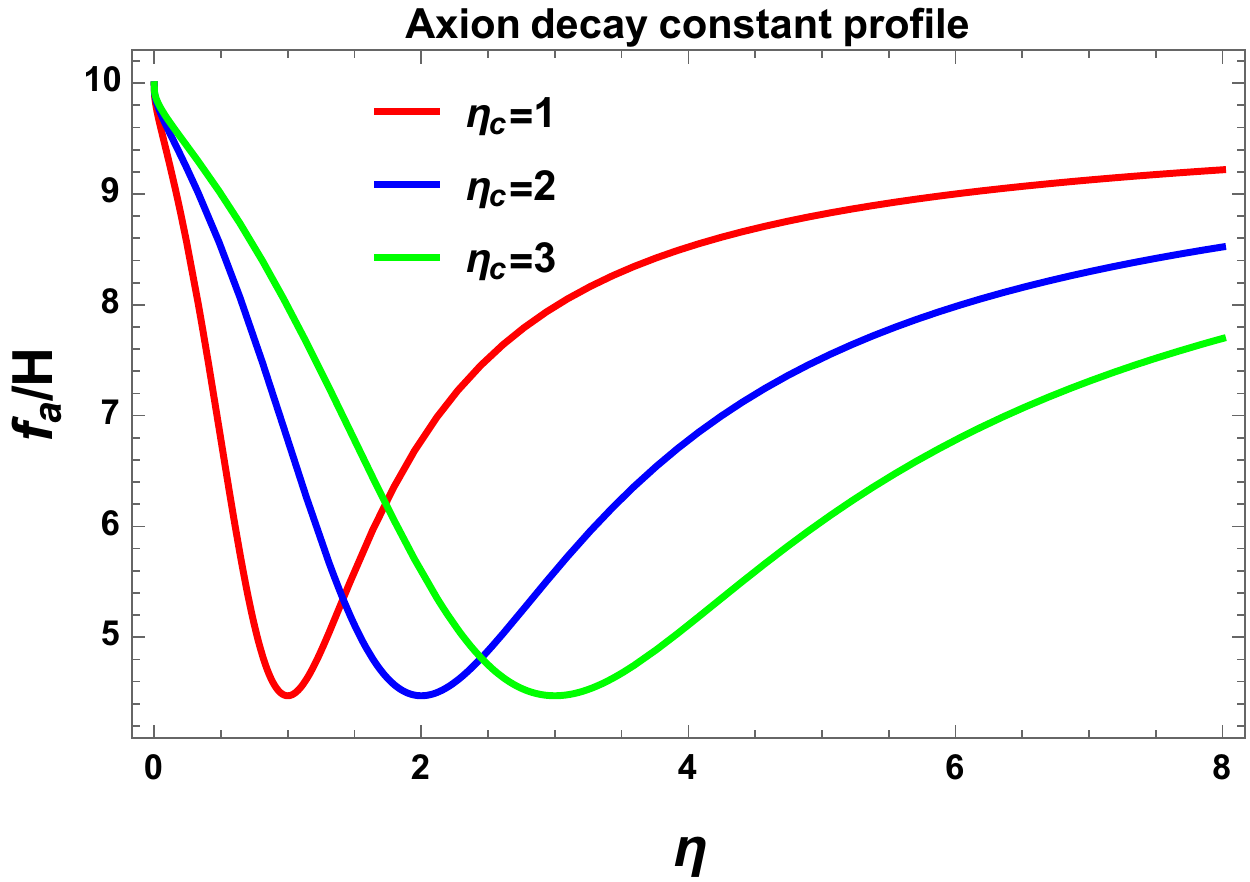}
    \label{fig1}
}
\subfigure[Axion mass profile for $\eta_{c}=-1,-2,-3$.]{
    \includegraphics[width=7.2cm,height=6cm] {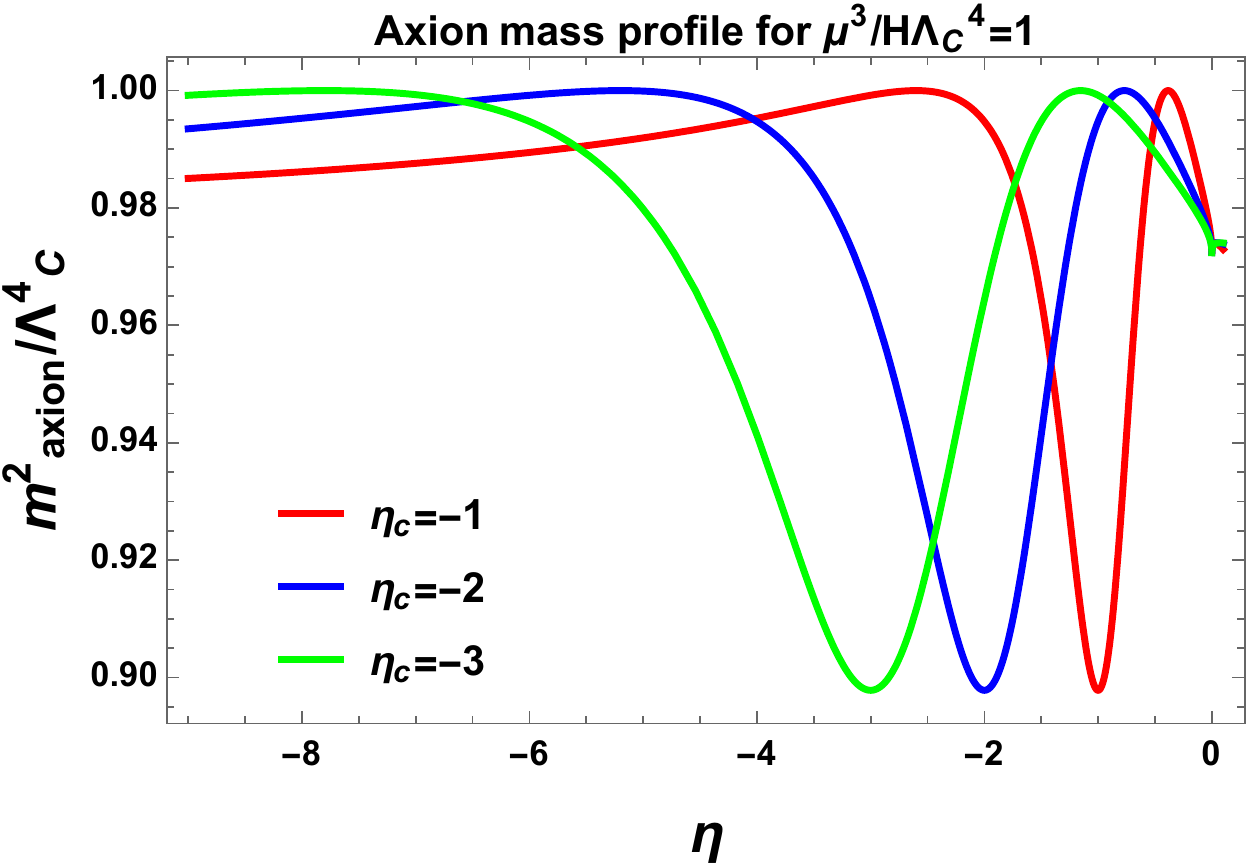}
    \label{fig2}
}
\subfigure[Axion mass profile for $\eta_{c}=1,2,3$.]{
    \includegraphics[width=7.2cm,height=6cm] {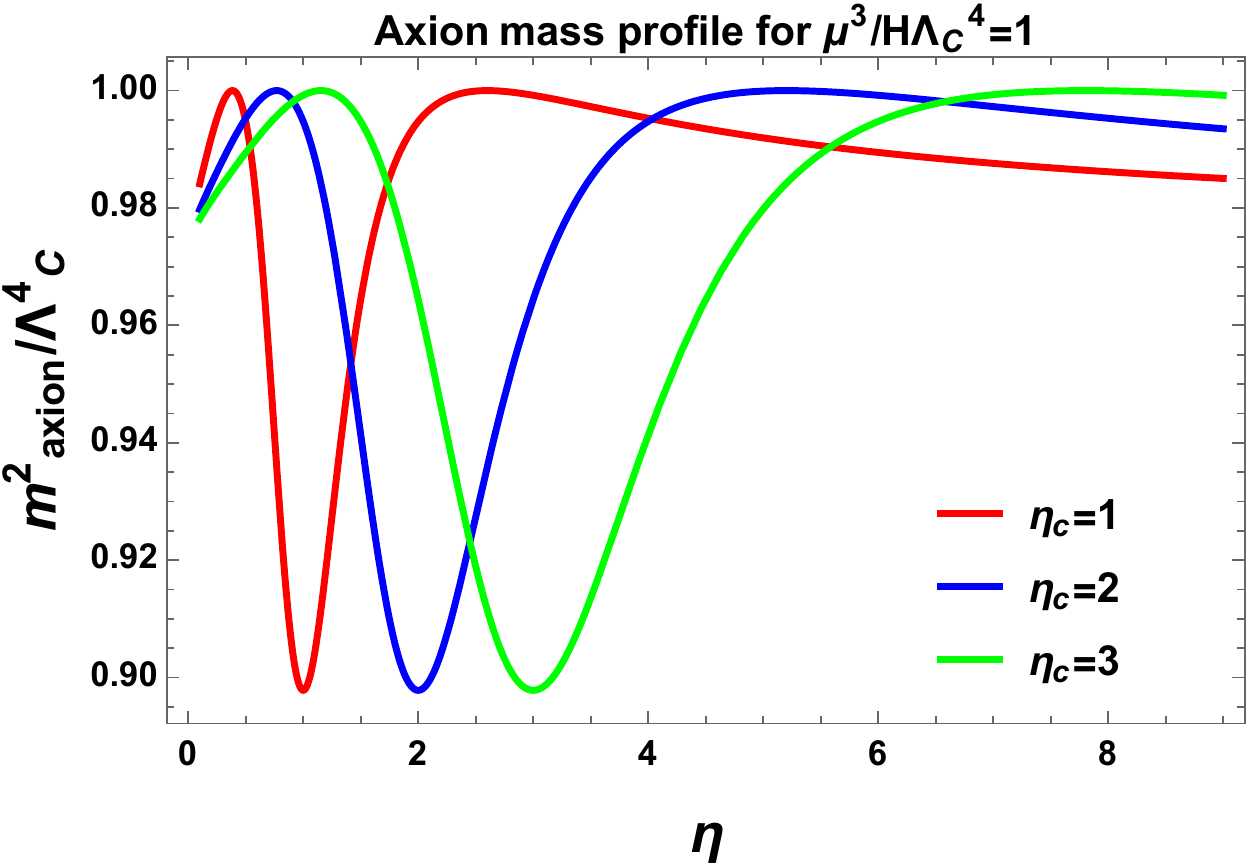}
    \label{fig2}
}
\subfigure[Axion mass parameter profile for $\eta_{c}=-1,-2,-3$.]{
    \includegraphics[width=7.2cm,height=6cm] {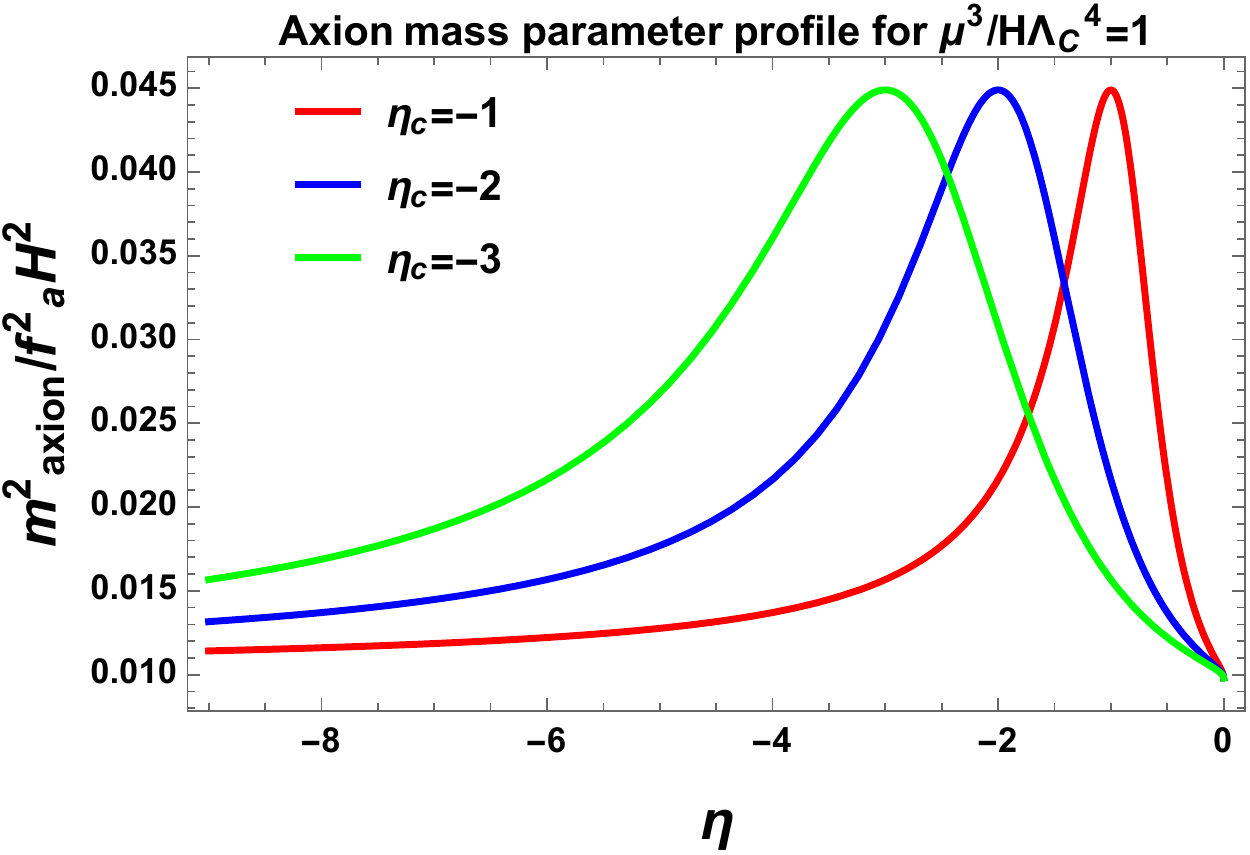}
    \label{fig3}
}
\subfigure[Axion mass parameter profile for $\eta_{c}=1,2,3$.]{
    \includegraphics[width=7.2cm,height=6cm] {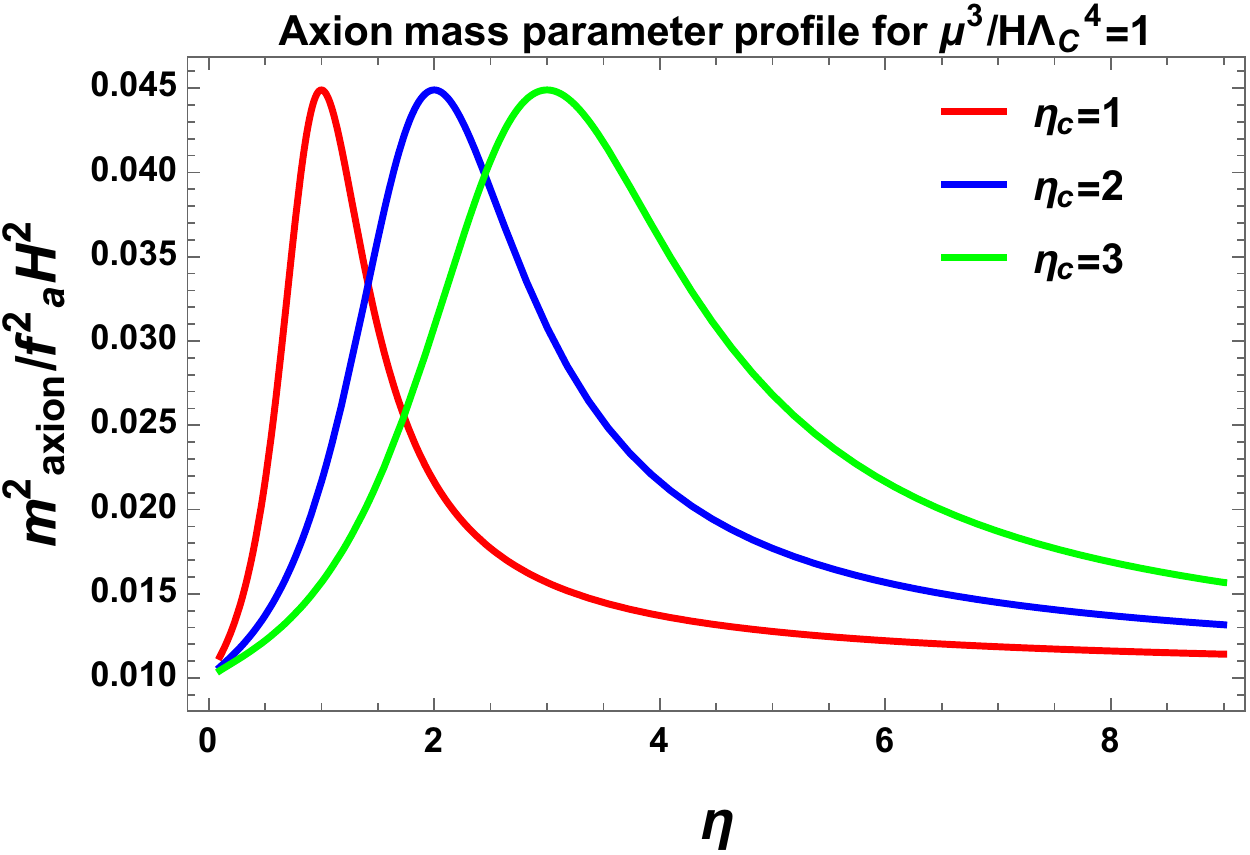}
    \label{fig3}
}
\caption[Optional caption for list of figures]{Conformal time scale dependent behaviour of axion decay constant $f_{a}/H$, axion mass $m^2_{axion}/\Lambda^4_C$ and axion mass parameter $m^2_{axion}/f^2_a H^2$ for a given profile.} 
\label{fza}
\end{figure*}
For this specific choice of the axion decay constant one can find the following characteristic features:
\begin{itemize}
 \item For very early times this is constant and we get $f_{a}\approx 10 H$.
 \item For very late times this is constant and we get $f_{a}\approx 10 H$.
 \item At conformal time scale $\eta \sim \eta_{c}$, it takes smaller value, $f_{a}\approx 2\sqrt{5} H$.
 \item For this specific choice Mukhanov-Sasaki variable can be computed as:
 \bea \frac{\partial^2_{\eta}\left(\frac{f^2_{a}}{H^2\eta^2}\right)}{\left(\frac{f^2_{a}}{H^2\eta^2}\right)}\approx 
\left\{\begin{array}{ll}
                    \displaystyle \frac{6}{\eta^2}~~~~ &
 \mbox{\small {\bf for ~$\eta\sim \eta_{c}$, early~\&~late~$\eta$}}  \\ 
	\displaystyle \frac{6+\Delta_{c}}{\eta^2} ~~~~ & \mbox{\small {\bf for~$\eta<\eta_{c}$}}.
          \end{array}
\right.\eea
\item Also for this case the axion mass parameter can be expressed as:
\bea \frac{m^2_{axion}}{f^2_a H^2}= -\frac{\Lambda^4_C }{\left[100-\frac{80}{1+\left(\ln\frac{\eta}{\eta_{c}}\right)^2}
\right]^2 H^4} \times \varSigma_{C}\eea
where $\varSigma_{C}$ is defined as:
\bea\varSigma_{C}=\left\{\begin{array}{ll}
                    \displaystyle\cos\left(\sin^{-1}\left(\frac{\mu^3 H}{
                    \Lambda^4_{C}}\sqrt{100-\frac{80}{1+\left(\ln\frac{\eta}{\eta_{c}}\right)^2}}\right)\right)~~~~ &
 \mbox{\small {\bf for ~total~$U(a)$}}  \\ 
	\displaystyle (-1)^{m} ~~~~ & \mbox{\small {\bf for~osc.~$U(a)$}}.
          \end{array}
\right.\eea
In this context we are interested in the following situations:
\begin{enumerate}
 \item At very early time scale.
 \item At very late time scale.
 \item At intermediate scale $\eta<\eta_c$.
 \item At the characteristic scale $\eta\sim \eta_{c}$.
\end{enumerate}
For all these four physical situations the axion mass parameter can be recast as:
\bea \underline{\bf For~total~U(a)}&& \nonumber\\
\frac{m^2_{axion}}{f^2_a H^2}:&\approx& 
\left\{\begin{array}{ll}
                    \displaystyle -\frac{\Lambda^4_C \times \cos\left(\sin^{-1}\left(\frac{10\mu^3 H}{
                    \Lambda^4_{C}}\right)\right)}{10000H^4} ~~~~ &
 \mbox{\small {\bf for ~early~\&~late~$\eta$}}  \\ 
	\displaystyle-\frac{\Lambda^4_C \times \cos\left(\sin^{-1}\left(\frac{\mu^3 H}{
                    \Lambda^4_{C}}\sqrt{100-\frac{80}{1+\omega_C}}\right)\right)}{\left[100-\frac{80}{1+\omega_C}
\right]^2 H^4}  ~~~~ & \mbox{\small {\bf for~$\eta<\eta_{c}$}}\\ 
	\displaystyle-\frac{\Lambda^4_C \times \cos\left(\sin^{-1}\left(\frac{2\sqrt{5}\mu^3 H}{
                    \Lambda^4_{C}}\right)\right)}{400H^4}  ~~~~ & \mbox{\small {\bf for~$\eta\sim\eta_{c}$}}.
          \end{array}
\right.\eea 
\bea \underline{\bf For~osc.~U(a)}&& \nonumber\\
\frac{m^2_{axion}}{f^2_a H^2}:&\approx& 
\left\{\begin{array}{ll}
                    \displaystyle -\frac{\Lambda^4_C }{10000 H^4} \times (-1)^{m}~~~~ &
 \mbox{\small {\bf for ~early~\&~late~$\eta$}}  \\ 
	\displaystyle-\frac{\Lambda^4_C }{\left[100-\frac{80}{1+\omega_C}
\right]^2 H^4} \times (-1)^{m} ~~~~ & \mbox{\small {\bf for~$\eta<\eta_{c}$}}\\ 
	\displaystyle-\frac{\Lambda^4_C }{400 H^4} \times (-1)^{m} ~~~~ & \mbox{\small {\bf for~$\eta\sim\eta_{c}$}}.
          \end{array}
\right.\eea
\end{itemize}
where $\omega_C$ is a small contribution. As for all the cases the axion mass parameter can be treated as constant factor one can recast the mode equations into the following form:
\bea
\vartheta^{''}_{\bf k}+
\left(k^2
+\frac{\left(\frac{m^2_{axion}}{f^2_a H^2}-6\right)}{\eta^2}\right)\vartheta_{\bf k} &=& 0~~~~~~~{\bf for~\eta\sim \eta_{c}, early~\&~late~\eta}\\ 
\vartheta^{''}_{\bf k}+
\left(k^2
+\frac{\left(\frac{m^2_{axion}}{f^2_a H^2}-6-\Delta_{c}\right)}{\eta^2}\right)\vartheta_{\bf k} &=& 0~~~~~~~{\bf for~\eta<\eta_{c}}.
\eea
The solutions for the mode function for the above mentioned cases can be expressed as:
\be\begin{array}{lll}\label{yusdsd1}
 \displaystyle \vartheta_{\bf k} (\eta) =\frac{f^2_{a}}{H^2\eta^2M^2_p}\bar{a}_{\bf k}\footnotesize=\left\{\begin{array}{ll}
                    \displaystyle   \sqrt{-\eta}\left[C_1  H^{(1)}_{\sqrt{\frac{25}{4}-\frac{m^2_{axion}}{f^2_a H^2}}} \left(-k\eta\right) 
+ C_2 H^{(2)}_{\sqrt{\frac{25}{4}-\frac{m^2_{axion}}{f^2_a H^2}}} \left(-k\eta\right)\right]~~~~ &
 \mbox{\small {\bf for~$\eta\sim \eta_{c}$, early~\&~late~$\eta$}}  \\ 
	\displaystyle \sqrt{-\eta}\left[C_1  H^{(1)}_{\sqrt{\frac{25}{4}+\Delta_{c}-\frac{m^2_{axion}}{f^2_a H^2}}} \left(-k\eta\right) 
+ C_2 H^{(2)}_{\sqrt{\frac{25}{4}+\Delta_{c}-\frac{m^2_{axion}}{f^2_a H^2}}} \left(-k\eta\right)\right]~~~~ & \mbox{\small {\bf for~$\eta<\eta_{c}$}}.
          \end{array}
\right.
\end{array}\ee
Here $C_{1}$ and $C_{2}$ are the arbitrary integration constants and the numerical value depend on the choice of the 
initial condition or more precisely the vacuum.

In the standard WKB approximation the total solution can be recast in the following form:
\bea\label{df3}
\vartheta_{\bf k} (\eta)&=& \frac{f^2_{a}}{H^2\eta^2M^2_p}\bar{a}_{\bf k}=\left[D_{1}u_{k}(\eta) + D_{2} \bar{u}_{k}(\eta)\right],\eea
where $D_{1}$ and and $D_{2}$ are two arbitrary integration constants, which depend on the 
choice of the initial condition during WKB approximation at early and late time scale. 
In the present context $u_{k}(\eta)$ and $\bar{u}_{k}(\eta)$ are defined as:
\bea 
u_{k}(\eta) &=&
                    \displaystyle  
                    \left\{\begin{array}{ll}
                    \displaystyle  \frac{\exp\left[i\int^{\eta}
                    d\eta^{\prime} \sqrt{k^2+\left(\frac{m^2_{axion}}{f^2_a H^2}-6\right)
                    \frac{1}{\eta^{'2}}} )\right]}
                    {\sqrt{2\sqrt{k^2+\left(\frac{m^2_{axion}}{f^2_a H^2}-6\right)\frac{1}{\eta^2}} }}\nonumber
                    ~~~~ &
 \mbox{\small {\bf for~$\eta\sim \eta_{c}$, early~\&~late~$\eta$}}  \\ 
	 \displaystyle  \frac{\exp\left[i\int^{\eta}
                    d\eta^{\prime} \sqrt{k^2+\left(\frac{m^2_{axion}}{f^2_a H^2}-6-\Delta_{c}\right)
                    \frac{1}{\eta^{'2}}} )\right]}
                    {\sqrt{2\sqrt{k^2+\left(\frac{m^2_{axion}}{f^2_a H^2}-6-\Delta_{c}\right)\frac{1}{\eta^2}} }}
                    
	 ~~~~ & \mbox{\small {\bf for~$\eta<\eta_{c}$}}.
          \end{array}
\right.\eea 
\bea 
u_{k}(\eta) &=&
                    \displaystyle  
                    \left\{\begin{array}{ll}
                    \displaystyle  \frac{\exp\left[i\int^{\eta}
                    d\eta^{\prime} \sqrt{k^2+\left(\frac{m^2_{axion}}{f^2_a H^2}-6\right)
                    \frac{1}{\eta^{'2}}} )\right]}
                    {\sqrt{2\sqrt{k^2+\left(\frac{m^2_{axion}}{f^2_a H^2}-6\right)\frac{1}{\eta^2}} }}\nonumber
                    ~~~~ &
 \mbox{\small {\bf for~$\eta\sim \eta_{c}$, early~\&~late~$\eta$}}  \\ 
	 \displaystyle  \frac{\exp\left[i\int^{\eta}
                    d\eta^{\prime} \sqrt{k^2+\left(\frac{m^2_{axion}}{f^2_a H^2}-6-\Delta_{c}\right)
                    \frac{1}{\eta^{'2}}} )\right]}
                    {\sqrt{2\sqrt{k^2+\left(\frac{m^2_{axion}}{f^2_a H^2}-6-\Delta_{c}\right)\frac{1}{\eta^2}} }}
                    
	 ~~~~ & \mbox{\small {\bf for~$\eta<\eta_{c}$}}.
          \end{array}
\right.\eea 
 Here in the most generalized situation
the new conformal time dependent factor $p(\eta)$ is defined as:
\be\begin{array}{lll}\label{soladfasx2}
 \displaystyle p(\eta) =\footnotesize\left\{\begin{array}{ll}
                    \displaystyle   \sqrt{k^2+\left(\frac{m^2_{axion}}{f^2_a H^2}-6\right)\frac{1}{\eta^2}}~~~~ &
 \mbox{\small {\bf for~$\eta\sim \eta_{c}$, early~\&~late~$\eta$}}  \\ 
	\displaystyle  \sqrt{k^2+\left(\frac{m^2_{axion}}{f^2_a H^2}-6-\Delta_{c}\right)\frac{1}{\eta^2}}~~~~ & \mbox{\small {\bf for~$\eta<\eta_{c}$}}.
          \end{array}
\right.
\end{array}\ee
which we use thoroughly in our computation. Here it is important to mention the expressions for the controlling 
factor $p(\eta)$ in different regime of solution:
\be\begin{array}{lll}\label{yu2dfvqasx3}\small
 \displaystyle \underline{\bf For~\eta\sim \eta_{c}, early~\&~late~\eta:}~~~~~~~~~p(\eta) \approx\footnotesize\left\{\begin{array}{ll}
                    \displaystyle  \sqrt{k^2-\frac{5}{\eta^2}}~~~~ &
 \mbox{\small {\bf for ~$m_{axion}/f_{a}\approx H$}}  \\ 
	\displaystyle \sqrt{k^2-
	\frac{6}{\eta^2}}~~~~ & \mbox{\small {\bf for ~$m_{axion}/f_{a}<< H$}}\\ 
	\displaystyle \sqrt{k^2+\left(\lambda^2-6\right)\frac{1}{\eta^2}}~~~~
	& \mbox{\small {\bf for ~$m_{axion}/f_{a}>> H$}}.
          \end{array}
\right.
\end{array}\ee
\\
\be\begin{array}{lll}\label{yu2dfveasx4}\small
 \displaystyle \underline{ \bf For~\eta<\eta_{c}:}~~~p(\eta) \approx\footnotesize\left\{\begin{array}{ll}
                    \displaystyle  \sqrt{k^2-\left(5+\Delta_{c}\right)\frac{1}{\eta^2}} &
 \mbox{\small {\bf for ~$m_{axion}/f_{a}\approx H$}}  \\ 
	\displaystyle \sqrt{k^2-\left(6+\Delta_{c}\right)\frac{1}{\eta^2}} 
	& \mbox{\small {\bf for ~$m_{axion}/f_{a}<< H$}}\\ 
	\displaystyle \sqrt{k^2+\left(\lambda^2-6-\Delta_{c}\right)\frac{1}{\eta^2}} 
	& \mbox{\small {\bf for ~$m_{axion}/f_{a}>> H$}}.
          \end{array}
\right.
\end{array}\ee
where for $m_{axion}/f_{a}>> H$ case we introduce a new parameter $\lambda$ as, $m_{axion}/f_{a}=\lambda H$ with $\lambda>>1$.
It is important to mention that the general expression for the Bogoliubov coefficient $\beta$ in Fourier 
space is given by the following approximation:
\be\begin{array}{lll}\label{soladfasx2}
 \displaystyle \beta(k) =\footnotesize\left\{\begin{array}{ll}
                    \displaystyle    \displaystyle   \int^{0}_{-\infty}d\eta~\frac{\left(\frac{m^2_{axion}}{f^2_a H^2}-6\right)^2\exp\left[2i\int^{\eta}_{-\infty}
d\eta^{'}\sqrt{k^2+\left(\frac{m^2_{axion}}{f^2_a H^2}-6\right)\frac{1}{\eta^{'2}}}\right]}{
                    4\eta^6\left(k^2+\left(\frac{m^2_{axion}}{f^2_a H^2}-6\right)\frac{1}{\eta^{2}}\right)^{5/2}}~~~~ &
 \mbox{\small {\bf for~$\eta\sim \eta_{c}$, early~\&~late~$\eta$}}  \\ 
	\displaystyle   \displaystyle    \int^{0}_{-\infty}d\eta~\frac{\left(\frac{m^2_{axion}}{f^2_a H^2}-6-\Delta_{c}\right)^2\exp\left[2i\int^{\eta}_{-\infty}
d\eta^{'}\sqrt{k^2+\left(\frac{m^2_{axion}}{f^2_a H^2}-6-\Delta_{c}\right)\frac{1}{\eta^{'2}}}\right]}{
                    4\eta^6\left(k^2+\left(\frac{m^2_{axion}}{f^2_a H^2}-6-\Delta_{c}\right)\frac{1}{\eta^{2}}\right)^{5/2}}~~~~ & \mbox{\small {\bf for~$\eta<\eta_{c}$}}.
          \end{array}
\right.
\end{array}\ee
One can use another equivalent way to define the the Bogoliubov coefficient $\beta$ in Fourier 
space by implementing instantaneous Hamiltonian diagonalization method in the present context. 
Using this diagonalized representation the regularized Bogoliubov coefficient $\beta$ in Fourier 
space can be written as:
\be\begin{array}{lll}\label{soladfasx2}
 \displaystyle \beta_{diag}(k;\tau,\tau^{'}) =\footnotesize\left\{\begin{array}{ll}
                    \displaystyle    \displaystyle   \int^{\tau}_{\tau^{'}}d\eta~\frac{-\left(\frac{m^2_{axion}}{f^2_a H^2}-6\right)\exp\left[-2i\int^{\eta}
d\eta^{'}\sqrt{k^2+\left(\frac{m^2_{axion}}{f^2_a H^2}-6\right)\frac{1}{\eta^{'2}}}\right]}{
                    2\eta^3\left(k^2+\left(\frac{m^2_{axion}}{f^2_a H^2}-6\right)\frac{1}{\eta^{2}}\right)}~~~~ &
 \mbox{\small {\bf for~$\eta\sim \eta_{c}$, early~\&~late~$\eta$}}  \\ 
	\displaystyle   \displaystyle    \int^{\tau}_{\tau^{'}}d\eta~\frac{-\left(\frac{m^2_{axion}}{f^2_a H^2}-6-\Delta_{c}\right)\exp\left[-2i\int^{\eta}
d\eta^{'}\sqrt{k^2+\left(\frac{m^2_{axion}}{f^2_a H^2}-6-\Delta_{c}\right)\frac{1}{\eta^{'2}}}\right]}{
                    2\eta^3\left(k^2+\left(\frac{m^2_{axion}}{f^2_a H^2}-6-\Delta_{c}\right)\frac{1}{\eta^{2}}\right)}~~~~ & \mbox{\small {\bf for~$\eta<\eta_{c}$}}.
          \end{array}
\right.
\end{array}\ee
where $\tau$ and $\tau^{'}$ introduced as the conformal time regulator in the present context.
We will also derive the expressions using Eq~(\ref{soladfasxsd3}) in the next two subsections.
In the next two subsection we will explicitly discuss two physical possibilities which captures the effect of 
massive particles in our computation.

Further using the expressions for Bogoliubov co-efficient $\beta$ in two different representations,
and substituting them in Eq~(\ref{sdq1}) we get the following
expressions for the Bogoliubov co-efficient $\alpha$ in two different representations as given by:
\be\begin{array}{lll}\label{soladfasx2}
 \displaystyle \alpha(k) =\tiny\left\{\begin{array}{ll}
                    \displaystyle    \displaystyle   \sqrt{\left[ 1+\left| \int^{0}_{-\infty}d\eta~\frac{\left(\frac{m^2_{axion}}{f^2_a H^2}-6\right)^2\exp\left[2i\int^{\eta}_{-\infty}
d\eta^{'}\sqrt{k^2+\left(\frac{m^2_{axion}}{f^2_a H^2}-6\right)\frac{1}{\eta^{'2}}}\right]}{
                    4\eta^6\left(k^2+\left(\frac{m^2_{axion}}{f^2_a H^2}-6\right)\frac{1}{\eta^{2}}\right)^{5/2}}\right|^2\right]}~e^{i\phi}&
 \mbox{\small {\bf for~$\eta\sim \eta_{c}$, early~\&~late~$\eta$}}  \\ 
	\displaystyle   \displaystyle     \sqrt{\left[ 1+\left| \int^{0}_{-\infty}d\eta~\frac{\left(\frac{m^2_{axion}}{f^2_a H^2}-6-\Delta_{c}\right)^2\exp\left[2i\int^{\eta}_{-\infty}
d\eta^{'}\sqrt{k^2+\left(\frac{m^2_{axion}}{f^2_a H^2}-6-\Delta_{c}\right)\frac{1}{\eta^{'2}}}\right]}{
                    4\eta^6\left(k^2+\left(\frac{m^2_{axion}}{f^2_a H^2}-6-\Delta_{c}\right)\frac{1}{\eta^{2}}\right)^{5/2}}\right|^2\right]}~e^{i\phi} & \mbox{\small {\bf for~$\eta<\eta_{c}$}}.
          \end{array}
\right.
\end{array}\ee
\be\begin{array}{lll}\label{soladfasx2}
 \displaystyle \alpha_{diag}(k;\tau,\tau^{'}) =\tiny\left\{\begin{array}{ll}
                    \displaystyle    \displaystyle   \sqrt{\left[ 1+\left| \int^{\tau}_{\tau^{'}}d\eta~\frac{-\left(\frac{m^2_{axion}}{f^2_a H^2}-6\right)\exp\left[-2i\int^{\eta}
d\eta^{'}\sqrt{k^2+\left(\frac{m^2_{axion}}{f^2_a H^2}-6\right)\frac{1}{\eta^{'2}}}\right]}{
                    2\eta^3\left(k^2+\left(\frac{m^2_{axion}}{f^2_a H^2}-6\right)\frac{1}{\eta^{2}}\right)}\right|^2\right]}~e^{i\phi}&
 \mbox{\small {\bf for~$\eta\sim \eta_{c}$, early~\&~late~$\eta$}}  \\ 
	\displaystyle   \displaystyle     \sqrt{\left[ 1+\left| \int^{\tau}_{\tau^{'}}d\eta~\frac{-\left(\frac{m^2_{axion}}{f^2_a H^2}-6-\Delta_{c}\right)\exp\left[-2i\int^{\eta}
d\eta^{'}\sqrt{k^2+\left(\frac{m^2_{axion}}{f^2_a H^2}-6-\Delta_{c}\right)\frac{1}{\eta^{'2}}}\right]}{
                    2\eta^3\left(k^2+\left(\frac{m^2_{axion}}{f^2_a H^2}-6-\Delta_{c}\right)\frac{1}{\eta^{2}}\right)}\right|^2\right]}~e^{i\phi} & \mbox{\small {\bf for~$\eta<\eta_{c}$}}.
          \end{array}
\right.
\end{array}\ee
where $\phi$ and $\phi_{diag}$ are the associated phase factors in two different representations.
Further using the expressions for Bogoliubov co-efficient $\alpha$ in two different representations as mentioned in
Eq~(\ref{soladfasxv1}) and Eq~(\ref{soladfasxsd3v2}), and substituting them in Eq~(\ref{sdq2}) we get the following
expressions for the reflection and transmission co-efficient in two different representations as given by:
\be\begin{array}{lll}\label{soladfasx2}
 \displaystyle {\cal R} =\tiny\left\{\begin{array}{ll}
                    \displaystyle      \frac{\displaystyle\int^{0}_{-\infty}d\eta~\frac{\left(\frac{m^2_{axion}}{f^2_a H^2}-6\right)^2\exp\left[2i\int^{\eta}_{-\infty}
d\eta^{'}\sqrt{k^2+\left(\frac{m^2_{axion}}{f^2_a H^2}-6\right)\frac{1}{\eta^{'2}}}\right]}{
                    4\eta^6\left(k^2+\left(\frac{m^2_{axion}}{f^2_a H^2}-6\right)\frac{1}{\eta^{2}}\right)^{5/2}}}{\sqrt{\displaystyle\left[ 1+\left| \int^{0}_{-\infty}d\eta~\frac{\left(\frac{m^2_{axion}}{f^2_a H^2}-6\right)^2\exp\left[2i\int^{\eta}_{-\infty}
d\eta^{'}\sqrt{k^2+\left(\frac{m^2_{axion}}{f^2_a H^2}-6\right)\frac{1}{\eta^{'2}}}\right]}{
                    4\eta^6\left(k^2+\left(\frac{m^2_{axion}}{f^2_a H^2}-6\right)\frac{1}{\eta^{2}}\right)^{5/2}}\right|^2\right]}}~e^{-i\phi} &
 \mbox{\small {\bf for~$\eta\sim \eta_{c}$, early~\&~late~$\eta$}}  \\ 
	\displaystyle   \displaystyle    \frac{\displaystyle\int^{0}_{-\infty}d\eta~\frac{\left(\frac{m^2_{axion}}{f^2_a H^2}-6-\Delta_{c}\right)^2\exp\left[2i\int^{\eta}_{-\infty}
d\eta^{'}\sqrt{k^2+\left(\frac{m^2_{axion}}{f^2_a H^2}-6-\Delta_{c}\right)\frac{1}{\eta^{'2}}}\right]}{
                    4\eta^6\left(k^2+\left(\frac{m^2_{axion}}{f^2_a H^2}-6-\Delta_{c}\right)\frac{1}{\eta^{2}}\right)^{5/2}}}{\sqrt{\displaystyle\left[ 1+\left| \int^{0}_{-\infty}d\eta~\frac{\left(\frac{m^2_{axion}}{f^2_a H^2}-6-\Delta_{c}\right)^2\exp\left[2i\int^{\eta}_{-\infty}
d\eta^{'}\sqrt{k^2+\left(\frac{m^2_{axion}}{f^2_a H^2}-6-\Delta_{c}\right)\frac{1}{\eta^{'2}}}\right]}{
                    4\eta^6\left(k^2+\left(\frac{m^2_{axion}}{f^2_a H^2}-6-\Delta_{c}\right)\frac{1}{\eta^{2}}\right)^{5/2}}\right|^2\right]}}~e^{-i\phi}  & \mbox{\small {\bf for~$\eta<\eta_{c}$}}.
          \end{array}
\right.
\end{array}\ee
\be\begin{array}{lll}\label{soladfasx2}
 \displaystyle {\cal T} =\tiny\left\{\begin{array}{ll}
                    \displaystyle      \frac{1}{\sqrt{\displaystyle\left[ 1+\left| \int^{0}_{-\infty}d\eta~\frac{\left(\frac{m^2_{axion}}{f^2_a H^2}-6\right)^2\exp\left[2i\int^{\eta}_{-\infty}
d\eta^{'}\sqrt{k^2+\left(\frac{m^2_{axion}}{f^2_a H^2}-6\right)\frac{1}{\eta^{'2}}}\right]}{
                    4\eta^6\left(k^2+\left(\frac{m^2_{axion}}{f^2_a H^2}-6\right)\frac{1}{\eta^{2}}\right)^{5/2}}\right|^2\right]}}~e^{-i\phi} &
 \mbox{\small {\bf for~$\eta\sim \eta_{c}$, early~\&~late~$\eta$}}  \\ 
	\displaystyle   \displaystyle    \frac{1}{\sqrt{\displaystyle\left[ 1+\left| \int^{0}_{-\infty}d\eta~\frac{\left(\frac{m^2_{axion}}{f^2_a H^2}-6-\Delta_{c}\right)^2\exp\left[2i\int^{\eta}_{-\infty}
d\eta^{'}\sqrt{k^2+\left(\frac{m^2_{axion}}{f^2_a H^2}-6-\Delta_{c}\right)\frac{1}{\eta^{'2}}}\right]}{
                    4\eta^6\left(k^2+\left(\frac{m^2_{axion}}{f^2_a H^2}-6-\Delta_{c}\right)\frac{1}{\eta^{2}}\right)^{5/2}}\right|^2\right]}}~e^{-i\phi}  & \mbox{\small {\bf for~$\eta<\eta_{c}$}}.
          \end{array}
\right.
\end{array}\ee
and 
\be\begin{array}{lll}\label{soladfasx2}
 \displaystyle {\cal R}_{diag}(k;\tau,\tau^{'}) =\tiny\left\{\begin{array}{ll}
                    \displaystyle      \frac{\displaystyle\int^{\tau}_{\tau^{'}}d\eta~\frac{-\left(\frac{m^2_{axion}}{f^2_a H^2}-6\right)\exp\left[-2i\int^{\eta}
d\eta^{'}\sqrt{k^2+\left(\frac{m^2_{axion}}{f^2_a H^2}-6\right)\frac{1}{\eta^{'2}}}\right]}{
                    2\eta^3\left(k^2+\left(\frac{m^2_{axion}}{f^2_a H^2}-6\right)\frac{1}{\eta^{2}}\right)}}{\sqrt{\displaystyle\left[ 1+\left| \int^{\tau}_{\tau^{'}}d\eta~\frac{-\left(\frac{m^2_{axion}}{f^2_a H^2}-6\right)\exp\left[-2i\int^{\eta}
d\eta^{'}\sqrt{k^2+\left(\frac{m^2_{axion}}{f^2_a H^2}-6\right)\frac{1}{\eta^{'2}}}\right]}{
                    2\eta^3\left(k^2+\left(\frac{m^2_{axion}}{f^2_a H^2}-6\right)\frac{1}{\eta^{2}}\right)}\right|^2\right]}}~e^{-i\phi} &
 \mbox{\small {\bf for~$\eta\sim \eta_{c}$, early~\&~late~$\eta$}}  \\ 
	\displaystyle   \displaystyle    \frac{\displaystyle\int^{\tau}_{\tau^{'}}d\eta~\frac{-\left(\frac{m^2_{axion}}{f^2_a H^2}-6-\Delta_{c}\right)\exp\left[-2i\int^{\eta}
d\eta^{'}\sqrt{k^2+\left(\frac{m^2_{axion}}{f^2_a H^2}-6-\Delta_{c}\right)\frac{1}{\eta^{'2}}}\right]}{
                    2\eta^3\left(k^2+\left(\frac{m^2_{axion}}{f^2_a H^2}-6-\Delta_{c}\right)\frac{1}{\eta^{2}}\right)}}{\sqrt{\displaystyle\left[ 1+\left| \int^{\tau}_{\tau^{'}}d\eta~\frac{-\left(\frac{m^2_{axion}}{f^2_a H^2}-6-\Delta_{c}\right)\exp\left[-2i\int^{\eta}
d\eta^{'}\sqrt{k^2+\left(\frac{m^2_{axion}}{f^2_a H^2}-6-\Delta_{c}\right)\frac{1}{\eta^{'2}}}\right]}{
                    2\eta^3\left(k^2+\left(\frac{m^2_{axion}}{f^2_a H^2}-6-\Delta_{c}\right)\frac{1}{\eta^{2}}\right)}\right|^2\right]}}~e^{-i\phi}  & \mbox{\small {\bf for~$\eta<\eta_{c}$}}.
          \end{array}
\right.
\end{array}\ee
\be\begin{array}{lll}\label{soladfasx2}
 \displaystyle {\cal T}_{diag}(k;\tau,\tau^{'}) =\tiny\left\{\begin{array}{ll}
                    \displaystyle      \frac{1}{\sqrt{\displaystyle\left[ 1+\left|\int^{\tau}_{\tau^{'}}d\eta~\frac{-\left(\frac{m^2_{axion}}{f^2_a H^2}-6-\Delta_{c}\right)\exp\left[-2i\int^{\eta}
d\eta^{'}\sqrt{k^2+\left(\frac{m^2_{axion}}{f^2_a H^2}-6-\Delta_{c}\right)\frac{1}{\eta^{'2}}}\right]}{
                    2\eta^3\left(k^2+\left(\frac{m^2_{axion}}{f^2_a H^2}-6-\Delta_{c}\right)\frac{1}{\eta^{2}}\right)}\right|^2\right]}}~e^{-i\phi} &
 \mbox{\small {\bf for~$\eta\sim \eta_{c}$, early~\&~late~$\eta$}}  \\ 
	\displaystyle   \displaystyle    \frac{1}{\sqrt{\displaystyle\left[ 1+\left| \int^{\tau}_{\tau^{'}}d\eta~\frac{-\left(\frac{m^2_{axion}}{f^2_a H^2}-6-\Delta_{c}\right)\exp\left[-2i\int^{\eta}
d\eta^{'}\sqrt{k^2+\left(\frac{m^2_{axion}}{f^2_a H^2}-6-\Delta_{c}\right)\frac{1}{\eta^{'2}}}\right]}{
                    2\eta^3\left(k^2+\left(\frac{m^2_{axion}}{f^2_a H^2}-6-\Delta_{c}\right)\frac{1}{\eta^{2}}\right)}\right|^2\right]}}~e^{-i\phi}  & \mbox{\small {\bf for~$\eta<\eta_{c}$}}.
          \end{array}
\right.
\end{array}\ee
Next the expression for the
number of produced particles at time $\tau$ can be calculated in the two representations using 
from the following formula as:
\be\begin{array}{lll}\label{soladfasx2}
 \displaystyle  {\cal N}(\tau,\tau^{'}) =\tiny\int \frac{d^{3}{\bf k}}{(2\pi a)^3}~\left\{\begin{array}{ll}
                   \left| \displaystyle    \displaystyle   \int^{0}_{-\infty}d\eta~\frac{\left(\frac{m^2_{axion}}{f^2_a H^2}-6\right)^2\exp\left[2i\int^{\eta}_{-\infty}
d\eta^{'}\sqrt{k^2+\left(\frac{m^2_{axion}}{f^2_a H^2}-6\right)\frac{1}{\eta^{'2}}}\right]}{
                    4\eta^6\left(k^2+\left(\frac{m^2_{axion}}{f^2_a H^2}-6\right)\frac{1}{\eta^{2}}\right)^{5/2}}\right|^2~~~~ &
 \mbox{\small {\bf for~$\eta\sim \eta_{c}$, early~\&~late~$\eta$}}  \\ 
	\left|\displaystyle   \displaystyle    \int^{0}_{-\infty}d\eta~\frac{\left(\frac{m^2_{axion}}{f^2_a H^2}-6-\Delta_{c}\right)^2\exp\left[2i\int^{\eta}_{-\infty}
d\eta^{'}\sqrt{k^2+\left(\frac{m^2_{axion}}{f^2_a H^2}-6-\Delta_{c}\right)\frac{1}{\eta^{'2}}}\right]}{
                    4\eta^6\left(k^2+\left(\frac{m^2_{axion}}{f^2_a H^2}-6-\Delta_{c}\right)\frac{1}{\eta^{2}}\right)^{5/2}}\right|^2~~~~ & \mbox{\small {\bf for~$\eta<\eta_{c}$}}.
          \end{array}
\right.
\end{array}\ee
\be\begin{array}{lll}\label{soladfasx2}
 \displaystyle {\cal N}_{diag}(\tau,\tau^{'}) =\tiny\int \frac{d^{3}{\bf k}}{(2\pi a)^3}\left\{\begin{array}{ll}
                    \left|\displaystyle    \displaystyle   \int^{\tau}_{\tau^{'}}d\eta~\frac{-\left(\frac{m^2_{axion}}{f^2_a H^2}-6\right)\exp\left[-2i\int^{\eta}
d\eta^{'}\sqrt{k^2+\left(\frac{m^2_{axion}}{f^2_a H^2}-6\right)\frac{1}{\eta^{'2}}}\right]}{
                    2\eta^3\left(k^2+\left(\frac{m^2_{axion}}{f^2_a H^2}-6\right)\frac{1}{\eta^{2}}\right)}\right|^2~~~~ &
 \mbox{\small {\bf for~$\eta\sim \eta_{c}$, early~\&~late~$\eta$}}  \\ 
	\left|\displaystyle   \displaystyle    \int^{\tau}_{\tau^{'}}d\eta~\frac{-\left(\frac{m^2_{axion}}{f^2_a H^2}-6-\Delta_{c}\right)\exp\left[-2i\int^{\eta}
d\eta^{'}\sqrt{k^2+\left(\frac{m^2_{axion}}{f^2_a H^2}-6-\Delta_{c}\right)\frac{1}{\eta^{'2}}}\right]}{
                    2\eta^3\left(k^2+\left(\frac{m^2_{axion}}{f^2_a H^2}-6-\Delta_{c}\right)\frac{1}{\eta^{2}}\right)}\right|^2~~~~ & \mbox{\small {\bf for~$\eta<\eta_{c}$}}.
          \end{array}
\right.
\end{array}\ee
Finally, one can define the total energy density of the produced particles using the following expression:
\be\begin{array}{lll}\label{soladfasx2}
   \rho(\tau,\tau^{'}) =\tiny\int \frac{d^{3}{\bf k}}{(2\pi a)^3a}~\tiny\left\{\begin{array}{ll}
                   \sqrt{k^2+\left(\frac{m^2_{axion}}{f^2_a H^2}-6\right)\frac{1}{\eta^2}}\\ \times
                   \left| \displaystyle    \displaystyle   \int^{0}_{-\infty}d\eta~\frac{\left(\frac{m^2_{axion}}{f^2_a H^2}-6\right)^2\exp\left[2i\int^{\eta}_{-\infty}
d\eta^{'}\sqrt{k^2+\left(\frac{m^2_{axion}}{f^2_a H^2}-6\right)\frac{1}{\eta^{'2}}}\right]}{
                    4\eta^6\left(k^2+\left(\frac{m^2_{axion}}{f^2_a H^2}-6\right)\frac{1}{\eta^{2}}\right)^{5/2}}\right|^2 &
 \mbox{\small {\bf for~$\eta\sim \eta_{c}$, early~\&~late~$\eta$}}  \\ 
	\sqrt{k^2+\left(\frac{m^2_{axion}}{f^2_a H^2}-6-\Delta_{c}\right)\frac{1}{\eta^2}}\\ \times\left|\displaystyle   \displaystyle    \int^{0}_{-\infty}d\eta~\frac{\left(\frac{m^2_{axion}}{f^2_a H^2}-6-\Delta_{c}\right)^2\exp\left[2i\int^{\eta}_{-\infty}
d\eta^{'}\sqrt{k^2+\left(\frac{m^2_{axion}}{f^2_a H^2}-6-\Delta_{c}\right)\frac{1}{\eta^{'2}}}\right]}{
                    4\eta^6\left(k^2+\left(\frac{m^2_{axion}}{f^2_a H^2}-6-\Delta_{c}\right)\frac{1}{\eta^{2}}\right)^{5/2}}\right|^2 & \mbox{\small {\bf for~$\eta<\eta_{c}$}}.
          \end{array}
\right.
\end{array}\ee
\be\begin{array}{lll}\label{soladfasx2}
 \rho_{diag}(\tau,\tau^{'},\eta^{'}) =\tiny\int \frac{d^{3}{\bf k}}{(2\pi a)^3a}\left\{\begin{array}{ll}
                    \sqrt{k^2+\left(\frac{m^2_{axion}}{f^2_a H^2}-6\right)\frac{1}{\eta^2}}\\ \times\left|\displaystyle    \displaystyle   \int^{\tau}_{\tau^{'}}d\eta~\frac{-\left(\frac{m^2_{axion}}{f^2_a H^2}-6\right)\exp\left[-2i\int^{\eta}
d\eta^{'}\sqrt{k^2+\left(\frac{m^2_{axion}}{f^2_a H^2}-6\right)\frac{1}{\eta^{'2}}}\right]}{
                    2\eta^3\left(k^2+\left(\frac{m^2_{axion}}{f^2_a H^2}-6\right)\frac{1}{\eta^{2}}\right)}\right|^2&
 \mbox{\small {\bf for~$\eta\sim \eta_{c}$, early~\&~late~$\eta$}}  \\ 
	\sqrt{k^2+\left(\frac{m^2_{axion}}{f^2_a H^2}-6-\Delta_{c}\right)\frac{1}{\eta^2}}\\ \times\left|\displaystyle   \displaystyle    \int^{\tau}_{\tau^{'}}d\eta~\frac{-\left(\frac{m^2_{axion}}{f^2_a H^2}-6-\Delta_{c}\right)\exp\left[-2i\int^{\eta}
d\eta^{'}\sqrt{k^2+\left(\frac{m^2_{axion}}{f^2_a H^2}-6-\Delta_{c}\right)\frac{1}{\eta^{'2}}}\right]}{
                    2\eta^3\left(k^2+\left(\frac{m^2_{axion}}{f^2_a H^2}-6-\Delta_{c}\right)\frac{1}{\eta^{2}}\right)}\right|^2 & \mbox{\small {\bf for~$\eta<\eta_{c}$}}.
          \end{array}
\right.
\end{array}\ee
\subsubsection{\bf Case I: $m_{axion}/f_{a} \approx H$}
Equation of motion for the axion field for $m_{axion}/f_{a} \approx H$ case is given by:
\bea
\vartheta''_k + \left\{k^2 - \frac{5}{\eta^2} \right\} \vartheta_k &=& 0~~~~~~~{\bf for~\eta\sim \eta_{c}, early~\&~late~\eta}\\
\vartheta''_k + \left\{k^2 - \left[5+\Delta_{c}\right]
\frac{1}{\eta^2} \right\} \vartheta_k &=& 0~~~~~~~{\bf for~\eta<\eta_{c}}.
\eea
The solution for the mode function for de Sitter and quasi de Sitter space can be expressed as: 
\be\begin{array}{lll}\label{yu2zx1}
 \displaystyle \vartheta_k (\eta) =\left\{\begin{array}{ll}
                    \displaystyle   \sqrt{-\eta}\left[C_1  H^{(1)}_{\sqrt{21}/2} \left(-k\eta\right) 
+ C_2 H^{(2)}_{\sqrt{21}/2} \left(-k\eta\right)\right]~~~~ &
 \mbox{\small {\bf for ~$\eta\sim \eta_{c}$, early~\&~late~$\eta$}}  \\ 
	\displaystyle \sqrt{-\eta}\left[C_1  H^{(1)}_{\sqrt{\frac{21}{4}+\Delta_{c}}} \left(-k\eta\right) 
+ C_2 H^{(2)}_{\sqrt{\frac{21}{4}+\Delta_{c}}} \left(-k\eta\right)\right]~~~~ & \mbox{\small {\bf for~$\eta<\eta_{c}$}}.
          \end{array}
\right.
\end{array}\ee
where $C_{1}$ and and $C_{2}$ are two arbitrary integration constant, which depend on the 
choice of the initial condition.

After taking $k\eta\rightarrow -\infty$, $k\eta\rightarrow 0$ and $|k\eta|\approx 1-\Delta(\rightarrow 0)$ limit the most general 
solution as stated in Eq~(\ref{yu2zx1}) can be recast as:
\bea\label{yu2zxx}
 \displaystyle \vartheta_k (\eta) &\stackrel{|k\eta|\rightarrow-\infty}{=}&\left\{\begin{array}{ll}
                    \displaystyle   \sqrt{\frac{2}{\pi k}}\left[C_1  e^{ -ik\eta}
e^{-\frac{i\pi}{2}\left(\frac{\sqrt{21}+1}{2}\right)} 
+ C_2 e^{ ik\eta}
e^{\frac{i\pi}{2}\left(\frac{\sqrt{21}+1}{2}\right)}\right]~~~~ &
 \mbox{\small {\bf for ~$\eta\sim \eta_{c}$, early~\&~late~$\eta$}} \\ 
	\displaystyle \sqrt{\frac{2}{\pi k}}\left[C_1  e^{ -ik\eta}
e^{-\frac{i\pi}{2}\left(\sqrt{\frac{21}{4}+\Delta_{c}}+\frac{1}{2}\right)} 
+ C_2 e^{ ik\eta}
e^{\frac{i\pi}{2}\left(\sqrt{\frac{21}{4}+\Delta_{c}}+\frac{1}{2}\right)}\right]~~~~ & \mbox{\small {\bf for~$\eta<\eta_{c}$}}.
          \end{array}
\right.\eea 
\bea
\label{yu2xxx}
 \displaystyle \vartheta_k (\eta) &\stackrel{|k\eta|\rightarrow 0}{=}&\left\{\begin{array}{ll}
                    \displaystyle  \frac{i\sqrt{-\eta}}{\pi}\Gamma\left(\frac{\sqrt{21}}{2}\right)
                    \left(-\frac{k\eta}{2}\right)^{-\frac{\sqrt{21}}{2}}\left[C_1   
- C_2 \right]~ &
 \mbox{\small {\bf for ~$\eta\sim \eta_{c}$, early~\&~late~$\eta$}}  \\ 
	\displaystyle\frac{i\sqrt{-\eta}}{\pi}\Gamma\left(\sqrt{\frac{21}{4}+\Delta_{c}}\right)\displaystyle
	\left(-\frac{k\eta}{2}\right)^{-\sqrt{\frac{21}{4}+\Delta_{c}}}\left[C_1   
- C_2 \right]~ & \mbox{\small {\bf for~$\eta<\eta_{c}$}}.
          \end{array}
\right.
\eea
\bea
\label{yu2}
 \displaystyle \vartheta_k (\eta) &\stackrel{|k\eta|\approx 1-\Delta(\rightarrow 0)}{=}&\left\{\begin{array}{ll}
                    \displaystyle  \frac{i}{\pi}\sqrt{-\eta}\left[ \frac{2}{\sqrt{21}}-\gamma+\frac{\sqrt{21}}{4}
                   \left(\gamma^2+\frac{\pi^2}{6}\right)\right.\\ \displaystyle \left.
                   \displaystyle~~~~~~~~-\frac{7}{8}\left(\gamma^3+\frac{\gamma \pi^2}{2}+2\zeta(3)\right)
                    +\cdots\right]\\ \displaystyle ~~~~~~~~~~~~\times\left(\frac{1+\Delta}{2}\right)^{-\frac{\sqrt{21}}{2}}\left[C_1   
- C_2 \right]~&
 \mbox{\small {\bf for ~$\eta\sim \eta_{c}$, early~\&~late~$\eta$}}  \\ 
	\displaystyle\frac{i}{\pi}\sqrt{-\eta}\left[ \frac{1}{ \left\{\sqrt{\frac{21}{4}+\Delta_{c}}\right\}}
  -\gamma  \displaystyle +\frac{1}{2}\left(\gamma^2+\frac{\pi^2}{6}\right) \left\{\sqrt{\frac{21}{4}+\Delta_{c}}\right\}
  \right.\\ \displaystyle \left. \displaystyle~-\frac{1}{6}\left(\gamma^3+\frac{\gamma \pi^2}{2}+2\zeta(3)\right) \left\{\sqrt{\frac{21}{4}+\Delta_{c}}\right\}^2
  +\cdots\right]\\ \displaystyle ~~~~~~~~~~~~\times\left(\frac{1+\Delta}{2}\right)^{-\sqrt{\frac{21}{4}+\Delta_{c}}}\left[C_1   
- C_2 \right]~ & \mbox{\small {\bf for~$\eta<\eta_{c}$}}.
          \end{array}
\right.
\eea
Next we assume that the WKB approximation is approximately valid for all times for the solution for the mode function $\vartheta_k$.
In the standard WKB approximation the total solution can be recast in the following form:
\bea\label{df3}
\vartheta_k (\eta)&=& \left[D_{1}u_{k}(\eta) + D_{2} \bar{u}_{k}(\eta)\right],\eea
where $D_{1}$ and and $D_{2}$ are two arbitrary integration constant, which depend on the 
choice of the initial condition during WKB approximation at early and late time scale.
In the present context $u_{k}(\eta)$ and $\bar{u}_{k}(\eta)$ are defined as:
\be\begin{array}{lll}\label{solp}
 \displaystyle\small u_{k}(\eta) =\footnotesize\left\{\begin{array}{ll}
                    \displaystyle   \frac{1}{\sqrt{2\sqrt{k^2 - \frac{5}{\eta^2} }}}
\exp\left[i\int^{\eta} d\eta^{\prime} \sqrt{k^2-\frac{5}{\eta^{'2}}}\right]\\
= \displaystyle\frac{1}{\sqrt{2\sqrt{k^2 - \frac{5}{\eta^2} }}} 
\exp\left[i\left(\eta\sqrt{k^2-\frac{5}{\eta^{2}}}+tan^{-1}\left[\frac{\sqrt{5}}{\eta\sqrt{k^2-\frac{5}{\eta^{2}}}}\right]
\right)\right] &
 \mbox{\small {\bf for ~$\eta\sim \eta_{c}$, early~\&~late~$\eta$}}  \\
	\displaystyle \frac{1}{\sqrt{2\sqrt{k^2-\frac{\left[5+\Delta_{c}\right]}{\eta^2}}}}
\exp\left[i\int^{\eta} d\eta^{\prime} \sqrt{k^2-\frac{\left[5+\Delta_{c}\right]}{\eta^{'2}}}\right]\\
= \displaystyle\frac{1}{\sqrt{2\sqrt{k^2-\frac{\left[5+\Delta_{c}\right]}{\eta^2}}}} 
\exp\left[i\left(\eta\sqrt{k^2-\frac{\left[5+\Delta_{c}\right]}{\eta^{2}}}
\displaystyle +\sqrt{5+\Delta_{c}}~tan^{-1}\left[\frac{\sqrt{5+\Delta_{c}}}{\eta\sqrt{k^2-\frac{\left[5+\Delta_{c}\right]}{\eta^{2}}}}\right]
\right)\right] & \mbox{\small {\bf for~$\eta<\eta_{c}$}}.
          \end{array}
\right.
\end{array}\ee
\\
\be\begin{array}{lll}\label{solp}
 \displaystyle\small \bar{u}_{k}(\eta) =\footnotesize\left\{\begin{array}{ll}
                    \displaystyle   \frac{1}{\sqrt{2\sqrt{k^2 - \frac{5}{\eta^2} }}}
\exp\left[i\int^{\eta} d\eta^{\prime} \sqrt{k^2-\frac{5}{\eta^{'2}}}\right]\\
= \displaystyle\frac{1}{\sqrt{2\sqrt{k^2 - \frac{5}{\eta^2} }}} 
\exp\left[i\left(\eta\sqrt{k^2-\frac{5}{\eta^{2}}}+tan^{-1}\left[\frac{\sqrt{5}}{\eta\sqrt{k^2-\frac{5}{\eta^{2}}}}\right]
\right)\right] &
 \mbox{\small {\bf for ~$\eta\sim \eta_{c}$, early~\&~late~$\eta$}}  \\
	\displaystyle \frac{1}{\sqrt{2\sqrt{k^2-\frac{\left[5+\Delta_{c}\right]}{\eta^2}}}}
\exp\left[i\int^{\eta} d\eta^{\prime} \sqrt{k^2-\frac{\left[5+\Delta_{c}\right]}{\eta^{'2}}}\right]\\
= \displaystyle\frac{1}{\sqrt{2\sqrt{k^2-\frac{\left[5+\Delta_{c}\right]}{\eta^2}}}} 
\exp\left[i\left(\eta\sqrt{k^2-\frac{\left[5+\Delta_{c}\right]}{\eta^{2}}}
\displaystyle +\sqrt{5+\Delta_{c}}~tan^{-1}\left[\frac{\sqrt{5+\Delta_{c}}}{\eta\sqrt{k^2-\frac{\left[5+\Delta_{c}\right]}{\eta^{2}}}}\right]
\right)\right] & \mbox{\small {\bf for~$\eta<\eta_{c}$}}.
          \end{array}
\right.
\end{array}\ee
In the present context the Bogoliubov coefficient $\beta(k)$ in Fourier space
can be computed approximately using the following expression:
\be\begin{array}{lll}\label{soladf129}
 \displaystyle \beta(k,\tau,\tau^{'},\eta^{'}) =\left\{\begin{array}{ll}
                    \displaystyle   \int^{\tau}_{\tau^{'}}d\eta~\frac{25}{
                    4\eta^{6}\left(k^2-\frac{5}{\eta^{2}}\right)^{\frac{5}{2}}}\exp\left[2i\int^{\eta}_{\eta^{'}}
d\eta^{''}\sqrt{k^2-\frac{5}{\eta^{''2}}}\right]&
 \mbox{\small {\bf for ~$\eta\sim \eta_{c}$, early~\&~late~$\eta$}}  \\ 
	\displaystyle   \int^{\tau}_{\tau^{'}}d\eta~\frac{\left[5+\Delta_{c}\right]^2}{
                    4\eta^{6}\left(k^2-\frac{\left[5+\Delta_{c}\right]}{\eta^{2}}\right)^{\frac{5}{2}}}
                    \exp\left[2i\int^{\eta}_{\eta^{'}}
d\eta^{''}\sqrt{k^2-\frac{\left[5+\Delta_{c}\right]}{\eta^{''2}}}\right]& \mbox{\small {\bf for~$\eta<\eta_{c}$}}.
          \end{array}
\right.
\end{array}\ee
which is not exactly analytically computable. To study the behaviour of this integral we consider here three 
consecutive physical situations-$|k\eta|<<1$, $|k\eta|\approx 1-\Delta(\rightarrow 0)$ and $|k\eta|>>1$ for de Sitter and quasi de Sitter case. 
In three cases we have:
\be\begin{array}{lll}\label{yu2dfvqasx3}\small
 \displaystyle \underline{\rm \bf For~\eta\sim \eta_{c}, early~\&~late~\eta:}~~~~~~~~~\sqrt{\left\{k^2 - \frac{5}{\eta^2} \right\}} \approx\left\{\begin{array}{ll}
                    \displaystyle  \frac{i\sqrt{5}}{\eta}~~~~ &
 \mbox{\small {\bf for ~$|k\eta|<<1$}}  \\ 
	\displaystyle \frac{i\sqrt{2\Delta+4}}{\eta}~~~~ & \mbox{\small
	{\bf for ~$|k\eta|\approx 1-\Delta(\rightarrow 0)$}}\\ 
	\displaystyle k~~~~ & \mbox{\small {\bf for ~$|k\eta|>>1$}}.
          \end{array}
\right.
\end{array}
\ee
\be\begin{array}{lll}\label{yu2dfvqasx3zxxbbb}
\small
 \displaystyle\underline{\rm \bf For~\eta<\eta_{c}:}~~~~~~~~~\sqrt{\left\{k^2 - \left[5+\Delta_{c}\right]
\frac{1}{\eta^2} \right\}} \approx\left\{\begin{array}{ll}
                    \displaystyle \frac{i\sqrt{\left[5+\Delta_{c}\right]}}{\eta} ~~~~ &
 \mbox{\small {\bf for ~$|k\eta|<<1$}}  \\ 
	\displaystyle \frac{i\sqrt{2\Delta+4+\Delta_{c}}}{\eta}~~~~ 
	& \mbox{\small {\bf for ~$|k\eta|\approx 1-\Delta(\rightarrow 0)$}}\\ 
	\displaystyle k~~~~ & \mbox{\small {\bf for ~$|k\eta|>>1$}}.
          \end{array}
\right.
\end{array}\ee
and further using this result Bogoliubov coefficient $\beta(k)$ in Fourier space can be expressed as:
\bea &&\underline{\bf For~\eta\sim \eta_{c}, early~\&~late~\eta:}\\
\small\beta(k,\tau,\tau^{'},\eta^{'})&=& \footnotesize\left\{\begin{array}{ll}
                    \displaystyle  \frac{5^{3/4}\eta^{'2\sqrt{5}}}{
                    4\left(\frac{5}{2}+2\sqrt{5}\right)i}\left[\frac{1}{\tau^{'\frac{5}{2}+2\sqrt{5}}}-\frac{1}{\tau^{\frac{5}{2}+2\sqrt{5}}}\right]
                     &
 \mbox{\small {\bf for ~$|k\eta|<<1$}}  \\ 
 \displaystyle  \frac{25\eta^{'2\sqrt{2\Delta+4}}}{
                    4(2\Delta+4)^{5/4}\left(\frac{5}{2}+2\sqrt{2\Delta+4}\right)i}\left[\frac{1}{\tau^{'\frac{5}{2}+2\sqrt{2\Delta+4}}}-\frac{1}{\tau^{\frac{5}{2}+2\sqrt{2\Delta+4}}}\right] &
 \mbox{\small {\bf for ~$|k\eta|\approx 1-\Delta(\rightarrow 0)$}}  \\ 
	\displaystyle 25\left[i\frac{\text{Ei}(2 ik \eta)e^{-2ik\eta^{'}}}{15}
	\displaystyle-\frac{e^{2 i k (\eta-\eta^{'})} }{120 k^5 \eta^5}\left(
	4 k^4 \eta^4-2 i k^3 \eta^3
	-2 k^2 \eta^2+3 i k \eta+6\right)\right]^{\tau}_{\tau^{'}}& \mbox{\small {\bf for ~$|k\eta|>>1$}}.
          \end{array}
\right.\nonumber\eea
\bea
&&\underline{\bf For~\eta<\eta_{c}:}\\
\small\beta(k,\tau,\tau^{'},\eta^{'})&=& \footnotesize\left\{\begin{array}{ll}
                     \frac{(5+\Delta_{c})^2\eta^{'2\sqrt{5+\Delta_{c}}}}{
                    4i\left(5+\Delta_{c}\right)^{5/4}\left(\frac{5}{2}+2\sqrt{5+\Delta_{c}}\right)}\left[\frac{1}{\tau^{'\frac{5}{2}+2\sqrt{5+\Delta_{c}}}}-\frac{1}{\tau^{\frac{5}{2}+2\sqrt{5+\Delta_{c}}}}\right]&
 \mbox{\small {\bf for ~$|k\eta|<<1$}}  \\ 
 \frac{(5+\Delta_{c})^2\eta^{'2\sqrt{2\Delta+4+\Delta_{c}}}}{
                    4(2\Delta+4+\Delta_{c})^{5/4}\left(\frac{5}{2}+2\sqrt{2\Delta+4+\Delta_{c}}\right)i}\left[\frac{1}{\tau^{'\frac{5}{2}+2\sqrt{2\Delta+4+\Delta_{c}}}}-\frac{1}{\tau^{\frac{5}{2}+2\sqrt{2\Delta+4+\Delta_{c}}}}\right]
                    &
 \mbox{\small {\bf for ~$|k\eta|\approx 1-\Delta(\rightarrow 0)$}} \\
	\displaystyle \left[5+\Delta_{c}\right]^2 \left[i\frac{\text{Ei}(2 ik \eta)e^{-2ik\eta^{'}}}{15}
	-\frac{e^{2 i k (\eta-\eta^{'})} }{120 k^5 \eta^5}\left(
	4 k^4 \eta^4-2 i k^3 \eta^3
	-2 k^2 \eta^2+3 i k \eta+6\right)\right]^{\tau}_{\tau^{'}} & \mbox{\small {\bf for ~$|k\eta|>>1$}}.
          \end{array}
\right.\nonumber
\eea
As mentioned earlier here one can use another equivalent way
to define the the Bogoliubov coefficient $\beta$ in Fourier 
space by implementing instantaneous Hamiltonian
diagonalization method to interpret the results. 
Using this diagonalized representation the
regularized Bogoliubov coefficient $\beta$ in Fourier 
space can be written as:
\be\begin{array}{lll}\label{soladfasxsd3cv129}
 \displaystyle \beta_{diag}(k;\tau,\tau^{'}) =\left\{\begin{array}{ll}
                    \displaystyle   \int^{\tau}_{\tau^{'}}d\eta~\frac{5\exp\left[-2i\int^{\eta}
d\eta^{'}\sqrt{k^2-\frac{5}{\eta^{'2}}}\right]}{
                    2\eta^{3}\left(k^2-\frac{5}{\eta^{2}}\right)}~~~~ &
 \mbox{\small {\bf for ~$\eta\sim \eta_{c}$, early~\&~late~$\eta$}}  \\ 
	\displaystyle  \int^{\tau}_{\tau^{'}}d\eta~\frac{\left[5+\Delta_{c}\right]\exp\left[-2i\int^{\eta}
d\eta^{'}\sqrt{k^2-\frac{\left[5+\Delta_{c}\right]}{\eta^{'2}}}\right]}{
                    2\eta^{3}\left(k^2-\frac{\left[5+\Delta_{c}\right]}{\eta^{2}}\right)}
                    ~~~~ & \mbox{\small {\bf for~ $\eta<\eta_{c}$}}.
          \end{array}
\right.
\end{array}\ee
where $\tau$ and $\tau^{'}$ introduced as the conformal time regulator in the present context.
In this case as well the Bogoliubov coefficient is not exactly analytically computable. 
To study the behaviour of this integral we consider here three similar
consecutive physical situations for de Sitter and quasi de Sitter case as discussed earlier.
\bea &&\underline{\bf For~\eta\sim \eta_{c}, early~\&~late~\eta:}\\
\small\beta_{diag}(k;\tau,\tau^{'}) &=& \footnotesize\left\{\begin{array}{ll}
                    \displaystyle  \frac{\sqrt{5}}{
                    2(2\sqrt{5}-1)}\left[\tau^{'2\sqrt{5}-1}-\tau^{2\sqrt{5}-1}\right]
                     &
 \mbox{\small {\bf for ~$|k\eta|<<1$}}  \\ 
 \displaystyle   \frac{5}{
                    2\sqrt{2\Delta+4}(2\sqrt{2\Delta+4}-1)}\left[\tau^{'~2\sqrt{2\Delta+4}-1}-\tau^{2\sqrt{2\Delta+4}-1}\nonumber
                    \right] &
 \mbox{\small {\bf for ~$|k\eta|\approx 1-\Delta(\rightarrow 0)$}}  \\ 
	\displaystyle \frac{5}{2}\left[\frac{e^{-2 i k\eta} \left(2ik\eta-1\right)}{4 k^2 \eta^2}-\text{Ei}(-2 i k\eta)\right]^{\tau}_{\tau^{'}} & \mbox{\small {\bf for ~$|k\eta|>>1$}}.
          \end{array}
\right.\eea
\bea
&&\underline{\bf For~\eta<\eta_{c}:}\\
\small\beta_{diag}(k;\tau,\tau^{'}) &=&\footnotesize \left\{\begin{array}{ll}
                    \displaystyle  \frac{\sqrt{5+\Delta_{c}}}{
                    2(2\sqrt{5+\Delta_{c}}-1)}\left[\tau^{'2\sqrt{5+\Delta_{c}}-1}-\tau^{2\sqrt{5+\Delta_{c}}-1}\right]
                     &
 \mbox{\small {\bf for ~$|k\eta|<<1$}}  \\ 
 \displaystyle   \frac{(5+\Delta_{c})\left[\tau^{'~2\sqrt{2\Delta+4+\Delta_{c}}-1}-\tau^{2\sqrt{2\Delta+4+\Delta_{c}}-1}\nonumber
                    \right]}{
                    2\sqrt{2\Delta+4+\Delta_{c}}(2\sqrt{2\Delta+4+\Delta_{c}}-1)} &
 \mbox{\small {\bf for ~$|k\eta|\approx 1-\Delta(\rightarrow 0)$}}  \\ 
	\displaystyle \frac{(5+\Delta_{c})}{2}\left[\frac{e^{-2 i k\eta} \left(2ik\eta-1\right)}{4 k^2 \eta^2}-\text{Ei}(-2 i k\eta)\right]^{\tau}_{\tau^{'}} & \mbox{\small {\bf for ~$|k\eta|>>1$}}.
          \end{array}
\right.\eea
Further using the regularized expressions for Bogoliubov co-efficient $\beta$ in two different representations as mentioned in
Eq~(\ref{soladf129}) and Eq~(\ref{soladfasxsd3cv129}), and substituting them in Eq~(\ref{sdq1}) we get the following
regularized expressions for the Bogoliubov co-efficient $\alpha$ in two different representations as given by:
\bea &&\underline{\bf For~\eta\sim \eta_{c}, early~\&~late~\eta:}\\
\small\alpha(k,\tau,\tau^{'},\eta^{'})&=& \tiny\left\{\begin{array}{ll}
                    \displaystyle  \sqrt{\left[ 1+ \left|\frac{5^{3/4}\eta^{'2\sqrt{5}}}{
                    4\left(\frac{5}{2}+2\sqrt{5}\right)i}\left[\frac{1}{\tau^{'\frac{5}{2}+2\sqrt{5}}}-\frac{1}{\tau^{\frac{5}{2}+2\sqrt{5}}}\right]\right|^2\right]}~e^{i\phi}
                     &
 \mbox{\small {\bf for ~$|k\eta|<<1$}}  \\ 
 \displaystyle   \sqrt{\left[ 1+ \left|\frac{25\eta^{'2\sqrt{2\Delta+4}}}{
                    4(2\Delta+4)^{5/4}\left(\frac{5}{2}+2\sqrt{2\Delta+4}\right)i}\left[\frac{1}{\tau^{'\frac{5}{2}+2\sqrt{2\Delta+4}}}-\frac{1}{\tau^{\frac{5}{2}+2\sqrt{2\Delta+4}}}\right]\right|^2\right]}~e^{i\phi} &
 \mbox{\small {\bf for ~$|k\eta|\approx 1-\Delta(\rightarrow 0)$}}  \\ 
	\displaystyle \sqrt{\left[ 1+ \left|25\left[i\frac{\text{Ei}(2 ik \eta)e^{-2ik\eta^{'}}}{15}
	-\frac{e^{2 i k (\eta-\eta^{'})} }{120 k^5 \eta^5}\left(
	4 k^4 \eta^4-2 i k^3 \eta^3
	-2 k^2 \eta^2+3 i k \eta+6\right)\right]^{\tau}_{\tau^{'}}\right|^2\right]}~e^{i\phi}& \mbox{\small {\bf for ~$|k\eta|>>1$}}.
          \end{array}
\right.\nonumber\eea
\bea
&&\underline{\bf For~\eta<\eta_{c}:}\\
\small\alpha(k,\tau,\tau^{'},\eta^{'})&=& \tiny\left\{\begin{array}{ll}
                       \sqrt{\left[ 1+ \left|\frac{(5+\Delta_{c})^2\eta^{'2\sqrt{5+\Delta_{c}}}}{
                    4i\left(5+\Delta_{c}\right)^{5/4}\left(\frac{5}{2}+2\sqrt{5+\Delta_{c}}\right)}\left[\frac{1}{\tau^{'\frac{5}{2}+2\sqrt{5+\Delta_{c}}}}-\frac{1}{\tau^{\frac{5}{2}+2\sqrt{5+\Delta_{c}}}}\right]\right|^2\right]}~e^{i\phi}&
 \mbox{\small {\bf for ~$|k\eta|<<1$}}  \\ 
   \sqrt{\left[ 1+ \left|\frac{(5+\Delta_{c})^2\eta^{'2\sqrt{2\Delta+4+\Delta_{c}}}}{
                    4(2\Delta+4+\Delta_{c})^{5/4}\left(\frac{5}{2}+2\sqrt{2\Delta+4+\Delta_{c}}\right)i}\left[\frac{1}{\tau^{'\frac{5}{2}+2\sqrt{2\Delta+4+\Delta_{c}}}}-\frac{1}{\tau^{\frac{5}{2}+2\sqrt{2\Delta+4+\Delta_{c}}}}\right]\right|^2\right]}~e^{i\phi}
                    &
 \mbox{\small {\bf for ~$|k\eta|\approx 1-\Delta(\rightarrow 0)$}} \\ 
	\displaystyle \sqrt{\left[ 1+ \left|\left[5+\Delta_{c}\right]^2 \left[i\frac{\text{Ei}(2 ik \eta)e^{-2ik\eta^{'}}}{15}
	-\frac{e^{2 i k (\eta-\eta^{'})} }{120 k^5 \eta^5}\left(
	4 k^4 \eta^4-2 i k^3 \eta^3
	-2 k^2 \eta^2+3 i k \eta+6\right)\right]^{\tau}_{\tau^{'}}\right|^2\right]}~e^{i\phi} & \mbox{\small {\bf for ~$|k\eta|>>1$}}.
          \end{array}
\right.\nonumber
\eea
and
\bea &&\underline{\bf For~\eta\sim \eta_{c}, early~\&~late~\eta:}\\
\small\alpha_{diag}(k;\tau,\tau^{'}) &=& \tiny\left\{\begin{array}{ll}
                    \displaystyle   \sqrt{\left[ 1+\left|\frac{\sqrt{5}}{
                    2(2\sqrt{5}-1)}\left[\tau^{'2\sqrt{5}-1}-\tau^{2\sqrt{5}-1}\right]\right|^2\right]}~e^{i\phi_{diag}}
                     &
 \mbox{\small {\bf for ~$|k\eta|<<1$}}  \\ 
 \displaystyle    \sqrt{\left[ 1+\left|\frac{5}{
                    2\sqrt{2\Delta+4}(2\sqrt{2\Delta+4}-1)}\left[\tau^{'~2\sqrt{2\Delta+4}-1}-\tau^{2\sqrt{2\Delta+4}-1}\nonumber
                    \right]\right|^2\right]}~e^{i\phi_{diag}} &
 \mbox{\small {\bf for ~$|k\eta|\approx 1-\Delta(\rightarrow 0)$}}  \\ 
	\displaystyle  \sqrt{\left[ 1+\left|\frac{5}{2}\left[\frac{e^{-2 i k\eta} \left(2ik\eta-1\right)}{4 k^2 \eta^2}-\text{Ei}(-2 i k\eta)\right]^{\tau}_{\tau^{'}}\right|^2\right]}~e^{i\phi_{diag}} & \mbox{\small {\bf for ~$|k\eta|>>1$}}.
          \end{array}
\right.\eea
\bea
&&\underline{\bf For~\eta<\eta_{c}:}\\
\small\alpha_{diag}(k;\tau,\tau^{'}) &=& \tiny\left\{\begin{array}{ll}
                    \displaystyle   \sqrt{\left[ 1+\left|\frac{\sqrt{5+\Delta_{c}}}{
                    2(2\sqrt{5+\Delta_{c}}-1)}\left[\tau^{'2\sqrt{5+\Delta_{c}}-1}-\tau^{2\sqrt{5+\Delta_{c}}-1}\right]\right|^2\right]}~e^{i\phi_{diag}}
                     &
 \mbox{\small {\bf for ~$|k\eta|<<1$}}  \\ 
 \displaystyle   \sqrt{\left[ 1+\left| \frac{(5+\Delta_{c})\left[\tau^{'~2\sqrt{2\Delta+4+\Delta_{c}}-1}-\tau^{2\sqrt{2\Delta+4+\Delta_{c}}-1}\nonumber
                    \right]}{
                    2\sqrt{2\Delta+4+\Delta_{c}}(2\sqrt{2\Delta+4+\Delta_{c}}-1)}\right|^2\right]}~e^{i\phi_{diag}} &
 \mbox{\small {\bf for ~$|k\eta|\approx 1-\Delta(\rightarrow 0)$}}  \\ 
	\displaystyle  \sqrt{\left[ 1+\left|\frac{(5+\Delta_{c})}{2}\left[\frac{e^{-2 i k\eta} \left(2ik\eta-1\right)}{4 k^2 \eta^2}-\text{Ei}(-2 i k\eta)\right]^{\tau}_{\tau^{'}}\right|^2\right]}~e^{i\phi_{diag}} & \mbox{\small {\bf for ~$|k\eta|>>1$}}.
          \end{array}
\right.\eea
where $\phi$ and $\phi_{diag}$ are the associated phase factors in two different representations.

Further using the expressions for Bogoliubov co-efficient $\alpha$ in two different representations
we get the following
expressions for the reflection and transmission co-efficient in two different representations 
for three consecutive physical situations-$|k\eta|<<1$, $|k\eta|\approx 1-\Delta(\rightarrow 0)$ and
$|k\eta|>>1$ for de Sitter and quasi de Sitter case as given by:
\bea &&\underline{\bf For~\eta\sim \eta_{c}, early~\&~late~\eta:}\\
{\cal R}&=& \footnotesize\left\{

\right.\nonumber\eea
 Next the expression for the
number of produced particles at time $\tau$ can be expressed in the two representations as:
\be%
\ee
which are not exactly analytically computable. To study the behaviour of this integral we consider here three 
consecutive physical situations-$|k\eta|<<1$, $|k\eta|\approx 1-\Delta(\rightarrow 0)$ and $|k\eta|>>1$ for de Sitter and quasi de Sitter case. 
In three cases we have:
\bea &&\underline{\bf For~\eta\sim \eta_{c}, early~\&~late~\eta:}\\
\small{\cal N}(\tau,\tau^{'},\eta^{'})&=& \footnotesize\left\{%
\right.\nonumber
\eea
where the symbol $V$ is defined earlier. 

Finally one can define the total energy density of the produced particles using the following expression:
\be%
\ee
which is not exactly analytically computable. To study the behaviour of this integral we consider here three 
consecutive physical situations-$|k\eta|<<1$, $|k\eta|\approx 1-\Delta(\rightarrow 0)$ and $|k\eta|>>1$ for de Sitter and quasi de Sitter case. 
In three cases we have:
\bea &&\underline{\bf For~\eta\sim \eta_{c}, early~\&~late~\eta:}\\
\small\rho(\tau,\tau^{'},\eta^{'})&=& \footnotesize\left\{%
\right.\nonumber
\eea
Throughout the discussion of total energy density of the produced particles 
we have introduced a symbol $J$ defined as:
\be%
\ee
which physically signifies the total finite volume weighted by $p(\eta)$ 
in momentum space within which the produced particles are occupied. To study the behaviour of this integral we consider here three 
consecutive physical situations-$|k\eta|<<1$, $|k\eta|\approx 1-\Delta(\rightarrow 0)$ and $|k\eta|>>1$ for de Sitter and quasi de Sitter case. 
In three cases we have:
\bea &&\underline{\bf For~\eta\sim \eta_{c}, early~\&~late~\eta:}\\
\small J&=& \left\{%
\right.\nonumber
\eea
\subsubsection{\bf Case II: $m_{axion}/f_{a} << H$}
Equation of motion for the axion field for $m_{axion}/f_{a} \approx H$ case is given by:
\bea
\vartheta''_k + \left\{k^2 - \frac{6}{\eta^2} \right\} \vartheta_k &=& 0~~~~~~~{\bf for~\eta\sim \eta_{c}, early~\&~late~\eta}\\
\vartheta''_k + \left\{k^2 - \left[6+\Delta_{c}\right]
\frac{1}{\eta^2} \right\} \vartheta_k &=& 0~~~~~~~{\bf for~\eta<\eta_{c}}.
\eea
The solution for the mode function for de Sitter and quasi de Sitter space can be expressed as: 
\be%
\ee
where $C_{1}$ and and $C_{2}$ are two arbitrary integration constant, which depend on the 
choice of the initial condition.

After taking $k\eta\rightarrow -\infty$, $k\eta\rightarrow 0$ and $|k\eta|\approx 1-\Delta(\rightarrow 0)$ limit the most general 
solution as stated in Eq~(\ref{yu2zx36}) can be recast as:
\bea\label{yu2zxx}
 \displaystyle \vartheta_k (\eta) &\stackrel{|k\eta|\rightarrow-\infty}{=}&\left\{%
\right.
\eea
Next we assume that the WKB approximation is approximately valid for all times for the solution for the mode function $\vartheta_k$.
In the standard WKB approximation the total solution can be recast in the following form:
\bea\label{df3}
\vartheta_k (\eta)&=& \left[D_{1}u_{k}(\eta) + D_{2} \bar{u}_{k}(\eta)\right],\eea
where $D_{1}$ and and $D_{2}$ are two arbitrary integration constant, which depend on the 
choice of the initial condition during WKB approximation at early and late time scale.
In the present context $u_{k}(\eta)$ and $\bar{u}_{k}(\eta)$ are defined as:
\be\begin{array}{lll}\label{solp}
 \displaystyle\small u_{k}(\eta) =\footnotesize \left\{\begin{array}{ll}
                    \displaystyle   \frac{1}{\sqrt{2\sqrt{k^2 - \frac{6}{\eta^2} }}}
\exp\left[i\int^{\eta} d\eta^{\prime} \sqrt{k^2-\frac{6}{\eta^{'2}}}\right]\\
= \displaystyle\frac{1}{\sqrt{2\sqrt{k^2 - \frac{6}{\eta^2} }}} 
\exp\left[i\left(\eta\sqrt{k^2-\frac{6}{\eta^{2}}}+tan^{-1}\left[\frac{\sqrt{6}}{\eta\sqrt{k^2
-\frac{6}{\eta^{2}}}}\right]
\right)\right] &
 \mbox{\small {\bf for ~$\eta\sim \eta_{c}$, early~\&~late~$\eta$}}  \\
	\displaystyle \frac{1}{\sqrt{2\sqrt{k^2-\frac{\left[6+\Delta_{c}\right]}{\eta^2}}}}
\exp\left[i\int^{\eta} d\eta^{\prime} \sqrt{k^2-\frac{\left[6+\Delta_{c}\right]}{\eta^{'2}}}\right]\\
= \displaystyle\frac{1}{\sqrt{2\sqrt{k^2-\frac{\left[6+\Delta_{c}\right]}{\eta^2}}}} 
\exp\left[i\left(\eta\sqrt{k^2-\frac{\left[6+\Delta_{c}\right]}{\eta^{2}}}
\displaystyle +\sqrt{6+\Delta_{c}}
~tan^{-1}\left[\frac{\sqrt{6+\Delta_{c}}}{\eta\sqrt{k^2-\frac{\left[6+\Delta_{c}\right]}{\eta^{2}}}}\right]
\right)\right] & \mbox{\small {\bf for~$\eta<\eta_{c}$}}.
          \end{array}
\right.
\end{array}\ee
\\
\be\begin{array}{lll}\label{solp}
 \displaystyle\small \bar{u}_{k}(\eta) =\footnotesize \left\{\begin{array}{ll}
                    \displaystyle   \frac{1}{\sqrt{2\sqrt{k^2 - \frac{6}{\eta^2} }}}
\exp\left[i\int^{\eta} d\eta^{\prime} \sqrt{k^2-\frac{6}{\eta^{'2}}}\right]\\
= \displaystyle\frac{1}{\sqrt{2\sqrt{k^2 - \frac{6}{\eta^2} }}} 
\exp\left[i\left(\eta\sqrt{k^2-\frac{6}{\eta^{2}}}
+tan^{-1}\left[\frac{\sqrt{6}}{\eta\sqrt{k^2-\frac{6}{\eta^{2}}}}\right]
\right)\right] &
 \mbox{\small {\bf for ~$\eta\sim \eta_{c}$, early~\&~late~$\eta$}}  \\
	\displaystyle \frac{1}{\sqrt{2\sqrt{k^2-\frac{\left[6+\Delta_{c}\right]}{\eta^2}}}}
\exp\left[i\int^{\eta} d\eta^{\prime} \sqrt{k^2-\frac{\left[6+\Delta_{c}\right]}{\eta^{'2}}}\right]\\
= \displaystyle\frac{1}{\sqrt{2\sqrt{k^2-\frac{\left[6+\Delta_{c}\right]}{\eta^2}}}} 
\exp\left[i\left(\eta\sqrt{k^2-\frac{\left[6+\Delta_{c}\right]}{\eta^{2}}}
\displaystyle +\sqrt{6+\Delta_{c}}
~tan^{-1}\left[\frac{\sqrt{6+\Delta_{c}}}{\eta\sqrt{k^2-\frac{\left[6+\Delta_{c}\right]}{\eta^{2}}}}\right]
\right)\right] & \mbox{\small {\bf for~$\eta<\eta_{c}$}}.
          \end{array}
\right.
\end{array}\ee
In the present context the Bogoliubov coefficient $\beta(k)$ in Fourier space
can be computed approximately using the following expression:
\be\begin{array}{lll}\label{soladf100}
 \displaystyle \beta(k,\tau,\tau^{'},\eta^{'}) =\left\{\begin{array}{ll}
                    \displaystyle   \int^{\tau}_{\tau^{'}}d\eta~\frac{36}{
                    4\eta^{6}\left(k^2-\frac{6}{\eta^{2}}\right)^{\frac{5}{2}}}\exp\left[2i\int^{\eta}_{\eta^{'}}
d\eta^{''}\sqrt{k^2-\frac{6}{\eta^{''2}}}\right]&
 \mbox{\small {\bf for ~$\eta\sim \eta_{c}$, early~\&~late~$\eta$}}  \\ 
	\displaystyle   \int^{\tau}_{\tau^{'}}d\eta~\frac{\left[6+\Delta_{c}\right]^2}{
                    4\eta^{6}\left(k^2-\frac{\left[6+\Delta_{c}\right]}{\eta^{2}}\right)^{\frac{5}{2}}}
                    \exp\left[2i\int^{\eta}_{\eta^{'}}
d\eta^{''}\sqrt{k^2-\frac{\left[6+\Delta_{c}\right]}{\eta^{''2}}}\right]& \mbox{\small {\bf for~$\eta<\eta_{c}$}}.
          \end{array}
\right.
\end{array}\ee
which is not exactly analytically computable. To study the behaviour of this integral we consider here three 
consecutive physical situations-$|k\eta|<<1$, $|k\eta|\approx 1-\Delta(\rightarrow 0)$ and $|k\eta|>>1$ for de Sitter and quasi de Sitter case. 
In three cases we have:
\be\begin{array}{lll}\label{yu2dfvqasx3}\small
 \displaystyle \underline{\rm \bf For~\eta\sim \eta_{c}, early~\&~late~\eta:}~~~~~~~~~\sqrt{\left\{k^2 - \frac{6}{\eta^2} \right\}} \approx\left\{\begin{array}{ll}
                    \displaystyle  \frac{i\sqrt{6}}{\eta}~~~~ &
 \mbox{\small {\bf for ~$|k\eta|<<1$}}  \\ 
	\displaystyle \frac{i\sqrt{2\Delta+5}}{\eta}~~~~ & \mbox{\small
	{\bf for ~$|k\eta|\approx 1-\Delta(\rightarrow 0)$}}\\ 
	\displaystyle k~~~~ & \mbox{\small {\bf for ~$|k\eta|>>1$}}.
          \end{array}
\right.
\\
\small
 \displaystyle \underline{\rm \bf For~\eta<\eta_{c}:}~~~~~~~~~\sqrt{\left\{k^2 - \left[6+\Delta_{c}\right]
\frac{1}{\eta^2} \right\}} \approx\left\{\begin{array}{ll}
                    \displaystyle \frac{i\sqrt{\left[6+\Delta_{c}\right]}}{\eta} ~~~~ &
 \mbox{\small {\bf for ~$|k\eta|<<1$}}  \\ 
	\displaystyle \frac{i\sqrt{2\Delta+5+\Delta_{c}}}{\eta}~~~~ 
	& \mbox{\small {\bf for ~$|k\eta|\approx 1-\Delta(\rightarrow 0)$}}\\ 
	\displaystyle k~~~~ & \mbox{\small {\bf for ~$|k\eta|>>1$}}.
          \end{array}
\right.
\end{array}\ee
and further using this result Bogoliubov coefficient $\beta(k)$ in Fourier space can be expressed as:
\bea &&\underline{\bf For~\eta\sim \eta_{c}, early~\&~late~\eta:}\\
\small\beta(k,\tau,\tau^{'},\eta^{'})&=& \footnotesize \left\{\begin{array}{ll}
                    \displaystyle  \frac{6^{3/4}\eta^{'2\sqrt{6}}}{
                    4\left(\frac{5}{2}+2\sqrt{6}\right)i}\left[\frac{1}{\tau^{'\frac{5}{2}+2\sqrt{6}}}-\frac{1}{\tau^{\frac{5}{2}+2\sqrt{6}}}\right]
                     &
 \mbox{\small {\bf for ~$|k\eta|<<1$}}  \\ 
 \displaystyle  \frac{36\eta^{'2\sqrt{2\Delta+5}}}{
                    4(2\Delta+5)^{5/4}\left(\frac{5}{2}+2\sqrt{2\Delta+5}\right)i}\left[\frac{1}{\tau^{'\frac{5}{2}+2\sqrt{2\Delta+5}}}-\frac{1}{\tau^{\frac{5}{2}+2\sqrt{2\Delta+5}}}\right] &
 \mbox{\small {\bf for ~$|k\eta|\approx 1-\Delta(\rightarrow 0)$}}  \\ 
	\displaystyle 36\left[i\frac{\text{Ei}(2 ik \eta)e^{-2ik\eta^{'}}}{15}
	\displaystyle-\frac{e^{2 i k (\eta-\eta^{'})} }{120 k^5 \eta^5}\left(
	4 k^4 \eta^4-2 i k^3 \eta^3
	-2 k^2 \eta^2+3 i k \eta+6\right)\right]^{\tau}_{\tau^{'}}& \mbox{\small {\bf for ~$|k\eta|>>1$}}.
          \end{array}
\right.\nonumber\eea
\bea
&&\underline{\bf For~\eta<\eta_{c}:}\\
\small\beta(k,\tau,\tau^{'},\eta^{'})&=& \footnotesize \left\{\begin{array}{ll}
                     \frac{(6+\Delta_{c})^2\eta^{'2\sqrt{6+\Delta_{c}}}}{
                    4i\left(6+\Delta_{c}\right)^{5/4}\left(\frac{5}{2}+2\sqrt{6+\Delta_{c}}\right)}\left[\frac{1}{\tau^{'\frac{5}{2}+2\sqrt{6+\Delta_{c}}}}-\frac{1}{\tau^{\frac{5}{2}+2\sqrt{6+\Delta_{c}}}}\right]&
 \mbox{\small {\bf for ~$|k\eta|<<1$}}  \\ 
 \frac{(6+\Delta_{c})^2\eta^{'2\sqrt{2\Delta+5+\Delta_{c}}}}{
                    4(2\Delta+5+\Delta_{c})^{5/4}\left(\frac{5}{2}+2\sqrt{2\Delta+5+\Delta_{c}}\right)i}\left[\frac{1}{\tau^{'\frac{5}{2}+2\sqrt{2\Delta+5+\Delta_{c}}}}-\frac{1}{\tau^{\frac{5}{2}+2\sqrt{2\Delta+5+\Delta_{c}}}}\right]
                    &
 \mbox{\small {\bf for ~$|k\eta|\approx 1-\Delta(\rightarrow 0)$}} \\
	\displaystyle \left[6+\Delta_{c}\right]^2 \left[i\frac{\text{Ei}(2 ik \eta)e^{-2ik\eta^{'}}}{15}
	-\frac{e^{2 i k (\eta-\eta^{'})} }{120 k^5 \eta^5}\left(
	4 k^4 \eta^4-2 i k^3 \eta^3
	-2 k^2 \eta^2+3 i k \eta+6\right)\right]^{\tau}_{\tau^{'}} & \mbox{\small {\bf for ~$|k\eta|>>1$}}.
          \end{array}
\right.\nonumber
\eea
As mentioned earlier here one can use another equivalent way
to define the the Bogoliubov coefficient $\beta$ in Fourier 
space by implementing instantaneous Hamiltonian
diagonalization method to interpret the results. 
Using this diagonalized representation the
regularized Bogoliubov coefficient $\beta$ in Fourier 
space can be written as:
\be\begin{array}{lll}\label{soladfasxsd3cv100}
 \displaystyle \beta_{diag}(k;\tau,\tau^{'}) =\left\{\begin{array}{ll}
                    \displaystyle   \int^{\tau}_{\tau^{'}}d\eta~\frac{6\exp\left[-2i\int^{\eta}
d\eta^{'}\sqrt{k^2-\frac{6}{\eta^{'2}}}\right]}{
                    2\eta^{3}\left(k^2-\frac{6}{\eta^{2}}\right)}~~~~ &
 \mbox{\small {\bf for ~$\eta\sim \eta_{c}$, early~\&~late~$\eta$}}  \\ 
	\displaystyle  \int^{\tau}_{\tau^{'}}d\eta~\frac{\left[6+\Delta_{c}\right]\exp\left[-2i\int^{\eta}
d\eta^{'}\sqrt{k^2-\frac{\left[6+\Delta_{c}\right]}{\eta^{'2}}}\right]}{
                    2\eta^{3}\left(k^2-\frac{\left[6+\Delta_{c}\right]}{\eta^{2}}\right)}
                    ~~~~ & \mbox{\small {\bf for~ $\eta<\eta_{c}$}}.
          \end{array}
\right.
\end{array}\ee
where $\tau$ and $\tau^{'}$ introduced as the conformal time regulator in the present context.
In this case as well the Bogoliubov coefficient is not exactly analytically computable. 
To study the behaviour of this integral we consider here three similar
consecutive physical situations for de Sitter and quasi de Sitter case as discussed earlier.
\bea &&\underline{\bf For~\eta\sim \eta_{c}, early~\&~late~\eta:}\\
\small\beta_{diag}(k;\tau,\tau^{'}) &=& \footnotesize \left\{\begin{array}{ll}
                    \displaystyle  \frac{\sqrt{6}}{
                    2(2\sqrt{6}-1)}\left[\tau^{'2\sqrt{6}-1}-\tau^{2\sqrt{6}-1}\right]
                     &
 \mbox{\small {\bf for ~$|k\eta|<<1$}}  \\ 
 \displaystyle   \frac{6}{
                    2\sqrt{2\Delta+5}(2\sqrt{2\Delta+5}-1)}\left[\tau^{'~2\sqrt{2\Delta+5}-1}-\tau^{2\sqrt{2\Delta+5}-1}\nonumber
                    \right] &
 \mbox{\small {\bf for ~$|k\eta|\approx 1-\Delta(\rightarrow 0)$}}  \\ 
	\displaystyle \frac{6}{2}\left[\frac{e^{-2 i k\eta} \left(2ik\eta-1\right)}{4 k^2 \eta^2}-\text{Ei}(-2 i k\eta)\right]^{\tau}_{\tau^{'}} & \mbox{\small {\bf for ~$|k\eta|>>1$}}.
          \end{array}
\right.\eea
\bea
&&\underline{\bf For~\eta<\eta_{c}:}\\
\small\beta_{diag}(k;\tau,\tau^{'}) &=&\footnotesize  \left\{\begin{array}{ll}
                    \displaystyle  \frac{\sqrt{6+\Delta_{c}}}{
                    2(2\sqrt{6+\Delta_{c}}-1)}\left[\tau^{'2\sqrt{6+\Delta_{c}}-1}-\tau^{2\sqrt{6+\Delta_{c}}-1}\right]
                     &
 \mbox{\small {\bf for ~$|k\eta|<<1$}}  \\ 
 \displaystyle   \frac{(6+\Delta_{c})\left[\tau^{'~2\sqrt{2\Delta+5+\Delta_{c}}-1}-\tau^{2\sqrt{2\Delta+5+\Delta_{c}}-1}\nonumber
                    \right]}{
                    2\sqrt{2\Delta+5+\Delta_{c}}(2\sqrt{2\Delta+5+\Delta_{c}}-1)} &
 \mbox{\small {\bf for ~$|k\eta|\approx 1-\Delta(\rightarrow 0)$}}  \\ 
	\displaystyle \frac{(6+\Delta_{c})}{2}\left[\frac{e^{-2 i k\eta} \left(2ik\eta-1\right)}{4 k^2 \eta^2}-\text{Ei}(-2 i k\eta)\right]^{\tau}_{\tau^{'}} & \mbox{\small {\bf for ~$|k\eta|>>1$}}.
          \end{array}
\right.\eea
Further using the regularized expressions for Bogoliubov co-efficient $\beta$ in two different representations as mentioned in
Eq~(\ref{soladf100}) and Eq~(\ref{soladfasxsd3cv100}), and
substituting them in Eq~(\ref{sdq1}) we get the following
regularized expressions for the Bogoliubov co-efficient $\alpha$ in two different representations as given by:
\bea &&\underline{\bf For~\eta\sim \eta_{c}, early~\&~late~\eta:}\\
\small\alpha(k,\tau,\tau^{'},\eta^{'})&=& \tiny\left\{\begin{array}{ll}
                    \displaystyle  \sqrt{\left[ 1+ \left|\frac{6^{3/4}\eta^{'2\sqrt{6}}}{
                    4\left(\frac{5}{2}+2\sqrt{6}\right)i}\left[\frac{1}{\tau^{'\frac{5}{2}+2\sqrt{6}}}-\frac{1}{\tau^{\frac{5}{2}+2\sqrt{6}}}\right]\right|^2\right]}~e^{i\phi}
                     &
 \mbox{\small {\bf for ~$|k\eta|<<1$}}  \\ 
 \displaystyle   \sqrt{\left[ 1+ \left|\frac{36\eta^{'2\sqrt{2\Delta+5}}}{
                    4(2\Delta+5)^{5/4}\left(\frac{5}{2}+2\sqrt{2\Delta+5}\right)i}\left[\frac{1}{\tau^{'\frac{5}{2}+2\sqrt{2\Delta+5}}}-\frac{1}{\tau^{\frac{5}{2}+2\sqrt{2\Delta+5}}}\right]\right|^2\right]}~e^{i\phi} &
 \mbox{\small {\bf for ~$|k\eta|\approx 1-\Delta(\rightarrow 0)$}}  \\ 
	\displaystyle \sqrt{\left[ 1+ \left|36\left[i\frac{\text{Ei}(2 ik \eta)e^{-2ik\eta^{'}}}{15}
	-\frac{e^{2 i k (\eta-\eta^{'})} }{120 k^5 \eta^5}\left(
	4 k^4 \eta^4-2 i k^3 \eta^3
	-2 k^2 \eta^2+3 i k \eta+6\right)\right]^{\tau}_{\tau^{'}}\right|^2\right]}~e^{i\phi}& \mbox{\small {\bf for ~$|k\eta|>>1$}}.
          \end{array}
\right.\nonumber\eea
\bea
&&\underline{\bf For~\eta<\eta_{c}:}\\
\small\alpha(k,\tau,\tau^{'},\eta^{'})&=& \tiny\left\{\begin{array}{ll}
                       \sqrt{\left[ 1+ \left|\frac{(6+\Delta_{c})^2\eta^{'2\sqrt{5+\Delta_{c}}}}{
                    4i\left(6+\Delta_{c}\right)^{5/4}\left(\frac{5}{2}+2\sqrt{6+\Delta_{c}}\right)}\left[\frac{1}{\tau^{'\frac{5}{2}+2\sqrt{6+\Delta_{c}}}}-\frac{1}{\tau^{\frac{5}{2}+2\sqrt{6+\Delta_{c}}}}\right]\right|^2\right]}~e^{i\phi}&
 \mbox{\small {\bf for ~$|k\eta|<<1$}}  \\ 
   \sqrt{\left[ 1+ \left|\frac{(6+\Delta_{c})^2\eta^{'2\sqrt{2\Delta+5+\Delta_{c}}}}{
                    4(2\Delta+5+\Delta_{c})^{5/4}\left(\frac{5}{2}+2\sqrt{2\Delta+5+\Delta_{c}}\right)i}\left[\frac{1}{\tau^{'\frac{5}{2}+2\sqrt{2\Delta+5+\Delta_{c}}}}-\frac{1}{\tau^{\frac{5}{2}+2\sqrt{2\Delta+5+\Delta_{c}}}}\right]\right|^2\right]}~e^{i\phi}
                    &
 \mbox{\small {\bf for ~$|k\eta|\approx 1-\Delta(\rightarrow 0)$}} \\ 
	\displaystyle \sqrt{\left[ 1+ \left|\left[6+\Delta_{c}\right]^2 \left[i\frac{\text{Ei}(2 ik \eta)e^{-2ik\eta^{'}}}{15}
	-\frac{e^{2 i k (\eta-\eta^{'})} }{120 k^5 \eta^5}\left(
	4 k^4 \eta^4-2 i k^3 \eta^3
	-2 k^2 \eta^2+3 i k \eta+6\right)\right]^{\tau}_{\tau^{'}}\right|^2\right]}~e^{i\phi} & \mbox{\small {\bf for ~$|k\eta|>>1$}}.
          \end{array}
\right.\nonumber
\eea
and
\bea &&\underline{\bf For~\eta\sim \eta_{c}, early~\&~late~\eta:}\\
\small\alpha_{diag}(k;\tau,\tau^{'}) &=& \tiny\left\{\begin{array}{ll}
                    \displaystyle   \sqrt{\left[ 1+\left|\frac{\sqrt{6}}{
                    2(2\sqrt{6}-1)}\left[\tau^{'2\sqrt{6}-1}-\tau^{2\sqrt{6}-1}\right]\right|^2\right]}~e^{i\phi_{diag}}
                     &
 \mbox{\small {\bf for ~$|k\eta|<<1$}}  \\ 
 \displaystyle    \sqrt{\left[ 1+\left|\frac{6}{
                    2\sqrt{2\Delta+5}(2\sqrt{2\Delta+5}-1)}\left[\tau^{'~2\sqrt{2\Delta+5}-1}-\tau^{2\sqrt{2\Delta+5}-1}\nonumber
                    \right]\right|^2\right]}~e^{i\phi_{diag}} &
 \mbox{\small {\bf for ~$|k\eta|\approx 1-\Delta(\rightarrow 0)$}}  \\ 
	\displaystyle  \sqrt{\left[ 1+\left|\frac{6}{2}\left[\frac{e^{-2 i k\eta} \left(2ik\eta-1\right)}{4 k^2 \eta^2}-\text{Ei}(-2 i k\eta)\right]^{\tau}_{\tau^{'}}\right|^2\right]}~e^{i\phi_{diag}} & \mbox{\small {\bf for ~$|k\eta|>>1$}}.
          \end{array}
\right.\eea
\bea
&&\underline{\bf For~\eta<\eta_{c}:}\\
\small\alpha_{diag}(k;\tau,\tau^{'}) &=& \tiny\left\{\begin{array}{ll}
                    \displaystyle   \sqrt{\left[ 1+\left|\frac{\sqrt{6+\Delta_{c}}}{
                    2(2\sqrt{6+\Delta_{c}}-1)}\left[\tau^{'2\sqrt{6+\Delta_{c}}-1}-\tau^{2\sqrt{6+\Delta_{c}}-1}\right]\right|^2\right]}~e^{i\phi_{diag}}
                     &
 \mbox{\small {\bf for ~$|k\eta|<<1$}}  \\ 
 \displaystyle   \sqrt{\left[ 1+\left| \frac{(6+\Delta_{c})\left[\tau^{'~2\sqrt{2\Delta+5+\Delta_{c}}-1}-\tau^{2\sqrt{2\Delta+5+\Delta_{c}}-1}\nonumber
                    \right]}{
                    2\sqrt{2\Delta+5+\Delta_{c}}(2\sqrt{2\Delta+5+\Delta_{c}}-1)}\right|^2\right]}~e^{i\phi_{diag}} &
 \mbox{\small {\bf for ~$|k\eta|\approx 1-\Delta(\rightarrow 0)$}}  \\ 
	\displaystyle  \sqrt{\left[ 1+\left|\frac{(6+\Delta_{c})}{2}\left[\frac{e^{-2 i k\eta} \left(2ik\eta-1\right)}{4 k^2 \eta^2}-\text{Ei}(-2 i k\eta)\right]^{\tau}_{\tau^{'}}\right|^2\right]}~e^{i\phi_{diag}} & \mbox{\small {\bf for ~$|k\eta|>>1$}}.
          \end{array}
\right.\eea
where $\phi$ and $\phi_{diag}$ are the associated phase factors in two different representations.

Further using the expressions for Bogoliubov co-efficient $\alpha$ in two different representations
we get the following
expressions for the reflection and transmission co-efficient in two different representations 
for three consecutive physical situations-$|k\eta|<<1$, $|k\eta|\approx 1-\Delta(\rightarrow 0)$ and
$|k\eta|>>1$ for de Sitter and quasi de Sitter case as given by:
\bea &&\underline{\bf For~\eta\sim \eta_{c}, early~\&~late~\eta:}\\
{\cal R}&=& \footnotesize \left\{

\right.\nonumber\eea
 Next the expression for the
number of produced particles at time $\tau$ can be expressed in the two representations as:
\be%
\ee
which are not exactly analytically computable. To study the behaviour of this integral we consider here three 
consecutive physical situations-$|k\eta|<<1$, $|k\eta|\approx 1-\Delta(\rightarrow 0)$ and $|k\eta|>>1$ for de Sitter and quasi de Sitter case. 
In three cases we have:
\bea &&\underline{\bf For~\eta\sim \eta_{c}, early~\&~late~\eta:}\\
\small{\cal N}(\tau,\tau^{'},\eta^{'})&=& \footnotesize \left\{%
\right.\nonumber
\eea
where the symbol $V$ is defined earlier. 

Finally one can define the total energy density of the produced particles using the following expression:
\be%
\ee
which is not exactly analytically computable. To study the behaviour of this integral we consider here three 
consecutive physical situations-$|k\eta|<<1$, $|k\eta|\approx 1-\Delta(\rightarrow 0)$ and $|k\eta|>>1$ for de Sitter and quasi de Sitter case. 
In three cases we have:
\bea &&\underline{\bf For~\eta\sim \eta_{c}, early~\&~late~\eta:}\\
\small\rho(\tau,\tau^{'},\eta^{'})&=&\footnotesize \left\{%
\right.\nonumber
\eea
Throughout the discussion of total energy density of the produced particles 
we have introduced a symbol $J$ defined as:
\be%
\ee
which physically signifies the total finite volume weighted by $p(\eta)$ 
in momentum space within which the produced particles are occupied. To study the behaviour of this integral we consider here three 
consecutive physical situations-$|k\eta|<<1$, $|k\eta|\approx 1-\Delta(\rightarrow 0)$ and $|k\eta|>>1$ for de Sitter and quasi de Sitter case. 
In three cases we have:
\bea &&\underline{\bf For~\eta\sim \eta_{c}, early~\&~late~\eta:}\\
\small J&=&\left\{%
\right.\nonumber
\eea

\subsubsection{\bf Case III: $m_{axion}/f_{a} >> H$}
Equation of motion for the axion field for $m_{axion}/f_{a} \approx H$ case is given by:
\bea
\vartheta''_k + \left\{k^2 + \frac{\lambda^2-6}{\eta^2} \right\} \vartheta_k &=& 0~~~~~~~{\bf for~\eta\sim \eta_{c}, early~\&~late~\eta}\\
\vartheta''_k + \left\{k^2 + \left[\lambda^2-6-\Delta_{c}\right]
\frac{1}{\eta^2} \right\} \vartheta_k &=& 0~~~~~~~{\bf for~\eta<\eta_{c}}.
\eea
The solution for the mode function for de Sitter and quasi de Sitter space can be expressed as: 
\be%
\ee
where $C_{1}$ and and $C_{2}$ are two arbitrary integration constant, which depend on the 
choice of the initial condition.

After taking $k\eta\rightarrow -\infty$, $k\eta\rightarrow 0$ and $|k\eta|\approx 1-\Delta(\rightarrow 0)$ limit the most general 
solution as stated in Eq~(\ref{yu2zx}) can be recast as:
\bea\label{yu2zxx}
 \displaystyle \vartheta_k (\eta) &\stackrel{|k\eta|\rightarrow-\infty}{=}&\left\{%
\right.
\eea
Next we assume that the WKB approximation is approximately valid for all times for the solution for the mode function $\vartheta_k$.
In the standard WKB approximation the total solution can be recast in the following form:
\bea\label{df3}
\vartheta_k (\eta)&=& \left[D_{1}u_{k}(\eta) + D_{2} \bar{u}_{k}(\eta)\right],\eea
where $D_{1}$ and and $D_{2}$ are two arbitrary integration constant, which depend on the 
choice of the initial condition during WKB approximation at early and late time scale.
In the present context $u_{k}(\eta)$ and $\bar{u}_{k}(\eta)$ are defined as:
\be\begin{array}{lll}\label{solp}
 \tiny u_{k}(\eta) =\footnotesize\left\{\begin{array}{ll}
                    \displaystyle   \frac{1}{\sqrt{2\sqrt{k^2 + \frac{\lambda^2-6}{\eta^2}}}}
\exp\left[i\int^{\eta} d\eta^{\prime} \sqrt{k^2 + \frac{\lambda^2-6}{\eta^{'2}}}\right]\\
= \frac{1}{\sqrt{2\sqrt{k^2 + \frac{\lambda^2-6}{\eta^2}}}} 
\exp\left[i\left(\eta\sqrt{k^2 + \frac{\lambda^2-6}{\eta^2}}
+\sqrt{\lambda^2-6}\ln\left[\frac{2}{\eta\sqrt{\lambda^2-6}}+\frac{2\sqrt{k^2 + \frac{\lambda^2-6}{\eta^{2}}}}{(\lambda^2-6)}\right]
\right)\right] &
 \mbox{\small {\bf for ~$\eta\sim \eta_{c}$, early~\&~late~$\eta$}}  \\ 
	\displaystyle \frac{1}{\sqrt{2\sqrt{k^2 + 
\frac{\left[\lambda^2-6-\Delta_{c}\right]}{\eta^2}}}}
\exp\left[i\int^{\eta} d\eta^{\prime} \sqrt{k^2 + 
\frac{\left[\lambda^2-6-\Delta_{c}\right]}{\eta^{'2}}}\right]\\
= \displaystyle\frac{1}{\sqrt{2\sqrt{k^2 + 
\frac{\left[\lambda^2-6-\Delta_{c}\right]}{\eta^2}}}} 
\exp\left[i\left(\eta\sqrt{k^2 + 
\frac{\left[\lambda^2-6-\Delta_{c}\right]}{\eta^2}}
\right.\right.\\ \left.\left. \displaystyle +\sqrt{\lambda^2-6-\Delta_{c}}\ln\left[\frac{2}{\eta\sqrt{\lambda^2-6-\Delta_{c}}}
+\frac{2\sqrt{k^2 + \frac{\lambda^2-6-\Delta_{c}}{\eta^{2}}}}{(\lambda^2-6-\Delta_{c})}\right]
\right)\right]& \mbox{\small {\bf for~$\eta<\eta_{c}$}}.
          \end{array}
\right.
\end{array}\ee
\be\begin{array}{lll}\label{sola}
 \tiny \bar{u}_{k}(\eta) =\footnotesize\left\{\begin{array}{ll}
                    \displaystyle   \frac{1}{\sqrt{2\sqrt{k^2 + \frac{\lambda^2-6}{\eta^2}}}}
\exp\left[-i\int^{\eta} d\eta^{\prime} \sqrt{k^2 + \frac{\lambda^2-6}{\eta^{'2}}}\right]\\
= \frac{1}{\sqrt{2\sqrt{k^2 + \frac{\lambda^2-6}{\eta^2}}}} 
\exp\left[-i\left(\eta\sqrt{k^2 + \frac{\lambda^2-6}{\eta^2}}
+\sqrt{\lambda^2-6}\ln\left[\frac{2}{\eta\sqrt{\lambda^2-6}}+\frac{2\sqrt{k^2 + \frac{\lambda^2-6}{\eta^{2}}}}{(\lambda^2-6)}\right]
\right)\right] &
 \mbox{\small {\bf for ~$\eta\sim \eta_{c}$, early~\&~late~$\eta$}}  \\ 
	\displaystyle \frac{1}{\sqrt{2\sqrt{k^2 + 
\frac{\left[\lambda^2-6-\Delta_{c}\right]}{\eta^2}}}}
\exp\left[-i\int^{\eta} d\eta^{\prime} \sqrt{k^2 + 
\frac{\left[\lambda^2-6-\Delta_{c}\right]}{\eta^{'2}}}\right]\\
= \displaystyle\frac{1}{\sqrt{2\sqrt{k^2 + 
\frac{\left[\lambda^2-6-\Delta_{c}\right]}{\eta^2}}}} 
\exp\left[-i\left(\eta\sqrt{k^2 + 
\frac{\left[\lambda^2-6-\Delta_{c}\right]}{\eta^2}}
\right.\right.\\ \left.\left. \displaystyle +\sqrt{\lambda^2-6-\Delta_{c}}\ln\left[\frac{2}{\eta\sqrt{\lambda^2-6-\Delta_{c}}}
+\frac{2\sqrt{k^2 + \frac{\lambda^2-6-\Delta_{c}}{\eta^{2}}}}{(\lambda^2-6-\Delta_{c})}\right]
\right)\right] & \mbox{\small {\bf for~$\eta<\eta_{c}$}}.
          \end{array}
\right.
\end{array}\ee
In the present context the Bogoliubov coefficient $\beta(k)$ in Fourier space
can be computed approximately using the following expression:
\be\begin{array}{lll}\label{soladf}
 \displaystyle \beta(k,\tau,\tau^{'},\eta^{'}) =\left\{\begin{array}{ll}
                    \displaystyle   \int^{\tau}_{\tau^{'}}d\eta~\frac{(\lambda^2-6)^2}{
                    4\eta^{6}\left(k^2+\frac{(\lambda^2-6)}{\eta^{2}}\right)^{\frac{5}{2}}}\exp\left[2i\int^{\eta}_{\eta^{'}}
d\eta^{''}\sqrt{k^2+\frac{(\lambda^2-6)}{\eta^{''2}}}\right]&
 \mbox{\small {\bf for ~$\eta\sim \eta_{c}$, early~\&~late~$\eta$}}  \\ 
	\displaystyle   \int^{\tau}_{\tau^{'}}d\eta~\frac{(\lambda^2-6-\Delta_{c})^2}{
                    4\eta^{6}\left(k^2+\frac{(\lambda^2-6-\Delta_{c})}{\eta^{2}}\right)^{\frac{5}{2}}}
                    \exp\left[2i\int^{\eta}_{\eta^{'}}
d\eta^{''}\sqrt{k^2+\frac{(\lambda^2-6-\Delta_{c})}{\eta^{''2}}}\right]& \mbox{\small {\bf for~$\eta<\eta_{c}$}}.
          \end{array}
\right.
\end{array}\ee
which is not exactly analytically computable. To study the behaviour of this integral we consider here three 
consecutive physical situations-$|k\eta|<<1$, $|k\eta|\approx 1-\Delta(\rightarrow 0)$ and $|k\eta|>>1$ for de Sitter and quasi de Sitter case. 
In three cases we have:
\be\begin{array}{lll}\label{yu2dfvqasx3}\small
 \displaystyle \underline{\rm \bf For~\eta\sim \eta_{c}, early~\&~late~\eta:}~~~~~~~~~\sqrt{\left\{k^2 + \frac{\lambda^2-6}{\eta^2} \right\}} \approx\left\{\begin{array}{ll}
                    \displaystyle  \frac{\sqrt{\lambda^2-6}}{\eta}~~~~ &
 \mbox{\small {\bf for ~$|k\eta|<<1$}}  \\ 
	\displaystyle \frac{\sqrt{\lambda^2-2\Delta-5}}{\eta}~~~~ & \mbox{\small
	{\bf for ~$|k\eta|\approx 1-\Delta(\rightarrow 0)$}}\\ 
	\displaystyle k~~~~ & \mbox{\small {\bf for ~$|k\eta|>>1$}}.
          \end{array}
\right.
\\
\small
 \displaystyle \underline{\rm \bf For~\eta<\eta_{c}:}~~~~~~~~~\sqrt{\left\{k^2 + \left[\lambda^2-6-\Delta_{c}\right]
\frac{1}{\eta^2} \right\}} \approx\left\{\begin{array}{ll}
                    \displaystyle \frac{\sqrt{\lambda^2-6-\Delta_{c}}}{\eta} ~~~~ &
 \mbox{\small {\bf for ~$|k\eta|<<1$}}  \\ 
	\displaystyle \frac{\sqrt{\lambda^2-2\Delta-5-\Delta_{c}}}{\eta}~~~~ 
	& \mbox{\small {\bf for ~$|k\eta|\approx 1-\Delta(\rightarrow 0)$}}\\ 
	\displaystyle k~~~~ & \mbox{\small {\bf for ~$|k\eta|>>1$}}.
          \end{array}
\right.
\end{array}\ee
and further using this result Bogoliubov coefficient $\beta(k)$ in Fourier space can be expressed as:
\bea &&\underline{\bf For~\eta\sim \eta_{c}, early~\&~late~\eta:}\\
\small\beta(k,\tau,\tau^{'},\eta^{'})&=& \footnotesize\left\{\begin{array}{ll}
                    \displaystyle   \frac{\left[\tau^{2i\sqrt{\lambda^2-6}}-\tau^{'2i\sqrt{\lambda^2-6}}\right]}{
                    8i(\lambda^2-6)\eta^{'2i\sqrt{\lambda^2-6}}}
                     &
 \mbox{\small {\bf for ~$|k\eta|<<1$}}  \\ 
 \displaystyle  \frac{(\lambda^2-6)^2\left[\tau^{2i\sqrt{\lambda^2-2\Delta-5}}-\tau^{'2i\sqrt{\lambda^2-2\Delta-5}}\right]}{
                    8i(\lambda^2-2\Delta-5)^3\eta^{'2i\sqrt{\lambda^2-2\Delta-5}}} &
 \mbox{\small {\bf for ~$|k\eta|\approx 1-\Delta(\rightarrow 0)$}}  \\ 
	\displaystyle (\lambda^2-6)^2\left[i\frac{\text{Ei}(2 ik \eta)e^{-2ik\eta^{'}}}{15}
	\displaystyle-\frac{e^{2 i k (\eta-\eta^{'})} }{120 k^5 \eta^5}\left(
	4 k^4 \eta^4-2 i k^3 \eta^3
	-2 k^2 \eta^2+3 i k \eta+6\right)\right]^{\tau}_{\tau^{'}}& \mbox{\small {\bf for ~$|k\eta|>>1$}}.
          \end{array}
\right.\nonumber\eea
\bea
&&\underline{\bf For~\eta<\eta_{c}:}\\
\small\beta(k,\tau,\tau^{'},\eta^{'})&=& \footnotesize\left\{\begin{array}{ll}
                    \frac{\left[\tau^{2i\sqrt{\lambda^2-6-\Delta_{c}}}-\tau^{'2i\sqrt{\lambda^2-6-\Delta_{c}}}\right]}{
                    8i\left[\lambda^2-6-\Delta_{c}\right]\eta^{'2i\sqrt{\lambda^2-6-\Delta_{c}}}}&
 \mbox{\small {\bf for ~$|k\eta|<<1$}}  \\ 
 \frac{\left[\lambda^2-6-\Delta_{c}\right]^2\left[\tau^{2i\sqrt{\lambda^2-2\Delta-5-\Delta_{c}}}-\tau^{'2i\sqrt{\lambda^2-2\Delta-5-\Delta_{c}}}\right]}{
                    8i\left[\lambda^2-2\Delta-5-\Delta_{c}\right]^3\eta^{'2i\sqrt{\lambda^2-2\Delta-5-\Delta_{c}}}}
                    &
 \mbox{\small {\bf for ~$|k\eta|\approx 1-\Delta(\rightarrow 0)$}} \\
	\displaystyle  \left[\lambda^2-6-\Delta_{c}\right]^2\left[i\frac{\text{Ei}(2 ik \eta)e^{-2ik\eta^{'}}}{15}
	-\frac{e^{2 i k (\eta-\eta^{'})} }{120 k^5 \eta^5}\left(
	4 k^4 \eta^4-2 i k^3 \eta^3
	-2 k^2 \eta^2+3 i k \eta+6\right)\right]^{\tau}_{\tau^{'}} & \mbox{\small {\bf for ~$|k\eta|>>1$}}.
          \end{array}
\right.\nonumber
\eea
As mentioned earlier here one can use another equivalent way
to define the the Bogoliubov coefficient $\beta$ in Fourier 
space by implementing instantaneous Hamiltonian
diagonalization method to interpret the results. 
Using this diagonalized representation the
regularized Bogoliubov coefficient $\beta$ in Fourier 
space can be written as:
\be\begin{array}{lll}\label{soladfasxsd3cv}
 \displaystyle \beta_{diag}(k;\tau,\tau^{'}) =\left\{\begin{array}{ll}
                    \displaystyle   \int^{\tau}_{\tau^{'}}d\eta~\frac{(\lambda^2-6)\exp\left[-2i\int^{\eta}
d\eta^{'}\sqrt{k^2
+\frac{\lambda^2-6}{\eta^{'2}}}\right]}{
                    2\eta^{3}\left(k^2+\frac{\lambda^2-6}{\eta^{2}}\right)}~~~~ &
 \mbox{\small {\bf for ~$\eta\sim \eta_{c}$, early~\&~late~$\eta$}}  \\ 
	\displaystyle  \int^{\tau}_{\tau^{'}}d\eta~\frac{\left[\lambda^2-6-\Delta_{c}\right]\exp\left[-2i\int^{\eta}
d\eta^{'}\sqrt{k^2+\frac{\left[\lambda^2-6-\Delta_{c}\right]}{\eta^{'2}}}\right]}{
                    2\eta^{3}\left(k^2+\frac{\left[\lambda^2-6-\Delta_{c}\right]}{\eta^{2}}\right)}
                    ~~~~ & \mbox{\small {\bf for~ $\eta<\eta_{c}$}}.
          \end{array}
\right.
\end{array}\ee
where $\tau$ and $\tau^{'}$ introduced as the conformal time regulator in the present context.
In this case as well the Bogoliubov coefficient is not exactly analytically computable. 
To study the behaviour of this integral we consider here three similar
consecutive physical situations for de Sitter and quasi de Sitter case as discussed earlier.
\bea &&\underline{\bf For~\eta\sim \eta_{c}, early~\&~late~\eta:}\\
\small\beta_{diag}(k;\tau,\tau^{'}) &=& \footnotesize\left\{\begin{array}{ll}
                    \displaystyle  \frac{1}{
                    4i\sqrt{\lambda^2-6}}\left[\frac{1}{\tau^{'2i\sqrt{\lambda^2-6}}}-
                    \frac{1}{\tau^{2i\sqrt{\lambda^2-6}}}\right]
                     &
 \mbox{\small {\bf for ~$|k\eta|<<1$}} \nonumber \\ 
 \displaystyle  \frac{(\lambda^2-6)}{
                    4i(\lambda^2-2\Delta-5)^{3/2}}\left[\frac{1}{\tau^{'2i\sqrt{\lambda^2-2\Delta-5}}}-
                    \frac{1}{\tau^{2i\sqrt{\lambda^2-2\Delta-5}}}\right] &
 \mbox{\small {\bf for ~$|k\eta|\approx 1-\Delta(\rightarrow 0)$}}  \\ 
	\displaystyle (\lambda^2-6)\left[\frac{e^{-2 i k\eta} \left(2ik\eta-1\right)}{4 k^2 \eta^2}-\text{Ei}(-2 i k\eta)\right]^{\tau}_{\tau^{'}} & \mbox{\small {\bf for ~$|k\eta|>>1$}}.
          \end{array}
\right.\eea
\bea
&&\underline{\bf For~\eta<\eta_{c}:}\\
\small\beta_{diag}(k;\tau,\tau^{'}) &=&\footnotesize \left\{\begin{array}{ll}
                    \displaystyle   \frac{1}{
                    4i\sqrt{\lambda^2-6-\Delta_{c}}}\left[\frac{1}{\tau^{'2i\sqrt{\lambda^2-6-\Delta_{c}}}}-
                    \frac{1}{\tau^{2i\sqrt{\lambda^2-6-\Delta_{c}}}}\right]
                     &
 \mbox{\small {\bf for ~$|k\eta|<<1$}} \nonumber \\ 
 \displaystyle    \frac{\left[\lambda^2-6-\Delta_{c}\right]}{
                    4i\left[\lambda^2-2\Delta-5-\Delta_{c}\right]^{3/2}}\left[\frac{1}{\tau^{'2i\sqrt{\lambda^2-2\Delta-5-\Delta_{c}}}}-
                    \frac{1}{\tau^{2i\sqrt{\lambda^2-2\Delta-5-\Delta_{c}}}}\right] &
 \mbox{\small {\bf for ~$|k\eta|\approx 1-\Delta(\rightarrow 0)$}}  \\ 
	\displaystyle \left[\lambda^2-6-\Delta_{c}\right]\left[\frac{e^{-2 i k\eta} \left(2ik\eta-1\right)}{4 k^2 \eta^2}-\text{Ei}(-2 i k\eta)\right]^{\tau}_{\tau^{'}} & \mbox{\small {\bf for ~$|k\eta|>>1$}}.
          \end{array}
\right.\eea
Further using the regularized expressions for Bogoliubov co-efficient $\beta$ in two different representations as mentioned in
Eq~(\ref{soladf}) and Eq~(\ref{soladfasxsd3cv}), and substituting them in Eq~(\ref{sdq1}) we get the following
regularized expressions for the Bogoliubov co-efficient $\alpha$ in two different representations as given by:
\bea &&\underline{\bf For~\eta\sim \eta_{c}, early~\&~late~\eta:}\\
\small\alpha(k,\tau,\tau^{'},\eta^{'})&=& \tiny\left\{\begin{array}{ll}
                    \displaystyle  \sqrt{\left[ 1+ \left|\frac{\left[\tau^{2i\sqrt{\lambda^2-6}}-\tau^{'2i\sqrt{\lambda^2-6}}\right]}{
                    8i(\lambda^2-6)\eta^{'2i\sqrt{\lambda^2-6}}}\right|^2\right]}~e^{i\phi}
                     &
 \mbox{\small {\bf for ~$|k\eta|<<1$}}  \\ 
 \displaystyle   \sqrt{\left[ 1+ \left|\frac{(\lambda^2-6)^2\left[\tau^{2i\sqrt{\lambda^2-2\Delta-5}}-\tau^{'2i\sqrt{\lambda^2-2\Delta-5}}\right]}{
                    8i(\lambda^2-2\Delta-5)^3\eta^{'2i\sqrt{\lambda^2-2\Delta-5}}}\right|^2\right]}~e^{i\phi} &
 \mbox{\small {\bf for ~$|k\eta|\approx 1-\Delta(\rightarrow 0)$}}  \\ 
	\displaystyle \sqrt{\left[ 1+ \left|(\lambda^2-6)^2\left[i\frac{\text{Ei}(2 ik \eta)e^{-2ik\eta^{'}}}{15}
	\displaystyle-\frac{e^{2 i k (\eta-\eta^{'})} }{120 k^5 \eta^5}\left(
	4 k^4 \eta^4-2 i k^3 \eta^3
	-2 k^2 \eta^2+3 i k \eta+6\right)\right]^{\tau}_{\tau^{'}}\right|^2\right]}~e^{i\phi}& \mbox{\small {\bf for ~$|k\eta|>>1$}}.
          \end{array}
\right.\nonumber\eea
\bea
&&\underline{\bf For~\eta<\eta_{c}:}\\
\small\alpha(k,\tau,\tau^{'},\eta^{'})&=& \tiny\left\{\begin{array}{ll}
                       \sqrt{\left[ 1+ \left|\frac{\left[\tau^{2i\sqrt{\lambda^2-6-\Delta_{c}}}-\tau^{'2i\sqrt{\lambda^2-6-\Delta_{c}}}\right]}{
                    8i\left[\lambda^2-6-\Delta_{c}\right]\eta^{'2i\sqrt{\lambda^2-6-\Delta_{c}}}}\right|^2\right]}~e^{i\phi}&
 \mbox{\small {\bf for ~$|k\eta|<<1$}}  \\ 
   \sqrt{\left[ 1+ \left|\frac{\left[\lambda^2-6-\Delta_{c}\right]^2\left[\tau^{2i\sqrt{\lambda^2-2\Delta-5-\Delta_{c}}}-\tau^{'2i\sqrt{\lambda^2-2\Delta-5-\Delta_{c}}}\right]}{
                    8i\left[\lambda^2-2\Delta-5-\Delta_{c}\right]^3\eta^{'2i\sqrt{\lambda^2-2\Delta-5-\Delta_{c}}}}\right|^2\right]}~e^{i\phi}
                    &
 \mbox{\small {\bf for ~$|k\eta|\approx 1-\Delta(\rightarrow 0)$}} \\ 
	\displaystyle \sqrt{\left[ 1+ \left|\left[\lambda^2-6-\Delta_{c}\right]^2\left[i\frac{\text{Ei}(2 ik \eta)e^{-2ik\eta^{'}}}{15}
	-\frac{e^{2 i k (\eta-\eta^{'})} }{120 k^5 \eta^5}\left(
	4 k^4 \eta^4-2 i k^3 \eta^3
	-2 k^2 \eta^2+3 i k \eta+6\right)\right]^{\tau}_{\tau^{'}}\right|^2\right]}~e^{i\phi} & \mbox{\small {\bf for ~$|k\eta|>>1$}}.
          \end{array}
\right.\nonumber
\eea
and
\bea &&\underline{\bf For~\eta\sim \eta_{c}, early~\&~late~\eta:}\\
\small\alpha_{diag}(k;\tau,\tau^{'}) &=& \tiny\left\{\begin{array}{ll}
                    \displaystyle   \sqrt{\left[ 1+\left|\frac{1}{
                    4i\sqrt{\lambda^2-6}}\left[\frac{1}{\tau^{'2i\sqrt{\lambda^2-6}}}-
                    \frac{1}{\tau^{2i\sqrt{\lambda^2-6}}}\right]\right|^2\right]}~e^{i\phi_{diag}}
                     &
 \mbox{\small {\bf for ~$|k\eta|<<1$}}  \\ 
 \displaystyle    \sqrt{\left[ 1+\left|\frac{(\lambda^2-6)}{
                    4i(\lambda^2-2\Delta-5)^{3/2}}\left[\frac{1}{\tau^{'2i\sqrt{\lambda^2-2\Delta-5}}}-
                    \frac{1}{\tau^{2i\sqrt{\lambda^2-2\Delta-5}}}\right]\right|^2\right]}~e^{i\phi_{diag}} &
 \mbox{\small {\bf for ~$|k\eta|\approx 1-\Delta(\rightarrow 0)$}} \nonumber \\ 
	\displaystyle  \sqrt{\left[ 1+\left|(\lambda^2-6)\left[\frac{e^{-2 i k\eta} \left(2ik\eta-1\right)}{4 k^2 \eta^2}-\text{Ei}(-2 i k\eta)\right]^{\tau}_{\tau^{'}}\right|^2\right]}~e^{i\phi_{diag}} & \mbox{\small {\bf for ~$|k\eta|>>1$}}.
          \end{array}
\right.\eea
\bea
&&\underline{\bf For~\eta<\eta_{c}:}\\
\small\alpha_{diag}(k;\tau,\tau^{'}) &=& \tiny\left\{\begin{array}{ll}
                    \displaystyle   \sqrt{\left[ 1+\left|\frac{1}{
                    4i\sqrt{\lambda^2-6-\Delta_{c}}}\left[\frac{1}{\tau^{'2i\sqrt{\lambda^2-6-\Delta_{c}}}}-
                    \frac{1}{\tau^{2i\sqrt{\lambda^2-6-\Delta_{c}}}}\right]\right|^2\right]}~e^{i\phi_{diag}}
                     &
 \mbox{\small {\bf for ~$|k\eta|<<1$}}  \\ 
 \displaystyle   \sqrt{\left[ 1+\left| \frac{\left[\lambda^2-6-\Delta_{c}\right]}{
                    4i\left[\lambda^2-2\Delta-5-\Delta_{c}\right]^{3/2}}\left[\frac{1}{\tau^{'2i\sqrt{\lambda^2-2\Delta-5-\Delta_{c}}}}-
                    \frac{1}{\tau^{2i\sqrt{\lambda^2-2\Delta-5-\Delta_{c}}}}\right]\right|^2\right]}~e^{i\phi_{diag}} &
 \mbox{\small {\bf for ~$|k\eta|\approx 1-\Delta(\rightarrow 0)$}}  \nonumber\\ 
	\displaystyle  \sqrt{\left[ 1+\left|\left[\lambda^2-6-\Delta_{c}\right]\left[\frac{e^{-2 i k\eta} \left(2ik\eta-1\right)}{4 k^2 \eta^2}-\text{Ei}(-2 i k\eta)\right]^{\tau}_{\tau^{'}}\right|^2\right]}~e^{i\phi_{diag}} & \mbox{\small {\bf for ~$|k\eta|>>1$}}.
          \end{array}
\right.\eea
where $\phi$ and $\phi_{diag}$ are the associated phase factors in two different representations.

Further using the expressions for Bogoliubov co-efficient $\alpha$ in two different representations
we get the following
expressions for the reflection and transmission co-efficient in two different representations 
for three consecutive physical situations-$|k\eta|<<1$, $|k\eta|\approx 1-\Delta(\rightarrow 0)$ and
$|k\eta|>>1$ for de Sitter and quasi de Sitter case as given by:
\bea &&\underline{\bf For~\eta\sim \eta_{c}, early~\&~late~\eta:}\\
{\cal R}&=& \footnotesize\left\{

\right.\nonumber\eea
 Next the expression for the
number of produced particles at time $\tau$ can be expressed in the two representations as:
\be%
\ee
which are not exactly analytically computable. To study the behaviour of this integral we consider here three 
consecutive physical situations-$|k\eta|<<1$, $|k\eta|\approx 1-\Delta(\rightarrow 0)$ and $|k\eta|>>1$ for de Sitter and quasi de Sitter case. 
In three cases we have:
\bea &&\underline{\bf For~\eta\sim \eta_{c}, early~\&~late~\eta:}\\
\small{\cal N}(\tau,\tau^{'},\eta^{'})&=& \footnotesize\left\{%
\right.\nonumber
\eea
where the symbol $V$ is defined earlier. 

Finally one can define the total energy density of the produced particles using the following expression:
\be%
\ee
which is not exactly analytically computable. To study the behaviour of this integral we consider here three 
consecutive physical situations-$|k\eta|<<1$, $|k\eta|\approx 1-\Delta(\rightarrow 0)$ and $|k\eta|>>1$ for de Sitter and quasi de Sitter case. 
In three cases we have:
\bea &&\underline{\bf For~\eta\sim \eta_{c}, early~\&~late~\eta:}\\
\small\rho(\tau,\tau^{'},\eta^{'})&=&\footnotesize \left\{%
\right.\nonumber
\eea
\begin{figure*}[htb]
\centering
\subfigure[]{
    \includegraphics[width=7.2cm,height=6cm] {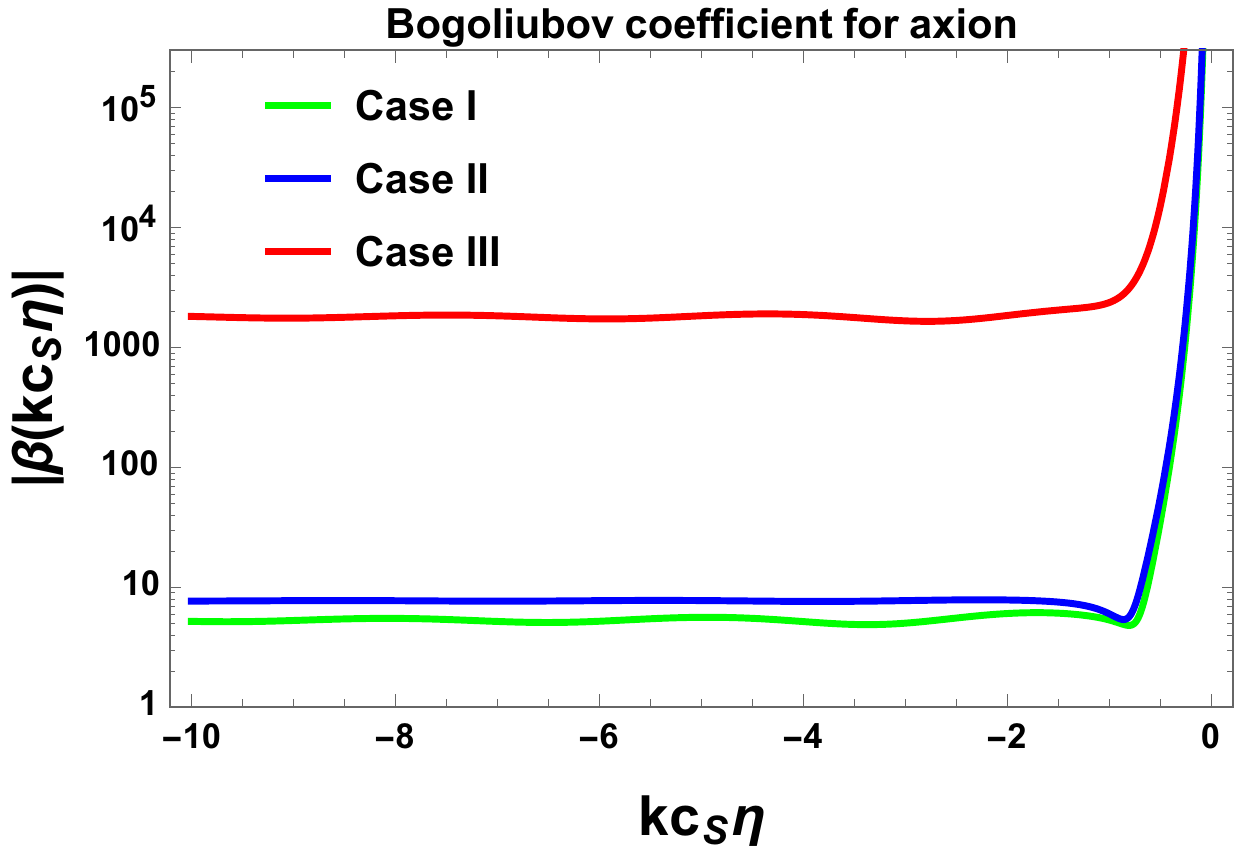}
    \label{fig1xcc}
}
\subfigure[]{
    \includegraphics[width=7.2cm,height=6cm] {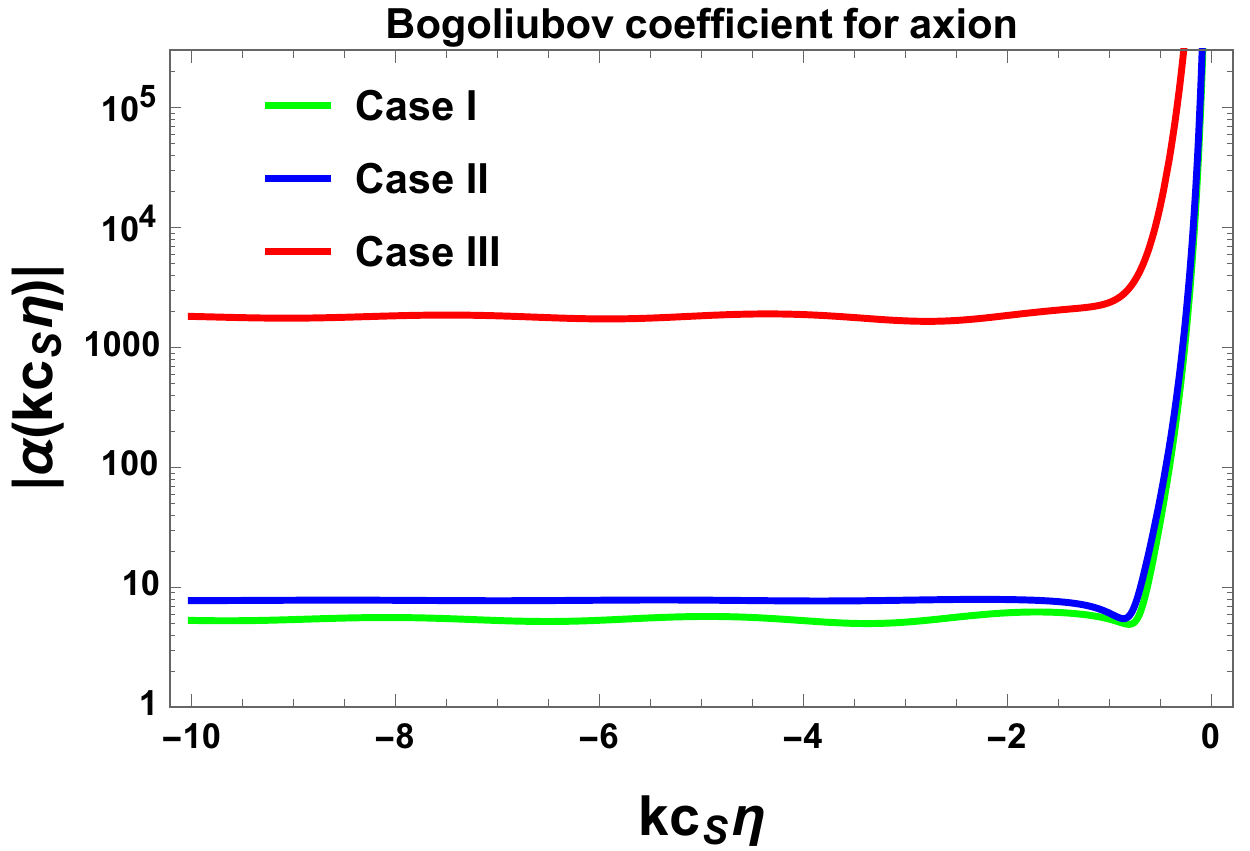}
    \label{fig2xcc}
}
\subfigure[]{
    \includegraphics[width=7.2cm,height=6cm] {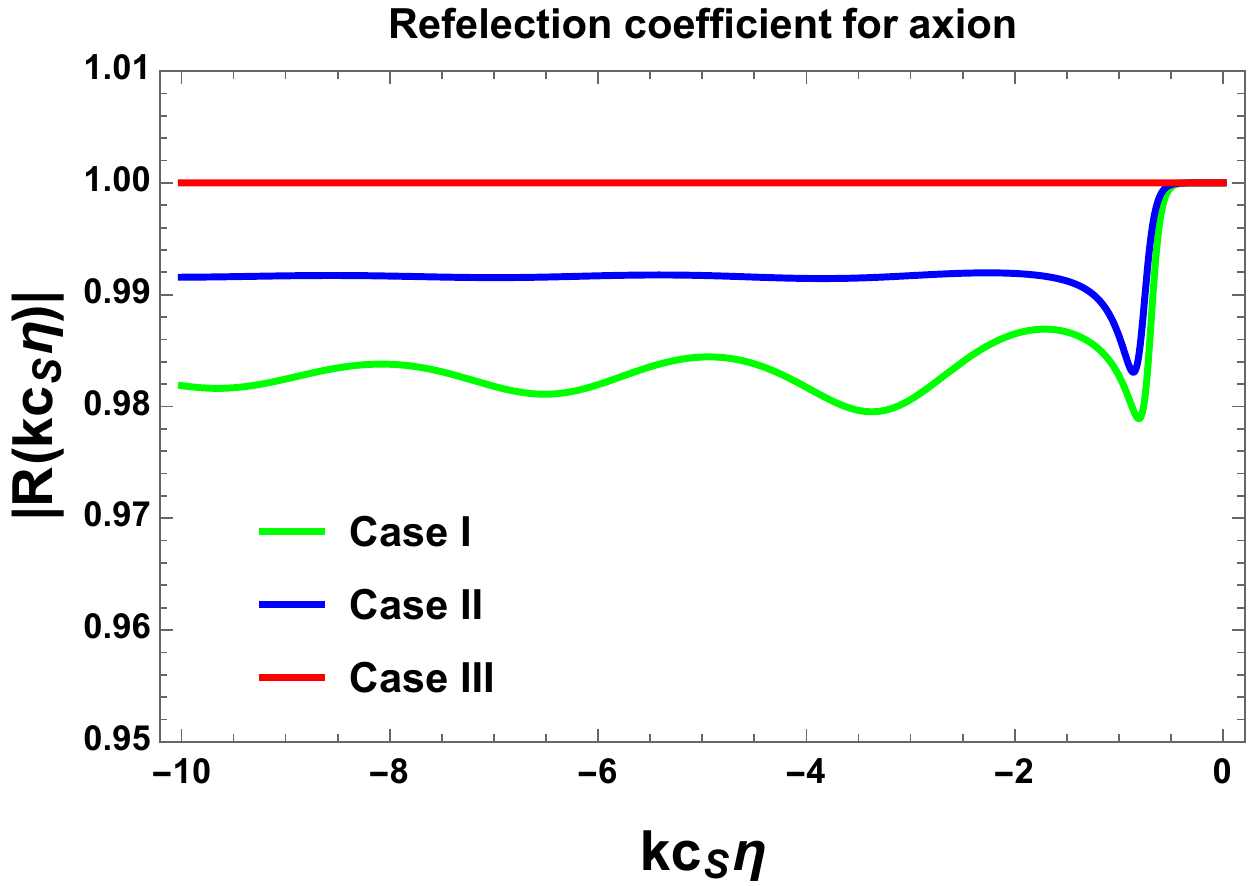}
    \label{fig3xcc}
}
\subfigure[]{
    \includegraphics[width=7.2cm,height=6cm] {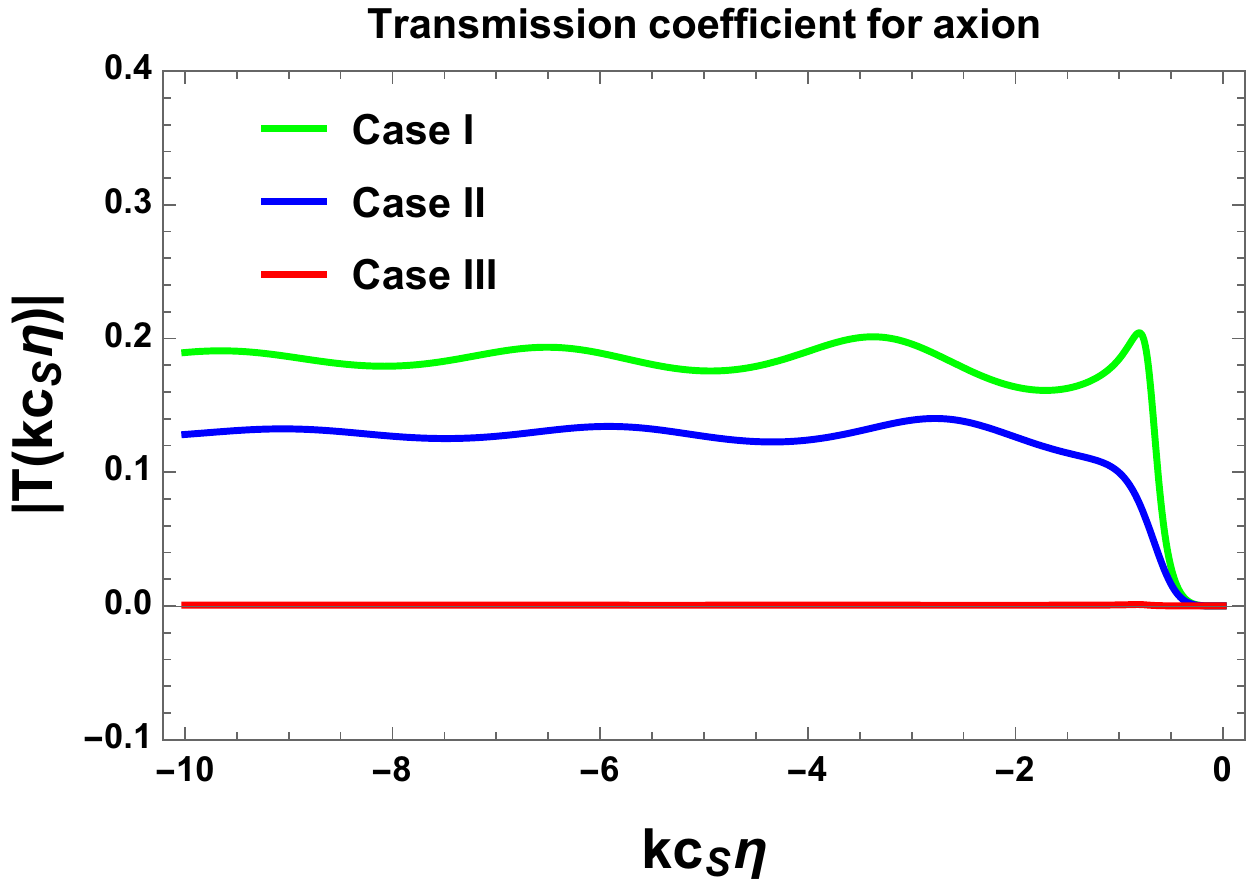}
    \label{fig4xcc}
}
\subfigure[]{
    \includegraphics[width=7.2cm,height=6.1cm] {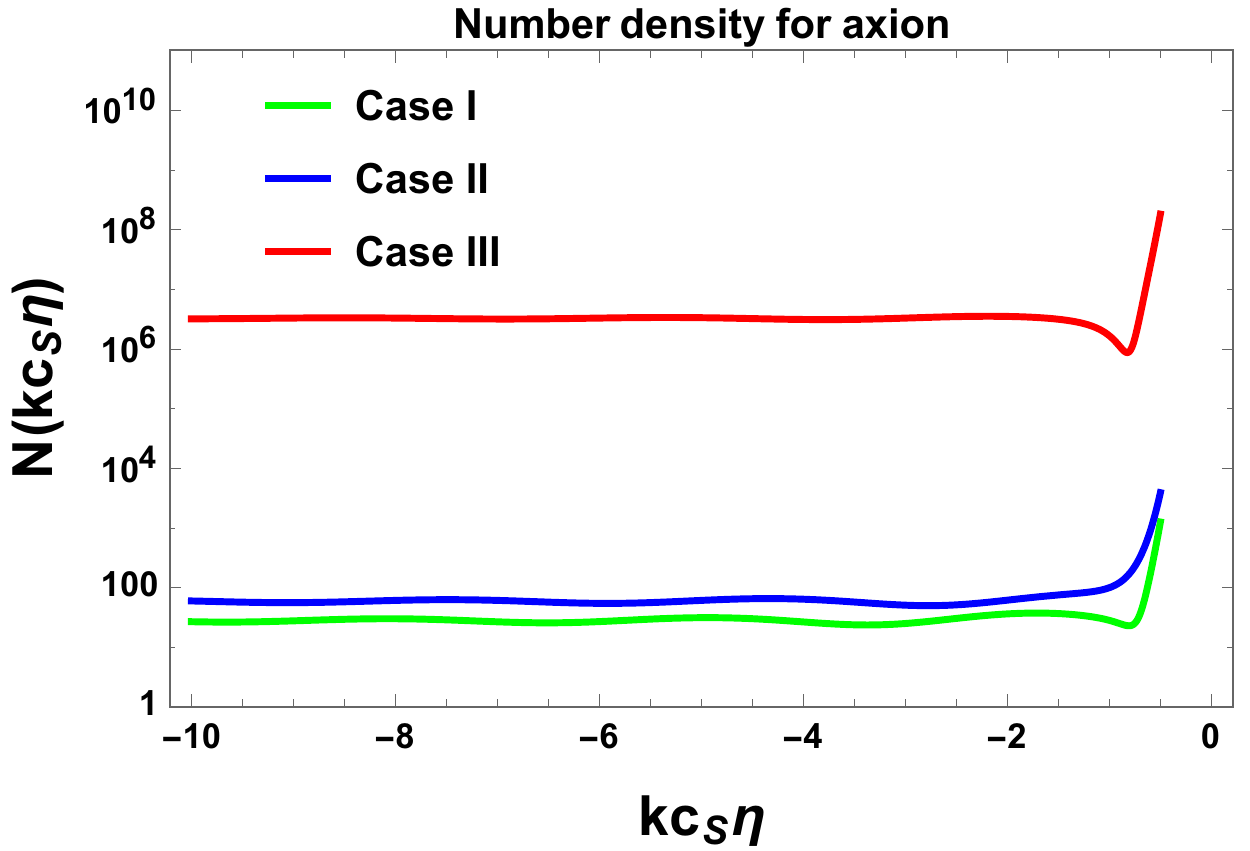}
    \label{fig3xcc}
}
\subfigure[]{
    \includegraphics[width=7.2cm,height=6.1cm] {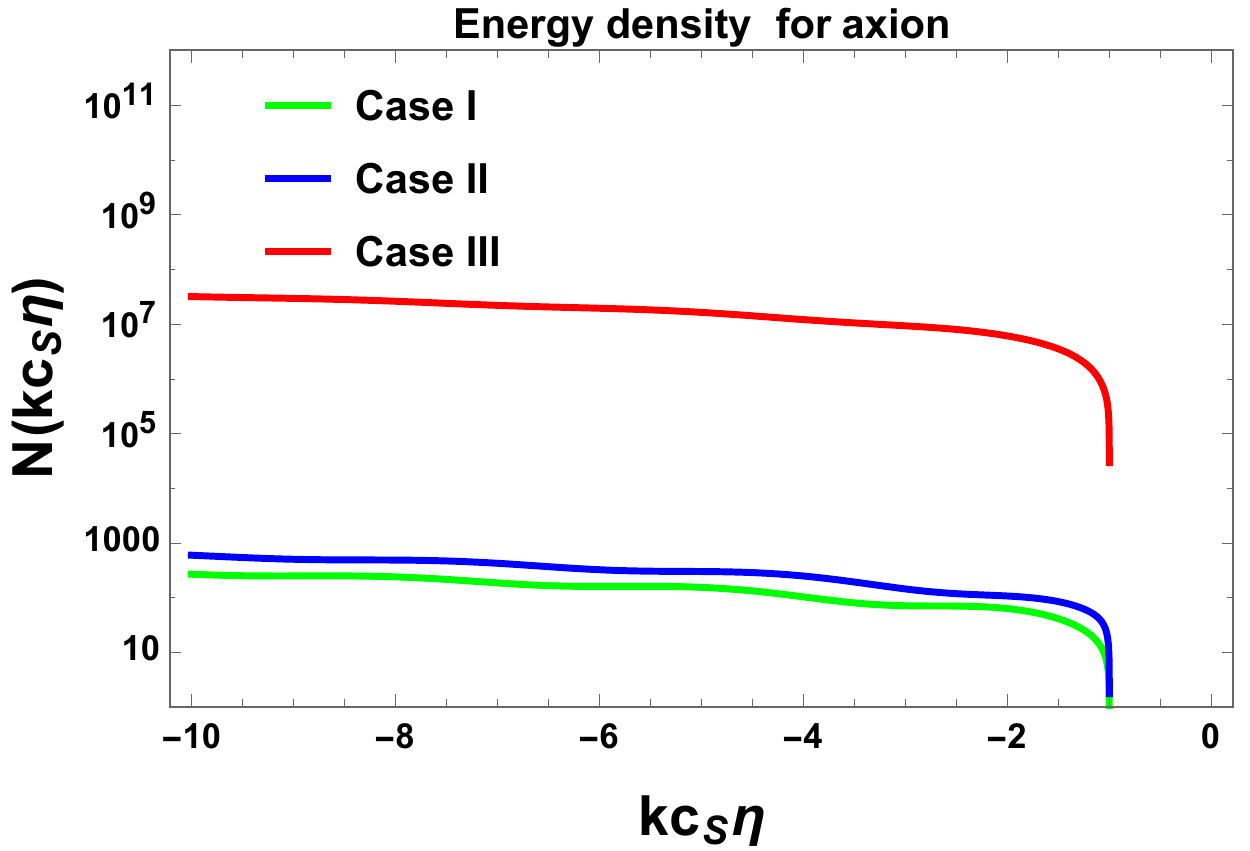}
    \label{fig4xcc}
}
\caption[Optional caption for list of figures]{Particle creation profile for {\bf Case I}, {\bf Case II} and {\bf Case III}.} 
\label{bog5xxx}
\end{figure*}
Throughout the discussion of total energy density of the produced particles 
we have introduced a symbol $J$ defined as:
\be\begin{array}{lll}\label{kghjk}
 \displaystyle J=\int d^3 {\bf k}~p(\tau)=\left\{\begin{array}{ll}
                    \displaystyle   \int d^3 {\bf k}~\sqrt{k^2+\frac{\lambda^2-6}{\tau^2}}~~~~ &
 \mbox{\small {\bf for ~dS}}  \\ 
	\displaystyle  \int d^3 {\bf k}~\sqrt{k^2+\frac{\left[\lambda^2-6-\Delta_{c}\right]}{\tau^2}}~~~~ & \mbox{\small {\bf for~ qdS}}.
          \end{array}
\right.
\end{array}\ee
which physically signifies the total finite volume weighted by $p(\eta)$ 
in momentum space within which the produced particles are occupied. To study the behaviour of this integral we consider here three 
consecutive physical situations-$|k\eta|<<1$, $|k\eta|\approx 1-\Delta(\rightarrow 0)$ and $|k\eta|>>1$ for de Sitter and quasi de Sitter case. 
In three cases we have:
\bea &&\underline{\bf For~\eta\sim \eta_{c}, early~\&~late~\eta:}\\
\small J&=& \left\{\begin{array}{ll}
                    \displaystyle  \int d^3 {\bf k}~\frac{\lambda^2-6}{\tau}=\frac{(\lambda^2-6)V}{\tau}
                    ~~&
 \mbox{\small {\bf for ~$|k\eta|<<1$}}  \\ 
 \displaystyle  \int d^3 {\bf k}~\frac{\sqrt{\lambda^2-2\Delta-5}}{\tau}=\frac{\sqrt{\lambda^2-2\Delta-5}~V}{\tau}~~~~ &
 \mbox{\small {\bf for ~$|k\eta|\approx 1-\Delta(\rightarrow 0)$}}  \\ 
	\displaystyle\int d^3 {\bf k}~k & \mbox{\small {\bf for ~$|k\eta|>>1$}}.~~~~~~~~
          \end{array}
\right.\nonumber\eea
\bea
&&\underline{\bf For~\eta<\eta_{c}:}\\
\small J&=&\left\{\begin{array}{ll}
                    \displaystyle  \int d^3 {\bf k}~\frac{\sqrt{\left[\lambda^2-6-\Delta_{c}\right]}}{\tau}=\frac{V\sqrt{\left[\lambda^2-6-\Delta_{c}\right]}}{\tau}~~ &
 \mbox{\small {\bf for ~$|k\eta|<<1$}}  \\ 
 \displaystyle   \int d^3 {\bf k}~\frac{\sqrt{\lambda^2-2\Delta-5-\Delta_{c}}}{\tau}=\frac{\sqrt{\lambda^2-2\Delta-5-\Delta_{c}}~V}{\tau}
                    ~~&
 \mbox{\small {\bf for ~$|k\eta|\approx 1-\Delta(\rightarrow 0)$}} \\ 
	\displaystyle \int d^3 {\bf k}~k~~ & \mbox{\small {\bf for ~$|k\eta|>>1$}}.~~~~~~~~
          \end{array}
\right.\nonumber
\eea
In fig.~(\ref{bog5xxx}), we have explicitly shown the particle creation profile for {\bf Case I}, {\bf Case II} and {\bf Case III}.
\subsection{Axion-massive particle correspondence}
\label{sec4d}

\begin{table*}
\centering
\footnotesize\begin{tabular}{|c|c|c|
}
\hline
\hline
{\bf Characteristics}& {\bf New particle} & {\bf Axion}
\\
\hline\hline\hline
 {\bf Action}&$S_{new}={\frac{1}{2}} \int d\eta d^3x \frac{2 \epsilon M^2_{p}}{\tilde{c}^2_{S}H^2} \left[\frac{(\partial_\eta \zeta)^2 
- c^2_{S}(\partial_i \zeta)^2}{\eta^2}\right]$ &  $S_{axion}=\int d\eta~d^{3}x \left[\frac{f^2_{a}(\eta)}{2H^2}\frac{\left[(\partial_{\eta}a)^2-(\partial_{i}a)^2\right]}{\eta^2} -\frac{U(a)}{H^4 \eta^4}\right].$
\\
  & $- \int \frac{d\eta}{\tilde{c}_{S}H} m(\eta) \partial_\eta \zeta(\eta,{\bf x}=0)$ &  $- \int \frac{d\eta}{f_a H} m_{axion} \partial_\eta \bar{a}(\eta,{\bf x}=0)$
\\
\hline
{\bf Sound speed} & $c_{S}\leq 1,\tilde{c}_{S}\leq 1$ & $c_{S}=1,\tilde{c}_{S}= 1$\\
\hline
 {\bf Coefficient }&$\frac{\epsilon M^2_{p}}{\tilde{c}^2_{S}}$ &  $\frac{f^2_{a}(\eta)}{2}$
\\
{\bf  parameter }& &  \\
\hline
{\bf Time }&$(\partial_\eta \zeta)^2$  &  $(\partial_\eta a)^2$ \\
{\bf derivative}& &  \\
\hline 
 {\bf Spatial } &$c^2_{S}(\partial_i \zeta)^2$ & $(\partial_{i}a)^2$
\\ {\bf derivative} & & 
\\ \hline
{\bf Additional}  &$\frac{m(\eta)}{\tilde{c}_{S}}  \partial_\eta \zeta(\eta,{\bf x}=0)$ &   $\frac{m_{axion} }{f_a H} \partial_\eta \bar{a}(\eta,{\bf x}=0)$\\
{\bf contribution}  & &  \\
\hline
{\bf Mass parameter}  &$m(\eta)$ &   $\frac{m_{axion}}{f_{a}}=\tiny\left\{\begin{array}{ll}
                    \displaystyle \sqrt{-\frac{\Lambda^4_C}{f^2_a} \cos\left(\sin^{-1}\left(\frac{\mu^3 f_{a}}{\Lambda^4_{C}}\right)\right)}~~~~ &
 \mbox{\small {\bf for ~total~$U(a)$}}  \\ \\
	\displaystyle \sqrt{\frac{\Lambda^4_C}{f^2_a}(-1)^{m+1}} ~~~~ & \mbox{\small {\bf for~osc.~$U(a)$}}.
          \end{array}
\right.$\\
\hline 
{\bf Rescaled mode}  &$h_{\bf k}=zM_{p}\zeta_{\bf k}$ &   $\vartheta_{\bf k}=\frac{f^2_a}{H^2\eta^2M^2_p}\bar{a}_{\bf k}$\\
{\bf function} & & \\
\hline
{\bf Mukhanov-Sasaki}  &$z=\frac{a\sqrt{2\epsilon}}{\tilde{c}_{S}}=\frac{\sqrt{2\epsilon}}{H\eta\tilde{c}_{S}}$
&   $\frac{f^2_a}{H^2\eta^2 M^2_p}$\\
{\bf variable} & & \\
\hline
{\bf Scalar mode }  &$h^{''}_{\bf k}+
\left(c^2_{S}k^2+\frac{\left(\frac{m^2}{H^2}-\delta\right)}{\eta^2}\right)h_{\bf k}=0$
&   $\partial^{2}_{\eta}\vartheta_{\bf k}+
\left(k^2-\frac{\partial^2_{\eta}\left(\frac{f^2_{a}}{H^2\eta^2}\right)}{\left(\frac{f^2_{a}}{H^2\eta^2}\right)}
+\frac{m^{2}_{axion}}{f^2_{a}H^2\eta^2}\right)\vartheta_{\bf k}=0$\\
{\bf equation} & where $\delta=\frac{z^{''}}{z}=\small\left\{\begin{array}{ll}
                    \displaystyle 2~~~~ &
 \mbox{\small {\bf for ~dS}}  \\ 
	\displaystyle \left(\nu^2-\frac{1}{4}\right) ~~~~ & \mbox{\small {\bf for~qdS}}.
          \end{array}
\right.$& where $\frac{\partial^2_{\eta}\left(\frac{f^2_{a}}{H^2\eta^2}\right)}{\left(\frac{f^2_{a}}{H^2\eta^2}\right)}
\approx \tiny
\left\{\begin{array}{ll}
                    \displaystyle \frac{6}{\eta^2}~~~~ &
 \mbox{\small {\bf for ~$\eta\sim \eta_{c}$}}  \\ 
 \displaystyle \frac{6}{\eta^2}~~~~ &
 \mbox{\small {\bf for ~early~\&~late~$\eta$}}  \\ 
	\displaystyle \frac{6+\Delta_{c}}{\eta^2} ~~~~ & \mbox{\small {\bf for~$\eta<\eta_{c}$}}.
          \end{array}
\right.$\\ 
\hline 
{\bf Parametrization}  &$\frac{m^2}{H^2}=\tiny
\left\{\begin{array}{ll}
                    \displaystyle \gamma\left(\frac{\eta}{\eta_0} - 1\right)^2 + \delta~~~~ &
 \mbox{\small {\bf ~Case~I}}  \\ 
 \displaystyle \frac{m^2_0}{2H^2}\left[1-\tanh\left(\rho\frac{\ln(-H\eta)}{H}\right)\right]~~~~ &
 \mbox{\small {\bf ~Case~II}}  \\ 
	\displaystyle \frac{m^2_0}{H^2}{\rm sech}^2\left(\rho\frac{\ln(-H\eta)}{H}\right) ~~~~ 
	& \mbox{\small {\bf ~Case~III}}.
          \end{array}
\right.
$ &   $\frac{m^2_{axion}}{f^2_a H^2}=\frac{m^2_{axion}}{\left[100-\frac{80}{1+\left(\ln\frac{\eta}{\eta_{c}}\right)^2} \right]H^4}$\\
\hline
\hline
\hline
\end{tabular}
\caption{ Table showing the analogy between the quantum fluctuation obtained from new particle and axion in string theory.}\label{fig:hysteresis}
\vspace{.4cm}
\end{table*}
Now if here we identify the above axion fluctuation equations with the scalar fluctuations originated from the new particles as mentioned in the earlier section then one can write:
\bea \underline{\bf For~\eta\sim \eta_{c}, early~\&~late~\eta}&&\nonumber\\ 
\frac{m^2_{axion}}{f^2_a H^2}-6 &\equiv&\footnotesize\left\{\begin{array}{ll}
                    \displaystyle   \frac{m^2}{H^2}-2~~~~ &
 \mbox{\small {\bf for~dS}}  \\ 
	 \displaystyle\frac{m^2}{H^2}-\left(\nu^2-\frac{1}{4}\right)~~~~ & \mbox{\small {\bf for~qdS}}.
          \end{array}
\right.\\
\tiny p(\eta):=\sqrt{k^2+\left(\frac{m^2_{axion}}{f^2_a H^2}-6\right)\frac{1}{\eta^2}} &\equiv&\footnotesize\left\{\begin{array}{ll}
                    \displaystyle  \sqrt{k^2+\left( \frac{m^2}{H^2}-2\right)\frac{1}{\eta^2}}~~~~ &
 \mbox{\small {\bf for~dS}}  \\ 
	 \displaystyle \sqrt{k^2+\left( \frac{m^2}{H^2}-\left(\nu^2-\frac{1}{4}\right)\right)\frac{1}{\eta^2}}~~~~ & \mbox{\small {\bf for~qdS}}.
          \end{array}
\right.\\ 
 \underline{\bf For~\eta<\eta_{c}}&&\nonumber\\\frac{m^2_{axion}}{f^2_a H^2}-6-\Delta_{c} &\equiv&\footnotesize\left\{\begin{array}{ll}
                    \displaystyle   \frac{m^2}{H^2}-2~~~~ &
 \mbox{\small {\bf for~dS}}  \\ 
	 \displaystyle\frac{m^2}{H^2}-\left(\nu^2-\frac{1}{4}\right)~~~~ & \mbox{\small {\bf for~qdS}}.
          \end{array}
\right.\\
p(\eta):=\sqrt{k^2+\left(\frac{m^2_{axion}}{f^2_a H^2}-6-\Delta_{c}\right)\frac{1}{\eta^2}} &\equiv&\footnotesize\left\{\begin{array}{ll}
                    \displaystyle  \sqrt{k^2+\left( \frac{m^2}{H^2}-2\right)\frac{1}{\eta^2}}~~~~ &
 \mbox{\small {\bf for~dS}}  \\ 
	 \displaystyle \sqrt{k^2+\left( \frac{m^2}{H^2}-\left(\nu^2-\frac{1}{4}\right)\right)\frac{1}{\eta^2}}~~~~ & \mbox{\small {\bf for~qdS}}.
          \end{array}
\right.\eea
Consequently the actual solution and WKB solution for axion fluctuation can be recast for $\eta\sim \eta_{c}$, early~\&~late~$\eta$ and $\eta<\eta_{c}$ as:
\be\begin{array}{lll}\label{yuriii2}
 \displaystyle \vartheta_{\bf k}  (\eta) =\frac{f^2_{a}}{H^2\eta^2M^2_p}\bar{a}_{\bf k}=\footnotesize\left\{\begin{array}{ll}
                    \displaystyle   \sqrt{-\eta}\left[C_1  H^{(1)}_{\sqrt{\frac{9}{4}-\frac{m^2}{H^2}}} \left(-k\eta\right) 
+ C_2 H^{(2)}_{\sqrt{\frac{9}{4}-\frac{m^2}{H^2}}} \left(-k\eta\right)\right]~~~~ &
 \mbox{\small {\bf for ~dS}}  \\ 
	\displaystyle \sqrt{-\eta}\left[C_1  H^{(1)}_{\sqrt{\nu^2-\frac{m^2}{H^2}}} \left(-k\eta\right) 
+ C_2 H^{(2)}_{\sqrt{\nu^2-\frac{m^2}{H^2}}} \left(-k\eta\right)\right]~~~~ & \mbox{\small {\bf for~ qdS}}.
          \end{array}
\right.
\end{array}\ee
In the standard WKB approximation the total solution can be recast in the following form:
\bea\label{df3}
\vartheta_{\bf k} (\eta)&=& \frac{f^2_{a}}{H^2\eta^2M^2_p}\bar{a}_{\bf k}=\tiny\left[D_{1}u_{k}(\eta) + D_{2} \bar{u}_{k}(\eta)\right],\eea
where $u_{k}(\eta)$ and $\bar{u}_{k}(\eta)$ are identified as:
\bea &&\underline{\bf For~\eta\sim \eta_{c}, early~\&~late~\eta}\nonumber\\ 
u_{k}(\eta) &=&\tiny
                    \displaystyle   \frac{\exp\left[i\int^{\eta}
                    d\eta^{\prime} \sqrt{k^2+\left(\frac{m^2_{axion}}{f^2_a H^2}-6\right)
                    \frac{1}{\eta^{'2}}} )\right]}
                    {\sqrt{2\sqrt{k^2+\left(\frac{m^2_{axion}}{f^2_a H^2}-6\right)\frac{1}{\eta^2}} }}
                    \equiv\footnotesize\left\{\begin{array}{ll}
                    \displaystyle  \frac{\exp\left[i\int^{\eta}
                    d\eta^{\prime} \sqrt{k^2+\left( \frac{m^2}{H^2}-2\right)\frac{1}{\eta^{'2}}}\right]}
                    {\sqrt{2\sqrt{k^2+\left( \frac{m^2}{H^2}-2\right)\frac{1}{\eta^2}} }}\nonumber
                    ~~~~ &
 \mbox{\small {\bf for~dS}}  \\ 
	 \displaystyle  \frac{\exp\left[i\int^{\eta}
                    d\eta^{\prime} \sqrt{k^2+\left( \frac{m^2}{H^2}-\left(\nu^2-\frac{1}{4}\right)\right)\frac{1}{\eta^{'2}}}\right]}
                    {\sqrt{2\sqrt{k^2+\left( \frac{m^2}{H^2}-\left(\nu^2-\frac{1}{4}\right)\right)\frac{1}{\eta^2}} }}
                    
	 ~~~~ & \mbox{\small {\bf for~qdS}}.
          \end{array}
\right.\eea 
\bea
\bar{u}_{k}(\eta) &=&\footnotesize
                    \displaystyle   \frac{\exp\left[-i\int^{\eta} d\eta^{\prime} 
                    \sqrt{k^2+\left(\frac{m^2_{axion}}{f^2_a H^2}-6\right)
                    \frac{1}{\eta^{'2}}} )\right]}{\sqrt{2\sqrt{k^2+
                    \left(\frac{m^2_{axion}}{f^2_a H^2}-6\right)\frac{1}{\eta^2}} }}
\equiv\footnotesize\left\{\begin{array}{ll}
                    \displaystyle  \frac{\exp\left[-i\int^{\eta}
                    d\eta^{\prime} \sqrt{k^2+\left( \frac{m^2}{H^2}-2\right)\frac{1}{\eta^{'2}}}\right]}
                    {\sqrt{2\sqrt{k^2+\left( \frac{m^2}{H^2}-2\right)\frac{1}{\eta^2}} }}\nonumber
                    ~~~~ &
 \mbox{\small {\bf for~dS}}  \\ 
	 \displaystyle  \frac{\exp\left[-i\int^{\eta}
                    d\eta^{\prime} \sqrt{k^2+\left( \frac{m^2}{H^2}-\left(\nu^2-\frac{1}{4}\right)\right)\frac{1}{\eta^{'2}}}\right]}
                    {\sqrt{2\sqrt{k^2+\left( \frac{m^2}{H^2}-\left(\nu^2-\frac{1}{4}\right)\right)\frac{1}{\eta^2}} }}
                    
	 ~~~~ & \mbox{\small {\bf for~qdS}}.
          \end{array}
\right.\eea
\bea
 \underline{\bf For~\eta<\eta_{c}}&&\nonumber\\
 u_{k}(\eta) &=&
                    \displaystyle   \frac{\exp\left[i\int^{\eta}
                    d\eta^{\prime} \sqrt{k^2+\left(\frac{m^2_{axion}}{f^2_a H^2}-6-\Delta_{c}\right)
                    \frac{1}{\eta^{'2}}} )\right]}
                    {\sqrt{2\sqrt{k^2+\left(\frac{m^2_{axion}}{f^2_a H^2}-6-\Delta_{c}\right)\frac{1}{\eta^2}} }}
                    \equiv\footnotesize\left\{\begin{array}{ll}
                    \displaystyle  \frac{\exp\left[i\int^{\eta}
                    d\eta^{\prime} \sqrt{k^2+\left( \frac{m^2}{H^2}-2\right)\frac{1}{\eta^{'2}}}\right]}
                    {\sqrt{2\sqrt{k^2+\left( \frac{m^2}{H^2}-2\right)\frac{1}{\eta^2}} }}\nonumber
                    ~~~~ &
 \mbox{\small {\bf for~dS}}  \\ 
	 \displaystyle  \frac{\exp\left[i\int^{\eta}
                    d\eta^{\prime} \sqrt{k^2+\left( \frac{m^2}{H^2}-\left(\nu^2-\frac{1}{4}\right)\right)\frac{1}{\eta^{'2}}}\right]}
                    {\sqrt{2\sqrt{k^2+\left( \frac{m^2}{H^2}-\left(\nu^2-\frac{1}{4}\right)\right)\frac{1}{\eta^2}} }}
                    
	 ~~~~ & \mbox{\small {\bf for~qdS}}.
          \end{array}
\right.\\
\bar{u}_{k}(\eta) &=&
                    \displaystyle   \frac{\exp\left[-i\int^{\eta}
                    d\eta^{\prime} \sqrt{k^2+
                    \left(\frac{m^2_{axion}}{f^2_a H^2}-6-\Delta_{c}\right)\frac{1}{\eta^{'2}}} )
                    \right]}{\sqrt{2\sqrt{k^2+\left(\frac{m^2_{axion}}{f^2_a H^2}-6-\Delta_{c}\right)\frac{1}{\eta^2}} }}
\equiv\footnotesize\left\{\begin{array}{ll}
                    \displaystyle  \frac{\exp\left[-i\int^{\eta}
                    d\eta^{\prime} \sqrt{k^2+\left( \frac{m^2}{H^2}-2\right)\frac{1}{\eta^{'2}}}\right]}
                    {\sqrt{2\sqrt{k^2+\left( \frac{m^2}{H^2}-2\right)\frac{1}{\eta^2}} }}\nonumber
                    ~~~~ &
 \mbox{\small {\bf for~dS}}  \\ 
	 \displaystyle  \frac{\exp\left[-i\int^{\eta}
                    d\eta^{\prime} \sqrt{k^2+\left( \frac{m^2}{H^2}-\left(\nu^2-\frac{1}{4}\right)\right)\frac{1}{\eta^{'2}}}\right]}
                    {\sqrt{2\sqrt{k^2+\left( \frac{m^2}{H^2}-\left(\nu^2-\frac{1}{4}\right)\right)\frac{1}{\eta^2}} }}
                    
	 ~~~~ & \mbox{\small {\bf for~qdS}}.
          \end{array}
\right.\eea
which is exactly same as the mode function obtained for the new particles in earlier section with effective sound speed $c_{S}=1$. 
So one can naturally expect that even we use WKB approximation method, as we have properly identified the exact connection between new
particle and axion through various parametrization of new particle mass parameter $m(\eta)$ as mentioned earlier, the final 
results obtained for Bogoliubov coefficients, reflection and transmission coefficients for two different representation can also be reproduced,
if we set effective sound speed $c_{S}=1$. But things will change once we consider the interaction term 
or more precisely the effective potential term as appearing in the context of axion. As we have already pointed the mathematical structure of the interaction terms 
are different for the new particle and for the axion in the present context. But we are interested in this possibility as in both theories mass parameters are conformal time dependent
in general. As we have already pointed in the last section that such contribution is solely responsible for the violation of Bell inequality in the cosmological experimental setup. 
In this section now we will explicitly investigate the cosmological consequences from the axion effective potential term in violating Bell inequality.
Before going to the further details to check the consistency of this statement let us explicitly four cases in the following where we give the exact estimate of the axion mass parameter for a given structure of new particle mass parameter:
\begin{enumerate}
 \item \underline{\bf Case I: $m\approx H$}\bea \underline{\bf For~\eta\sim \eta_{c}, early~\&~late~\eta}&&\nonumber\\ 
\frac{m_{axion}}{f_a H}&\equiv&\tiny\left\{\begin{array}{ll}
                    \displaystyle   \sqrt{5}~~~~ &
 \mbox{\small {\bf for~dS}}  \\ 
	 \displaystyle\sqrt{\frac{29}{4}-\nu^2}~~~~ & \mbox{\small {\bf for~qdS}}.
          \end{array}
\right.\\
p(\eta):=\sqrt{k^2+\left(\frac{m^2_{axion}}{f^2_a H^2}-6\right)\frac{1}{\eta^2}} &\equiv&\tiny\left\{\begin{array}{ll}
                    \displaystyle  \sqrt{k^2-\frac{1}{\eta^2}}~~~~ &
 \mbox{\small {\bf for~dS}}  \\ 
	 \displaystyle \sqrt{k^2-\left(\nu^2-\frac{5}{4}\right)\frac{1}{\eta^2}}~~~~ & \mbox{\small {\bf for~qdS}}.
          \end{array}
\right.\\
 \underline{\bf For~\eta<\eta_{c}}&&\nonumber\\\frac{m_{axion}}{f_a H}&\equiv&\tiny\left\{\begin{array}{ll}
                    \displaystyle   \sqrt{5+\Delta_{c}}~~~~ &
 \mbox{\small {\bf for~dS}}  \\ 
	 \displaystyle\sqrt{\frac{29}{4}+\Delta_{c}-\nu^2}~~~~ & \mbox{\small {\bf for~qdS}}.
          \end{array}
\right.\\
p(\eta):=\sqrt{k^2+\left(\frac{m^2_{axion}}{f^2_a H^2}-6-\Delta_{c}\right)\frac{1}{\eta^2}} &\equiv&\tiny\left\{\begin{array}{ll}
                    \displaystyle  \sqrt{k^2-\frac{1}{\eta^2}}~~~~ &
 \mbox{\small {\bf for~dS}}  \\ 
	 \displaystyle \sqrt{k^2-\left(\nu^2-\frac{5}{4}\right)\frac{1}{\eta^2}}~~~~ & \mbox{\small {\bf for~qdS}}.
          \end{array}
\right.\eea
 Consequently the solution for axion fluctuation can be recast for dS and qdS case as:
\be\begin{array}{lll}\label{yuriii2}
 \displaystyle \vartheta_{\bf k}  (\eta) =\frac{f^2_{a}}{H^2\eta^2M^2_p}\bar{a}_{\bf k}=\footnotesize\left\{\begin{array}{ll}
                    \displaystyle   \sqrt{-\eta}\left[C_1  H^{(1)}_{\sqrt{5}/2} \left(-k\eta\right) 
+ C_2 H^{(2)}_{\sqrt{5}/2} \left(-k\eta\right)\right]~~~~ &
 \mbox{\small {\bf for ~dS}}  \\ 
	\displaystyle \sqrt{-\eta}\left[C_1  H^{(1)}_{\sqrt{\nu^2-1}} \left(-k\eta\right) 
+ C_2 H^{(2)}_{\sqrt{\nu^2-1}} \left(-k\eta\right)\right]~~~~ & \mbox{\small {\bf for~ qdS}}.
          \end{array}
\right.
\end{array}\ee
In the standard WKB approximation the total solution can be recast in the following form:
\bea\label{vbcx}
\vartheta_{\bf k}  (\eta) =\frac{f^2_{a}}{H^2\eta^2M^2_p}\bar{a}_{\bf k}&=& \left[D_{1}u_{k}(\eta) + D_{2} \bar{u}_{k}(\eta)\right],\eea
where $D_{1}$ and and $D_{2}$ are two arbitrary integration constants, which depend on the 
choice of the initial condition during WKB approximation at early and late time scale. 
In the present context $u_{k}(\eta)$ and $\bar{u}_{k}(\eta)$ are defined as:
\bea &&\underline{\bf For~\eta\lesssim \eta_{c}, early~\&~late~\eta}\nonumber\\ 
u_{k}(\eta) &=&
                    \footnotesize\left\{\begin{array}{ll}
                    \displaystyle  \frac{\exp\left[i\int^{\eta}
                    d\eta^{\prime} \sqrt{k^2-\frac{1}{\eta^{'2}}}\right]}
                    {\sqrt{2\sqrt{k^2-\frac{1}{\eta^2}} }}\nonumber
                    ~~~~ &
 \mbox{\small {\bf for~dS}}  \\ 
	 \displaystyle  \frac{\exp\left[i\int^{\eta}
                    d\eta^{\prime} \sqrt{k^2-\left(\nu^2-\frac{5}{4}\right)\frac{1}{\eta^{'2}}}\right]}
                    {\sqrt{2\sqrt{k^2-\left(\nu^2-\frac{5}{4}\right)\frac{1}{\eta^2}} }}
                    
	 ~~~~ & \mbox{\small {\bf for~qdS}}.
          \end{array}
\right.\eea 
\bea
\bar{u}_{k}(\eta) &=&
\tiny\left\{\begin{array}{ll}
                    \footnotesize\frac{\exp\left[-i\int^{\eta}
                    d\eta^{\prime} \sqrt{k^2-\frac{1}{\eta^{'2}}}\right]}
                    {\sqrt{2\sqrt{k^2-\frac{1}{\eta^2}} }}\nonumber
                    ~~~~ &
 \mbox{\small {\bf for~dS}}  \\ 
	 \displaystyle  \frac{\exp\left[-i\int^{\eta}
                    d\eta^{\prime} \sqrt{k^2-\left(\nu^2-\frac{5}{4}\right)\frac{1}{\eta^{'2}}}\right]}
                    {\sqrt{2\sqrt{k^2-\left(\nu^2-\frac{5}{4}\right)\frac{1}{\eta^2}} }}
                    
	 ~~~~ & \mbox{\small {\bf for~qdS}}.
          \end{array}
\right.\eea 
 \item \underline{\bf Case II: $m\approx \sqrt{\gamma\left(\frac{\eta}{\eta_0} - 1\right)^2 + \delta}~H$}\bea \underline{\bf For~\eta\sim \eta_{c}, early~\&~late~\eta}&&\nonumber\\ 
\frac{m_{axion}}{f_a H}&\equiv&\tiny\left\{\begin{array}{ll}
                    \displaystyle   \sqrt{4+\gamma\left(\frac{\eta}{\eta_0} - 1\right)^2 + \delta}~~~~ &
 \mbox{\small {\bf for~dS}}  \\ 
	 \displaystyle\sqrt{\frac{25}{4}-\nu^2+\gamma\left(\frac{\eta}{\eta_0} - 1\right)^2 + \delta}~~~~ & \mbox{\small {\bf for~qdS}}.
          \end{array}
\right.\eea \bea
p(\eta):=\sqrt{k^2+\left(\frac{m^2_{axion}}{f^2_a H^2}-6\right)\frac{1}{\eta^2}} &\equiv&\tiny\left\{\begin{array}{ll}
                    \displaystyle  \sqrt{k^2+\left(\gamma\left(\frac{\eta}{\eta_0} - 1\right)^2 + \delta-2\right)\frac{1}{\eta^2}}~~~~ &
 \mbox{\small {\bf for~dS}}  \\ 
	 \displaystyle \sqrt{k^2+\left(\gamma\left(\frac{\eta}{\eta_0} - 1\right)^2 + \delta-\left[\nu^2-\frac{1}{4}\right]\right)\frac{1}{\eta^2}}~~~~ & \mbox{\small {\bf for~qdS}}.
          \end{array}
\right.\eea \bea
 \underline{\bf For~\eta<\eta_{c}}&&\nonumber\\\frac{m_{axion}}{f_a H}&\equiv&\tiny\left\{\begin{array}{ll}
                    \displaystyle   \sqrt{4+\gamma\left(\frac{\eta}{\eta_0} - 1\right)^2 + \delta+\Delta_{c}}~~~~ &
 \mbox{\small {\bf for~dS}}  \\ 
	 \displaystyle\sqrt{\frac{25}{4}-\nu^2+\gamma\left(\frac{\eta}{\eta_0} - 1\right)^2 + \delta+\Delta_{c}}~~~~ & \mbox{\small {\bf for~qdS}}.
          \end{array}
\right.\\ 
p(\eta):=\sqrt{k^2+\left(\frac{m^2_{axion}}{f^2_a H^2}-6-\Delta_{c}\right)\frac{1}{\eta^2}} &\equiv&\tiny\left\{\begin{array}{ll}
                    \displaystyle  \sqrt{k^2+\left(\gamma\left(\frac{\eta}{\eta_0} - 1\right)^2 + \delta-2\right)\frac{1}{\eta^2}}~~~~ &
 \mbox{\small {\bf for~dS}}  \\ 
	 \displaystyle \sqrt{k^2+\left(\gamma\left(\frac{\eta}{\eta_0} - 1\right)^2 + \delta-\left[\nu^2-\frac{1}{4}\right]\right)\frac{1}{\eta^2}}~~~~ & \mbox{\small {\bf for~qdS}}.
          \end{array}
\right.\eea
 Consequently the solution for axion fluctuation can be recast for dS and qdS case as:
\be\begin{array}{lll}\label{yu2aazuu}
 \displaystyle  \footnotesize \vartheta_{\bf k}  (\eta) =\frac{f^2_{a}}{H^2\eta^2M^2_p}\bar{a}_{\bf k} =
                   (-\eta)^{\frac{3}{2}}e^{-\frac{P\eta}{2}} (P\eta)^{A+\frac{d}{P}}
                    \left[G_{1}  ~_1F_1\left(A; B;P \eta\right)+G_2 ~ U\left(A; B; P\eta\right)\right]
\end{array}\ee
where $A$, $B$ and $P$ is defined as:
\bea\label{yu2aaz}
 \displaystyle  A&=&\footnotesize\left\{\begin{array}{ll}
                   \displaystyle -\frac{d}{2i \sqrt{k^2+C}}+i \sqrt{(\gamma+\delta) -\frac{9}{4}}+\frac{1}{2}~~~~~~~~~~~~~~~~~~~&
 \mbox{\small {\bf for ~dS}}  \\ 
	\displaystyle -\frac{d}{2i \sqrt{k^2+C}}+i \sqrt{(\gamma+\delta) -\nu^2}+\frac{1}{2}~~~~~~~~~~~~~~~~~~~ & \mbox{\small {\bf for~ qdS}}.
          \end{array}
\right.\\
\label{yu2bbz}
 \displaystyle  B &=&\footnotesize\left\{\begin{array}{ll}
                   \displaystyle 2i \sqrt{(\gamma+\delta) -\frac{9}{4}}+1~~~~~~~~~~~~~~~~~~~~~~~~~~~~~~~~~~~~~~~~~~~~~~&
 \mbox{\small {\bf for ~dS}}  \\ 
	\displaystyle 2i \sqrt{(\gamma+\delta) -\nu^2}+1~~~~~~~~~~~~~~~~~~~~~~~~~~~~~~~~~~~~~~~~~~~~~~& \mbox{\small {\bf for~ qdS}}.
          \end{array}
\right.\\
P&=&2 i\sqrt{k^2+C}.
\eea
In the standard WKB approximation the total solution can be recast in the following form:
\bea\label{vbcx}
\vartheta_{\bf k}  (\eta) =\frac{f^2_{a}}{H^2\eta^2M^2_p}\bar{a}_{\bf k}&=& \left[D_{1}u_{k}(\eta) + D_{2} \bar{u}_{k}(\eta)\right],\eea
where $D_{1}$ and and $D_{2}$ are two arbitrary integration constants, which depend on the 
choice of the initial condition during WKB approximation at early and late time scale. 
In the present context $u_{k}(\eta)$ and $\bar{u}_{k}(\eta)$ are defined as:
\bea &&\underline{\bf For~\eta\lesssim \eta_{c}, early~\&~late~\eta}\nonumber\\ 
u_{k}(\eta) &=&
                   \footnotesize\left\{\begin{array}{ll}
                    \displaystyle  \frac{\exp\left[i\int^{\eta}
                    d\eta^{\prime} \sqrt{k^2+\left(\gamma\left(\frac{\eta}{\eta_0} - 1\right)^2 + \delta-2\right)\frac{1}{\eta^{'2}}}\right]}
                    {\sqrt{2\sqrt{k^2+\left(\gamma\left(\frac{\eta}{\eta_0} - 1\right)^2 + \delta-2\right)\frac{1}{\eta^2}} }}\nonumber
                    ~~~~ &
 \mbox{\small {\bf for~dS}}  \\ 
	 \displaystyle  \frac{\exp\left[i\int^{\eta}
                    d\eta^{\prime} \sqrt{k^2+\left(\gamma\left(\frac{\eta}{\eta_0} - 1\right)^2 + \delta-\left[\nu^2-\frac{1}{4}\right]\right)\frac{1}{\eta^{'2}}}\right]}
                    {\sqrt{2\sqrt{k^2+\left(\gamma\left(\frac{\eta}{\eta_0} - 1\right)^2 + \delta-\left[\nu^2-\frac{1}{4}\right]\right)\frac{1}{\eta^2}} }}
                    
	 ~~~~ & \mbox{\small {\bf for~qdS}}.
          \end{array}
\right.\eea 
\bea
\bar{u}_{k}(\eta) &=&
\footnotesize\left\{\begin{array}{ll}
                    \displaystyle  \frac{\exp\left[-i\int^{\eta}
                    d\eta^{\prime} \sqrt{k^2+\left(\gamma\left(\frac{\eta}{\eta_0} - 1\right)^2 + \delta-2\right)\frac{1}{\eta^{'2}}}\right]}
                    {\sqrt{2\sqrt{k^2+\left(\gamma\left(\frac{\eta}{\eta_0} - 1\right)^2 + \delta-2\right)\frac{1}{\eta^2}} }}\nonumber
                    ~~~~ &
 \mbox{\small {\bf for~dS}}  \\ 
	 \displaystyle  \frac{\exp\left[-i\int^{\eta}
                    d\eta^{\prime} \sqrt{k^2+\left(\gamma\left(\frac{\eta}{\eta_0} - 1\right)^2 + \delta-\left[\nu^2-\frac{1}{4}\right]\right)\frac{1}{\eta^{'2}}}\right]}
                    {\sqrt{2\sqrt{k^2+\left(\gamma\left(\frac{\eta}{\eta_0} - 1\right)^2 + \delta-\left[\nu^2-\frac{1}{4}\right]\right)\frac{1}{\eta^2}} }}
                    
	 ~~~~ & \mbox{\small {\bf for~qdS}}.
          \end{array}
\right.\eea 
 \item \underline{\bf Case III: $m>> H$}\bea \underline{\bf For~\eta\sim \eta_{c}, early~\&~late~\eta}&&\nonumber\\ 
\frac{m_{axion}}{f_a H}&\equiv&\tiny\left\{\begin{array}{ll}
                    \displaystyle   \sqrt{4+\Upsilon^2}~~~~ &
 \mbox{\small {\bf for~dS}}  \\ 
	 \displaystyle\sqrt{\frac{25}{4}-\nu^2+\Upsilon^2}~~~~ & \mbox{\small {\bf for~qdS}}.
          \end{array}
\right.\\ 
p(\eta):=\sqrt{k^2+\left(\frac{m^2_{axion}}{f^2_a H^2}-6\right)\frac{1}{\eta^2}} &\equiv&\tiny\left\{\begin{array}{ll}
                    \displaystyle  \sqrt{k^2+\left( \Upsilon^2-2\right)\frac{1}{\eta^2}}~~~~ &
 \mbox{\small {\bf for~dS}}  \\ 
	 \displaystyle \sqrt{k^2+\left( \Upsilon^2-\left(\nu^2-\frac{1}{4}\right)\right)\frac{1}{\eta^2}}~~~~ & \mbox{\small {\bf for~qdS}}.
          \end{array}
\right.\\ 
 \underline{\bf For~\eta<\eta_{c}}&&\nonumber\\\frac{m_{axion}}{f_a H}&\equiv&\tiny\left\{\begin{array}{ll}
                    \displaystyle   \sqrt{4+\Upsilon^2+\Delta_{c}}~~~~ &
 \mbox{\small {\bf for~dS}}  \\ 
	 \displaystyle\sqrt{\frac{25}{4}-\nu^2+\Upsilon^2+\Delta_{c}}~~~~ & \mbox{\small {\bf for~qdS}}.
          \end{array}
\right.\\
p(\eta):=\sqrt{k^2+\left(\frac{m^2_{axion}}{f^2_a H^2}-6-\Delta_{c}\right)\frac{1}{\eta^2}} &\equiv&\tiny\left\{\begin{array}{ll}
                    \displaystyle  \sqrt{k^2+\left( \Upsilon^2-2\right)\frac{1}{\eta^2}}~~~~ &
 \mbox{\small {\bf for~dS}}  \\ 
	 \displaystyle \sqrt{k^2+\left( \Upsilon^2-\left(\nu^2-\frac{1}{4}\right)\right)\frac{1}{\eta^2}}~~~~ & \mbox{\small {\bf for~qdS}}.
          \end{array}
\right.\\ \eea
 Consequently the solution for axion fluctuation can be recast for dS and qdS case as:
\be\begin{array}{lll}\label{yuriii2}
 \displaystyle \footnotesize\vartheta_{\bf k}  (\eta) =\frac{f^2_{a}}{H^2\eta^2M^2_p}\bar{a}_{\bf k}=\left\{\begin{array}{ll}
                    \displaystyle   \sqrt{-\eta}\left[C_1  H^{(1)}_{i\sqrt{\Upsilon^2-\frac{9}{4}}} \left(-k\eta\right) 
+ C_2 H^{(2)}_{i\sqrt{\Upsilon^2-\frac{9}{4}}} \left(-k\eta\right)\right]~~~~ &
 \mbox{\small {\bf for ~dS}}  \\ 
	\displaystyle \sqrt{-\eta}\left[C_1  H^{(1)}_{i\sqrt{\Upsilon^2-\nu^2}} \left(-k\eta\right) 
+ C_2 H^{(2)}_{i\sqrt{\Upsilon^2-\nu^2}} \left(-k\eta\right)\right]~~~~ & \mbox{\small {\bf for~ qdS}}.
          \end{array}
\right.
\end{array}\ee
 In the standard WKB approximation the total solution can be recast in the following form:
\bea\label{vbcx}
\vartheta_{\bf k}  (\eta) =\frac{f^2_{a}}{H^2\eta^2M^2_p}\bar{a}_{\bf k}&=& \left[D_{1}u_{k}(\eta) + D_{2} \bar{u}_{k}(\eta)\right],\eea
where $D_{1}$ and and $D_{2}$ are two arbitrary integration constants, which depend on the 
choice of the initial condition during WKB approximation at early and late time scale. 
In the present context $u_{k}(\eta)$ and $\bar{u}_{k}(\eta)$ are defined as:
\bea &&\underline{\bf For~\eta\lesssim \eta_{c}, early~\&~late~\eta}\nonumber\\ 
u_{k}(\eta) &=&
                    \footnotesize\left\{\begin{array}{ll}
                    \displaystyle  \frac{\exp\left[i\int^{\eta}
                    d\eta^{\prime} \sqrt{k^2+\frac{\Upsilon^2-2}{\eta^{'2}}}\right]}
                    {\sqrt{2\sqrt{k^2+\frac{\Upsilon^2-2}{\eta^2}} }}\nonumber
                    ~~~~ &
 \mbox{\small {\bf for~dS}}  \\ 
	 \displaystyle  \frac{\exp\left[i\int^{\eta}
                    d\eta^{\prime} \sqrt{k^2+\left[\Upsilon^2-\left(\nu^2-\frac{1}{4}\right)\right]\frac{1}{\eta^{'2}}}\right]}
                    {\sqrt{2\sqrt{k^2+\left[\Upsilon^2-\left(\nu^2-\frac{1}{4}\right)\right]\frac{1}{\eta^2}} }}
                    
	 ~~~~ & \mbox{\small {\bf for~qdS}}.
          \end{array}
\right.\eea 
\bea
\bar{u}_{k}(\eta) &=&
\footnotesize\left\{\begin{array}{ll}
                    \displaystyle  \frac{\exp\left[-i\int^{\eta}
                    d\eta^{\prime} \sqrt{k^2+\frac{\Upsilon^2-2}{\eta^{'2}}}\right]}
                    {\sqrt{2\sqrt{k^2+\frac{\Upsilon^2-2}{\eta^2}} }}\nonumber
                    ~~~~ &
 \mbox{\small {\bf for~dS}}  \\ 
	 \displaystyle  \frac{\exp\left[-i\int^{\eta}
                    d\eta^{\prime} \sqrt{k^2-\left(\nu^2-\frac{5}{4}\right)\frac{1}{\eta^{'2}}}\right]}
                    {\sqrt{2\sqrt{k^2-\left(\nu^2-\frac{5}{4}\right)\frac{1}{\eta^2}} }}
                    
	 ~~~~ & \mbox{\small {\bf for~qdS}}.
          \end{array}
\right.\eea 
 \item \underline{\bf Case IV: $m<< H$}\bea \underline{\bf For~\eta\sim \eta_{c}, early~\&~late~\eta}&&\nonumber\\ 
\frac{m_{axion}}{f_a H}&\equiv&\tiny\left\{\begin{array}{ll}
                    \displaystyle   2~~~~ &
 \mbox{\small {\bf for~dS}}  \\ 
	 \displaystyle\sqrt{\frac{25}{4}-\nu^2}~~~~ & \mbox{\small {\bf for~qdS}}.
          \end{array}
\right.\\ 
p(\eta):=\sqrt{k^2+\left(\frac{m^2_{axion}}{f^2_a H^2}-6\right)\frac{1}{\eta^2}} &\equiv&\tiny\left\{\begin{array}{ll}
                    \displaystyle  \sqrt{k^2-\frac{2}{\eta^2}}~~~~ &
 \mbox{\small {\bf for~dS}}  \\ 
	 \displaystyle \sqrt{k^2-\left(\nu^2-\frac{1}{4}\right)\frac{1}{\eta^2}}~~~~ & \mbox{\small {\bf for~qdS}}.
          \end{array}
\right.\\ 
 \underline{\bf For~\eta<\eta_{c}}&&\nonumber\\\frac{m_{axion}}{f_a H}&\equiv&\tiny\left\{\begin{array}{ll}
                    \displaystyle   \sqrt{4+\Delta_{c}}~~~~ &
 \mbox{\small {\bf for~dS}}  \\ 
	 \displaystyle\sqrt{\frac{25}{4}-\nu^2+\Delta_{c}}~~~~ & \mbox{\small {\bf for~qdS}}.
          \end{array}
\right.\\
p(\eta):=\sqrt{k^2+\left(\frac{m^2_{axion}}{f^2_a H^2}-6-\Delta_{c}\right)\frac{1}{\eta^2}} &\equiv&\tiny\left\{\begin{array}{ll}
                    \displaystyle  \sqrt{k^2-\frac{2}{\eta^2}}~~~~ &
 \mbox{\small {\bf for~dS}}  \\ 
	 \displaystyle \sqrt{k^2-\left(\nu^2-\frac{1}{4}\right)\frac{1}{\eta^2}}~~~~ & \mbox{\small {\bf for~qdS}}.
          \end{array}
\right.\\ \eea
Consequently the solution for axion fluctuation can be recast for dS and qdS case as:
\be\begin{array}{lll}\label{yuriii2}
 \displaystyle \footnotesize\vartheta_{\bf k}  (\eta) =\frac{f^2_{a}}{H^2\eta^2M^2_p}\bar{a}_{\bf k}=\left\{\begin{array}{ll}
                    \displaystyle   \sqrt{-\eta}\left[C_1  H^{(1)}_{3/2} \left(-k\eta\right) 
+ C_2 H^{(2)}_{3/2} \left(-k\eta\right)\right]~~~~ &
 \mbox{\small {\bf for ~dS}}  \\ 
	\displaystyle \sqrt{-\eta}\left[C_1  H^{(1)}_{\nu} \left(-k\eta\right) 
+ C_2 H^{(2)}_{\nu} \left(-k\eta\right)\right]~~~~ & \mbox{\small {\bf for~ qdS}}.
          \end{array}
\right.
\end{array}\ee
In the standard WKB approximation the total solution can be recast in the following form:
\bea\label{vbcx}
\vartheta_{\bf k}  (\eta) =\frac{f^2_{a}}{H^2\eta^2M^2_p}\bar{a}_{\bf k}&=& \left[D_{1}u_{k}(\eta) + D_{2} \bar{u}_{k}(\eta)\right],\eea
where $D_{1}$ and and $D_{2}$ are two arbitrary integration constants, which depend on the 
choice of the initial condition during WKB approximation at early and late time scale. 
In the present context $u_{k}(\eta)$ and $\bar{u}_{k}(\eta)$ are defined as:
\bea &&\underline{\bf For~\eta\lesssim \eta_{c}, early~\&~late~\eta}\nonumber\\ 
u_{k}(\eta) &=&
                    \footnotesize\left\{\begin{array}{ll}
                    \displaystyle  \frac{\exp\left[i\int^{\eta}
                    d\eta^{\prime} \sqrt{k^2-\frac{2}{\eta^{'2}}}\right]}
                    {\sqrt{2\sqrt{k^2-\frac{2}{\eta^2}} }}\nonumber
                    ~~~~ &
 \mbox{\small {\bf for~dS}}  \\ 
	 \displaystyle  \frac{\exp\left[i\int^{\eta}
                    d\eta^{\prime} \sqrt{k^2-\left(\nu^2-\frac{1}{4}\right)\frac{1}{\eta^{'2}}}\right]}
                    {\sqrt{2\sqrt{k^2-\left(\nu^2-\frac{1}{4}\right)\frac{1}{\eta^2}} }}
                    
	 ~~~~ & \mbox{\small {\bf for~qdS}}.
          \end{array}
\right.\eea 
\bea
\bar{u}_{k}(\eta) &=&
\footnotesize\left\{\begin{array}{ll}
                    \displaystyle  \frac{\exp\left[-i\int^{\eta}
                    d\eta^{\prime} \sqrt{k^2-\frac{2}{\eta^{'2}}}\right]}
                    {\sqrt{2\sqrt{k^2-\frac{2}{\eta^2}} }}\nonumber
                    ~~~~ &
 \mbox{\small {\bf for~dS}}  \\ 
	 \displaystyle  \frac{\exp\left[-i\int^{\eta}
                    d\eta^{\prime} \sqrt{k^2-\left(\nu^2-\frac{1}{4}\right)\frac{1}{\eta^{'2}}}\right]}
                    {\sqrt{2\sqrt{k^2-\left(\nu^2-\frac{1}{4}\right)\frac{1}{\eta^2}} }}
                    
	 ~~~~ & \mbox{\small {\bf for~qdS}}.
          \end{array}
\right.\eea 
\end{enumerate}
As we have seen that the structure of the WKB solution for the axion fluctuation is exactly same with the scalar mode fluctuation for different mass parametrization for the new particle 
as appearing in the previous section, the final 
results obtained for Bogoliubov coefficients, reflection and transmission coefficients for two different representation can also be the same,
if we set effective sound speed $c_{S}=1$ in the previously obtained results. 
\subsection{Cosmological implication from axion fluctuation}
\label{sec4e}
Here things will change once we consider the interaction term 
or more precisely the effective potential term as appearing in the context of axion. This is because of the fact that the mathematical structure of the interaction terms 
are different for the new particle and for the axion in the present context. Apart from this fact we are here interested in this possibility as in both the theories mass parameters are conformal time dependent
in general and this is the key point of this work. As we have already pointed in the last section that such type of contribution in the effective action 
is solely responsible for the violation of Bell inequality in the cosmological
experimental setup. In this section we will explicitly investigate this possibility for axion fluctuation.

Here our prime objective is to explicitly compute the expression for VEV or one point function of the axion fluctuation in momentum space in presence of the axion mass parameter present in the axion interaction term
with respect to the arbitrary choice of vacuum, 
which leads to important contribution to the Bell's inequalities or violation in the context of primordial cosmology. Before computing this effect one has to remember the fact that initially we have assumed 
that effect of back reaction are very small and so that can easily be neglected from our analysis. Because of this reason we will only concentrate only in the axion mass contribution in the effective interaction or more precisely in the 
axion potential. Using the interaction picture in curved background in the present context one point function of the axion fluctuation in momentum space can be expressed as:  
\begin{equation}\label{opiocvcv}
\langle \bar{a}_k(\eta=0)\rangle = -i \int\limits_{-\infty}^0 {d\eta}~ a(\eta)~ \langle0|\left[\bar{a}_{\bf{k}} \left(0\right) , H^{axion}_{int} \left(\eta\right)\right]|0\rangle,
\end{equation}
where $a(\eta)$ is the scale factor defined in the earlier section in terms of Hubble parameter $H$ and conformal time scale $\eta$. In the interaction picture the Hamiltonian can written as:
\begin{equation}\label{hamilcvcv}
H^{axion}_{int} = - \frac{ m_{axion} }{f_a H}\partial_\eta \bar{a}(\eta,{\bf x}=0)
\end{equation}
After applying Fourier transform in Eq~(\ref{hamilcvcv}) we get the following expression:
\begin{equation}\label{uytdcvcv}
H^{axion}_{int} = -\int \frac{d^3k}{(2\pi)^3}\frac{M^2_pH^2\eta^2}{f^2_a} \frac{ m_{axion} }{f_a H} \left[{\vartheta}^\prime \left(\eta , \bf{k}\right) a\left(\bf{k}\right) + {\vartheta^{\dagger}}^\prime\left(\eta,-\bf{k}\right) a^\dagger 
\left(-\bf{k}\right)\right] 
\end{equation}
and further substituting Eq~(\ref{uytdcvcv}) in Eq~(\ref{opiocvcv}) finally we get:
\be\begin{array}{llll}\label{dfincvcv}
\displaystyle \langle\bar{a}_{\bf k}(\eta=0)\rangle = -i \int\limits_{-\infty}^0 {d\eta}~ 
\frac{M^4_pH^4\eta^4a(\eta)}{f^4_a}
\frac{m_{axion}}{f_{a}} \left({\vartheta}_{\bf{k}}
\left(0\right) {\vartheta^\dagger}^{\prime}_{\bf{k}}\left(\eta\right) - 
{\vartheta^\dagger}_{-\bf{k}}\left(0\right) {\vartheta}^{\prime}_{-\bf{k}} \left(\eta\right)\right)
\end{array}\ee
where $h_{\bf k}(\eta)$ is the exact solution or the WKB solution of the mode function as explicitly computed
in the earlier section of this paper. Now sometimes it happens that the exact solution of mode function is not 
exactly defined at $\eta=0$ point. To avoid such complicacy in the present computation for the sake of clarity 
here we introduce a Infra-Red (IR) cut-off regulator $\xi$ in the conformal time integral and consequently Eq~(\ref{dfin}) can be 
recast as:
\be\begin{array}{llll}\label{dfinv2}
\displaystyle \langle\bar{a}_{\bf k}(\eta=\xi\rightarrow 0)\rangle = -i \lim_{kc_{S}\xi\rightarrow 0}\int\limits_{-\infty}^{\xi} {d\eta}
~\frac{M^4_pH^4\eta^4a(\eta)}{f^4_a}
\frac{m_{axion}}{f_{a}} \left({\vartheta}_{\bf{k}}
\left(0\right) {\vartheta^\dagger}^{\prime}_{\bf{k}}\left(\eta\right) - 
{\vartheta^\dagger}_{-\bf{k}}\left(0\right) {\vartheta}^{\prime}_{-\bf{k}} \left(\eta\right)\right).
\end{array}\ee
Further substituting the explicit form of the scalar mode functions computed from the 
exact solution or from the WKB approximated solution we get the following simplified expression for 
the VEV of the curvature fluctuation in momentum space:
\be\begin{array}{llll}\label{dfinv3}
\displaystyle \langle\bar{a}_{\bf k}(\eta=\xi\rightarrow 0)\rangle = -i \lim_{k\xi\rightarrow 0}
\int\limits_{-\infty}^{\xi} {d\eta}
~\frac{M^4_pH^4\eta^4a(\eta)}{f^4_a}
\frac{m_{axion}}{f_{a}} \sum^{2}_{i=1}\sum^{2}_{j=1}C^{*}_{i}C_{j}\left[{\cal A}_{ij}\right]_{c_{S}=1}.
\end{array}\ee
where the conformal time dependent functions ${\cal A}_{ij}\forall i,j=1,2$ in momentum space is defined in the earlier section.

 For further simplification we consider here two limiting cases $|k\eta|\rightarrow -\infty$,
 $|k\eta|\rightarrow 0$ and $|k\eta|\approx 1$
 which are physically acceptable in the present context. First of we consider the
 results for $|k\eta|\rightarrow -\infty$. In this case we get:
 \bea\label{dfinv4bnbncv}
\displaystyle \langle\bar{a}_{\bf k}(\eta=0)\rangle_{|k\eta|\rightarrow -\infty} &=& \frac{4}{M^2_{p}\pi}
\int\limits_{-\infty}^{0} {d\eta}
~\frac{H}{a(\eta) k}\frac{M^4_pH^2\eta^2}{f^4_a} \frac{m_{axion}}{f_{a}H}\left[|C_{2}|^2 
e^{-ik\eta}-|C_{1}|^2 e^{ik\eta}\right.\nonumber \\ &&\left. 
\displaystyle~~~~~~~~~~~-i\left(C^{*}_{1}C_{2}e^{i\pi\left(\Lambda+\frac{1}{2}\right)}
+C_{1}C^{*}_{2}e^{-i\pi\left(\Lambda+\frac{1}{2}\right)}\right)\sin {\it k\eta}\right]\\&=&\left\{\begin{array}{ll}
                    \displaystyle  -\frac{2}{M^2_{p}}
\int\limits_{-\infty}^{0} {d\eta}
~\frac{H}{a(\eta) k}\frac{M^4_pH^2\eta^2}{f^4_a} \frac{m_{axion}}{f_{a}H}e^{ik\eta}~~~~ &
 \mbox{\small {\bf for ~BD vacua}}  \\ 
	\displaystyle  \frac{4}{M^2_{p}\pi}
\int\limits_{-\infty}^{0} {d\eta}
~\frac{H}{a(\eta) k}\frac{M^4_pH^2\eta^2}{f^4_a} \frac{m_{axion}}{f_{a}H}\left[\sinh^2\alpha~ 
e^{ik\eta}-\cosh^2\alpha~ e^{-ik\eta}\right.\nonumber \\ \left. 
\displaystyle~~~~~~~~~~~+i~\sinh2\alpha\cos\left({\it \pi\left(\Lambda+\frac{1}{2}\right)}+\delta\right)\sin{\it 
k\eta}\right]
                    ~~~~ & \mbox{\small {\bf for~$\alpha$~vacua~Type-I}}\\ 
	\displaystyle  \frac{4|N_{\alpha}|^2}{M^2_{p}\pi}
\int\limits_{-\infty}^{0} {d\eta}
~\frac{H}{a(\eta) k}\frac{M^4_pH^2\eta^2}{f^4_a} \frac{m_{axion}}{f_{a}H}\left[e^{\alpha+\alpha^{*}} 
e^{ik\eta}-e^{-ik\eta}\right.\nonumber \\ \left. 
\displaystyle~~~~~~~~~~~+i\left(e^{\alpha}e^{i\pi\left(\Lambda+\frac{1}{2}\right)}
+e^{\alpha^{*}}e^{-i\pi\left(\Lambda+\frac{1}{2}\right)}\right)\sin{\it 
k\eta}\right]
                    ~~~~ & \mbox{\small {\bf for~$\alpha$~vacua~Type-II}}\\ 
	\displaystyle  \frac{16i|C|^2}{M^2_{p}\pi}
\int\limits_{-\infty}^{0} {d\eta}
~\frac{H}{a(\eta) k}\frac{M^4_pH^2\eta^2}{f^4_a} \frac{m_{axion}}{f_{a}H}\sin{\it 
k\eta}\cos^2{\it \frac{\pi}{2}\left(\Lambda+\frac{1}{2}\right)}
                    ~~~~ & \mbox{\small {\bf for~special~vacua}}.
          \end{array}
\right.
\eea
On the other hand for $|k\eta|\rightarrow 0$ we get the following simplified expression:
\bea\label{dfinv4bnbnxcxcx}
\displaystyle \langle\bar{a}_{\bf k}(\eta=\xi\rightarrow 0)\rangle_{|k\eta|\rightarrow 0} &=& 
\frac{4\pi\sqrt{\xi}}{M^2_{p}}\left(C^{*}_{1}C_{2}
+C_{1}C^{*}_{2}-|C_{1}|^2-|C_{2}|^2\right)
\int\limits_{-\infty}^{\xi} {d\eta}
~\frac{H}{a(\eta)(2\pi)^3\sqrt{-\eta}} \frac{M^4_pH^2\eta^2}{f^4_a}\frac{m_{axion}}{f_{a}H}~~~~~~~~~\\ &&\displaystyle~~~~~~~~~~~~~~~~ 
\left(\Lambda-\frac{1}{2}\right)\left[\left(-\frac{k\eta}{2}\right)^{-\Lambda}\left(-\frac{k\xi}{2}\right)^{-\Lambda}
+\left(\frac{k\eta}{2}\right)^{-\Lambda}\left(\frac{kc_{S}\xi}{2}\right)^{-\Lambda}\right]\nonumber
\\&=&I_{\xi}\times\left\{\begin{array}{ll}
                    \displaystyle  -\frac{2\pi^2\sqrt{\xi}}{M^2_{p}}~~~~ &
 \mbox{\small {\bf for ~Bunch Davies vacua}}  \\ 
	\displaystyle  \frac{4\pi\sqrt{\xi}}{M^2_{p}}\left(\cos\delta\sinh 2\alpha-\cosh^2\alpha-\sinh^2\alpha\right)
                    ~~~~ & \mbox{\small {\bf for~$\alpha$~vacua~Type-I}}\\ 
	\displaystyle  \frac{4\pi\sqrt{\xi}|N_{\alpha}|^2}{M^2_{p}}\left(e^{\alpha}+e^{\alpha^{*}}
-e^{\alpha+\alpha^{*}}-1\right)
\nonumber
                    ~~~~ & \mbox{\small {\bf for~$\alpha$~vacua~Type-II}}\\ 
	\displaystyle  0
~~~ & \mbox{\small {\bf for~special~vacua}}.
          \end{array}
\right.
\eea
where the integral $I_{\xi}$ is defined as:
\bea I_{\xi}&=&\int\limits_{-\infty}^{\xi} {d\eta}
~\frac{H}{a(\eta)(2\pi)^3\sqrt{-\eta}}  \frac{M^4_pH^2\eta^2}{f^4_a}\frac{m_{axion}}{f_{a}H}
\left(\Lambda-\frac{1}{2}\right)\left[\left(-\frac{k\eta}{2}\right)^{-\Lambda}\left(-\frac{k\xi}{2}\right)^{-\Lambda}
+\left(\frac{k\eta}{2}\right)^{-\Lambda}\left(\frac{k\xi}{2}\right)^{-\Lambda}\right].~~~~~~~~~~~~\eea
Finally the other hand for $|k\eta|\approx 1$ we get the following simplified expression:
\bea\label{dfinv4bnbnxcxcx}
\displaystyle \langle\bar{a}_{\bf k}(\eta=\xi\rightarrow 0)\rangle_{|k\eta|\approx 1} &=& 
\frac{4\pi\sqrt{\xi}}{M^2_{p}}\left(C^{*}_{1}C_{2}
+C_{1}C^{*}_{2}-|C_{1}|^2-|C_{2}|^2\right)
\int\limits_{-\infty}^{\xi} {d\eta}
~\frac{H}{a(\eta)(2\pi)^3\sqrt{-\eta}}  \frac{M^4_pH^2\eta^2}{f^4_a}\frac{m_{axion}}{f_{a}H}~~~~~~~~~~~~~\\ &&
\displaystyle~~~~~~~~~~~~~~~~ 
\left(\Lambda-\frac{1}{2}\right)\left(\frac{1}{2}\right)^{-\Lambda}\left[\left(-\frac{k\xi}{2}\right)^{-\Lambda}
+\left(-1\right)^{-\Lambda}\left(\frac{k\xi}{2}\right)^{-\Lambda}\right]\nonumber
\\&=&J_{\xi}\times\left\{\begin{array}{ll}
                    \displaystyle  -\frac{2\pi^2\sqrt{\xi}}{M^2_{p}}~~~~ &
 \mbox{\small {\bf for ~Bunch Davies vacua}}  \\ 
	\displaystyle  \frac{4\pi\sqrt{\xi}}{M^2_{p}}\left(\cos\delta\sinh 2\alpha-\cosh^2\alpha-\sinh^2\alpha\right)
                    ~~~~ & \mbox{\small {\bf for~$\alpha$~vacua~Type-I}}\\ 
	\displaystyle  \frac{4\pi\sqrt{\xi}|N_{\alpha}|^2}{M^2_{p}}\left(e^{\alpha}+e^{\alpha^{*}}
-e^{\alpha+\alpha^{*}}-1\right)
\nonumber
                    ~~~~ & \mbox{\small {\bf for~$\alpha$~vacua~Type-II}}\\ 
	\displaystyle  0
~~~ & \mbox{\small {\bf for~special~vacua}}.
          \end{array}
\right.
\eea
where the integral $J_{\xi}$ is defined as:
\bea J_{\xi}&=&\int\limits_{-\infty}^{\xi} {d\eta}
~\frac{H}{a(\eta)(2\pi)^3\sqrt{-\eta}}  \frac{M^4_pH^2\eta^2}{f^4_a}\frac{m_{axion}}{f_{a}H}
\left(\Lambda-\frac{1}{2}\right)\left(\frac{1}{2}\right)^{-\Lambda}
\left[\left(-\frac{k\xi}{2}\right)^{-\Lambda}
+\left(-1\right)^{-\Lambda}\left(\frac{k\xi}{2}\right)^{-\Lambda}\right].~~~~~~~~~~~~\eea
Now to analyze the behaviour of the expectation value of scalar curvature perturbation in position space 
we need to take the Fourier transform of the expectation value of scalar curvature perturbation already computed 
in momentum space. For the most generalized solution we get the following result:
\be\begin{array}{llll}\label{dfinsd}
\displaystyle \langle\bar{a}({\bf x},\eta=0)\rangle = -i \int \frac{d^{3}k}{(2\pi)^3}\int\limits_{-\infty}^0 {d\eta}~
\frac{M^4_pH^4\eta^4a(\eta)}{f^4_a}
\frac{m_{axion}}{f_{a}} \left({\vartheta}_{\bf{k}}
\left(0\right) {\vartheta^\dagger}^{\prime}_{\bf{k}}\left(\eta\right) - 
{\vartheta^\dagger}_{-\bf{k}}\left(0\right) {\vartheta}^{\prime}_{-\bf{k}} \left(\eta\right)\right)
\end{array}\ee
where $h_{\bf k}(\eta)$ is the exact solution or the WKB solution of the mode function as explicitly computed
in the earlier section of this paper. Following the previous methodology here we also
introduce a Infra-Red (IR) cut-off regulator $\xi$ in the conformal time integral and 
consequently Eq~(\ref{dfinsd}) can be 
recast in the following form as:
\be\begin{array}{llll}\label{dfinv2}
\displaystyle \langle\bar{a}({\bf x},\eta=0)\rangle = -i \lim_{k\xi\rightarrow 0}\int \frac{d^{3}k}{(2\pi)^3}\int\limits_{-\infty}^{\xi} {d\eta}
~ \frac{M^4_pH^4\eta^4a(\eta)}{f^4_a}
\frac{m_{axion}}{f_{a}} \left({\vartheta}_{\bf{k}}
\left(0\right) {\vartheta^\dagger}^{\prime}_{\bf{k}}\left(\eta\right) - 
{\vartheta^\dagger}_{-\bf{k}}\left(0\right) {\vartheta}^{\prime}_{-\bf{k}} \left(\eta\right)\right).
\end{array}\ee
Further substituting the explicit form of the scalar mode functions computed from the 
exact solution or from the WKB approximated solution we get the following simplified expression for 
the VEV of the curvature fluctuation in position space:
\be\begin{array}{llll}\label{dfinv3}
\displaystyle \langle\bar{a}({\bf x},\eta=0)\rangle = -i \lim_{k\xi\rightarrow 0}
\int \frac{d^{3}k}{(2\pi)^3}\int\limits_{-\infty}^{\xi} {d\eta}
~\frac{M^4_pH^4\eta^4a(\eta)}{f^4_a}
\frac{m_{axion}}{f_{a}}\sum^{2}_{i=1}\sum^{2}_{j=1}C^{*}_{i}C_{j}{\cal A}_{ij}e^{i{\bf k}.{\bf x}}.
\end{array}\ee
where the conformal time dependent functions ${\cal A}_{ij}\forall i,j=1,2$ in momentum
space is already defined earlier.

Similarly in the position space the representative expressions for the expectation value of the 
scalar curvature perturbation along with three
limiting cases $|k\eta|\rightarrow -\infty$, $|k\eta|\rightarrow 0$ and $|k\eta|\approx 1$
are given by:
\bea\label{dfinv4bnbncv}
\displaystyle \langle\bar{a}({\bf x},\eta=0)\rangle_{|k\eta|\rightarrow -\infty} &\approx& 
-\frac{4H^3 M^2_p}{\pi f^4_a}
\left[|C_{2}|^2 O_{1}
-|C_{1}|^2 O_{2}\right.\\ &&\left.\nonumber~~~~~~~~~~~~~~~~~~~~~~-i\left(C^{*}_{1}C_{2}e^{i\pi\left(\Lambda+\frac{1}{2}\right)}
+C_{1}C^{*}_{2}e^{-i\pi\left(\Lambda+\frac{1}{2}\right)}\right)O_{3}\right]\\&=&\left\{\begin{array}{ll}
                  \displaystyle    \frac{2H^3 M^2_p}{f^4_a}
O_{2}\nonumber &
 \mbox{\small {\bf for ~Bunch Davies}}  \\ 
	 \displaystyle  -\frac{4H^3 M^2_p}{\pi f^4_a}
\left[\sinh^2\alpha~ 
O_{1}-\cosh^2\alpha~O_{2}\right.\nonumber \\ \left.  \displaystyle
-i~\sinh2\alpha\cos\left({\it \pi\left(\Lambda+\frac{1}{2}\right)}+\delta\right)O_{3}\right]
                     & \mbox{\small {\bf for~$\alpha$~vacua~Type-I}}\\ 
	  \displaystyle -\frac{4H^3 M^2_p|N_{\alpha}|^2}{\pi f^4_a}
\left[e^{\alpha+\alpha^{*}} 
O_{1}-O_{2}\right.\nonumber \\ \left.  \displaystyle
-i\left(e^{\alpha}e^{i\pi\left(\Lambda+\frac{1}{2}\right)}
+e^{\alpha^{*}}e^{-i\pi\left(\Lambda+\frac{1}{2}\right)}\right)O_{3}\right]
                     & \mbox{\small {\bf for~$\alpha$~vacua~Type-II}}\\ 
	 \displaystyle - \frac{16i|C|^2H^3M^2_p}{\pi f^4_a}
\cos^2{\it \frac{\pi}{2}\left(\Lambda+\frac{1}{2}\right)}O_{3}
                    & \mbox{\small {\bf for~special~vacua}}.
          \end{array}
\right.
\eea
\bea\label{dfinv4bnbnxcxcx}
\small\displaystyle \langle\bar{a}({\bf x},\eta=\xi\rightarrow 0)\rangle_{|k\eta|\rightarrow 0} &=& 
\frac{H^3M^2_p\sqrt{\xi}}{2f^4_a}\left(C^{*}_{1}C_{2}
+C_{1}C^{*}_{2}-|C_{1}|^2-|C_{2}|^2\right)
O^{\xi\theta}_{4}
\\&=&O^{\xi\theta}_{4}\times\left\{\begin{array}{ll}
                    \displaystyle  -\frac{2H^3M^2_p\pi^2\sqrt{\xi}}{f^4_a}~~~~ &
 \mbox{\small {\bf for ~Bunch Davies vacua}}  \\ 
	\displaystyle  \frac{H^3M^2_p\sqrt{\xi}}{2f^4_a}\left(\cos\delta\sinh 2\alpha-\cosh^2\alpha-\sinh^2\alpha\right)
                    ~~~~ & \mbox{\small {\bf for~$\alpha$~vacua~Type-I}}\\ 
	\displaystyle  \frac{H^3M^2_p\sqrt{\xi}|N_{\alpha}|^2}{2f^4_a}\left(e^{\alpha}+e^{\alpha^{*}}
-e^{\alpha+\alpha^{*}}-1\right)
\nonumber
                    ~~~~ & \mbox{\small {\bf for~$\alpha$~vacua~Type-II}}\\ 
	\displaystyle  0
~~~ & \mbox{\small {\bf for~special~vacua}}.
          \end{array}
\right.
\eea
\bea\label{dfinv4bnbnxcxcx}
\small\displaystyle \langle\bar{a}({\bf x},\eta=\xi\rightarrow 0)\rangle_{|kc_{S}\eta|\approx 1} &=& 
\frac{H^3M^2_p\sqrt{\xi}}{2f^4_a}\left(C^{*}_{1}C_{2}
+C_{1}C^{*}_{2}-|C_{1}|^2-|C_{2}|^2\right)
O^{\xi\theta}_{5}
\\&=&O^{\xi\theta}_{5}\times\left\{\begin{array}{ll}
                    \displaystyle  -\frac{2H^3M^2_p\pi^2\sqrt{\xi}}{f^4_a}~~~~ &
 \mbox{\small {\bf for ~Bunch Davies vacua}}  \\ 
	\displaystyle  \frac{H^3M^2_p\sqrt{\xi}}{2f^4_a}\left(\cos\delta\sinh 2\alpha-\cosh^2\alpha-\sinh^2\alpha\right)
                    ~~~~ & \mbox{\small {\bf for~$\alpha$~vacua~Type-I}}\\ 
	\displaystyle  \frac{H^3M^2_p\sqrt{\xi}|N_{\alpha}|^2}{2f^4_a}\left(e^{\alpha}+e^{\alpha^{*}}
-e^{\alpha+\alpha^{*}}-1\right)
\nonumber
                    ~~~~ & \mbox{\small {\bf for~$\alpha$~vacua~Type-II}}\\ 
	\displaystyle  0
~~~ & \mbox{\small {\bf for~special~vacua}}.
          \end{array}
\right.
\eea
where we introduce the following integrals $O_{1}, O_{2}, O_{3}, O^{\xi\theta}_{4}, O^{\xi\theta}_{5}$ are given by:
\bea O_{1}&=& \int \frac{d^{3}k}{(2\pi)^3}~e^{i{\bf k}.{\bf x}}\int\limits_{-\infty}^{0} {d\eta}~
~\frac{\eta^3}{k}~\frac{m_{axion}}{f_a}
e^{-ik\eta},\\ 
 O_{2}&=&  \int \frac{d^{3}k}{(2\pi)^3}~e^{i{\bf k}.{\bf x}}\int\limits_{-\infty}^{0} {d\eta}
~\frac{\eta^3}{k}~\frac{m_{axion}}{f_a}~e^{ik\eta},\\ 
 O_{3}&=& \int \frac{d^{3}k}{(2\pi)^3}~e^{i{\bf k}.{\bf x}}\int\limits_{-\infty}^{0} {d\eta}
~\frac{\eta^3}{k}~\frac{m_{axion}}{f_a}\sin {\it k\eta},\\
 O^{\xi\theta}_{4}&=&\int \frac{d^{3}k}{(2\pi)^3}~e^{i{\bf k}.{\bf x}}\int\limits_{-\infty}^{\xi} {d\eta}
~\frac{\left(\Lambda-\frac{1}{2}\right)\eta^2}{a(\eta)\sqrt{-\eta}} \frac{m_{axion}}{f_aH}
\left[\left(-\frac{k\eta}{2}\right)^{-\Lambda}\left(-\frac{k\xi}{2}\right)^{-\Lambda}
+\left(\frac{k\eta}{2}\right)^{-\Lambda}\left(\frac{k\xi}{2}\right)^{-\Lambda}\right],~~~~~~~\\
 O^{\xi\theta}_{5}&=&\int \frac{d^{3}k}{(2\pi)^3}~e^{i{\bf k}.{\bf x}}\int\limits_{-\infty}^{\xi} {d\eta}
~\frac{\left(\Lambda-\frac{1}{2}\right)\eta^2}{a(\eta)\sqrt{-\eta}} \frac{m_{axion}}{f_aH}
\left(\frac{1}{2}\right)^{-\Lambda}
\left[\left(-\frac{k\xi}{2}\right)^{-\Lambda}
+\left(-1\right)^{-\Lambda}\left(\frac{k\xi}{2}\right)^{-\Lambda}\right].~~\eea
 Now to compute these momentum integrals we follow a few sets of assumptions that we have mentioned for the new massive particle in the last section. two situations where ${\bf k}$ and ${\bf x}$ are parallel and having an angle $\Theta$ in between them. For the 
  first case ${\bf k}.{\bf x}=kx$ and for the second case we have ${\bf k}.{\bf x}=kx\cos\Theta,$ where the range of the 
  angular parameter is lying within the window $\Theta_{1}\leq \Theta\leq \Theta_{2}$, where 
  $\Theta_{1}$ and $\Theta_{2}$ are two cut-off in the angular coordinate which are introduced 
  to regularize the momentum integrals in the present context.
 For Case I and Case II we get the following results: 
\bea {\bf For~Case~I:}\nonumber\\
O_{1}&=& \frac{1}{2\pi^2}\int^{\infty}_{0} dk~\int\limits_{-\infty}^{0} {d\eta}~\eta^3 k~\frac{m_{axion}}{f_a}~e^{ik\left(x-\eta\right)}
=-\frac{i|{\bf x}|^2}{2\pi^2}\left(\frac{m_{axion}}{f_{a}}\right)_{\eta=-|{\bf x}|},\\ 
 O_{2}&=&  \frac{1}{2\pi^2}\int^{\infty}_{0} dk~\int\limits_{-\infty}^{0} {d\eta}~\eta^3 k~\frac{m_{axion}}{f_a}~e^{ik\left(x+\eta\right)}
=\frac{i|{\bf x}|^2}{2\pi^2}\left(\frac{m_{axion}}{f_{a}}\right)_{\eta=-|{\bf x}|},\\ 
 O_{3}&=&  \frac{1}{2\pi^2}\int^{\infty}_{0} dk\int\limits_{-\infty}^{0} {d\eta}
~\eta^3 k~\frac{m_{axion}}{f_a}~e^{ikx}\sin {\it k\eta}=\frac{|{\bf x}|^2}{2\pi^2}\left(\frac{m_{axion}}{f_{a}}\right)_{\eta=-|{\bf x}|},\\
 O^{\xi\theta}_{4}&=& \frac{1}{2\pi^2}\int^{\infty}_{0} dk~k^2~e^{ikx}\int\limits_{-\infty}^{\xi} {d\eta}
~\frac{\left(\Lambda-\frac{1}{2}\right)\eta^2}{a(\eta)\sqrt{-\eta}} \frac{m_{axion}}{f_aH}
\left[\left(-\frac{k\eta}{2}\right)^{-\Lambda}\left(-\frac{k\xi}{2}\right)^{-\Lambda}
+\left(\frac{k\eta}{2}\right)^{-\Lambda}\left(\frac{k\xi}{2}\right)^{-\Lambda}\right],~~~~~~~~\\
 O^{\xi\theta}_{5}&=& \frac{1}{2\pi^2}\int^{\infty}_{0} dk~e^{ikx}~k^{2-\Lambda}\int\limits_{-\infty}^{\xi} {d\eta}
~\frac{\left(\Lambda-\frac{1}{2}\right)\eta^2}{a(\eta)\sqrt{-\eta}} \frac{m_{axion}}{f_aH}
\left(\frac{1}{2}\right)^{-\Lambda}
\left[\left(-\frac{\xi}{2}\right)^{-\Lambda}
+\left(-1\right)^{-\Lambda}\left(\frac{\xi}{2}\right)^{-\Lambda}\right].\eea
 \bea {\bf For~Case~II:}\nonumber\\
 O_{1}&=&  \frac{1}{4\pi^2}\int^{\infty}_{0} dk~\int\limits_{-\infty}^{0} {d\eta}\int^{\Theta_{2}}_{\Theta_{1}}d\Theta~\eta^3 k~\frac{m_{axion}}{f_{a}}
 ~e^{ik\left(x\cos\Theta-\eta\right)}\nonumber\\
&=&-i\int^{\Theta_{2}}_{\Theta_{1}}d\Theta \frac{|{\bf x}|^2\left(\frac{m_{axion}}{f_{a}}\right)_{\eta=-|{\bf x}|\cos\Theta}}{4\pi^2c^2_{S}}\cos^2\Theta,~~~~~~~~~~~~~\\ 
 O_{2}&=& \frac{1}{4\pi^2}\int^{\infty}_{0} dk~\int\limits_{-\infty}^{0} {d\eta}\int^{\Theta_{2}}_{\Theta_{1}}d\Theta~\eta^3 k~\frac{m_{axion}}{f_{a}}~
 e^{ik\left(x\cos\Theta+\eta\right)}
\nonumber\\
&=&i\int^{\Theta_{2}}_{\Theta_{1}}d\Theta \frac{|{\bf x}|^2\left(\frac{m_{axion}}{f_{a}}\right)_{\eta=-|{\bf x}|\cos\Theta}}{4\pi^2c^2_{S}}\cos^2\Theta,\\ 
 O_{3}&=& \frac{1}{4\pi^2}\int^{\infty}_{0} dk\int\limits_{-\infty}^{0} {d\eta}\int^{\Theta_{2}}_{\Theta_{1}}d\Theta
~\eta^3 k~\frac{m_{axion}}{f_{a}}~e^{ikx\cos\Theta}\sin {\it k\eta}\nonumber\\
&=&\int^{\Theta_{2}}_{\Theta_{1}}d\Theta \frac{|{\bf x}|^2\left(\frac{m_{axion}}{f_{a}}\right)_{\eta=-|{\bf x}|\cos\Theta}}{4\pi^2c^2_{S}}\cos^2\Theta,\\
 O^{\xi\theta}_{4}&=& \frac{1}{4\pi^2}\int^{\infty}_{0} dk~\int\limits_{-\infty}^{\xi} {d\eta}~\int^{\Theta_{2}}_{\Theta_{1}}~d\Theta~k^2~e^{ikx\cos\Theta}
~\frac{\eta^2\left(\Lambda-\frac{1}{2}\right)}{a(\eta)\sqrt{-\eta}}  \frac{m_{axion}}{f_aH}
\\
\nonumber&&~~~~~~~~~~~~~~~~~~~\left[\left(-\frac{k\eta}{2}\right)^{-\Lambda}\left(-\frac{k\xi}{2}\right)^{-\Lambda}
+\left(\frac{k\eta}{2}\right)^{-\Lambda}\left(\frac{k\xi}{2}\right)^{-\Lambda}\right],~~~~~~~~~~~~\\
 O^{\xi\theta}_{5}&=&\frac{1}{4\pi^2}\int^{\infty}_{0} dk~\int^{\Theta_{2}}_{\Theta_{1}}~d\Theta~e^{ikx\cos \Theta}~k^{2-\Lambda}\int\limits_{-\infty}^{\xi} {d\eta}
~\frac{\eta^2\left(\Lambda-\frac{1}{2}\right)}{a(\eta)\sqrt{-\eta}}  \frac{m_{axion}}{f_aH}
\left(\frac{1}{2}\right)^{-\Lambda}\\
\nonumber&&~~~~~~~~~~~~~~~~~~~\left[\left(-\frac{\xi}{2}\right)^{-\Lambda}
+\left(-1\right)^{-\Lambda}\left(\frac{\xi}{2}\right)^{-\Lambda}\right].~~~~~~~~~~~~\eea
 where $\Theta_{1}$ and $\Theta_{2}$ plays the role of angular regulator in the present context.
 
 Now our objective is to compute the expression for the 
 two point correlation function from scalar curvature perturbation. Following the previously mentioned 
 computational technique of in-in formalism, which is commonly known as the Swinger-Keyldish formalism here we 
 compute the expression for the two point correlation function from scalar curvature perturbation.
 Using the interaction picture the two point correlation function
 of the curvature fluctuation in momentum space can be expressed as:  
\begin{equation}\label{opiovbvb}
\langle\bar{a}_{\bf k}(\eta)\bar{a}_{\bf q}(\eta)\rangle = (2\pi)^3 \delta^{3}({\bf k}+{\bf q})\frac{2\pi^2}{k^3}
\Delta_{\bar{a}}(k),
\end{equation}
where the primordial power spectrum for scalar mode at any arbitrary momentum scale can be written as:
\bea\label{dfinv4bnbncv}
\displaystyle \Delta_{\bar{a}}(k) &=& \frac{k^3|\vartheta_{\bf k}|^2}{2\pi^2 (f^2_a/H^2\eta^2)}\\
&=&\frac{(-k\eta)^3 H^2}{2\pi^2 f^2_a}\sum^{2}_{i=1}\sum^{2}_{j=1} C^{*}_{i}C_{j} 
\left[U_{ij}\right]_{c_{S}=1}\nonumber.
\eea
where $U_{ij}\forall i,j=1,2$ are defined for new massive particles in the earlier section.

For further simplification we consider here three limiting cases $|k\eta|\rightarrow -\infty$,
$|k\eta|\rightarrow 0$ and $|k\eta|\approx 1$,
 which are physically acceptable in the present context. First of we consider the
 results for $|k\eta|\rightarrow -\infty$. In this case we get:
 \begin{equation}\label{opiovbvb}
\langle\bar{a}_{\bf k}(\eta)\bar{a}_{\bf q}(\eta)\rangle_{|k\eta|\rightarrow -\infty} =
(2\pi)^3 \delta^{3}({\bf k}+{\bf q})\frac{2\pi^2}{k^3}
\left[\Delta_{\bar{a}}(k)\right]_{{|k\eta|\rightarrow -\infty}},
\end{equation}
where the primordial power spectrum for scalar mode at $|k\eta|\rightarrow -\infty$ can be written as:
 \bea\label{dfinv4bnbncv}
\displaystyle \left[\Delta_{\bar{a}}(k)\right]_{{|k\eta|\rightarrow -\infty}} &\approx& 
\frac{H^2}{f^2_a}\frac{k^2\eta^2}{\pi^3}
\left[|C_{2}|^2+|C_{1}|^2\right.\\ &&\left.~~~~~~~\nonumber+\left(C^{*}_{1}C_{2}e^{2ikc_{S}\eta}e^{i\pi\left(\Lambda+\frac{1}{2}\right)}
+C_{1}C^{*}_{2}e^{-2ikc_{S}\eta}e^{-i\pi\left(\Lambda+\frac{1}{2}\right)}\right)\right]\\&=&\left\{\begin{array}{ll}
                     \displaystyle  \frac{H^2}{f^2_a}\frac{(-k\eta)^2}{2\pi^2}\nonumber &
 \mbox{\small {\bf for ~Bunch Davies}}  \\ 
	 \displaystyle  \frac{H^2}{f^2_a}\frac{(-k\eta)^2}{\pi^3}
\left[\sinh^2\alpha +\cosh^2\alpha
\right.\\ \left. \displaystyle ~~~~~~~~~~~~~+\sinh2\alpha\cos{\it \left(2kc_{S}\eta+\pi\left(\Lambda+\frac{1}{2}\right)+\delta\right)}\right]
                     & \mbox{\small {\bf for~$\alpha$~vacua~Type-I}}\\ 
	  \displaystyle  \frac{H^2}{f^2_a}\frac{4(-k\eta)^2 |N_{\alpha}|^2}{\pi^3}
\cos^2{\it \left(kc_{S}\eta+\frac{\pi}{2}\left(\Lambda+\frac{1}{2}\right)-i\frac{\alpha}{2}\right)}
                     & \mbox{\small {\bf for~$\alpha$~vacua~Type-II}}\\ 
	 \displaystyle \frac{H^2}{f^2_a}\frac{4(-k\eta)^2}{\pi^3}
\cos^2{\it \left(kc_{S}\eta+\frac{\pi}{2}\left(\Lambda+\frac{1}{2}\right)\right)}
                    & \mbox{\small {\bf for~special~vacua}}.
          \end{array}
\right.
\eea
Next we consider the
 results for $|k\eta|\rightarrow 0$. In this case we get:
 \begin{equation}\label{opiovbvb}
\langle\bar{a}_{\bf k}(\eta)\bar{a}_{\bf q}(\eta)\rangle_{|k\eta|\rightarrow 0} =
(2\pi)^3 \delta^{3}({\bf k}+{\bf q})\frac{2\pi^2}{k^3}
\left[\Delta_{\bar{a}}(k)\right]_{{|k\eta|\rightarrow 0}},
\end{equation}
where the primordial power spectrum for scalar mode at $|k\eta|\rightarrow 0$ can be written as:
 \bea\label{dfinv4bnbncv}
\displaystyle \left[\Delta_{\bar{a}}(k)\right]_{{|k\eta|\rightarrow 0}} &\approx& 
\frac{H^2}{f^2_a}\frac{(-k\eta)^3}{2\pi^4}\Gamma^2(\Lambda)\left(-\frac{k\eta}{2}\right)^{-2\Lambda}
\left[|C_{2}|^2+|C_{1}|^2-\left(C^{*}_{1}C_{2}
+C_{1}C^{*}_{2}\right)\right]\\&=&\left\{\begin{array}{ll}
                    \displaystyle   \frac{H^2}{f^2_a}\frac{(-k\eta)^{3-2\Lambda}}{2^{2(2-\Lambda)}
                      \pi^2}\left|\frac{\Gamma(\Lambda)}{\Gamma\left(\frac{3}{2}\right)}\right|^2
                     \nonumber &
 \mbox{\small {\bf for ~Bunch Davies}}  \\ 
	 \displaystyle \frac{H^2}{f^2_a}\frac{(-k\eta)^{3-2\Lambda}}{2^{(3-2\Lambda)}
                      \pi^3}\left|\frac{\Gamma(\Lambda)}{\Gamma\left(\frac{3}{2}\right)}\right|^2
\left[\sinh^2\alpha +\cosh^2\alpha
-\sinh2\alpha\cos{\it \delta}\right]
                     & \mbox{\small {\bf for~$\alpha$~vacua~Type-I}}\\ 
 \displaystyle	\frac{H^2}{f^2_a}\frac{(-k\eta)^{3-2\Lambda}|N_{\alpha}|^2}{2^{(1-2\Lambda)}
                      \pi^3}\left|\frac{\Gamma(\Lambda)}
                      {\Gamma\left(\frac{3}{2}\right)}\right|^2
\sin^2\frac{\alpha}{2}
                     & \mbox{\small {\bf for~$\alpha$~vacua~Type-II}}\\ 
	 0
                    & \mbox{\small {\bf for~special~vacua}}.
          \end{array}
\right.
\eea
Finally we consider the
 results for $|k\eta|\approx 1$. In this case we get:
 \begin{equation}\label{opiovbvb}
\langle\bar{a}_{\bf k}(\eta=0)\bar{a}_{\bf q}(\eta=0)\rangle_{|k\eta|\approx 1} =
(2\pi)^3 \delta^{3}({\bf k}+{\bf q})\frac{2\pi^2}{k^3}
\left[\Delta_{\bar{a}}(k)\right]_{{|k\eta|\approx 1}},
\end{equation}
where the primordial power spectrum for scalar mode at $|k\eta|\approx 1$ can be written as:
 \bea\label{dfinv4bnbncv}
\displaystyle \left[\Delta_{\bar{a}}(k)\right]_{{|k\eta|\approx 1}} &\approx& 
\frac{H^2}{f^2_a}\frac{1}{2^{2(1-\Lambda)}\pi^3}\left|\frac{\Gamma(\Lambda)}{\Gamma
\left(\frac{3}{2}\right)}\right|^2
\left[|C_{2}|^2+|C_{1}|^2-\left(C^{*}_{1}C_{2}
+C_{1}C^{*}_{2}\right)\right]\\&=&\left\{\begin{array}{ll}
                     \displaystyle  \frac{H^2}{f^2_a}\frac{1}{2^{2(2-\Lambda)}
                      \pi^2}\left|\frac{\Gamma(\Lambda)}{\Gamma\left(\frac{3}{2}\right)}\right|^2
                     \nonumber &
 \mbox{\small {\bf for ~Bunch Davies}}  \\ 
	 \displaystyle \frac{H^2}{f^2_a}\frac{1}{2^{(3-2\Lambda)}
                      \pi^3}\left|\frac{\Gamma(\Lambda)}{\Gamma\left(\frac{3}{2}\right)}\right|^2
\left[\sinh^2\alpha +\cosh^2\alpha
\right.\\ \left. ~~~~~~~~~~~~~~~~~~~~~~~~~~~~~~~~~~~~~ \displaystyle-\sinh2\alpha\cos{\it \delta}\right]
                     & \mbox{\small {\bf for~$\alpha$~vacua~Type-I}}\\ 
	 \displaystyle \frac{H^2}{f^2_a}\frac{|N_{\alpha}|^2}{2^{(1-2\Lambda)}
                      \pi^3}\left|\frac{\Gamma(\Lambda)}
                      {\Gamma\left(\frac{3}{2}\right)}\right|^2
\sin^2\frac{\alpha}{2}
                     & \mbox{\small {\bf for~$\alpha$~vacua~Type-II}}\\ 
	 0
                    & \mbox{\small {\bf for~special~vacua}}.
          \end{array}
\right.
\eea
Now to analyze the behaviour of the two point correlation function of scalar curvature perturbation in position space 
we need to take the Fourier transform of the two point correlation function of scalar curvature perturbation already computed 
in momentum space. For the most generalized solution we get the following result:
\be\begin{array}{llll}\label{dfinv3}
\displaystyle \langle\bar{a}({\bf x},\eta)\bar{a}({\bf y},\eta)\rangle = \int \frac{d^{3}k}{(2\pi)^3 }
~e^{i{\bf k}.({\bf x}-{\bf y})}\frac{(-\eta)^3H^2}{f^2_a}\sum^{2}_{i=1}\sum^{2}_{j=1}C^{*}_{i}C_{j}\left[U_{ij}\right]_{c_{S}=1}.
\end{array}\ee
where the conformal time dependent functions $U_{ij}\forall i,j=1,2$ in momentum
space is already defined earlier.

Similarly in the position space the representative expressions for the expectation value of the 
scalar curvature perturbation along with 
limiting case $|k\eta|\rightarrow -\infty$ is given by:
 \bea\label{dfinv4bnbncv}
\displaystyle \langle\bar{a}({\bf x},\eta)\bar{a}({\bf y},\eta)\rangle_{|k\eta|\rightarrow -\infty}&\approx& 
\frac{1}{4\pi^4}\frac{H^2}{f^2_a}\eta^2
\left[\left(|C_{2}|^2+|C_{1}|^2\right)J_{1}\right.\\ &&\left.~~~~~~~\nonumber+\left(C^{*}_{1}C_{2}
e^{i\pi\left(\Lambda+\frac{1}{2}\right)}J_{2}
+C_{1}C^{*}_{2}e^{-i\pi\left(\Lambda+\frac{1}{2}\right)}J_{3}\right)\right]\\&=&\left\{\begin{array}{ll}
                    \displaystyle   \frac{H^2}{f^2_a}\frac{\eta^2}{(2\pi)^3}J_{1}\nonumber &
 \mbox{\small {\bf for ~Bunch Davies}}  \\ 
	 \displaystyle  \frac{H^2}{f^2_a}\frac{\eta^2}{4\pi^4}
\left[\left(\sinh^2\alpha +\cosh^2\alpha\right)J_{1}
\right.\\ \left. ~~~~~~~~~~~~~ \displaystyle+\frac{1}{2}\sinh2\alpha\left( e^{
i\left(\pi\left(\Lambda+\frac{1}{2}\right)+\delta\right)}J_{2}+e^{
-i\left(\pi\left(\Lambda+\frac{1}{2}\right)+\delta\right)}\right)J_{3}\right]
                     & \mbox{\small {\bf for~$\alpha$~vacua~Type-I}}\\ 
	 \displaystyle \frac{H^2}{f^2_a}\frac{\eta^2 |N_{\alpha}|^2}{2\pi^4}
\left[J_{1}+\left(e^{i\pi\left(\Lambda+\frac{1}{2}\right)}e^{\alpha}J_{2}+
e^{-i\pi\left(\Lambda+\frac{1}{2}\right)}e^{-\alpha}J_{3}\right)\right]
                     & \mbox{\small {\bf for~$\alpha$~vacua~Type-II}}\\ 
	 \displaystyle \frac{H^2}{f^2_a}\frac{\eta^2}{2\pi^4}
\left[J_{1}+\left(e^{i\pi\left(\Lambda+\frac{1}{2}\right)}e^{\alpha}J_{2}+
e^{-i\pi\left(\Lambda+\frac{1}{2}\right)}e^{-\alpha}J_{3}\right)\right]
                    & \mbox{\small {\bf for~special~vacua}}.
          \end{array}
\right.
\eea
where the momentum integrals $J_{1}, J_{2}$ and $J_{3}$ are defined in the previous section.
Here by setting ${\bf y}=0$ one can derive the results for $\langle\bar{a}({\bf x},\eta)\bar{a}(0,\eta)\rangle$
at conformal time scale $\eta$ with $|k\eta|\rightarrow -\infty$.

Next we consider the
 results for $|k\eta|\rightarrow 0$. In this case we get:
 \bea\label{dfinv4bnbncv}
\displaystyle \langle\bar{a}({\bf x},\eta)\bar{a}({\bf y},\eta)\rangle_{|k\eta|\rightarrow 0}&\approx& 
\frac{H^2}{f^2_a}\frac{(-\eta)^{3-2\Lambda}}{2^{2(2-\Lambda)}\pi^4}\left|\frac{\Gamma(\Lambda)}{\Gamma
\left(\frac{3}{2}\right)}\right|^2
\left[(|C_{2}|^2+|C_{1}|^2)-\left(C^{*}_{1}C_{2}
+C_{1}C^{*}_{2}\right)\right]K_{I}~~~~~~~~~~~~~~~~\\&=& K_{I}\times\left\{\begin{array}{ll}
                     \displaystyle  \frac{H^2}{f^2_a}\frac{(-\eta)^{3-2\Lambda}}{2^{2(3-\Lambda)}
                      \pi^3}\left|\frac{\Gamma(\Lambda)}{\Gamma\left(\frac{3}{2}\right)}\right|^2
                     \nonumber &
 \mbox{\small {\bf for ~Bunch Davies}}  \\ 
	 \displaystyle \frac{H^2}{f^2_a}\frac{(-\eta)^{3-2\Lambda}}{2^{(5-2\Lambda)}
                      \pi^4}\left|\frac{\Gamma(\Lambda)}{\Gamma\left(\frac{3}{2}\right)}\right|^2
\left[\sinh^2\alpha +\cosh^2\alpha
\right.\\ \left. ~~~~~~~~~~~~~~~~~~~~~~~~~~~~~~~~~~~~~ \displaystyle
-\sinh2\alpha\cos{\it \delta}\right]
                     & \mbox{\small {\bf for~$\alpha$~vacua~Type-I}}\\ 
	 \displaystyle \frac{H^2}{f^2_a}\frac{(-\eta)^{3-2\Lambda}|N_{\alpha}|^2}{2^{(3-2\Lambda)}
                      \pi^4}\left|\frac{\Gamma(\Lambda)}
                      {\Gamma\left(\frac{3}{2}\right)}\right|^2
\sin^2\frac{\alpha}{2}
                     & \mbox{\small {\bf for~$\alpha$~vacua~Type-II}}\\ 
	 0
                    & \mbox{\small {\bf for~special~vacua}}.
          \end{array}
\right.
\eea
where the momentum integrals $K_{I}$ is defined in the previous section.
Here by setting ${\bf y}=0$ one can derive the results for $\langle\bar{a}({\bf x},\eta)\bar{a}(0,\eta)\rangle$
at conformal time scale $\eta$ with $|k\eta|\rightarrow 0$.
 
Finally we consider the
 results for $|k\eta|\approx 1$. In this case we get:
  \bea\label{dfinv4bnbncv}
\displaystyle \langle\bar{a}({\bf x},\eta=0)\bar{a}({\bf y},\eta=0)\rangle_{|k\eta|\approx 1}&\approx& 
\frac{H^2}{f^2_a}\frac{1}{2^{2(2-\Lambda)}\pi^4}\left|\frac{\Gamma(\Lambda)}{\Gamma
\left(\frac{3}{2}\right)}\right|^2
\left[(|C_{2}|^2+|C_{1}|^2)-\left(C^{*}_{1}C_{2}
+C_{1}C^{*}_{2}\right)\right]Z_{I}~~~~~~~~~~~~~~~~~~~~~~~~\\&=& Z_{I}\times\left\{\begin{array}{ll}
                      \displaystyle \frac{H^2}{f^2_a}\frac{1}{2^{2(3-\Lambda)}
                      \pi^3}\left|\frac{\Gamma(\Lambda)}{\Gamma\left(\frac{3}{2}\right)}\right|^2
                     \nonumber &
 \mbox{\small {\bf for ~Bunch Davies}}  \\ 
	 \displaystyle \frac{H^2}{f^2_a}\frac{1}{2^{(5-2\Lambda)}
                      \pi^4}\left|\frac{\Gamma(\Lambda)}{\Gamma\left(\frac{3}{2}\right)}\right|^2
\left[\sinh^2\alpha +\cosh^2\alpha
\right.\\ \left. ~~~~~~~~~~~~~~~~~~~~~~~~~~~~~~~~~~~~~ \displaystyle
-\sinh2\alpha\cos{\it \delta}\right]
                     & \mbox{\small {\bf for~$\alpha$~vacua~Type-I}}\\ 
	 \displaystyle \frac{H^2}{f^2_a}\frac{|N_{\alpha}|^2}{2^{(3-2\Lambda)}
                      \pi^4}\left|\frac{\Gamma(\Lambda)}
                      {\Gamma\left(\frac{3}{2}\right)}\right|^2
\sin^2\frac{\alpha}{2}
                     & \mbox{\small {\bf for~$\alpha$~vacua~Type-II}}\\ 
	 0
                    & \mbox{\small {\bf for~special~vacua}}.
          \end{array}
\right.
\eea
where the momentum integrals $Z_{I}$ is defined in the previous section.
Here by setting ${\bf y}=0$ one can derive the results for $\langle\bar{a}({\bf x},\eta=0)\bar{a}(0,\eta=0)\rangle$
at conformal time scale $\eta$ with $|kc_{S}\eta|\rightarrow -\infty$.

Now to derive a exact connection between the computed 
 VEV and two point function of the scalar curvature perturbation in presence of axion fluctuation 
 in momentum space we write down the following sets of new consistency relations in primordial cosmology
for the three limiting cases $|k\eta|\rightarrow -\infty$, $|k\eta|\rightarrow 0$
and $|k\eta|\approx 1$:
\bea\langle \bar{a}_{\bf k}(\eta=0) \rangle_{|k\eta|
\rightarrow -\infty}&=&\langle \bar{a}_{\bf k}(\eta=0)\bar{a}_{\bf q}(\eta=0)
\rangle^{'}_{|kc_{S}\eta|\rightarrow -\infty}\times\frac{2 k }{\eta^2H}{\cal I}_{1},\\
\langle \bar{a}_{\bf k}(\eta=\xi\rightarrow 0) \rangle_{|kc_{S}\eta|
\rightarrow 0}&=&\langle \bar{a}_{\bf k}(\eta=\xi\rightarrow 0)\bar{a}_{\bf q}(\eta=\xi\rightarrow 0)
\rangle^{'}_{|k\eta|\rightarrow 0}\nonumber\\
&&~~~~~\times\frac{2^{2-2\Lambda} k^{2\Lambda}}{(-\eta)^{3-2\Lambda} H\pi}
\left|\frac{\Gamma\left(\frac{3}{2}\right)}{\Gamma\left(\Lambda\right)}\right|^2{\cal I}_{2},\\
\langle \bar{a}_{\bf k}(\eta=0) \rangle_{|k\eta|
\approx 1}&=&\langle \bar{a}_{\bf k}(\eta=0)\bar{a}_{\bf q}(\eta=0)
\rangle^{'}_{|k\eta|\approx 1}\nonumber\\
&&~~~~~\times\frac{2^{2-2\Lambda} k^{3}}{ H\pi}
\left|\frac{\Gamma\left(\frac{3}{2}\right)}{\Gamma\left(\Lambda\right)}\right|^2{\cal I}_{3},
\eea
where $\langle \bar{a}_{\bf k}(\eta=0)\bar{a}_{\bf q}(\eta=0)
\rangle^{'}_{|k\eta|\rightarrow -\infty}$, $\langle \bar{a}_{\bf k}(\eta=0)\bar{a}_{\bf q}(\eta=0)
\rangle^{'}_{|k\eta|\rightarrow 0}$ and $\langle \bar{a}_{\bf k}(\eta=0)\bar{a}_{\bf q}(\eta=0)
\rangle^{'}_{|k\eta|\approx 1}$ are defined as:
\bea \langle \bar{a}_{\bf k}(\eta=0)\bar{a}_{\bf q}(\eta=0)
\rangle^{'}_{|k\eta|\rightarrow -\infty}&=&\frac{\langle \bar{a}_{\bf k}(\eta=0)\bar{a}_{\bf q}(\eta=0)
\rangle_{|k\eta|\rightarrow -\infty}}{(2\pi)^3\delta^{3}({\bf k}+{\bf q})},\\ 
\langle \bar{a}_{\bf k}(\eta=0)\bar{a}_{\bf q}(\eta=0)
\rangle^{'}_{|k\eta|\rightarrow 0}&=&\frac{\langle \bar{a}_{\bf k}(\eta=0)\bar{a}_{\bf q}(\eta=0)
\rangle_{|k\eta|\rightarrow 0}}{(2\pi)^3\delta^{3}({\bf k}+{\bf q})},\\
\langle \bar{a}_{\bf k}(\eta=0)\bar{a}_{\bf q}(\eta=0)
\rangle^{'}_{|k\eta|\approx 1}&=&\frac{\langle \bar{a}_{\bf k}(\eta=0)\bar{a}_{\bf q}(\eta=0)
\rangle_{|k\eta|\approx 1}}{(2\pi)^3\delta^{3}({\bf k}+{\bf q})}.
\eea
Let us now define a new cosmological observable $\hat{ O}_{obs}$ in the three limiting cases $|k\eta|\rightarrow -\infty$, $|k\eta|\rightarrow 0$
and $|k\eta|\approx 1$ as:
\bea \hat{ O}_{obs}&\stackrel{|k\eta|\rightarrow -\infty}{=}&\frac{2}{(-k\eta)^2H}{\cal I}_{1},\\ 
\hat{ O}_{obs}&\stackrel{|k\eta|\rightarrow 0}{=}&\frac{2^{3-2\Lambda}\pi
}{(-k\eta)^{3-2\Lambda} H}
\left|\frac{\Gamma\left(\frac{3}{2}\right)}{\Gamma\left(\Lambda\right)}\right|^2{\cal I}_{2},\\ 
\hat{ O}_{obs}&\stackrel{|k\eta|\approx 1}{=}&\frac{2^{3-2\Lambda} \pi }{ H}
\left|\frac{\Gamma\left(\frac{3}{2}\right)}{\Gamma\left(\Lambda\right)}\right|^2{\cal I}_{3},\eea
where ${\cal I}_{1}, {\cal I}_{2}$ and ${\cal I}_{3}$ are defined as:
\be\begin{array}{lll}\label{dfinv4bnbncv}
 \displaystyle{\cal I}_{1} = -\frac{
\int\limits_{-\infty}^{0} {d\eta}
~\frac{\eta}{k}\frac{m_{axion}}{f_{a}}\left[|C_{2}|^2 
e^{-ik\eta}-|C_{1}|^2 e^{ik\eta} 
-i\left(C^{*}_{1}C_{2}e^{i\pi\left(\Lambda+\frac{1}{2}\right)}
+C_{1}C^{*}_{2}e^{-i\pi\left(\Lambda+\frac{1}{2}\right)}\right)\sin {\it k\eta}\right]}
{\left[|C_{2}|^2+|C_{1}|^2
+\left(C^{*}_{1}C_{2}e^{2ik\eta}e^{i\pi\left(\Lambda+\frac{1}{2}\right)}
+C_{1}C^{*}_{2}e^{-2ik\eta}e^{-i\pi\left(\Lambda+\frac{1}{2}\right)}\right)\right]}
\\=\left\{\begin{array}{ll}
                    \displaystyle 
\int\limits_{-\infty}^{0} {d\eta}
~\frac{\eta}{k} \frac{m_{axion}}{f_{a}}~e^{ik\eta} &
 \mbox{\small {\bf for ~Bunch Davies}}  \\ 
	\displaystyle  -\frac{
\int\limits_{-\infty}^{0} {d\eta}
~\frac{\eta}{k}\frac{m_{axion}}{f_{a}}\left[\sinh^2\alpha~ 
e^{ik\eta}-\cosh^2\alpha~ e^{-ik\eta}
+i~\sinh2\alpha\cos\left({\it \pi\left(\Lambda+\frac{1}{2}\right)}+\delta\right)\sin{\it 
k\eta}\right]}{\left[\sinh^2\alpha +\cosh^2\alpha
+\sinh2\alpha\cos{\it \left(2k\eta+\pi\left(\Lambda+\frac{1}{2}\right)+\delta\right)}\right]}
                     & \mbox{\small {\bf for~$\alpha$~vacua~Type-I}}\\ 
	\displaystyle  -\frac{
\int\limits_{-\infty}^{0} {d\eta}
~\frac{\eta}{k} \frac{m_{axion}}{f_{a}}\left[e^{\alpha+\alpha^{*}} 
e^{ik\eta}-e^{-ik\eta}
+i\left(e^{\alpha}e^{i\pi\left(\Lambda+\frac{1}{2}\right)}
+e^{\alpha^{*}}e^{-i\pi\left(\Lambda+\frac{1}{2}\right)}\right)\sin{\it 
k\eta}\right]}{4
\cos^2{\it \left(k\eta+\frac{\pi}{2}\left(\Lambda+\frac{1}{2}\right)-i\frac{\alpha}{2}\right)}}
                     & \mbox{\small {\bf for~$\alpha$~vacua~Type-II}}\\ 
	\displaystyle  -\frac{i\pi^2
\int\limits_{-\infty}^{0} {d\eta}
~\frac{\eta}{k} \frac{m_{axion}}{f_{a}}\sin{\it 
k\eta}\cos^2{\it \frac{\pi}{2}\left(\Lambda+
\frac{1}{2}\right)}}{\cos^2{\it \left(k\eta+\frac{\pi}{2}\left(\Lambda+\frac{1}{2}\right)\right)}}
                     & \mbox{\small {\bf for~special~vacua}}.
          \end{array}
\right.
\end{array}\ee
\bea\label{dfinv4bnbncv}
 \displaystyle{\cal I}_{2} &=& \sqrt{\xi}\int\limits_{-\infty}^{\xi} {d\eta}
~\sqrt{-\eta}~\frac{m_{axion}}{f_{a}}~
\left(\Lambda-\frac{1}{2}\right)\left[\left(-\frac{k\eta}{2}\right)^{-\Lambda}\left(-\frac{k\xi}{2}\right)^{-\Lambda}
+\left(\frac{k\eta}{2}\right)^{-\Lambda}\left(\frac{k\xi}{2}\right)^{-\Lambda}\right],~~~~~~~~\eea
\bea\label{dfinv4bnbncv}
 \displaystyle{\cal I}_{3} &=& \sqrt{\xi}\int\limits_{-\infty}^{\xi} {d\eta}
~\sqrt{-\eta}~\frac{m_{axion}}{f_{a}}~
\left(\Lambda-\frac{1}{2}\right)2^{1+\Lambda}\left(\frac{k\xi}{2}\right)^{-\Lambda}
\left(-1\right)^{-\Lambda},~~~~~~~~\eea
Next we write down the expression for new cosmological observable $\hat{ O}_{obs}$ 
in terms of all other known cosmological 
\bea\label{fgdfzxzx}\hat{ O}_{obs}&\stackrel{|k\eta|\approx 1}{=}&\frac{2^{n_{\bar{a}}-1}\pi }{ H}
\left|\frac{\Gamma\left(\frac{3}{2}\right)}{\Gamma\left(\frac{4-n_{\bar{a}}}{2}\right)}\right|^2{\cal I}_{3}\left(\Lambda=
\frac{4-n_{\bar{a}}}{2}\right),\eea
where the the spectral tilt for scalar fluctuations is given by: 
\be\begin{array}{ll}\label{dfgsaaaa} n_{\bar{a}}-1\equiv\left(\frac{d\ln \Delta_{\bar{a}}(k)}{d\ln k}\right)_{|k\eta|\approx 1}
\approx 3-2\Lambda=\left\{\begin{array}{ll}
                    \displaystyle 3-2\sqrt{\frac{9}{4}-\frac{m^2_{axion}}{f^2_{a}H^2}}
 &
 \mbox{\small {\bf for ~de~Sitter}}  \\ 
	\displaystyle  3-2\sqrt{\nu^2-\frac{m^2_{axion}}{f^2_{a}H^2}}
                     & \mbox{\small {\bf for~quasi~de~Sitter}}.
          \end{array}
\right.
\end{array}\ee 
and substituting the explicit expression for axion mass parameter we get:
\bea \underline{\bf For~total~U(a)}&& \nonumber\\
\underline{\bf A.~de~Sitter}&&\nonumber\\
n_{\bar{a}}-1\equiv\left(\frac{d\ln \Delta_{\bar{a}}(k)}{d\ln k}\right)_{|k\eta|\approx 1}
&\approx& 
\footnotesize\left\{\begin{array}{ll}
                    \displaystyle 3-2\sqrt{\frac{9}{4}+\frac{\Lambda^4_C \times \cos\left(\sin^{-1}\left(\frac{10\mu^3 H}{
                    \Lambda^4_{C}}\right)\right)}{10000H^4}} ~~~~ &
 \mbox{\small {\bf for ~early~\&~late~$\eta$}}  \\ 
	\displaystyle3-2\sqrt{\frac{9}{4}+\frac{\Lambda^4_C \times \cos\left(\sin^{-1}\left(\frac{\mu^3 H}{
                    \Lambda^4_{C}}\sqrt{100-\frac{80}{1+\omega_C}}\right)\right)}{\left[100-\frac{80}{1+\omega_C}
\right]^2 H^4}}  ~~~~ & \mbox{\small {\bf for~$\eta<\eta_{c}$}}\\ 
	\displaystyle3-2\sqrt{\frac{9}{4}+\frac{\Lambda^4_C \times \cos\left(\sin^{-1}\left(\frac{2\sqrt{5}\mu^3 H}{
                    \Lambda^4_{C}}\right)\right)}{400H^4}}   ~~~~ & \mbox{\small {\bf for~$\eta\sim\eta_{c}$}}.
          \end{array}
\right.\eea \bea
\underline{\bf B.~quasi~de~Sitter}&&\nonumber\\
n_{\bar{a}}-1\equiv\left(\frac{d\ln \Delta_{\bar{a}}(k)}{d\ln k}\right)_{|k\eta|\approx 1}&\approx& 
\footnotesize\left\{\begin{array}{ll}
                    \displaystyle  3-2\sqrt{\nu^2+\frac{\Lambda^4_C \times \cos\left(\sin^{-1}\left(\frac{10\mu^3 H}{
                    \Lambda^4_{C}}\right)\right)}{10000H^4} }~~~~ &
 \mbox{\small {\bf for ~early~\&~late~$\eta$}}  \\ 
	\displaystyle3-2\sqrt{\nu^2+\frac{\Lambda^4_C \times \cos\left(\sin^{-1}\left(\frac{\mu^3 H}{
                    \Lambda^4_{C}}\sqrt{100-\frac{80}{1+\omega_C}}\right)\right)}{\left[100-\frac{80}{1+\omega_C}
\right]^2 H^4}  } ~~~~ & \mbox{\small {\bf for~$\eta<\eta_{c}$}}\\ 
	\displaystyle3-2\sqrt{\nu^2+\frac{\Lambda^4_C \times \cos\left(\sin^{-1}\left(\frac{2\sqrt{5}\mu^3 H}{
                    \Lambda^4_{C}}\right)\right)}{400H^4}}   ~~~~ & \mbox{\small {\bf for~$\eta\sim\eta_{c}$}}.
          \end{array}
\right.
\eea 
\bea \underline{\bf For~osc.~U(a)}&& \nonumber\\
\underline{\bf A.~de~Sitter}&&\nonumber\\
n_{\bar{a}}-1\equiv\left(\frac{d\ln \Delta_{\bar{a}}(k)}{d\ln k}\right)_{|k\eta|\approx 1}
&\approx& 
\footnotesize\left\{\begin{array}{ll}
                    \displaystyle 3-2\sqrt{\frac{9}{4}+\frac{\Lambda^4_C }{10000 H^4} \times (-1)^{m}} ~~~~ &
 \mbox{\small {\bf for ~early~\&~late~$\eta$}}  \\ 
	\displaystyle3-2\sqrt{\frac{9}{4}+\frac{\Lambda^4_C }{\left[100-\frac{80}{1+\omega_C}
\right]^2 H^4} \times (-1)^{m}}  ~~~~ & \mbox{\small {\bf for~$\eta<\eta_{c}$}}\\ 
	\displaystyle3-2\sqrt{\frac{9}{4}+\frac{\Lambda^4_C }{400 H^4} \times (-1)^{m}}   ~~~~ & \mbox{\small {\bf for~$\eta\sim\eta_{c}$}}.
          \end{array}
\right.\eea\bea
\underline{\bf B.~quasi~de~Sitter}&&\nonumber\\
n_{\bar{a}}-1\equiv\left(\frac{d\ln \Delta_{\bar{a}}(k)}{d\ln k}\right)_{|k\eta|\approx 1}&\approx& 
\left\{\begin{array}{ll}
                    \displaystyle  3-2\sqrt{\nu^2+\frac{\Lambda^4_C }{10000 H^4} \times (-1)^{m}}~~~~ &
 \mbox{\small {\bf for ~early~\&~late~$\eta$}}  \\ 
	\displaystyle3-2\sqrt{\nu^2+\frac{\Lambda^4_C }{\left[100-\frac{80}{1+\omega_C}
\right]^2 H^4} \times (-1)^{m}} ~~~~ & \mbox{\small {\bf for~$\eta<\eta_{c}$}}\\ 
	\displaystyle3-2\sqrt{\nu^2+\frac{\Lambda^4_C }{400 H^4} \times (-1)^{m}}   ~~~~ & \mbox{\small {\bf for~$\eta\sim\eta_{c}$}}.
          \end{array}
\right.
\eea 
where $\omega_C$ is a small contribution as pointed earlier.
 
 It is important to mention here that if we use the constraint on scalar spectral tilt as obtained from Planck 2015 data we get the following $2\sigma$ bound on the magnitude of the 
 axion mass parameter:
 \bea
 &&\underline{\bf For ~dS:}\nonumber\\
 && 0.23<\left|\frac{m_{axion}}{f_{a}H}\right|_{kc_{S}\eta\approx 1}<0.28,\\
 &&\underline{\bf For~qdS:}\nonumber\\
 && 0.23\times\left|\sqrt{1-56.18\left(\epsilon+\frac{\eta}{2}+\frac{s}{2}\right)}\right|<
 \left|\frac{m_{axion}}{f_{a}H}\right|_{kc_{S}\eta\approx 1}<0.28\times\left|\sqrt{1-39.06\left(\epsilon+\frac{\eta}{2}+\frac{s}{2}\right)}\right|.~~~~\eea 
 Using this bound on the axion mass parameter one can restrict the value of $\Lambda_{C}/H$ as well from this calculation.

\subsection{Role of isospin breaking interaction}
\label{sec4f}
\begin{figure*}[htb]
\centering
\subfigure[Here we set $\gamma_{\bf even}=1=\delta_{\bf even}$, $\gamma_{\bf odd}=0.5=\delta_{\bf odd}$.]{
    \includegraphics[width=12.2cm,height=6cm] {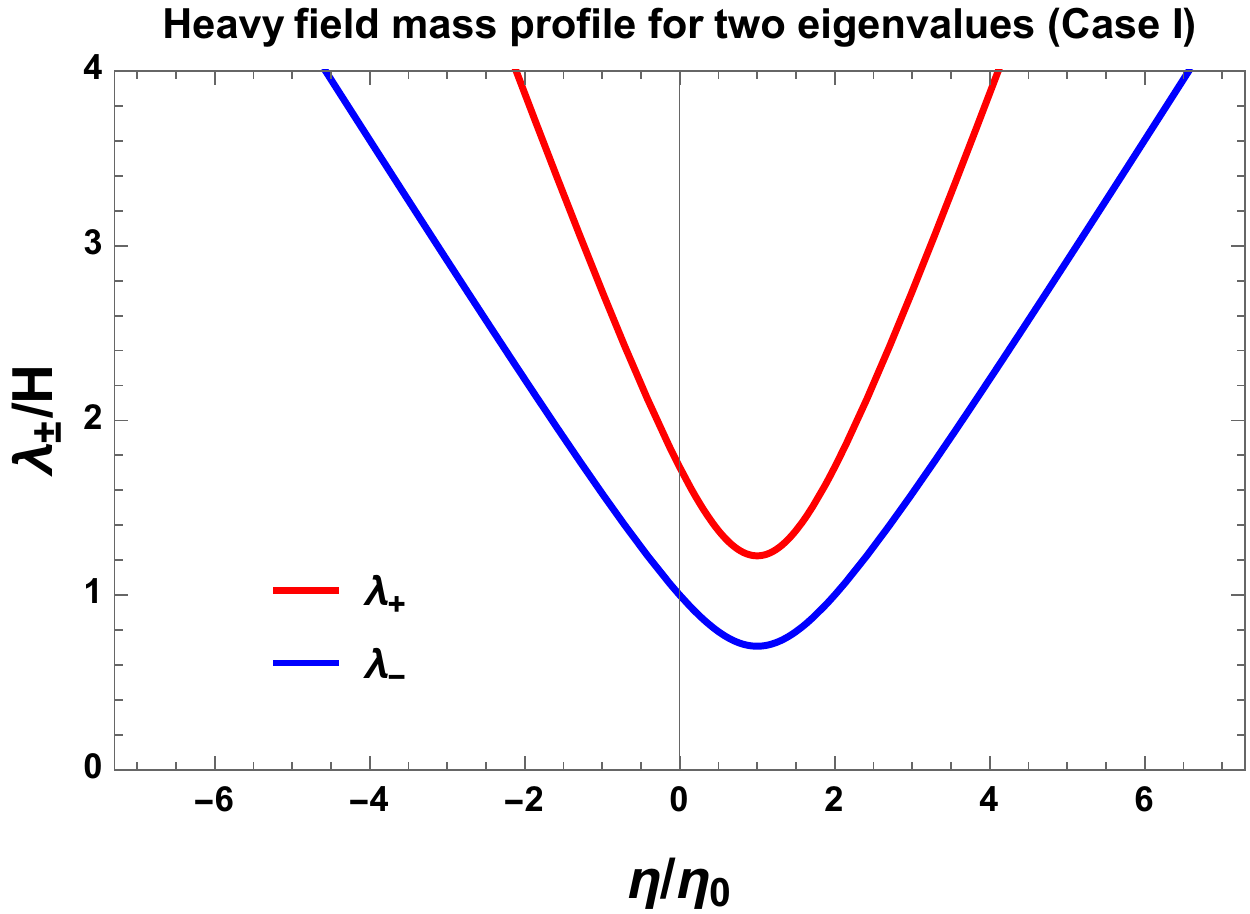}
    \label{fig1}
}
\subfigure[Here we set $m_{\bf even}=3$, $m_{\bf odd}=1.5$ and $\rho/H=1$.]{
    \includegraphics[width=12.2cm,height=6cm] {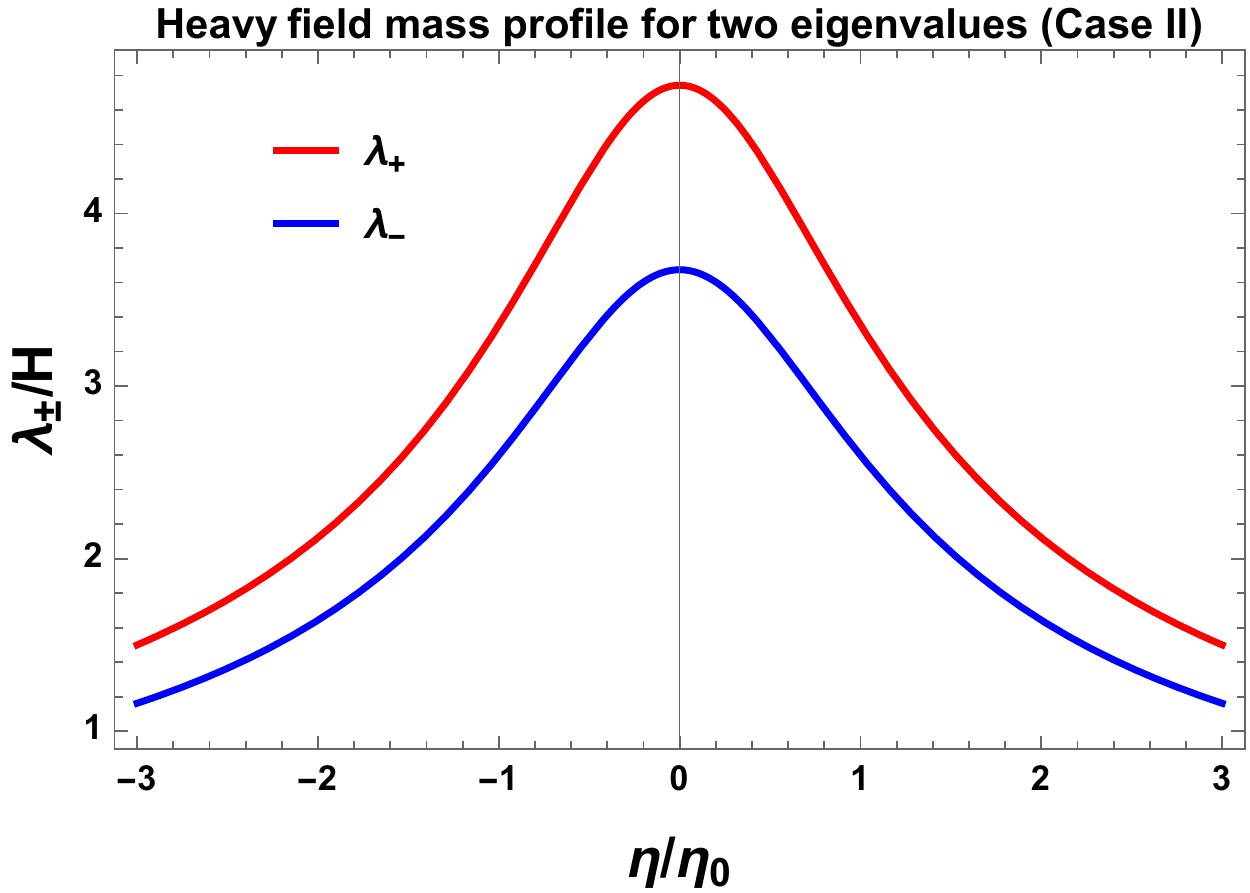}
    \label{fig1}
}
\subfigure[Here we set $m_{\bf even}=2$, $m_{\bf odd}=1.5$ and $\rho/H=1$.]{
    \includegraphics[width=12.2cm,height=6cm] {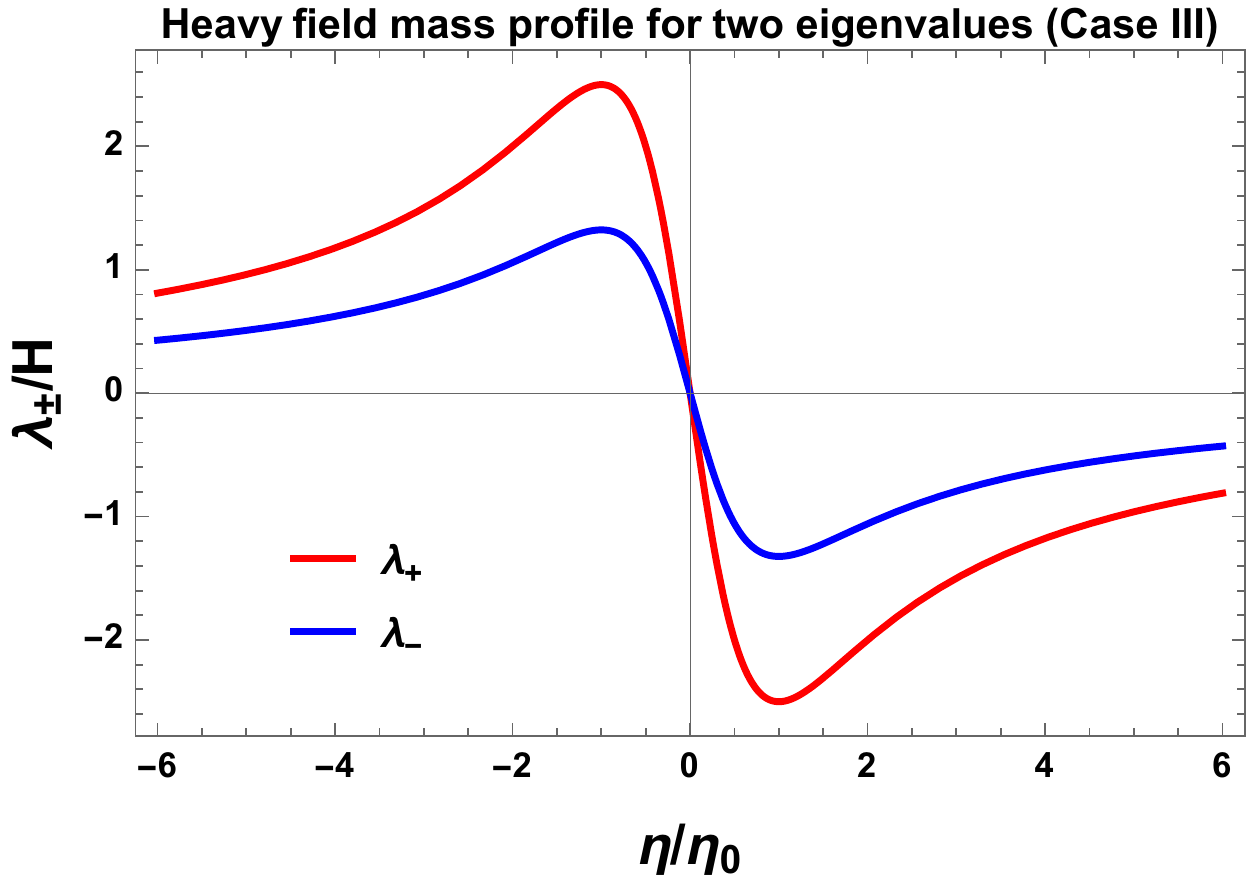}
    \label{fig2}
}
\caption[Optional caption for list of figures]{Conformal time scale dependent behaviour of heavy particle mass profile for two eigenstates.} 
\label{hmassiso}
\end{figure*}

To construct a cosmological Bell violating experimental setup specific role of isospin is very significant. See ref \cite{juan:2015ja} for details on this important issue. 
To setup such cosmological experiment we need here isospin breaking interactions in the present context, that can be implemented in a phenomenological way.
Here one can start with any type of isospin breaking interactions which involves the mass term of the heavy field. To star with let us consider the following 
situation where the effective Lagrangian for the inflaton and heavy scalar field interaction is given by:
\be\label{axhev} S=\int d^{4}x\sqrt{-g}\left[\frac{1}{2}g^{\mu\nu}(\partial_{\mu}{\cal H})^{\dagger}(\partial_{\nu}H)+\frac{1}{2}g^{\mu\nu}(\partial_{\mu}\phi)(\partial_{\nu}\phi)
-{\cal H}^{\dagger}\left(\sum^{\infty}_{n=0}{\cal M}^{2}_{n}(\phi)(\sigma.{\bf n})^{n}\right){\cal H}+\cdots\right],\ee
where $\cdots$ contains all other possible isospin breaking interactions in the effective Lagrangian. Here $\phi$ is inflaton field and ${\cal H}$ is the heavy field, which is a isospin $SU(2)$ doublet and structure of ${\cal H}$ is given by,
${\cal H}=({\cal H}_{1},{\cal H}_{2})$. Here ${\cal M}^{2}_{n}(\phi)$ represents the mass term for every quadratic operator labeled by $n$. In the above mentioned interaction for $n=0$ it is a isospin preserving interaction and physically represents the mass term of the heavy scalar field $H$, provided the other interactions are absent or sub dominant 
in the effective Lagrangian. One can also identify this specific type of interaction as the leading order effect in the effective Lagrangian. Here terms for $n\geq 1$ take care of all the isospin breaking interaction in the
present discussion. We here study the effect of these type of isospin breaking interactions explicitly. Here it is important to mention that $\sigma$ is the Pauli spin matrices and ${\bf n}$ is the unit vector along which we are 
taking isospin projection of the heavy field $H$. Before going the further details it is important to note that in our computation we can use the following results:
\be\begin{array}{lll}
 \displaystyle  (\sigma.{\bf n})^{n}=(\sigma_{x}\cos m\theta+\sigma_{y} \sin m\theta)^{n}=\left\{\begin{array}{ll}
                    \displaystyle   I,~~~~&
 \mbox{\small {\bf for ~even~$n$}}  \\ 
	\displaystyle (\sigma.{\bf n}),~~~~& \mbox{\small {\bf for ~odd~$n$}}.
          \end{array}
\right.
\end{array}\ee
where $I$ is a identity matrix, $m$ is an integer and $\theta$ is the angular dependence of isospin projection along the axis of measurement. Specially the role of the integer $m$ is very important for the 
present discussion as it amplifies the effect of quantum fluctuations. Now to understand the behaviour of the interactions in a more comprehensive manner let us investigate explicitly first few terms in the above mentioned series:
\begin{itemize}
 \item For $n=0$, the interaction term is given by the following expression:
 \be {\cal M}^{2}_{0}(\phi){\cal H}^{\dagger}{\cal H}={\cal M}^{2}_{0}(\phi)\left[|{\cal H}_{1}|^2+|{\cal H}_{2}|^2\right].\ee
 In the case of axion one can identify ${\cal M}^{2}_{0}(\phi)=m^2_{axion}$.
 \item For $n=1$, the interaction term is given by the following expression:
 \be {\cal M}^{2}_{1}(\phi){\cal H}^{\dagger}(\sigma.{\bf n}){\cal H}={\cal M}^{2}_{1}(\phi)\left[\exp(-im\theta){\cal H}^{*}_{1}{\cal H}_{2}+\exp(im\theta){\cal H}^{*}_{2}{\cal H}_{1}\right].\ee
 \item For $n=2$, the interaction term is given by the following expression:
 \be {\cal M}^{2}_{2}(\phi){\cal H}^{\dagger}(\sigma.{\bf n})^2{\cal H}={\cal M}^{2}_{2}(\phi){\cal H}^{\dagger}{\cal H}={\cal M}^{2}_{2}(\phi)\left[|{\cal H}_{1}|^2+|{\cal H}_{2}|^2\right].\ee
 \item For $n=3$, the interaction term is given by the following expression:
 \be {\cal M}^{2}_{3}(\phi){\cal H}^{\dagger}(\sigma.{\bf n})^3{\cal H}={\cal M}^{2}_{3}(\phi){\cal H}^{\dagger}(\sigma.{\bf n}){\cal H}={\cal M}^{2}_{3}(\phi)\left[\exp(-im\theta){\cal H}^{*}_{1}{\cal H}_{2}+\exp(im\theta){\cal H}^{*}_{2}{\cal H}_{1}\right].\ee
\end{itemize}
This implies that:
\be\begin{array}{lll}
 \displaystyle  {\cal H}^{\dagger}(\sigma.{\bf n})^{n}{\cal H} =\left\{\begin{array}{ll}
                    \displaystyle  {\cal H}^{\dagger}{\cal H}=\left[|{\cal H}_{1}|^2+|{\cal H}_{2}|^2\right],~~~~&
 \mbox{\small {\bf for ~even~$n$}}  \\ 
	\displaystyle {\cal H}^{\dagger}(\sigma.{\bf n}){\cal H}=\left[\exp(-im\theta){\cal H}^{*}_{1}{\cal H}_{2}+\exp(im\theta){\cal H}^{*}_{2}{\cal H}_{1}\right],~~~~& \mbox{\small {\bf for ~odd~$n$}}.
          \end{array}
\right.
\end{array}\ee
Consequently the effective Lagrangian as stated in Eq~(\ref{axhev}) can be recast as:
\bea\label{iopb} S&=&\int d^{4}x\sqrt{-g}\left[\frac{1}{2}g^{\mu\nu}(\partial_{\mu}{\cal H})^{\dagger}(\partial_{\nu}{\cal H})
+\frac{1}{2}g^{\mu\nu}(\partial_{\mu}\phi)(\partial_{\nu}\phi)
\right.\nonumber\\ &&\left.~~~~~ -\sum^{\infty}_{n=0,2,4}{\cal H}^{\dagger}\left({\cal M}^{2}_{n}(\phi)(\sigma.{\bf n})^{n}\right){\cal H}
-\sum^{\infty}_{n=1,3,5}{\cal H}^{\dagger}\left({\cal M}^{2}_{n}(\phi)(\sigma.{\bf n})^{n}\right){\cal H}+\cdots\right],\nonumber\\
&=&\int d^{4}x\sqrt{-g}\left[\frac{1}{2}g^{\mu\nu}(\partial_{\mu}{\cal H}_{1})^{*}(\partial_{\nu}{\cal H}_{1})+\frac{1}{2}g^{\mu\nu}(\partial_{\mu}{\cal H}_{2})^{*}(\partial_{\nu}{\cal H}_{2})+\frac{1}{2}g^{\mu\nu}(\partial_{\mu}\phi)(\partial_{\nu}\phi)
\right.\nonumber\\ &&\left.~~~~~~~ -\left[|{\cal H}_{1}|^2+|{\cal H}_{2}|^2\right]
\sum^{\infty}_{n=0,2,4}{\cal M}^{2}_{n}(\phi)
-\left[\exp(-im\theta){\cal H}^{*}_{1}{\cal H}_{2}+\exp(im\theta){\cal H}^{*}_{2}{\cal H}_{1}\right]
\sum^{\infty}_{n=1,3,5}{\cal M}^{2}_{n}(\phi)+\cdots\right],\nonumber\\
&=&\int d^{4}x\sqrt{-g}\left[\frac{1}{2}g^{\mu\nu}(\partial_{\mu}{\cal H}_{1})^{*}(\partial_{\nu}{\cal H}_{1})+\frac{1}{2}g^{\mu\nu}(\partial_{\mu}{\cal H}_{2})^{*}(\partial_{\nu}{\cal H}_{2})+\frac{1}{2}g^{\mu\nu}(\partial_{\mu}\phi)(\partial_{\nu}\phi)
\right.\nonumber\\ &&\left.~~~~~ -\left[|{\cal H}_{1}|^2+|{\cal H}_{2}|^2\right]
{\cal M}^{2}_{\bf even}(\phi)-\left[\exp(-im\theta){\cal H}^{*}_{1}{\cal H}_{2}+\exp(im\theta){\cal H}^{*}_{2}H_{1}\right]
{\cal M}^{2}_{\bf odd}(\phi)+\cdots\right],~~~~~~~~~\eea
where the mass-term ${\cal M}^{2}_{\bf even}(\phi)$ and ${\cal M}^{2}_{\bf odd}(\phi)$ are defined as:
\bea {\cal M}^{2}_{\bf even}(\phi)&=&\sum^{\infty}_{n=0,2,4}{\cal M}^{2}_{n}(\phi),\\
{\cal M}^{2}_{\bf odd}(\phi)&=&\sum^{\infty}_{n=1,3,5}{\cal M}^{2}_{n}(\phi).\eea
From Eq.~(\ref{iopb}), one can construct a mass matrix as given by~\footnote{Here the determinant and trace of the mass matrix is given by:
\bea {\it Det}\left[ {\cal M}^{2}_{\bf even}(\phi)\right]&=&\lambda^2_{\pm}(\phi)\left(\lambda_{\pm}(\phi)-\sqrt{2}{\cal M}_{\bf odd}(\phi)\right)\left(\lambda_{\pm}(\phi)+\sqrt{2}{\cal M}_{\bf odd}(\phi)\right)\nonumber\\
&=&\lambda^2_{\pm}(\phi)\left(\sqrt{2}{\cal M}_{\bf even}(\phi)-\lambda_{\pm}(\phi)\right)\left(\sqrt{2}{\cal M}_{\bf even}(\phi)+\lambda_{\pm}(\phi)\right).\\
{\it Tr}\left[ {\cal M}^{2}_{\bf even}(\phi)\right]&=&{\cal M}^{2}_{\bf even}(\phi)+{\cal M}^{2}_{\bf odd}(\phi)=\lambda^2_{\pm}(\phi).\eea}:
\bea {\cal M}^2(\phi)&=&\left( \begin{array}{ccc}
{\cal M}^{2}_{\bf even}(\phi) & \exp(-im\theta){\cal M}^{2}_{\bf odd}(\phi) \\
\exp(im\theta){\cal M}^{2}_{\bf odd}(\phi) & \displaystyle{\cal M}^{2}_{\bf even}(\phi) \end{array} \right)\nonumber\\
&=&\left( \begin{array}{ccc}
\displaystyle\sum^{\infty}_{n=0,2,4}{\cal M}^{2}_{n}(\phi) & \displaystyle\exp(-im\theta)\sum^{\infty}_{n=1,3,5}{\cal M}^{2}_{n}(\phi) \\
\exp(im\theta)\displaystyle\sum^{\infty}_{n=1,3,5}{\cal M}^{2}_{n}(\phi) 
& \displaystyle\displaystyle\sum^{\infty}_{n=0,2,4}{\cal M}^{2}_{n}(\phi) \end{array} \right).\eea
and one can construct a physical mass eigen basis from the following eigenvalues:
\bea\label{eigen1} \lambda_{\pm}(\phi)&=&\sqrt{{\cal M}^{2}_{\bf even}(\phi)\pm {\cal M}^{2}_{\bf odd}(\phi)}
=\sqrt{\sum^{\infty}_{n=0,2,4}{\cal M}^{2}_{n}(\phi)\pm \sum^{\infty}_{n=1,3,5}{\cal M}^{2}_{n}(\phi)}.\eea
One can also express the eigenvalues presented in Eq~(\ref{eigen1}) as:
\bea\label{eigen2} \lambda_{\pm}(\phi)&=&\sqrt{{\cal M}^{2}_{\bf even}(\phi)+{\bf sign}(\sigma.{\bf n}) {\cal M}^{2}_{\bf odd}(\phi)}
=\sqrt{\sum^{\infty}_{n=0,2,4}{\cal M}^{2}_{n}(\phi)+{\bf sign}(\sigma.{\bf n}) \sum^{\infty}_{n=1,3,5}{\cal M}^{2}_{n}(\phi)},\eea
where ${\bf sign}(\sigma.{\bf n})=\pm 1$. As the inflaton field $\phi$ is function of the conformal time $\eta$, one can express the eigenvalues in terms of $\eta$ as well. Consequently the physical mass parameter for these two eigenstates can be written as:
\bea \frac{ \lambda_{\pm}(\eta)}{H}&=&\sqrt{\left(\frac{{\cal M}_{\bf even}(\eta)}{H}\right)^2\pm \left(\frac{{\cal M}^{2}_{\bf odd}(\eta)}{H}\right)^2}\nonumber\\
&=&\sqrt{\left(\frac{\sum^{\infty}_{n=0,2,4}{\cal M}^{2}_{n}(\eta)}{H}\right)^2\pm \left(\frac{\sum^{\infty}_{n=1,3,5}{\cal M}^{2}_{n}(\eta)}{H}\right)^2}.\eea
Similarly in terms of {\bf sign} function the physical mass parameter can be expressed as:
\bea \frac{ \lambda_{\pm}(\eta)}{H}&=&\sqrt{\left(\frac{{\cal M}_{\bf even}(\eta)}{H}\right)^2+{\bf sign}(\sigma.{\bf n}) \left(\frac{{\cal M}^{2}_{\bf odd}(\eta)}{H}\right)^2}\nonumber\\
&=&\sqrt{\left(\frac{\sum^{\infty}_{n=0,2,4}{\cal M}^{2}_{n}(\eta)}{H}\right)^2+{\bf sign}(\sigma.{\bf n}) \left(\frac{\sum^{\infty}_{n=1,3,5}{\cal M}^{2}_{n}(\eta)}{H}\right)^2}.\eea
Now to show the time dependent behaviour of the mass parameters constructed out of the {\bf even} and {\bf odd} contributions in the interaction picture here we chose few time dependent mass profile. For example:
\bea {\cal M}_{\bf even}(\eta)
&=& \left\{\begin{array}{ll}
                    \displaystyle \sqrt{\gamma_{\bf even}\left(\frac{\eta}{\eta_0} - 1\right)^2 + \delta_{\bf even}}~H~~~~ &
 \mbox{\small {\bf Case~I}}  \\ 
	\displaystyle \frac{m_{\bf even}}{\sqrt{2}}\sqrt{\left[1-\tanh\left(\rho\frac{\ln(-H\eta)}{H}\right)\right]}~~~~ & \mbox{\small {\bf Case~II}}\\ 
	\displaystyle m_{\bf even}~{\rm sech}\left(\rho\frac{\ln(-H\eta)}{H}\right)  ~~~~ & \mbox{\small {\bf Case~III}}.
          \end{array}
\right.\\
 {\cal M}_{\bf odd}(\eta)
&=&\left\{\begin{array}{ll}
                    \displaystyle \sqrt{\gamma_{\bf odd}\left(\frac{\eta}{\eta_0} - 1\right)^2 + \delta_{\bf odd}}~H~~~~ &
 \mbox{\small {\bf Case~I}}  \\ 
	\displaystyle \frac{m_{\bf odd}}{\sqrt{2}}\sqrt{\left[1-\tanh\left(\rho\frac{\ln(-H\eta)}{H}\right)\right]}~~~~ & \mbox{\small {\bf Case~II}}\\ 
	\displaystyle m_{\bf odd}~{\rm sech}\left(\rho\frac{\ln(-H\eta)}{H}\right)  ~~~~ & \mbox{\small {\bf Case~III}}.
          \end{array}
\right. \eea
Consequently the physical mass parameter for these two eigenstates can be written as:
\bea  \frac{ \lambda_{\pm}(\eta)}{H}&=&\sqrt{\left(\frac{{\cal M}_{\bf even}(\eta)}{H}\right)^2\pm \left(\frac{{\cal M}^{2}_{\bf odd}(\eta)}{H}\right)^2}\nonumber\\
&=& \left\{\begin{array}{ll}
                    \displaystyle \sqrt{\left(\gamma_{\bf even}\pm\gamma_{\bf odd}\right) \left(\frac{\eta}{\eta_0} - 1\right)^2 + \left(\delta_{\bf even}\pm\delta_{\bf odd}\right)}~~~~ &
 \mbox{\small {\bf Case~I}}  \\ 
	\displaystyle \frac{1}{\sqrt{2}H}\sqrt{\left\{{m}^2_{\bf even}\pm {m}^2_{\bf even}\right\}\left[1-\tanh\left(\rho\frac{\ln(-H\eta)}{H}\right)\right]}~~~~ & \mbox{\small {\bf Case~II}}\\ 
	\displaystyle \frac{1}{H}\sqrt{\left\{{m}^2_{\bf even}\pm {m}^2_{\bf even}\right\}}{\rm sech}\left(\rho\frac{\ln(-H\eta)}{H}\right)~~~~ & \mbox{\small {\bf Case~III}}.
          \end{array}
\right.\eea
Similarly in terms of {\bf sign} function the physical mass parameter for these two eigenstates can be expressed as:
\bea  \frac{ \lambda_{\pm}(\eta)}{H}&=&\sqrt{\left(\frac{{\cal M}_{\bf even}(\eta)}{H}\right)^2\pm \left(\frac{{\cal M}^{2}_{\bf odd}(\eta)}{H}\right)^2}\nonumber\\
&=& \left\{\begin{array}{ll}
                    \displaystyle \sqrt{\left(\gamma_{\bf even}+{\bf sign}(\sigma.{\bf n})\gamma_{\bf odd}\right) \left(\frac{\eta}{\eta_0} - 1\right)^2 + \left(\delta_{\bf even}+{\bf sign}(\sigma.{\bf n})\delta_{\bf odd}\right)}~~~~ &
 \mbox{\small {\bf Case~I}}  \\ 
	\displaystyle \frac{1}{\sqrt{2}H}\sqrt{\left\{{m}^2_{\bf even}+{\bf sign}(\sigma.{\bf n}){m}^2_{\bf even}\right\}\left[1-\tanh\left(\rho\frac{\ln(-H\eta)}{H}\right)\right]}~~~~ & \mbox{\small {\bf Case~II}}\\ 
	\displaystyle \frac{1}{H}\sqrt{\left\{{m}^2_{\bf even}+{\bf sign}(\sigma.{\bf n}) {m}^2_{\bf even}\right\}}{\rm sech}\left(\rho\frac{\ln(-H\eta)}{H}\right)~~~~ & \mbox{\small {\bf Case~III}}.
          \end{array}
\right.\eea
For the sake of simplicity here we introduce new parameters defined as:
\bea \gamma_{\pm}&=& \gamma_{\bf even}\pm \gamma_{\bf odd}=\gamma_{\bf even}+{\bf sign}(\sigma.{\bf n}) \gamma_{\bf odd},\\
\delta_{\pm}&=&\delta_{\bf even}\pm \delta_{\bf odd}=\delta_{\bf even}+{\bf sign}(\sigma.{\bf n}) \delta_{\bf odd},\\
m_{0\pm}&=&\sqrt{{m}^2_{\bf even}\pm {m}^2_{\bf even}}=\sqrt{{m}^2_{\bf even}+{\bf sign}(\sigma.{\bf n}) {m}^2_{\bf even}}.\eea
Then in terms of these new parameters physical mass parameter for these two eigenstates can be recast as:
\bea  \frac{ \lambda_{\pm}(\eta)}{H}&=&\sqrt{\left(\frac{{\cal M}_{\bf even}(\eta)}{H}\right)^2\pm \left(\frac{{\cal M}^{2}_{\bf odd}(\eta)}{H}\right)^2}\nonumber\\
&=& \left\{\begin{array}{ll}
                    \displaystyle \sqrt{\gamma_{\pm}\left(\frac{\eta}{\eta_0} - 1\right)^2 +\delta_{\pm}}~~~~ &
 \mbox{\small {\bf Case~I}}  \\ \\
	\displaystyle \frac{m_{0\pm}}{\sqrt{2}H}\sqrt{\left[1-\tanh\left(\rho\frac{\ln(-H\eta)}{H}\right)\right]}~~~~ & \mbox{\small {\bf Case~II}}\\ \\
	\displaystyle \frac{m_{0\pm}}{H}{\rm sech}\left(\rho\frac{\ln(-H\eta)}{H}\right)~~~~ & \mbox{\small {\bf Case~III}}.
          \end{array}
\right.\eea
In fig.~(\ref{hmassiso}) we have explicitly shown the conformal time scale dependent behaviour of heavy particle mass profile for two eigenstates.

Let us give the physical interpretation of both the eigen values obtained in this context. According to the phenomenological construction of the above mentioned effective Lagrangian ${\cal H}$ is a complex $SU(2)$ isospin 
doublet. During the particle production ${\cal H}=({\cal H}_{1},{\cal H}_{2})$ and its complex conjugate part ${\cal H}^{*}=({\cal H}^{*}_{1},{\cal H}^{*}_{2})=(\tilde{{\cal H}}_{1},-\tilde{{\cal H}}_{2})$ both of them participate in the creation of particle and antiparticle production 
in the present context. 

Here we need the following requirements:
\begin{itemize}
 \item To avoid any instability in the eigen basis, eigen values of the mass matrix are always positive definite.
Consequently we get the following constraint condition:
\bea {\cal M}^{2}_{\bf even}(\phi) \geq {\cal M}^{2}_{\bf odd}(\phi)\Rightarrow
\sum^{\infty}_{n=0,2,4}{\cal M}^{2}_{n}(\phi)\geq\sum^{\infty}_{n=1,3,5}{\cal M}^{2}_{n}(\phi). \eea
For the specific three mass profile one can rewrite this constraint condition as:
\bea 0
&\leq & \left\{\begin{array}{ll}
                    \displaystyle \left[\gamma_{-}\left(\frac{\eta}{\eta_0} - 1\right)^2 + \delta_{-}\right]~H^2~~~~ &
 \mbox{\small {\bf Case~I}}  \\ 
	\displaystyle \frac{m^2_{0-}}{2}\left[1-\tanh\left(\rho\frac{\ln(-H\eta)}{H}\right)\right]~~~~ & \mbox{\small {\bf Case~II}}\\ 
	\displaystyle m^2_{0-}~{\rm sech}^2\left(\rho\frac{\ln(-H\eta)}{H}\right)  ~~~~ & \mbox{\small {\bf Case~III}}.
          \end{array}
\right.\eea
\item At late time scales:
\bea {\cal M}^{2}_{\bf even}(\phi) \sim {\cal M}^{2}_{\bf odd}(\phi)\Rightarrow
\sum^{\infty}_{n=0,2,4}{\cal M}^{2}_{n}(\phi)\sim\sum^{\infty}_{n=1,3,5}{\cal M}^{2}_{n}(\phi). \eea
Here if the value of ${\cal M}_{\bf even}(\phi)$ increases then the magnitude of ${\cal M}_{\bf odd}(\phi)$ also increases as they are of the same order. Consequently, the eigen value of the mass-matrix 
also increases. For the specific three mass profile one can rewrite this constraint condition as:
\bea 0
&= & \left\{\begin{array}{ll}
                    \displaystyle \left[\gamma_{-}\left(\frac{\eta}{\eta_0} - 1\right)^2 + \delta_{-}\right]~H^2~~~~ &
 \mbox{\small {\bf Case~I}}  \\ 
	\displaystyle \frac{m^2_{0-}}{2}\left[1-\tanh\left(\rho\frac{\ln(-H\eta)}{H}\right)\right]~~~~ & \mbox{\small {\bf Case~II}}\\ 
	\displaystyle m^2_{0-}~{\rm sech}^2\left(\rho\frac{\ln(-H\eta)}{H}\right)  ~~~~ & \mbox{\small {\bf Case~III}}.
          \end{array}
\right.\eea
\item Another crucial requirement is eigen values of the mass matrix is of the order of UV cut-off of the EFT $\Lambda_{UV}$. 
In our purpose one can choose $\Lambda_{UV}\sim M_p=2.43\times 10^{18}{\rm GeV}$. Which implies that:
\bea \lambda^2_{\pm}\sim {\cal M}^{2}_{\bf even}(\phi)\pm {\cal M}^{2}_{\bf odd}(\phi)=\left[
\sum^{\infty}_{n=0,2,4}{\cal M}^{2}_{n}(\phi)\pm 
\sum^{\infty}_{n=1,3,5}{\cal M}^{2}_{n}(\phi)\right]\approx \Lambda^2_{UV}\sim M^2_p.
\eea
This is necessarily required to measure the eigenvalues with the given Bell violating cosmological setup. For the specific three mass profile one can rewrite this constraint condition as:
\bea \lambda^2_{\pm}\sim\Lambda^2_{UV}
&= & \left\{\begin{array}{ll}
                    \displaystyle \left[\gamma_{\pm}\left(\frac{\eta}{\eta_0} - 1\right)^2 + \delta_{\pm}\right]~H^2~~~~ &
 \mbox{\small {\bf Case~I}}  \\ \\
	\displaystyle \frac{m^2_{0\pm}}{2}\left[1-\tanh\left(\rho\frac{\ln(-H\eta)}{H}\right)\right]~~~~ & \mbox{\small {\bf Case~II}}\\ \\
	\displaystyle m^2_{0\pm}~{\rm sech}^2\left(\rho\frac{\ln(-H\eta)}{H}\right)  ~~~~ & \mbox{\small {\bf Case~III}}.
          \end{array}
\right.\eea
\item When all the isospin breaking interactions are absent from the effective Lagrangian in that case the eigen value
of the mass matrix is given by, \be \lambda(\phi)\sim {\cal M}_{\bf even}(\phi).\ee This is another very crucial criteria
which can be observed during the time of heavy mass particle creation with $SU(2)$ isospin singlet state. For the specific three mass profile one can rewrite this constraint condition as:
\bea \lambda(\eta)
&=& \left\{\begin{array}{ll}
                    \displaystyle \sqrt{\gamma_{\bf even}\left(\frac{\eta}{\eta_0} - 1\right)^2 + \delta_{\bf even}}~H~~~~ &
 \mbox{\small {\bf Case~I}}  \\ \\
	\displaystyle \frac{m_{\bf even}}{\sqrt{2}}\sqrt{\left[1-\tanh\left(\rho\frac{\ln(-H\eta)}{H}\right)\right]}~~~~ & \mbox{\small {\bf Case~II}}\\ \\
	\displaystyle m_{\bf even}~{\rm sech}\left(\rho\frac{\ln(-H\eta)}{H}\right)  ~~~~ & \mbox{\small {\bf Case~III}}.
          \end{array}
\right.\eea
\item Here signature of the angular parameter $\theta$ and its functional dependence on the background plays very crucial role to setup the Bell violating setup in the context of cosmology.
For a very simplified case one can assume that $\theta$ is a constant. In this case if we identify the particle mass eigen value as $\lambda_{\pm}(\phi)$, then the antiparticle mass eigen values also characterized by 
$\lambda_{\pm}(\phi)$. Sign flip in the eigen value of the antiparticle mass eigen state may happen if the angular parameter $\theta$ is background dependent and not a constant quantity. 
\end{itemize}
Now we will include additional quartic contributions in the phenomenological effective Lagrangian to describe isospin breaking interactions through self integrations between the heavy fields. In that case Eq~(\ref{axhev}) can be modified as:
\bea\label{axhev2} S&=&\int d^{4}x\sqrt{-g}\left[\frac{1}{2}g^{\mu\nu}(\partial_{\mu}{\cal H})^{\dagger}(\partial_{\nu}{\cal H})+\frac{1}{2}g^{\mu\nu}(\partial_{\mu}\phi)(\partial_{\nu}\phi)
\right.\nonumber \\&& \left.~~~~~~~~~~~~~~~~~~~-{\cal H}^{\dagger}\left(\sum^{\infty}_{n=0}{\cal M}^{2}_{n}(\phi)(\sigma.{\bf n})^{n}\right){\cal H}-\left({\cal H}^{\dagger}\left(\sum^{\infty}_{n=0}{G}_{n}(\phi)(\sigma.{\bf n})^{n}\right){\cal H}\right)^2
+\cdots\right].\eea
where $G_{n}(\phi)$ is the quartic coupling for each values of $n$. Let us analyze the effect of only last term first and then we will comment on the total combined effect. Here in this context one can write down only the quartic contribution as:
\bea\label{axhev3} S_{quartic}&=&-\int d^{4}x\sqrt{-g}\left({\cal H}^{\dagger}\left(\sum^{\infty}_{n=0}{G}_{n}(\phi)(\sigma.{\bf n})^{n}\right){\cal H}\right)^2,\nonumber\\
&=&-\int d^{4}x\sqrt{-g}\left[{\cal H}^{\dagger}\left(\sum^{\infty}_{n=0,2,4}{G}_{n}(\phi)(\sigma.{\bf n})^{n}\right){\cal H}+{\cal H}^{\dagger}\left(\sum^{\infty}_{n=1,3,5}{G}_{n}(\phi)(\sigma.{\bf n})^{n}\right){\cal H}\right]^2,\nonumber\\
&=&-\int d^{4}x\sqrt{-g}\left[\left(|{\cal H}_{1}|^2+|{\cal H}_{2}|^2\right)\sum^{\infty}_{n=0,2,4}{G}_{n}(\phi)\right.\nonumber\\ && \left.~~~~~~~~~~~~~~~~~~~~~~~~~~~+\left[\exp(-im\theta){\cal H}^{*}_{1}{\cal H}_{2}+\exp(im\theta){\cal H}^{*}_{2}{\cal H}_{1}\right]
\sum^{\infty}_{n=1,3,5}G_{n}(\phi)\right]^2,\nonumber\\
&=&-\int d^{4}x\sqrt{-g}\left[\left(|{\cal H}_{1}|^2+|{\cal H}_{2}|^2\right){G}_{\bf even}(\phi)\right.\nonumber\\ && \left.~~~~~~~~~~~~~~~~~~~~~~~~~~~+\left[\exp(-im\theta){\cal H}^{*}_{1}{\cal H}_{2}+\exp(im\theta){\cal H}^{*}_{2}{\cal H}_{1}\right]
G_{\bf odd}(\phi)\right]^2,\eea
where the couplings ${G}_{\bf even}(\phi)$ and $G_{\bf odd}(\phi)$ are defined by the following expressions:
\bea {G}_{\bf even}(\phi)&=& \sum^{\infty}_{n=0,2,4}{G}_{n}(\phi),\\
G_{\bf odd}(\phi)&=& \sum^{\infty}_{n=1,3,5}{G}_{n}(\phi).\eea
Here further we assume that all the higher mass dimensional non-re-normalizable operators of the following structure:
\bea {\cal O}_{NR}&=&\sum^{\infty}_{d>2}\frac{1}{\Lambda^{2d-4}_{UV}}\left[{\cal H}^{\dagger}\left(\sum^{\infty}_{n=0}{Z}_{n}(\phi)(\sigma.{\bf n})^{n}\right){\cal H}\right]^d,\eea
are highly suppressed by the EFT cut-off scale $\Lambda_{UV}$, so that one neglect all such contributions in the EFT Lagrangian.

Further combining the effect of quartic and quadratic interaction the re-normalizable part of the total effective potential can be written as:
\bea V({\cal H}_{1},{\cal H}_{2},\theta)&=&V_{\bf IP}({\cal H}_{1},{\cal H}_{2})+V_{\bf IB}({\cal H}_{1},{\cal H}_{2},\theta),\eea
where the isospin preserving and isospin breaking re-normalizable contributions are given by:
\bea V_{\bf IP}({\cal H}_{1},{\cal H}_{2})&=&{\cal M}^{2}_{\bf even}(\phi)\left(|{\cal H}_{1}|^2+|{\cal H}_{2}|^2\right)+{G}^2_{\bf even}(\phi)\left(|{\cal H}_{1}|^2+|{\cal H}_{2}|^2\right)^2,\\
V_{\bf IB}({\cal H}_{1},{\cal H}_{2},\theta)&=&{\cal M}^{2}_{\bf odd}(\phi)\left[\exp(-im\theta){\cal H}^{*}_{1}{\cal H}_{2}+\exp(im\theta){\cal H}^{*}_{2}{\cal H}_{1}\right]\nonumber\\
&&+{G}^{2}_{\bf odd}(\phi)\left[\exp(-2im\theta)({\cal H}^{*}_{1}{\cal H}_{2})^2+\exp(2im\theta)({\cal H}^{*}_{2}{\cal H}_{1})^2+|{\cal H}_{1}|^2|{\cal H}_{2}|^2\right]\nonumber\\
&&+{G}_{\bf even}(\phi){G}_{\bf odd}(\phi)\left(|{\cal H}_{1}|^2+|{\cal H}_{2}|^2\right)
\left[\exp(-im\theta){\cal H}^{*}_{1}{\cal H}_{2}+\exp(im\theta){\cal H}^{*}_{2}{\cal H}_{1}\right].\eea
Now exactly identify the specific role of self interactions in the present context one can consider the following 
simplest  but special
physical situation where the self coupling can be connected with the mass parameter for the {\bf even} 
type of interactions as:
\bea G_{\bf even}(\phi)={\cal M}^2_{\bf even}(\phi),\eea
According to our three previous choice of mass parameter here one can re-express the self coupling parameter 
for the {\bf even} type of contributions in the isospin preserving interactions as:
\bea G_{\bf even}(\eta)={\cal M}^2_{\bf even}(\eta)
&=& \left\{\begin{array}{ll}
                    \displaystyle \left[\gamma_{\bf even}\left(\frac{\eta}{\eta_0} - 1\right)^2 
                    + \delta_{\bf even}\right]~H^2~~~~ &
 \mbox{\small {\bf Case~I}}  \\ \\
	\displaystyle \frac{m^2_{\bf even}}{2}\left[1-\tanh\left(\rho\frac{\ln(-H\eta)}{H}\right)\right]~~~~ & \mbox{\small {\bf Case~II}}\\ \\
	\displaystyle m^2_{\bf even}~{\rm sech}^2\left(\rho\frac{\ln(-H\eta)}{H}\right)  ~~~~ & \mbox{\small {\bf Case~III}}.
          \end{array}
\right.\eea
Similarly following the same argument one can also express the connection between self coupling and mass parameter for the 
{\bf odd} type of interactions as:
\bea G_{\bf odd}(\phi)={\cal M}^2_{\bf odd}(\phi).\eea
According to our three previous choice of mass parameter here one can re-express the self coupling parameter 
for the {\bf odd} type of contributions in the isospin violating interactions as:
\bea G_{\bf odd}(\eta)={\cal M}^2_{\bf odd}(\eta)
&=& \left\{\begin{array}{ll}
                    \displaystyle \left[\gamma_{\bf odd}\left(\frac{\eta}{\eta_0} - 1\right)^2 
                    + \delta_{\bf odd}\right]~H^2~~~~ &
 \mbox{\small {\bf Case~I}}  \\ \\
	\displaystyle \frac{m^2_{\bf odd}}{2}\left[1-\tanh\left(\rho\frac{\ln(-H\eta)}{H}\right)\right]~~~~ & \mbox{\small {\bf Case~II}}\\ \\
	\displaystyle m^2_{\bf odd}~{\rm sech}^2\left(\rho\frac{\ln(-H\eta)}{H}\right)  ~~~~ & \mbox{\small {\bf Case~III}}.
          \end{array}
\right.\eea
Because of this identification one can also express the two eigenvalues of the mass eigenstates
in terms of the self coupling parameters as:
\be \lambda_{\pm}=\sqrt{G_{\bf even}\pm G_{\bf odd}}=\sqrt{G_{\bf even}
+{\bf sign}(\sigma.{\bf n}) G_{\bf odd}}={\cal G}_{\pm}(\eta).\ee
where ${\cal G}_{\pm}(\eta)$ is defined as:
\bea{\cal G}_{\pm}(\eta)
&=& \left\{\begin{array}{ll}
                    \displaystyle \sqrt{\gamma_{\pm}\left(\frac{\eta}{\eta_0} - 1\right)^2 +\delta_{\pm}}~H~~~~ &
 \mbox{\small {\bf Case~I}}  \\ \\
	\displaystyle \frac{m_{0\pm}}{\sqrt{2}}\sqrt{\left[1-\tanh\left(\rho\frac{\ln(-H\eta)}{H}\right)\right]}~~~~ & \mbox{\small {\bf Case~II}}\\ \\
	\displaystyle m_{0\pm}{\rm sech}\left(\rho\frac{\ln(-H\eta)}{H}\right)~~~~ & \mbox{\small {\bf Case~III}}.
          \end{array}
\right.\eea
Here it is important to mention here that the isospin preserving re-normalizable interaction can be identified with the axion interaction as discussed in the earler section. In case axion interaction we have neglected the quartic interaction due to very very small 
back reaction. But in case of axion also one can include the effects of isospin breaking interaction to measure the mass eigenvalues in the eigenstate as discussed here for any heavy fields. During the post inflationary era to avoid 
the problem of
formation of domain wall one can set the axion potential to be zero at that period. As the signatures of these heavy fields or the axion have not observed yet through any observational probes, one can treat such heavy fields or the axions correspond to a component of dark matter and the corresponding density fluctuations 
can be treated as isocurvature fluctuations. 
\section{Conclusion}
\label{sec5}
To summarize, in the present article, we have addressed the following points:
\begin{itemize}
\item Firstly we have briefly reviewed Bell's inequality in quantum mechanics and its implications. For this we reviewed the proof of Bell's inequality.
Further we have discussed the violation of Bell's inequality in the context of quantum mechanics. Also we have given the explanation for such violation, which finally give rise to new physical concepts and phenomena.

\item Next we have briefly discussed about the setup for Bell's inequality violating test experiment in the context of primordial cosmology. Further we have studied creation of new massive particles
as introduced in the context of inflationary paradigm for various choice of time dependent mass profile. We have also presented the calculation for the three limiting
situations-{\bf (1)} $m\approx H$, {\bf (2)} $m>>H$ and {\bf (3)} $m<<H$. To describe a very small 
fraction of particle creation after inflation we have computed the expression for Bogoliubov coefficient $\beta$ in FLRW space-time, which characterizes the amount of mixing between the two types of WKB
solutions. Next using the results for Bogoliubov co-efficients we have further calculated reflection and transmission co-efficients, number density and energy density of the created particles for
various mass profiles. Here we have provided the results for three specific cases:-\begin{enumerate}
                                                                              \item  $|kc_{S}\eta|=c_{S}k/aH<<1$ (\textcolor{blue}{super horizon}),
                                                                              \item $|kc_{S}\eta|=c_{S}k/aH\approx 1$ (\textcolor{blue}{horizon crossing}),
                                                                              \item  $|kc_{S}\eta|=c_{S}k/aH>>1$ (\textcolor{blue}{sub horizon}).
                                                                             \end{enumerate}
Further we have studied cosmological scalar curvature fluctuations in presence of new massive particles for arbitrary choice of initial condition and also for any arbitrary mass profile. Here 
we have explicitly derived the expression for one point and two point correlation function using in-in formalism. Then we have quoted the results 
for the three limiting
situations-{\bf (1)} $m\approx H$, {\bf (2)} $m>>H$ and {\bf (3)} $m<<H$ in \textcolor{blue}{super horizon}, \textcolor{blue}{sub horizon} and \textcolor{blue}{horizon crossing}.
Here in our computation we have introduced a new cosmological observable which captures the effect of Bell's inequality violation in cosmology.
Further we have expressed the scale of inflation in terms of the amount of Bell's inequality violation in cosmology experimental setup using model independent prescription like EFT. Additionally we have derived
 a model independent expression for first Hubble slow roll parameter $\epsilon=-\dot{H}/H^2$ and tensor-to-scalar ratio in terms of the Bell's inequality violating observable
 within the framework of EFT. Additionally,  we have given an estimate of heavy field mass parameter $m/H$
to violate Bell's inequality within cosmological setup.

\item It is important to note that when all the EFT interactions are absent in that case both $c_{S}\sim \tilde{c}_{S}=1$ and one can get back the results for canonical slow-roll models.
On the other hand when the EFT interactions are switched on within the present description, one can able to accommodate the non-canonical as well 
 as non-minimal interactions within this framework. In that case both $c_{S}$ and $\tilde{c}_{S}$ are less than
 unity and in such a situation one can always constraint the sound speed parameter as well the strength 
 of the EFT interactions using observational probes (Planck 2015 data).
 One can easily compare the present setup with effective time varying mass 
 parameter with the axions with time varying decay constant.
 For $m<<H$ case the last term in the effective action is absent and in that case the reduced form of the action will able to explain
 the EFT of inflation in presence of previously mentioned non-trivial effective interactions. Once we switch off all such interactions the above action mimics the case for single field slow-roll inflation.

\item Further we have given an example of axion model with time dependent decay constant as appearing in the context of string theory. Hence we have mentioned the effective axion interaction of axion fields.
Now to give a analogy between the newly introduced massive particle and the axion we have further discussed the creation of axion in early universe.
Next we have established the one to one correspondence between heavy field and axion by comparing the particle creation mechanism, one and two point correlation functions.
Additionally, we have given an estimate of axion mass parameter $m_{axion}/f_{a}H$
to violate Bell's inequality within cosmological setup. Finally, we have discussed the specific role of 
 isospin breaking interaction for axion type of heavy fields to measure the effect of Bell's inequality violation in primordial cosmology.
 
\item  Next we have explicitly shown the role of quantum decoherence in cosmological setup to violate Bell's inequality. Additionally here we have also 
 mentioned a possibility to enhance the value of primordial non-Gaussianity from Bell's inequality violating setup in presence of massive time dependent field profile. Further we have discussed the role of 
 three specific time dependent mass profile for producing massive particles and to generate quantum fluctuations. Finally, we have discussed the role of arbitrary spin heavy field to violate Bell's inequality. 
 Here we have provided a bound on the mass parameter for massive scalar with spin ${\cal S}=0$, axion with spin ${\cal S}=0$, graviton with spin ${\cal S}=2$ and for particles with high spin ${\cal S}>2$ in \textcolor{blue}{horizon crossing}, \textcolor{blue}{super horizon}
 and \textcolor{blue}{sub horizon} regime. 
 \end{itemize}
The future prospects of our work are appended below:
\begin{itemize}
 \item In this work we have not explored in detail the possibility of enhancing the primordial non-Gaussianity from the violation of Bell's inequality and the exact role of time dependent mass profile for heavy fields. 
 In appendix we have pointed on some of the issues but not given detailed calculation on this issue. In near future we are planning to address this important issue.
 
 \item One can also comment on the dependence on time dependent mass profile for the heavy field to derive the consistency relation in the context of inflationary cosmology. Due to the enhancement of the primordial 
 non-Gaussian amplitude it is expected from the basic understanding 
 that due to the presence of such non-negligible Bell violating contribution, all the inflationary consistency relations will modify significantly. In future we are also planning to derive all such modified consistency relations from this work.
 
 \item In this work we have implemented the idea of Bell violation in the context of inflationary cosmology. But the explicit role of alternatives idea of inflation to design a Bell's inequality violating experiment in cosmology is not studied yet.
 One can check whether this can be done or not. If this is possible, then one can also study the consequences from this, including non-Gaussianities.
 
 \item The explicit role of entanglement entropy is very important to understand the underlying physical principles in the present context for de Sitter and quasi de Sitter case. In this paper we have not
 addressed this issue in detail, which one can address 
 in future.

 \item One can also carry forward our analysis in the context of higher derivative gravity set up which nobody have addressed yet.
 
 \item To give a model independent bound on the scale of inflation and primordial gravitational waves we have defined a new cosmological observable which explicitly captures the effect of violation of Bell's inequality in cosmology.
 But for this we need a prior knowledge of such observables. We have only mentioned the explicit role of isospin breaking interactions in this context. But in this work we have not studied the exact connection between such isospin breaking interactions for heavy fields 
 and newly defined Bell violating cosmological observable. One can also address this issue to comment on the measurement of such observables through various observational probes.
\end{itemize}

\section{Appendix}
\label{sec6}
\subsection{Role of quantum decoherence in Bell violating cosmological setup}
\label{sec6a}
Before going to the next section let us briefly discuss about the some more issues related to the cosmological setup in which we want to study
the violation of Bell inequality. In this context the basis field eigenstates can be identified with $|\zeta({\bf x})\rangle$, which satisfies the following eigenvalue equation:
\be \hat{\zeta}({\bf x})|\zeta({\bf x})\rangle = \zeta({\bf x})|\zeta({\bf x})\rangle,\ee
where the quantum operator $\hat{\zeta}({\bf x})$ is specified in any point in the space-time and the eigenvalue $\zeta({\bf x})$ representing the classical configuration in the present context. 
Here for our discussion we start with a Gaussian arbitrary vacuum state which can be expressed in terms of the coherent superposition of scalar curvature fluctuation field eigenstates:
\be |\Psi_{\zeta}\rangle\equiv |\Psi[\zeta({\bf x})]\rangle =\sum_{\zeta}{\cal A}_{\zeta({\bf x})}|\zeta({\bf x})\rangle,
\ee
where the Gaussian functional co-efficient ${\cal A}_{\zeta({\bf x})}$ can be defined as:
\be {\cal A}_{\zeta({\bf x})}=\langle \zeta({\bf x})|\Psi_{\zeta}\rangle\equiv \langle \zeta({\bf x})|\Psi[\zeta({\bf x})]\rangle.\ee
Within this theoretical setup to accommodate quantum decoherence phenomena we consider a general environment source
functional $G({\bf x})$ as we have already introduced earlier. In general, this may account any type of non-linear interaction 
with the environment. But specific structure of the interaction actually decides the behaviour of the response to the classical configuration characterized by $\zeta({\bf x})$ and consequently the 
response of the interaction with environment can be characterized by the following quantum entanglement:
\be |\Psi[G({\bf x})]\rangle \otimes|\zeta({\bf x})\rangle=|\Psi[G({\bf x})]\rangle_{\zeta({\bf x})}\otimes |\zeta({\bf x})\rangle,\ee
where one can physically interpret $|\Psi[G({\bf x})]\rangle_{\zeta({\bf x})}$ as the conditional state or pointer state in the present context which satisfies the following condition:
\be \langle \Psi[G({\bf x})|\Psi[G({\bf x})\rangle =1.\ee
Further using this expression one can express the configuration space joint state vector as:
\bea |\Psi[G({\bf x})]\rangle \otimes|\Psi[\zeta({\bf x})]\rangle&=&|\Psi[G({\bf x})]\rangle \otimes\left\{\sum_{\zeta}{\cal A}_{\zeta({\bf x})}|\zeta({\bf x})\rangle\right\}\nonumber\\
&=&\sum_{\zeta}{\cal A}_{\zeta({\bf x})}\left(|\Psi[G({\bf x})]\rangle\otimes |\zeta({\bf x})\rangle\right)\nonumber\\
&=&\sum_{\zeta}{\cal A}_{\zeta({\bf x})}\left(|\Psi[G({\bf x})]\rangle_{\zeta({\bf x})}\otimes |\zeta({\bf x})\rangle\right),\eea
 which in turn destroy the effect of coherent superposition of the eigenstates $|\zeta({\bf x})\rangle$. Additionally it is important to mention here that, the reduced density matrix for the system can be represented by:
 \bea \rho_{\bf Reduced}[\zeta({\bf x}),\alpha({\bf x})]&=&\underbrace{\Psi[\zeta({\bf x})]
 \left(\Psi[\alpha({\bf x})]\right)^{\dagger}}_{\bf Interference ~term}\left\{\int {\cal D}G~ \Psi[G({\bf x})]_{\zeta({\bf x})}\left(
 \Psi[G({\bf x})]_{\zeta({\bf x})}\right)^{\dagger}\right\}\nonumber \\
 &=&\underbrace{\Psi[\zeta({\bf x})]
 \left(\Psi[\alpha({\bf x})]\right)^{\dagger}}_{\bf Interference ~term}\langle\Psi[G({\bf x})]_{\zeta({\bf x})}|\Psi[G({\bf x})]_{\zeta({\bf x})}\rangle, \eea
 where we use:
 \be \int {\cal D}G~ \Psi[G({\bf x})]_{\zeta({\bf x})}\Psi[G({\bf x})]^{\dagger}_{\zeta({\bf x})}=\langle\Psi[G({\bf x})]_{\zeta({\bf x})}|\Psi[G({\bf x})]_{\zeta({\bf x})}\rangle.\ee
 Here the most general structure of the conditional or pointer wave function for the environment can be expressed as:
 \be \Psi[G({\bf x})]_{\zeta({\bf x})}={\cal N}_{G\zeta}
 \exp\left[\zeta({\bf x})\star G({\bf x}) \star G({\bf x}) \star \left({\rm Re }~{\cal M}({\bf x})+i{\rm Im }~{\cal M}({\bf x})\right)\right],\ee
 where ${\cal N}_{G\zeta}$ represents the normalization constant for conditional or pointer wave
 function and $\star$ characterize the convolution operation in the present context. From this specific
 structure of the conditional or pointer wave function it is clearly observed that:
 \begin{itemize}
  \item ${\rm Re }~{\cal M}({\bf x})$ and ${\rm Im }~{\cal M}({\bf x})$ is the unknown function in the present context which one needs to compute 
  for a specified structure of the interaction between system and environment,
  \item Here contribution from ${\rm Im }~{\cal M}({\bf x})$ is large compared to ${\rm Re }~{\cal M}({\bf x})$,
  \item The phase factor is rapidly oscillating in conditional or pointer wave function and consequently the following integral vanishes:
   \be \int {\cal D}G~ \Psi[G({\bf x})]_{\zeta({\bf x})}
   \Psi[G({\bf x})]^{\dagger}_{\zeta({\bf x})}=\langle
   \Psi[G({\bf x})]_{\zeta({\bf x})}|\Psi[G({\bf x})]_{\zeta({\bf x})}\rangle\sim 0.\ee
   \item Finally the off diagonal component of the reduced density matrix vanishes.
 \end{itemize}
Now in Schr$\ddot{o}$dinger picture of quantum mechanics we define a configuration space eigenstate by $|G({\bf x}),\zeta({\bf x})\rangle$ which satisfy the following sets of eigenvalue equation:
\bea \hat{\zeta}({\bf x})|G({\bf x}),\zeta({\bf x})\rangle=\zeta({\bf x})|G({\bf x}),\zeta({\bf x})\rangle,\\
 \hat{\zeta}({\bf x})|G({\bf x}),\zeta({\bf x})\rangle=\zeta({\bf x})|G({\bf x}),\zeta({\bf x})\rangle.\eea
 Further using configuration space eigenstate $|G({\bf x}),\zeta({\bf x})\rangle$ one can also define the wave functional of the joint system and environment at conformal time $\eta$ as:
 \be \langle G({\bf x}),\zeta({\bf x})|\Psi(\eta)\rangle = \Psi[G({\bf x}),\zeta({\bf x})](\eta).\ee
 Also one can express the the wave functional of the joint system and environment at time $\eta$ in the following convenient product form:
 \be \Psi[G({\bf x}),\zeta({\bf x})](\eta)=\underbrace{\Psi_{\bf Gaussian}[\zeta({\bf x})](\eta)\Psi_{\bf Gaussian}[G({\bf x})](\eta)}_{\bf Gaussian~component}
 \underbrace{\Psi_{\bf Non-Gaussian}[G({\bf x}),\zeta({\bf x})](\eta)}_{\bf Non-Gaussian~component},\ee
 where each components of the wave functional can be written as:
 \bea \Psi_{\bf Gaussian}[\zeta({\bf x})](\eta)&=&{\cal N}_{\zeta}(\eta)\exp\left[-\int \frac{d^3{\bf k}}{(2\pi)^3}\zeta_{\bf k} \zeta^{\dagger}_{\bf k}\Theta_{\zeta}(k,\eta)\right],\\
 \Psi_{\bf Gaussian}[G({\bf x})](\eta)&=&{\cal N}_{G}(\eta)\exp\left[-\int \frac{d^3{\bf k}}{(2\pi)^3}G_{\bf k} G^{\dagger}_{\bf k}\Theta_{G}(k,\eta)\right],\\
 \Psi_{\bf Non-Gaussian}[G({\bf x}),\zeta({\bf x})](\eta)&=&{\cal N}_{G\zeta}(\eta)\exp\left[\int_{\bf k} \int_{\bf p}\int_{\bf q}\frac{d^3{\bf k}}{(2\pi)^3} \frac{d^3{\bf p}}{(2\pi)^3} \frac{d^3{\bf q}}{(2\pi)^3}\nonumber\right.\\ &&\left.~~~~~~~~~~~~~~~~~~~~~~~~~~~ (2\pi)^3\delta^3({\bf k}+{\bf p}+{\bf q})
 G_{\bf k} G_{\bf p}\zeta_{\bf q}{\cal M}_{{\bf k},{\bf p},{\bf q}}(\eta)\right].~~~~~~~~~~~
 \eea
 Here ${\cal N}_{\zeta}(\eta)$, ${\cal N}_{G}(\eta)$ and ${\cal N}_{G\zeta}(\eta)$ characterize the conformal time dependent normalization constant for system, environment and the joint system-environment in the present context.
 Hence by explicitly studying the time evolution using the master equation from the generic cubic type of interaction one can write down the model independent expressions for the complex functions
 $\Theta_{\zeta}(k,\eta)$, $\Theta_{G}(k,\eta)$ and ${\cal M}_{{\bf k},{\bf p},{\bf q}}(\eta)$
 in the present context. By knowing such structures for arbitrary vacuum state one can further study various issues related to theoretical and observational part of cosmology. 
 
 Further the reduced density matrix can be written as:
 \be \rho_{\bf Reduced}\equiv {\bf Tr}_{G}(\rho_{\bf Global})=\int {\cal D}G \langle G|\rho_{\bf Global}| G\rangle,\ee
 where we have integrated or tracing over the environment $G$ here. Additionally, it is important to note that $\rho_{\bf Global}$ is the global density matrix which is defined as:
 \be \rho_{\bf Global} \equiv |\Psi\rangle \langle \Psi|.\ee
 In the field basis the reduced density matrix can be re-expressed as:
 \bea \rho_{\bf Reduced}[\zeta({\bf x}),\alpha({\bf x})]&=&\langle \zeta({\bf x})|\rho_{\bf Reduced}| \alpha({\bf x})\rangle\nonumber\\
 &=&\underbrace{\Psi[\zeta({\bf x})]
 \left(\Psi[\alpha({\bf x})]\right)^{\dagger}}_{\bf Interference ~term}\left\{\int {\cal D}G~ \Psi[G({\bf x})]_{\zeta({\bf x})}\left(
 \Psi[G({\bf x})]_{\zeta({\bf x})}\right)^{\dagger}\right\}\nonumber \\
 &=&\int {\cal D}G~ \Psi[\zeta({\bf x}),G({\bf x})]_{\zeta({\bf x})}\left(
 \Psi[\alpha({\bf x}),G({\bf x})]_{\zeta({\bf x})}\right)^{\dagger}\nonumber \\
 &=&\int {\cal D}G~ \underbrace{\Psi_{\bf Gaussian}[\zeta({\bf x})](\eta)\Psi_{\bf Gaussian}[G({\bf x})](\eta)}_{\bf Gaussian~component}
 \underbrace{\Psi_{\bf Non-Gaussian}[G({\bf x}),\zeta({\bf x})](\eta)}_{\bf Non-Gaussian~component}\nonumber\\
 &&~~~~~~~~~~~~~~~~~~~~\left(\underbrace{\Psi_{\bf Gaussian}[\alpha({\bf x})](\eta)\Psi_{\bf Gaussian}[G({\bf x})](\eta)}_{\bf Gaussian~component}
 \underbrace{\Psi_{\bf Non-Gaussian}[G({\bf x}),\zeta({\bf x})](\eta)}_{\bf Non-Gaussian~component}\right)^{\dagger}\nonumber\\
 &=&\Psi_{\bf Gaussian}[\zeta({\bf x})](\eta)\left(\Psi_{\bf Gaussian}[\alpha({\bf x})](\eta)\right)^{\dagger}\underbrace{{\LARGE \bf D}_{\bf Decoherence}[\zeta({\bf x}),\alpha({\bf x})]}_{\bf Decoherence~factor},\eea
 where the decoherence factor is defined as:
 \bea {\LARGE \bf D}_{\bf Decoherence}[\zeta({\bf x}),\alpha({\bf x})]
 &=&\frac{|\rho_{\bf Reduced}[\zeta({\bf x}),\alpha({\bf x})]|}{\sqrt{\rho_{\bf Reduced}[\zeta({\bf x}),\zeta({\bf x})]\rho_{\bf Reduced}[\alpha({\bf x}),\alpha({\bf x})]}}\nonumber\\
 &=&\int {\cal D}G|\Psi_{\bf Gaussian}[G({\bf x})](\eta)|^2|\Psi_{\bf Non-Gaussian}[G({\bf x}),\zeta({\bf x})](\eta)|^2\nonumber\\
 &=&|{\cal N}_{G\zeta}(\eta)|^2\int {\cal D}G|\Psi_{\bf Gaussian}[G({\bf x})](\eta)|^2\nonumber\\
 &&~~~~~~~~~~~~~\exp\left[\int_{\bf k} \int_{\bf p}\int_{\bf q}\frac{d^3{\bf k}}{(2\pi)^3} \frac{d^3{\bf p}}{(2\pi)^3} \frac{d^3{\bf q}}{(2\pi)^3}\nonumber\right.\\ &&\left.~~~~~~~~~~~~~~~ (2\pi)^3\delta^3({\bf k}+{\bf p}+{\bf q})
 G_{\bf k} G_{\bf p}\left\{\zeta_{\bf q}{\cal M}_{{\bf k},{\bf p},{\bf q}}(\eta)+\alpha_{\bf q}{\cal M}^{\dagger}_{{\bf k},{\bf p},{\bf q}}(\eta)\right\}\right]\nonumber\\
 &=&|{\cal N}_{G\zeta}(\eta)|^2|{\cal N}_{G}(\eta)|^2\int {\cal D}G\exp\left[-\int \frac{d^3{\bf k}}{(2\pi)^3}G_{\bf k} G^{\dagger}_{\bf k}\left(\Theta_{G}(k,\eta)+\Theta_{G}^{\dagger}(k,\eta)\right)\right]\nonumber\\
 &&~~~~~~~~~~~~~\exp\left[\int_{\bf k} \int_{\bf p}\int_{\bf q}\frac{d^3{\bf k}}{(2\pi)^3} \frac{d^3{\bf p}}{(2\pi)^3} \frac{d^3{\bf q}}{(2\pi)^3}\nonumber\right.\\ &&\left.~~~~~~~~~~~~~~~ (2\pi)^3\delta^3({\bf k}+{\bf p}+{\bf q})
 G_{\bf k} G_{\bf p}\left\{\zeta_{\bf q}{\cal M}_{{\bf k},{\bf p},{\bf q}}(\eta)+\alpha_{\bf q}{\cal M}^{\dagger}_{{\bf k},{\bf p},{\bf q}}(\eta)\right\}\right]\nonumber\\
 &=&|{\cal N}_{G\zeta}(\eta)|^2|{\cal N}_{G}(\eta)|^2\int {\cal D}G\exp\left[-2\int \frac{d^3{\bf k}}{(2\pi)^3}G_{\bf k} G^{\dagger}_{\bf k}{\rm Re}\left[\Theta_{G}(k,\eta)\right]\right]\nonumber\\
 &&~~~~~~~~~~~~~\exp\left[\int_{\bf k} \int_{\bf p}\int_{\bf q}\frac{d^3{\bf k}}{(2\pi)^3} \frac{d^3{\bf p}}{(2\pi)^3} \frac{d^3{\bf q}}{(2\pi)^3}\nonumber\right.\\ &&\left.~~~~~~~(2\pi)^3\delta^3({\bf k}+{\bf p}+{\bf q})
 G_{\bf k} G_{\bf p}\left\{{\bf \Delta}_{q+}{\rm Re}\left[{\cal M}_{{\bf k},{\bf p},{\bf q}}(\eta)\right]+i{\bf \Delta}_{q-}{\rm Im}\left[{\cal M}_{{\bf k},{\bf p},{\bf q}}(\eta)\right]\right\}\right],~~~~~~~~~ \eea
 which takes care of off diagonal amplitudes as well as non-Gaussian contributions in the present context. Here the newly defined functions ${\bf \Delta}_{q+}$ and ${\bf \Delta}_{q-}$ are defined as:
 \bea {\bf \Delta}_{q+}&=&\zeta_{\bf q}+\alpha_{\bf q},\\
 {\bf \Delta}_{q-}&=&\zeta_{\bf q}-\alpha_{\bf q}.\eea
 Further absorbing the normalization factors in the definition of decoherence factor one can write:
 \bea {\LARGE \bf\cal D}_{\bf Decoherence}[\zeta({\bf x}),\alpha({\bf x})]
 &=&\frac{{\LARGE \bf D}_{\bf Decoherence}[\zeta({\bf x}),\alpha({\bf x})]}{|{\cal N}_{G\zeta}(\eta)|^2|{\cal N}_{G}(\eta)|^2}\nonumber\\
 &=&\int {\cal D}G\exp\left[-2\int \frac{d^3{\bf k}}{(2\pi)^3}G_{\bf k} G^{\dagger}_{\bf k}{\rm Re}\left[\Theta_{G}(k,\eta)\right]+\int_{\bf k} \int_{\bf p}\int_{\bf q}\frac{d^3{\bf k}}{(2\pi)^3} \frac{d^3{\bf p}}{(2\pi)^3} \frac{d^3{\bf q}}{(2\pi)^3}\nonumber\right.\\ &&\left.~~~~~~~(2\pi)^3\delta^3({\bf k}+{\bf p}+{\bf q})
 G_{\bf k} G_{\bf p}\left\{{\bf \Delta}_{q+}{\rm Re}\left[{\cal M}_{{\bf k},{\bf p},{\bf q}}(\eta)\right]+i{\bf \Delta}_{q-}{\rm Im}\left[{\cal M}_{{\bf k},{\bf p},{\bf q}}(\eta)\right]\right\}\right]\nonumber\\
 &=&\langle\exp\left[{ \cal Z}\right]\rangle\nonumber\\
 &=&\exp\left[\underbrace{\sum^{\infty}_{n=1}\frac{1}{(2n)!}\langle {\cal Z}^{2n}\rangle_{\bf C}}_{\bf Connected~even~cumulants}+\underbrace{\sum^{\infty}_{m=0}\frac{1}{(2m+1)!}\langle {\cal Z}^{2m+1}\rangle_{\bf DC}}_{\bf Disconnected~odd~cumulants}\right],~~~~~~~~~ \eea
 where the factor ${\cal Z}$ is defined as:
 \bea {\cal Z}&=&-2\int \frac{d^3{\bf k}}{(2\pi)^3}G_{\bf k} G^{\dagger}_{\bf k}{\rm Re}\left[\Theta_{G}(k,\eta)\right]\nonumber\\ &&~~~+\int_{\bf k} \int_{\bf p}\int_{\bf q}\frac{d^3{\bf k}}{(2\pi)^3} \frac{d^3{\bf p}}{(2\pi)^3} \frac{d^3{\bf q}}{(2\pi)^3}(2\pi)^3\delta^3({\bf k}+{\bf p}+{\bf q})
 G_{\bf k} G_{\bf p}\left\{{\bf \Delta}_{q+}{\rm Re}\left[{\cal M}_{{\bf k},{\bf p},{\bf q}}(\eta)\right]+i{\bf \Delta}_{q-}{\rm Im}\left[{\cal M}_{{\bf k},{\bf p},{\bf q}}(\eta)\right]\right\}~~~~~~~~~~~\eea
 and the subscript ${\bf C}$ and ${\bf DC}$ indicates the connected and disconnected contribution of the correlation function.
 To further simplify the form of rescaled decoherence factor ${\LARGE \bf\cal D}_{\bf Decoherence}[\zeta({\bf x}),\alpha({\bf x})]$ one can introduce the environment two point correlation function as:
 \bea \langle G_{\bf k}G_{\bf q}\rangle_{\eta}&=&(2\pi)^2\delta^{3}({\bf k}+{\bf q})P_{G}(k,\eta),\eea
 where $P_{G}(k,\eta)$ is the power spectrum of the environment and can be expressed in terms of the variance of the wave function. Here due to the presence of the additional non-Gaussian part in the wave function it is expected that 
 decoherence effect cannot be negligible in the present context, provided the structure of complex functions $\Theta_{\zeta}(k,\eta)$, $\Theta_{G}(k,\eta)$ and ${\cal M}_{{\bf k},{\bf p},{\bf q}}(\eta)$ involves 
 the time dependent mass parameter used for violating Bell's inequality in the cosmological setup. Due to the presence of time dependent mass parameter in the interaction part of the Hamiltonian one can expect that 
 the contribution from the disconnected odd cumulants in the correlation function is non negligible. Specifically due to Bell violation $\langle {\cal Z} \rangle$ contributes to the rescaled decoherence factor ${\LARGE \bf\cal D}_{\bf Decoherence}[\zeta({\bf x}),\alpha({\bf x})]$ and in the 
 expression for reduced density matrix $\rho_{\bf Reduced}[\zeta({\bf x}),\alpha({\bf x})]$. Also $\langle {\cal Z}^2 \rangle$, $\langle {\cal Z}^3 \rangle$ and $\langle {\cal Z}^4 \rangle$ captures the effect of power spectrum, 
 non-Gaussian bi-spectrum and tri-spectrum in the present context. See ref.~\cite{Burgess:2006jn,Nelson:2016kjm} for more details. After knowing the specific structure of the reduced density matrix one can define Wigner function as:
 \bea W&=&\frac{1}{(2\pi)^2}\int dx \int dy \exp[-i(\pi_{\bf k}x+\pi_{-\bf k}y)]\langle q_{\bf k}+x/2,q_{\bf -k}+y/2|\rho_{\bf Reduced}|q_{\bf k}-x/2,q_{\bf -k}-y/2\rangle\nonumber\\
 &=&\frac{1}{(2\pi)^2}\int dx \int dy \exp[-i(\pi_{\bf k}x+\pi_{-\bf k}y)]\nonumber\\
 &&~~~~~\Psi_{\bf Gaussian}[\zeta(q_{\bf k}+x/2,q_{\bf -k}+y/2)]\left(\Psi_{\bf Gaussian}[\alpha(q_{\bf k}-x/2,q_{\bf -k}-y/2)]\right)^{\dagger}\nonumber\\
 &&~~~~~~~~~~~~~~~~~~~~~~~~~~~~~~~~~~~~~~~~~~~~
 \times\underbrace{{\LARGE \bf D}_{\bf Decoherence}[\zeta(q_{\bf k}+x/2,q_{\bf -k}+y/2),\alpha(q_{\bf k}-x/2,q_{\bf -k}-y/2)]}_{\bf Decoherence~factor}.~~~~~~~~~~~
 \eea
 To understand the structure of the Wigner function we need to substitute the specific form of the wave functions as mentioned earlier. Here after substitution the wave function and absorbing the normalization constants one 
 can be recast the Wigner function into the following rescaled form: 
\bea {\cal W}&=&\frac{W}{|{\cal N}_{G\zeta}(\eta)|^2|{\cal N}_{G}(\eta)|^2}\nonumber\\
&=&\frac{1}{(2\pi)^2}\int dx \int dy \exp[-i(\pi_{\bf k}x+\pi_{-\bf k}y)]\nonumber\\
 &&~~~~~\Psi_{\bf Gaussian}[\zeta(q_{\bf k}+x/2,q_{\bf -k}+y/2)]\left(\Psi_{\bf Gaussian}[\alpha(q_{\bf k}-x/2,q_{\bf -k}-y/2)]\right)^{\dagger}\nonumber\\
 &&~~~~~~~~~~~~~~~~~~~~~~~~~~~~~~~~~~~~~~~~~~~~
 \times\langle\exp\left({ \cal Z}[\zeta(q_{\bf k}+x/2,q_{\bf -k}+y/2),\alpha(q_{\bf k}-x/2,q_{\bf -k}-y/2)]\right)\rangle.~~~~~~~~\eea
 As our motivation of this paper is restricted to setup the cosmological experiment where 
 Bell's inequality violation can be tested, we have not compute further results in this paper from arbitrary vacuum state. 
 But we will report soon on such computation in the continuation part of this paper, where we will also comment on the primordial non-Gaussianity for this cosmological setup as well.

\subsection{Time dependent mass profile for heavy field}
\label{sec6b}
\subsubsection{\bf Profile A:~$ m= \sqrt{\gamma\left(\frac{\eta}{\eta_0} - 1\right)^2 + \delta}~H$}
Equation of motion for the massive field is:
\bea
h''_k + \left(c^{2}_{S}k^2 + C + \frac{\left(\gamma+\delta-2\right)}{\eta^2} - \frac{d}{\eta} \right) h_k &=& 0~~~~~~~{\bf for~ dS}\\ 
h''_k + \left(c^{2}_{S}k^2 + C + \frac{\left(\gamma+\delta-\left[\nu^2-\frac{1}{4}\right]\right)}{\eta^2} - \frac{d}{\eta} \right) h_k &=& 0~~~~~~~{\bf for~ qdS}.
\eea
where we introduce two new parameter:
\bea
C &=& \frac{\gamma}{\eta_0^2},~~~ d =\frac{2\gamma}{\eta_0}.
\eea
The solution for the mode function for de Sitter and quasi de Sitter space can be expressed as: 
\be\begin{array}{lll}\label{yu2}
 \displaystyle  h_k (\eta) =\footnotesize\left\{\begin{array}{ll}
                    \displaystyle   G_1 M_{-\frac{d}{2i \sqrt{c^{2}_{S}k^2+C}},i \sqrt{(\gamma+\delta) -\frac{9}{4}}}
\left[2 i\sqrt{c^{2}_{S}k^2+C} \eta\right]+G_2 W_{-\frac{d}{2i \sqrt{c^{2}_{S}k^2+C}},
i \sqrt{(\gamma+\delta) -\frac{9}{4}}}\left[2i \sqrt{k^2+C} \eta\right]&
 \mbox{\small {\bf for ~dS}}  \\ 
	\displaystyle G_1 M_{-\frac{d}{2i \sqrt{c^{2}_{S}k^2+C}}, i \sqrt{\gamma +\delta-\nu^2}}
\left[2 i\sqrt{c^{2}_{S}k^2+C} \eta\right]+G_2 W_{-\frac{d}{2i \sqrt{c^{2}_{S}k^2+C}},
 i \sqrt{\gamma +\delta-\nu^2}}\left[2i \sqrt{c^{2}_{S}k^2+C} \eta\right]& \mbox{\small {\bf for~ qdS}}.
          \end{array}
\right.
\end{array}\ee
where $G_{1}$ and $G_{2}$ are two arbitrary integration constant, which depend on the 
choice of the initial condition. For the sake of simplicity one can recast these solution as:
\be\begin{array}{lll}\label{yu2aazuu}
 \displaystyle  h_k (\eta) =
                   (-\eta)^{\frac{3}{2}}e^{-\frac{P\eta}{2}} (P\eta)^{A+\frac{d}{P}}
                    \left[G_{1}  ~_1F_1\left(A; B;P \eta\right)+G_2 ~ U\left(A; B; P\eta\right)\right]
\end{array}\ee
where $A$, $B$ and $P$ is defined as:
\bea\label{yu2aaz}
 \displaystyle  A&=&\left\{\begin{array}{ll}
                   \displaystyle -\frac{d}{2i \sqrt{c^{2}_{S}k^2+C}}+i \sqrt{(\gamma+\delta) -\frac{9}{4}}+\frac{1}{2}~~~~~~~~~~~~~~~~~~~&
 \mbox{\small {\bf for ~dS}}  \\ 
	\displaystyle -\frac{d}{2i \sqrt{c^{2}_{S}k^2+C}}+i \sqrt{(\gamma+\delta) -\nu^2}+\frac{1}{2}~~~~~~~~~~~~~~~~~~~ & \mbox{\small {\bf for~ qdS}}.
          \end{array}
\right.\\ \nonumber \\
\label{yu2bbz}
 \displaystyle  B &=&\left\{\begin{array}{ll}
                   \displaystyle 2i \sqrt{(\gamma+\delta) -\frac{9}{4}}+1~~~~~~~~~~~~~~~~~~~~~~~~~~~~~~~~~~~~~~~~~~~~~~&
 \mbox{\small {\bf for ~dS}}  \\ 
	\displaystyle 2i \sqrt{(\gamma+\delta) -\nu^2}+1~~~~~~~~~~~~~~~~~~~~~~~~~~~~~~~~~~~~~~~~~~~~~~& \mbox{\small {\bf for~ qdS}}.
          \end{array}
\right.\\ \nonumber \\ 
P&=&2 i\sqrt{c^{2}_{S}k^2+C}.
\eea
After taking $kc_{S}\eta\rightarrow -\infty$ and $kc_{S}\eta\rightarrow 0$ limit for the arbitrary choice of the initial condition or vacuum we get the following results:
\begin{eqnarray}
\label{sol1c1}\lim_{kc_{S}\eta\rightarrow -\infty} U\left(A; B; Qkc_{S}\eta\right)&\approx&
(Qkc_{S}\eta)^{-A},\\
\label{sol1c2}\lim_{kc_{S}\eta\rightarrow -\infty} ~_1F_1\left(A; B;Q kc_{S}\eta\right)&\approx&
\frac{\Gamma(B)(-Qkc_{S}\eta)^{-A}}{\Gamma(B-A)}+\frac{\Gamma(B)e^{Qkc_{S}\eta}(Qkc_{S}\eta)^{A-B}}{\Gamma(A)},\\
\label{sol1cc}\lim_{kc_{S}\eta\rightarrow 0} U\left(A; B; Qkc_{S}\eta\right)&\approx&
\frac{\Gamma(1-B)}{\Gamma(1+A-B)}\left(1+\frac{AQ kc_{S}\eta}{B}\right)+\frac{(Qkc_{S}\eta)^{1-B}\Gamma(B-1)}{\Gamma(A)},~~~~~~~~~\\
\lim_{kc_{S}\eta\rightarrow 0} ~_1F_1\left(A; B;Qkc_{S}\eta\right)&\approx&
\left(1+\frac{AQkc_{S}\eta}{B}\right).
\end{eqnarray}
where the parameter $Q$ is defined as:
 \be\begin{array}{lll}\label{yu2dfvqdfdfdf}\small
 \displaystyle Q=\frac{P}{kc_{S}}.
\end{array}\ee
One can also consider the following approximations to simplify the final derived form of the solution for arbitrary vacuum with $|kc_{S}\eta|= 1$ or equivalently  $|kc_{S}\eta|\approx 1-\Delta$ with $\Delta\rightarrow 0$ case:
\begin{enumerate}
 \item We start with the {\it Laurent expansion} of the Gamma function:
       \bea \label{r1zax} \Gamma(X) &=& \frac{1}{X}-\gamma+\frac{1}{2}\left(\gamma^2+\frac{\pi^2}{6}\right)X
       -\frac{1}{6}\left(\gamma^3+\frac{\gamma \pi^2}{2}+2\zeta(3)\right)X^2 +{\cal O}(X^3)~~~~~~~~~~~~~~~~~~
\eea
where $\gamma$ being the Euler Mascheroni constant and $\zeta(3)$ characterizing the Reimann zeta function of 
       order $3$ originating in the expansion of the gamma function. Here the parameter $X$ is defined as:
       \be X=A, B-1, 1-B, A-B+1,\ee
       where for de Sitter and quasi de Sitter case we get:
  \bea\label{yu2aaz}
 \displaystyle  A&=&\left\{\begin{array}{ll}
                   \displaystyle -\frac{d}{2i \sqrt{c^{2}_{S}k^2+C}}+i \sqrt{(\gamma+\delta) -\frac{9}{4}}+\frac{1}{2}~~~~~~~~~~~~~~~~~~~&
 \mbox{\small {\bf for ~dS}}  \\ 
	\displaystyle -\frac{d}{2i \sqrt{c^{2}_{S}k^2+C}}+i \sqrt{(\gamma+\delta) -\nu^2}+\frac{1}{2}~~~~~~~~~~~~~~~~~~~ & \mbox{\small {\bf for~ qdS}}.
          \end{array}
\right.\\ 
\label{yu2bbz}
 \displaystyle  B-1 &=&\left\{\begin{array}{ll}
                   \displaystyle 2i \sqrt{(\gamma+\delta) -\frac{9}{4}}~~~~~~~~~~~~~~~~~~~~~~~~~~~~~~~~~~~~~~~~~~~~~~&
 \mbox{\small {\bf for ~dS}}  \\ 
	\displaystyle 2i \sqrt{(\gamma+\delta) -\nu^2}~~~~~~~~~~~~~~~~~~~~~~~~~~~~~~~~~~~~~~~~~~~~~~& \mbox{\small {\bf for~ qdS}}.
          \end{array}
\right.\\ 
\label{yu2bbz}
 \displaystyle  1-B &=&\left\{\begin{array}{ll}
                   \displaystyle -2i \sqrt{(\gamma+\delta) -\frac{9}{4}}~~~~~~~~~~~~~~~~~~~~~~~~~~~~~~~~~~~~~~~~~~~~~~&
 \mbox{\small {\bf for ~dS}}  \\ 
	\displaystyle 2i \sqrt{(\gamma+\delta) -\nu^2}~~~~~~~~~~~~~~~~~~~~~~~~~~~~~~~~~~~~~~~~~~~~~~& \mbox{\small {\bf for~ qdS}}.
          \end{array}
\right.\\ 
\label{yu2bbz}
 \displaystyle  A-B+1 &=&\left\{\begin{array}{ll}
                   \displaystyle  -\frac{d}{2i \sqrt{c^{2}_{S}k^2+C}}
                   -i \sqrt{(\gamma+\delta) -\frac{9}{4}}+\frac{1}{2}
                   ~~~~~~~~~~~~~~~&
 \mbox{\small {\bf for ~dS}}  \\ 
	\displaystyle -\frac{d}{2i \sqrt{c^{2}_{S}k^2+C}}-i \sqrt{(\gamma+\delta) -\nu^2}+\frac{1}{2}
	~~~~~~~~~~~~~~~& \mbox{\small {\bf for~ qdS}}.
          \end{array}
\right.
\eea     
\item In this case the solution Hyper-geometric functions of first and second kind can re-expressed into the following simplified form  as:
\begin{eqnarray}
\label{sol1cccc}\lim_{|kc_{S}\eta|\approx 1-\Delta(\rightarrow 0)}U\left(A; B; Qkc_{S}\eta\right)&\approx&
f_{1}(A,B)\left(1-\frac{AQ(1+\Delta)}{B}\right)
+f_{2}(A,B)(-Q(1+\Delta))^{1-B},~~~~~~~~~~~~\\
\lim_{|kc_{S}\eta|\approx 1-\Delta(\rightarrow 0)}~_1F_1\left(A; B;Qkc_{S}\eta\right)&\approx&\left(1-\frac{AQ(1+\Delta)}{B}\right).~~~~~~~~~~~~
\end{eqnarray}
where $f_{1}(A,B)$ and $f_{2}(A,B)$ is defined as:
\bea f_{1}(A,B)&=& \tiny\frac{\Gamma(1-B)}{\Gamma(1+A-B)}
\approx \left[\frac{(1+A-B)}{(1-B)}-\gamma+\frac{1}{2}\left(\gamma^2+\frac{\pi^2}{6}\right)(1-B)(1+A-B)+\cdots\right],~~~~~~\\
     f_{2}(A,B)&=&\tiny\frac{\Gamma(B-1)}{\Gamma(A)}
\approx \left[\frac{A}{(B-1)}-\gamma+\frac{1}{2}\left(\gamma^2+\frac{\pi^2}{6}\right)(B-1)A+\cdots\right].\eea
\end{enumerate}
After taking $kc_{S}\eta\rightarrow -\infty$, $kc_{S}\eta\rightarrow 0$ and $|kc_{S}\eta|\approx 1-\Delta(\rightarrow 0)$ limit the most general 
solution as stated in Eq~(\ref{yu2aazuu}) can be recast as:
\bea\label{yu2zzx}
 \displaystyle h_k (\eta) &\stackrel{|kc_{S}\eta|\rightarrow-\infty}{=}&
 (-\eta)^{\frac{3}{2}}e^{-\frac{P\eta}{2}} (P\eta)^{A+\frac{d}{P}}
                    \left[G_{1}  \left\{\frac{\Gamma(B)(-Qkc_{S}\eta)^{-A}}{\Gamma(B-A)}\right.\right.\\ && \left.\left. \nonumber
                    ~~~~~~~~~~~~~~~~~~~~~~~~~~~~~~+\frac{\Gamma(B)e^{Qkc_{S}\eta}(Qkc_{S}\eta)^{A-B}}{\Gamma(A)}\right\}+G_2 ~ (Qkc_{S}\eta)^{-A}\right]\\  \nonumber\\&=&\footnotesize\left\{\begin{array}{ll}
                    \displaystyle  (-\eta)^{\frac{3}{2}}e^{-i\sqrt{c^{2}_{S}k^2+C}\eta} \left(2 i\sqrt{c^{2}_{S}k^2+C}\eta\right)^{i \sqrt{(\gamma+\delta) -\frac{9}{4}}+\frac{1}{2}}
                    \\ \displaystyle \left[G_{1}  \left\{\frac{\Gamma\left( 2i \sqrt{(\gamma+\delta) -\frac{9}{4}}+1\right)
                    }{\Gamma\left(i \sqrt{(\gamma+\delta) -\frac{9}{4}}+\frac{1}{2}+\frac{d}{2i \sqrt{c^{2}_{S}k^2+C}}\right)}\right.\right.\nonumber\\ \left.\left. 
                    \displaystyle ~~~~~~~~~~~~~~~~~~~~~~~~~~~~~\times\left(-2 i\sqrt{c^{2}_{S}k^2+C}\eta\right)^{\frac{d}{2i \sqrt{c^{2}_{S}k^2+C}}-i \sqrt{(\gamma+\delta) -\frac{9}{4}}-
                    \frac{1}{2}}\right.\right.\\ \displaystyle \left.\left. \nonumber
                    +\frac{\Gamma\left( 2i \sqrt{(\gamma+\delta) -\frac{9}{4}}+1\right)}
                    {\Gamma\left(-\frac{d}{2i \sqrt{c^{2}_{S}k^2+C}}+i \sqrt{(\gamma+\delta) -\frac{9}{4}}+\frac{1}{2}\right)}\right.\right.\nonumber\\ \left. \left.\displaystyle  ~~~~~~~~~~~~~~~~~~~~~~~~~~~~~\times
                    \frac{e^{2 i\sqrt{c^{2}_{S}k^2+C}\eta}}{\left(2 i\sqrt{c^{2}_{S}k^2+C}\eta\right)^{i \sqrt{(\gamma+\delta) -\frac{9}{4}}+\frac{1}{2}+
                    \frac{d}{2i \sqrt{c^{2}_{S}k^2+C}}}}\right\}\right.\nonumber\\ \left. \displaystyle +G_2 ~ \left(2 i\sqrt{c^{2}_{S}k^2+C}\eta\right)^{\frac{d}{2i \sqrt{c^{2}_{S}k^2+C}}-i \sqrt{(\gamma+\delta) -\frac{9}{4}}-
                    \frac{1}{2}}\right]~~~~ &
 \mbox{\small {\bf for ~dS}}  \nonumber\\  \nonumber\\ \nonumber\\
	\displaystyle  (-\eta)^{\frac{3}{2}}e^{-i\sqrt{c^{2}_{S}k^2+C}\eta} \left(2 i\sqrt{c^{2}_{S}k^2+C}\eta\right)^{i \sqrt{(\gamma+\delta) -\nu^2}+\frac{1}{2}}
                    \\ \displaystyle \left[G_{1}  \left\{\frac{\Gamma\left( 2i \sqrt{(\gamma+\delta) -\nu^2}+1\right)
                    }{\Gamma\left(i \sqrt{(\gamma+\delta) -\nu^2}+\frac{1}{2}+\frac{d}{2i \sqrt{c^{2}_{S}k^2+C}}\right)}\right.\right.\nonumber\\ \left.\left. 
                    \displaystyle ~~~~~~~~~~~~~~~~~~~~~~~~~~~~~\times\left(-2 i\sqrt{c^{2}_{S}k^2+C}\eta\right)^{\frac{d}{2i \sqrt{c^{2}_{S}k^2+C}}-i \sqrt{(\gamma+\delta) -\nu^2}-
                    \frac{1}{2}}\right.\right.\\ \displaystyle \left.\left. \nonumber
                    +\frac{\Gamma\left( 2i \sqrt{(\gamma+\delta) -\nu^2}+1\right)}
                    {\Gamma\left(-\frac{d}{2i \sqrt{c^{2}_{S}k^2+C}}+i \sqrt{(\gamma+\delta) -\nu^2}+\frac{1}{2}\right)}\right.\right.\nonumber\\ \left. \left.\displaystyle  ~~~~~~~~~~~~~~~~~~~~~~~~~~~~~\times
                    \frac{e^{2 i\sqrt{c^{2}_{S}k^2+C}\eta}}{\left(2 i\sqrt{c^{2}_{S}k^2+C}\eta\right)^{i \sqrt{(\gamma+\delta) -\nu^2}+\frac{1}{2}+
                    \frac{d}{2i \sqrt{c^{2}_{S}k^2+C}}}}\right\}\right.\nonumber\\ \left. \displaystyle +G_2 ~ \left(2 i\sqrt{c^{2}_{S}k^2+C}\eta\right)^{\frac{d}{2i \sqrt{c^{2}_{S}k^2+C}}-i \sqrt{(\gamma+\delta) -\nu^2}-
                    \frac{1}{2}}\right]~~~~ & \mbox{\small {\bf for~ qdS}}.
          \end{array}
\right.\nonumber\eea 
\bea
\label{yu2}
 \displaystyle h_k (\eta) &\stackrel{|kc_{S}\eta|\rightarrow 0}{=}& (-\eta)^{\frac{3}{2}}e^{-\frac{P\eta}{2}} (P\eta)^{A+\frac{d}{P}}
                    \left[G_{1}  \left(1+\frac{AQkc_{S}\eta}{B}\right)\right.\\ \nonumber &&\left.+G_2 \left\{\frac{\Gamma(1-B)}{\Gamma(1+A-B)}
                    \left(1+\frac{AQ kc_{S}\eta}{B}\right)+\frac{(Qkc_{S}\eta)^{1-B}\Gamma(B-1)}{\Gamma(A)}\right\}\right]\\  \nonumber\\&=&\left\{\begin{array}{ll}
                    \displaystyle (-\eta)^{\frac{3}{2}}e^{-i\sqrt{c^{2}_{S}k^2+C}\eta} \left(2 i\sqrt{c^{2}_{S}k^2+C}\eta\right)^{i \sqrt{(\gamma+\delta) -\frac{9}{4}}+\frac{1}{2}}\\ \displaystyle 
                    \left[G_{1}  \left(1+\frac{\left(2 i\sqrt{c^{2}_{S}k^2+C}\eta\right)\left(-\frac{d}{2i \sqrt{c^{2}_{S}k^2+C}}+i \sqrt{(\gamma+\delta) -\frac{9}{4}}+\frac{1}{2}\right)}{
                    \left(2i \sqrt{(\gamma+\delta) -\frac{9}{4}}+1\right)}\right)\right.\\ \nonumber \left.\displaystyle 
                    +G_2 \left\{\frac{\Gamma\left(-2i \sqrt{(\gamma+\delta) -\frac{9}{4}}\right)}{\Gamma\left(-\frac{d}{2i \sqrt{c^{2}_{S}k^2+C}}
                   -i \sqrt{(\gamma+\delta) -\frac{9}{4}}+\frac{1}{2}\right)}\right.\right.\\ \left. \left. \displaystyle 
                    ~~~~~~\times \left(1+\frac{\left(2 i\sqrt{c^{2}_{S}k^2+C}\eta\right)\left(-\frac{d}{2i \sqrt{c^{2}_{S}k^2+C}}+i \sqrt{(\gamma+\delta) -\frac{9}{4}}+\frac{1}{2}\right)}{
                    \left(2i \sqrt{(\gamma+\delta) -\frac{9}{4}}+1\right)}\right)\right. \right. \\ \left. \left. \displaystyle +\frac{\Gamma\left(2i \sqrt{(\gamma+\delta) 
                    -\frac{9}{4}}\right)}{\Gamma\left(-\frac{d}{2i \sqrt{c^{2}_{S}k^2+C}}+i \sqrt{(\gamma+\delta) -\frac{9}{4}}+\frac{1}{2}\right)}\right.\right. \\ \left. \left. \displaystyle 
                    ~~~~~~~~~~~~~~~~~~~~~~~~~\times \left(2 i\sqrt{c^{2}_{S}k^2+C}\eta\right)^{-2i \sqrt{(\gamma+\delta) -\frac{9}{4}}}\right\}\right]~ &
 \mbox{\small {\bf for ~dS}} \nonumber \\ \\ \nonumber\\
	\displaystyle (-\eta)^{\frac{3}{2}}e^{-i\sqrt{c^{2}_{S}k^2+C}\eta} \left(2 i\sqrt{c^{2}_{S}k^2+C}\eta\right)^{i \sqrt{(\gamma+\delta) -\nu^2}+\frac{1}{2}}\\ \displaystyle 
                    \left[G_{1}  \left(1+\frac{\left(2 i\sqrt{c^{2}_{S}k^2+C}\eta\right)\left(-\frac{d}{2i \sqrt{c^{2}_{S}k^2+C}}+i \sqrt{(\gamma+\delta) -\nu^2}+\frac{1}{2}\right)}{
                    \left(2i \sqrt{(\gamma+\delta) -\nu^2}+1\right)}\right)\right.\\ \nonumber \left.\displaystyle 
                    +G_2 \left\{\frac{\Gamma\left(-2i \sqrt{(\gamma+\delta) -\nu^2}\right)}{\Gamma\left(-\frac{d}{2i \sqrt{c^{2}_{S}k^2+C}}
                   -i \sqrt{(\gamma+\delta) -\nu^2}+\frac{1}{2}\right)}\right.\right.\\ \left. \left. \displaystyle 
                    ~~~~~~\times \left(1+\frac{\left(2 i\sqrt{c^{2}_{S}k^2+C}\eta\right)\left(-\frac{d}{2i \sqrt{c^{2}_{S}k^2+C}}+i \sqrt{(\gamma+\delta) -\nu^2}+\frac{1}{2}\right)}{
                    \left(2i \sqrt{(\gamma+\delta) -\nu^2}+1\right)}\right)\right. \right. \\ \left. \left. \displaystyle +\frac{\Gamma\left(2i \sqrt{(\gamma+\delta) 
                    -\nu^2}\right)}{\Gamma\left(-\frac{d}{2i \sqrt{c^{2}_{S}k^2+C}}+i \sqrt{(\gamma+\delta) -\nu^2}+\frac{1}{2}\right)}\right.\right. \\ \left. \left. \displaystyle 
                    ~~~~~~~~~~~~~~~~~~~~~~~~~\times \left(2 i\sqrt{c^{2}_{S}k^2+C}\eta\right)^{-2i \sqrt{(\gamma+\delta) -\nu^2}}\right\}\right]~ & \mbox{\small {\bf for~ qdS}}.
          \end{array}
\right.
\eea
\bea
\label{yu2}
 \displaystyle h_k (\eta) &\stackrel{|kc_{S}\eta|\approx 1-\Delta(\rightarrow 0)}{=}&(-\eta)^{\frac{3}{2}}e^{-\frac{P\eta}{2}} (P\eta)^{A+\frac{d}{P}}
                    \left[G_{1}  \left(1-\frac{AQ(1+\Delta)}{B}\right)\right.\\ \nonumber &&\left.+G_2 \left\{f_{1}(A,B)
                    \left(1-\frac{AQ(1+\Delta)}{B}\right)+f_{2}(A,B)(-Q(1+\Delta))^{1-B}\right\}\right]\\  \nonumber\\&=&\footnotesize\left\{\begin{array}{ll}
                    \displaystyle (-\eta)^{\frac{3}{2}}e^{-i\sqrt{c^{2}_{S}k^2+C}\eta} \left(2 i\sqrt{c^{2}_{S}k^2+C}\eta\right)^{i \sqrt{(\gamma+\delta) -\frac{9}{4}}+\frac{1}{2}}\\ \displaystyle 
                    \left[G_{1}  \left(1-\frac{\left(\frac{2 i}{kc_{S}}\sqrt{c^{2}_{S}k^2+C}(1+\Delta)\right)\left(-\frac{d}{2i \sqrt{c^{2}_{S}k^2+C}}+i \sqrt{(\gamma+\delta) -\frac{9}{4}}+\frac{1}{2}\right)}{
                    \left(2i \sqrt{(\gamma+\delta) -\frac{9}{4}}+1\right)}\right)\right.\\ \nonumber \left.\displaystyle 
                    +G_2 \left\{\left[\frac{\left(-\frac{d}{2i \sqrt{c^{2}_{S}k^2+C}}
                   -i \sqrt{(\gamma+\delta) -\frac{9}{4}}+\frac{1}{2}\right)}{\left(-2i \sqrt{(\gamma+\delta) -\frac{9}{4}}\right)}-\gamma+\cdots\right]\right.\right.\\ \left. \left. \displaystyle 
                    ~\times \left(1-\frac{\left(\frac{2 i}{kc_{S}}\sqrt{c^{2}_{S}k^2+C}(1+\Delta)\right)\left(-\frac{d}{2i \sqrt{c^{2}_{S}k^2+C}}+i \sqrt{(\gamma+\delta) -\frac{9}{4}}+\frac{1}{2}\right)}{
                    \left(2i \sqrt{(\gamma+\delta) -\frac{9}{4}}+1\right)}\right)\right. \right. \\ \left. \left. \displaystyle +\left[\frac{\left(-\frac{d}{2i \sqrt{c^{2}_{S}k^2+C}}
                   +i \sqrt{(\gamma+\delta) -\frac{9}{4}}+\frac{1}{2}\right)}{\left(2i \sqrt{(\gamma+\delta) -\frac{9}{4}}\right)}-\gamma+\cdots\right]\right.\right. \\ \left. \left. \displaystyle 
                    ~~~~~~~~~~\times \left(2 i\sqrt{c^{2}_{S}k^2+C}\eta\right)^{-2i \sqrt{(\gamma+\delta) -\frac{9}{4}}}\right\}\right] ~~~~~~~~~~~~~~
 \mbox{\small {\bf for ~dS}} \nonumber \\  \nonumber\\ \nonumber\\
	\displaystyle (-\eta)^{\frac{3}{2}}e^{-i\sqrt{c^{2}_{S}k^2+C}\eta} \left(2 i\sqrt{c^{2}_{S}k^2+C}\eta\right)^{i \sqrt{(\gamma+\delta) -\nu^2}+\frac{1}{2}}\\ \displaystyle 
                    \left[G_{1}  \left(1-\frac{\left(\frac{2 i}{kc_{S}}\sqrt{c^{2}_{S}k^2+C}(1+\Delta)\right)\left(-\frac{d}{2i \sqrt{c^{2}_{S}k^2+C}}+i \sqrt{(\gamma+\delta) -\nu^2}+\frac{1}{2}\right)}{
                    \left(2i \sqrt{(\gamma+\delta) -\nu^2}+1\right)}\right)\right.\\ \nonumber \left.\displaystyle 
                    +G_2 \left\{\left[\frac{\left(-\frac{d}{2i \sqrt{c^{2}_{S}k^2+C}}
                   -i \sqrt{(\gamma+\delta) -\nu^2}+\frac{1}{2}\right)}{\left(-2i \sqrt{(\gamma+\delta) -\nu^2}\right)}-\gamma+\cdots\right]\right.\right.\\ \left. \left. \displaystyle 
                    ~\times \left(1-\frac{\left(\frac{2 i}{kc_{S}}\sqrt{c^{2}_{S}k^2+C}(1+\Delta)\right)\left(-\frac{d}{2i \sqrt{c^{2}_{S}k^2+C}}+i \sqrt{(\gamma+\delta) -\nu^2}+\frac{1}{2}\right)}{
                    \left(2i \sqrt{(\gamma+\delta) -\nu^2}+1\right)}\right)\right. \right. \\ \left. \left. \displaystyle +\left[\frac{\left(-\frac{d}{2i \sqrt{c^{2}_{S}k^2+C}}
                   +i \sqrt{(\gamma+\delta) -\nu^2}+\frac{1}{2}\right)}{\left(2i \sqrt{(\gamma+\delta) -\nu^2}\right)}-\gamma+\cdots\right]\right.\right. \\ \left. \left. \displaystyle 
                    ~~~~~~~~~~\times \left(2 i\sqrt{c^{2}_{S}k^2+C}\eta\right)^{-2i \sqrt{(\gamma+\delta) -\nu^2}}\right\}\right]~ ~~~~~~~~~~~~~ \mbox{\small {\bf for~ qdS}}.
          \end{array}
\right.
\eea
Next we assume that the WKB approximation is approximately valid for all times
for the solution for the mode function $h_{k}$. In the standard WKB approximation the total solution can be recast in the following form:
\bea\label{df3}
\large h_k (\eta)&=& \left[D_{1}u_{k}(\eta) + D_{2} \bar{u}_{k}(\eta)\right],\eea
where $D_{1}$ and and $D_{2}$ are two arbitrary integration constant, which depend on the 
choice of the initial condition during WKB approximation at early and late time scale.
In the present context $u_{k}(\eta)$ and $\bar{u}_{k}(\eta)$ are defined as:
\be\begin{array}{lll}\label{solp}
 \displaystyle   u_{k}(\eta) =\footnotesize\left\{\begin{array}{ll}
                    \displaystyle   \frac{\exp\left[i\int^{\eta} d\eta^{\prime} \sqrt{c^{2}_{S}k^2 + C + \frac{\left(\gamma+\delta-2\right)}{\eta^{'2}} - \frac{d}{\eta^{'}}}\right]}{\sqrt{2\sqrt{c^{2}_{S}k^2 + C + \frac{\left(\gamma+\delta-2\right)}{\eta^2} - \frac{d}{\eta}}}}
&
 \mbox{\small {\bf for ~dS}}  \\ 
	\displaystyle  \frac{\exp\left[i\int^{\eta} d\eta^{\prime} \sqrt{c^{2}_{S}k^2 + C + \frac{\left(\gamma+\delta-\left[\nu^2-\frac{1}{4}\right]\right)}{\eta^{'2}} - \frac{d}{\eta^{'}}}\right]}{\sqrt{2\sqrt{c^{2}_{S}k^2 + C + \frac{\left(\gamma+\delta-\left[\nu^2-\frac{1}{4}\right]\right)}{\eta^2} - \frac{d}{\eta}}}}
&
 \mbox{\small {\bf for ~qdS}}.
 \end{array}
\right.
\end{array}\ee
\be\begin{array}{lll}\label{sola}
 \displaystyle  \bar{u}_{k}(\eta) =\footnotesize\left\{\begin{array}{ll}
                    \displaystyle   \frac{\exp\left[-i\int^{\eta} d\eta^{\prime} \sqrt{c^{2}_{S}k^2 + C + \frac{\left(\gamma+\delta-2\right)}{\eta^{'2}} - \frac{d}{\eta^{'}}}\right] }{\sqrt{2\sqrt{c^{2}_{S}k^2 + C + \frac{\left(\gamma+\delta-2\right)}{\eta^2} - \frac{d}{\eta}}}}
&
 \mbox{\small {\bf for ~dS}}  \\ 
	\displaystyle \frac{\exp\left[-i\int^{\eta} d\eta^{\prime}
	\sqrt{c^{2}_{S}k^2 + C + \frac{\left(\gamma+\delta-\left[\nu^2-\frac{1}{4}\right]\right)}{\eta^{'2}} - \frac{d}{\eta^{'}}}\right]}{\sqrt{2\sqrt{c^{2}_{S}k^2 + C + \frac{\left(\gamma+\delta-\left[\nu^2-\frac{1}{4}\right]\right)}{\eta^2} - \frac{d}{\eta}}}}
 & \mbox{\small {\bf for~ qdS}}.
          \end{array}
\right.
\end{array}\ee
where 
\bea&&\underline{\bf For~dS:}\nonumber\\
\displaystyle &&\int^{\eta} d\eta^{\prime} 
\displaystyle\sqrt{c^{2}_{S}k^2 + C + \frac{\left(\gamma+\delta-2\right)}{\eta^{'2}} - \frac{d}{\eta^{\prime}}}\nonumber\\
 ~~~~~~&=&\frac{1}{2 \sqrt{C+c^{2}_{S}k^2}}\left\{2 \sqrt{C+c^{2}_{S}k^2} \sqrt{\gamma +C \eta ^2-d \eta +\delta +\eta ^2 c^{2}_{S}k^2-2}
\right.\nonumber\\ &&\left.\displaystyle~~~~~~~~~~-d \ln \left[2 \sqrt{C+c^{2}_{S}k^2} 
\sqrt{\gamma +C \eta ^2-d \eta +\delta +\eta ^2 c^{2}_{S}k^2-2}
+2 C \eta -d+2 \eta  c^{2}_{S}k^2\right]\right.\nonumber\\&& \left.-2 \sqrt{\gamma+\delta -2} \sqrt{C+c^{2}_{S}k^2}
\ln \left[\frac{2 \sqrt{\gamma +C \eta ^2-d \eta +\delta +\eta ^2 c^{2}_{S}k^2-2}
+\frac{i (2 \gamma -d \eta +2 \delta -4)}{\sqrt{-\gamma -\delta +2}}}{\eta
(\gamma +\delta -2)}\right]\right\},~~~~~~~~~~\eea
\bea
&&\underline{\bf For~qdS:}\nonumber\\
 &&\int^{\eta} d\eta^{\prime} 
\sqrt{c^{2}_{S}k^2 + C + \frac{\left(\gamma+\delta-\left[\nu^2-\frac{1}{4}\right]\right)}{\eta^{'2}} - \frac{d}{\eta^{\prime}}}\nonumber\\
 ~~~~~~&=&\displaystyle\frac{1}{2 \sqrt{C+c^{2}_{S}k^2}}\left\{2 \sqrt{C+k^2} \sqrt{\gamma +C \eta ^2-d \eta +\delta +\eta ^2 c^{2}_{S}k^2-\left[\nu^2-\frac{1}{4}\right]}
\right.\nonumber\\ &&\left.\displaystyle-d \ln \left[2 \sqrt{C+k^2} 
\sqrt{\gamma +C \eta ^2-d \eta +\delta +\eta ^2 c^{2}_{S}k^2-\left[\nu^2-\frac{1}{4}\right]}
+2 C \eta -d+2 \eta  c^{2}_{S}k^2\right]\right.\nonumber\\&& \left.\small-2 \sqrt{\gamma+\delta -\left[\nu^2-\frac{1}{4}\right]}\right.\nonumber\\&&\left. \times\sqrt{C+c^{2}_{S}k^2}
\ln \left[\frac{2 \sqrt{\gamma +C \eta ^2-d \eta +\delta +\eta ^2 c^{2}_{S}k^2-\left[\nu^2-\frac{1}{4}\right]}
+\frac{i (2 \gamma -d \eta +2 \delta -2\left[\nu^2-\frac{1}{4}\right])}{\sqrt{-\gamma -\delta +\left[\nu^2-\frac{1}{4}\right]}}}{\eta
(\gamma +\delta -\left[\nu^2-\frac{1}{4}\right])}\right]\right\},~~~~~~~~
\eea
where we have written the total solution for the mode $h_k$ in terms of 
two linearly independent solutions. Here it is important to note that the both of the 
solutions are hermitian conjugate of each other. If in the present context the exact solution of the mode $h_k$ 
is expanded with respect to these two linearly independent solutions then particle creation is absent in our EFT 
setup. In the present context correctness of 
WKB approximation is guarantee at very early and very late time scales. In this discussion $u_{k}(\eta)$
is valid at very early time scale and $\bar{u}_{k}(\eta)$ perfectly works in the late time scale.

\subsubsection{\bf Profile B:~$ m= \frac{m_0}{\sqrt{2}}\sqrt{\left[1
 -\tanh\left(\frac{\rho}{H}\ln(-H\eta)\right)\right]}$}
 Equation of motion for the massive field is:
\bea
h''_k + \left(c^{2}_{S}k^2 +\left\{\frac{m^2_0}{2H^2}~\left[1-{\rm tanh}\left(\frac{\rho}{H}\ln(-H\eta)\right)\right]-2\right\}\frac{1}{\eta^2} \right) h_k &=& 0~~~~~~~{\bf for~ dS}\\ 
h''_k + \left(c^{2}_{S}k^2 + \left\{\frac{m^2_0}{2H^2}~\left[1-{\rm tanh}\left(\frac{\rho}{H}\ln(-H\eta)\right)\right]-\left[\nu^2-\frac{1}{4}\right]\right\}\frac{1}{\eta^2}\right) h_k &=& 0~~~~~~~{\bf for~ qdS}.
\eea
Which is not at all possible to solve exactly. But if we assume that:
\be \rho << \frac{H}{\ln(-H\eta)},\ee then in that limiting situation one can recast the equation of motions as:
\bea
h''_k + \left(c^{2}_{S}k^2 +\left\{\frac{m^2_0}{H^2}-2\right\}\frac{1}{\eta^2} \right) h_k &=& 0~~~~~~~{\bf for~ dS}\\ 
h''_k + \left(c^{2}_{S}k^2 + \left\{\frac{m^2_0}{H^2}-\left[\nu^2-\frac{1}{4}\right]\right\}\frac{1}{\eta^2}\right) h_k &=& 0~~~~~~~{\bf for~ qdS}.
\eea
which exactly similar to equation of motions obtained for the cases where the conformal time dependent mass function varying slowly. 
The solution for the mode function for de Sitter and quasi de Sitter space in this limiting cases can be expressed as: 
\be\begin{array}{lll}\label{yu2}
 \displaystyle h_k (\eta) =\footnotesize\left\{\begin{array}{ll}
                    \displaystyle   \sqrt{-\eta}\left[C_1  H^{(1)}_{\sqrt{\frac{9}{4}-\frac{m^2_0}{H^2}}} \left(-kc_{S}\eta\right) 
+ C_2 H^{(2)}_{\sqrt{\frac{9}{4}-\frac{m^2_0}{H^2}}} \left(-kc_{S}\eta\right)\right]~~~~ &
 \mbox{\small {\bf for ~dS}}  \\ 
	\displaystyle \sqrt{-\eta}\left[C_1  H^{(1)}_{\sqrt{\nu^2-\frac{m^2_0}{H^2}}} \left(-kc_{S}\eta\right) 
+ C_2 H^{(2)}_{\sqrt{\nu^2-\frac{m^2_0}{H^2}}} \left(-kc_{S}\eta\right)\right]~~~~ & \mbox{\small {\bf for~ qdS}}.
          \end{array}
\right.
\end{array}\ee
where $C_{1}$ and and $C_{2}$ are two arbitrary integration constant, which depend on the 
choice of the initial condition. From this solution one can study $m_{0}\approx H$, $m_{0}>>H$ and $m_{0}<<H$ physical situations as studied before.

To solve this we assume that the WKB approximation is approximately valid for all times
for the solution for the mode function $h_{k}$. In the standard WKB approximation the total solution can be recast in the following form:
\bea\label{df3}
\large h_k (\eta)&=& \left[D_{1}u_{k}(\eta) + D_{2} \bar{u}_{k}(\eta)\right],\eea
where $D_{1}$ and and $D_{2}$ are two arbitrary integration constant, which depend on the 
choice of the initial condition during WKB approximation at early and late time scale.
In the present context $u_{k}(\eta)$ and $\bar{u}_{k}(\eta)$ are defined as:
\be\begin{array}{lll}\label{solpzxzxzxzx}
 \displaystyle   u_{k}(\eta) =\left\{\begin{array}{ll}
                    \displaystyle   \frac{\exp\left[i\int^{\eta} d\eta^{\prime} 
                    \sqrt{c^{2}_{S}k^2 +\left\{\frac{m^2_0}{2H^2}~\left[1-{\rm tanh}\left(\frac{\rho}{H}\ln(-H\eta^{'})\right)\right]-2\right\}\frac{1}{\eta^{'2}}  }\right]}
                    {\sqrt{2\sqrt{c^{2}_{S}k^2 +\left\{\frac{m^2_0}{2H^2}~\left[1-{\rm tanh}\left(\frac{\rho}{H}\ln(-H\eta)\right)\right]-2\right\}\frac{1}{\eta^2}  }}}
&
 \mbox{\small {\bf for ~dS}}  \\ 
	\displaystyle  \frac{\exp\left[i\int^{\eta} d\eta^{\prime} \sqrt{c^{2}_{S}k^2 + \left\{\frac{m^2_0}{2H^2}~\left[1-{\rm tanh}\left(\frac{\rho}{H}\ln(-H\eta^{'})\right)\right]-\left[\nu^2-\frac{1}{4}\right]\right\}\frac{1}{\eta^{'2}}}\right]}
	{\sqrt{2\sqrt{c^{2}_{S}k^2 + \left\{\frac{m^2_0}{2H^2}~\left[1-{\rm tanh}\left(\frac{\rho}{H}\ln(-H\eta)\right)\right]-\left[\nu^2-\frac{1}{4}\right]\right\}\frac{1}{\eta^2}}}}
&
 \mbox{\small {\bf for ~qdS}}.
 \end{array}
\right.
\end{array}\ee
\be\begin{array}{lll}\label{sola}
 \displaystyle   \bar{u}_{k}(\eta) =\left\{\begin{array}{ll}
                    \displaystyle   \frac{\exp\left[-i\int^{\eta} d\eta^{\prime} 
                    \sqrt{c^{2}_{S}k^2 +\left\{\frac{m^2_0}{2H^2}~\left[1-{\rm tanh}\left(\frac{\rho}{H}\ln(-H\eta^{'})\right)\right]-2\right\}\frac{1}{\eta^{'2}}  }\right]}
                    {\sqrt{2\sqrt{c^{2}_{S}k^2 +\left\{\frac{m^2_0}{2H^2}~\left[1-{\rm tanh}\left(\frac{\rho}{H}\ln(-H\eta)\right)\right]-2\right\}\frac{1}{\eta^2}  }}}
&
 \mbox{\small {\bf for ~dS}}  \\ 
	\displaystyle  \frac{\exp\left[-i\int^{\eta} d\eta^{\prime} \sqrt{c^{2}_{S}k^2 + \left\{\frac{m^2_0}{2H^2}~\left[1-{\rm tanh}\left(\frac{\rho}{H}\ln(-H\eta^{'})\right)\right]-\left[\nu^2-\frac{1}{4}\right]\right\}\frac{1}{\eta^{'2}}}\right]}
	{\sqrt{2\sqrt{c^{2}_{S}k^2 + \left\{\frac{m^2_0}{2H^2}~\left[1-{\rm tanh}\left(\frac{\rho}{H}\ln(-H\eta)\right)\right]-\left[\nu^2-\frac{1}{4}\right]\right\}\frac{1}{\eta^2}}}}
&
 \mbox{\small {\bf for ~qdS}}.
          \end{array}
\right.
\end{array}\ee
Again if we assume $\rho << \frac{H}{\ln(-H\eta)}$ then these WKB solutions can be recast as:
\be\begin{array}{lll}\label{solp}
 \tiny u_{k}(\eta) =\footnotesize\left\{\begin{array}{ll}
                    \displaystyle  \displaystyle\frac{1}{\sqrt{2\sqrt{c^{2}_{S}k^2 + \frac{\frac{m^2_0}{2H^2}-2}{\eta^2}}}} 
\exp\left[i\left(\eta\sqrt{c^{2}_{S}k^2 + \frac{\frac{m^2_0}{2H^2}-2}{\eta^2}}
+\sqrt{\frac{m^2_0}{2H^2}-2}\ln\left[\frac{2}{\eta\sqrt{\frac{m^2_0}{2H^2}-2}}+\frac{2\sqrt{c^{2}_{S}k^2 + \frac{\frac{m^2_0}{2H^2}-2}{\eta^{2}}}}{\left(\frac{m^2_0}{2H^2}-2\right)}\right]
\right)\right]~~~~ &
 \mbox{\small {\bf for ~dS}}  \\ 
	\displaystyle\frac{1}{\sqrt{2\sqrt{c^{2}_{S}k^2 + 
\frac{\left[\frac{m^2_0}{2H^2}-\left(\nu^2-\frac{1}{4}\right)\right]}{\eta^2}}}} 
\exp\left[i\left(\eta\sqrt{c^{2}_{S}k^2 + 
\frac{\left[\frac{m^2_0}{2H^2}-\left(\nu^2-\frac{1}{4}\right)\right]}{\eta^2}}
\right.\right.\\ \left.\left. \displaystyle +\sqrt{\frac{m^2_0}{2H^2}-\left(\nu^2-\frac{1}{4}\right)}\ln\left[\frac{2}{\eta\sqrt{\frac{m^2_0}{2H^2}-\left(\nu^2-\frac{1}{4}\right)}}
+\frac{2\sqrt{c^{2}_{S}k^2 + \frac{\frac{m^2_0}{2H^2}-\left(\nu^2-\frac{1}{4}\right)}{\eta^{2}}}}{\left(\frac{m^2_0}{2H^2}-\left(\nu^2-\frac{1}{4}\right)\right)}\right]
\right)\right]~~~~ & \mbox{\small {\bf for~ qdS}}.
          \end{array}
\right.
\end{array}\ee
\be\begin{array}{lll}\label{sola}
 \tiny \bar{u}_{k}(\eta) =\footnotesize\left\{\begin{array}{ll}
                    \displaystyle  \displaystyle\frac{1}{\sqrt{2\sqrt{c^{2}_{S}k^2 + \frac{\frac{m^2_0}{2H^2}-2}{\eta^2}}}} 
\exp\left[-i\left(\eta\sqrt{c^{2}_{S}k^2 + \frac{\frac{m^2_0}{2H^2}-2}{\eta^2}}
+\sqrt{\frac{m^2_0}{2H^2}-2}\ln\left[\frac{2}{\eta\sqrt{\frac{m^2_0}{2H^2}-2}}+\frac{2\sqrt{c^{2}_{S}k^2 + \frac{\frac{m^2_0}{2H^2}-2}{\eta^{2}}}}{\left(\frac{m^2_0}{2H^2}-2\right)}\right]
\right)\right]~~~~ &
 \mbox{\small {\bf for ~dS}}  \\
	\displaystyle \displaystyle\frac{1}{\sqrt{2\sqrt{c^{2}_{S}k^2 + 
\frac{\left[\frac{m^2_0}{2H^2}-\left(\nu^2-\frac{1}{4}\right)\right]}{\eta^2}}}} 
\exp\left[-i\left(\eta\sqrt{c^{2}_{S}k^2 + 
\frac{\left[\frac{m^2_0}{2H^2}-\left(\nu^2-\frac{1}{4}\right)\right]}{\eta^2}}
\right.\right.\\ \left.\left. \displaystyle +\sqrt{\frac{m^2_0}{2H^2}-\left(\nu^2-\frac{1}{4}\right)}\ln\left[\frac{2}{\eta\sqrt{\frac{m^2_0}{2H^2}-\left(\nu^2-\frac{1}{4}\right)}}
+\frac{2\sqrt{c^{2}_{S}k^2 + \frac{\frac{m^2_0}{2H^2}-\left(\nu^2-\frac{1}{4}\right)}{\eta^{2}}}}{\left(\frac{m^2_0}{2H^2}-\left(\nu^2-\frac{1}{4}\right)\right)}\right]
\right)\right]~~~~ & \mbox{\small {\bf for~ qdS}}.
          \end{array}
\right.
\end{array}\ee
 
 \subsubsection{\bf Profile C:~$m= m_0~{\rm sech}\left(\frac{\rho}{H}\ln(-H\eta)\right) $}
Equation of motion for the massive field is:
\bea
h''_k + \left(c^{2}_{S}k^2 +\left\{\frac{m^2_0}{H^2}~{\rm sech}^2\left(\frac{\rho}{H}\ln(-H\eta)\right)-2\right\}\frac{1}{\eta^2} \right) h_k &=& 0~~~~~~~{\bf for~ dS}\\ 
h''_k + \left(c^{2}_{S}k^2 + \left\{\frac{m^2_0}{H^2}~{\rm sech}^2\left(\frac{\rho}{H}\ln(-H\eta)\right)-\left[\nu^2-\frac{1}{4}\right]\right\}\frac{1}{\eta^2}\right) h_k &=& 0~~~~~~~{\bf for~ qdS}.
\eea
Which is not at all possible to solve exactly. But if we assume that:
\be \rho << \frac{H}{\ln(-H\eta)},\ee then in that limiting situation the solutions are exactly same as appearing for Profile B. Only here we have to change $m^2_0/2H^2\rightarrow m^2_0/H^2$.

To solve this we assume that the WKB approximation is approximately valid for all times
for the solution for the mode function $h_{k}$. In the standard WKB approximation the total solution can be recast in the following form:
\bea\label{df3}
\large h_k (\eta)&=& \left[D_{1}u_{k}(\eta) + D_{2} \bar{u}_{k}(\eta)\right],\eea
where $D_{1}$ and and $D_{2}$ are two arbitrary integration constant, which depend on the 
choice of the initial condition during WKB approximation at early and late time scale.
In the present context $u_{k}(\eta)$ and $\bar{u}_{k}(\eta)$ are defined as:
\be\begin{array}{lll}\label{solpzxzxzxzx}
 \displaystyle   u_{k}(\eta) =\left\{\begin{array}{ll}
                    \displaystyle   \frac{\exp\left[i\int^{\eta} d\eta^{\prime} 
                    \sqrt{c^{2}_{S}k^2 +\left\{\frac{m^2_0}{H^2}~{\rm sech}^2\left(\frac{\rho}{H}\ln(-H\eta^{'})\right)-2\right\}\frac{1}{\eta^{'2}} }\right]}
                    {\sqrt{2\sqrt{c^{2}_{S}k^2 +\left\{\frac{m^2_0}{H^2}~{\rm sech}^2\left(\frac{\rho}{H}\ln(-H\eta)\right)-2\right\}\frac{1}{\eta^2} }}}
&
 \mbox{\small {\bf for ~dS}}  \\ \\
	\displaystyle  \frac{\exp\left[i\int^{\eta} d\eta^{\prime} \sqrt{c^{2}_{S}k^2 + \left\{\frac{m^2_0}{H^2}~{\rm sech}^2\left(\frac{\rho}{H}\ln(-H\eta^{'})\right)-\left[\nu^2-\frac{1}{4}\right]\right\}\frac{1}{\eta^{'2}}}\right]}
	{\sqrt{2\sqrt{c^{2}_{S}k^2 + \left\{\frac{m^2_0}{H^2}~{\rm sech}^2\left(\frac{\rho}{H}\ln(-H\eta)\right)-\left[\nu^2-\frac{1}{4}\right]\right\}\frac{1}{\eta^2}}}}
&
 \mbox{\small {\bf for ~qdS}}.
 \end{array}
\right.
\end{array}\ee
\be\begin{array}{lll}\label{sola}
 \displaystyle   \bar{u}_{k}(\eta) =\left\{\begin{array}{ll}
                    \displaystyle   \frac{\exp\left[-i\int^{\eta} d\eta^{\prime} 
                    \sqrt{c^{2}_{S}k^2 +\left\{\frac{m^2_0}{H^2}~{\rm sech}^2\left(\frac{\rho}{H}\ln(-H\eta^{'})\right)-2\right\}\frac{1}{\eta^{'2}} }\right]}
                    {\sqrt{2\sqrt{c^{2}_{S}k^2 +\left\{\frac{m^2_0}{H^2}~{\rm sech}^2\left(\frac{\rho}{H}\ln(-H\eta)\right)-2\right\}\frac{1}{\eta^2} }}}
&
 \mbox{\small {\bf for ~dS}}  \\ \\
	\displaystyle  \frac{\exp\left[-i\int^{\eta} d\eta^{\prime} \sqrt{c^{2}_{S}k^2 + \left\{\frac{m^2_0}{H^2}~{\rm sech}^2\left(\frac{\rho}{H}\ln(-H\eta^{'})\right)-\left[\nu^2-\frac{1}{4}\right]\right\}\frac{1}{\eta^{'2}}}\right]}
	{\sqrt{2\sqrt{c^{2}_{S}k^2 + \left\{\frac{m^2_0}{H^2}~{\rm sech}^2\left(\frac{\rho}{H}\ln(-H\eta)\right)-\left[\nu^2-\frac{1}{4}\right]\right\}\frac{1}{\eta^2}}}}
&
 \mbox{\small {\bf for ~qdS}}.
          \end{array}
\right.
\end{array}\ee
Again if we assume $\rho << \frac{H}{\ln(-H\eta)}$ then these WKB solutions is exactly same as appearing for Profile B. Only here we have to change $m^2_0/2H^2\rightarrow m^2_0/H^2$.

 \subsection{Role of spin for heavy field}
 \label{sec6c}
 Let us consider the situation for a dynamical massive field with arbitrary spin ${\cal S}$. In this case we assume that the dynamics of all such arbitrary spin fields with spin ${\cal S}>2$ is similar with the graviton. For this case 
the classical time dependence of the high spin modes leads to a time dependent mass $m_{\cal S}\left(\eta\right)$ for the spin field. The equation of motion for the massive field with arbitrary spin ${\cal S}$ is given by:
\bea
h''_k + \left\{c^2_{S}k^2 + \left(\frac{m^2_{\cal S}}{H^2} - \left[\nu^2_{\cal S}-\frac{1}{4}\right]
\right) \frac{1}{\eta^2} \right\} h_k &=& 0.
\eea
where in the de Sitter and quasi de Sitter case the parameter $\nu_{\cal S}$ can be written as:
\be\begin{array}{lll}\label{yu2dfvqdfdfdf}\small
 \displaystyle \nu_{\cal S}=\left\{\begin{array}{ll}
                    \displaystyle \left({\cal S}-\frac{1}{2}\right) ~~~~ &
 \mbox{\small {\bf for ~dS}}  \\ \\
	\displaystyle \left({\cal S}-\frac{1}{2}\right)+\epsilon+\frac{\eta}{2}+\frac{s}{2} ~~~~ & \mbox{\small {\bf for~ qdS}}.
          \end{array}
\right.
\end{array}\ee

The most general solution of the mode function for de Sitter and quasi de Sitter case can be written as:
\be\begin{array}{lll}\label{yu2}
 \displaystyle h_k (\eta) =
	\displaystyle \sqrt{-\eta}\left[C_1  H^{(1)}_{\sqrt{\nu^2_{\cal S}-\frac{m^2_{\cal S}}{H^2}}} \left(-kc_{S}\eta\right) 
+ C_2 H^{(2)}_{\sqrt{\nu^2_{\cal S}-\frac{m^2_{\cal S}}{H^2}}} \left(-kc_{S}\eta\right)\right].
\end{array}\ee
Here $C_{1}$ and $C_{2}$ are the arbitrary integration constants and the numerical value depend on the choice of the 
initial condition or more precisely the vacuum. It is important to note that, solution for spin ${\cal S}=2$ exactly matches with our previously obtained results for massive scalar fields. This implies that 
tensor fluctuations for massive graviton field mimics the role of massive scalar field with spin ${\cal S}=0$ at the level of equation of motion if we identify:
\be\begin{array}{lll}
 \displaystyle \frac{m^2_{\cal S}}{H^2} =\left\{\begin{array}{ll}
                    \displaystyle    \frac{m^2}{H^2}-2~~~~ &
 \mbox{\small {\bf for ~dS}}  \\ \\
	\displaystyle  \frac{m^2}{H^2}-\left[\nu^2-\frac{1}{4}\right]~~~~ & \mbox{\small {\bf for~ qdS}}.
          \end{array}
\right.
\end{array}\ee
But here for arbitrary spin (mostly for high spin with spin 
${\cal S}>2$) we get a generic result which may be different from the previously mentioned massive scalar field as well as useful to study the effects of other high spin massive fields in the present context.

Here it is important to mention that the argument in the Hankel function for the solution of the $h_k$ takes the following 
values in different regime:
\be\begin{array}{lll}\label{yu2dfvq}\small
 \displaystyle \underline{\rm \bf For~dS:}~~~~~~~~~\sqrt{\nu^2_{\cal S}-\frac{m^2_{\cal S}}{H^2}} \approx\left\{\begin{array}{ll}
                    \displaystyle  \sqrt{\left({\cal S}-\frac{1}{2}\right)^2-1} ~~~~ &
 \mbox{\small {\bf for ~$m_{\cal S}\approx H$}}  \\ 
	\displaystyle \left({\cal S}-\frac{1}{2}\right) ~~~~ & \mbox{\small {\bf for ~$m_{\cal S}<< H$}}\\ 
	\displaystyle i\sqrt{\Upsilon^2_{\cal S}-\left({\cal S}-\frac{1}{2}\right)^2}~~~~ & \mbox{\small {\bf for ~$m_{\cal S}>> H$}}.
          \end{array}
\right.
\end{array}\ee
\be\begin{array}{lll}\label{yu2dfve}\small
 \displaystyle \underline{\rm \bf For~qdS:}~~~~~~~~~\sqrt{\nu^2_{\cal S}-\frac{m^2_{\cal S}}{H^2}} \approx\left\{\begin{array}{ll}
                    \displaystyle  \sqrt{\left(\left({\cal S}-\frac{1}{2}\right)+\epsilon+\frac{\eta}{2}+\frac{s}{2}\right)^2-1}~~~~ &
 \mbox{\small {\bf for ~$m_{\cal S}\approx H$}}  \\ 
	\displaystyle \left({\cal S}-\frac{1}{2}\right)+\epsilon+\frac{\eta}{2}+\frac{s}{2}~~~~ & \mbox{\small {\bf for ~$m_{\cal S}<< H$}}\\ 
	\displaystyle i\sqrt{\Upsilon^2_{\cal S}-\left(\left({\cal S}-\frac{1}{2}\right)+\epsilon+\frac{\eta}{2}+\frac{s}{2}\right)^2}~~~~ & \mbox{\small {\bf for ~$m_{\cal S}>> H$}}.
          \end{array}
\right.
\end{array}\ee
Here we set $m_{\cal S}=\Upsilon_{\cal S} H$, where the parameter $\Upsilon_{\cal S}>>1$ for $m_{\cal S}>>H$ case.
 
After taking $kc_{S}\eta\rightarrow -\infty$, $kc_{S}\eta\rightarrow 0$ and $|kc_{S}\eta|\approx 1-\Delta(\rightarrow 0)$ limit the most general 
solution as stated in Eq~(\ref{yu2}) can be recast as:
\bea\label{yu2}
 \displaystyle h_k (\eta) &\stackrel{|kc_{S}\eta|\rightarrow-\infty}{=}&
  \left\{\begin{array}{ll}
                    \displaystyle   \sqrt{\frac{2}{\pi kc_{S}}}\left[C_1  e^{ -ikc_{S}\eta}
e^{-\frac{i\pi}{2}\left(\sqrt{\left({\cal S}-\frac{1}{2}\right)^2-\frac{m^2_{\cal S}}{H^2}}+\frac{1}{2}\right)} 
+ C_2 e^{ ikc_{S}\eta}
e^{\frac{i\pi}{2}\left(\sqrt{\left({\cal S}-\frac{1}{2}\right)^2-\frac{m^2_{\cal S}}{H^2}}+\frac{1}{2}\right)}\right]~~~~ &
 \mbox{\small {\bf for ~dS}}  \nonumber\\ 
	\displaystyle \sqrt{\frac{2}{\pi kc_{S}}}\left[C_1  e^{ -ikc_{S}\eta}
e^{-\frac{i\pi}{2}\left(\sqrt{\left(\left({\cal S}-\frac{1}{2}\right)+\epsilon+\frac{\eta}{2}+\frac{s}{2}\right)^2-\frac{m^2_{\cal S}}{H^2}}+\frac{1}{2}\right)} 
\right.\\ \left.\displaystyle~~~~~~~~~~~~~~~~~~~~~~~~~ + C_2 e^{ ikc_{S}\eta}
e^{\frac{i\pi}{2}\left(\sqrt{\left(\left({\cal S}-\frac{1}{2}\right)+\epsilon+\frac{\eta}{2}+\frac{s}{2}\right)^2-\frac{m^2_{\cal S}}{H^2}}+\frac{1}{2}\right)}\right]~~~~ & \mbox{\small {\bf for~ qdS}}.
          \end{array}
\right.\\ &&\eea 
\bea
\label{yu2}
 \displaystyle h_k (\eta) &\stackrel{|kc_{S}\eta|\rightarrow 0}{=}&\footnotesize\left\{\begin{array}{ll}
                    \displaystyle  \frac{i\sqrt{-\eta}}{\pi}\Gamma\left(\sqrt{\left({\cal S}-\frac{1}{2}\right)^2-\frac{m^2_{\cal S}}{H^2}}\right)
                    \left(-\frac{kc_{S}\eta}{2}\right)^{-\sqrt{\left({\cal S}-\frac{1}{2}\right)^2-\frac{m^2_{\cal S}}{H^2}}}\left[C_1   
- C_2 \right]~ &
 \mbox{\small {\bf for ~dS}} \nonumber \\ 
	\displaystyle\frac{i\sqrt{-\eta}}{\pi}\Gamma\left(\sqrt{\left(\left({\cal S}-\frac{1}{2}\right)+\epsilon+\frac{\eta}{2}+\frac{s}{2}\right)^2-\frac{m^2_{\cal S}}{H^2}}\right)\left(-\frac{kc_{S}\eta}{2}\right)^{-\sqrt{\left(\left({\cal S}-\frac{1}{2}\right)+\epsilon+\frac{\eta}{2}+\frac{s}{2}\right)^2-\frac{m^2_{\cal S}}{H^2}}}\left[C_1   
- C_2 \right]~ & \mbox{\small {\bf for~ qdS}}.
          \end{array}
\right.\\ &&\eea 
\bea
\label{yu2}
 \displaystyle h_k (\eta) &\stackrel{|kc_{S}\eta|\approx 1-\Delta(\rightarrow 0)}{=}&\left\{\begin{array}{ll}
                    \displaystyle  \frac{i}{\pi}\sqrt{-\eta}\left[ \frac{1}{\left(\sqrt{\left({\cal S}-\frac{1}{2}\right)^2-\frac{m^2_{\cal S}}{H^2}}\right)}-\gamma+\frac{1}{2}
                   \left(\gamma^2+\frac{\pi^2}{6}\right)\left(\sqrt{\left({\cal S}-\frac{1}{2}\right)^2-\frac{m^2_{\cal S}}{H^2}}\right)\right.\\ \displaystyle \left.
                   \displaystyle~~-\frac{1}{6}\left(\gamma^3+\frac{\gamma \pi^2}{2}+2\zeta(3)\right)
                   \left(\sqrt{\left({\cal S}-\frac{1}{2}\right)^2-\frac{m^2_{\cal S}}{H^2}}\right)^2 +\cdots\right]\\ \displaystyle ~~~~~~~~~~~~\times
                   \left(\frac{1+\Delta}{2}\right)^{-\sqrt{\left({\cal S}-\frac{1}{2}\right)^2-\frac{m^2_{\cal S}}{H^2}}}\left[C_1   
- C_2 \right]~&
 \mbox{\small {\bf for ~dS}}\\ \\
	\displaystyle\frac{i}{\pi}\sqrt{-\eta}\left[ \frac{1}{ \left\{\sqrt{\left(\left({\cal S}-\frac{1}{2}\right)+\epsilon+\frac{\eta}{2}+\frac{s}{2}\right)^2-\frac{m^2_{\cal S}}{H^2}}\right\}}
  -\gamma  \displaystyle\right.\\ \left. \displaystyle +\frac{1}{2}\left(\gamma^2+\frac{\pi^2}{6}\right) \left\{\sqrt{\left(\left({\cal S}-\frac{1}{2}\right)+\epsilon+\frac{\eta}{2}+\frac{s}{2}\right)^2-\frac{m^2_{\cal S}}{H^2}}\right\}
  \right.\\ \displaystyle \left. \displaystyle~-\frac{1}{6}\left(\gamma^3+\frac{\gamma \pi^2}{2}+2\zeta(3)\right) \left\{\sqrt{\left(\left({\cal S}-\frac{1}{2}\right)+\epsilon+\frac{\eta}{2}+\frac{s}{2}\right)^2-\frac{m^2_{\cal S}}{H^2}}\right\}^2
  +\cdots\right]\\ \displaystyle ~~~~~~~~~~~~\times
  \left(\frac{1+\Delta}{2}\right)^{-\sqrt{\left(\left({\cal S}-\frac{1}{2}\right)+\epsilon+\frac{\eta}{2}+\frac{s}{2}\right)^2-\frac{m^2_{\cal S}}{H^2}}}\left[C_1   
- C_2 \right]~ & \mbox{\small {\bf for~ qdS}}.
          \end{array}
\right.
\eea
In the standard WKB approximation the total solution can be recast in the following form:
\bea\label{df3}
h_k (\eta)&=& \left[D_{1}u_{k}(\eta) + D_{2} \bar{u}_{k}(\eta)\right],\eea
where $D_{1}$ and and $D_{2}$ are two arbitrary integration constants, which depend on the 
choice of the initial condition during WKB approximation at early and late time scale.
In the present context $u_{k}(\eta)$ and $\bar{u}_{k}(\eta)$ are defined as:
\bea\label{solpzxzx}
 \displaystyle u_{k}(\eta) &=&
                    \displaystyle   \frac{1}{\sqrt{2p(\eta)}}
\exp\left[i\int^{\eta} d\eta^{\prime} p(\eta^{'})\right]\\
\label{solazxzx}
 \displaystyle\small \bar{u}_{k}(\eta) &=&
                    \displaystyle   \frac{1}{\sqrt{2p(\eta)}}
\exp\left[-i\int^{\eta} d\eta^{\prime} p(\eta^{'})\right]
\eea
where we have written the total solution for the mode $h_k$ in terms of 
two linearly independent solutions. Here in the most generalized situation
the new conformal time dependent factor $p(\eta)$ is defined as:
\be\begin{array}{lll}\label{soladfasx2}
 \displaystyle p(\eta) =\left\{\begin{array}{ll}
                    \displaystyle   \sqrt{\left\{c^{2}_{S}k^2 + \left(\frac{m^2_{\cal S}}{H^2} - \left\{\left({\cal S}-\frac{1}{2}\right)^2-\frac{1}{4}\right\}\right) \frac{1}{\eta^2} \right\}}~~~~ &
 \mbox{\small {\bf for ~dS}}  \\ \\
	\displaystyle  \sqrt{\left\{c^{2}_{S}k^2 + \left(\frac{m^2_{\cal S}}{H^2} - \left\{\left(\left({\cal S}-\frac{1}{2}\right)+\epsilon+\frac{\eta}{2}+\frac{s}{2}\right)^2-\frac{1}{4}\right\}
\right) \frac{1}{\eta^2} \right\}}~~~~ & \mbox{\small {\bf for~ qdS}}.
          \end{array}
\right.
\end{array}\ee
using which one can calculate the Bogoliubov coefficients and other components as we have computed for the other cases.

In this context one can compare the dynamical equations for scalar mode fluctuations with the well known 
Schr$\ddot{o}$dinger scattering problem in one spatial dimension then one can write down the following expression for the spin dependent potential and energy:
\bea
 \displaystyle V(t)&=&\left\{\begin{array}{ll}
                    \displaystyle -\frac{1}{2m}\left(\frac{m^2_{\cal S}}{H^2} - \left[\left({\cal S}-\frac{1}{2}\right)^2-\frac{1}{4}\right]\right) H^2~e^{2Ht} ~~~~ &
 \mbox{\small {\bf for ~dS}}  \\ \\
	\displaystyle -\frac{1}{2m}\left(\frac{m^2_{\cal S}}{H^2} - \left[\left(\left({\cal S}-\frac{1}{2}\right)+\epsilon+\frac{\eta}{2}+\frac{s}{2}\right)^2-\frac{1}{4}\right]
\right)  H^2~e^{2Ht} ~~~~ & \mbox{\small {\bf for~ qdS}}.
          \end{array}
\right.\\ \nonumber\\
E&=&\frac{1}{2m}c^{2}_{S}k^2.
\eea
Now we have already shown in the earlier section the detailed study of cosmological scalar curvature fluctuations from new massive particles, where we have computed the expression for the one point and two point functions 
in terms of the parameter $\Lambda$. Here in presence of arbitrary spin (mostly for high spin with spin 
${\cal S}>2$) this $\Lambda$ parameter is replaced by new parameter $\Lambda_{\cal S}$, where $\Lambda_{\cal S}$ is defined as:
 \be\begin{array}{lll}
 \displaystyle \Lambda_{\cal S}=\left\{\begin{array}{ll}
                    \displaystyle \sqrt{\left({\cal S}-\frac{1}{2}\right)^2-\frac{m^2_{\cal S}}{H^2}} ~~~~ &
 \mbox{\small {\bf for ~dS}}  \\ \\
	\displaystyle \sqrt{\left(\left({\cal S}-\frac{1}{2}\right)+\epsilon+\frac{\eta}{2}+\frac{s}{2}\right)^2-\frac{m^2_{\cal S}}{H^2}} ~~~~ & \mbox{\small {\bf for~ qdS}}.
          \end{array}
\right.
\end{array}\ee
For ${\cal S}=2$ only $\Lambda_{\cal S}=\Lambda$ both for de Sitter and quasi de Sitter case. Otherwise the rest of the computation of one point and two point correlation function is exactly same. If we replace $\Lambda$ by 
$\Lambda_{\cal S}$ then the spectral tilt for scalar fluctuations with the horizon crossing $|kc_{S}\eta| \approx 1$ can be expressed in presence of arbitrary spin contribution as: 
\be\begin{array}{ll}\label{dfgs} n_{\zeta}-1
\approx 3-2\Lambda_{\cal S}=\left\{\begin{array}{ll}
                    \displaystyle 3-2\sqrt{\left({\cal S}-\frac{1}{2}\right)^2-\frac{m^2_{\cal S}}{H^2}}
 &
 \mbox{\small {\bf for ~dS}}  \\ \\
	\displaystyle  3-2\sqrt{\left(\left({\cal S}-\frac{1}{2}\right)+\epsilon+\frac{\eta}{2}+\frac{s}{2}\right)^2-\frac{m^2_{\cal S}}{H^2}}
                     & \mbox{\small {\bf for~qdS}}.
          \end{array}
\right.
\end{array}\ee 
As the value of scalar spectral tilt $n_{\zeta}$ is known from observation, one can easily give the estimate of the
value of the ratio of the mass parameter $m$ with Hubble scale during inflation $H$ as:
\be\begin{array}{ll}\label{dfgs1} \displaystyle \left|\frac{m_{\cal S}}{H}\right|=\left\{\begin{array}{ll}
                    \displaystyle \left|\sqrt{\left({\cal S}-\frac{1}{2}\right)^2-\frac{(4-n_{\zeta})^2}{4}}\right|
 &
 \mbox{\small {\bf for ~dS}}  \\ \\
	\displaystyle   \left|\sqrt{\left(\left({\cal S}-\frac{1}{2}\right)+\epsilon+\frac{\eta}{2}+\frac{s}{2}\right)^2-\frac{(4-n_{\zeta})^2}{4}}\right|
                     & \mbox{\small {\bf for~qdS}}.
          \end{array}
\right.
\end{array}\ee 
 It is important to mention here that if we use the constraint on scalar spectral tilt as obtained from Planck 2015 data we get the following $2\sigma$ bound on the magnitude of the 
 mass parameter of the new heavy particles:
 \bea
 &&\underline{\bf For ~dS:}\nonumber\\
 && \left|\sqrt{\left({\cal S}-\frac{1}{2}\right)^2-2.33}\right|<\left|\frac{m_{\cal S}}{H}\right|_{|kc_{S}\eta| \approx 1}<\left|\sqrt{\left({\cal S}-\frac{1}{2}\right)^2-2.30}\right|,\\ \nonumber\\
 &&\underline{\bf For~qdS:}\nonumber\\
 && \left|\sqrt{\left(\left({\cal S}-\frac{1}{2}\right)+\epsilon+\frac{\eta}{2}+\frac{s}{2}\right)^2-2.33}\right|<
 \left|\frac{m_{\cal S}}{H}\right|_{|kc_{S}\eta| \approx 1}<\left|\sqrt{\left(\left({\cal S}-\frac{1}{2}\right)+\epsilon+\frac{\eta}{2}+\frac{s}{2}\right)^2-2.30}\right|,~~~~~~~~~~~\eea 
 and for $|kc_{S}\eta|<<1$ and $|kc_{S}\eta|>>1$ the allowed lower bound on the magnitude of the 
 mass parameter of the new heavy particle is given by:
 \be\begin{array}{ll} \displaystyle \left|\frac{m_{\cal S}}{H}\right|_{|kc_{S}\eta|<<1,|kc_{S}\eta|>>1}\geq  \sqrt{\nu^2_{\cal S}-\frac{1}{4}}=\left\{\begin{array}{ll}
                    \displaystyle  \sqrt{\left({\cal S}-\frac{1}{2}\right)^2-\frac{1}{4}}
 &
 \mbox{\small {\bf for ~dS}}  \\ \\
	\displaystyle   \sqrt{\left(\left({\cal S}-\frac{1}{2}\right)+\epsilon+\frac{\eta}{2}+\frac{s}{2}\right)^2-\frac{1}{4}}
                     & \mbox{\small {\bf for~qdS}}.
          \end{array}
\right.
\end{array}\ee 
 For graviton with spin ${\cal S}=2$ the  bound on the graviton mass parameter for $|kc_{S}\eta|\approx 1$, $|kc_{S}\eta|<<1$ and $|kc_{S}\eta|>>1$ is given by:
 \bea
 &&\underline{\bf For ~dS:}\nonumber\\
 && 0.23<\left|\frac{m_{{\cal S}=2}}{H}\right|_{kc_{S}\eta\approx 1}<0.28,\\ \nonumber\\
 &&\underline{\bf For~qdS:}\nonumber\\
 && 0.23\times\left|\sqrt{1-56.18\left(\epsilon+\frac{\eta}{2}+\frac{s}{2}\right)}\right|<
 \left|\frac{m_{{\cal S}=2}}{H}\right|_{kc_{S}\eta\approx 1}<0.28\times\left|\sqrt{1-39.06\left(\epsilon+\frac{\eta}{2}+\frac{s}{2}\right)}\right|.~~~~\eea 
 \be\begin{array}{ll} \displaystyle \left|\frac{m_{{\cal S}=2}}{H}\right|_{|kc_{S}\eta|<<1,|kc_{S}\eta|>>1}\geq \left\{\begin{array}{ll}
                    \displaystyle  \sqrt{2}
 &
 \mbox{\small {\bf for ~dS}}  \\ \\
	\displaystyle   \sqrt{2}\sqrt{1+\frac{3}{2}\left(\epsilon+\frac{\eta}{2}+\frac{s}{2}\right)+\frac{1}{2}\left(\epsilon+\frac{\eta}{2}+\frac{s}{2}\right)^2}
                     & \mbox{\small {\bf for~qdS}}.
          \end{array}
\right.
\end{array}\ee 
 On the other hand for massive scalar field with ${\cal S}=0$ we get the following lower bound on the mass parameter:
 \be\begin{array}{ll} \displaystyle \left|\frac{m}{H}\right|_{|kc_{S}\eta|<<1,|kc_{S}\eta|>>1}\geq \left\{\begin{array}{ll}
                    \displaystyle  \sqrt{2}
 &
 \mbox{\small {\bf for ~dS}}  \\ \\
	\displaystyle   \sqrt{2}\sqrt{1+\frac{3}{2}\left(\epsilon+\frac{\eta}{2}+\frac{s}{2}\right)+\frac{1}{2}\left(\epsilon+\frac{\eta}{2}+\frac{s}{2}\right)^2}
                     & \mbox{\small {\bf for~qdS}}.
          \end{array}
\right.
\end{array}\ee 
which is exactly similar to graviton. Similarly for axion field with ${\cal S}=0$ we get the following lower bound on the mass parameter:
 \be\begin{array}{ll} \displaystyle \left|\frac{m_{axion}}{f_{a}H}\right|_{|kc_{S}\eta|<<1,|kc_{S}\eta|>>1}\geq \left\{\begin{array}{ll}
                    \displaystyle  \sqrt{6}~~~~~~~~~~~~~
 &
 \mbox{\small {\bf for ~$\eta\sim \eta_{c}$, early~\&~late~$\eta$}}  \\ \\
	\displaystyle   \sqrt{6+\Delta_{C}}~~~~~~~~~~~~~
                     & \mbox{\small {\bf for~$\eta<\eta_{c}$}}.
          \end{array}
\right.
\end{array}\ee 
where $\Delta_{C}$ is a slowly varying quantity as introduced earlier in context of axion fluctuation.

\subsection{More on Bell Inequalities}
\subsubsection{CHSH inequality}

Theory of locality of Einstein are based on another
form of the Bell inequality, which applies to a situation in which Aace can
measure either one of two observables $A_1$ and $A_2$ , while Bace can measure
either $B_1$ or $B_2$.
\\ Now suppose that the observables $A_1$, $A_2$, $B_1$ and $B_2$ take values in $\pm 1$, and are functions of hidden random variables.
If $A_1$, $A_2 = \pm 1$, therefore either $A_1 + A_2 = 0$, in which case $A_1 − A_2 = \pm 2$,
or else $A_1 − A_2 = 0$, in which case $A_1 + A_2 = \pm 2$; therefore
\begin{equation}
C = \left(A_{1} + A_{2}\right)B_{1} + \left(A_{1} - A_{2}\right)B_{2} = \pm 2
\end{equation}
This is where the local hidden-variable assumption comes in. Here it is assumed
that values in $\pm 1$ can be assigned simultaneously to all four
observables, even though it is impossible to measure both of $A_1$ and $A_2$ , or
both of $B_1$ and $b_2$.
Therefore,
\bea
\lvert\langle C \rangle\vert \leq \langle\lvert C \rvert\rangle = 2
\\ \lvert\langle A_1 B_1 \rangle + \langle A_1 B_2 \rangle + \langle A_2 B_1 \rangle - \langle A_2 B_2 \rangle \rvert \leq 2
\eea
This inequality is called the CHSH (Clauser-Horne-Shimony-
Holt) inequality which is introduced by John Clauser, Michael Horne, Abner Shimony and R. A. Holt.
See ref~ \cite{Ekert:1991zz,Adesso:2016ygq,Beaudry:2015jna,Fujikawa:2013ama,Fujikawa:2012jd,Horodecki:2009zz,
Toner:2007kea,Gisin:2002zz,Jennewein:2000zz,Hughes:1999pd,Kempe:1999vk,Lomonaco:1998we}
for more details. One can also test the future consequences for CHSH inequality violation in the cosmological setup. After Aspect's second experiment in 1982, many test experiments were done which
used the CHSH inequality.

\subsubsection{Consequences of Bell Inequality Violation}

\begin{itemize}

\item Bell's inequalities violation due to entanglement gave a solid evidence that theory of quantum mechanics cannot be represented by any theory of classical physics.

\item Elements which are compatible with classical theory are complementarity and wave function collapse.
\item The `Einstein, Rosen and Podolsky' paper pointed some properties of entangled states which are unusual which
in turn is the fundamental foundation for the applications of quantum physics we use today in daily lives
such as \textcolor{blue}{{\it quantum cryptography}} and \textcolor{blue}{{\it quantum non-locality}}.

\item Theorem of John Bell proved that the quantum mechanical property \textcolor{blue}{{\it entanglement}}
has a degree of non­locality which can't be explained by any local theory.
\item No phenomena which is predicted by the theory of quantum mechanics can be reproduce by any
combination of
local deterministic and local random variables and this is also observed experimentally.
\item There are various applications of entanglement in quantum information theory.
Many impossible work can be done using the concept of entanglement theory.
\item The most well known applications are \textcolor{blue}{{\it superdense coding}} and \textcolor{blue}{{\it
quantum teleportation}}.
\item \textcolor{blue}{{\it Quantum Computing}} and \textcolor{blue}{{\it Quantum Cryptography}} \cite{Ekert:1991zz,Adesso:2016ygq,Beaudry:2015jna,Fujikawa:2013ama,Fujikawa:2012jd,Horodecki:2009zz,
Toner:2007kea,Gisin:2002zz,Jennewein:2000zz,Hughes:1999pd,Kempe:1999vk,Lomonaco:1998we}
are very well known disciplines of physics where
entanglement theory is used. Entanglement ­based quantum cryptography is very useful to detect the presence of any third party
between two communication parties.
\end{itemize}

	\section*{Acknowledgments}
	SC would like to thank Department of Theoretical Physics, Tata Institute of Fundamental
Research, Mumbai for providing me Visiting (Post-Doctoral) Research Fellowship. SC would like to thank Gautam Mandal and Guruprasad Kar for useful discussions and suggestions. SC take this opportunity to thank
sincerely to Sandip P. Trivedi and Shiraz Minwalla for their constant support
and inspiration. SC additionally take this
opportunity to thank the organizers of STRINGS, 2015,
International Centre for Theoretical Science, Tata Institute of Fundamental Research (ICTS,TIFR),
Indian Institute of Science (IISC) and specially Shiraz Minwalla for giving the opportunity
to participate in STRINGS, 2015 and also providing the local hospitality during the work. SC also thank 
the organizers of National String Meet 2015 and International Conference on Gravitation and Cosmology, IISER, Mohali 
and COSMOASTRO 2015, Institute of Physics, Bhubaneswar for providing the local hospitality during the work. 
SC also thanks the organizers of School and Workshop on Large Scale Structure: From Galaxies to Cosmic Web, 
The Inter-University Centre for Astronomy and Astrophysics (IUCAA), Pune, India
and specially Aseem Paranjape and Varun Sahni for providing the academic visit during the work. A very special thanks to The Inter-University Centre for Astronomy and Astrophysics (IUCAA), Pune, India, where the problem was formulated 
and Institute of Physics, Bhubaneswar where the part of the work was done. SC also would like to thank Department of Theoretical Physics, Indian Association for the Cultivation of Science, Kolkata and specially Somitra SenGupta for 
providing the academic visit during the work. RS would like to thank Department of Theoretical Physics, Tata Institute of Fundamental Research. 
and Department of Physics, Savitribai Phule Pune University for providing a platform to work on this topic and for their constant support and resources.
Last but not the least, we would all like to acknowledge our debt to the people of India for their generous and steady support for research in natural
sciences, especially for theoretical high energy physics, string theory and cosmology.

\end{document}